\documentclass[fleqn, twoside, A4]{scrreprt}

\usepackage{nomencl}

\makenomenclature

\usepackage{setspace}
\usepackage{amsmath}
\usepackage{fancyhdr}
\usepackage{enumerate}
\usepackage[T1]{fontenc}
\usepackage[latin1]{inputenc}
\usepackage{graphicx}
\usepackage{graphics}
\usepackage{tabularx}
\usepackage{array}
\usepackage{here}
\usepackage{youngtab}
\usepackage[normal,font=small,labelfont=bf]{caption}
\usepackage{wasysym}
\usepackage{amsfonts}
\usepackage{amssymb}
\usepackage{slashed}
\usepackage[mathscr]{eucal}
\usepackage{feynmp}
\usepackage{xspace}
\usepackage{rotating}
\usepackage{sidecap}
\usepackage[Bjarne]{fncychap}
\usepackage[colorlinks=false,hyperindex=false,linkcolor=blue]{hyperref}
\usepackage[ngerman,english]{babel}
\usepackage{subfigure}
\usepackage{wallpaper}

\usepackage{rotating}
\usepackage[numbers, square, comma, sort&compress]{natbib}

\makeatletter
\ChNameUpperCase
\ChTitleUpperCase
\ChNameVar{\raggedright \bfseries \normalsize}
\ChNumVar{\raggedright \bfseries \normalsize }
\ChRuleWidth{1pt}
\ChTitleVar{\raggedright \Large\rm}

\newcommand{\TheAlphaChapterTwo}{%
      \setcounter{AlphaCnt}{\c@chapter}
        \AlphaNo
    }  

\renewcommand{\DOCH}{%
\vskip 80\p@
\setlength{\fboxrule}{\RW}
 \CNV\FmN{\@chapapp}\space \CNoV\TheAlphaChapterTwo\par\nobreak
\vskip 4\p@ \hrule}

\renewcommand{\DOTI}[1]{%
\vskip 5 \p@
\CTV\FmTi{#1}\par\nobreak
\vskip 5 \p@
\vskip 40\p@}
\renewcommand{\DOTIS}[1]{%
\CTV\FmTi{#1}\par\nobreak
\vskip 5 \p@ 
\vskip 40\p@}
\makeatother

\graphicspath{{./Abbildungen/}}
\newcommand{\minusOne}{$-1$}

\fancypagestyle{plain} 
{
    \fancyhf{}

 \fancyfoot[RO,LE]{\thepage} 
}

\renewcommand{\footnoterule}{\rule{0pt}{0pt}\vspace{0pt}} 

\newenvironment{longtable*}{%
  \addtocounter{table}{-1}%
  \longtable
}{%
  \endlongtable
}

\newcommand\SQ{{\tilde{q}}}
\newcommand\SU{{\tilde{u}}}
\newcommand\SL{{\tilde{l}}}
\newcommand\SD{{\tilde{d}}}
\newcommand\SE{{\tilde{e}}}
\newcommand\SG{{\tilde{g}}}

\newcommand\FG{{\tilde{G}}}
\newcommand\SHE{{H_d}}
\newcommand\SHZ{{H_u}}

\newcommand\Neu{\tilde{\chi}^0}
\newcommand\Cha{\tilde{\chi}^-}
\newcommand\bCha{\tilde{\chi}^+}

\newcommand\GeV{{\text{GeV}}}
\newcommand\eV{{\text{eV}}}
\newcommand\TeV{{\text{TeV}}}
\newcommand\MeV{{\text{MeV}}}
\newcommand\keV{{\text{keV}}}

\newcommand\sign{{\text{sign}}}

\newcommand\DR{{\overline{\text{DR}}}}

\newcommand\FeynArts{\texttt{FeynArts}\xspace}
\newcommand\FormCalc{\texttt{FormCalc}\xspace}
\newcommand\CalcHep{\texttt{CalcHep}\xspace}
\newcommand\CompHep{\texttt{CompHep}\xspace}
\newcommand\SARAH{\texttt{SARAH}\xspace}
\newcommand\SPheno{\texttt{SPheno}\xspace}
\newcommand\micrOmegas{\texttt{micrOMEGAs}\xspace}

\newcommand\bra{\langle}
\newcommand\ket{\rangle}
\newcommand\halb{\frac{1}{2}}
\newcommand\La{\mathscr{L}}
\newcommand\Ord{\mathscr{O}}

\newcommand{\eq}[1]{eq.~(\ref{eq:#1})}

\begin{document}
\ThisTileWallPaper{\paperwidth}{\paperheight}{wallpaper}

 \pagenumbering{roman}
 \makeatletter
\def\thickhrulefill{\leavevmode \leaders \hrule height 1pt\hfill \kern \z@}
\renewcommand{\maketitle}{\begin{titlepage}%
    \let\footnotesize\small
    \let\footnoterule\relax
    \parindent \z@
    \reset@font
    \null\vfil
    \vspace{-3cm}
    \begin{flushleft}
      \includegraphics[width=4cm]{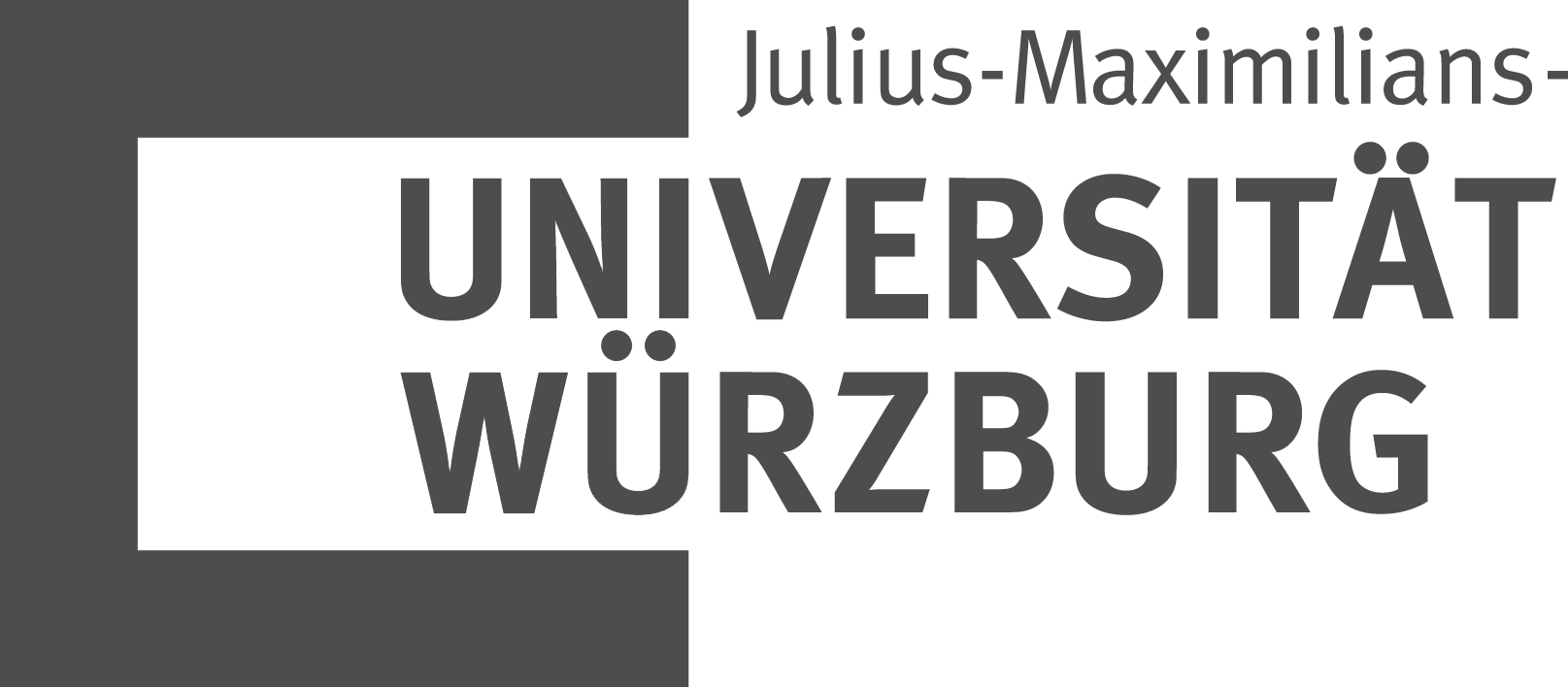} 
      \hspace{11.2cm}
      \includegraphics[width=2cm]{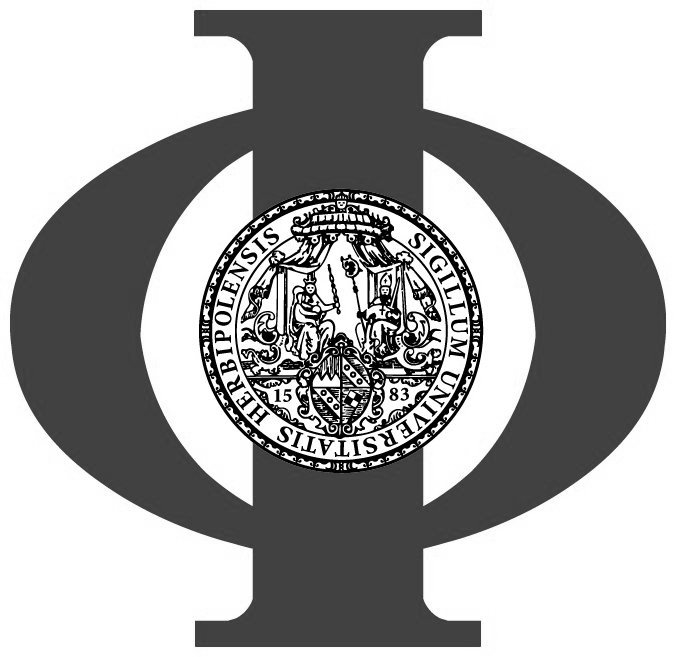} \\
\vspace{0.2cm}
 \hrule height 1pt
      \vspace{2.5cm}
 \end{flushleft}
\begin{center}
      \huge \@title
\end{center}
    \par
    \par
      \large \@author \par
    \begin{center}
\vskip 2,cm
\Large \ Dissertation zur Erlangung des \\
naturwissenschaftlichen Doktorgrades \\
der Bayerischen Julius-Maximilians-Universit\"at W\"urzburg \\
    \vskip 9.5cm
    \normalsize \bf vorgelegt von \\
    \vskip 0.2cm
    \Large \Large Florian Roland Albert Staub\\
    \vskip 0.2cm
\normalsize aus Aschaffenburg a. Main
\vskip 1.cm
W\"urzburg 2010
    \end{center}
  \end{titlepage}%
   \setcounter{footnote}{0}%
}

\title{Considerations on supersymmetric Dark Matter beyond the MSSM}
 \date{}
 
\vspace{-9cm}
\hspace{-0.5cm} 
\parbox{17.5cm}{ \maketitle}

\newpage
\thispagestyle{empty}
\section*{}

\newpage

\hspace{1cm}\\
 \thispagestyle{empty}
{\onehalfspacing 

\vspace{5cm}
Eingereicht am:  \begin{picture}(100,5)  \line(1,0){170}  \end{picture} \\
bei der Fakult\"at f\"ur Physik und Astronomie \\

\vspace{1cm}

1. Gutachter:  \begin{picture}(100,5)  \line(1,0){170}  \end{picture} \\
2. Gutachter:  \begin{picture}(100,5)  \line(1,0){170}  \end{picture} \\
3. Gutachter:  \begin{picture}(100,5)  \line(1,0){170}  \end{picture} \\
der Dissertation \\

\vspace{1cm}

1. Pr\"ufer:  \begin{picture}(100,5)  \line(1,0){170}  \end{picture} \\
2. Pr\"ufer:  \begin{picture}(100,5)  \line(1,0){170}  \end{picture} \\
3. Pr\"ufer:  \begin{picture}(100,5)  \line(1,0){170}  \end{picture} \\
im Promotionskolloquium

\vspace{1cm}

Tag des Promotionskolloquiums:  \begin{picture}(100,5)  \line(1,0){170}  \end{picture} 
\vspace{0.5cm} \\
Doktorurkunde ausgeh\"andigt am:  \begin{picture}(100,5)  \line(1,0){170}  \end{picture} \\
}

\newpage
\thispagestyle{empty}
\section*{}

\newpage
\thispagestyle{empty}
\vspace*{\fill}
\hspace*{\fill} \parbox{7cm}{
Der Spa\ss{} f\"angt erst dann an, wenn man die Regeln kennt. Im Universum aber sind wir momentan noch dabei, die Spielanleitung zu lesen. \\
\begin{flushright}
{\em Richard P. Feynman}
\end{flushright}
}
\newpage

\newpage
\thispagestyle{empty}
\section*{}

\newpage

\chapter*{Zusammenfassung}
Das Standardmodell der Teilchenphysik ist seit drei Jahrzehnten eine \"uberaus erfolgreiche \linebreak Beschreibung der Eigenschaften und Wechselwirkungen der bekannten Elementarteilchen. Derzeit wird es durch die ersten Kollisionen des Large Hadron Colliders (LHC) erneut auf die Probe gestellt. Es wird weitl\"aufig erwartet, dass am LHC neue Physik entdeckt wird und somit das \linebreak Standardmodell erweitert werden muss. Die am meisten untersuchte Erweiterung des\linebreak  Standardmodells ist Supersymmetrie (SUSY). In SUSY k\"onnen nicht nur intrinsische Probleme des Standardmodells wie das Hierarchieproblem gel\"ost werden, sondern es werden auch Teilchen postuliert, welche die gemessene Dunkle Materie im Universum erkl\"aren k\"onnen. Der Gro{\ss}teil der bis\-herigen Studien \"uber Dunkle Materie in SUSY hat sich hierbei auf die minimale supersymmetrische Erweiterung des Standardmodells, das MSSM, beschr\"ankt. Das Ziel dieser Arbeit ist es, Szenarien zu betrachten, die dar\"uber hinaus gehen. Hierbei handelt es sich um zwei Modelle, mit denen auch Neutrinomassen erkl\"art werden k\"onnen: Das Gravitino als Dunkle \linebreak  Materie im Rahmen von Gauge Mediated SUSY Breaking (GMSB) mit $R$-Parit\"atsverletzung sowie Seesaw-Modelle mit einem Neutralino als leichtestem SUSY Teilchen. Weiterhin betrachten wir das \glqq Next-to-Minimal Supersymmetric Standard Model\grqq{} (NMSSM), welches das \(\mu\)-Problem des MSSM l\"ost, und diskutieren dort das leichteste Neutralino als Dunkle Materie Kandidaten. \\
Im Rahmen von leichten Gravitinos als Dunkle Materie wird das kosmologische Gravitino \linebreak Problem betrachtet. Es wird gezeigt, dass die in der Literatur vorgeschlagene L\"osung gegen die \"Uberbev\"olkerung des Universums durch solche  Gravitinos, n\"amlich die Entropieproduktion durch Zerf\"alle der GMSB-Messenger, nur in ausgew\"ahlten Modellen und kleinen Regionen des Para\-meterraums funktioniert. Die Ursache hierf\"ur sind zwei Faktoren, die bislang au{\ss}er Acht gelassen wurden: M\"ogliche Zerf\"alle der neutralen Messenger in massive Vektorbosonen sowie der Einfluss geladener Messenger. Beide Aspekte bewirken zusammen ein Wechselspiel von verschiedenen, kosmologischen Randbedingungen, welches zu starken Bedingungen an die zu Grunde liegenden Parameter f\"uhrt.\\
Als n\"achstes werden Modelle im Rahmen minimaler Supergravitation (mSugra) untersucht, welche bei sehr hohen Energien \"uber zus\"atzliche chirale Superfelder verf\"ugen. Diese zus\"atzlichen Teilchen sind in kompletten $SU(5)$ Multiplets angeordnet, um  Eichvereinheitlichung nicht zu gef\"ahrden. Die neuen Teilchen erzeugen durch den so genannten Seesaw-Mechanismus einen \linebreak Dimension~5 Operator, welcher Neutrinodaten erkl\"aren kann. Dar\"uber hinaus erzeugen sie aber durch das ge\"anderte Laufen der Renormierungsgruppengleichungen Unterschiede im Massenspektrum der SUSY Teilchen, was nat\"urlich auch die Eigenschaften des Neutralinos als Dunkle Materie Kandidaten ver\"andert. Wir diskutieren den Parameterraum aller drei m\"oglichen Seesaw-Szenarien im Hinblick auf Dunkle Materie sowie die Auswirkungen auf Leptonflavor verletzende Prozesse.  Wir werden sehen, dass insbesondere in Typ~III aber auch in Typ~II sowohl gro{\ss}e Unterschiede im Massenspektrum als auch in den Parameter\-be\-reichen, welche konsistent mit Dunkler Materie sind, im Vergleich zu einem gew\"ohnlichen mSugra-Szenario bestehen. Dar\"uber hinaus f\"uhren vor allem die oberen, experimentellen Schranken der  Verzweigungsverh\"altnisse von \(l_i \rightarrow l_j \gamma\) zu starken Bedingungen an die zu Grunde liegenden Seesaw-Parameter. \\
Abschlie{\ss}end wird das Neutralino im Rahmen des NMSSM untersucht. In dieser Erweiterung des MSSM ist zwar das Neutralino immer noch der beste Kandidat f\"ur Dunkle Materie, kann sich jedoch auf Grund der Anteile eines Eichsinglets sehr unterschiedlich im Vergleich zum MSSM verhalten. Wir zeigen nicht nur die Unterschiede zum MSSM auf, sondern berechnen auch die Dichte Dunkler Materie im NMSSM mit der gleichen Pr\"azision wie im MSSM. F\"ur diesen Zweck ist es notwendig, eine komplette Einschleifenrenormierung des elektroschwachen Sektors des NMSSM durchzuf\"uhren. Es wird sich zeigen, dass insbesondere die Strahlungskorrekturen zu den Massen der Staus gro{\ss}e Auswirkung auf die Neutralinodichte in der Koannihilationsregion haben. Weiterhin ist der so genannte Higgs-Funnel, also Bereiche im Parameterraum, in denen die Masse eines Higgs Bosons in etwa der zweifachen Masse des leichtesten Neutralinos entspricht, sehr sensitiv auf die Ein- und Zweischleifenkorrekturen im pseudoskalaren Sektor. \\
Im Rahmen dieser Projekte wurde ein Mathematica Package namens \SARAH entwickelt, um super\-symmetrische Modelle schnell, effektiv und mit sehr hoher Pr\"azision untersuchen zu k\"onnen. \SARAH berechnet f\"ur ein gegebenes Modell alle analytischen Ausdr\"ucke f\"ur die Massen, Wechselwirkungen, Selbstenergien auf Einschleifenniveau sowie Renormierungsgruppengleichungen auf Ein- und Zweischleifenniveau. Eine gro{\ss}e Bandbreite von SUSY Modellen kann analysiert und auch von dem Benutzer intuitiv ver\"andert werden. Die berechneten Ausdr\"ucke k\"onnen dazu benutzt werden, um neue Modelle in Programme zum diagrammatischen Berechnen von Prozessen (\FeynArts/\FormCalc bzw. \CalcHep/\CompHep) zu implementieren oder das gesamte Spektrum und alle Parameter des neuen SUSY Modells mit Hilfe von \SPheno berechnen zu lassen. Die sich durch \SARAH bietenden M\"oglichkeiten gehen hierbei \"uber reine Studien zur Dunkle Materie weit hinaus. 
\chapter*{Abstract}
The standard model (SM) of particle physics is for the last three decades a very successful description of the properties and interactions of all known elementary particles. Currently, it is again probed with the first collisions at the Large Hadron Collider (LHC). It is widely expected that new physics will be detected at the LHC and the SM has to be extended. The most exhaustive analyzed extension of the SM is supersymmetry (SUSY). SUSY can not only solve intrinsic problems of the SM like the hierarchy problem, but it also postulates new particles which might explain the nature of dark matter in the universe. The majority of all studies about dark matter in the framework of SUSY has focused on the minimal supersymmetric standard model (MSSM). The aim of this work is to consider scenarios beyond that scope. We consider two models which explain not only dark matter but also neutrino masses: the gravitino as dark matter in gauge mediated SUSY breaking (GMSB) with bilinear broken $R$-parity as well as different seesaw scenarios with the neutralino as dark matter candidate. Furthermore, we also study  the next-to-minimal supersymmetric standard model (NMSSM) which solves the \(\mu\)-problem of the MSSM and discuss the properties of the neutralino as dark matter candidate. \\
In case of $R$-parity violation, light gravitinos are often the only remaining candidate for dark matter in SUSY because of their very long life time. We reconsider the cosmological gravitino problem arising for this kind of models. It will be shown that the proposed solution for the overclosure of the universe by light gravitinos, namely the entropy production by decays of GMSB messenger, just works in a small subset of models and in fine-tuned regions of the parameter space. This is a consequence of two effects so far overlooked: the enhanced decay channels in massive vector bosons and the impact of charged messenger particles. Both aspects cause an interplay between different cosmological restrictions which lead to strong constraints on the parameters of GMSB models. \\
Afterwards, a minimal supergravity (mSugra) scenario with additional chiral superfields at high energy scales is considered. These fields are arranged in complete $SU(5)$ multiplets in order to maintain gauge unification. The new fields generate a dimension 5 operator to explain neutrino data. Furthermore, they cause large differences in mass spectrum of MSSM fields because of the different evaluation of the renormalization group equations what changes also the properties of the lightest neutralino as dark matter candidate. We discuss the parameter space of all three possible seesaw scenarios with respect to dark matter and the impact on rare lepton flavor violating processes. As we will see, especially in seesaw type~III but also in type~II the mass spectrum and regions of parameter space consistent with dark matter differ significantly in comparison to a common mSugra scenario. Moreover, the experimental bounds, in particular of branching ratios like \(l_i \rightarrow l_j \gamma\), cause large constraints on the seesaw parameters.   \\
Finally, we study dark matter in the NMSSM. In this extension of the MSSM the neutralino is still a valid dark matter candidate. However, the properties of the lightest neutralino can be quite different to the MSSM due to the contributions of a gauge singlet. We point not only  out these differences to the MSSM, but also calculate the relic density in the NMSSM with the same precision known for the MSSM. For this purpose, it is necessary to perform a complete one-loop calculation of the electroweak sector of the NMSSM. As we will see in particular, the one-loop corrections to staus have significant influence on the correct amount of dark matter in the coannihilation region. Furthermore, the so called Higgs funnel, i.e. the region in parameter space with a Higgs mass close to twice the mass of the lightest neutralino, is very sensitive to the one- and two-loop corrections in the Higgs sector.   \\
During this work, a Mathematica package called \SARAH was developed for the fast, effective and precise  analysis of SUSY models. \SARAH calculates for a given model all analytical expressions for the masses, vertices, one-loop self-energies and one- and two-loop RGEs. A large variety of models can be handled and changed by the user in an intuitive way. The expressions can be used to produce model files for diagram calculators (\FeynArts/\FormCalc or \CalcHep/\CompHep) or to calculate the spectrum and parameters of a new SUSY model with \SPheno. The possibilities in this context offered by \SARAH go far beyond the scope of pure dark matter studies. 

\newpage

\tableofcontents 

\newpage
 \pagenumbering{arabic}

\chapter{Introduction}
With the first collisions at the  Large Hadron Collider (LHC), a new era of high energy physics has started. The LHC is designed to get new and deeper insights into the fundamental principles of our world. It is not only supposed  to find the last missing particle of the Standard Model (SM) of particle physics, the Higgs boson but also discover physics beyond the SM. Although the SM is  a very successful and precise description of all experiments in particle physics for last 30 years, it lacks on some theoretical shortcomings.  The SM suffers from the hierarchy problem and it can't provide an explanation for some observations. One open question is: what is the invisible, 'dark' matter in the universe whose amount is five times the amount of all visible stars, clouds and planets? We consider in this work the most studied extension of the SM: Supersymmetry (SUSY). Supersymmetric models provide not only candidates for dark matter, but also solve many intrinsic problems of the SM. We will discuss these appealing properties of SUSY in the first section of this introduction. Afterwards, we concentrate on the astrophysical properties, evidences and constraints on dark matter, before we briefly present the two main candidates for supersymmetric dark matter. Finally, we discuss the necessity to extend the studies of SUSY dark matter to scenarios beyond the MSSM.
\section{Supersymmetry}
\subsection{Motivation}
\label{introduction:motivation}
\begin{figure}[t]
\centering
\includegraphics[scale=1.]{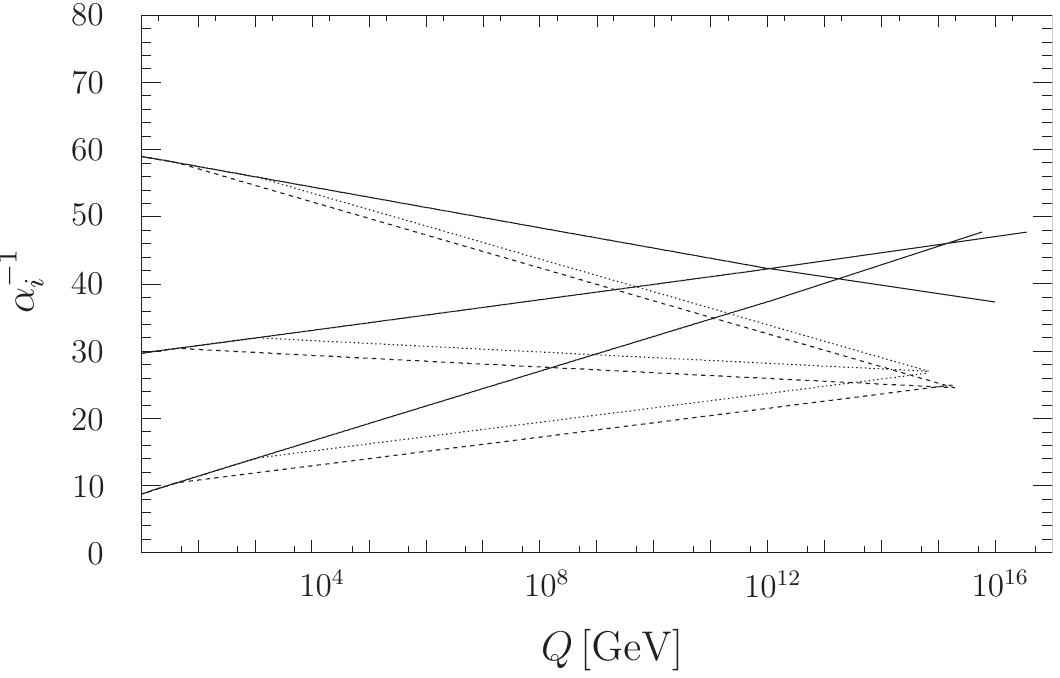}
\caption[Gauge unification in SUSY]{Unification of the gauge couplings in the MSSM. The dotted lines are for a SUSY scale of 1~TeV and the dashed lines belong to a SUSY scale of  10~TeV. The straight lines are for the SM.}
\label{fig:unficiation}
\end{figure}

As already mentioned, the SM must be extended in order to explain some experimental observations and to get rid of theoretically drawbacks. These extensions take most likely place at the TeV scale and can hopefully be probed at the LHC in the near future. SUSY is still the most prominent and promising extension and was studied very well during the last decades.  It overcomes not only many shortcomings of the SM but has also additional and very attracting features. \\
\paragraph*{SUSY algebra} The name 'Supersymmetry' is based on its unique role as the largest possible symmetry group. SUSY describes the unification of internal and space-time symmetries. For a long time, it was assumed that such an unification isn't possible according to the no-go-theorem of Coleman and Mandula \cite{Coleman:1967ad}. However, it was found that the generators of both types of symmetry fulfill the commutator and anti-commutator relations \cite{Ramond:1971gb,Wess:1974tw,Volkov:1973ix}
\begin{equation}
\label{susygen}
\{Q_\alpha, Q_{\dot{\alpha}}^\dagger\} = -2 \sigma_{\alpha \dot{\alpha}}^\mu P^\mu\thickspace, \hspace{1cm} \{Q_\alpha,Q_\beta\} = \{Q_{\dot{\alpha}}^\dagger, Q_{\dot{\beta}}^\dagger\} = 0\thickspace, \hspace{1cm} [Q_\alpha,P^\mu] = [Q_{\dot{\alpha}}^\dagger,P^\mu] = 0\thickspace ,
\end{equation}
with \(\alpha, \dot{\alpha}, \beta, \dot{\beta} = 1,2\). Here, \(P^\mu\) is the four dimensional operator of space-time translation and the Weyl spinor \(Q_\alpha\) is the generator of a supersymmetric transformation. Such a transformation describes the conversion of bosons in fermions and vice versa:
\begin{equation}
\label{susygen2}
Q_\alpha |\text{Boson} \ket = |\text{Fermion} \ket\thickspace, \hspace{1cm} Q_\alpha |\text{Fermion} \ket = |\text{Boson} \ket \thickspace.
\end{equation}
There are models with several generators \(Q^i_\alpha\) but we focus in this work on so called \(\mathscr{N}=1\) supersymmetric models with only one generator \(Q_\alpha\). Using eqs.~(\ref{susygen}) and (\ref{susygen2}), it can be shown that every bosonic state has a fermionic (super-)partner and the other way round. Both are connected by the operator \(Q\). The consequence is that for every SM particle a supersymmetric partner must exist. Both have the same properties apart from their spin which differs by \(\frac{1}{2}\). However, this is not sufficient and such a model would suffer from chiral anomalies: the hypercharge of the fermionic partner of the Higgs would not be canceled by another contribution. This is the reason, why the Higgs sector of the smallest possible supersymmetric extension of the SM, the Minimal Supersymmetric Standard Model (MSSM), has to be extended. It consists of two scalar Higgs doublets with opposite hypercharge and of their fermionic superpartners, the Higgsinos. We will discuss the MSSM and its particle content in more detail in sec.~\ref{MSSM_teilchen}. Since no fundamental scalar field was detected so far, SUSY has to be broken. SUSY breaking is discussed in sec.~\ref{sec:SUSY_Breaking}. 
\begin{figure}[t]
\begin{minipage}{16cm}
\begin{center}
\begin{tabular}{ c }
\parbox{15cm}{
\begin{center}
\begin{fmffile}{Feynmangraphen/cancel7x}
\fmfframe(5,5)(5,5){
\begin{fmfgraph*}(100,100)
\fmfkeep{fermion}
\fmfleft{i}
\fmfright{o}
\fmflabel{$h$}{i}
\fmflabel{$h$}{o}
\fmf{dashes}{i,v1}
\fmf{dashes}{v2,o}
\fmf{fermion,left,tension=.3, label=$\bar{t}$}{v1,v2}
\fmf{fermion,left,tension=.3, label=$t$}{v2,v1}
\end{fmfgraph*}
}
\end{fmffile}
\quad
\begin{picture}(10,10)(0,0)
\put(0,50){+}
\end{picture}
\quad
\begin{fmffile}{Feynmangraphen/cancel7y}
\fmfframe(5,5)(5,5){
\begin{fmfgraph*}(100,100)
\fmfleft{i} 
\fmfright{o} 
\fmftop{k} 
\fmf{dashes}{i,v,o}
\fmffreeze
\fmf{dashes,left,tension=0.3}{k,v}
\fmf{dashes,right,tension=-0.3}{k,v}
\fmflabel{$h$}{i}
\fmflabel{$h$}{o}
\fmflabel{$\tilde{t}_i$}{k}
\end{fmfgraph*}
}
\end{fmffile}
\end{center}}
\\
\(\Delta m_h^2 \sim \ln\Lambda_{UV}^2 \)\\
\end{tabular}
\end{center}
\caption[Solution of the hierarchy problem in SUSY]{Solution of the hierarchy problem in SUSY: the quadratic divergence due to contribution of fermions is canceled by bosonic contributions. Only a logarithmic divergence remains. If SUSY is unbroken also this divergence is canceled by another scalar loop.}
\label{hierarchie}
\end{minipage}
\end{figure} 
\paragraph*{Hierarchy problem} One of the main advantages of SUSY and its extended particle content is the elegant solution of the hierarchy problem. The hierarchy problem results from the fact that the mass of the Higgs boson is not protected by any symmetry in the SM.  Therefore, it receives large loop corrections proportional to some cut-off scale \(\Lambda^2\). \(\Lambda\) is normally assumed to be a very high scale of \(\Ord(10^{16})\)~GeV, so the Higgs mass will be pushed up to this scale. However, the Higgs mass is demanded to be of the order of the electroweak scale. The only loophole in the SM is the cancellation of different loop contributions, but this demands an extreme fine-tuning of parameters \cite{Weinberg:1975gm,Weinberg:1979bn}. This problem is absent in supersymmetric theories because every loop correction caused by SM particles is in a natural way completely canceled by supersymmetric contributions if SUSY is unbroken. This exact cancellation is based on the observation that fermions obey the Fermi statistics while bosons belong to the Bose statistics. Hence, the contributions connected by SUSY differ exactly by one sign. Such a cancellation is depicted in Fig.~\ref{hierarchie}. SUSY can be broken in a way that there are no quadratical divergences introduced again. This case is referred to as 'softly broken' SUSY and leads only to logarithmic divergences depending on the cut-off scale.  
\paragraph*{Gauge unification} The extension of the particle content by supersymmetric partners influences the scale dependence of all gauge couplings. This dependence is described by the renormalization group equations (RGE) \cite{Wilson:1973jj}. We will discuss the behavior of the RGEs depending on the particle content of a model in more detail in chapter~\ref{chapter:SU5}. Here, we just state: while the RGE running of the gauge couplings in the standard model doesn't lead to an unification at any scale, the couplings seem to meet in SUSY a few orders below the Planck scale (see Fig.~\ref{fig:unficiation}). This is a further motivation for the assumption that SUSY is the next step towards a grand unified description of particle physics \cite{Langacker:1991an,Ellis:1990wk,Giunti:1991ta,Hall:1980kf}.
\paragraph*{Electroweak symmetry breaking} Another unanswered question in the SM is, how the electroweak symmetry gets broken in a natural way. In SUSY, the negative mass squared of a Higgs doublet and consequently electroweak symmetry breaking (EWSB) are related to the large top mass: the evaluation of the scalar masses from the GUT scale down to the electroweak scale is strongly affected by the Yukawa couplings. While the gauge contributions increase the mass while running down, the Yukawa couplings decrease it. The top Yukawa coupling is big enough to cause a negative mass squared for the up-type Higgs  \cite{Martin:2001vx,Ibanez:1982fr}. 
\paragraph*{Dark matter} For completeness, we mention briefly that many supersymmetric models provide a particle which might explain the measurement of 23~\% non-baryonic matter in the universe \cite{Goldberg:1983nd, Ellis:1983ew}. This aspect is discussed in great detail in sec.~\ref{section:SUSY_DM}.  \\

We continue this introductory section about SUSY with a short summary of the superspace formalism and the motivation of the Wess-Zumino gauge. We will go on with a description of the MSSM, its particle content and its Lagrangian, before we come to possible origins of SUSY breaking. Afterwards, there will be a short section on supersymmetric grand unified theories. At the end of this section, we discuss different phenomenological aspects and constraints on SUSY models. 
\subsection{Superspace formalism}
\label{sec:superspace}
The superspace formalism is a very compact and elegant language to describe supersymmetric theories and their properties \cite{Ferrara:1974ac,Friedman:1977pf,Dine:2007zp, Gates:1983nr}. The basic idea is to extended the four space-time coordinates \(x_\mu\) by four Grassmann coordinates  \(\Theta_\alpha\) and \(\bar{\Theta}_{\dot{\alpha}}\) which anti-commute. That's why, these coordinates are called the 'fermionic coordinates' while the coordinates of Minkowski space are the 'bosonic coordinates'. Useful features of Grassmann numbers are: the squared of a number is always zero and integration and differentiation with respect to Grassmann numbers are identical operations. Using this formalism, the description of SUSY operators in superspace is similar to Poincare operators in Minkowski space. Keeping this in mind, many results can be motivated in analogy to the known results of gauge theories. We can write the SUSY generator as
\begin{equation}
Q_\alpha = \frac{\partial}{\partial \Theta_\alpha} - i \sigma^\mu_{\alpha \dot{\alpha}} \bar{\Theta}^{\dot{\alpha}} \partial_\mu \thickspace, \hspace{1cm} Q^\dagger_{\dot{\alpha}} = - \frac{\partial}{\partial \bar{\Theta}_{\dot{\alpha}}} + i \Theta^{\alpha} \sigma^\mu_{\alpha \dot{\alpha}}  \partial_\mu \thickspace .
\end{equation}
When we consider \(Q\) as generator of an infinitesimal translation in superspace and call the parameter of this translation \(\epsilon\), a supersymmetric transformation acting on a superfield \(\Phi\) reads
\begin{equation} 
\exp \left(\epsilon Q + \epsilon^* Q^\dagger \right)
\Phi(x^\mu, \Theta, \bar{\Theta}) = \Phi(x^\mu - i \epsilon
\sigma^\mu \bar{\Theta} + i \Theta \sigma^\mu \epsilon^*, \Theta +
\epsilon,
 \bar{\Theta}+\epsilon^*) \thickspace.
\end{equation}
Because of the property of Grassmann numbers, each Taylor series truncates after a finite number of terms of fermionic coordinates. We can decompose these terms in two irreducible representations: one describes a chiral superfield, the other a vector superfield. To discuss the properties of these representations,  we define the covariant derivative \(D_\alpha\) as
\begin{equation} 
D_\alpha = \partial_\alpha + i \sigma^\mu_{\alpha \dot{\alpha}} \bar{\Theta}^{\dot{\alpha}} \partial_\mu \thickspace .
\end{equation}
The covariant derivative and the SUSY generator commute, thus 
\begin{equation}
\bar{D}_{\dot{\alpha}} \Phi = 0
\end{equation}
is invariant under a supersymmetric transformation. The fields \(\Phi\) fulfilling this relation are called left-chiral superfields and we can express them using of  \(y^\mu = x^\mu + i \Theta \sigma^\mu \bar{\Theta}\) by
\begin{equation}
\label{taylor_chiral}
\Phi = \phi(y) + \sqrt{2} \Theta \Psi(y) + \Theta^2 F(y) \thickspace.
\end{equation}
The component \(F\) is an auxiliary field and doesn't propagate, \(\phi\) is a scalar and \(\Psi\) a fermion. \\
Vector superfields, which form the other irreducible representation of the algebra, satisfy
\begin{equation}
V = V^\dagger \thickspace .
\end{equation}
This relation is also invariant under a supersymmetric translation. If we consider a left-chiral superfield \(\xi (\Theta)\) with mass dimensions 0, we can expand \(V\) as
\begin{equation}
\label{taylor_vektor}
V = i \xi(\Theta) - i \xi^\dagger(\Theta) - \Theta \sigma^\mu \bar{\Theta} A_\mu + i \Theta^2 \bar{\Theta} \bar{\lambda} - i \bar{\Theta}^2 \Theta \lambda + \halb \Theta^2 \bar{\Theta}^2 D \thickspace .
\end{equation}
\(D\) is again an auxiliary field, \(\lambda\) is the fermionic component of the vector superfield and \(A_\mu\) is a gauge boson.  Since \(V\) is massless, it can be shifted by a gauge transformation. The gauge in which the fields \(\xi\) disappear is called Wess-Zumino  gauge. \\
To sum up, we have found two different kind of superfields which are invariant under a SUSY transformation in superspace: the complex chiral superfield \(\Phi\) and the real vector superfield \(V\). Both can in the Wess-Zumino gauge be expressed in component fields as
\begin{align}
\Phi &= \phi(y) + \sqrt{2} \Theta \Psi(y) + \Theta^2 F(y) \thickspace , \\
V &= \Theta \sigma^\mu \bar{\Theta} A_\mu + i \Theta^2 \bar{\Theta} \bar{\lambda} - i \bar{\Theta}^2 \Theta \lambda + \halb \Theta^2 \bar{\Theta}^2 D \thickspace .
\end{align}
The matter interactions of a supersymmetric models can written in a short form using the superfield formalism. The result is the so called superpotential. Generally spoken, the superpotential is a holomorphic function of chiral superfields \(\Phi_i\). The most general form of the superpotential for a renormalizable model is  
\begin{equation}
W = L^i \Phi_i + \halb \mu^{ij} \Phi_i \Phi_j + \frac{1}{3} Y^{ijk} \Phi_i \Phi_j \Phi_k \thickspace .
\end{equation}
\subsection{The Minimal Supersymmetric Standard Model}
\subsubsection{Particle content of the MSSM}
\label{MSSM_teilchen}
As already stated, the MSSM is the smallest possible supersymmetric extensions of the SM without gauge anomalies \cite{Inoue:1982pi,Haber:1984rc,Djouadi:1998di}. The superpartners of the quarks and leptons are the spin 0 squarks and sleptons, e.g. the stop, the sup or the selectron. The superpartners of the gauge bosons are the fermionic gauginos. These are called bino, wino and gluino. The Higgs sector of the MSSM has to be more extended: there are two scalar Higgs-doublets and two fermionic doublets called Higgsinos.  \\
The left-chiral fields are arranged in doublets and they transform under the fundamental representation. The right-chiral fields are  singlets and transform in the conjugated representation. The vector bosons and gauginos transform under the adjoint representation. The names of the different particles as well as their quantum numbers with respect to the SM gauge groups  \(SU(3)_C\times SU(2)_L\times U(1)_Y\) are shown in Table~\ref{chirales_superfeld} and Table~\ref{Vektorsuperfeld}.
\begin{table}[t]
\begin{center}
\begin{tabular}{|c|c|c|c|c|}
\hline Name & SF & spin 0 & spin \(\frac{1}{2}\) & \(SU(3)_C \times SU(2)_L \times U(1)_Y \)\\
\hline \hline squarks, quarks & \(\hat{q}\) & \((\tilde{u}_L \thinspace
\tilde{d}_L)\) & \((u_L \thinspace d_L)\) &
\((\textbf{3},\textbf{2},\frac{1}{6})\)\\
(3 generations)& \(\hat{u}\) & \(\tilde{u}_R^*\) & \(u_R^\dagger\) &
\((\bar{\textbf{3}},\textbf{1},-\frac{2}{3})\)\\
& \(\hat{d}\) & \(\tilde{d}_R^*\) & \(d_R^\dagger\) &
\((\bar{\textbf{3}},\textbf{1},\frac{1}{3})\)\\
\hline sleptons, leptons & \(\hat{l}\) & \((\tilde{\nu} \thinspace
\tilde{e}_L)\) & \((\nu \thinspace e_L)\) &
\((\textbf{1},\textbf{2},-\frac{1}{2})\)\\
(3 generations)& \(\hat{e}\) & \(\tilde{e}_R^*\) & \(e_R^\dagger\) &
\((\textbf{1},\textbf{1},1)\)\\
\hline Higgs, Higgsinos & \(\hat{H}_u\) & \((H_u^+ \thinspace H_u^0)\) &
\((\tilde{H}_u^+ \thinspace \tilde{H}_u^0)\) &
\((\textbf{1},\textbf{2},\frac{1}{2})\)\\
& \(\hat{H}_d\) & \((H_d^0 \thinspace H_d^-)\) & \((\tilde{H}_d^0
\thinspace \tilde{H}_d^-)\) &
\((\textbf{1},\textbf{2},-\frac{1}{2})\)\\ \hline \hline
\end{tabular}
\end{center}
\caption[Left-chiral superfields of the MSSM]{Left-chiral superfields of the MSSM and the quantum numbers with respect to SM gauge group. The names of the scalar superpartners of the SM fermions and the Higgsinos are written with a tilde. Superfields are assigned by a hat and the hermitian conjugated is assigned by a dagger.} 
\label{chirales_superfeld}
\end{table}
%
\begin{table}[t]
\begin{center}
\begin{tabular}{|c|c|c|c|c|}
\hline Name &  SF & spin \(\frac{1}{2}\) & spin 1 & \(SU(3)_C, SU(2)_L, U(1)_Y \) \\
\hline \hline gluino, gluon &\(\hat{g}_\alpha\) & \(\tilde{g}_\alpha\) & \(g_\alpha\) &
\((\textbf{8},\textbf{1},0)\) \\
\hline winos, W bosons&\(\hat{W}_i\) & \(\tilde{W}_i\) &
\(W_i\) &
\((\textbf{1},\textbf{3},0)\) \\
bino, B boson&\(\hat{B}\) & \(\tilde{B}^0\) & \(B^0\) &
\((\textbf{1},\textbf{1},0)\) \\
\hline \hline
\end{tabular}
\end{center}
\caption[{Vector superfields of the MSSM}]{Vector superfields of the MSSM and the quantum numbers with respect to the SM gauge group. The fermionic superpartners of the SM gauge bosons are written with a tilde, superfields are assigned by a hat.}
\label{Vektorsuperfeld}
\end{table}
%
\subsubsection{Lagrangian of the MSSM}
The superpotential of the MSSM is
\begin{equation}
\label{superpotential}
W= Y^{a b}_{e}\,\hat{l}_a^j\,\hat{e}_b\,\hat{H}_d^i\,\epsilon_{ij}+Y^{a b}_{d}\,\hat{q}^{j \alpha}_a\,\hat{d}_{\alpha b}\,\hat{H}_d^i\,\epsilon_{ij} +
Y^{a b}_{u} \hat{q}^{i \alpha}_a\,\hat{u}_{\alpha b}\,\hat{H}_u^j\,\epsilon_{ij} + \mu\,\hat{H}_u^i\,\hat{H}_d^j \epsilon_{ij} \thickspace .
\end{equation}
Here, \(i,j\) are \(SU(2)_L\) indices, \(\alpha\) is a \(SU(3)_C\) color index and \(\epsilon_{i j}\) is the antisymmetric Levi-Civita tensor in two dimensions. \(a,b\) are flavor indices. The three Yukawa couplings  \(Y_e, Y_d\) and \(Y_u\) are complex \(3 \times 3\) matrices while \(\mu\) is a dimensionful mass parameter. \\
Generally spoken, the Lagrangian is a sum of gauge interactions, kinetic terms, matter interactions and terms stemming from the auxiliary fields when they are replaced by the equations of motion
\begin{equation}
\La = \La_{\rm gauge } + \La_{\rm kin} + \La_{\rm W}  + \La_D + \La_F \thickspace . 
\end{equation}
Furthermore, it is necessary to specify the gauge by adding gauge fixing terms \(\La_{GF}\) to the Lagrangian. A method to calculate the complete Lagrangian from the superpotential and the gauge structure for any renormalizable supersymmetric model is described in app.~\ref{sec:SARAH_Lag}. The complete Lagrangian for the MSSM is published for example in \cite{Kuroda:1999ks}. Here, we show only the scalar potential of the MSSM. This potential is formed by the F-Terms, the D-Terms and the soft breaking terms of the MSSM. The F- and D-terms are 
{\allowdisplaybreaks
\begin{eqnarray}
\nonumber V_F = - \La_F &=& |\sum_{a,b} (Y_{d,ab}\, \SQ_a^{j \alpha}\, \SD^*_{b \alpha} + Y_{e,a b}\, \SL_a^j\, \SE_b^*)\epsilon_{ij} + \mu H_u^j \epsilon_{ij}|^2 + \sum_{a,b}|Y_{e,a b}\, H_d^j\, \SE_b^*\, \epsilon_{ij}|^2\\
\nonumber && + |\sum_{a,b} (Y_{u,a b}\, \SQ_a^{j \alpha}\, \SU^*_{b \alpha}) \epsilon_{ij} - \mu H_d^j\, \epsilon_{ij}|^2 + \sum_{a,b} |Y_{e,a b} \, H_d^i \, \SL_a^j \, \epsilon_{ij}|^2 \\
\nonumber && + \sum_{a,b} | Y_{d,a b} \, H_d^j \, \epsilon_{ji} \, \SD^*_{b \alpha} + Y_{u,a b} \, H_u^j \,
\epsilon_{ji} \, \SU^*_{b \alpha}|^2
 + \sum_{a,b} | Y_{d, a b} \, H_d^i \, \epsilon_{ij} \, \SQ_a^{j\alpha} |^2 \nonumber \\ && + \sum_{a,b} | Y_{u,a b} \, H_u^j \, \SQ_a^{j\alpha} \, \epsilon_{ij}|^2 \label{F} \thickspace , \\
 \nonumber V_D = - \La_D &=& \halb g^{'2} \left(-\halb |\SHE|^2 + \halb |\SHZ|^2 + \sum_a \left(-\halb |\SL_a|^2 + |\SE_a^*|^2
 -\frac{2}{3} |\SU_a^*|^2 + \frac{1}{3} |\SD_a^*|^2+\frac{1}{6} |\SQ_a|^2\right)\right)^2 \\
 \nonumber &&+ \halb g_3^2 \left( \sum_a \left(\SQ_a^\dagger \frac{\lambda^A}{2} \SQ_a
 -\SD_a \frac{\lambda^A}{2} \SD_a^* - \SU_a \frac{\lambda^A}{2} \SU_a^*\right)\right)^2 \\
 \nonumber &&+ \frac{1}{8} g^2 \Big( |\SHE|^4 + |\SHZ|^4 + (\sum_a |\SQ_a|^2)^2 + (\sum_a |\SL_a|^2)^2
 - 2 |\SHE| |\SHZ|^2 \\
 \nonumber &&- 2|\SHE|^2 \sum_a|\SQ_a|^2 - 2|\SHZ|^2 \sum_a|\SQ_a|^2 - 2 |\SHE|^2 \sum_a|\SL_a|^2 -
 2|\SHZ|^2 \sum_a|\SL_a|^2 \\
 \label{D} &&+ 4 \sum_a |H_u^\dagger \, \SL_a|^2 + 4 \sum_a|H_d^\dagger \, \SQ_a|^2 + 4 \sum_a|H_u^\dagger \, \SQ_a|^2\Big).
\end{eqnarray}
The breaking of SUSY can be parametrized by adding so called soft breaking terms to the Lagrangian. They consist of mass terms for the scalars and for the gauginos as well as of scalar couplings related to the superpotential couplings \cite{Girardello:1981wz}:
 \begin{eqnarray}
 \label{eq:MSSM_SB}
 \nonumber \La_{SB} &=& - m_{\SHE}^2 |\SHE|^2 - m_{\SHZ}^2 |\SHZ|^2 -
 \SQ^\dagger \, m_\SQ^2 \, \SQ - \SD^\dagger \, m_\SD^2 \, \SD - \SU^\dagger \, m_\SU^2 \, \SU - \SL^\dagger \, m_\SL^2 \, \SL - \SE^\dagger \, m_\SE^2 \, \SE  \\
\nonumber && - \frac{1}{2}\left(M_1 \, \tilde{B} \tilde{B} + M_2 \, \tilde{W}_a \tilde{W}^a  + M_3 \, \tilde{g}_\alpha \tilde{g}^\alpha + \mbox{h.c.} \right) \\
\nonumber && - \left(T_e \, \SL^j \, \SE \, H_d^i \, \epsilon_{ij} + T_d \, \SQ^{j \alpha} \SD_\alpha \, H_d^i \epsilon_{ij} - T_u \, \SQ^{j \alpha} \, \SU_\alpha \, H_u^i \epsilon_{ij} + \mbox{h.c.} \right)\\
&&  - \left( B_\mu \, H_d^i \, H_u^j \, \epsilon_{ij} + \mbox{h.c.} \right) \thickspace .
\label{soft_breaking}
 \end{eqnarray}
\(T_i\) as well as \(m_\SQ^2\), \(m_\SL^2\), \(m_\SD^2\), \(m_\SU^2\) and \(m_\SE^2\)  are complex \(3\times 3\)-matrices while \(m_{\SHE}\) and \(m_{\SHZ}\) are real parameters. \(B_\mu\) is complex. } \\
\subsubsection{Electroweak symmetry breaking in the MSSM}
\label{sec:SUSY_EWSB}
For successful EWSB in the MSSM, the Higgs sector has to fulfill several conditions \cite{Kim:1983dt,Casas:1992mk,Moxhay:1984im}. This can easily be seen when we simplify the scalar potential of the Higgs fields: the charged component \(H_u^+\) can be rotated away by a \(SU(2)_L\) gauge transformation. In this case, we find that the minimum of the potential satisfying \(\frac{\partial V}{\partial H_u^+} = 0\) fulfills also \(H_d^-=0\). Thus, we can restrict the following discussion to neutral components of the Higgs fields.  The terms of eqs.(\ref{F})~-~(\ref{eq:MSSM_SB}) which involve only neutral Higgs fields are
\begin{eqnarray}
 \nonumber V &=& (|\mu|^2+m_\SHZ^2)|H_u^0|^2 + (|\mu|^2 + m_\SHE^2)|H_d^0|^2 - (B_\mu H_u^0 H_d^0 + \mbox{h.c.}) + \\
  && \frac{1}{8} (g^2 + g^{'2})(|H_u^0|^2-|H_d^0|^2)^2 \thickspace.
\end{eqnarray}
With  \(m_{1,2}^2 = m_{\SHE,\SHZ}^2 + \mu^2\).  We see that EWSB takes places if
\begin{equation}
m_1^2 m_2^2 -  B_\mu^2 < 0 \thickspace . 
\end{equation}
The one-loop corrections \(\Delta V_1\) to the potential are in an effective potential approach calculable by \cite{Coleman:1973jx} 
\begin{equation}
\Delta V_1 = \frac{1}{64 \pi^2} \mbox{STr} \left(M^4 \left(\ln \frac{M^2}{Q^2} - \frac{3}{2} \right) \right) \thickspace .
\end{equation}
Here, \(STr(x)\) is the spin weighted supertrace. We get the subsequent conditions for EWSB at one-loop:
\begin{equation}
\label{eq:MuProb}
 M_Z^2 = 2 \frac{\bar{m}_1^2 - \bar{m}_2^2 \tan^2 \beta}{\tan^2 \beta - 1}, \hspace{1cm}
 \sin 2 \beta = - \frac{2 B_\mu}{\bar{m}_1^2 + \bar{m}_2^2} 
\end{equation}
with \(\bar{m}_{1,2}^2 = m_{1,2}^2 + \frac{\partial \Delta V_1}{\partial v_{d,u}^2}\). These relations point to the so called \(\mu\)-problem in the MSSM: the SUSY parameters \(\mu, B_\mu, m_\SHZ\) and \(m_\SHE\) have to be of \(\Ord(M_Z)\) to prevent the necessity of fine-tuning to fulfill eq.~(\ref{eq:MuProb}). In contrast, the natural choice of the \(\mu\)-term would be of order of the GUT scale if it is not forbidden by a symmetry or exactly zero otherwise. A solution to this problem is to generate the \(\mu\)-term dynamically after SUSY breaking in order to relate it to the SUSY breaking scale. This is the underlying idea of the Next-to-minimal Supersymmetric Standard Model (NMSSM) considered in chapter~\ref{chapter:NMSSM}. 
\subsubsection{Mass eigenstates}
\label{section:masseigenstates}
The gauge eigenstates in Table~\ref{chirales_superfeld} and Table~\ref{Vektorsuperfeld} with the same quantum numbers mix to the mass eigenstates of the MSSM after EWSB. These eigenstates are formed by the massive vector bosons like in the SM and by the neutralinos, charginos and the mixtures of sleptons, squarks and Higgs fields. Of course, there is also a mixing of the SM fermions when we consider off-diagonal elements of the Yukawa couplings.  A complete overview of all mixings is given in app.~\ref{chapter:MSSM}. We summarize here the main results. A collection of the gauge and corresponding mass eigenstates is shown in Table~\ref{coll}.  \\
\begin{table}[h!]
\begin{center}
\begin{center} \begin{tabular}{|c|c|c|}
\hline Names & Electroweak eigenstates & Mass eigenstates \\
\hline \hline
Higgs & \(H_u^0, H_d^0, H_u^+, H_d^-\) & \(h_i, A^0, H^\pm\) \\
\hline 
 {} & \(\tilde{u}_L, \tilde{u}_R, \tilde{d}_L, \tilde{d}_R \) & \(\tilde{u}_1, \tilde{u}_2, \tilde{d}_1, \tilde{d}_2\) \\
squarks & \(\tilde{s}_L, \tilde{s}_R, \tilde{c}_L, \tilde{c}_R \) & \(\tilde{s}_1, \tilde{s}_2, \tilde{c}_1, \tilde{c}_2\) \\
{} & \(\tilde{t}_L, \tilde{t}_R, \tilde{b}_L, \tilde{b}_R \) & \(\tilde{t}_1, \tilde{t}_2, \tilde{b}_1, \tilde{b}_2\)\\
\hline
{} & \(\tilde{e}_L, \tilde{e}_R, \tilde{v}_e \) & \(\tilde{e}_1, \tilde{e}_2, \tilde{v}_e\) \\
sleptons & \(\tilde{\mu}_L, \tilde{\mu}_R, \tilde{v}_\mu \) & \(\tilde{\mu}_1, \tilde{\mu}_2, \tilde{v}_\mu\) \\
{} & \(\tilde{\tau}_L, \tilde{\tau}_R, \tilde{v}_\tau \) & \(\tilde{\tau}_1, \tilde{\tau}_2, \tilde{v}_\tau\)\\
\hline
neutralinos & \(\tilde{B}, \tilde{W}, \tilde{H}_d^0, \tilde{H}_u^0\) & \(\Neu_1, \Neu_2, \Neu_3, \Neu_4\) \\
\hline
charginos & \(\tilde{W}^\pm, \tilde{H}^+_u, \tilde{H}_d^.\) & \(\tilde{\chi}^\pm_1, \tilde{\chi}^\pm_2\) \\
\hline
gauge bosons & \(W^1, W^2, W^3, B\) & \(W^\pm, Z, \gamma\) \\
\hline
gluon \& gluino  & \(g, \tilde{g}\) & same \\
\hline \hline
\end{tabular} \end{center}
\end{center}
\caption[Gauge and Mass Eigenstates of the MSSM]{Gauge and Mass Eigenstates of the MSSM. For the squarks and sleptons is no flavor violation assumed. If flavor is violated all six down- and up-time squarks and all six charged sleptons mix separately to six mass eigenstates. Also the SM fermions would mix in that case. } 
\label{coll}
\end{table} 
\paragraph*{Charginos and neutralinos} The first and second wino \(\tilde{W}_{1,2}\) mix after EWSB to charged fermions \(\tilde{W}^\pm\). This is analog to the mixing of the charged vector bosons in the SM.  Furthermore, the charged components of the Higgsinos, \(H_u^+,H_d^-\), and the charged winos \(\tilde{W}^\pm\) mix to new mass eigenstates, called charginos \(\bCha\), while the neutral components of the Higgsinos, the bino and the third wino mix to four Majorana fermions called neutralinos \(\Neu_i\).
\paragraph*{Sleptons, sneutrinos and squarks}
The three generations of the fields \(\tilde{e}_L\) and \(\tilde{e}_R\) combine to six charged eigenstates
 \(\tilde{e}_{1} \dots \tilde{e}_{6}\) while the three sneutrinos \(\tilde{\nu}_L\) just rotate to the mass eigenstates \(\tilde{\nu}\) if flavor symmetry is broken. The up-type squarks \(\tilde{u}_L\) and \(\tilde{u}_R\) mix to
\(\tilde{u}_i\), and the down-type squarks  \(\tilde{d}_L\) and \(\tilde{d}_R\) mix to 
 \(\tilde{d}_i\). If flavor symmetry is conserved, only the fields of one generation mix among each other. 
\paragraph*{Leptons and quarks}
If the Yukawa couplings  are not assumed to be diagonal, the SM leptons and quarks rotate, too. The charged lepton mass matrix is diagonalized by two matrices \(Z^{E,L}\) and \(Z^{E,R}\) while in the quark sector the matrices \(Z^{U,L}\), \(Z^{U,R}\), \(Z^{D,L}\) and \(Z^{D,R}\) are used. The symmetric mass matrix of neutrinos is diagonalized by the unitary matrix \(Z^V\).  It is common to redefine the parameters in a way that \(Z^{U,R}\), \(Z^{D,R}\) and \(Z^{E,R}\) disappear from the Lagrangian. Only the products
 \begin{eqnarray}
  (Z^{U,L})^\dagger Z^{D,L}  & = & V_{\rm CKM} \thickspace ,\\
  (Z^{E,L})^\dagger Z^V  & = & V_{\rm PMNS}
 \end{eqnarray}
remain. \(V_{\rm CKM}\) is the well known Cabibbo-Kobayashi-Maskawa matrix \cite{Kobayashi:1973fv} and \(V_{\rm PMNS}\) is the Pontecorvo-Maki-Nakagawa-Sakata matrix \cite{Maki:1962mu}. 
\paragraph*{Higgs}
After EWSB, the Higgs fields are expressed by their scalar and pseudo scalar component as well as by their  vacuum expectation value (VEV)
\begin{equation}
\label{higgsvev}
H_d^0 \rightarrow \frac{1}{\sqrt{2}}( v_d + i \sigma_d + \phi_d) \thickspace, \hspace{1cm} H_u^0 \rightarrow  \frac{1}{\sqrt{2}}(v_u + i \sigma_u +  \phi_u) \thickspace . 
\end{equation}
Here \(v_u\) and \(v_d\) are the VEVs of the Higgs fields and they are connected to the mass of the \(W\) boson and \(Z\) boson by 
\begin{eqnarray}
\label{eq:VB_Masses}
M_W = \frac{e}{2 \sin\Theta_W} \sqrt{v_d^2 + v_u^2} \thickspace, \hspace{1cm}
M_Z = \frac{e}{2 \sin\Theta_W \cos\Theta_W} \sqrt{v_d^2 + v_u^2} \thickspace .
\end{eqnarray}
\(\sin \Theta_W\) is the well known Weinberg angle resulting from EWSB. The ratio of the VEVs is parametrized by  \(\tan\beta = \frac{v_d}{v_u}\). The scalar components \(\phi_d\) and \(\phi_u\) mix to the CP even mass eigenstates \(h_i\) (i=1,2), the pseudo scalars \(\sigma_d\) and \(\sigma_u\) mix to \(G^0\) and \(A^0\) and the charged components \(\phi_u^+\) and \(\phi_d^-\) build the eigenstates \(G^\pm\) and \(H^\pm\). \(G^0\) and \(G^\pm\) are the Goldstone bosons of EWSB. Thus, they are absorbed by the massive vector bosons. Hence, the physical spectrum of Higgs fields consists of two CP even (\(h_i\)), one CP odd (\(A^0\)) and two charged scalar bosons (\(H^-/H^+\)) with \(H^+=(H^-)^*\). \\
\subsubsection{Parameter space of the MSSM}
A careful counting of the parameters in the MSSM leads to 107 free parameters. Such an huge amount of parameters is not only impossible to handle in numerical studies but also very disappointing from the theoretical point of view: it just parametrizes our ignorance about the mechanism of SUSY breaking. That's why it is natural to search for underlying principles which relate these parameters among each other and reduce the number of free parameters. The basic idea is to embed the MSSM in a more fundamental theory which lives at a higher scale, normally assumed to be the Planck scale or a little bit below. In the fundamental theory, all SUSY parameters can be expressed by a limited number of variables (see sec.~\ref{sec:SUSY_Breaking}). However, these relations are just valid at the high scale. The values at the electroweak scale are related to those at the GUT scale by the RGEs. \\
For instance, the boundary conditions are at the GUT scale with regard to minimal supergravity (mSugra) that the squark and slepton masses are proportional to the unit matrix. A stronger demand is that all scalar squared masses are identical, i.e.
\begin{equation}
 \label{MNull}
 m_\SQ^2 = m_\SU^2 = m_\SD^2 = m_\SL^2 = m_\SE^2  \equiv m_0^2 {\bf 1}, \hspace{1cm}
 m_\SHE^2 = m_\SHZ^2 \equiv m_0^2  \thickspace .
\end{equation}
In addition, all gaugino masses are assumed to be the same
\begin{equation}
 \label{Mhalb}
 M_3 = M_2 = M_1 \equiv M_{1/2} .
\end{equation}
and all phases of the SUSY couplings are \(0\) or \(\pi\). Finally, the cubic, soft breaking couplings are proportional to the corresponding Yukawa couplings
\begin{equation}
\label{ANull}
T_u = A_0 Y_u \thickspace, \hspace{1cm} T_d = A_0 Y_d \thickspace, \hspace{1cm} T_e = A_0 Y_e \thickspace .
\end{equation}
In extensions of the MSSM, mSugra-like boundary conditions are often used, too. We will see this in chapters~\ref{chapter:SU5} and \ref{chapter:NMSSM}. Another set of boundary conditions is presented in sec.~\ref{sec:Intro_GMSB} for the case of gauge mediated SUSY breaking. 
\subsection{\texorpdfstring{$R$}{R}-parity, \texorpdfstring{$R$}{R}-parity violation and neutrino data}
\label{sec:R-parity}
Adding SUSY partners to all SM particles without further restrictions to their interactions would open decay channels for the proton. The reason is that interactions like
\begin{equation}
\label{eq:RpV}
W_{\slashed{R}} = \frac{1}{2} \lambda_{ijk} \hat{u}_i \hat{d}_j \hat{d}_k + \frac{1}{2}  \lambda^{'}_{ijk} \hat{q}_i \hat{d}_j \hat{l}_k+
\frac{1}{2}  \lambda^{''}_{ijk} \hat{l}_i \hat{l}_k \hat{e}_k + \epsilon_i \hat{l}_i \hat{H}_u 
\end{equation}
are allowed by gauge invariance. These interactions enable the decay of a proton, for instance, into \(e^+\) and \(\pi^0\). The life time of the proton would be the fraction of a second if the couplings were of \(\Ord(1)\). Consequently, these couplings must either extremely small or completely forbidden. One possibility to forbid all interactions of eq.~(\ref{eq:RpV}) at once is to demand that $R$-parity is conserved. $R$-parity is a multiplicative quantum number defined as \cite{Farrar:1978xj,Dimopoulos:1981zb,Weinberg:1981wj,Sakai:1981pk,Dimopoulos:1981dw}
\begin{equation}
R= (-1)^{3B+L+2S} \thickspace. 
\end{equation}
\(S\) is the spin of the particle. \(B\) and \(L\) are the baryon and lepton number, respectively. Obviously, all SUSY particle have  $R$-parity -1. Thus, there can't be any interaction with an odd number of SUSY particles and all terms of eq.~(\ref{eq:RpV}) are forbidden. As another consequence, the  lightest supersymmetric particles (LSP) must be stable if $R$-parity is conserved. This is the origin of many supersymmetric dark matter candidates. \\
On the other side, it would be sufficient to forbid only some interactions of eq.~(\ref{eq:RpV}) to get proton decay under control. One possibility is to allow just lepton or just baryon number violating operators. A often studied assumption is that only the baryon number violating terms of eq.~(\ref{eq:RpV}) are forbidden while the lepton number violating terms are allowed. This can be motivated by a symmetry called baryon triality \cite{Dreiner:1997uz}. Another ansatz is to assume that $R$-parity is conserved at the high scale and the bilinear term is created dynamically at the low scale. Bilinear $R$-parity violation has interesting, phenomenological aspects: it leads to a mixing between neutrinos and neutralinos. In this way, neutrino masses are generated \cite{Romao:1999up,Hirsch:2000ef,AristizabalSierra:2008ye}. While neutrinos are massless in the MSSM, oscillation experiments have shown that the neutrinos have tiny, but non-zero masses \cite{Fukuda:1998mi,Eguchi:2002dm,Ahmad:2002jz}. In addition, the mass eigenstates of neutrinos are not equal to the gauge eigenstates \cite{Schwetz:2008er}. This leads to several observables which can be related to the $R$-parity violating parameters. For example,  
\begin{equation}
\label{eq:BiRpV_neutrino}
\tan \Theta_{23} = \left| \frac{\Lambda_2}{\Lambda_3} \right|, \hspace{1cm} \tan\Theta_{1 2} \simeq \left| \frac{\tilde{\epsilon}_1}{\tilde{\epsilon}_2} \right| \thickspace. 
\end{equation}
\(\Theta_{ij}\) is the observed mixing angle between the neutrino generations \(i\) and \(j\). We used in eq.~(\ref{eq:BiRpV_neutrino}) the definitions \(\tilde{\epsilon}_i = Z^{\nu}_{i j} \epsilon_j\) and \(\Lambda_i = \epsilon_i v_d + \mu v_i\). \(v_i\) are the VEVs of the sneutrinos and \(Z^{\nu}\) is the sneutrino mixing matrix. Similar relations for the neutrinos masses can be found in \cite{Hirsch:2003fe}. The measured values for the neutrino observables are 
\begin{eqnarray*}
\tan^2 \Theta_{12} = 0.47 \pm 0.05\thickspace,\hspace{1cm}
\tan^2 \Theta_{23} = 0.83 \pm 0.35\thickspace,\hspace{1cm}
\sin^2 \Theta_{13} < 0.019 \thickspace ,\\
\Delta m_{21}^2 = 7.67 \pm 0.22 \times 10^{-5}\, \eV^2 \thickspace,\hspace{1cm}
\Delta m_{31}^2 = 2.58 \pm 0.15 \times 10^{-3}\, \eV^2 \thickspace .
\end{eqnarray*}
\(\Delta m^2_{ij}\) is the measured mass difference between the generation \(i\) and \(j\). The bilinear term doesn't lead to fast proton decay, but the neutralino wouldn't be stable any longer. It could decay via \(W, Z\) or Higgs exchange for instance into \(\nu \nu \nu\), \(l^\pm l^\mp \nu\) or \(l^\pm q q'\). This poses the question if there can be SUSY dark matter when $R$-parity is broken. We will check this in case of the gravitino as LSP in chapter~\ref{chapter:GMSB}.   
\subsection{Supersymmetry breaking}
The easiest way to break a symmetry is to add breaking terms to the Lagrangian. This happens in the MSSM when writing down the soft breaking terms in eq.~(\ref{eq:MSSM_SB}). However, as already mentioned, this parametrizes only the ignorance about the fundamental origin of SUSY breaking. Therefore, it is necessary to search for a mechanism to break SUSY spontaneously. We discuss first SUSY breaking at tree level and show the need for hidden sector SUSY breaking. This is discussed in the second part.  
\subsubsection{Breaking of supersymmetry at tree level}
\label{sec:SUSY_Breaking}
So far, we have introduced in eq.~(\ref{eq:MSSM_SB}) an explicit breaking of SUSY by the soft breaking terms. Now we want to generate such a breaking dynamically in order to explain the relations in eqs.~(\ref{Mhalb})~-~(\ref{ANull}). If SUSY is spontaneously broken, the VEV of the SUSY generator is
\begin{equation}
 \bra 0 | H | 0 \ket = \frac{1}{4} \left( ||Q_1^\dagger|0\ket||^2+ ||Q_1|0\ket||^2+ ||Q_2^\dagger|0\ket||^2+
  ||Q_2|0\ket||^2 \right) > 0 \thickspace .
\end{equation}
Furthermore, it holds \(\bra | H | \ket = \bra | V | \ket\), i.e. SUSY gets broken if the F- and/or D-Terms don't vanish in the ground state. Consequently, it is sufficient to search for theories in which the equations  
\begin{equation}
 F_i = 0 \thickspace, \hspace{1cm}
 D^a = 0 \thickspace.
\end{equation}
can't be fulfilled simultaneously for all \(i\) and \(a\). Unfortunately, the easiest ans\"atze for a spontaneous breaking of SUSY are not sufficient. 
\begin{enumerate}
 \item \textbf{Fayet-Illopoulus D-Term breaking} \cite{Fayet:1974jb,Dvali:1996rj}: \\
The D-terms receive in that case a VEV due to a term linear in the auxiliary fields
\begin{equation}
\label{eq:DTermBreaking}
 \La_{\text{Fayet}} = - \kappa D
\end{equation}
with a constant \(\kappa\). Such a term is not gauge invariant for a non-abelian group. Therefore only \(U(1)_Y\) D-terms are possible. However, it has been shown that eq.~(\ref{eq:DTermBreaking}) causes VEVs of sfermions and would break electromagnetism or color but not SUSY.
\item  \textbf{O`Raifeartaight F-Term breaking}  \cite{O'Raifeartaigh:1975pr,Shih:2007av}:\\
In this case, a set of chiral superfields \(\Phi_i\) and a superpotential \(W\) are chosen in a way that the equations
\begin{equation}
F_i = - \frac{\delta W^*}{\delta \phi^{*i}} = 0 
\end{equation}
can't be solved simultaneously. The simplest example for such a superpotential is
\begin{equation}
 W = - k \Phi_1 + m \Phi_2 \Phi_3 + \frac{y}{2} \Phi_1 \Phi_3^2 \thickspace.
\end{equation}
SUSY breaking gets fixed by the parameter \(k\). One problem in this context is to explain  \(k \ll M_p^2\) what is necessary for correct phenomenology.  
\end{enumerate}
Even more complicated models of D- and F-term SUSY breaking have another problem in common, namely, to fulfill the supertrace mass sum rule \cite{Ferrara:1979wa}
\begin{equation}
\mbox{STr} m^2 \equiv  \sum_{J=0}^{1/2} (-1)^{2 J} (2 J + 1) \mbox{Tr}(m_J^2) = 0 \thickspace .
\end{equation}
This means that the spin weighted trace of the squared mass matrix taken over all chiral superfields has to vanish. This constraint arises naturally for tree level SUSY breaking and leads to the prediction of scalars lighter than down- or up-quark \cite{Luty:2005sn}. A possibility to circumvent this problem is to break SUSY at a very high scale in a hidden sector by superfields which are singlets under the SM gauge groups and couple this sector weakly to the sector containing the SUSY fields.
\subsubsection{Hidden sector supersymmetry breaking}
\label{section:introdcution_SUSY_Breaking}
\begin{figure}[t]
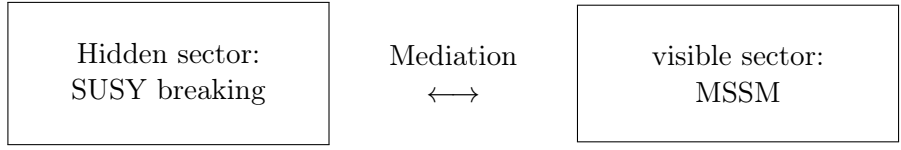

\begin{center}
\fbox{\parbox{4cm}{
\begin{center}
Hidden sector:\\
SUSY breaking
\end{center} }}
\parbox{3cm}{
\begin{center}
Mediation \\
\(\longleftrightarrow\)
\end{center}}
\fbox{\parbox{4cm}{
\begin{center}
visible sector:\\
MSSM
\end{center} }}
\end{center}
\caption[Hidden sector SUSY breaking]{SUSY is broken in a hidden sector and the breaking is transmitted via gauge or gravitational interactions to the visible sector}
\label{hiddensector}
\end{figure}

The discussion in the last section has shown that it is necessary to extend the framework in order to get phenomenological correct SUSY breaking. The idea is to break SUSY in a 'secluded' or 'hidden' sector spontaneously. This sector consists of superfields which are uncharged under the SM gauge group. Interactions between the secluded and visible sector transmit SUSY breaking afterwards to our 'visible' sector. This happens either by non-renormalizable terms in the K\"ahler potential or by loop corrections. The scenario is sketched in Fig.~\ref{hiddensector}.  
The two main approaches for the mediation of SUSY breaking work with gravitational or gauge interactions.
\begin{enumerate}
 \item {\bf Gravitational mediation of SUSY breaking} \cite{Chamseddine:1982jx,Ellis:1982wr}:\\
We assume that SUSY is broken in the hidden sector by a VEV \(\bra F \ket\) of some new spurion field. The soft breaking terms can be approximated to
\begin{equation}
 m_{\rm soft} \simeq \frac{\bra F \ket}{M_P} \thickspace ,
\end{equation}
if SUSY breaking is mediated by gravity. \(M_P\) is the Planck scale of \(2.4 \times 10^{18}\)~\GeV. This approximation can be motivated by the observation that \(m_{\rm soft}\) has to vanish in the limits  \(\bra F \ket \rightarrow 0\) and \(M_P \rightarrow \infty \) but is also the result of a rigorous calculation:  in case of minimal supergravity (mSugra) the results are \cite{Nilles:1983ge}
\begin{equation}
 M_{1/2} = f \frac{\bra F \ket}{M_P} \thickspace, \hspace{1cm} m_0^2 = k \frac{|\bra F \ket |^2}{M_P^2} \thickspace, \hspace{1cm} A_0 = \alpha \frac{\bra F \ket}{M_P} 
\end{equation}
with parameters \(f, k\) and \(\alpha\) depending on the underlying model. Obviously, this explains the relations eqs.~(\ref{MNull}) - (\ref{ANull}). 
\item {\bf Gauge mediated SUSY breaking} \cite{Dine:1981gu,Dine:1993yw}: \\
A possibility for a flavor independent mediation of SUSY breaking are interactions involving the SM gauge couplings. In these scenarios, the soft breaking terms are generated at the loop level due to radiative contributions by messenger fields. These messengers are chiral superfields charged under \(SU(3)_C\times SU(2)_L \times U(1)_Y\). This leads to soft breaking masses of the order 
\begin{equation}
m_{\rm soft} \simeq \frac{g^2}{16 \pi^2} \frac{\bra F \ket}{M} \thickspace . 
\end{equation}
Therefore, if the messenger masses \(M\) and the breaking parameter \(\sqrt{\bra F \ket}\) are of the same order, the SUSY breaking is of the order \(\sqrt{\bra F \ket}\). We will discuss GMSB especially with regard to the messenger sector in chapter~\ref{chapter:GMSB}. 
\end{enumerate}
\subsection{Supersymmetric grand unified theories}
\subsubsection{Introduction}
\label{sec:introduction_GUT}
In sec.~\ref{introduction:motivation},  we have already seen that SUSY leads to an unification of gauge couplings and might therefore be a further step towards a grand unified theory (GUT). Another reason to search for a GUT theory is the particle content of the SM. The quantum numbers of the known particles with respect to the SM gauge groups  look quite random at first glance :
\begin{eqnarray*}
e^c = \left({\bf 1},{\bf 1}\right)_{1}\thickspace, \hspace{0.5cm}  d^c = \left({\bf \bar{3}},{\bf 1}\right)_{-\frac{2}{6}}\thickspace, \hspace{0.5cm} u^c  \left({\bf \bar{3}},{\bf 1}\right)_{\frac{2}{3}}\thickspace, \hspace{0.5cm} l = \left({\bf 1},{\bf 2}\right)_{-\frac{1}{2}}\thickspace, \hspace{0.5cm} q = \left({\bf 3},{\bf 2}\right)_{\frac{1}{6}} \thickspace.
\end{eqnarray*}
For a long time it was  a time an open question how these numbers, especially the \(U(1)_Y\) hypercharges, can be explained from a more fundamental principle. Glashow and Georgi proposed the solution that the SM gauge groups are embedded in a \(SU(5)\) group that gets broken at a high scale \cite{Georgi:1974sy}
\begin{equation}
\label{eq:SU5breakingToSM}
  SU(5) \rightarrow SU(3)_C \times SU(2)_L \times U(1)_Y \thickspace .
\end{equation}
Using this assumption, all SM particles of one generation can be described by a 5- and 10-plet under \(SU(5)\) and all quantum numbers are fixed up to a normalization of the \(U(1)_Y\). This can be proofed with the technique of Young Tableaux which we briefly summarize in app.~\ref{chapter:young}. \\
The embedding in one gauge group leads inevitably to the demand that all gauge couplings have to unify at a high scale. While this is not the case for the particle content of the SM, this aim can be reached in the context of supersymmetric models as already shown. Even if there might be no strict unification by just extrapolating the measured gauge couplings, the other particles of the complete \(SU(5)\) theory with masses just below the GUT scale, like the additional gauge bosons and the colored Higgs triplets, lead to threshold corrections. Taking all these factors into account, the possibility of a  gauge unification in minimal supersymmetric \(SU(5)\) has already be shown at three-loop level \cite{Martens:2010nm}. The main uncertainty is still the experimental error of the strong interaction.  \\
On the other hand, exactly the same particles which might lead to an unification in the framework of a GUT theory cause some severe constraints on the symmetry breaking scale: the \(SU(5)\) gauge bosons which get high masses after breaking to the SM gauge group are charged under \(SU(2)_L\) and \(SU(3)_C\). Therefore, they enable interactions which can lead to proton decay. According to the decoupling theorem of Appelquist and Carazzone \cite{Appelquist:1974tg} those processes are suppressed by the high mass of the particles. This suppression might not be sufficient to circumvent experimental bounds on the life-time of the proton. Therefore, \(SU(5)\) GUT theories are under big pressure \cite{Nath:1985ub}. However, there are new models arising which circumvent the problem with proton decay, e.g. dual \(SU(5)\) \cite{Abel:2009bj}.\\
Of course,  larger groups have also been explored in the scope of grand unification. The \(SO(10)\) is the smallest group in which all particles of one generation of SM fermions can be described by just one multiplet, the {\bf 16} \cite{Fritzsch:1974nn}. Another group motivated by string theory is the exceptional group \(E_6\) \cite{Gursey:1975ki}. 
\subsubsection{Supersymmetric \texorpdfstring{$SU(5)$}{SU5} }
A pedagogical introduction to the topic of supersymmetric GUT theories can be found in \cite{Mohapatra:1999vv}. In this section, we write down the simplest \(SU(5)\)-invariant superpotential which reproduces the superpotential of the MSSM as low energy limit. Moreover, we  decompose this superpotential in case of the symmetry breaking of eq.~(\ref{eq:SU5breakingToSM}). We start with the superpotential
\begin{equation}
\label{SU5superpotential}
 W^{SU(5)} = \sqrt{2}\, \overline{5}_{M,i}\, Y^5\, 10_M^{ij}\, \overline{5}_{H,j} - \epsilon_{ijklm} \frac{1}{4}\, 10_M^{ij}\, Y^{10}\, 10_M^{kl}\, 5_H^{m}\, + M_5\, 5_H^i\, \overline{5}_{H,i} \thickspace . 
\end{equation}
For clearness, we have written here once explicitly the \(SU(5)\) indices \(i,j,k,l,m=1,\dots 5\). We will skip them in the following expressions. The used numbers as name for the superfields assign the transformation properties with respect to \(SU(5)\) and a lower index \(M\) refers to matter fields while \(H\) refers to Higgs fields.  \(Y^{10}\) is a symmetric matrix and \(Y^5\) is a generic one. The origin of the numerical coefficients of the different terms is the normalization of the multiplets
\begin{equation}
\label{eq:SU5matter}
 10_M = \frac{1}{\sqrt{2}} \left( \begin{array}{cc} \epsilon_{abc} \hat{u}_c & - \hat{q}_{\alpha \beta}^T \\ \hat{q}_{\alpha \beta} & -\epsilon_{\alpha \beta} \hat{e} \end{array} \right) \thickspace, \hspace{1cm} 5_M = \left( \begin{array}{c} \hat{d}_a \\ (\epsilon \hat{l})_\alpha \end{array} \right)
\end{equation}
with \(a,b,c \in \{1,2,3\}\) and \(\alpha,\beta \in \{1,2\}\). The factor \(\frac{1}{\sqrt{2}}\) guarantees a correct normalization of the kinetic terms.\\
Further Higgs fields have to be introduced to break \(SU(5)\)
\begin{equation}
\label{eq:SU5breaking}
W^{SU(5),H} = \lambda_5\, 5_h\, 24_H\, \bar{5}_H + \frac{1}{2}\, M_{24}\, (24_H)^2 + \frac{1}{6}\, \lambda_{24}\, (24_H)^3 \thickspace. 
\end{equation}
To write these interactions in a \(SU(3)_C\times SU(2)_L \times U(1)_Y \) invariant form, we can use the identifications of eq.~(\ref{eq:SU5matter}) and
\begin{eqnarray}
\label{eq:SU5Higgs}
 5_H &=& \left(\hat{H}_u^C, \hat{H}_u \right), \hspace{1cm}  \bar{5}_H = \left(\hat{H}_d^C, \hat{H}_d \right), \hspace{1cm}  24_H = \left(\hat{H}_G, \hat{H}_W, \hat{H}_B, \hat{H}_{X}, \hat{H}_{\bar{X}}\right) \thickspace .
\end{eqnarray}
These identifications are based on the quantum numbers with regard to the broken gauge group and can again be proofed with the technique of Young Tableaux shown in app.~\ref{chapter:young}. \(\hat{H}_u^C\) and \(\hat{H}_d^C\) are Higgs fields which carry a color charge. Thus, they can also trigger proton decay and have to be much heavier than the other components \(\hat{H}_d\) and \(\hat{H}_u\). Satisfying this multiplet-splitting is one of the great difficulties in \(SU(5)\) model building. We take this in the following as given. \\
The fields \(\hat{H}_G, \hat{H}_W, \hat{H}_B\) have the same quantum numbers as the vector superfields \(\hat{g}\), \(\hat{W}\) and \(\hat{B}\). \(\hat{H}_X\) and \(\hat{H}_{\bar{X}}\) have the quantum numbers of the additional \(SU(5)\) gauge bosons, which lead to proton decay: they are triplet/anti-triplet under \(SU(3)_C\) and doublets under \(SU(2)_L\). \\
We will concentrate in the following on the matter sector of \(SU(5)\) and don't discuss the interactions stemming from eq.~(\ref{eq:SU5breaking}). The different terms of eq.~(\ref{eq:SU5matter}) can be decomposed as
\begin{eqnarray*}
 \sqrt{2}\, \bar{5}_M\, Y^5\, 10_M\, \bar{5}_H &=& - \hat{d}\, Y^5\, \hat{q}\, \hat{H}_d - \hat{e}\, \left(Y^5\right)^T \, \hat{l} \, \hat{H}_d - \hat{l} \, Y^5 \, \hat{q}  \, \hat{H}_d^C - \hat{d} \, Y^5 \, \hat{u} \, \hat{H}_d^C \thickspace , \\
-\frac{1}{4}\, 10_M\, Y^{10}\, 10_M 5_H &=& \hat{u} \, Y^{10} \, \hat{q} \, \hat{H}_u + \frac{1}{2} \hat{q} \, Y^{10} \hat{q} \, \hat{H}_u^C + \hat{u} \, Y^{10} \, \hat{e} \, \hat{H}_u^C  \thickspace , \\
M_5\, 5_H\, \bar{5}_H &=& M_5 \, \hat{H}_u^C \, \hat{H}_d^C + M_5 \, \hat{H}_u \, \hat{H}_d \thickspace .
\end{eqnarray*}
After integrating out the colored Higgs fields, the superpotential of the MSSM remains. We will extend this minimal model of \(SU(5)\) in chapter~\ref{chapter:SU5} by additional chiral superfields. 
\subsection{Constraints from precision data} 
\label{sec:LowE}
As we will show in sec.~\ref{sec:DM_constraints}, a potential candidate for dark matter has to fulfill several constraints in order to be a valid aspirant for building up 23~\% of the matter in the universe. Besides, there are also constraints on the parameter space of a SUSY model by precision measurements: the particles postulated by a supersymmetric extension can contribute to rare processes like \(b\rightarrow s\gamma\), or open new channels which are not possible in the SM without neutrino masses like \(\tau \rightarrow 3 e\).  We give a short overview of the observables which provide the most stringent bounds for new physics beyond the SM. 

\paragraph*{$\tau$ and $\mu$ decays}
In SUSY models, large contributions to lepton flavor violation can be a result of off-diagonal elements in the soft breaking masses as well as of $R$-parity violation. This could lead to enhanced branching ratios of processes like \(l_i \rightarrow 3 l_j\) or \(l_i \rightarrow \gamma l_j\). A detailed derivation of the analytical formulas as well as numerical studies can be found in \cite{Arganda:2005ji}. So far, none of these processes have been measured, but there are severe upper limits from experiment \cite{Aubert:2003pc,Bellgardt:1987du,Aubert:2005ye,Aubert:2005wa,Brooks:1999pu} 
\begin{eqnarray*}
 \mbox{Br}\left(\tau \rightarrow 3 \mu \right) <  1.9\cdot 10^{-7} \thickspace, \hspace{1cm}  
\mbox{Br}\left(\tau \rightarrow 3 e \right) <  2.0\cdot 10^{-7} \thickspace, \hspace{1cm} 
 \mbox{Br}\left(\mu \rightarrow 3 e\right) <  1.0\cdot 10^{-12} \thickspace . \\
 \mbox{Br}\left(\tau \rightarrow \mu \gamma \right) <  6.8\cdot 10^{-8} \thickspace, \hspace{1cm} 
 \mbox{Br}\left(\tau \rightarrow e \gamma \right) <  1.1\cdot 10^{-7}\thickspace, \hspace{1cm}
 \mbox{Br}\left(\mu \rightarrow e \gamma \right) <  1.2\cdot 10^{-11} \thickspace . \\
\end{eqnarray*}

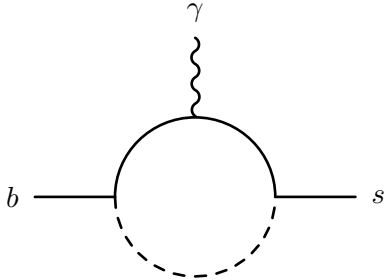
\begin{figure}[t]
\centering
\begin{fmffile}{Feynmangraphen/BToSgamma}
 \fmfframe(0,0)(0,0){
   \begin{fmfgraph*}(120,120)
    \fmfleft{l1}
   \fmfright{r1}
    \fmftop{t1}
   \fmf{plain}{l1,v1}
    \fmf{plain}{v2,r1}
    \fmf{dashes,right,tension=0.5}{v1,v2}
    \fmffreeze
    \fmfforce{60,90}{v3}
    \fmf{plain,left,tension=0.5}{v1,v2}
    \fmf{wiggly}{v3,t1}
    \fmflabel{$b$}{l1}
    \fmflabel{$s$}{r1}
    \fmflabel{$\gamma$}{t1}
\end{fmfgraph*}}
\end{fmffile}
\vspace{-0.8cm}
\caption[One loop contribution to the branching ratio $b\rightarrow s \gamma$]{One loop contribution to the branching ratio $b\rightarrow s \gamma$. In the case of SUSY, the heavy fields in the loop can be a squark together with a neutralino, a charginos or a gluino, or a charged Higgs together with a quark. }
\label{bsgamma}
\end{figure}

\paragraph*{Flavor changing neutral and charged currents in the B-sector} 
Today, some of the most strict constraints on flavor physics come from the measurements of rare processes in the B-sector. Details of the calculations of the subsequent  processes can e.g. be found in \cite{Buras:2002vd} and \cite{Baek:2001kh}. \\
Maybe, the best known of those constraints is the branching ratio of \(\bar{B} \rightarrow X_s \gamma\). The main contribution to this decay stems from the partonic process \(b \rightarrow s \gamma\). The effective Lagrangian for this process which violates bottom number by 1 can be written as  
\begin{equation}
 \La^{\Delta B=1} = - \frac{G_F}{\sqrt{2}} \left( C_7^{'} \left(\bar{s}_R \sigma^{\mu\nu} b_L\right) F_{\mu\nu} + C_7 \left(\bar{s}_L \sigma^{\mu\nu} b_R \right) F_{\mu\nu} \right) \thickspace .
\end{equation}
The Wilson coefficients \(C_7\) and \(C_7^{'}\) are zero at tree level, but can at one-loop level be generated by diagrams like the one shown in Fig.~\ref{bsgamma}. The branching ratio is already unequal to zero in the SM because of the off-diagonal elements in the CKM matrix. Today, the dominant standard model contributions to these coefficient are known up to three loop level \cite{Schutzmeier:2007xm}.\\ 
The combined value of the measurements by BaBar, Belle and CLEO of the Heavy Flavor Averaging Group (HFAG) leads to \cite{Abe:2008sxa, Chen:2001fja,Aubert:2006gg,Barberio:2007cr}
\begin{equation}
 \mbox{Br}\left(\bar{B} \rightarrow X_s \gamma \right) = \left(3.55 \pm 0.24 \right)\cdot 10^{-4} \thickspace .
\end{equation}
With the discovery of the mixings in the neutral B-meson sector, new strong constraints on flavor physics have been set. Similar constraints were already known for years coming from the neutral Kaon sector. These mixings violate the corresponding flavor number by 2 and can again be expressed by an effective Lagrangian. In the case of B-mesons the Lagrangian reads
\begin{eqnarray}
\nonumber
\La^{\Delta B = 2} &=& - \frac{8 G_F}{\sqrt{2}} \Bigg(\frac{1}{2} g_R^V \left(\bar{d}_R^\alpha \gamma^\mu b_{R,\alpha} \right) \left(\bar{d}_R^\beta \gamma^\mu b_{R,\beta} \right) +  \frac{1}{2} g_L^V \left(\bar{d}_L^\alpha \gamma^\mu b_{L,\alpha} \right) \left(\bar{d}_L^\beta \gamma^\mu b_{L,\beta} \right) +\\
\nonumber &&  \frac{1}{2} g_{RR}^S \left(\bar{d}_L^\alpha b_{R,\alpha} \right) \left(\bar{d}_L^\beta b_{R,\beta} \right) +  \frac{1}{2} g_{LL}^S \left(\bar{d}_R^\alpha b_{L,\alpha} \right) \left(\bar{d}_R^\beta  b_{L,\beta} \right) \\
\nonumber &&  \frac{1}{2} g_{RR}^{S'} \left(\bar{d}_L^\alpha b_{R,\beta} \right) \left(\bar{d}_L^\beta b_{R,\alpha} \right) +  \frac{1}{2} g_{LL}^{S'} \left(\bar{d}_R^\alpha b_{L,\beta} \right) \left(\bar{d}_R^\beta  b_{L,\alpha} \right) \\
&&  \frac{1}{2} g_{RL}^S \left(\bar{d}_L^\alpha b_{R,\alpha} \right) \left(\bar{d}_R^\beta b_{L,\beta} \right) +  \frac{1}{2} g_{RL}^{S'} \left(\bar{d}_L^\alpha b_{R,\beta} \right) \left(\bar{d}_R^\beta  b_{L,\alpha} \right) \Bigg)
\label{eq:EffMBs} 
\end{eqnarray}
with effective couplings \(g_i\). For Kaons, the bottom quark has to be replaced by a strange quark. Again, these couplings are  zero at tree level and can be generated by box diagrams or by double penguin diagrams at the loop level. We can express two observables \(\Delta M_{B_{s,d}}\) using the effective Lagrangians
\begin{equation}
 \Delta M_{B_{s,d}} = 2 |M_{12}(B_{s,d})| \thickspace, \hspace{1cm} M_{12}(B_{s,d}) = - \frac{\bra B_{s,d}^0 | \La^{\Delta B=2} | \bar{B}_{s,d}^0 \ket}{2 m_{B_{s,d}}} \thickspace .
\end{equation}
The measured values are \cite{Amsler:2008zzb}
\begin{equation}
 \Delta M_{B_s} = \left(117.0 \pm 0.8 \right)\cdot 10^{-10}\,\MeV \thickspace, \hspace{1cm}
 \Delta M_{B_d} = \left(3.337 \pm 0.033 \right)\cdot 10^{-10}\, \MeV \thickspace .
\end{equation}
\paragraph*{Electric dipole moments}
The values for a SM calculation of the electric dipole moment of the neutron \(d_n\) and electron \(d_e\) are
\begin{equation}
 d_n \simeq 10^{-32}\, \mbox{e cm}, \hspace{1cm} d_e \simeq 10^{-38}\, \mbox{e cm} \thickspace .
\end{equation}
These values are much smaller than the current experimental limits of
\begin{equation}
 d_n \leq 2.9\cdot 10^{-26}\, \mbox{e cm}, \hspace{1cm} d_e \leq 1.6\cdot 10^{-27}\, \mbox{e cm} \thickspace .
\end{equation}
In SUSY  many couplings can carry a CP phase and, consequently, contribute to the electric dipole moment. Either these phases have to be very small or the sfermions of the first two generations must be heavy. Otherwise, some cancellation between different contributions at one-loop level has to take place. 
\paragraph*{SUSY contributions to $\Delta\rho$}
If there is a large mass splitting of SUSY particles sitting in the same \(SU(2)_L\)-doublet, there can be large contributions to the self-energies \(\Sigma\) of the electroweak gauge bosons. Therefore  the \(\rho\) parameter defined as
\begin{equation}
\rho = \frac{1}{1 - \Delta \rho} \hspace{1cm} \mbox{with} \hspace{1cm} \Delta \rho = \frac{\Sigma^Z(0)}{M^2_Z} - \frac{\Sigma^W(0)}{M^2_W} 
\end{equation}
can be significantly changed. The allowed range from experiment is \cite{PDBook}
\begin{equation}
 1.00989 \simeq \rho \simeq 1.01026 \thickspace.
\end{equation}
\section{Dark Matter}
We can only touch here the main astrophysical aspects of dark matter. A good review about evidences, constraints and candidates  of dark matter is given in \cite{Bertone:2004pz}. Before we present the different topics concerning dark matter, we give a general introduction to the thermodynamics in the expanding universe: we start with a short introduction to the standard model of cosmology and discuss afterwards the freeze-out of species and late time decays of particles. 

\subsection{Thermodynamics in the expanding universe}
\subsubsection{Standard Big Bang cosmology}
\label{sec:SM_cosmology}
The standard model of cosmology is not a fixed term like the standard model of particle physics \cite{1931MNRAS..91..490L,1975seu..book.....Z}. Nevertheless, it contains some basic ingredients: everything started about \(10^{10}\)~years ago with the Big Bang. Since then, the universe expanded and cooled down. This expansion was in the first \(10^{-35}\)~-~\(10^{-33}\)~seconds exponential and this period is called the inflation. Inflation is the reason, why the universe looks today to a very high level homogeneous and isotropic. While the early universe was very hot and consisted of a soup of matter and energy, more and more particles left the thermal equilibrium with decreasing temperature in the universe. During that time, an asymmetry between matter and anti-matter must have appeared, which prevented a complete annihilation of all matter. This so called baryogenesis is not yet completely understood. When the universe was cooled down to \(10^{16}\)~K after \(10^{-12}\)~s, the electroweak interaction splits into electromagnetism and weak interaction. After about 10~s the protons and neutrons started to fuse to the light elements deuterium, helium and lithium. This process is called Big Bang nucleosynthesis (BBN). The universe cooled further down and free nuclei and electrons combined to atoms. This process is known as recombination and it is the first time that photons could travel over large distances. This happened when the universe was about 400.000~years old. \\
The quantitative description of this scenario is based on Einstein's field equation \cite{Einstein:1916:GARa}
\begin{equation}
\label{eq:Einstein}
 R_{\mu\nu} - \frac{1}{2} g_{\mu\nu} R = - \frac{8 \pi G_N}{c^4} T_{\mu\nu} + \Lambda g_{\mu\nu} \thickspace.
\end{equation}
Here, \(R_{\mu\nu}\) and \(R\) are the Ricci tensor and scalar which incorporate the geometry of the space time. \(g_{\mu\nu}\) is the metric tensor and \(T_{\mu\nu}\) the energy-momentum tensor. \(G_N\) is Newton's constant and \(\Lambda\) is the cosmological constant which can be used to describe the accelerated expansion of the universe. Eq.~(\ref{eq:Einstein}) is a set of second order differential equations which can analytically only be solved if some symmetries are assumed. Because of the isotropy and homogeneity of the space time, the line element can be approximated by the Friedman-Robertson-Walker metric as \cite{springerlink:10.1007/BF01332580,1935ApJ....82..284R,Walker01011937}
\begin{equation}
 d s^2 = - c^2 d t^2 + a(t)^2 \left(\frac{d r^2}{1-k r^2} + r^2 d \Omega^2 \right) \thickspace.
\end{equation}
\(a(t)\) is the so called scale factor and \(k\) describes the curvature of the space time. Solving eq.~(\ref{eq:Einstein}) using this metric leads to the Friedman equation \cite{springerlink:10.1007/BF01332580}
\begin{equation}
\label{eq:Friedmann}
 \left( \frac{\dot{a}}{a} \right)^2 + \frac{k}{a^2} = \frac{8 \pi G_N}{3} \rho_{tot} \thickspace.
\end{equation}
\(\rho_{tot}\) is the total averaged energy density of the universe. Obviously, this equation leads to a flat universe (\(k=0\)) if the total density is equal to the critical density defined as
\begin{equation}
 \rho_{crit} = \frac{3 H^2}{8 \pi G_N} \thickspace.
\end{equation}
Here we have introduced the Hubble parameter \(H(t) = \frac{\dot{a}(t)}{a(t)}\). The measured value of the Hubble parameter is \(H_0 = \left(70.4 \pm 1.4\right) \frac{\mbox{km}}{\mbox{s Mpc}}\) \cite{Jarosik:2010iu}. It is common to normalize the values for the matter and energy in the universe with respect to the critical density. This defines the quantity \(\Omega_i\) for different species \(i\) of matter and energy as
\begin{equation}
 \Omega_i = \frac{\rho_i}{\rho_{crit}} \thickspace.
\end{equation}
\subsubsection{The freeze out of species}
\label{sec:freezeout}
Weakly interacting massive particles (WIMPs) \(\chi\) with mass \(m_{\chi}\) were kept in thermal equilibrium in the hot, early universe \cite{Srednicki:1988ce}. As long as they stayed in equilibrium, their number density \(n_{eq}\) is exponentially suppressed with decreasing temperature. When the interaction rate became smaller than the expansion of the universe, i.e.
\begin{equation}
\label{eq:freezout}
 \Gamma < H \thickspace  \thickspace ,
\end{equation}
the particle left thermal equilibrium. After freeze out, the density of the WIMPs stayed approximatively  constant in a comoving volume while the photons were diluted and redshifted. We can calculate the current WIMP density normalized to today's temperature \(T_0=2.7\)~K and photon density  \(\rho_0 \simeq 422 \frac{\mbox{photons}}{\mbox{cm}^3}\) of the CMB by calculating the decoupling temperature \(T_f\) between WIMPs and the plasma defined by eq.~(\ref{eq:freezout}).\\
The starting point for the calculation of the relic density is the Boltzmann equation 
\begin{equation}
\label{eq:Boltzmann}
L[f] = C[f] \thickspace. 
\end{equation}
Here, \(L\) is the Liouville operator describing the evolution of the universe and \(C\) is the collision operator describing the interactions of particles.  The way to rewrite this equation to the better known Lee-Weinberg formula \cite{Lee:1977ua}
\begin{equation}
 \dot{n} + 3 H n = - \bra \sigma |v| \ket  \left(n^2 - n_{eq}^2 \right)
\end{equation}
is shown in \cite{Kolb:1998ki}.
\(n\) is the density of the WIMP, \(\bra \sigma |v| \ket\) is the thermal averaged cross section and \(H\) is the Hubble parameter. By introducing the yield \(Y=\frac{n}{s}\) and  freeze out parameter \(x = \frac{m_\chi}{T_f}\) we derive finally the Riccati equation 
\begin{equation}
\frac{d Y}{d x} = -\frac{x s}{H(T=m)} \bra \sigma v \ket \left(Y^2 - Y^2_{eq}\right) \thickspace.
\end{equation}
This equation can often be solved by a non relativistic expansion in \(x\): \(\bra \sigma  v \ket \simeq a + \frac{b}{x} \). Hence, the yield \(Y_\infty\) long after the freeze out is given by \cite{Gondolo:1990dk}
\begin{equation}
\label{eq:Riccati}
Y_\infty^{-1} = \frac{\pi g_*}{45} M_P m_\chi x^{-1} \left(a+3 \frac{b}{x}\right) \thickspace .
\end{equation}
Since  the present density of \(\chi\) is given by \(\rho_\chi = m_\chi s_0 Y_\infty\), the relic density can be easily calculated by eq.~(\ref{eq:Riccati}). A good first order estimation for the relic density of a WIMP is often
\begin{equation}
\label{eq:ApproxRelic}
\Omega_{\chi} h^2 \simeq \frac{0.1 \, \mbox{pb}}{\bra \sigma v \ket} \thickspace . 
\end{equation}

\begin{figure}[t]
\centering
\includegraphics[scale=1.0]{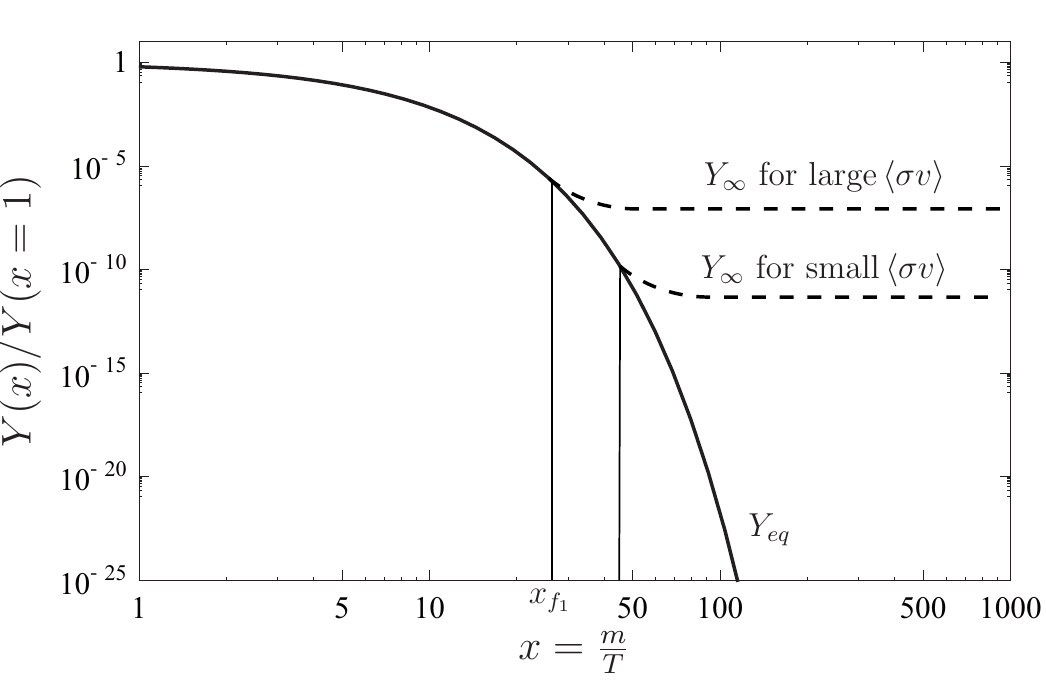}
\caption[Freeze out of species]{Dependence of the yield after freeze out $Y_\infty$ on the cross section $\langle \sigma v \rangle$. As long as the particles are in thermal equilibrium, their yield $Y_{eq}$ is exponentially decreasing with temperature $T$, but stays constant after freeze out.}
\label{fig:freeze_out}
\end{figure}

Here, we have inserted the today's entropy \(s_0 = 2970\, \mbox{cm}^{-3}\), the critical density \(\rho_c = 1.054 10^{-5}\, \frac{\GeV}{\mbox{cm}^3}\), the freeze out parameter \(x_f \simeq 20\), the number of degrees of freedom \(\sqrt{g_*} \simeq 10\) and the Hubble parameter \(h=\frac{H_0}{100}\, \frac{\mbox{km}}{\mbox{s Mpc}}\). The general dependence between the cross section and the yield is depicted in Fig.~\ref{fig:freeze_out}.  \\
If we use the approximation eq.~(\ref{eq:ApproxRelic}) for the relic density together with the bounds from CMB, we get an upper mass bound for a WIMP of
\begin{equation}
 m_{\chi} \leq 120~\mbox{TeV} \thickspace .
\end{equation}
As we will see below, this calculation doesn't hold necessarily  for all realistic scenarios: especially the gravitino is so weakly interacting that it probably has never been in thermal equilibrium as discussed in sec.~\ref{section:introdution_GravitinoDM}. Furthermore, the effects of coannihilation and resonances have been neglected so far. Both will be discussed for the case of a neutralino as dark matter in chapter~\ref{chapter:SU5} and chapter~\ref{chapter:NMSSM}. The decays of other particles after the freeze out can change the relic density of a WIMP, too, as we will see now.

\subsubsection{Late time decays}
\label{sec:LateTimeDecays}
A particle \(\Psi\) with mass \(m\) and life time \(\tau\) decays at time \(t \sim H^{-1} \sim \tau \) and at temperature \(T_D\). Here, we have neglected the freeze out time \(\tau_{FO}\), hence, the following discussion is valid for \(\tau_{FO} \ll \tau\). We assume that the energy content of the universe is dominated by \(\Psi\), i.e. the energy density of the universe is given by \(\rho \sim \rho_\Psi = s Y_i m\). The decay temperature \(T_D\) and life time \(\tau\) of \(\Psi\) are related by \cite{Kolb:1998ki}
\begin{equation}
H^2(T_D)  \sim Y_i T_D^3 \frac{m}{M_P^2}  \sim \tau^{-2} \thickspace.
\end{equation}
If the particles decay into relativistic particles, they rapidly thermalize and yield a post-decay radiation density \(\rho_R\) of
\begin{equation}
\rho_R \sim g_{\star} T^4_{RH} \thickspace.
\end{equation}
By energy conservation this radiation must be equal to the energy density of \(\Psi\) just before its demise: \(H_D^2 M_P^2\). Hence, the ratio of the entropy density before and after the decay is given by
\begin{equation}
\label{eq:entropyDecay}
\frac{s_{\text{after}}}{s_{\text{before}}} = \frac{g_\star a^3 T_{RH}^3}{g_\star a^3 T_D^3} \sim g^{1/4} \frac{Y_i m \sqrt{\tau}}{\sqrt{M_P}} \thickspace .
\end{equation}
This increment of the entropy dilutes the relic density of already frozen out particles by a factor
\begin{equation}
\label{eq:dilutionfactor}
 \Delta = 1 + \frac{4}{3} \frac{M Y}{T_D} \thickspace.
\end{equation}
Finally, the connection between the decay temperature \(T_D\) of a particle and its decay width \(\Gamma\) is given by
\begin{equation}
\label{DecayTime}
T_D = \left(g_* \frac{\pi^2}{90} \right)^{\frac{1}{4}} \sqrt{\Gamma M_P} \thickspace .
\end{equation}
This dilution is  very important for us when we try to solve the cosmological gravitino problem in chapter~\ref{chapter:GMSB}.

\subsection{Evidences for dark matter}
\label{sec:evidence}
Evidences for dark matter can be found at all astrophysical scales: from the rotation curves of galaxies, over the luminosity of galaxy clusters up to the cosmological scale of the microwave background, there are many different hints for the existence of additional matter in our universe.
\begin{figure}[ht]
\centering
\includegraphics[scale=0.55]{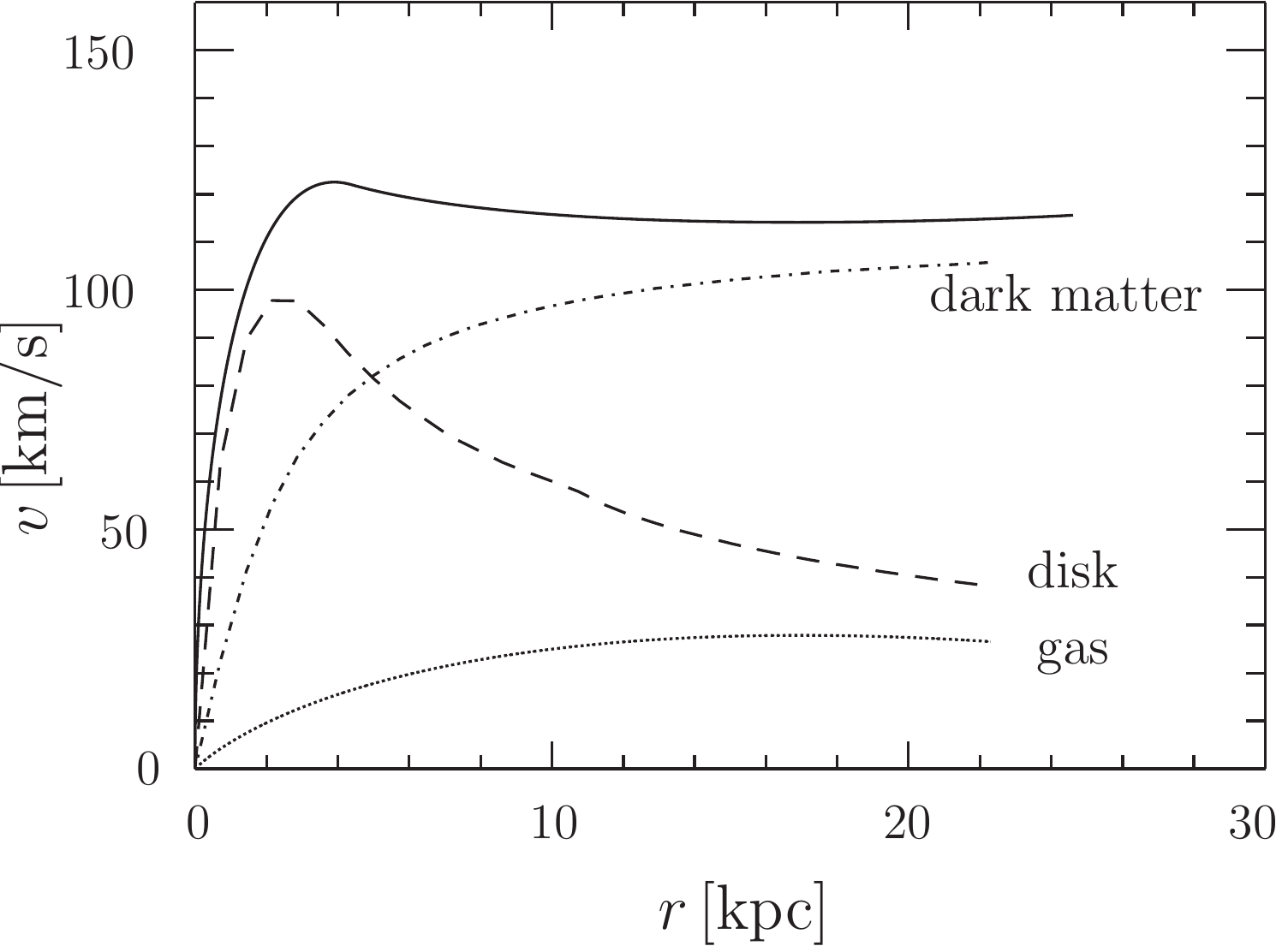}
\caption[Rotation curve of a galaxy]{Schematic picture of the rotation curve of a galaxy. The dotted line is the contribution of gas, the dashed line of the disk and the dash-dotted line of a dark matter halo. The sum of all three contributions reproduces the observed, flat curve (solid line).}
\label{fig:rotation}
\end{figure}
\paragraph*{Rotation curves} The first hint for the existence of 'invisible matter' came from the observation of the rotation curves of galaxies \cite{1937ApJ....86..217Z,1970ApJ...159..379R}. While the velocity \(v\) of stars should drop according to Newton's law like
\begin{equation}
 v(r) = \sqrt{\frac{G M(r)}{r}}
\end{equation}
as a function of the distance \(r\) to the center of the galaxy, it was observed that \(v\) is constant over a large range of \(r\).  This behavior is not an effect of general relativity and not explainable by the visible amount of matter in the galaxy without changing the physical laws. There are models trying to explain this observation with modified Newtonian dynamics (MOND) \cite{1983ApJ...270..365M}. However, such models have often severe problems explaining other observations. Another explanation for the flatness of the rotation curves are new sources of gravitational attraction in form of dark matter halo as depicted in Fig.~\ref{fig:rotation}.

\begin{figure}[!htb]
\centering
\includegraphics[scale=0.3,angle=90]{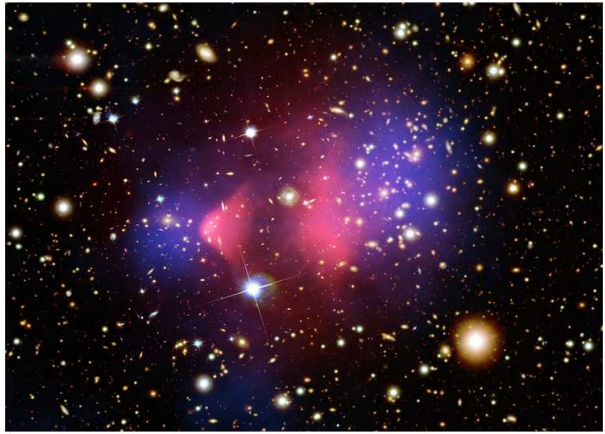}
\caption[Bullet Cluster]{Picture of the bullet cluster \cite{bulett}. The pink areas show X-ray emission observed by the Chandra X-Ray telescope. At this region the luminosity matter is  located. The blue areas are the region where most of the mass is located according to gravitational lensing. The explanation for this large separation of visible and dark matter is that dark matter can pass the collision area much faster because of its weak interaction.} 
\label{fig:bullett}
\end{figure}

\paragraph*{Cosmic microwave background} The most precise measurement of the amount of dark matter comes from the observation of the cosmic microwave background (CMB). The CMB is an echo of the decoupling of the photons from matter in the early universe. This effect was first predicted by Gamow in 1948 \cite{1948Natur.162..680G} and accidentally discovered by Penzias and Wilson 1965. While the CMB looks at first glance very isotropic with a temperature of \(T=2.726\)~K, satellite experiments have detected a distortion of these isotropy at the \(10^{-5}\) level. The anisotropies are usually parametrized by an expansion in spherical harmonics \(Y(\Theta,\Phi)\):
\begin{equation}
 \frac{\delta T}{T} = \sum_{l=2}^{+\infty}\sum_{m=-l}^{+l} a_{lm} Y_{l m}(\Theta,\Phi) \thickspace .
\end{equation}
All measurements show that the anisotropies are Gaussian-like distributed and the power spectrum can be expressed as a function of \(\frac{1}{2\pi} l(l+1)C_l\). \(C_l\) is the variance of \(a_{lm}\)
\begin{equation}
 C_l = \bra |a_{lm}|^2 \ket = \frac{1}{2 l +1} \sum_{m=-l}^{+l} |a_{lm}|^2 \thickspace.
\end{equation}

\begin{figure}[t]
\centering
\includegraphics[scale=1.0]{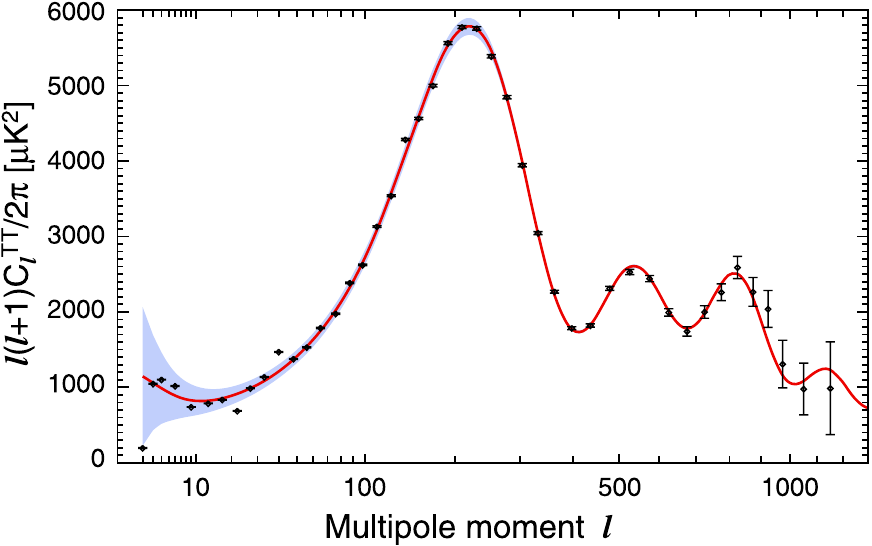}
\caption[7-year result of WMAP ]{7-year result of WMAP mission for the power spectrum of the universe \cite{Larson:2010gs}. The temperature anisotropies of the CMB are expressed as function of the multipole moment \(l\). Assuming a cosmological scenario, the best fit parameters to this curve are related to the amount of matter and energy in the universe at the time of decoupling.}
\label{fig:WMAP_5y}
\end{figure}

Fig.~\ref{fig:WMAP_5y} shows the result of the power spectrum based on the 7-year data of the Wilkinson Microwave Anisotropy Probe (WAMP). The different peaks are an effect of the different constituents of the universe building the potential of the photons at the time of decoupling. Therefore, this power spectrum can be translated into the matter and energy content of the universe. The result is normally expressed in units of \(\Omega h^2\). The best fit value for the amount of baryonic and dark matter from the 7-year data of WMAP together with results of Baryon Acoustic Oscillations (BAO) and Hubble constant measurements is \cite{Jarosik:2010iu}
\begin{equation}
 \Omega_b h^2 = 0.02260 \pm 0.00053\thickspace, \hspace{1cm} \Omega_{DM} h^2 = 0.1123 \pm 0.0035 \thickspace.
\end{equation}
Remarkably, the largest contribution to the energy density in the universe with 
\begin{equation}
\Omega_{\Lambda} = 0.728\pm 0.015
\end{equation}
comes from the so called 'Dark Energy', which is still poorly understood. 

\paragraph*{Bullet Cluster} Finally, we show a picture of the galaxy cluster 1E 0657-56, called 'the bullet cluster' in Fig.~\ref{fig:bullett}. This picture is often referred to as the first direct detection of dark matter \cite{Clowe:2006eq}: it shows the merging of two galaxies. The combination of the X-ray image taken by the Chandra X-Ray telescope  together with optical and weak lensing observation shows that the mass center of the collision doesn't coincide with the visible mass distribution. Thus, most of the matter has passed the collision area much faster. This can only happen if a high amount of matter in these galaxies is only weakly interacting.  \\ 

This was just a small extract of the clues and evidences for dark matter. We have here skipped, for instance, the results of weak lensing (see \cite{2008SSRv..134....7D} and references therein) or simulations of structure formation of the universe like the well-known millennium project \cite{2005Natur.435..629S}. 
\subsection{Constraints on dark matter}
\label{sec:DM_constraints}
There are different constraints for dark matter. All of them must be fulfilled by a particle in order to be a valid candidate. The authors of \cite{Taoso:2007qk} give a ten-point:
\begin{itemize}
 \item[ 1] Does it match the appropriate relic density?
 \item[ 2] Is it cold?
 \item[ 3] Is it neutral?
 \item[ 4] Is it consistent with BBN?
 \item[ 5] Does it leave stellar evolution unchanged?
 \item[ 6] Is it compatible with constraints on self-interactions?
 \item[ 7] It is consistent with direct dark matter searches?
 \item[ 8] It is compatible with gamma-ray constraints?
 \item[ 9] Is it compatible with other astrophysical bounds?
 \item[10] Can it be probed experimentally?
\end{itemize}
The last point is perhaps a little bit philosophical and the splitting between 8 and 9 seems arbitrary. On the other hand, the list gives an impression of the large number and big variety of constraints one has to consider when dealing with dark matter candidates. The first point was already addressed in sec.~\ref{sec:freezeout}. We discuss now the astrophysical bounds as well as the bounds from direct detection. 
\subsubsection{Astrophysical bounds}
\paragraph{Hot, cold or warm}
\label{introduction:WarmDM}
Today, it is a widely accepted fact that dark matter is directly involved in the creation of the large scale structures in the universe: when the universe became matter dominated, the density perturbations of dark matter started to grow. This caused an oscillation of the fluid of photons and baryons around the dark matter potential wells. When photons and baryons decoupled, the baryons remained trapped in the dark matter potential. Their density perturbations grew and formed the large scale structures observed today  \cite{Kolb:1998ki,Dodelson:2005tp}. \\
If the dark matter particles were still relativistic when they decoupled from the thermal equilibrium, they would have washed out small scale structures in the universe.  Particles with this property are called 'hot' \cite{Bond:1983hb}. Simulations of structure formation are not consistent with high amount of hot dark matter. The best known particle with this property is the SM neutrino and it was therefore ruled out as a dark matter candidate some time ago. Moreover, this result can be used to derive a bound on the summed mass of neutrinos of \(\sum m_\nu < 0.17\, \mbox{eV} \)  \cite{Seljak:2006qw,Fogli:2006yq}.\\
Particles which have been non-relativistic long time before their decoupling are called 'cold' and they are favored as dark matter. Simulations of an accelerating universe with cold dark matter (\(\Lambda\)CDM)  are in agreement with a wide range of observations like the galaxy-galaxy correlation functions \cite{Primack:2001ia,Green:2003un}. However, there are some observations contradicting the cold dark matter observations, the best known is the so called 'satellite problem': the number of observed dwarf galaxies of the Milky Way is much smaller than predicted by cold dark matter simulations \cite{Moore:2005jj,Klypin:1999uc}. Aside from that the density profile of cold dark matter would be a cusp while observations point to a flat distribution \cite{Flores:1994gz,McGaugh:1998tq,Gentile:2005tu, Gentile:2004tb}. \\
Those contradictions to observation in the context of cold dark matter have increased the effort to study the properties of so called 'warm dark matter', i.e. a particle with free streaming lengths in between hot and cold dark matter \cite{Bode:2000gq,SommerLarsen:1999jx}.  The mass of warm dark matter particles is limited from below by the observations of the Lyman-\(\alpha\) forest \cite{Viel:2005qj}. Current results rule out pure warm dark matter scenarios with particle masses below 8~keV for non resonantly produced dark matter \cite{Boyarsky:2008xj}. The main candidates for warm dark matter are sterile neutrinos and light gravitinos. We will study  the case of a light gravitino and the corresponding bounds from Lyman-\(\alpha\) forest in chapter~\ref{chapter:GMSB}.  \\
\paragraph{Big Bang nucleosynthesis}
\label{sec:BBN}
Big Bang nucleosynthesis (BBN) is one of the major successes of standard cosmology. It predicts to a very large accuracy the abundances of light elements in the universe. This prediction is based on simulations of chemical and nuclear processes 3~minutes after Big Bang. This complicated network of coupled Boltzmann equations can easily be perturbed, mainly through three effects. \\
First, additional energy density as consequence of the presence of new relativistic particles changes the rate of Hubble expansion in the early universe
\begin{equation}
 H = \frac{\pi T^2}{M_P} \sqrt{\frac{g_*}{90} }  \thickspace . 
\end{equation}
This leads to an earlier freeze out of certain nuclear reactions and changes the ratio of protons to neutrons. Thus, the observed abundance of \({}^4\)He and \({}^2\)H restricts the number of neutrino-like particles in the early universe to \cite{Cyburt:2004yc}
\begin{equation}
2.2 < N_\nu <4.4 \thickspace .
\end{equation}
Second, another problem could occur if there are unstable particles which decay after BBN in SM particles. This release of electromagnetic or hadronic energy could destroy the products of BBN \cite{Jedamzik:2004er,Cyburt:2002uv,Steffen:2006hw}. Especially in case of gravitino dark matter, this restricts often the life time of the next-to-lightest supersymmetric particle (NLSP): the neutralino as LSP decays dominantly in \(\FG \gamma\) and produces mostly electromagnetic energy. We can estimate this energy as
\begin{equation}
 \epsilon_{em}^{\Neu} = \frac{m_{\Neu}^2 - m^2_\FG}{2 m_{\Neu}}
\end{equation}
and the branching ratio is nearly one \(B_{em}^{\Neu} \simeq 1\). Above kinematic thresholds there are decays into gravitino and Higgs or Z boson possible which contribute to hadronic fluxes \cite{Cerdeno:2005eu}. The dominant process for a lighter stau NLSP is \(\tilde{\tau}_1 \rightarrow \FG \tau\) and releases also mostly electromagnetic energy. However, also hadronic showers in case of a stau NLSP are coming from three-body decays like \(\tilde{\tau}_1 \rightarrow \FG \tau Z\). \\
Third, there is another limit from the measured amount of \({}^6\)Li on the life time of the NLSP because catalyzed BBN via
\begin{equation}
 ({}^4 \mbox{He} X^-) + \mbox{D} \rightarrow {}^6 \mbox{Li} + X^-
\end{equation}
could result in an overproduction of light elements. For a detailed discussion how BBN could constrain the parameter space in case of gravitino  dark matter with masses in the GeV range see  \cite{Steffen:2008qp}. 
\paragraph{Other astrophysical constraints}
We just mention other astrophysical constraints and refer for details to literature. Constraints from stellar evolution come from the possible production of light and weakly interacting particles in the hot plasma of stars. These particles can escape without further interactions and cause a large energy loss of the stars \cite{Raffelt:1999tx,Raffelt:2006cw}. This leads especially for axionic dark matter to strong constraints on the Peccei-Quinn scale. \\
If dark matter particles annihilate, this might lead to characteristic signals in \(\gamma\)-ray spectra \cite{Bertone:2006nw}. Those signals haven't been observed so far even though there are some claims. For Instance, HESS and MAGIC have discovered a very high energy source spatially coincident with Sgr A*. However, the shape of the energy spectrum is close to a perfect power-law over two decades in energy and points towards ordinary astrophysical sources \cite{Aharonian:2006wh,Profumo:2005xd}. \\
The most severe bounds on self-interaction of dark matter are resulting from the bullet cluster observations described in sec.~\ref{sec:evidence}. It is obvious that parts of the halo are ahead the interaction area. That won't be possible if the self-interaction of dark matter is too large \cite{1995ApJ...444..532T,Markevitch:2003at}.      
\subsubsection{Direct search for dark matter}
The underlying idea of the direct search for dark matter is quite simple: the expected flux of dark matter on the earth is 
\begin{equation}
 \Phi \simeq 10^5 \left( \frac{100 \GeV}{m_{\chi}}\right) \frac{1}{\mbox{cm}^2 \mbox{s}} \thickspace ,
\end{equation}
if we assume a mean velocity of \(\bar{v} = 220\,\frac{\mbox{km}}{\mbox{s}}\) and a local density of \(\rho \simeq 0.3\, \frac{\GeV}{\mbox{cm}^3}\). Even if the dark matter particle is weakly interacting, there should be interactions with electrons and nuclei of atoms. There are many experiments searching for the recoil energy of a WIMP scattered at the nuclei using different material and different detector types. The main difficulty of these experiments is to suppress the background. The approximated rate \(R\) for an interaction between nucleus and WIMP is given by
\begin{equation}
 R \simeq \sum_i N_i n_\chi \bra \sigma_{i \chi} \ket \thickspace. 
\end{equation}

\begin{figure}[t]
\centering
\includegraphics[scale=0.55]{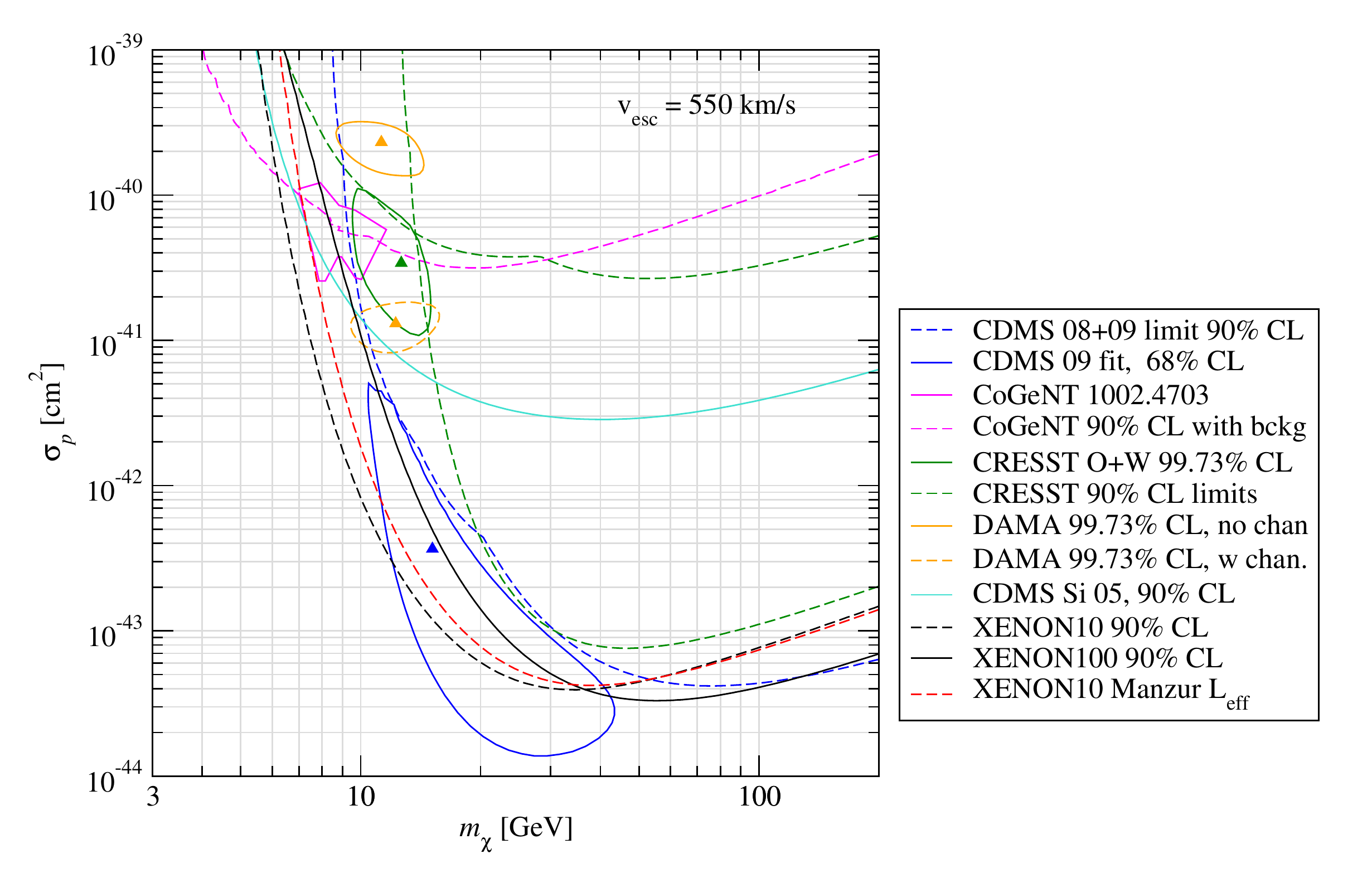}
\caption[Dark matter direct detection bounds]{Combined plot for several bounds on spin-independent WIMP-nucleon cross section: XENON10 \cite{Angle:2008we,Aprile:2010bt}, first results of XENON100 \cite{Aprile:2010um}, CRESST \cite{Altmann:2001ax,Seidel}, CoGeNT \cite{Aalseth:2010vx} and CDMS \cite{Akerib:2005zy}. Also the preferred regions of DAMA/Libra \cite{Bernabei:2008yi,Savage:2008er}, CoGeNT \cite{Aalseth:2010vx}, CDMS \cite{Ahmed:2009zw,Ahmed:2009rh} and CRESST \cite{Seidel}  are shown. Plot taken from \cite{schwetz,Kopp:2009qt}.}
\label{fig:DirectSearch}
\end{figure}

\(i\) is running over the nuclei species in the detector and \(N_i\) is the number of target nuclei of these species in the detector. \(n_\chi\) is the ratio of the WIMP energy density to its mass and \(\sigma_{i \chi}\) is the cross section of the WIMP with the nucleus of a certain species. \\
The types of scatterings are divided into elastic and inelastic or spin-dependent and spin-independent. Elastic scattering takes place between the WIMP and the nucleus resulting in a recoil energy while inelastic scattering is between the WIMP and an electron of the atom. This leads not to a recoil energy, but to an excitation or ionization of the atom. Spin-dependent scatterings are proportional to \(J(J+1)\) where \(J\) is the spin of the nucleus and depend barely on the mass of the target. In contrast, spin-independent cross sections increase dramatically with increasing mass of the target.\\
Current experiments are usually dominated by the spin-independent cross section. All of them but DAMA/Libra and CoGeNT have not yet reported a signal. DAMA claims that it has seen an annular modulation with 8.9\(\sigma\) confident level  \cite{Bernabei:2006mx}. CoGeNT has recently measured some events not immediately identifiable with background  \cite{Aalseth:2010vx}. In addition, CDMS has seen two events, but with very low statistical significance \cite{Ahmed:2009rh}. Also CRESST showed preliminary results with some event-candidates \cite{Seidel}. These experiments have in common that they point to quite low WIMP masses of \(\Ord(10)\)~GeV. These results are heavily discussed: they contradict for many beyond SM models the null results of the other experiments, especially the new bounds from XENON100 \cite{2009arXiv0902.4253A}. We give in Fig.~\ref{fig:DirectSearch} an overview of the upper limits derived for the mass and interaction of dark matter from the different experiments and show the preferred regions according to the signals. If the measured signals are really dark matter, they are hardly explainable in the standard MSSM \cite{Vasquez:2010ru}.

\subsection{Dark matter candidates}
In the last years many candidates for dark matter have been proposed with a large variation of their properties. Some of them already ruled out, others are  still in good shape. Besides SUSY, especially models with extra dimensions or additional gauge symmetries provide particles that have the correct properties and are well studied (see \cite{Hooper:2007qk,Perelstein:2005ka} and references therein). In \cite{Taoso:2007qk} several candidates have been probed with respect to their ten-point test. The result is given in Table~\ref{dark_matter_test}. We will focus in this work on the candidates arising in SUSY models.  
\begin{table}[t]
\begin{center}
{\tiny
\begin{tabular}{|p{3.8 cm}||*{11}{c|}|}
\hline
 & {\bf I.} & {\bf II.} & {\bf III.} & {\bf IV.} & {\bf V.} & {\bf VI.} & {\bf VII.} & {\bf VIII.} & {\bf IX.} & {\bf X.}& {\bf Result}\\
 {\bf {\it DM candidate}}  &   $\Omega h^2$   & Cold & Neutral & BBN & Stars & Self & Direct & $\gamma$-rays & Astro & Probed &  \\
                  \hline
\hline

SM Neutrinos & $\times$ & $\times$ & \checkmark & \checkmark & \checkmark & \checkmark & \checkmark & -- & -- & \checkmark &$\times$ \\
\hline
{\bf Sterile Neutrinos} & $\sim$ & $\sim$ & \checkmark & \checkmark & \checkmark & \checkmark & \checkmark & \checkmark & \checkmark ! & \checkmark & $\sim$ \\
\hline
{\bf Neutralino} & \checkmark & \checkmark & \checkmark & \checkmark & \checkmark & \checkmark & \checkmark ! & \checkmark ! & \checkmark ! & \checkmark & \checkmark \\
\hline
{\bf Gravitino} & \checkmark & \checkmark & \checkmark & $\sim$ & \checkmark & \checkmark & \checkmark & \checkmark & \checkmark & \checkmark & $\sim$ \\
\hline
{\bf Gravitino (broken \(R\)-parity)} & \checkmark & \checkmark & \checkmark & \checkmark & \checkmark & \checkmark & \checkmark & \checkmark & \checkmark & \checkmark & \checkmark \\
\hline
Sneutrino $\tilde{\nu}_{L}$ & $\sim$ & \checkmark & \checkmark & \checkmark & \checkmark & \checkmark & $\times$ &\checkmark  ! & \checkmark ! & \checkmark & $\times$ \\
\hline
{\bf Sneutrino} $\tilde{\nu}_{R}$ & \checkmark & \checkmark & \checkmark& \checkmark & \checkmark & \checkmark & \checkmark ! & \checkmark ! &\checkmark  ! & \checkmark & \checkmark \\
\hline
{\bf Axino} & \checkmark & \checkmark & \checkmark & \checkmark & \checkmark & \checkmark & \checkmark & \checkmark & \checkmark & \checkmark & \checkmark \\
\hline
{\bf SUSY Q-balls} & \checkmark & \checkmark & \checkmark & \checkmark & $\sim$& -- & \checkmark ! & \checkmark & \checkmark & \checkmark  & $\sim$\\
\hline
{\bf $B^1$ UED} & \checkmark & \checkmark & \checkmark & \checkmark & \checkmark & \checkmark & \checkmark ! & \checkmark !& \checkmark ! & \checkmark & \checkmark \\
\hline
First level graviton UED & \checkmark & \checkmark & \checkmark & \checkmark & \checkmark & \checkmark & \checkmark & $\times$ & $\times$ & \checkmark & $\times$$^a$
\\
\hline
{\bf Axion} & \checkmark & \checkmark & \checkmark & \checkmark & \checkmark & \checkmark & \checkmark ! & \checkmark & \checkmark & \checkmark & \checkmark \\
\hline
{\bf Heavy photon (Little Higgs)} & \checkmark & \checkmark & \checkmark & \checkmark & \checkmark & \checkmark & \checkmark & \checkmark !& \checkmark ! & \checkmark & \checkmark \\
\hline
{\bf Inert Higgs model} & \checkmark & \checkmark & \checkmark & \checkmark & \checkmark & \checkmark & \checkmark & \checkmark ! & -- & \checkmark & \checkmark \\
\hline
Champs & \checkmark & \checkmark & $\times$ & \checkmark & $\times$ & -- & -- & -- & -- & \checkmark & $\times$ \\
\hline
{\bf Wimpzillas} & \checkmark & \checkmark & \checkmark & \checkmark & \checkmark & \checkmark & \checkmark & \checkmark & \checkmark & $\sim$& $\sim$ \\
\hline
\end{tabular}
}
\end{center}
\caption[Check of dark matter candidates]{Check for different dark matter candidates if they fulfill all possible constraints. The \checkmark is used if the bounds are so far satisfied. A ! is added if upcoming data can probe significant areas of the parameter space. The bounds marked with $\sim$ are only satisfied in non-standard or unnatural scenarios while the bounds with $\times$ contradict the nature of the candidate. Taken from \cite{Taoso:2007qk}.}
\label{dark_matter_test}
\end{table}
\section{Supersymmetric dark matter}
\label{section:SUSY_DM}
SUSY provides a large variety of dark matter candidates. The most known and best examined one is the neutralino in case of the MSSM. It can be the LSP for mSugra scenarios and has all properties of a WIMP: its mass is of \(\Ord(100)\)~GeV and it is a weakly interacting particle (see sec.~\ref{section:neutralino_DM}). Another possible LSP in mSugra scenarios is the gravitino. In addition, the gravitino is always the LSP in GMSB models. The properties of the gravitino can vary a lot. The mass ranges from some eV to several hundred GeV. The interaction is purely gravitational, but can be enhanced depending on its mass by several orders (see sec.~\ref{section:introdution_GravitinoDM}). Other SUSY candidates for dark matter are sterile sneutrinos while the left-chiral sneutrinos in the correct mass range are already ruled out by direct detection. Heavier, weakly charged sneutrino escaping direct detection would annihilate too efficient to contribute significantly to dark matter \cite{Falk:1994es,Lee:2007mt}. Furthermore, if the Peccei-Quinn solution to the strong CP-problem is assumed, two more candidates are present: the pseudo scalar Axion and its superpartner, the Axino \cite{Covi:2009pq}. The properties of those particles can vary a lot depending on the underlying model. In the subsequent, we will discuss the two main candidates, the neutralino and the gravitino in the MSSM, and discuss afterwards the necessity to study these candidates in beyond-MSSM models. 
\subsection{Neutralino dark matter}
\label{section:neutralino_DM}
\begin{figure}[t]
\centering
\includegraphics[scale=0.66]{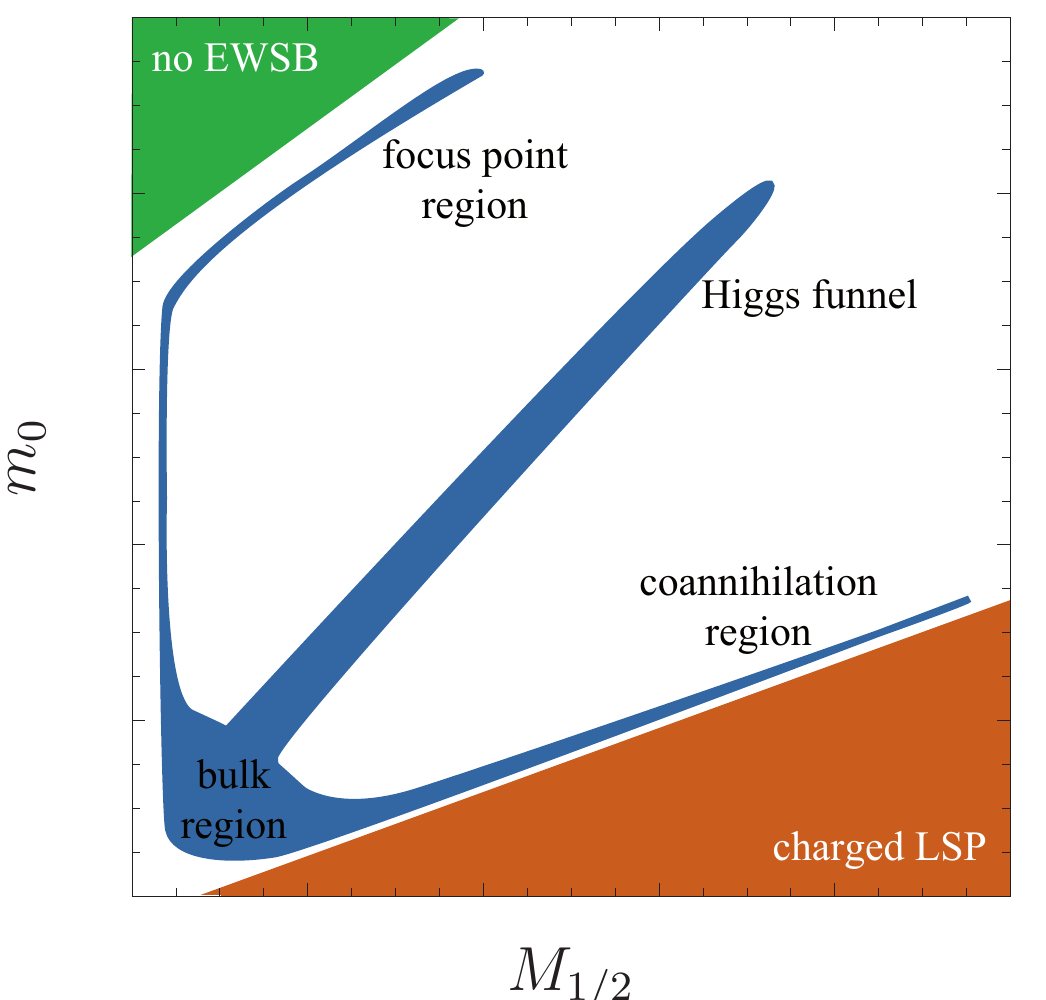}
\caption[Neutralino relic density]{Schematic picture of the allowed regions in the $(m_0,M_{1/2})$-plane for neutralino dark matter (blue area). In the bulk region the sleptons are light, in the coannihilation region the mass of the NLSP is close to the mass of the LSP, in the focus point region the Higgsino component of the lightest neutralino is enhanced and in the Higgs funnel the mass of a Higgs is close to twice the mass of the LSP.}
\label{fig:m0m12}
\end{figure}

The main candidate for dark matter in SUSY models is still the neutralino. As shown in sec.~\ref{section:masseigenstates}, the neutralino is a mixture of the neutral Higgsinos as well as the neutral gauginos after EWSB. The neutralino is, for example, the LSP in mSugra models and its mass is normally of \(\Ord(100)\)~GeV. This classifies the neutralino as a standard WIMP \cite{Jungman:1995df,Primack:1988zm,Rich:1987jd}. Nevertheless, in the minimal scenario of supergravity, the relic density of neutralinos is often larger than the upper bounds of WAMP. If we fix the parameters \(A_0\), \(\tan\beta\) and \(\sign\mu\) of mSugra, there are just tiny regions in the \((m_0,M_{1/2})\)-plane left which are consistent with the measured amount of dark matter. These regions are shown in Fig.~\ref{fig:m0m12} and can be categorized by the different effects which lead to sufficient annihilation \cite{Griest:1990kh}. 

\paragraph*{Bulk region} This region is characterized by light sleptons. Two neutralino can therefore annihilate sufficiently via a t-channel  exchange of a sfermion into two leptons:
\begin{equation}
 \Neu_1 \Neu_1 \rightarrow l^+ l^- \thickspace.
\end{equation}

\paragraph*{Focus point region} In this region the Higgsino part of the neutralino is in the percent range. This can lead to an efficient annihilation into vector bosons if kinematically allowed
\begin{equation}
 \Neu_1 \Neu_1 \rightarrow W^+ W^-, \hspace{1cm}  \Neu_1 \Neu_1 \rightarrow Z Z \thickspace .
\end{equation}
For lighter neutralinos, the s-channel annihilation via a \(Z\) boson into two fermions dominates
\begin{equation}
 \Neu_1 \Neu_1 \rightarrow Z \rightarrow l^+ l^- \thickspace , 
\end{equation}
while for neutralinos heavier than the top quark \(t\), the dominant process is
\begin{equation}
 \Neu_1 \Neu_1 \rightarrow t \bar{t}  \thickspace .
\end{equation}

\paragraph*{Coannihilation region} In this region,  the mass of the NLSP, normally a stau or chargino, is very close to the mass of the lightest neutralino \cite{Edsjo:1997bg}. This allows an efficient coannihilation of a neutralino and chargino into two SM fermions
\begin{equation}
 \Neu_1 \Cha_1 \rightarrow l^- \nu \thickspace .
\end{equation}
In the case of a stau NLSP, the final states are normally a SM fermion together with a vector boson or Higgs 
\begin{equation}
 \Neu_1 \tilde{\tau}_1 \rightarrow l^- \gamma/Z \thickspace, \hspace{1cm} \Neu_1 \tilde{\tau}_1 \rightarrow l^- h_i
\end{equation}

\paragraph*{Higgs funnel} This region is characterized by a pseudo scalar Higgs mass very close to twice the neutralino mass. Because of this resonance, the cross section of two neutralinos into two \(b\)~quarks is very large
\begin{equation}
 \Neu_1 \Neu_1 \rightarrow A^0 \rightarrow b \bar{b} \thickspace .
\end{equation}

\subsection{Gravitino dark matter}
\label{section:introdution_GravitinoDM}
As already mentioned, the mass \(m_{3/2}\) of the gravitino \(\FG\) can lie in the range of a fraction of an eV up to several TeV. In GMSB scenarios,  the mass is normally below 1 GeV while  in scenarios involving gravity it is in the GeV up to TeV range. The cosmological constraints for the gravitino depend on its mass, its life time and the reheating temperature \(T_R\). Since the gravitino is a singlet under the gauge groups of the SM and interacts purely gravitationally, all interactions are suppressed by the Planck scale \(M_P\). For light gravitinos those interactions are enhanced by the Super-Higgs-mechanism.\\
A light gravitino  gives severe restrictions to the mechanism of SUSY breaking. The reason is the existence of a critical energy scale \(E_{cr}\) at which unitarity gets broken \cite{Bhattacharya:1988zp,Bhattacharya:1988nk}
\begin{equation}
E_{cr} = 12 \sqrt{2 \pi} \left(\left(\frac{m_{\tilde{g}}}{m_{3/2}}\right)^2 - \frac{6}{5} \right)^{-\frac{1}{2}} M_P  \thickspace .
\end{equation}
This implies for scenarios with SUSY breaking transmitted by gravitational interactions an upper bound of \(\frac{m_{\tilde{g}}}{m_{3/2}} \simeq 21\) because \(E_{cr} > M_P\) must hold. 

\paragraph*{Very light gravitino}
First, the interactions of the longitudinal \(s_z = \pm \halb\) components of the gravitino is enhanced by a factor \(\frac{M_s}{m_{3/2}}\). \(M_s\) is of the order of the sparticles in the visible sector. In this case, the interactions of the spin-\(\frac{3}{2}\) components are negligible. The very light gravitino decouples at a temperature \( T_{3/2} \gg m_{3/2} \), so it was still relativistic. If the mass of the gravitino is very small, i.e. less then 1 eV, the contribution to the matter density of the universe is negligible. However, they were still relativistic at the time of nucleosynthesis. Hence, the Hubble parameter \( H = \frac{\pi T^2}{M_P} \sqrt{\frac{g_*}{90} } \) was increased at this time by the additional degrees of freedom. This leads to severe constraints from BBN observations as described in sec.~\ref{sec:BBN}. It has been shown that relativistic gravitinos are safe as long as they decouple before muons did \cite{Drees:2004jm}.

\paragraph*{Light gravitino}
Second, we calculate the relic density for a light gravitino, which is favored in GMSB scenarios \cite{Pagels:1981ke}: the gravitinos were in thermal equilibrium  in the early universe and decoupled at a temperature \(T_{3/2} \gg m_{3/2}\). Thus, they were still relativistic and their number density obeys
\begin{equation}
 n_{3/2} = \frac{3}{4} \frac{\zeta(3)}{\pi^2} g_*(T_{3/2}) T^3 \thickspace,
\end{equation}
while the number density for photons is
\begin{equation} 
\label{dof_gamma}
n_{\gamma} = \frac{\zeta(3)}{\pi^2} g_*(T) T^3 \thickspace.
\end{equation}
Particles with masses \(m < T_{3/2}\) contribute until their decoupling to the effective degrees of freedom of the photon, but not to those of the gravitino. Hence, the today's ratio of gravitinos to photons is
\begin{equation}
 \frac{n_{3/2}}{n_\gamma} = \frac{3}{4} \frac{g_*(T_0)}{g_*(T_{3/2})} \thickspace.
\end{equation}
With \(\Omega h^2 = \frac{\rho}{\rho_c} h^2 \) we obtain
\begin{equation}
 \Omega h^2 = \frac{ m_{3/2} n_{3/2}(T_0)}{\rho_c} h^2 = \frac{3 m_{3/2} n_{\gamma} g_*(T_0)}{4 g_*(T_{3/2}) \rho_c} h^2 \thickspace.
\end{equation}
Using eq.~(\ref{dof_gamma}), \(T_0 = 2.3 \cdot 10^{-12} \thinspace \mbox{GeV}\) and \(\frac{\rho_c}{h^2} = 8.1 \cdot 10^{-47} \thinspace \mbox{GeV}^4\), we have finally a compact expression for the relic density of light gravitinos
\begin{equation}
\label{eq:Omega32}
 \Omega_{3/2} h^2 \simeq \frac{m_{3/2}}{0.85 \, \mbox{keV}} \frac{100}{g_*(T_{3/2})} \thickspace .
\end{equation} 
Neglecting the electroweak gauge interactions we obtain 
\begin{equation}
 T_{3/2} = 0.62 \frac{m_{3/2}^2 M_P \sqrt{g_*}}{\alpha_s M_SG^2}
\end{equation}
as freeze out temperature for the gravitinos which have once been in thermal equilibrium. Gravitinos might  have never been in thermal equilibrium after inflation if the reheating temperature was too low. In Big Bang scenarios with inflation any initial population of gravitinos was diluted by the exponential expansion of the universe. After inflation, there was a thermal production of gravitinos depending on the reheating temperature \(T_R\) which can be approximated by \cite{Moroi:1993mb}
\begin{equation}
 \Omega_{3/2}^{TP} h^2 = 0.32 \left(\frac{10\,\GeV}{m_{3/2}} \right) \left(\frac{M_{1/2}}{1\, \TeV}\right)^2 \left(\frac{T_R}{10^8 \, \GeV}\right) \thickspace.
\end{equation}
Moreover, there was a non-thermal production by the decays of the NLSP of
\begin{equation}
 \Omega_{3/2}^{NTP} h^2 = m_{3/2} \, Y^{dec}_{NLSP} \, s(T_0) \, \frac{h^2}{\rho_c} \thickspace.
\end{equation}

\paragraph*{Heavy gravitino}
Finally, in case of a heavy gravitino, the longitudinal component is not enhanced by the Super-Higgs mechanism and the interactions of the \(\frac{3}{2}\) components are dominant. The most important processes for calculating the gravitino production cross section are 2 \(\rightarrow\) 2 reactions involving gauge supermultiplets. Since the decays of sparticles into gravitinos are negligible, we receive for the relic density 
\begin{equation}
 \Omega_{3/2} h^2 = \frac{m_{3/2}}{100\, \GeV} \frac{T_R}{10^{11}\, \GeV} \thickspace.
\end{equation}
The case of gravitino dark matter with \(m_{3/2}\) of \(\Ord(100)\, \GeV\) is elaborately studied in literature, e.g.
\cite{Steffen:2005cn, Pradler:2006hh,Steffen:2006hw}. We will concentrate in chapter~\ref{chapter:GMSB} on light gravitinos arising in GMSB with the relic density given by eq.~(\ref{eq:Omega32}).

\subsection{SUSY dark matter beyond the MSSM}
The features of dark matter build up by a gravitino or neutralino are very well explored today, but most studies were done in the framework of the MSSM. On the other hand, even if SUSY will be discovered at the LHC, it will most likely not be a pure MSSM. Although the MSSM suffers from intrinsic less  problems than the SM does, there are still open questions: how do the neutrinos get their masses and what's about the \(\mu\)-problem? If we also want to answer these questions, we have to go beyond the MSSM. Either we have to break \(R\)-parity or to introduce new particles to generate effective neutrino mass terms. The \(\mu\)-term in the superpotential might be generated in a dynamic way like in the NMSSM. All of these extensions of the MSSM have significant influence on the properties of the dark matter particle:\\
In the NMSSM, the neutralino is still the main candidate for dark matter but its properties can be quite different due to the contributions of a gauge singlet.  In seesaw scenarios heavy particles are added to the spectrum.  These fields can not only generate a dimension 5 operator to explain neutrino data but they lead also to changed properties of the neutralino as dark matter candidate because of the different evolution of the RGEs. In the case of \(R\)-parity violation, light gravitinos are often the only remaining candidate for dark matter in SUSY because of their long life time. However, their small mass leads normally to a thermal production of gravitinos which overcloses the universe. \\

The focus of this work is to determine the viability of SUSY dark matter beyond the pure MSSM case. For this purpose we have developed \SARAH, a tool for the efficient analysis of new supersymmetric models, which will be presented in the next chapter. Afterwards in chapter~\ref{chapter:GMSB}, we present the results for gravitino dark matter in GMSB scenarios with and without broken \(R\)-parity. In chapter~\ref{chapter:SU5}, we discuss neutralino dark matter and  the impact  of different high scale extensions of the MSSM which provide neutrino masses by miscellaneous seesaw scenarios.  Finally, we study neutralino dark matter in the NMSSM in chapter~\ref{chapter:NMSSM}.  In chapter~\ref{chapter:summary}, we summarize our results. 
\nomenclature[m32]{$m_{3/2}$}{Gravitino Mass} 
\nomenclature[T32]{$T_{3/2}$}{Freeze out temperature of gravitino} 
\nomenclature[Y]{$Y$}{Yield} 
\nomenclature[n]{$n$}{Number density} 
\nomenclature[TRH]{$T_{RH}$}{Reheating temperature} 
\nomenclature[g1g2g3]{$g_1,g_2,g_3$}{Standard Model gauge coupling constants} 
\nomenclature[eps]{$\epsilon_{ij}$}{Entries of Levi-Civita Tensor $\epsilon=\left( \begin{array}{cc} 0 & 1 \\ -1 & 0 \end{array} \right)$}  \nomenclature[Yi]{$Y_e,Y_d,Y_u$}{Yukawa couplings}  
\nomenclature[R]{$R_{\mu\nu}$}{Ricci Tensor} 
\nomenclature[R]{$R$}{Ricci Scalar} 
\nomenclature[gmu]{$g_{\mu\nu}$}{Minkowski Metrik, $diag(g) = (1,-1,-1,-1)$} 
\nomenclature[R]{$T_{\mu\nu}$}{Energy Momentum Tensor} 
\nomenclature[G]{$G_N$}{Newton's constant} 
\nomenclature[Lambda]{$\Lambda_{\mu\nu}$}{Cosmological constant} 
\nomenclature[Ti]{$T_e,T_d,T_u$}{Trilinear softbreaking couplings}
\nomenclature[m0]{$m_0$}{Universal scalar mass parameter} 
\nomenclature[M12]{$M_{1/2}$}{Universal gaugino mass parameter} 
\nomenclature[s0]{$s_0$}{Today's entropy density $s_0=2970\, \mbox{cm}^{-3}$} 
\nomenclature[rhoC]{$\rho_C$}{Critical density $\rho_C = 1.054 10^{-5}\, \frac{\GeV}{\mbox{cm}^3}$} 
\nomenclature[gS]{$g_*$}{effective degrees of freedom}
\nomenclature[s]{$s$}{Entropy density} 
\nomenclature[H]{$H$}{Hubble Constant $H=\left(73 \pm 3\right) \frac{\mbox{km}}{\mbox{s Mpc}}$}     
\nomenclature[Lambda]{$\lambda_a$}{Gell-Mann matrices} 
\nomenclature[Sigma]{$\sigma_a$}{Pauli matrices} 
\nomenclature[Ta]{$T^a$}{Generators of  $SU(N)$ gauge group} 
\nomenclature[fijk]{$f^{ijk}$}{Structure constants of $SU(N)$ gauge group}
\nomenclature[mSugra]{mSugra}{minimal Supergravity}%
\nomenclature[Sugra]{Sugra}{Supergravity}%
\nomenclature[GMSB]{GMSB}{Gauge Mediated SUSY Breaking}
\nomenclature[DR]{DR}{Dimensional Reduction}
\nomenclature[DRb]{$\overline{\mbox{DR}}$}{modified Dimensional Reduction}
\nomenclature[CMSSM]{CMSSM}{Constrained Minimal Supersymmetric Standard Model}
\nomenclature[FCNC]{FCNC}{Flavor changing neutral current}%
\nomenclature[GIM]{GIM}{Glashow-Iliopoulos-Maiani (Mechanism)}
\nomenclature[CKM]{CKM}{Cabibbo-Kobayashi-Maskawa (Matrix)}%
\nomenclature[Rp]{$R_P$}{$R$-parity}
\nomenclature[RpV]{$\slashed{R}_P$}{$R$-parity Violation}
\nomenclature[STr]{$STr$}{Supertrace}    
\nomenclature[SM]{SM}{Standard Model}
\nomenclature[LHC]{LHC}{Large Hadron Collider}
\nomenclature[DM]{DM}{Dark Matter} 
\nomenclature[SUSY]{SUSY}{Supersymmetry}
\nomenclature[MSSM]{MSSM}{Minimal Supersymmetric Standard Model}
\nomenclature[GUT]{GUT}{Grand Unified Theory}
 \nomenclature[WMAP]{WMAP}{Wilkinson Microwave Anisotropy Probe}
 \nomenclature[Q]{$Q$}{SUSY generator}
\nomenclature[RGE]{RGE}{Renormalization Group Equations} 
  \nomenclature[EWSB]{EWSB}{Electroweak Symmetry Breaking} 
  \nomenclature[W]{$W$}{Superpotential}
\nomenclature[NMSSM]{NMSSM}{Next-to-minimal supersymmetric standard model}
\nomenclature[VEV]{VEV}{Vacuum Expectation Value}
\nomenclature[LSP]{LSP}{Lightest Supersymmetric Particles}
   \nomenclature[OPE]{OPE}{Operator Product Expansion}
\nomenclature[C]{$C_7,C_7^{'}$}{Wilson Coefficients}
\nomenclature[HFAG]{HFAG}{Heavy Flavor Averaging Group} 
\nomenclature[CMB]{CMB}{Cosmic Microwave Background}
 \nomenclature[BBN]{BBN}{Big Bang Nucleosynthesis}
 \nomenclature[Omega]{$\Omega_i$}{Relic density of species $i$}
\nomenclature[WIMP]{WIMP}{Weakly Interacting Massive Particle} 
\nomenclature[MP]{$M_P$}{Planck Constant}
\nomenclature[NSLP]{NLSP}{Next-to-Lightest SUSY Particle}
\cleardoublepage
\chapter{\SARAH: A Tool for the Analysis of SUSY Models}
The way from the idea about a supersymmetric model to the first phenomenological results is usually elaborative and time consuming. One reason is that even in the smallest possible extension of the SM, the MSSM, the particle content is more than doubled and it provides a lot of new interactions which are related due to SUSY. Thus, the calculations of all necessary masses, vertices or renormalization group equations are very laborious. A more technical reason is that the implementation of new models in existing software tools demands some good knowledge about these programs. Of course, even with this knowledge writing and testing the new parts needs time. This was just a small excerpt of all  necessary steps needed for a comprehensive analysis of a new SUSY model. A complete list might at least incorporate the following tasks:
\begin{enumerate}
 \item {\bf The idea} The starting point of each new supersymmetric model is usually the particle content, the gauge structure and the superpotential. All further steps are needed to extract all important properties from this underlying information.
 \item {\bf Group properties} If chiral superfields are present which are not in the fundamental representation, it is necessary to calculate the corresponding generators. Moreover, the quadratic Casimir and the Dynkin index are needed for loop calculations.
 \item {\bf Gauge anomalies} Before more work is put into the analysis of a model, it should be checked that it is well defined at quantum level, i.e. free from gauge anomalies \cite{Adler:1969gk}.
 \item {\bf Calculating the Lagrangian} If the model passes the first check, the Lagrangian is needed. Although there are simple rules for calculating the complete Lagrangian of a supersymmetric model either using the superspace approach of sec.~\ref{sec:superspace}  or the method explained in app.~\ref{sec:SARAH_Lag}, the expressions are getting quickly very long. 
 \item {\bf Breaking of symmetries and rotations of particles} Normally, all phenomenological relevant models involve at least one breaking of  the fundamental gauge symmetries. This leads to a rotation of matter and gauge fields and the Lagrangian has to be rewritten with respect to the new basis.
 \item {\bf Masses, vertices and tadpole equations} The first information which can be extracted directly from the Lagrangian are the tree level relations for the masses and the minimum conditions of the vacuum. Furthermore, the interactions between the different fields can be derived.
 \item {\bf Renormalization group equations} The RGEs are needed to connect the values of parameters of the model at different energy scales. For instance, the embedding of the SUSY model in a more fundamental theory is often considered. This theory is defined at a high scale and RGEs have to be evaluated down to the SUSY scale. The RGEs at a certain loop level can be derived in a diagrammatic approach or with help of some generic formulas. 
 \item {\bf Loop corrections} For producing reliable results, the tree level masses are often not sufficient. For this reason the self-energies are needed to get the radiative correction to the masses.
 \item {\bf Calculating the mass spectrum} When all analytical formulas are calculated, they have to be used for numerical calculations of all masses and couplings of the model. At the moment, there exist some spectrum calculators for the MSSM and its extensions with the same particle content at the SUSY scale (\SPheno \cite{Porod:2003um}, {\tt Suspect} \cite{Djouadi:2002ze}, {\tt IsaJet} \cite{Baer:1993ae} and {\tt SoftSusy} \cite{Allanach:2001kg}). For the NMSSM only one spectrum calculator ({\tt NMSSM-Tools} \cite{Ellwanger:2006rn}) exists. All of these tools have in common that the supported models are hard coded. Hence, adding a new model to these programs demands a very good knowledge about the source code and a lot of work. 
 \item {\bf Model files for diagram calculators} In order to use the known programs for calculating Feynman diagrams like {\tt FeynArts/FormCalc}  \cite{Hahn:2000kx,Hahn:2010zi} or {\tt CalcHep/CompHep}  \cite{Pukhov:1999gg} with a new model, the corresponding model files have to be created. The peculiarities of the different programs make this step often tricky.
 \item {\bf Relic density calculation}: Dark matter puts very severe bounds on the allowed parameter space of supersymmetric models. There are some standard programs at the market to calculate the relic density, e.g. \micrOmegas  \cite{ Belanger:2006is} or {\tt DarkSusy} \cite{Gondolo:2000ee}. Especially \micrOmegas is not restricted to neutralino dark matter and also new models can be implemented without changing the source code by using \CalcHep model files. 
 \item {\bf Low energy constraints} There are often severe bounds on the parameter space of a new model stemming from precision data. Repeating all the calculations to check these constraints for a new model is a tedious challenge. 
 \item {\bf Monte-Carlo simulation} The last step of the analysis of a new model is usually to study the collider phenomenology. For that purpose, different Monte-Carlo (MC) \nomenclature[MC]{MC}{Monte-Carlo} generators like {\tt WHIZARD} \cite{Kilian:2007gr} or {\tt Phytia} \cite{Bengtsson:1986rz} exist. Of course, for a new model they must be fed with all necessary information.
\end{enumerate}
One of the main topics of this work was to provide a tool which can automatize the above mentioned steps as much as possible. The result is the Mathematica package \SARAH \cite{Staub:2008uz,Staub:2009bi,Staub:2010jh}. 
\section{Introduction to \SARAH}
\begin{figure}
\vspace{-2cm}
\centering
\includegraphics[scale=0.75]{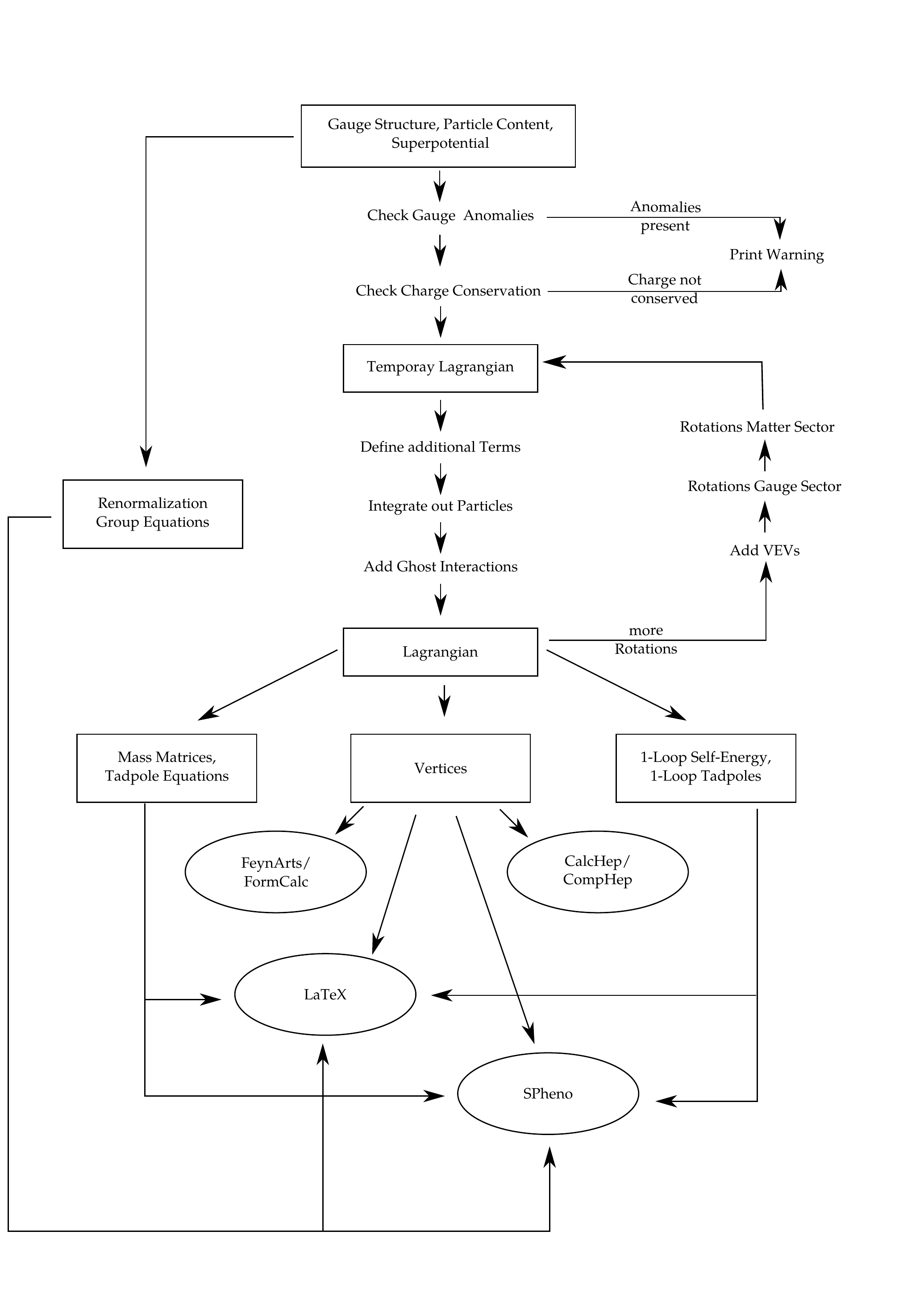}
\caption[Work-flow of \SARAH]{Schematic picture of the different steps performed by \SARAH. The user has access to the calculated information shown in boxes. The ellipses show the output which can be created by \SARAH.}
\label{fig:SARAH_workflow}
\end{figure}

\SARAH is a package for Mathematica  and was mainly written with version 5.2, but also tested with versions 6.0 and 7.0. In principle, \SARAH can handle every \(N=1\) SUSY theory with a direct product of \(SU(N)\) and/or \(U(1)\) gauge groups. The chiral superfields can be any arbitrary, irreducible representation with regard to these gauge groups, and all possible, renormalizable superpotential terms are supported. There is no restriction on either the number of gauge groups, the number for chiral superfields or the number of superpotential terms.  Furthermore, one can have any number of symmetry breakings or field rotations. A schematic picture of the different steps performed by \SARAH is given in Fig.~\ref{fig:SARAH_workflow}.  \\
We  give in this section an overview how \SARAH addresses the different tasks. In  app.~\ref{section:appendix_SARAH}, short examples for the work with \SARAH are given. A comprehensive manual of \SARAH can be found in Ref.~\cite{Staub:2008uz} and the package can be downloaded from 
\begin{verbatim}
http://theorie.physik.uni-wuerzburg.de/~fnstaub/sarah.html
\end{verbatim}
\subsection{Gauge groups and group factors}
\SARAH supports all \(SU(N)\) gauge groups. The gauge sector in \SARAH is defined by adding a vector superfield for each gauge group to the list {\tt Gauge} in the model file, e.g. 
\begin{verbatim}
Gauge[[3]]={G,  SU[3], color, g3, False};
\end{verbatim}
The different parts define the name of the superfield, of the gauge group and of the gauge coupling. In addition, the dimension of the gauge group is given. The last entry states, if the gauge indices should be implicit or explicit. \\
In \SARAH, chiral superfields are defined by using the list {\tt Field}, e.g. 
\begin{verbatim}
Fields[[1]] = {{uL, dL},  3, q, 1/6, 2,  3};  
...
Fields[[5]] = {conj[dR],  3, d, 1/3, 1, -3};
\end{verbatim}
The first entry defines the names used for the component fields, then the number of generation and the name for the superfield follows. Afterwards, the representation with respect to the gauge groups defined by {\tt Gauge} are assigned.  The transformation of an irreducible representation \(r\) under a given gauge group is in most cases fixed by its dimension \(D\). Therefore, it is sufficient to assign only \(D\) if it is unique.  Otherwise,the Dynkin labels of \(r\) have to be given as  additional input. \\
For all gauge groups the generators for all appearing representations are needed in order to write the kinetic part of the Lagrangian and the D-terms. All generators for non-fundamental representations are written as tensor product in \SARAH. Furthermore, the eigenvalues \(C_2(r)\) of the quadratic Casimir
\begin{equation}
T^a T^a \phi(r) = C_2(r) \phi(r)
\end{equation}
as well as the Dynkin index \(I_2(r)\)
\begin{equation}
Tr(T^a T^b) \phi(r) = I_2(r) \delta_{a b} \phi(r)
\end{equation}
for any field \(\phi\) transforming as irreducible representation \(r\) have to be derived. Those are needed for the calculation of the RGEs and one-loop self-energies. \(T^a\) are the fundamental generators of the gauge group. An introduction to such calculations for arbitrary Lie groups is given in \cite{Slansky:1981yr}. The calculations are performed in \SARAH by using the technique of Young Tableaux \cite{Georgi:1982jb} which is shortly presented in app.~\ref{chapter:group}.  The different steps are: the Young Tableaux corresponding to the dimension \(D\) and, if necessary, to the Dynkin labels is created. This is done by using the hook formula eq.~(\ref{eq:hook}). The number of rows and columns of the Young Tableaux defines the number of co- and contra-variant indices. This provides already all information for writing the generators. Afterwards, the vector for the highest weight \(\Lambda\) in Dynkin basis is extracted from the tableaux. The value of \(C_2(r)\) is calculated from the highest weight using the Weyl formula
\begin{equation}
C_2(r) = ( \Lambda, \Lambda + \rho) \thickspace. 
\end{equation}
Here, \(\rho\) is the Weyl vector and the highest weight is expressed in the basis of the fundamental weights of the gauge group. The Dynkin index \(I_2(r)\) is calculated from \(C_2(r)\) by
\begin{equation}
I_2(r) = C_2(r) \frac{D(r)}{D(G)} \thickspace,
\end{equation}
where \(D(G)\) is the dimension of the adjoint representation.  For the last step, the value of \(I_2(f)\) of the fundamental representation \(f\) was normalized to \(\frac{1}{2}\). \\
The user can calculate \(C_2(r), I_2(r)\) and the Dynkin labels independently of the current model for any irreducible representation of \(SU(N)\) with the function \verb"CalcIrrepSUN[r,N]" (see app.~\ref{calcIrrep} for examples). 
\subsection{Gauge anomalies}
Before \SARAH starts the calculation of the Lagrangian, it checks the model for the different triangle anomalies. These anomalies can involve diagrams with three external gauge bosons belonging to the same \(U(1)\) or \(SU(N)\) gauge group. To be anomaly free, all possible sums over  internal fermions have to vanish
\begin{eqnarray}
 U(1)^3_i &:&  \sum_n {Y^i_n}^3 = 0 \thickspace , \\
 SU(N)^3_i &:&  \sum_n \mbox{Tr}(T^i_n T^i_n T^i_n) = 0 \thickspace .
 \end{eqnarray}
We label the different gauge groups with the indices \(i,j,k\). \(Y^i_n\) is the charge of particle \(n\) under the abelian gauge group \(i\) while \(T^i_n\) is the generator with respect to a non-abelian gauge group.\\
Combinations of two different gauge groups are possible, if one group is an \(U(1)\). Hence, another condition for the absence of anomalies is
\begin{equation}
 U(1)_i\times SU(N)^2_j  :  \sum_n Y^i_n\, \mbox{Tr}(T^j_n T^j_n) = 0  \thickspace .
\end{equation}
If more than one \(U(1)\) gauge group are present, anomalies can be generated by  two or three different \(U(1)\) gauge bosons as external fields, too. Therefore, it has to be checked that 
{\allowdisplaybreaks
\begin{eqnarray}
 U(1)_i\times U(1)_j^2 &:& \thinspace  \sum_n  Y^i_n {Y^j_n}^2 = 0 \thickspace , \\
 U(1)_i\times U(1)_j\times U(1)_k &:& \thinspace  \sum_n  Y^i_n Y^j_n  Y^k_n= 0
\end{eqnarray}}
holds. In addition, it is checked that there is an even number of \(SU(2)\) doublets. This is necessary for a model in order to be free of the Witten anomaly \cite{Witten:1982fp}. If one condition is not fulfilled a warning is given by \SARAH but the model can be evaluated anyway.
\subsection{Superpotential and Lagrangian}
The superpotential is defined in a compact form using the variable {\tt SuperPotential}: 
\begin{verbatim}
SuperPotential = {{{Coefficient,Parameter,(Contraction)},
                      {Particle 1, Particle 2, Particle 3} }, ...}
\end{verbatim}
Each term of the superpotential is defined by two lists. The second list assigns all involved fields. The first list is two- or three-dimensional. It defines a numerical coefficient and the name of the coupling. The gauge and generation indices of the involved superfields are automatically contracted by \SARAH. The used contraction can be displayed via
\begin{verbatim}
ShowSuperpotentialContractions; 
\end{verbatim}
Sometimes, there are more possibilities to contract all indices. Therefore, it is possible to fix the contraction of each term using the third entry of the first list.  \\
\SARAH calculates the complete Lagrangian in component fields by performing the steps shown in app.~\ref{sec:SARAH_Lag}. Since all terms of the Lagrangian are  automatically generated, it is sufficient to check the given superpotential for charge conservation. The matter interactions and F-terms are derived from the superpotential while the kinetic terms, D-terms and gauge interactions are generated using the information about the gauge groups derived in the first step. \\
If the gauge fixing terms are defined in \(R_\xi\)-gauge, the ghost interactions are deduced. In addition, the soft breaking masses for scalars and gauginos as well as the soft breaking couplings corresponding to the superpotential couplings are affiliated to the Lagrangian. \\
Furthermore, it is possible to integrate fields out. During this process the effective operators up to dimension 6 are calculated. Otherwise, it is possible just to 'delete' fields. This performs the same steps as integrating them out but doesn't calculate the effective operators and, thus, saves time. This might be demanded, if a non-supersymmetric limit of a theory should be considered.  \\
Finally, it is possible to add non-canonical terms to the Lagrangian which originate not from the superpotential or the gauge interactions. This can be done for any eigenstate. These interactions are treated in the same way as all other terms, i.e. they are also rotated by a change of basis.  The definition of the MSSM is given as example in app.~\ref{sec:SARAH_Modelfiles}.
\subsection{Symmetry breaking and rotations}
Rotations for all matter and gauge fields as well as the decomposition of complex scalar fields into their scalar components, pseudo scalar components and VEVs can be performed. All appearing coefficients as well as the names of the rotation matrices to parametrize this change of the basis can be chosen individually. Besides, it is possible to decompose a field carrying a generation index into its different flavors in order to treat them separately. Afterwards, the complete Lagrangian for the new set of eigenstates is calculated. \\
In principle, these steps can be repeated as often as needed. Therefore, it is no problem to go first to the SCKM basis and afterwards to the mass eigenbasis. GUT theories involving several symmetry breakings can be treated in the same way. The information of all intermediate steps is saved. Hence, it is possible to calculate the vertices or masses of all eigenstates without the necessity of a new model file or a new evaluation of the model. 
\subsection{Masses and tadpole equations}
When the complete Lagrangian is calculated, tree level relations can easily be extracted. The masses and tadpole equations are derived automatically for each set of eigenstates during the evaluation of a model. In this regard, the masses or the entries of a mass matrix are calculated as second derivative of the Lagrangian
\begin{equation}
 m_{i j} = - \frac{\partial^2 \La}{\partial \phi_i \partial \phi_j^*}
\end{equation}
with respect to the considered fields \(\phi_{i}\). The tadpoles \(T_i\) are the first derivative of the scalar potential with respect to the different VEVs
\begin{equation}
\frac{\partial V}{\partial v_i} \equiv T_i \thickspace .
\end{equation}
The user has access to both information by using the command
\begin{verbatim}
 MassMatrix[Particle]
\end{verbatim}
for the mass matrix of  \verb"Particle" and 
\begin{verbatim}
 TadpoleEquation[VEV]
\end{verbatim}
for the tadpole equation corresponding to \verb"VEV". Examples for using this functions are given in app.~\ref{sec:SARAHmass}.\\
\subsection{Vertices}
The vertices are calculated as partial derivatives of the Lagrangian with respect to the involved fields and applying afterwards the vacuum conditions. The vertices can be calculated in two ways. Either it is possible to calculate the vertices for a specific choice of external particles or to calculate all vertices of the complete model at once. The former task is evolved by
\begin{verbatim}
 Vertex[{Particles},Options];
\end{verbatim}
The argument of this function is a list with the external particles. The options define the set of eigenstates and usage of relations among the parameters. In the results, the different parts of a vertex are ordered by their Lorentz structures.  If possible, the expressions are simplified by using the unitarity of rotation matrices, the properties of generators and, if defined, simplifying assumptions about parameters.\\
All vertices for a set of eigenstates are calculated at once  by
\begin{verbatim}
MakeVertexList[Eigenstates, Options];
\end{verbatim}
This searches for all possible interactions present in the Lagrangian and creates lists for the generic subclasses of interactions, e.g. {\tt VertexList[FFS]} or {\tt VertexList[SSVV]} for all two-fermion-one-scalar interactions and all two-scalar-two-vector-boson interactions, respectively. If effective theories are considered, six-particle interaction can be switched off during this calculation. Those interactions slow down  the computation and they are often not needed. 
\subsection{Renormalization group equations}
\label{RGEs}
\SARAH calculates the RGEs at one- and two-loop level for the high scale theory without broken gauge symmetries. This is done by using the generic formulas of \cite{Martin:1993zk}. The calculation of the RGEs can be started after the initialization of a model via
\begin{verbatim}
CalcRGEs[Options];
\end{verbatim}
The result of this calculation are the one- and two-loop anomalous dimensions for all chiral superfields and the one- and two-loop \(\beta\)-functions for all gauge couplings, all parameters of the superpotential, all soft breaking parameters and all VEVs. During this calculation, it is possible to treat the number of generations of specific chiral superfields as variable to make the dependence of the different parameters on these fields explicit. This might be helpful for models which include chiral superfields much heavier than the SUSY scale like GUT theories. Details about the output and the different options are given in app.~\ref{examplesRGEs}.
\subsection{One-loop tadpoles, self-energies and masses}
\label{sec:OneLoopSelf}
\SARAH calculates the analytical expressions for the one-loop corrections to the tadpoles and the one-loop self-energies for all particles. The executed steps are a generalization of the procedure applied in \cite{Pierce:1996zz}: the calculations are performed in \(\overline{\mbox{DR}}\)-scheme using 't Hooft gauge and they are started for the different eigenstates through
\begin{verbatim}
CalcLoopCorrections[Eigenstates];
\end{verbatim}
The results for the loop corrections are saved in two different ways. First as list containing the different loop contributions for each particle. Every entry of this list includes for each contribution the internal particles, generic type of the diagram, numerical factors coming from symmetry considerations and possible charges in the loop. The second output is a sum of all contributions. This sum is generated using the generic results of appendix~\ref{sec:Integrals}. The second form can, for example, be written as pdf-file using the \LaTeX{} output of \SARAH.\\
The self-energies can be used to calculate the radiative corrections to masses. We show this in sec.~\ref{sec:spectrum}, when we perform a complete one-loop corrections of the masses in the NMSSM.
\section{Output of \SARAH}
\subsection{Output for diagram calculators and  \LaTeX}
\paragraph*{CalcHep/CompHep} \CalcHep and \CompHep are well known  and widely used programs for calculating cross sections and decay widths via a diagrammatic approach at tree level. The model files produced by \SARAH for those programs support both Feynman gauge and unitarity gauge. Furthermore, \SARAH can split  interactions between four colored particles in a way that they can be handled by \CalcHep/\CompHep and also models with CP violation are possible. The model files for \CalcHep/\CompHep are created by 
\begin{verbatim}
 MakeCHep[Options];
\end{verbatim}
The options define, whether the Feynman gauge should be included. Also the splitting of specific four-scalar interactions can be suppressed as long as they are not colored. In addition, the running of the strong coupling constant can be included as it is usually  done in the standard \CalcHep files.   
\paragraph*{\FeynArts/\FormCalc} \FeynArts is a Mathematica package for creating Feynman diagrams and the corresponding amplitudes. This information is afterwards used by \FormCalc to simplify the amplitudes and square them by using {\tt FORM}. In contrast to \CalcHep/\CompHep{}, \FeynArts/\FormCalc can deal also with loop diagrams.  Beside the standard model file for \FeynArts, \SARAH writes a second file including supplementary information about the model: all defined dependences, the numerical values for the parameters and masses, if they are available, and special abbreviations to speed up the calculations with \FormCalc. The model files are generated via
\begin{verbatim}
 MakeFeynArts;
\end{verbatim}
\paragraph*{\LaTeX}
The generated \LaTeX\xspace files include all information about the model in a pleasant form: particle content, mixing matrices, tadpole equations, RGEs, one-loop self-energies and one-loop corrections to the tadpoles as well as all interactions. The \LaTeX\xspace output using the standard functions of Mathematica is not really readable when dealing with long expressions. With this in mind, special functions were developed for a better typesetting. For each vertex, the corresponding Feynman diagram is automatically drawn using \verb"FeynMF" \cite{Ohl:1995kr}. In addition, a batch file for Linux and for Windows is written by \SARAH to simplify the compilation of the different \LaTeX{} files including all Feynman diagrams. The command for producing the \LaTeX{} output is 
\begin{verbatim}
 MakeTeX[Options]; 
\end{verbatim}
One option is to disable the output of the Feynman diagrams. 
\subsection{Spectrum calculation: Combining \SARAH and \SPheno}
\label{sec:SARAH_SPheno}

\begin{figure}[t]
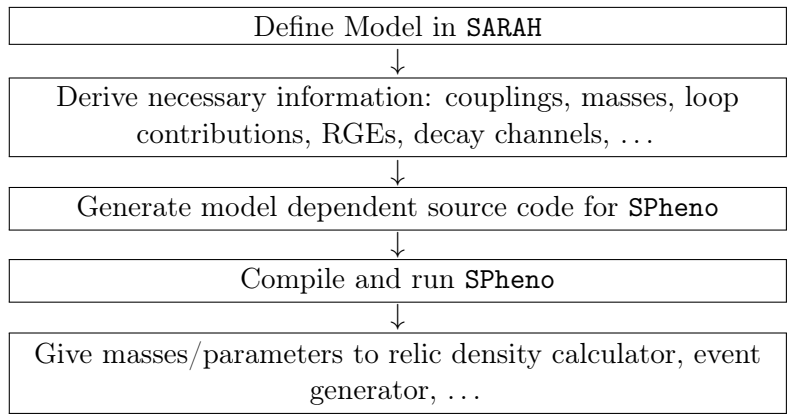

 \begin{center}
 \fbox{\parbox{10cm}{\centering
 Define Model in \SARAH
 }} \\
 \(\downarrow\) \\
 \fbox{\parbox{10cm}{\centering
 Derive necessary information: couplings, masses, loop contributions, RGEs, decay channels, \dots
 }}\\ 
\(\downarrow\) \\
\fbox{\parbox{10cm}{\centering
Generate model dependent source code for \SPheno
}} \\
\(\downarrow\) \\
\fbox{\parbox{10cm}{\centering
Compile and run \SPheno
}}\\
 \(\downarrow\) \\
 \fbox{\parbox{10cm}{\centering
 Give masses/parameters to relic density calculator, event generator, \dots
 }}
\end{center}
\caption[Automatized way from model building to phenomenology]{The model is defined in \SARAH. Afterwards, \SARAH generates all necessary files to implement this model in \SPheno, \micrOmegas and an event generators. This provides an completely automatized way from model building to phenomenology.}
\label{fig:SPheno_SARAH}
\end{figure}

\SARAH is based on Mathematica and therefore it is not sensible to do exhaustive numerical calculations in {\tt SARAH's} native environment. As opposed to that, there is \SPheno, a well tested spectrum calculator written in Fortran. \SPheno provides fast numerically routines for the evaluation of the RGEs, calculating the phase space of 2- and 3-body decays as well as Passarino Veltman integrals and much more. Since these routines are model independent, they can be used in principle for all SUSY models implemented in \SARAH. \\
Our idea for combining all advantages of \SPheno and \SARAH in order to create a very efficient and easy way from the model building to numerical results is depicted in Fig.~\ref{fig:SPheno_SARAH}. The model is defined in \SARAH in the usual way. \SARAH calculates all analytical expressions needed for a complete analysis of the model. This information is exported to Fortran code in a way that it can be included in \SPheno. This generates a fully functional version of \SPheno for the new model without any need to change the source code by hand. \\
Moreover, the user has control over the properties of the \SPheno version which is generated by \SARAH using a special input file: first, it is possible to define the free parameters of the model. Those build later on the Block {\tt MINPAR} in the LesHouches input file. Second, the boundary conditions at the GUT-, SUSY- and electroweak scale as well as at possible threshold scales can be set. Third, it can be defined what parameters are fixed by the solutions of the tadpoles equations. An approximated solution to the tadpoles can also be given, if there isn't an analytic one. \\
\SARAH produces replacements for all model dependent files of \SPheno. These files have to be copied in the {\tt src}-directory of \SPheno. A {\tt Makefile} for compiling the new model afterwards as well as a template for a LesHouches input file is written by \SARAH.\\
The routines for generating the source code for \SPheno will be included in the next main upgrade to version 3.0 of \SARAH. The command to calculate automatically all necessary information and to write the source code files is
\begin{verbatim}
 MakeSPheno[Options];
\end{verbatim}
As option, the name of the \SPheno specific input file of \SARAH can be given. This offers the possibility to create easily \SPheno versions for the same model with changed boundary conditions or another set of free parameters.

\subsection{Low energy constraints and Monte-Carlo generators}
It is possible to check the low energy constraints for new models, since the corresponding routines are added to the generated \SPheno version.  However, the used formulas are not yet derived in a fully generic way. They are based on the formulas already implemented in \SPheno and don't support particles in a  non-trivial representation not present in the MSSM. \\
Moreover, functions for the output of specific model files for a MC generator don't exist yet. At least, the first version of {\tt WHIZARD} \cite{Kilian:2007gr} can be used with model files for \CompHep. Both aspects should be improved in future. 
\section{Input and evaluation time}
\subsection{Input files}
All information about a model is saved in three different files: \verb"Model.m", \verb"parameters.m" and \verb"particles.m". Only the first one is absolutely necessary and contains the information about the gauge sector, particle content, superpotential and mixings. In \verb"parameters.m" the user can assign properties to all parameters of the model, give numerically values and define the \LaTeX{} names for each parameter. In \verb"particles.m" supplementary information about the particles can be given, which might be needed for an appropriate output: $R$-parity, mass, width, PDG code, \LaTeX{} name and output name. \\
All three files are written in an intuitive way and can easily and quickly be changed. The model file for the MSSM is explained in  app.~\ref{sec:SARAH_Modelfiles}. An example for changing the model file of the MSSM to a model file for the NMSSM is given in \cite{Staub:2009bi}.
\subsection{Existing models}
Besides different implementation of the standard MSSM (with/without flavor violation, in CKM basis and with CP violation), there exist models file for the MSSM with bi- and trilinear $R$-parity violation  and for the \(\mu \nu\)SSM \cite{Bartl:2009an}. Model files with and without CP violation are also available for the NMSSM \cite{Ellwanger:2009dp}. In addition, several other singlet extensions of the MSSM are already implemented \cite{Barger:2006dh}: the singlet extended MSSM (SMSSM), the $U(1)$-extended MSSM (UMSSM), the secluded $U(1)$-extended MSSM (sMSSM) and the near-to-MSSM (nMSSM). Also for the \(SU(5)\)-motivated models analyzed in chapter~\ref{chapter:SU5} model files have been created. Moreover, the next public version will contain model files for  models with an additional \(U(1)_{B-L}\) gauge group \cite{Khalil:2007dr} as well as a SUSY left-right model \cite{Esteves:2010si}. 
\begin{table}[t]
\begin{center}
\begin{tabular}{|c|cccc|}
\hline
Command & MSSM  & MSSM-CKM & NMSSM & \(\mu\nu\)SSM  \\
\hline \hline
\verb"Start" & 12.75 & 18.03 & 19.02 & 27.06 \\
\verb"ModelOutput[EWSB]" & 74.83 & 78.70 & 94.64 & 115.08  \\
\verb"MakeFeynArts" & 0.74 & 3.58 & 1.12 & 0.98 \\
\verb"MakeCalcHep[]" &  6.03 & 22.74 & 15.57 & 47.08  \\
\verb"MakeTeX[]" & 0.81 & 5.79  & 1.25 & 1.38  \\
\hline
\verb"CalcRGEs[]" & 50.72 & 50.8  & 91.07 & 265.29 \\
\verb"CalcLoopCorrections[EWSB]" & 7.07 & 28.44 & 8.14 & 7.98   \\
\hline \hline 
\end{tabular}
\caption[Evaluation time of \SARAH]{Needed time in seconds to evaluate several commands of \SARAH in Mathematica 5.2.}
\label{time}
\end{center}
\end{table}
\subsection{Evaluation time}
To give an impression for the needed evaluation time for different routines and models, we have collected some values in Table~\ref{time}. These times were measured with Mathematica 5.2 running on an Intel Quad CPU Q8200 with 2.33 GHz and 4GB RAM. 
\section{Verification of the output}
\label{section:SARAH_checks}
\begin{figure}[t]
\begin{minipage}{16cm}
\includegraphics[scale=1.]{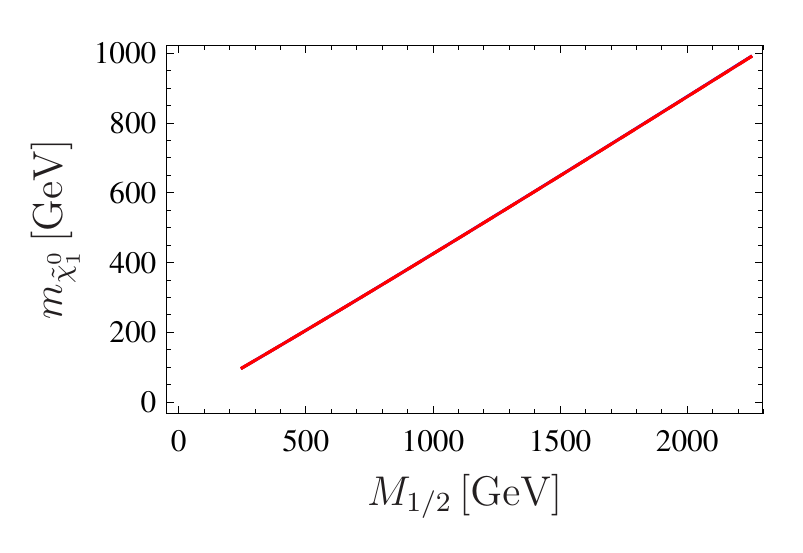}
\hfill
\includegraphics[scale=1.]{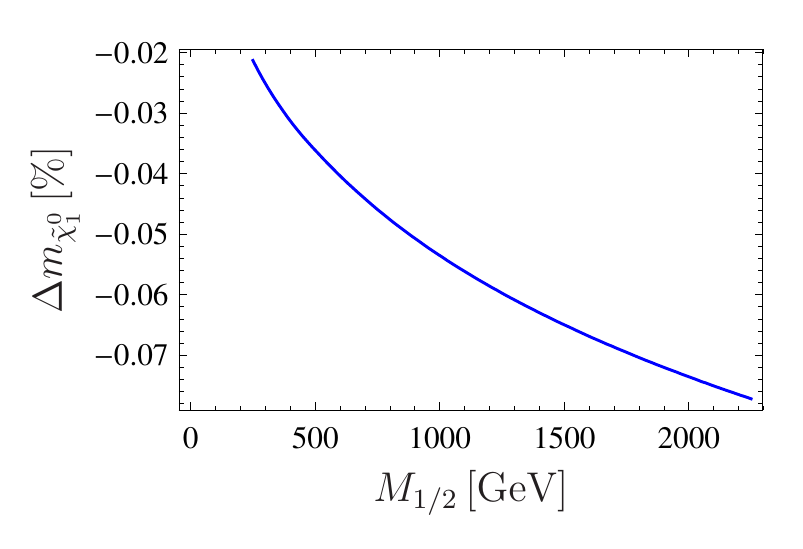}
\end{minipage}
\caption[Comparison \SARAH and \SPheno: lightest neutralino mass]{Comparison between an automatically generation version of \SARAH (blue line) and the official version \SPheno 3v48 (red) by a variation of $M_{1/2}$. Left: mass of the lightest neutralino $\Neu_1$. The lines are nearly identical. Right: relative difference.} 
\label{fig:SPheno_SARAH_m12}
\end{figure}

\paragraph*{Masses and interactions}
We have checked the model files generated with \SARAH  for the MSSM against the existing files of \FeynArts and \CalcHep. The checks happened at vertex level and for complete processes, too. We have compared the numerical value of each vertex for different sets of parameters and all possible combinations of generations (more than 5000). In addition, we have calculated several \(1\rightarrow 2\) and \(2\rightarrow 2\) processes with the old and new model files. During our tests, we have also calculated the relic density with \micrOmegas. An example for the good agreement is given in Table~\ref{SARAH:DM_comparison}. \\
\SARAH was also used in \cite{Staub:2009ww} and the results were cross checked in several ways. Furthermore, there was a detailed check of the expressions calculated by \SARAH for the $\mu\nu$SSM, the NMSSM \cite{Staub:2010ty} and a \(SU(2)_L \times SU(2)_R\)~-~model \cite{Vicente:2010wa}. 
\begin{table}[ht]
\begin{center}
\begin{tabular}{|l|ll|}
\hline 
 & CalcHep & SARAH \\
\hline \hline
\(\Omega h^2\) & 0.191    &  0.191  \\ 
\hline
Channels & 38.72~\%: \(\chi_1 \chi_1 \rightarrow e_3 \bar{e}_3\) & 38.73~\%: \(\chi_1 \chi_1 \rightarrow e_3 \bar{e}_3\) \\
&30.40~\%: \(\chi_1 \chi_1 \rightarrow e_2 \bar{e}_2\) & 30.39~\%: \(\chi_1 \chi_1 \rightarrow e_2 \bar{e}_2\) \\
&29.23~\%: \(\chi_1 \chi_1 \rightarrow e_1 \bar{e}_1\) & 29.23~\%: \(\chi_1 \chi_1 \rightarrow e_1 \bar{e}_1\) \\
& 0.31~\%: \(\chi_1 \chi_1 \rightarrow \nu_3 \bar{\nu}_3\) &  0.31~\%: \(\chi_1 \chi_1 \rightarrow \nu_3 \bar{\nu}_3\)  \\ 
& 0.30~\%: \(\chi_1 \chi_1 \rightarrow \nu_2 \bar{\nu}_2\) &  0.30~\%: \(\chi_1 \chi_1 \rightarrow \nu_2 \bar{\nu}_2\)  \\
& 0.30~\%:  \(\chi_1 \chi_1 \rightarrow \nu_1 \bar{\nu}_1\) &  0.30~\%:  \(\chi_1 \chi_1 \rightarrow \nu_1 \bar{\nu}_1\) \\
& 0.24~\%: \(\chi_1 \chi_1 \rightarrow u_2 \bar{u}_2\) &  0.24~\%: \(\chi_1 \chi_1 \rightarrow u_2 \bar{u}_2\) \\
 &0.23~\%: \(\chi_1 \chi_1 \rightarrow u_1 \bar{u}_1\) &  0.23~\%: \(\chi_1 \chi_1 \rightarrow u_1 \bar{u}_1\) \\
& 0.10~\%: \(\chi_1 \chi_1 \rightarrow d_3 \bar{d}_3\) &  0.10~\%: \(\chi_1 \chi_1 \rightarrow d_3 \bar{d}_3\) \\
&   0.07~\%: \(\chi_1 \chi_1 \rightarrow Z Z\) &  0.07~\%: \(\chi_1 \chi_1 \rightarrow Z Z\) \\
& 0.04~\%: \(\chi_1 \chi_1 \rightarrow d_2 \bar{d}_2\) &  0.04~\%: \(\chi_1 \chi_1 \rightarrow d_2 \bar{d}_2\) \\
& 0.04~\%: \(\chi_1 \chi_1 \rightarrow d_1 \bar{d}_1\) &  0.04~\%: \(\chi_1 \chi_1 \rightarrow d_1 \bar{d}_1\) \\ 
\hline \hline
\end{tabular}
\end{center}
\caption[Relic density calculation whit \SARAH model files]{Comparison of a relic density calculation with \micrOmegas 2.2.0. For the left column, the model file for the MSSM included in \micrOmegas was used. On the right side, the model file generated by \SARAH was implemented in \micrOmegas.}
\label{SARAH:DM_comparison}
\end{table}
\paragraph*{Renormalization group equations and loop corrections}
We have compared the analytical results for the one- and two-loop RGEs calculated by \SARAH for the MSSM with \cite{Martin:1993zk} and for the NMSSM with \cite{Ellwanger:2009dp}. The only difference has been in the NMSSM the two-loop RGE of \(A_\lambda\). A second calculation by authors of  \cite{Ellwanger:2009dp} has confirmed the result of \SARAH. Besides, we have checked a model containing non-fundamental representations, namely, the \(SU(5)\) inspired seesaw~II model of \cite{Borzumati:2009hu} and \cite{Rossi:2002zb}. It is known that the there are discrepancies of the RGEs given in these two papers. Our result fully agrees with  \cite{Borzumati:2009hu}. \\
The analytical expressions of the self-energies calculated by \SARAH for the MSSM were checked against the results of \cite{Pierce:1996zz}. In addition, the self-energies and the resulting loop corrected masses for the NMSSM are discussed in chapter \ref{chapter:NMSSM}.
Furthermore, numerically checks have been done by comparing the RGEs and self-energies for the MSSM separately with the functions implemented in \SPheno \cite{Porod:2003um}. Both sets of RGEs and self-energies are in full agreement. 
\begin{figure}[t]
\begin{minipage}{16cm}
\includegraphics[scale=1.]{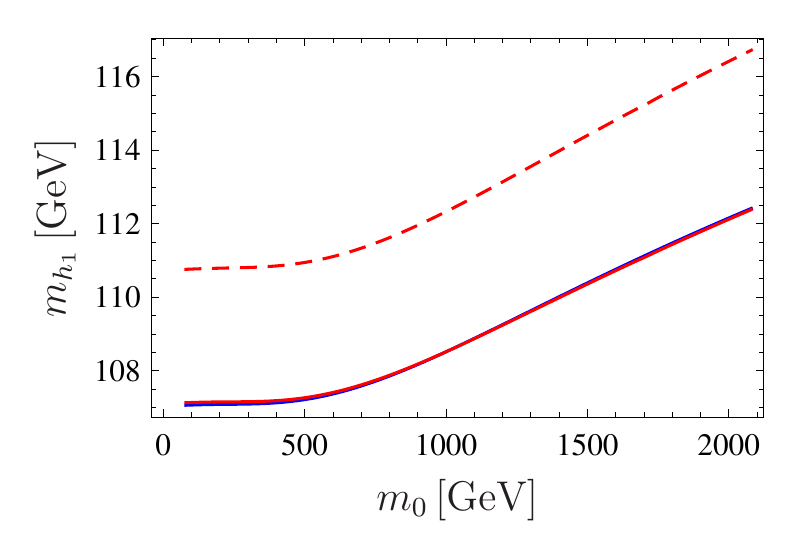}
\hfill
\includegraphics[scale=1.]{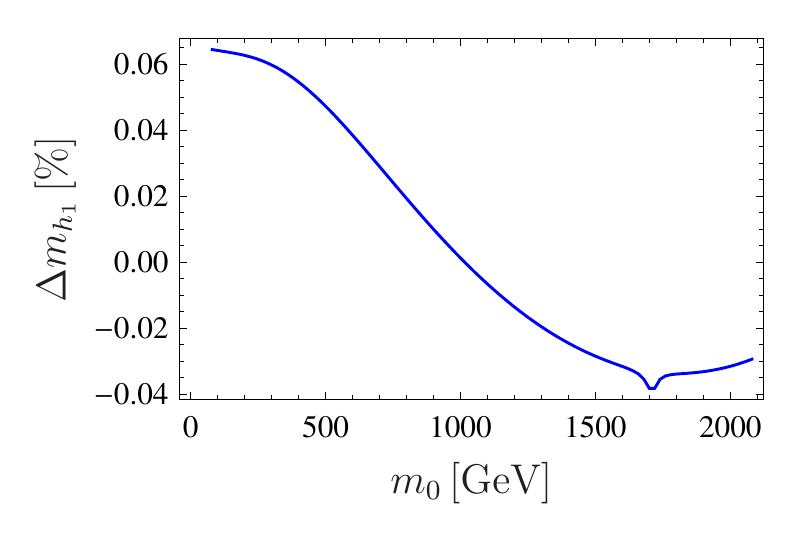}
\end{minipage}
\caption[Comparison \SARAH and \SPheno: lightest Higgs]{Comparison between an automatically generated version of \SPheno by \SARAH (blue line) and the official version \SPheno{\tt 3v48} with (red) and without (dashed red)  two-loop corrections in the Higgs sector  by a variation of $m_0$. Left: mass of the lightest Higgs $h_1$. Right: relative difference between the one-loop masses of both programs.} 
\label{fig:SPheno_SARAH_m0}
\end{figure} 
\paragraph*{Complete spectrum} Calculating the complete spectrum for the MSSM by using only formulas derived by \SARAH is a combined check for all vertices, RGEs and loop corrections. As mentioned, there are spectrum calculators for the MSSM and NMSSM to check the obtained results. For the discussion of the NMSSM, see sec.~\ref{sec:comparison}. Here, we show two examples for the comparison with the official \SPheno version. The chosen parameter point for a check of a standard mSugra scenario was
\begin{equation}
m_0 = 80\, \GeV\thickspace, \quad M_{1/2} = 250\, \GeV\thickspace, \quad A_0 = - 300\, \GeV\thickspace, \quad \tan\beta = 10\thickspace, \quad \sign\mu = 1 \thickspace .
\end{equation}
Fig.~\ref{fig:SPheno_SARAH_m0} and Fig.~\ref{fig:SPheno_SARAH_m12} show a comparison between the official \SPheno version 3.0v48 and a \SPheno version completely generated by \SARAH. We will call this version in the subsequent \SARAH-\SPheno. In Fig.~\ref{fig:SPheno_SARAH_m0}, the mass of the lightest Higgs for a variation of \(m_0\) between 80~GeV and 2~TeV is depicted. Since \SPheno calculates the Higgs masses and tadpoles at two-loop level, there is of course a difference to the result of \SARAH-\SPheno which calculated the masses at one-loop. After switching off these corrections in \SPheno, the agreement between both codes is nearly of order \(10^{-4}\), what was the used numerical precision for calculating the spectrum. This can be seen at the right plot of Fig.~\ref{fig:SPheno_SARAH_m0}, where the relative differences
\begin{equation}
 \Delta m = \frac{m_{\SPheno} - m_{\SARAH-\SPheno}}{m_{\SPheno}}
\end{equation}
is depicted. The origin of small numerical differences comes from the treatment of the loop corrections to the tadpole equations: \SPheno calculates them in an iterative way with tree level masses modified by the tadpoles equations while \SARAH-\SPheno calculates them only once to maintain gauge invariance. This leads to slightly different Higgs masses which cause a small shift in the GUT scale. Fig.~\ref{fig:SPheno_SARAH_m12} shows the mass of the lightest neutralino for both versions of \SPheno for a variation of \(M_{1/2}\) in the range 250~GeV~-~2250~GeV.

\begin{table}[t]
\begin{center}
\begin{tabular}{|c|c|c|}
\hline 
 &{\tt SPheno 3v50} & {\tt SARAH-SPheno} \\
\hline \hline
GUT Scale \([\GeV]\) & \(2.533\cdot 10^{16}\)  & \(2.504 \cdot 10^{16}\)   \\
\hline
\(m_{W_3} [\GeV]\) at  \(Q= 8 \cdot 10^{14} \, \GeV\) &  \(8.879\cdot 10^{14}\) & \(8.885\cdot 10^{14}\) \\
\(m_{W_2} [\GeV]\) at \(Q= 6 \cdot 10^{14} \, \GeV\) &  \(6.707\cdot 10^{14}\) & \(6.711\cdot 10^{14}\) \\
\(m_{W_1} [\GeV]\) at \(Q= 4 \cdot 10^{14} \, \GeV\) &  \(4.516\cdot 10^{14}\) & \(4.519\cdot 10^{14}\) \\
\hline   
\(m_{\Neu_1}[\GeV]\) & 93.0  & 92.8 \\
\(m_{\Neu_2}[\GeV]\) & 175.3 & 175.0 \\
 \(m_{\Neu_3}[\GeV]\) & 454.5 & 454.6 \\
 \(m_{\Neu_4}[\GeV]\) & 465.7 & 465.8 \\
\(m_{\Cha_1}[\GeV]\) & 175.2 & 175.0 \\
\(m_{\Cha_2}[\GeV]\) & 466.2 & 466.3 \\
\hline 
\(m_{h_1}[\GeV]\) &  98.7   & 98.7 \\
\(m_{h_2}[\GeV]\) &  467.8 &  467.9 \\
\(m_{A^0}[\GeV]\) &  467.9 &  468.9 \\
\(m_{H^+}[\GeV]\) &  474.2 &  474.8 \\
\hline \hline 
\end{tabular}
\end{center}
\caption[Comparison between {\tt SPheno 3v50} and {\tt SPheno SARAH} in case of Seesaw III]{Comparison between {\tt SPheno 3v50} and {\tt SPheno-SARAH} in case of seesaw~III and three threshold scales. In the official version {\tt SPheno 3v50}, the two-loop corrections to the Higgs masses and tadpoles were disabled. $m_{W_i}$ are the masses of the different generations of the $SU(2)_L$-triplet at their threshold scale.  }
\label{tab:SARAH_SeesawIII}
\end{table}

In Table~\ref{tab:SARAH_SeesawIII}, the result for a comparison between {\tt SPheno 3v50} and the the \SPheno version generated by \SARAH for the seesaw~III model is summarized. For this comparison, the two-loop self-energies for the Higgs and the two-loop tadpoles were again disabled  in \SPheno. The mSugra input parameters are chosen to be
\begin{equation}
m_0 = 90\, \GeV \thickspace, \quad M_{1/2} = 400\, \GeV \thickspace, \quad A_0 = 1500\, \GeV \thickspace, \quad \tan\beta = 10 \thickspace, \quad \sign\mu = 1 \thickspace. 
\end{equation}
In addition, the specific parameters for seesaw~III were set to
\begin{eqnarray}
\nonumber
& Y_{B,11} = 1\cdot 10^{-8} \thickspace, \quad Y_{B,22} = 2\cdot 10^{-8} \thickspace, \quad Y_{B,33} = 3\cdot 10^{-8} \thickspace, &\\
& M_{W,11} = 4\cdot 10^{14}\,\GeV, \quad M_{W,22} = 6\cdot 10^{14}\,\GeV, \quad M_{W,33} = 8\cdot 10^{14}\,\GeV \thickspace . &
\end{eqnarray}
The exact meaning of those parameters is explained in chapter~\ref{chapter:SU5}. This input leads to three different threshold scales. At each scale one generation of heavy fields is integrated out and the finite shifts for gauge couplings and gauginos are calculated. All necessary routines for these steps are written by \SARAH. This leads to an agreement similar to the case of the MSSM. 

\cleardoublepage
\chapter[Gravitino Dark Matter]{Gravitino Dark Matter in Gauge Mediated SUSY Breaking}
\label{chapter:GMSB}
As already discussed in the introduction, it is assumed that the SUSY breaking is transmitted from a hidden or secluded sector to the visible sector. Models in which the effects of SUSY breaking are communicated via the usual gauge interactions \cite{Dine:1981gu,Dine:1981za,Dimopoulos:1981au,Nappi:1982hm,AlvarezGaume:1981wy,Dine:1993yw,Dine:1993qm,Dine:1994vc,Dine:1995ag} to the visible sector have recently achieved considerable attention (see e.g.~\cite{Intriligator:2006dd,Murayama:2006yf,Murayama:2007fe,Carpenter:2008wi} and references therein), in particular in view of model building and understanding the mechanism of SUSY breaking.  An attractive feature of such models is the natural explanation for the smallness of SUSY contributions to flavor changing neutral current phenomena due to the strongly constrained SUSY spectrum. In gauge mediated SUSY breaking (GMSB) models, there are two new sectors \cite{Dine:1993yw,Dine:1993qm,Dine:1994vc,Dine:1995ag,Giudice:1998bp}:
\begin{enumerate}
\item the secluded sector: this is a strongly interacting sector in which SUSY is broken dynamically. We refer here to 'secluded' sector to differ it from the 'hidden' sector of models where SUSY breaking is transmitted via gravity. 
\item the messenger sector: it contains fields charged under   $SU(3)_C\times SU(2)_L \times U(1)_Y$ gauge interactions which communicate   SUSY breaking to the ordinary sparticles. Usually, it is assumed that  they come in complete $SU(5)$ representations or representations of larger groups containing $SU(5)$ as subgroup, so that the success of gauge coupling unification does not get spoiled.
\end{enumerate}
In these scenarios, the gravitino is usually very light, in the range between a few eV up to \(\Ord(1)\)~MeV. The masses of the messenger particles and some of the fields in the secluded sector can be as low as 100~TeV implying that they can act as cold dark matter if their masses are below the reheating temperature and if they are stable. The gravitino is the lightest SUSY particle in GMSB models and all MSSM particles decay into it in a cosmologically short time. Therefore, the gravitino forms the dark matter in these models. This aspect has been extensively discussed in the literature \cite{Dimopoulos:1996gy,Han:1997wn,Baltz:2001rq,Fujii:2002fv,Fujii:2002yx,Jedamzik:2005ir}.\\
However, in GMSB models there are serious problems with cosmology: first, if only the gravitino was responsible for dark matter assuming the standard history of the universe, it would be warm dark matter with a mass of about 100~eV. But, bounds from observations of the Lyman-\(\alpha\) forest rule warm dark matter with masses below 1.5~keV out. Gravitinos with a mass  consistent with these observations would overclose the universe. Second, if messenger particles are produced after inflation and if messenger number is conserved, one will in general obtain an exceedingly large contribution to $\Omega h^2$ which overcloses the universe, too. Both problems can, in principle, be solved by breaking messenger number. The idea is that decays of the lightest messenger particles give rise to sufficient additional entropy production once the gravitino has decoupled from the thermal bath. However, we will show below that this statement is incorrect  as only part of the possible messenger decay modes have been taken into account in the literature. \\
So far, we have implicitly assumed that $R$-parity is conserved implying a stable LSP. There is one experimental observation which can be explained by the breaking of $R$-parity, namely, neutrino masses and mixings. In the simplest model, one adds bilinear $R$-parity breaking terms to the MSSM superpotential and in this way neutrino data can be explained. Moreover, this class of models is also consistent with constraints from baryogenesis \cite{Akeroyd:2003pb}. Neutrino physics gives a lower bound on the $R$-parity breaking parameters such that the lightest MSSM particle will decay within a small fraction of a second. However, this does not apply to a light gravitino, which eventually decays, but neutrino physics now implies that its life time is much larger than the age of universe \cite{Borgani:1996ag,Takayama:2000uz,Hirsch:2005ag}.\\
After an introduction to GMSB, we discuss first the case of messenger decays with conserved $R$-parity in minimal GMSB. Afterwards, we show the effect of $R$-parity violation and extend the discussion to other messenger sectors. 
\section{Gauge mediated SUSY breaking}
\label{sec:Intro_GMSB}
\subsection{Messenger sector and SUSY masses}
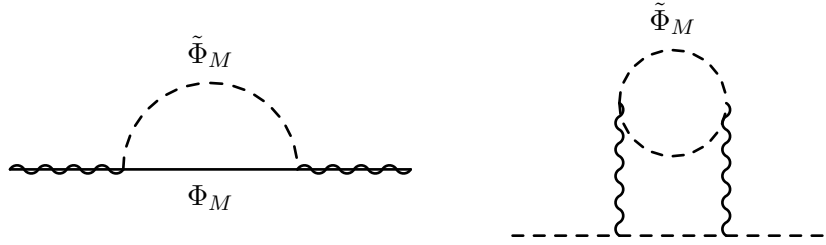
\begin{figure}
\begin{center}
\begin{minipage}{12cm}
\parbox{6cm}{
	\begin{fmffile}{Feynmangraphen/GMSB_fermion2}
 \fmfframe(25,25)(25,25){
		\begin{fmfgraph*}(150,50)
		\fmfleft{i}
		\fmfright{o}
		\fmf{plain}{i,v1}
		\fmf{plain}{v2,o}
		\fmf{plain,label=$\Phi_M$}{v1,v2}
		\fmf{wiggly}{i,v1}
		\fmf{wiggly}{v2,o}
		\fmf{dashes,label=$\tilde{\Phi}_M$,left,tension=.3}{v1,v2}
		\end{fmfgraph*}
}
	\end{fmffile}
}
\hfill
\parbox{6cm}{
	\begin{fmffile}{Feynmangraphen/GMSB_scalar2}
  \fmfframe(25,25)(25,25){
		\begin{fmfgraph*}(150,50)
		\fmfleft{i1,i2}
		\fmfright{o1,o2}
		\fmf{dashes}{i1,v1}
		\fmf{dashes}{v2,o1}
		\fmf{dashes}{v1,v2}
		\fmf{phantom}{i2,v3}
		\fmf{phantom}{o2,v4}
		\fmf{phantom}{v3,v4}
		\fmffreeze
		\fmf{wiggly}{v1,v3}
		\fmf{wiggly}{v2,v4}
		\fmf{dashes,label=$\tilde{\Phi}_M$,left,tension=.3}{v3,v4}
		\fmf{dashes,right,tension=.3}{v3,v4}
		\end{fmfgraph*}
}
	\end{fmffile}
}
\end{minipage}
\end{center}
\vspace{-1cm}
\caption{Generation of soft breaking masses at different loop levels for the scalars and gauginos due to messenger interactions.}
\label{fig:GMSB_soft}
\end{figure}

The mediation of the SUSY breaking from the secluded to the visible sector happens in GMSB by messenger particles charged under SM gauge groups. These messengers are described by chiral superfields of \(N_f\) flavors and come always in pairs which transform under the representation \(\textbf{r}+\bar{\textbf{r}}\) of the gauge group. In the minimal case, the messengers are  \(\textbf{5}+\bar{\textbf{5}}\) under \(SU(5)\) and only one flavor exists.  An important quantity for the phenomenology of GMSB models is the so called messenger index  \(N\) defined as 
\begin{equation}
\label{eq:MessengerIndex}
 N = \sum_{i=1}^{N_f} n_i \thickspace . 
\end{equation}
Here, \(n_i\) is twice the Dynkin index of the representation. In the case of \(SU(5)\), the values are \(n=1\) for messengers transforming as a {\bf 5} and \(n=3\) for a {\bf 10}. We will denote the messenger 5-plets with \(\Phi_M,\bar{\Phi}_M\). The interaction of the messengers with a secluded spurion-field \(S\) is described by the superpotential term 
\begin{equation}
\label{eq:mass_messenger}
W = \lambda S \Phi_{M} \bar{\Phi}_{M}  \thickspace . 
\end{equation}
\(S\) is a gauge singlet under SM gauge groups and acquires a VEV along its scalar and auxiliary component due to hidden sector interactions, which we leave here unspecified 
\begin{equation}
\label{eq:vevS}
\bra S \ket = M + \Theta^2 F \thickspace.
\end{equation}
The coupling \(\lambda\) of eq.~(\ref{eq:mass_messenger}) is often absorbed into a redefinition of \(M\) and \(F\)
\begin{equation}
\label{eq:reMF}
 M \rightarrow \lambda M \thickspace, \hspace{1cm}  F \rightarrow \lambda F \thickspace.
\end{equation}
The parameters \(M\) and \(F\) are the fundamental scales in this theory. \(F < M^2\) is required because of the positivity of the squared masses of all scalar messenger particles. This can be seen as follows: by inserting eq.~(\ref{eq:vevS}) in eq.~(\ref{eq:mass_messenger}) and applying eq.~(\ref{eq:reMF}), the fermionic component \(\phi\) is getting a mass \(M\) while the masses of the scalar components \(\tilde{\phi}_i\) are given by the diagonalization of the mass matrix 
\begin{equation}
\label{eq:MessengerMassMatrix}
\left(
\begin{array}{cc}
 M^2  & F  \\
F  & M^2
\end{array}
\right) 
\end{equation}
written with respect to the basis \(\left(\tilde{\phi}_M,\tilde{\bar{\phi}}_M\right)\). The eigenvalues and eigenstates of this matrix are
\begin{equation}
\label{eq:ScalarMessengerMass}
\tilde{\phi}_{+,-} = \frac{1}{\sqrt{2}} \left(\tilde{\phi}_M \pm \tilde{\bar{\phi}}_M\right) \thickspace, \hspace{1cm} m_{+,-} = \sqrt{M^2 \pm F} \thickspace.
\end{equation}
In phenomenological studies, the limit \(F \ll M^2\) is often  considered because the ratio \(\Lambda \equiv\frac{F}{M}\) defines the scale of the soft breaking parameters: when the spurion \(S\) has received its VEVs according to eq.~(\ref{eq:vevS}) and induces the mass splitting of the components of the messenger superfields, the soft breaking masses of the MSSM fields are generated via loop diagrams involving the messenger particles. The gauginos receive masses \(M_{\tilde{\lambda}}\) at one-loop level while the scalar masses \(m^2_{\tilde{f}}\) are generated at two-loop level due to diagrams like these depicted in Fig.~\ref{fig:GMSB_soft}. 
The leading approximations for the soft breaking masses are
\begin{equation}
\label{eq:SoftBreakingGMSB}
 M_{\tilde{\lambda}_i}(t) = k_i \frac{\alpha_i(t)}{4 \pi} \Lambda_G \thickspace , \hspace{1cm}
m_{\tilde{f}_i}^2(t) = 2 \sum_{r=1}^3 C_r(\tilde{f}) k_r \frac{\alpha(0)^2}{16 \pi^2} \left( \Lambda_S^2 + h_r \Lambda_G^2 \right) \thickspace .
\end{equation}
\(\alpha(t)_i\) are the running coupling constants. The values of \(k_i\) for the different gauge groups are  \(k_1 = \frac{5}{3},\, k_2=k_3 = 1\). Furthermore, we defined
\begin{equation}
h_r = \frac{k_r}{b_r} \left( 1 - \frac{\alpha_r^2(t)}{\alpha_r^2(0)}\right) 
\end{equation}
with the one-loop  \(\beta\)-coefficients \(b_1 = 11,\,b_2 = 1\) and \(b_3 = -3\). The SUSY soft breaking scales \(\Lambda_G\) and \(\Lambda_S\) depend  only on the ratio  \(\frac{F}{M}\) for a fixed messenger index
\begin{equation}
 \Lambda_G = N \frac{F}{M} \thickspace , \hspace{1cm} \Lambda_S^2 = N \frac{F^2}{M^2} \thickspace  .
\end{equation}
We have neglected terms of \(O\left(\frac{F^2}{M^4}\right)\). On the other hand, it is possible to get soft breaking masses of the correct order in the case \(F \simeq M^2\), too. While eq.~(\ref{eq:SoftBreakingGMSB}) keeps its form, the results for the SUSY breaking scales change to 
\begin{equation}
\Lambda_G = N g\left(\frac{F}{M}\right), \hspace{1cm} \Lambda_S^2 = N f\left(\frac{F^2}{M^2}\right)
\end{equation}
with
\begin{equation}
 g(x) \simeq 1 + \frac{x^2}{6} + \frac{x^4}{15} + \frac{x^6}{28} + \Ord(x^8) \thickspace, \hspace{0.5cm} f(x)\simeq 1 + \frac{x^2}{36} - \frac{11 x^4}{450} - \frac{319 x^6}{11760} + \Ord(x^8) \thickspace. 
\end{equation}
A detailed discussion of the SUSY spectrum in GMSB models is beyond the scope of this work, so we refer to \cite{Giudice:1998bp} and references therein regarding this topic.  In the subsequent, we will focus on the cosmological aspects of the messengers and of the LSP in GMSB, the gravitino.  
\subsection{The gravitino in GMSB}
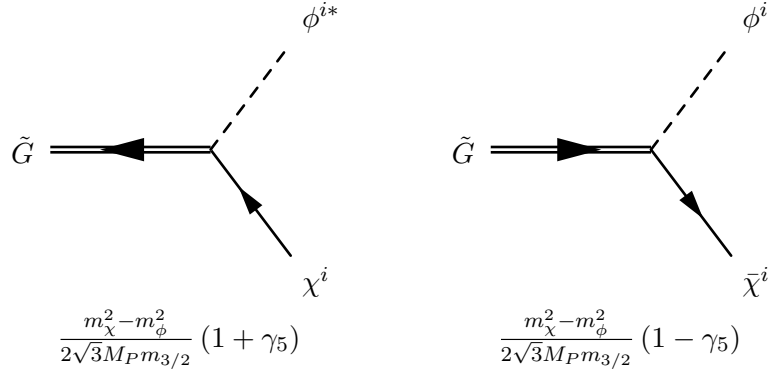
\begin{figure}[t]
\begin{center}
\begin{tabular}{ccc}
\begin{fmffile}{Feynmangraphen/grav1c}
  \fmfframe(15,15)(15,15){
    \begin{fmfgraph*}(100,80)
    \fmfleft{l1}
    \fmfright{r1,r2}
    \fmf{dbl_plain_arrow}{v1,l1}
    \fmf{plain_arrow}{r1,v1}
    \fmf{dashes}{v1,r2}
    \fmflabel{$\tilde{G}$}{l1}
    \fmflabel{$\chi^i$}{r1}
    \fmflabel{$\phi^{i *}$}{r2}
\end{fmfgraph*}}
\end{fmffile}
& {} &
\begin{fmffile}{Feynmangraphen/grav2c}
  \fmfframe(15,15)(15,15){
    \begin{fmfgraph*}(100,80)
    \fmfleft{l1}
    \fmfright{r1,r2}
    \fmf{dbl_plain_arrow}{l1,v1}
    \fmf{plain_arrow}{v1,r1}
    \fmf{dashes}{v1,r2}
    \fmflabel{$\tilde{G}$}{l1}
    \fmflabel{$\bar{\chi}^i$}{r1}
    \fmflabel{$\phi^{i}$}{r2}
\end{fmfgraph*}}
\end{fmffile} \\
\(\frac{m_\chi^2 - m_\phi^2}{2 \sqrt{3} M_P m_{3/2}} \left(1+\gamma_5\right)\) & &
\(\frac{m_\chi^2 - m_\phi^2}{2 \sqrt{3} M_P m_{3/2}} \left(1-\gamma_5\right)\) 
\end{tabular}
\end{center}
\caption[Interactions of the gravitino]{Vertices of the gravitino in case of small gravitino masses. The interactions are dominated by the Goldstino component and the contribution of the spin $\frac{3}{2}$-component was neglected here. }
\label{fig:FR_gravitino}
\end{figure}

\paragraph*{Mass of the gravitino} The gravitino is the LSP in GMSB  and receives its mass via the Super-Higgs mechanism by 'eating up' the Goldstino after SUSY breaking. The mass of the gravitino can therefore be related to  \(F\) which is responsible for SUSY breaking and to the Planck mass \(M_P\) by \cite{Volkov:1973jd,Deser:1977uq}
\begin{equation}
\label{eq:m32}
m_{3/2} = \frac{1}{\sqrt{3}} \left( \frac{\Lambda M}{k M_P} \right) \thickspace .
\end{equation}
Here, we introduced the parameter \(k\) defined as
\begin{equation}
k = \frac{F}{F_{\rm tot}} \thickspace ,
\end{equation}
where \(F_{\rm tot} = F + \sum_i F_{Z_i}\) is the sum of all F-terms in the secluded sector. \(k\) parametrizes how SUSY breaking is communicated to the messengers. In our case, it is normally of \(\Ord(1)\) because of the superpotential term eq.~(\ref{eq:mass_messenger}). Whereas,  it can be much smaller in case of radiative SUSY breaking in the messenger sector. Normally, \(M\) of \(\Ord(10^6)\)~GeV is needed to produce phenomenologically valid SUSY masses and not to introduce large flavor violation due to a high SUSY breaking scale and large RGE effects. In that case and for moderate values of \(k\), the gravitino mass is in the keV range.
\paragraph*{Interactions of the gravitino} The dominant interaction of the gravitino \(\tilde{G}\) is an effect from the coupling to the supercurrent. The corresponding terms in the Lagrangian are \cite{Moroi:1995fs}
\begin{eqnarray}
\nonumber \La_{\tilde{G} J } &=& - \frac{1}{\sqrt{2}M_P} D_\nu \phi^{* i} \bar{\tilde{G}}_\mu  \gamma^\nu \gamma^\mu \chi_R^i - \frac{1}{\sqrt{2} M_P} D_\nu \phi^i \bar{\chi}_L^i \gamma^\mu \gamma^\nu \tilde{G}_\mu \\
&& \hspace{1cm}- \frac{i}{8 M} \bar{\tilde{G}}_\mu [\gamma^\nu,\gamma^\rho] \gamma^\mu \lambda^{(a)} F_{\nu \rho}^{(a)} \thickspace.
\end{eqnarray}
As long as the gravitinos are lighter than the considered energy scale here, the interactions of the spin \(\frac{1}{2}\) are dominant, since they are given by the Goldstino component. In this limit, an effective Lagrangian  can be written as
\begin{eqnarray}
\nonumber \La_{\rm eff} &=& \left( \frac{i}{\sqrt{3} m_{3/2}} \left[ \left( \bar{\tilde{G}} \chi^i_R \right) \partial_\mu \partial^\mu \phi^{i *} - \left( \bar{\tilde{G}} \partial_\mu \partial^\mu \chi_R^i \right) \phi^{i *} \right] + \mbox{h.c.} \right) \\
&& \hspace{1cm}+ \frac{1}{4 \sqrt{6} M_P m_{3/2}} \bar{\tilde{G}} [\gamma^\mu,\gamma^\nu] \gamma^\rho \partial_\rho \lambda^{(a)} F^{(a)}_{\mu \rho} \label{eq:GravitinoInteractions}.
\end{eqnarray}
The Feynman rules  for the coupling to matter fields stemming from the first two terms of eq.~(\ref{eq:GravitinoInteractions}) are depicted in Fig.~\ref{fig:FR_gravitino}.
\paragraph*{Life time with broken $R$-parity}
If $R$-parity is broken, the LSP is no longer stable. To explain neutrino data with bilinear $R$-parity violation, the mixing parameters between neutrino and neutralinos are of a size to rule out the neutralino as dark matter completely because it will decay in cosmological short times. Just the gravitino might be still a dark matter candidate because its interaction is naturally suppressed by the Planck scale. The decay width for a gravitino into neutrino and photon was calculated to \cite{Hirsch:2005ag}
\begin{equation}
\Gamma \big(\tilde{G} \rightarrow \sum_i \gamma \nu_i\big)  \simeq \frac{1}{32 \pi} | U_{\gamma \nu} |^2 \frac{m^3_{3/2}}{M_P^2} \thickspace . 
\end{equation}
Here, \(|U_{\gamma \nu}|^2 = \sum_{i=1}^3 | \cos\Theta_W N_{i1} +\sin\Theta_W N_{i2}|^2\) is fixed by neutrino data. It can be approximated by
\begin{equation} 
| U_{\gamma \nu} |^2 = 3.5 \cdot 10^{-14} \frac{m_\nu}{0.05\,\mbox{eV}} \thickspace .
\end{equation}
If we assume a light gravitino with mass \(m_{3/2} \simeq 1\)~keV, this leads to a life time of about \(10^{28}\)~Hubble times, i.e. it is on cosmological scales still stable and contributes to the measured amount of dark matter in the universe.    
\section{The cosmological gravitino problem}
\subsection{The problem}
We have already discussed the cosmological aspects of the gravitino in sec.~\ref{section:introdution_GravitinoDM}, with this in mind, we just repeat the main statement. Using assumptions consistent with the standard thermal evolution of the early universe, the abundance of thermal produced gravitinos is   given by 
\begin{equation}
\label{eq:relic}
\Omega_{3/2} h^2 = \frac{m_{3/2}}{\mbox{keV}} \frac{100}{g_*} \thickspace .
\end{equation} 
Here \(g_*\) is the effective number of degrees of freedom at the time of gravitino decoupling. This implies that the gravitino forms warm dark matter in the mass range of $\Ord(100)$~eV. \nomenclature{WDM}{Warm Dark Matter} \\
However, as already mentioned, there are stringent constraints on the contribution of warm dark matter particles  with free-streaming lengths of the order of galaxy scales or larger to the total dark matter content. More precisely, if dark matter is assumed to consist of only one particle species, its mass is limited from below by the amplitude of the small scale power spectrum which, in turn, currently receives its tightest constraints from observations of the Lyman-\(\alpha\) forest \cite{Viel:2005qj}. This bound has recently been increased by an order of magnitude ruling out pure warm dark matter scenarios with particle masses below 8~keV for non resonantly produced dark matter \cite{Boyarsky:2008xj}.  Gravitinos with masses up to \(\Ord(\mbox{MeV})\) have once been in thermal equilibrium  as long as the reheating temperature is above \(10^6\, \mbox{GeV}\). For those thermal relics the bound is 1.5~keV while mixed dark matter scenarios dominated by a cold component allow a contribution of up to 60\% by a warm dark matter particle of any mass above 1.1~keV.  There remains a number of  systematic uncertainties in the interpretation of Lyman-\(\alpha\) observations, mostly related to the poorly understood thermal evolution of the intergalactic medium. Nevertheless, the overall result is fairly robust (for a detailed discussion, see \cite{Boyarsky:2008xj}), so that pure gravitino dark matter allowed by Lyman-\(\alpha\) bounds would have a relic density at least 15 times higher than the measured dark matter relic density.  \\
An additional difficulty stems from the fact that in typical GMSB models, the lightest messenger particle is stable as a result of the conservation of messenger quantum number. Its relic density is  calculable similarly to the case of a neutralino LSP and is found to scale as  \cite{Han:1997wn}
\begin{equation}
\label{eq:Relic_Messenger}
\Omega_M h^2 \simeq 10^5 \frac{m_-^2}{(10^3\,\mbox{TeV})^2} \thickspace ,
\end{equation}
where $m_-$ is the mass of the lightest messenger particle.  This overcloses the universe in most of the parameter space as discussed below. The situation becomes even worse because, as we will show below, there are usually several different types of stable messenger particles, one for each type of the corresponding SM gauge group representation. 
\subsection{Proposed solution: messenger number violation}
\paragraph*{Messenger decays} A possible solution to both problems might be additional entropy generated by decays of the lightest messengers into SM fermions \cite{Han:1997wn,Baltz:2001rq,Fujii:2002fv,Fujii:2002yx,Jedamzik:2005ir}. Those decays are motivated by the observation that, in general, gravitational interactions break global symmetries and thus one expects terms like 
\begin{equation}
\label{eq:deltaW}
\delta W = f \frac{\bra W \ket}{M^2_\star} \Phi_{MSSM} \bar{\Phi}_M  = f m_{\frac{3}{2}} \Phi_{MSSM} \bar{\Phi}_M
\end{equation}
in the superpotential \cite{Jedamzik:2005ir}. $\Phi_{MSSM}$ is a 5-plet containing the right down
quark superfield and the left lepton superfields of the MSSM. In the framework of supergravity, a possible origin for such a superpotential term can be highlighted by adding to the minimal K\"ahler potential 
\begin{equation}
K_0 = \sum_i \Phi_i^\dagger \Phi_i 
\end{equation}
a non minimal part \(\delta K\) given by
\begin{equation}
\delta K = \Phi_{MSSM} \bar{\Phi}_M + \mbox{h.c.} \thickspace.
\end{equation}
This term is allowed by gauge symmetries and possibly by an $R$-symmetry as well, for conveniently chosen $R$ charges. Then, making use of the usual invariance of the supergravity Lagrangian under K\"ahler transformation \(K \rightarrow K + F(\Phi) + F^*(\Phi^*) \) followed by superpotential \(W\rightarrow e^{-F} W \) scalings, the above term in the superpotential is obtained for \(F(\Phi) = - \delta K \) to the lowest order in \(\frac{1}{M_P}\) \cite{Jedamzik:2005ir}. \\
Owing to such interactions, the lightest messenger field decays into standard model fermions. It has been claimed in literature that it is sufficiently long lived to substantially produce entropy, diluting the gravitino abundance. This in turn would imply that heavier gravitinos with
masses above 8~keV  would be viable dark matter candidates. 
\paragraph*{Calculation in literature} We summarize briefly the calculation so far presented in literature \cite{Baltz:2001rq,Fujii:2002fv,Jedamzik:2005ir}. It is assumed that \(F \ll M^2\) and the lightest messenger decays due to the mixing term of eq.~(\ref{eq:deltaW}) predominantly into a neutralino and a SM fermion with a decay width of
\begin{equation}
\label{eq:gamma_neut}
 \Gamma \simeq \frac{g_2^2}{16 \pi} \left(\frac{f m_{3/2}}{M}\right)^2 M \thickspace.
\end{equation}
Applying this result, one can calculate the decay temperature to
\begin{equation}
T_D \simeq 68\, \mbox{MeV} \times \frac{f}{\sqrt{k}} \left(\frac{10}{g_*(T_D)}\right)^{1/4} \left(\frac{m_{3/2}}{10\, \keV}\right)^{1/2} \left(\frac{\Lambda}{10^5\, \GeV}\right)^{1/2} \thickspace.
\end{equation}
Furthermore, inserting eq.~(\ref{eq:Relic_Messenger}) leads to a yield of \(Y_M = \frac{\Omega_M \rho_c}{s_0 M} \simeq 3.65\cdot 10^{-10} \left(\frac{M}{\mbox{PeV}}\right)\). Taking the results together, the authors end up with a diluted relic density of the gravitinos of
\begin{eqnarray}
\nonumber
\Omega_{3/2} h^2 &=& \Omega_{3/2}\Big|_{\text{initial}} \times \frac{1}{\Delta_M} \\
&=& 0.14 \times f \left(\frac{10}{g_*(T_D)}\right)^{1/4} \left(\frac{230}{g_*(T_F)}\right)\left(\frac{\Lambda/k}{3 10^5\, \GeV}\right)^{5/2} \left(\frac{2\, \keV}{m_{3/2}}\right)^{1/2} .
\end{eqnarray}
In the last step the definition of the dilution factor of eq.~(\ref{eq:dilutionfactor}) was used. This would imply that the lightest messenger scalars decay after the freeze out of the gravitinos and before BBN. Furthermore, the dilution for a natural choice of parameters $f = \Ord(1)$ could lead to a relic density of the gravitinos of \(\Omega h^2 \simeq 0.1\). This conclusion would be correct in the absence of an event which has not been  taken into account so far: electroweak symmetry breaking which opens new decay channels, namely, into vector bosons. Also the effect of the other messenger particles was ignored so far. 
\section{Messenger decays}
\subsection{The considered model without \texorpdfstring{$R$}{R}-parity violation}
\paragraph*{Superpotential and soft breaking terms} We do a complete re-analysis of the scenario presented in literature. We start with the analysis in  case of conserved $R$-parity. The effects of bilinear $R$-parity violation are discussed in sec.~\ref{sec:GMSB_RpV}. First, we consider the minimal messenger sector by adding a {\bf 5} and \({\bf \bar{5}}\) messenger pair under \(SU(5)\) to the particle content. The mass term of the messengers in the superpotential is given by eq.~(\ref{eq:mass_messenger}) and we use the mixing between messengers and SUSY fields resulting from eq.~(\ref{eq:deltaW}). We decompose the messenger 5-plets according to SM representations by defining the superfields as
\begin{equation}
\Phi_M = \left(\begin{array}{c} \hat{D}^\alpha_R \\ \hat{L}^i \end{array} \right) \thickspace, \hspace{1cm} \bar{\Phi}_M = \left(\begin{array}{c} \hat{\bar{D}}^\alpha_R \\ \hat{\bar{L}}^i \end{array} \right)  \thickspace . 
\end{equation}
Furthermore, the fermionic components are called \(D_R\) and \(\bar{D}_R\) while the scalar components are written as \(\tilde{D}_R\) and \(\tilde{\bar{D}}_R\). The lepton-like messengers are named in the same way. Using these conventions, the superpotential for the considered scenario is
\begin{equation}
W = W_{MSSM} +  M \hat{D}_R \hat{\bar{D}}_R + M \hat{L} \hat{\bar{L}} + f_i m_{3/2} \left(\hat{D} \hat{d} + \hat{L} \hat{l} \right) \thickspace .
\end{equation}
\(W_{MSSM}\) is the superpotential of the MSSM of eq.~(\ref{superpotential}). We have suppressed here the color and \(SU(2)_L\) indices. The arising soft breaking terms are 
\begin{eqnarray}
\nonumber \La_{SB} &=& \La_{SB,MSSM} -  B_M \left(\tilde{D}_R \tilde{\bar{D}}^*_R + \tilde{L} \tilde{\bar{L}}^*\right) -  B_{f_i} m_{3/2} \left(\tilde{D}_R  \tilde{d}_R^* + \tilde{L} \tilde{l}^* \right)  \\
&& \hspace{1cm} - m^2_{\tilde{D}_R} \tilde{D}_R \tilde{D}_R^*  - m^2_{\tilde{\bar{D}}_R} \tilde{\bar{D}}_R \tilde{\bar{D}}^*_R - m^2_{\tilde{L}} \tilde{L} \tilde{L}^*  + m^2_{\tilde{\bar{L}}} \tilde{\bar{L}} \tilde{\bar{L}}^* \thickspace . 
\end{eqnarray}
\(L_{SB}\) are the soft breaking terms of the MSSM already given in eq.~(\ref{eq:MSSM_SB}). We will neglect in the following the soft breaking parameter \(B_M\). The reason for this is that such a mixing term is not induced at one- or two-loop level. This can be understood as follows: the amplitudes at one- or two-loop level contributing to this term are proportional to \(\cos(2\alpha)\), where \(\alpha\) is the mixing angle of the two scalar messenger particles. The corresponding mixing matrix is given by the diagonalization of eq.~(\ref{eq:MessengerMassMatrix}), i.e. \(\alpha\) must be equal to \(\frac{\pi}{4}\) and the amplitudes vanish. The soft breaking mass terms of the messengers \(m^2_{\tilde{\bar{L}}},\,m^2_{\tilde{L}},\,m^2_{\tilde{\bar{D}}_R}\) and \(m^2_{\tilde{D}_R}\) are given by the corresponding loop diagrams involving the messenger fields. We will calculate them in sec.~\ref{sec:MessengerMass}. 
\paragraph*{Effective Lagrangian} In the case of \(F \simeq  M^2\), there is a large mass splitting in the scalar messenger sector. This has the effect that at energy scales below \(M\) only the light scalar components of eq.~(\ref{eq:ScalarMessengerMass}) are degrees of freedom, since all interactions of the heavy fields are suppressed by inverse powers of its mass \cite{Appelquist:1974tg}. In such cases we can consider  an effective theory in which the heavy scalar messengers and the fermionic messengers are integrated out. The effective operators contributing to masses of the messengers are \(O\left(m^2_{3/2}\right)\), i.e. they are negligible. Furthermore, there are no large contributions to the decays of the messengers coming from higher dimensional operators, so there is no need to calculate them. Hence, the step from the complete to the effective theory can easily be done by just dropping the fermionic messengers. In the scalar sector, the rotation eq.~(\ref{eq:ScalarMessengerMass}) has to be performed and the heavy messenger can be removed from the spectrum afterwards. To this end, we define
\begin{align}
 \tilde{L} = \frac{1}{\sqrt{2}} \left(\tilde{L}_- - \tilde{L}_+ \right), \hspace{1cm} \tilde{\bar{L}} = \frac{1}{\sqrt{2}} \left(\tilde{L}^*_- + \tilde{L}^*_+ \right) \thickspace , \\
\tilde{D} = \frac{1}{\sqrt{2}} \left(\tilde{D}_- - \tilde{D}_+ \right), \hspace{1cm} \tilde{\bar{D}} = \frac{1}{\sqrt{2}} \left(\tilde{D}^*_- + \tilde{D}^*_+ \right) \thickspace .
\end{align}
\(\tilde{L}_-\) and \(\tilde{D}_-\) are the light messengers remaining in the spectrum. 
\paragraph*{Mass eigenstates} After EWSB, the messenger doublet \(\tilde{L}_-\) splits in sneutrino and selectron-like components called \(\tilde{\nu}_-\) and \(\tilde{E}_-\),  respectively. These particle mix with the MSSM sneutrinos and charged sleptons. In addition, the strongly interacting messenger \(\tilde{D}_-\) mixes with the d-squarks of the MSSM. We have summarized the mass eigenstates involving light scalar messenger after EWSB in Table~\ref{tab:mix_GSMB_S_no}. Since the heaviest mass eigenstates of \(\tilde{\nu}\), \(\tilde{e}\) and \(\tilde{d}\) are almost completely messenger-like, we use for them in the following the names \(\tilde{\nu}_M\), \(\tilde{e}_M\) and \(\tilde{d}_M\).
\begin{table}[t]
\begin{center}
\begin{center} \begin{tabular}{|c|c|c|}
\hline Electroweak eigenstates & Mass eigenstates \\
\hline \hline
 \(\tilde{\nu}_L, \tilde{\nu}_-\) & \(\tilde{\nu}_i\) \\
 \(\tilde{e}_{L,i}, \tilde{e}_{R,i}, \tilde{E}_- \) & \(\tilde{e}_i\) \\
 \(\tilde{d}_{L,i}, \tilde{d}_{R,i}, \tilde{D}_- \) & \(\tilde{d}_i\)\\
 \hline \hline
\end{tabular} \end{center}
\end{center}
\caption[Mass eigenstates involving light messengers]{Mass eigenstates after EWSB involving light scalar messengers. The other mass eigenstates are the same as in the MSSM shown in Table~\ref{coll}. We use for the heaviest mass eigenstates of charged sleptons, sneutrinos and down-type squarks a lower index M to assign their messenger nature: $\tilde{e}_M = \tilde{e}_7$, $\tilde{d}_M = \tilde{d}_7$, $\tilde{\nu}_M = \tilde{\nu}_4$. } 
\label{tab:mix_GSMB_S_no}
\end{table} 
\paragraph*{Masses and vertices} We have used \SARAH to calculate all vertices and mass matrices of the model. The mass matrices as well as the vertices responsible for the dominant decays are given apps.~\ref{Masses_GMSB} and \ref{Vertices_GMSB}. All other vertices like the lengthy but negligibly small, scalar interactions can easily be reproduced with the model file \verb"GMSB_eff" for the effective model, which is not a part of the official distribution but listed in app.~\ref{Modelfile_GMSB}. The minimum conditions of the vacuum are  calculated by
\begin{equation}
	T_i = \frac{\partial V}{\partial v_i} = 0 \thickspace, \hspace{1cm} i=d,u \thickspace .
	\label{EqTadPole}
\end{equation}
The results are the same as in the MSSM and read
{\allowdisplaybreaks
\begin{align}
\frac{\partial V}{\partial v_d} &= - v_u \mathrm{Re}\big\{B_{\mu}\big\} + v_d \Big( m_{H_d}^2  \
+  |\mu|^2  + \frac{1}{8} \Big(g_{1}^{2} + g_{2}^{2}\Big)\Big(- v_u^2  + v_d^2\Big)\Big) \thickspace , \\
\frac{\partial V}{\partial v_u} &= - v_d \mathrm{Re}\big\{B_{\mu}\big\} + v_u \Big( m_{H_u}^2  \
+ |\mu|^2  - \frac{1}{8} \Big(g_{1}^{2} + g_{2}^{2}\Big)\Big(- v_u^2  + v_d^2\Big)\Big) \thickspace .
\end{align}
We will solve these equations later on with respect to \(B_\mu\) and \(\mu\) like usually done in the MSSM. }
\paragraph*{Analytical approximation} Before we do a complete numerical analysis of this model, we want to get an estimation for the size of the mixing of the messengers with other fields. The mixing between the messenger fields and ordinary MSSM fields is induced by \eq{deltaW}. For simplicity, we can assume that the couplings of the messenger superfield is generation independent, i.e.~$f_1=f_2=f_3=f$. Relaxing this assumption does not change any of our conclusions. The mixing between the lightest messenger and the SUSY fields in the basis $(\tilde{\nu}_{L_i}, \tilde{\nu}_-)$ is given by
\begin{equation}
\left( \begin{array}{cc}
m_{\tilde{l},ij} + \left(D + f^2 m^2_{3/2}\right) \delta_{ij} & \frac{1}{\sqrt{2}} f m_{3/2} M \\
\frac{1}{\sqrt{2}} f m_{3/2} M &  M^2 - F + D' + \frac{3}{2}f^2 m^2_{3/2} 
\end{array} \right) \thickspace ,
\end{equation}
where $D$ and $D'$ denote  $D$-terms occurring after EWSB with $D = \Ord(M^2_Z)$ and $D'\simeq 0$. The latter holds because $D'$ is proportional to  $\cos 2\alpha$, i.e. such contributions vanish. The induced mixing between the sneutrinos and the neutral messenger scalar is then of the order
 \begin{equation}
 \delta \simeq \frac{f m_{3/2} M}{\sqrt{2} (M^2 - F)}
        \simeq \frac{f m_{3/2} M}{\sqrt{2} m^2_-} \thickspace .
 \end{equation}
\subsection{Corrections to messenger masses}
\label{sec:MessengerMass}
\paragraph*{One-loop mass} The dilution factor is sensitive to the the decay temperature. Therefore, it is of vital importance to know how small $m_-$ can be as this gives an upper bound on the  life-time. At tree level one might argue that \(\sqrt{M^2 - F}\) can be adjusted to arbitrary small values but it  turns out that this doesn't hold at one-loop level. 
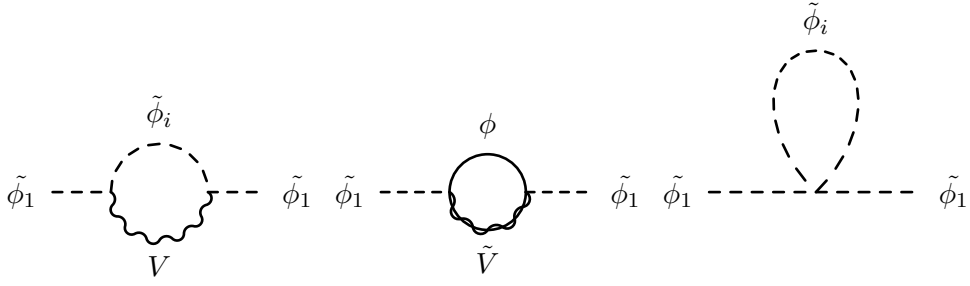
\begin{figure}[t]
\begin{center}
\vspace{0.5cm}
\begin{tabular}{ccc}
\begin{fmffile}{Feynmangraphen/Messenger1}
 \fmfframe(10,10)(10,10){
   \begin{fmfgraph*}(80,80)
    \fmfleft{l1}
   \fmfright{r1}
   \fmf{dashes}{l1,v1}
    \fmf{dashes}{v2,r1}
    \fmf{wiggly,right,tension=0.3,label=$V$}{v1,v2}
    \fmf{dashes,left,tension=0.3,label=$\tilde{\phi}_i$}{v1,v2}
    \fmflabel{$\tilde{\phi_1}$}{l1}
    \fmflabel{$\tilde{\phi_1}$}{r1}
\end{fmfgraph*}}
\end{fmffile}
&
\begin{fmffile}{Feynmangraphen/Messenger2}
  \fmfframe(10,10)(10,10){
    \begin{fmfgraph*}(80,80)
    \fmfleft{l1}
    \fmfright{r1}
    \fmf{dashes}{l1,v1}
    \fmf{plain,right,tension=0.3,label=$\tilde{V}$}{v1,v2}
    \fmf{wiggly,right,tension=0.3}{v1,v2}
    \fmf{plain,left,tension=0.3,label=$\phi$}{v1,v2}
    \fmf{dashes}{v2,r1}
    \fmflabel{$\tilde{\phi_1}$}{l1}
    \fmflabel{$\tilde{\phi_1}$}{r1}
\end{fmfgraph*}}
\end{fmffile}
&
\begin{fmffile}{Feynmangraphen/Messenger3}
  \fmfframe(10,10)(10,10){
    \begin{fmfgraph*}(80,80)
    \fmfleft{l1}
    \fmfright{r1}
    \fmf{dashes}{l1,v1,r1}
    \fmf{dashes, tension=0.5, label=$\tilde{\phi}_i$}{v1,v1}
    \fmflabel{$\tilde{\phi_1}$}{l1}
    \fmflabel{$\tilde{\phi_1}$}{r1}
\end{fmfgraph*}}
\end{fmffile}
\end{tabular}
\end{center}
\vspace{-1cm}
\caption[One-loop corrections to mass of the lightest, scalar messenger]{One-loop corrections to mass of the lightest, scalar messenger. We neglect possible contributions from spurion interactions as well as tiny corrections proportional to $m_{3/2}$. }
\label{graph_1loop}
\end{figure}
We have calculated the one-loop corrections to the mass parameter \(M^2\) and the mixing parameter \(F\) coming from the diagrams in Fig.~\ref{graph_1loop} in \(\overline{\mbox{DR}}\)-scheme. We neglected possible contributions from spurion interactions as well as tiny corrections proportional to $m_{3/2}$.  The amplitudes are given by the generic expressions eqs.~(\ref{eq:SE_FFS}), (\ref{eq:SE_SSSS}) and (\ref{eq:SE_SSV}). We can assume that all involved fields but the messengers are massless, so we can simplify the Passarino-Veltman integrals by
\begin{equation}
 A_0(m^2) = m^2 \left(1 + \Delta + \log \frac{Q^2}{m^2} \right) \thickspace, \hspace{1cm}  B_0(0,0,m^2) = \frac{A_0(m^2)}{m^2} \thickspace.
\end{equation}
\(Q\) is the renormalization scale. The dominant loop corrections are 
\begin{equation}
\delta M^2 \simeq \frac{1}{2 \pi^2} C_2^i g_i^2 M^2 \thickspace, \hspace{1cm} \delta F \simeq \frac{1}{4 \pi^2} C_2^i g_i^2 F  \thickspace ,
\end{equation}
where \(C_2^i\) are the Casimir operators of different gauge groups. The loop corrected mass matrix is calculated by eq.~(\ref{eq:MessengerMassMatrix}) using the shifts \(M^2 \rightarrow M^2 + \delta M^2\) and \(F \rightarrow F + \delta F\). Diagonalization leads to a smaller eigenvalue of
\begin{equation}
\label{eq:OneLoopMass}
m_-^2 = M^2 - F + \left(\delta M^2 - \delta F\right) = M^2 - F + \frac{1}{4 \pi^2} C_2^i g_i^2 \left(2 M^2 - F \right) .
\end{equation}
Hence, we are getting the smallest possible scalar mass in this model by taking the limit \(F \rightarrow M^2\)
\begin{equation}
m_{\text{min}} = \frac{1}{4 \pi^2} C_2^i g_i^2 M^2 \thickspace . 
\end{equation}
The final conclusion is that the lightest messengers has at least a mass of approximately
\begin{equation}
\label{eq:MinMass}
m_{\text{min}} \simeq  0.02 \cdot M^2   \thickspace .
\end{equation}
This large corrections can be easily understood by noting that at the scale $M$, the messenger fermions decouple from the spectrum and, thus, supersymmetry is broken as the number of bosonic degrees of freedom does not match the number of fermionic degrees of freedom leading to large radiative corrections. 

\paragraph*{Effect of RGE running} Additionally, the masses are increased by the RGE running from the SUSY breaking scale to the low energy scale. If we neglect contributions to the running coming from interactions with the spurion \(S\), it is possible to get an analytical expression for the RGEs for the scalar messengers from the SUSY breaking scale to the renormalization scale \(Q\). The starting point are the \(\beta\)-functions \cite{Huang:2000rn}
\begin{eqnarray}
 \label{eq:RGE1}
 \dot{\alpha}_i &=& - b \alpha_i \thickspace , \\
\label{eq:RGE2}
 \dot{m}^2 &=& - 2 \alpha_i C_2^i m^2 \thickspace .
\end{eqnarray}
Solving eq.~(\ref{eq:RGE1}) gives \(\alpha_i(Q) = \frac{\alpha_i(M)}{1+b \alpha_i(M) t} \) with \(t=\ln\frac{M^2}{Q^2}\). We can insert this in the solution of eq.~(\ref{eq:RGE2}) 
\begin{equation}
m(Q)^2 = \frac{m(M)^2}{1+2 \alpha_i(M) C_2^i t}  
\end{equation}
and obtain the final expression
\begin{equation}
\frac{d}{dt} m^2(Q) = \frac{m(M)^2}{1-\frac{1}{4 \pi} 2  C_2^i \alpha_i(Q)^2 t } \thickspace .
\end{equation}
Note, that the radiative corrections as well as the RGE running induces an additional mass splitting between the members of the 5-plet caused by the different Casimirs. 
\subsection{Decays of messengers}
\begin{figure}
\begin{center}
\includegraphics[scale=1.]{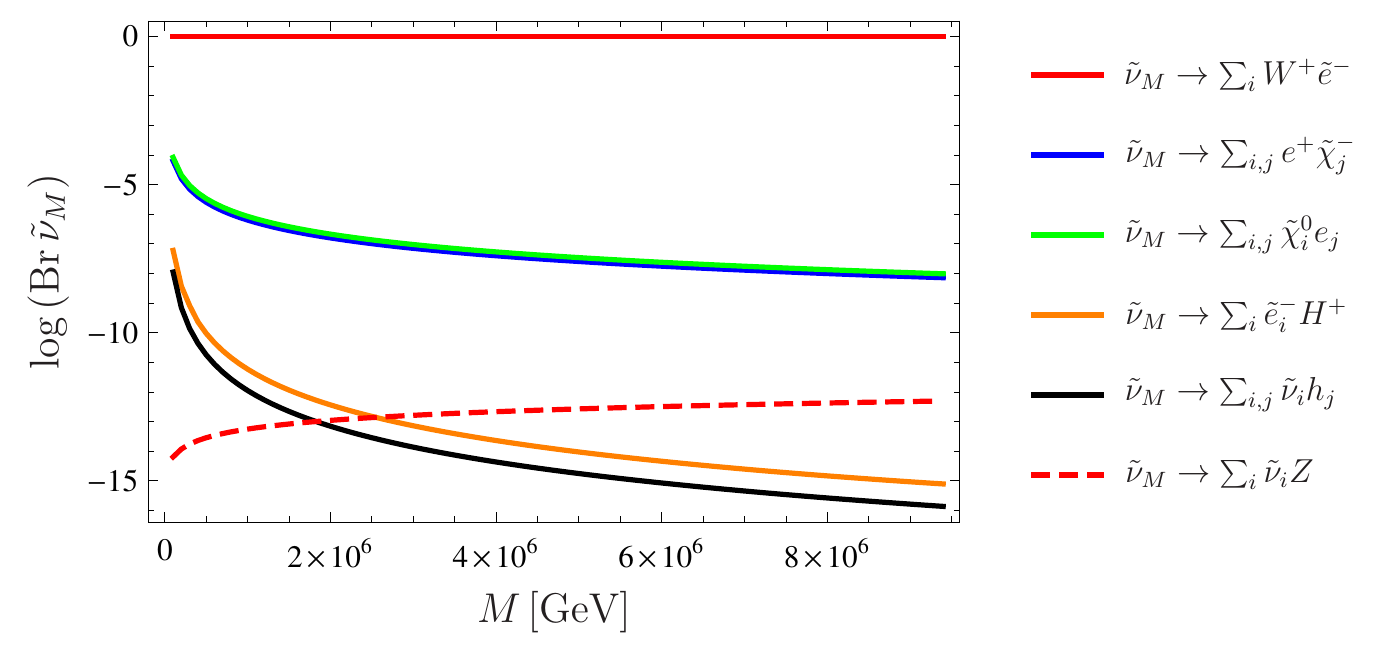} 
\end{center}
\caption[Branching ratios of sneutrino-like messengers]{Branching ratios for the decay of the lightest, sneutrino-like messenger. In this plot $\Lambda$ is fixed by the assumption that the messenger has the lightest possible mass and $f=1$. For the messenger mass the one-loop corrections are take into account.}
\label{fig:BrV}
\end{figure}
\begin{figure}[t]
\begin{center}
\includegraphics[scale=1.]{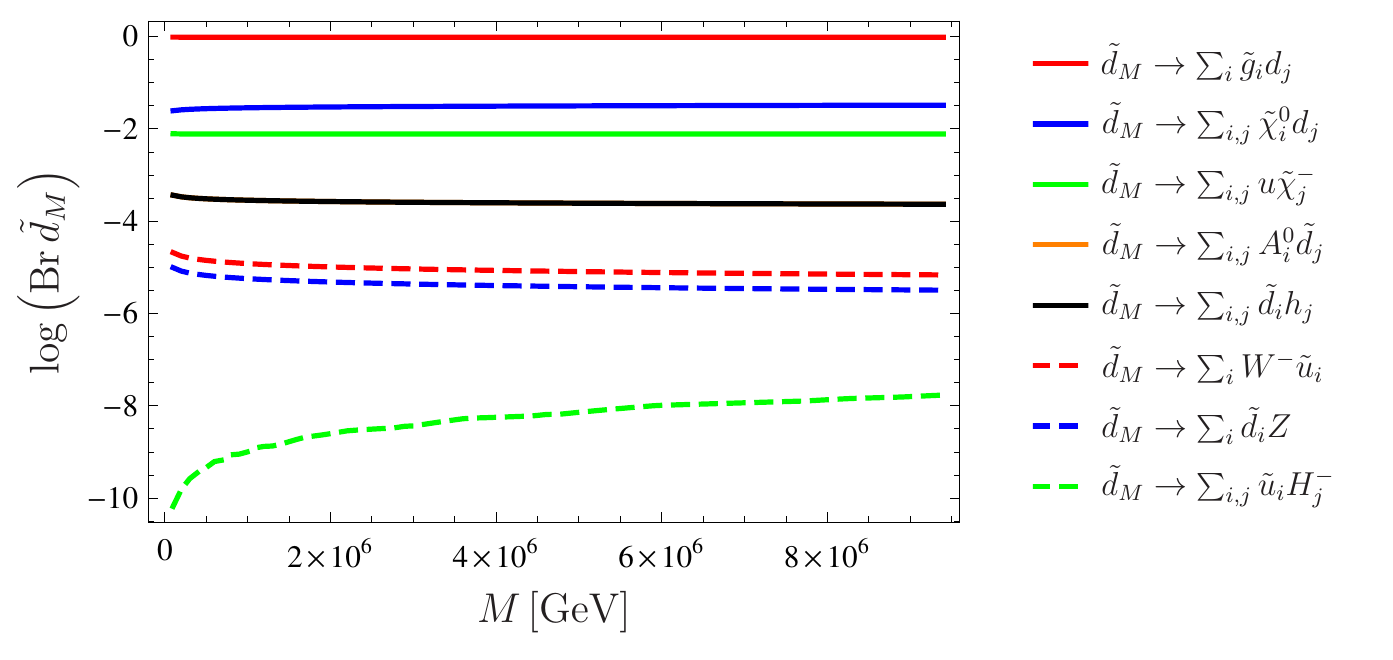} 
\end{center}
\caption[Branching ratios of squark-like messengers]{Branching ratios for the decay of the lighter, strongly interacting messenger. The same assumptions and input parameters as in Fig.~\ref{fig:BrV} are used.}
\label{fig:BrD}
\end{figure}

We  have now all necessary information for a complete numerical analysis of the messenger decays. We will focus here on the decays of the light sneutrino-like and squark-like messengers because they are the only ones which might be cosmologically relevant: in the model with conserved messenger number and conserved $R$-parity one has $m_+ > M > m_- \gg m_{SUSY} \gg m_{3/2}$. Both, $f m_{3/2}$ and $R$-parity violating parameters $\epsilon_i$, as we will see later,  are small compared to $m_{SUSY}$ implying that the induced mixing will only give small corrections to the various masses. This spectrum gives rise to decays of the following type: $\tilde \phi_+ \to \tilde B  \phi$ and $ \phi \to \tilde B \tilde \phi_-$, where $\phi$ is the messenger fermion. These decays are so fast that $\tilde \phi_+$ and $\phi$ will decay immediately after their decoupling from the thermal bath. The lightest selectron messenger \(\tilde{E}_-\) can decay into the sneutrino like messenger \(\tilde{\nu}_-\) and \(W^+\) if the induced mass splitting coming from D-terms is large enough. If this decay is kinematically forbidden, the decay $\tilde{E}_- \rightarrow \tilde{\nu}_- f \bar{f}$ due to an off-shell \(W^+\) takes place. Both happens instantaneously on cosmological scales  and long before the freeze out of the gravitino.\\
We turn now to the cosmologically relevant decays. The different branching ratios after EWSB for the  neutral messenger \(\tilde{\nu}_M\)  are shown in Fig.~\ref{fig:BrV} and for the strongly interacting messenger \(\tilde{d}_M\), they are shown in Fig.~\ref{fig:BrD}.  \\
Obviously, there is a big hierarchy in the decay widths of the lightest messengers. It has been shown  \cite{Bartl:1998xk,Bartl:1999bg} that decays into $W$  and $Z$ bosons can dominate the decays of supersymmetric scalar particles if there is sufficient phase space. It turns out that in our scenario, the decay into $Z$ bosons are suppressed via an extended  GIM mechanism. However, this is not the case for decays into $W$ bosons and one gets for the  corresponding  widths 
\begin{equation}
\label{eq:WidthW}
\Gamma \simeq \frac{g^2}{16 \pi} \frac{m_-^2}{M_W^2}\delta^2 m_- \thickspace.
\end{equation}  
The additional factor \(\frac{m_-^2}{M_W^2} \), a consequence of the polarization sum, implies that this  decay mode dominates once the lightest messenger scalar has a mass of a few TeV, hence, its life time gets reduced by the inverse factor.
\subsection{Entropy production by messenger decays}
\label{sec:entroy}
\begin{figure}[t]
\begin{minipage}{16cm}
\includegraphics[scale=0.95]{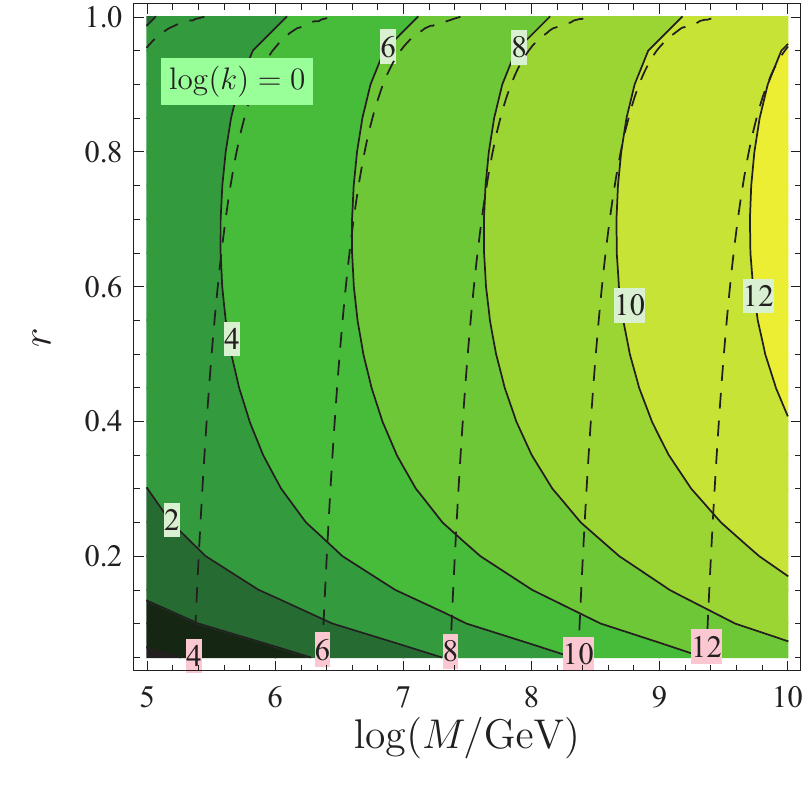} 
\hfill
\includegraphics[scale=0.95]{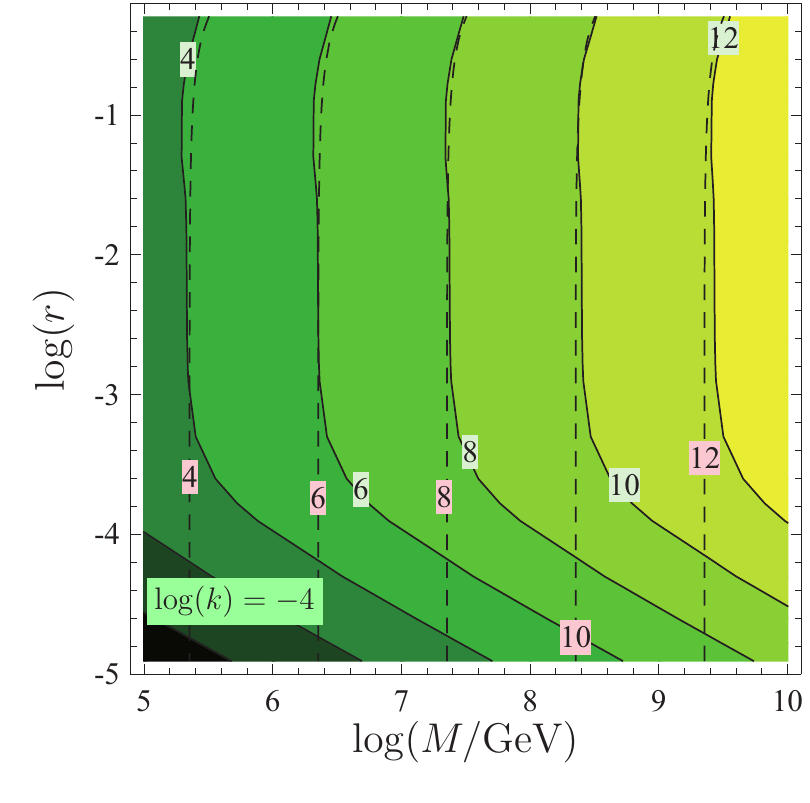} 
\end{minipage}
\caption[Relic density of sneutrino Messengers]{Relic density of the $\tilde{v}_L$-like messengers against the approximation of \cite{Han:1997wn}. The dashed lines are the approximation of \cite{Han:1997wn} while the contour plot shows the result of a calculation with \micrOmegas. Left: large values of $r=\frac{\Lambda}{M}$ and $k=1$. Right: small values of $r$ and $k=10^{-4}$. The large discrepancies are mostly an effect of Goldstino annihilation.}
\label{fig:RelicDensity_D1}
\end{figure}

\begin{figure}[t]
\begin{minipage}{16cm}
\includegraphics[scale=0.77]{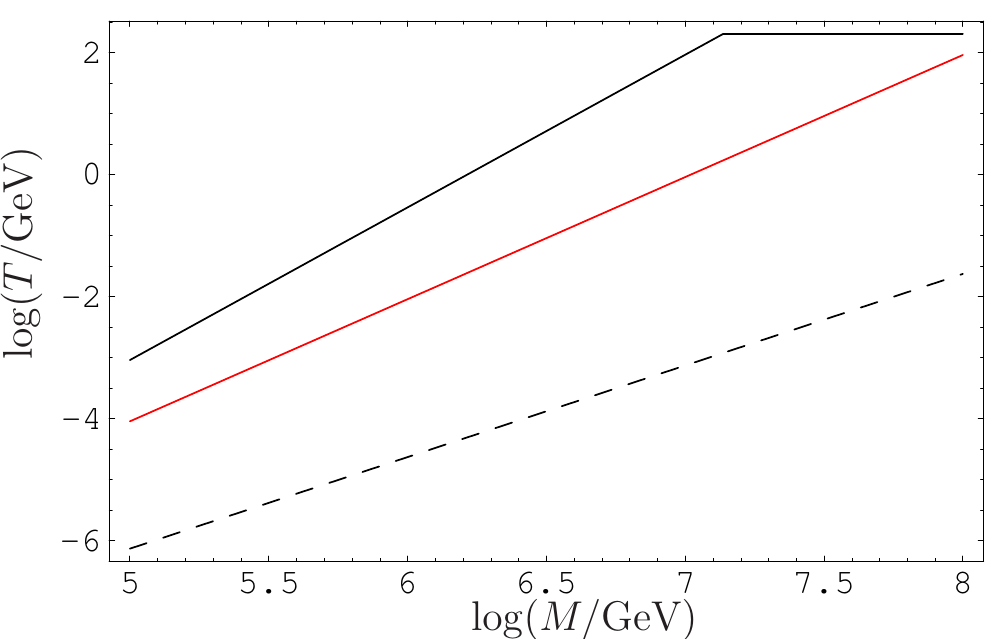} 
\hfill
\includegraphics[scale=0.77]{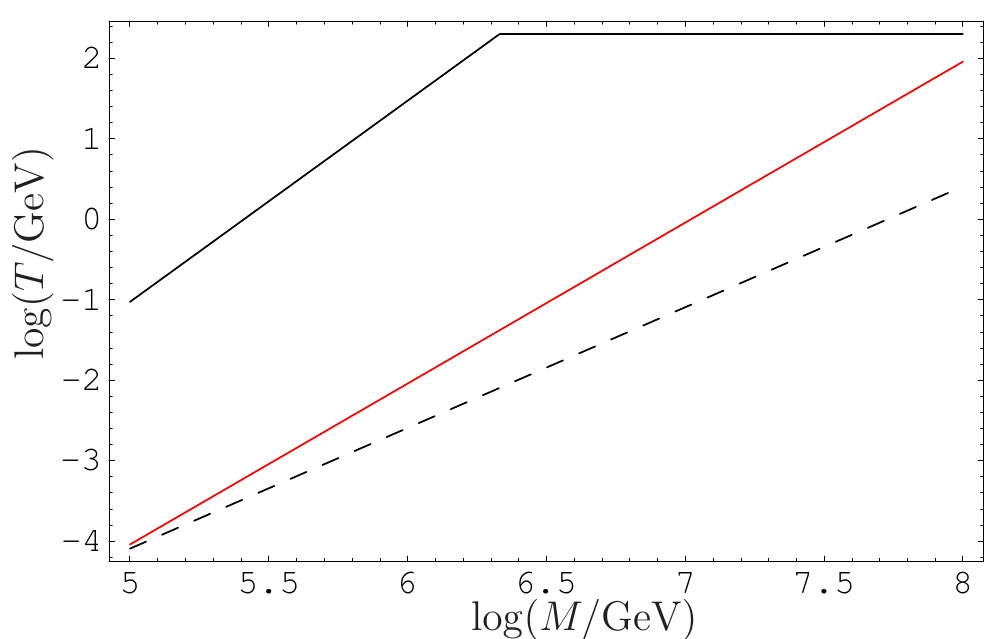} 
\end{minipage}
\caption[Decay temperature of messengers]{Comparison of the decay temperature of the messenger to the freeze out temperature of the gravitino (red line). The black, solid line shows the temperature including the decay in $W^-$, the dashed line the temperature without this channel. Left plot for $M^2 \gg F$ and right plot for $M^2 \cong F$, both for $k=1,f=1$. } 
\label{fig:plot_decay}
\end{figure}

\paragraph*{Relic density of the gravitinos} We have used eq.~(\ref{eq:relic}) for calculating the relic density of the gravitinos which have been once in thermal equilibrium. Since this formula is sensitive to the effective degrees of freedom, we have calculated those at the freeze out temperature of the gravitino \cite{Gondolo:2000ee}. This leads to sharp edges in the isocurves when the freeze out temperature crosses the mass threshold of particles. For analyzing the dependence on the reheating temperature, we have used the formula  \cite{Moroi:1993mb}
\begin{equation}
\label{eq:Omega32_TR}
\Omega_{3/2} h^2 = 10^7 \frac{T_{RH}}{1\, \TeV} \frac{M_3^2}{M_P\, m_{3/2}} 
\end{equation}
for gravitinos which have never been in thermal equilibrium.
\paragraph*{Yield of the messengers} It was already pointed out by the authors of \cite{Jedamzik:2005ir} that the approximation for the relic density of the messengers given in eq.~(\ref{eq:Relic_Messenger}) is not always valid.  Therefore, we will calculate for our studies the yield of the different messengers using \micrOmegas \cite{Belanger:2006is}. We show in Fig.~\ref{fig:RelicDensity_D1} a comparison between our numerical results and the approximation of eq.~(\ref{eq:Relic_Messenger}).  We see that the approximation is excellent in some parameter regions, but fails in others totally. This is the case for large values of $k$ because the Goldstino annihilation dominates, which have not been taken into account in \cite{Han:1997wn}. Also for large values of $r$, a larger discrepancy is obvious. This is based on the fact that the analytical calculation was done for the limit $M^2 \gg F$. 
\paragraph*{Decay temperature} We assume that the decay temperature \(T_D\) of the messengers depends only on the  decay width \(\Gamma\), i.e. we ignore the time which the messengers need for freezing out. That this is a very reasonable approximation can be seen as follows. The decay temperature is given by 
\begin{equation}
T_D  = \left(g_* \pi^2 \right)^{-1/4} \sqrt{\tilde{\Gamma} M_P} \thickspace .
\end{equation}
Here, we have defined \(\tilde{\Gamma}\) as
\begin{equation}
 \tilde{\Gamma} = \left(\tau_f + \tau_D\right)^{-1} \thickspace ,
\end{equation}
where \(\tau_f\) is the freeze out time and \(\tau_D = \Gamma^{-1}\) is the inverse decay width. The freeze out time is connected to the freeze out temperature by
\begin{equation}
\tau_f = \frac{M_P}{T_f^2} \sqrt{g_* \frac{\pi^2}{90}}  \thickspace . 
\end{equation}
Our numerical calculations with \micrOmegas show that \(X_f = \frac{M}{T_f}\) is of \(\Ord(10)\). This leads for a width \(\Gamma \simeq \frac{g^2}{16 \pi} \delta^2 m_-\) to an estimated ratio of decay to freeze out time of \(\frac{M_P M}{\Lambda^2} \gg 1\) if the decays don't involve massive vector bosons in the final states. Hence, \(\tilde{\Gamma} \simeq  \tau_D^{-1} = \Gamma \). This is later important for the decays of fields like \(\tilde{d}_M\) or \(\tilde{e}_M\) which also have to be considered as we will see. The decay of \(\tilde{\nu}_M\) is determined by the \(W\)-channel which just opens after EWSB. Therefore, that case is again independent of the freeze out time. 
\paragraph*{Entropy production in time} The freeze out temperature \(T_{3/2}\) of the gravitino is
\begin{equation}
\label{eq:T32}
T_{3/2} = 0.62 \frac{m_{3/2}^2 M_P \sqrt{g_* }}{\alpha_S M_\SG^2} \thickspace . 
\end{equation}
The gluino mass is according to eq.~(\ref{eq:SoftBreakingGMSB}) given by 
\(m_{\tilde{g}} \simeq \frac{\alpha_s}{4 \pi} \Lambda\).  As we  only need  a rough estimate of this temperature, we use eq.~(\ref{eq:m32}) and set all parameters of \(\Ord(1)\) to 1. As result, we obtain \(T_{3/2}\) as function of the messenger mass \(M\): \(T_{3/2} \simeq 10^{-13} \frac{M^2}{k^2\,\mbox{GeV}}\). Comparing this  to the decay temperature of the scalar messengers in Fig.~\ref{fig:plot_decay}, one sees that the neutral messenger always decays before the freeze-out of the gravitinos has taken place. Obviously, the simplest scenario proposed in literature doesn't work. We will now consider the possibilities of adjusting \(f\) or the reheating temperature and examine the effect of the other kinds of messenger fields.

\section{Results for minimal messenger sector}
\subsection{Weakly interacting messengers}
\begin{figure}[t]
\begin{minipage}{16cm}
\includegraphics[scale=0.95]{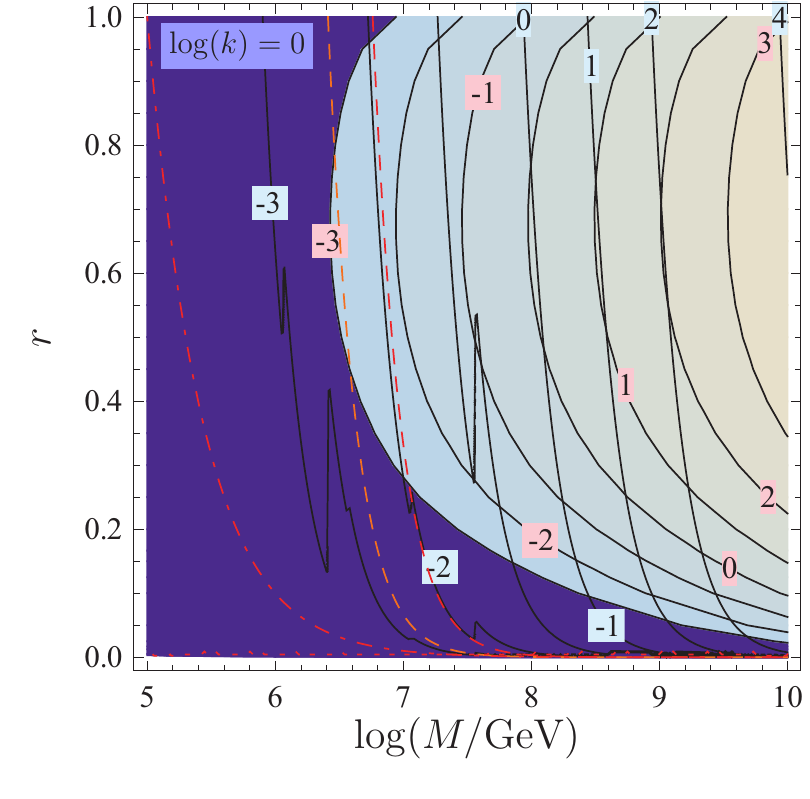} 
\hfill
\includegraphics[scale=0.95]{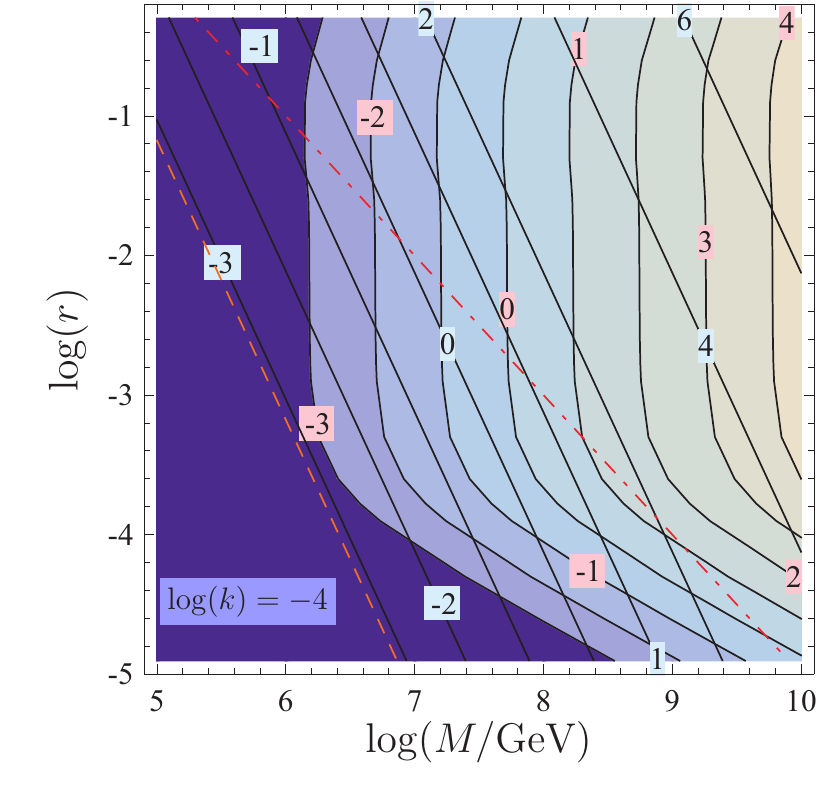} 
\end{minipage}
\caption[Yield of sneutrino-like messengers for sufficient dilution]{Calculated (black lines) against needed (contour plot) values of $\log(Y_{\tilde{\nu}_M} m_{\tilde{\nu}_M})$ in case of sneutrino-like messengers. The decay temperature is fixed to $T_D = 1$~MeV and a diluted relic density of gravitinos is demanded to be $\Omega_{3/2}=0.23$. The right plot is for values of $r$ of $\Ord(1)$, the left for smaller values. The exclusion lines are: $m_{3/2} < 1.6 $~keV (orange dashed line), $m_{3/2} < 8.0 $~keV (red dashed line), $\Lambda < 10^5 $~GeV (red dot dashed line).}
\label{fig:Nu_TD}
\end{figure}

We start with the lightest messenger, the only one that has so far been considered for entropy production in literature for the minimal messenger sector of GMSB. We have already seen in sec.~\ref{sec:entroy} (Fig.~\ref{fig:plot_decay}) that the proposed scenario fails for \(f=1\) because the messengers decay always before the gravitino has frozen out. On the other hand, \(f\) can still be considered as free parameter. Therefore, it is interesting to see if a change of \(f\) leads to a satisfying solution. The dilution is given by \(\Delta = 1 + \frac{4}{3} \frac{m_{\tilde{\nu}_M} Y_{\tilde{\nu}_M}}{T_D} \) (see eq.~(\ref{eq:dilutionfactor})), i.e. for fixed yield $Y_{\tilde{\nu}_M}$ and mass $m_{\tilde{\nu}_M}$ we get the maximal dilution for the smallest possible decay temperatures \(T_D\). The bounds from BBN on the life time of heavy particles are given in \cite{Kawasaki:2004qu,Kanzaki:2007pd,Kawasaki:2007xb}: the minimal temperature allowed for the messenger decays is roughly 1~MeV.  Therefore, we will test if it is in principle possible to reach a sufficient dilution due to the decay of sneutrino-like messengers by choosing $T_D=1$~MeV. 
The plots in Fig.~\ref{fig:Nu_TD} compare the calculated values of $m_{\tilde{\nu}_M} Y_{\tilde{\nu}_M}$ with the values needed for sufficient entropy production. The depicted exclusion lines are $m_{3/2} < 1.6 $~keV (orange dashed line) and $m_{3/2} < 8.0 $~keV (red dashed line) coming from Lyman-$\alpha$ forest observations, $T_D < 1 $~MeV (red dotted line) in order not to spoil BBN and $\Lambda < 10^5 $~GeV (red dot dashed line) to have SUSY masses fulfilling the LEP bounds. It seems that a tuning of \(f\) to smaller values can really lead to a successful dilution of the gravitinos. The preferred parameter range is roughly \(10^9\,\GeV > M > 10^8\, \GeV\) and \(0.8 > r > 0.4\). However, we have not yet considered the effect of the  other kinds of messenger fields. 

\subsection{Influence of the strongly interacting messenger}
\begin{figure}[!ht]
\begin{minipage}{16cm}
\includegraphics[scale=0.95]{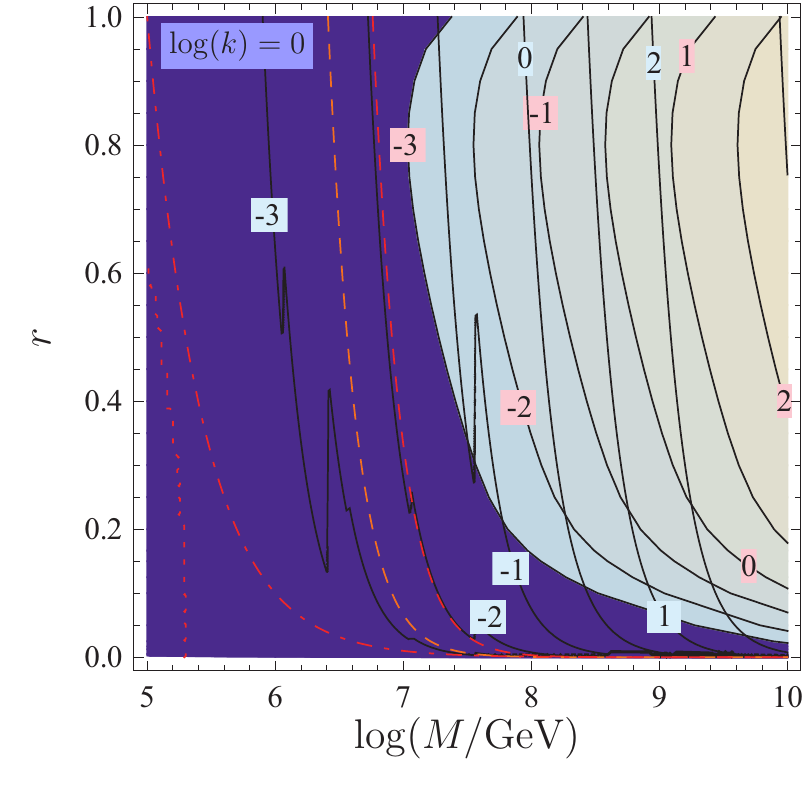} 
\hfill
\includegraphics[scale=0.95]{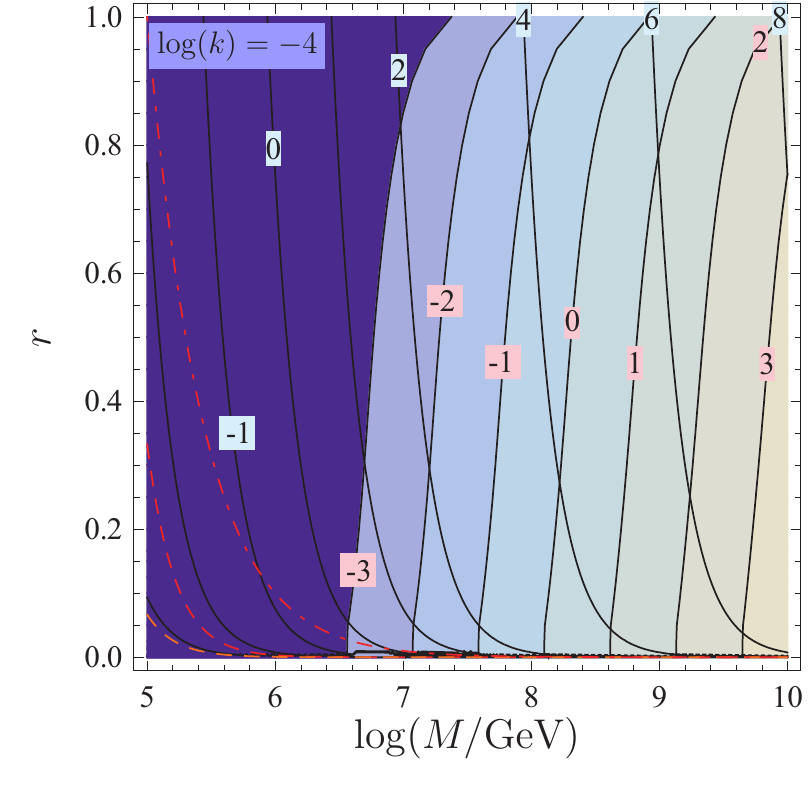}  \\
\includegraphics[scale=0.95]{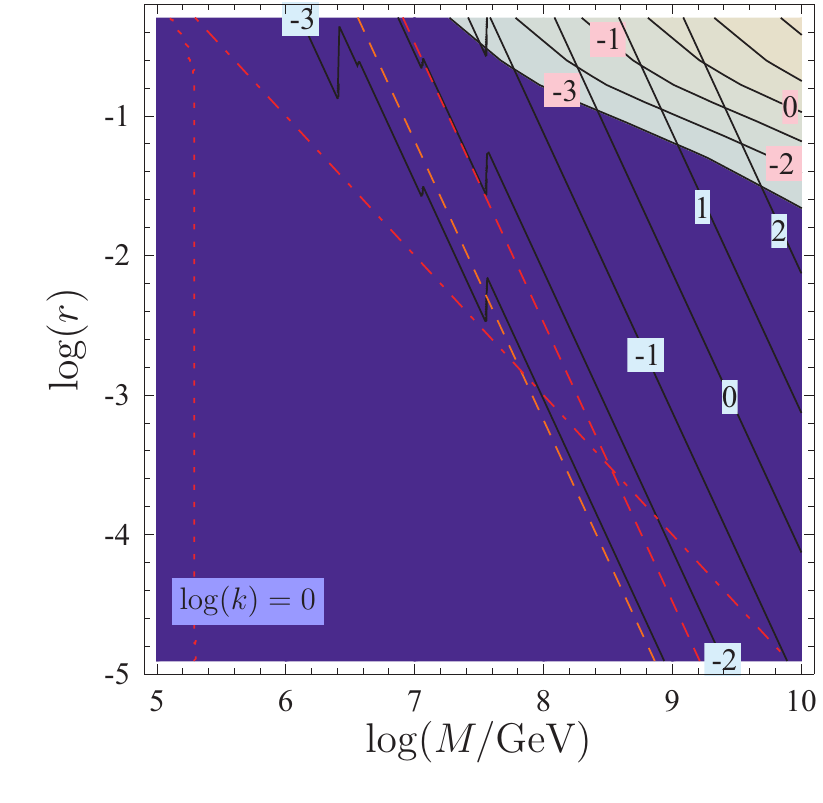} 
\hfill
\includegraphics[scale=0.95]{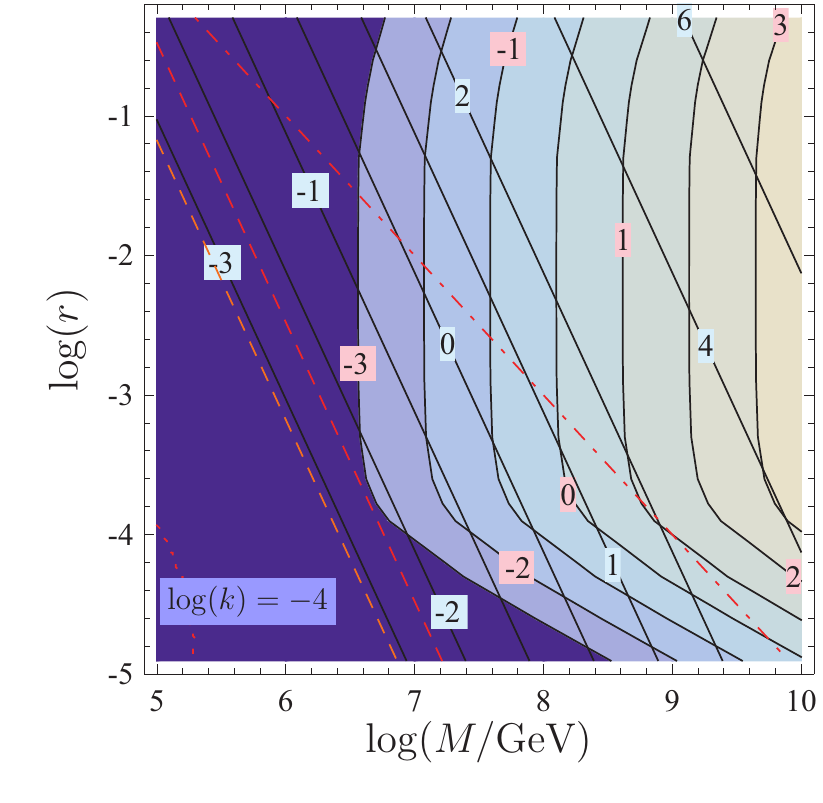} 
\end{minipage}
\caption[Tuning of $f$ in case of strongly interacting messengers]{Tuning of $f$ in case of the $SU(3)_C$-messengers. In these plots $T_D = 1\,\MeV$ is fixed. The contour plot shows the needed values of $\log(Y_{\tilde{d}_M} m_{\tilde{d}_M})$ to dilute the relic density to $\Omega_{3/2}=0.23$ while the blick lines give the calculated values of $\log(Y_{\tilde{d}_M} m_{\tilde{d}_M})$. Obviously, there is no solution. The exclusion lines are the same as in Fig.~\ref{fig:Nu_TD}. The kinks are an effect of the different degrees of freedom depending on the freeze out temperature of the gravitinos, see discussion in sec.~\ref{sec:entroy}.}
\label{fig:TD_D1}
\end{figure}
\begin{figure}[!ht]
\begin{minipage}{16cm}
\includegraphics[scale=0.95]{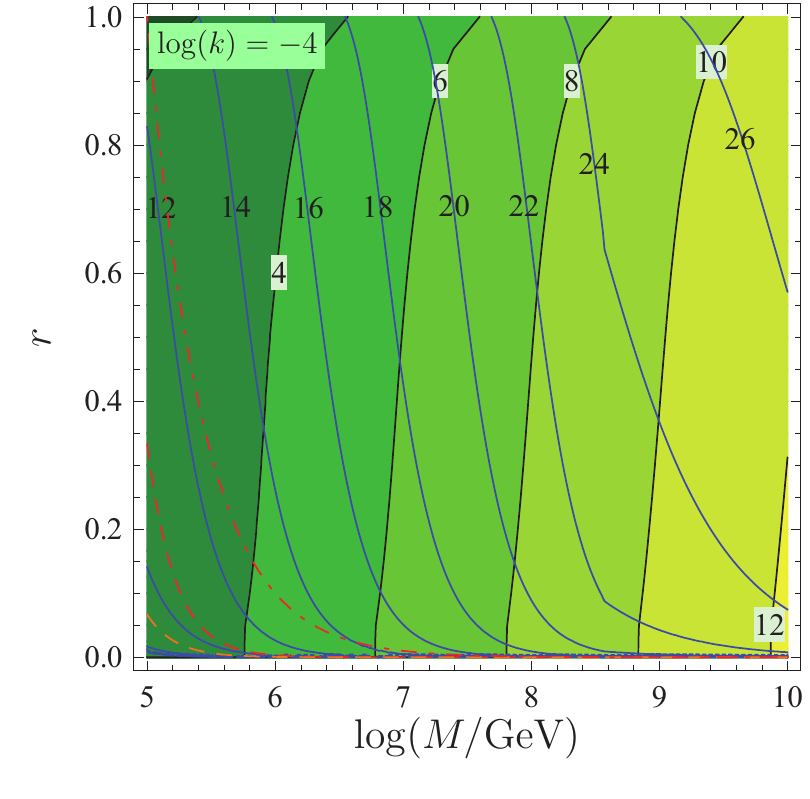} 
\hfill
\includegraphics[scale=0.95]{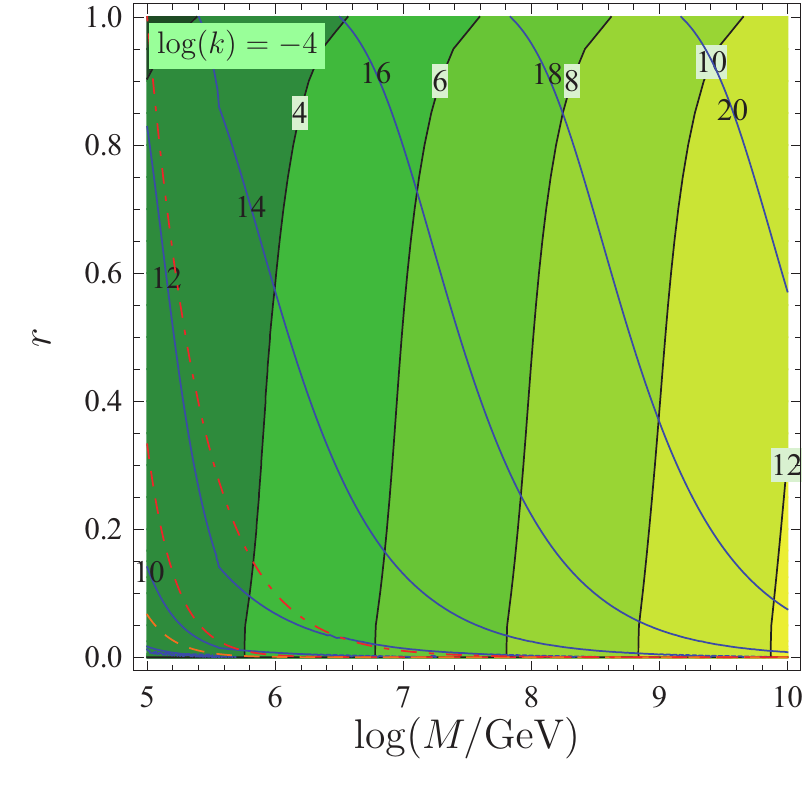} 
\end{minipage}
\caption[Dependence on reheating temperature of the dilution  by strongly interacting messengers]{The black lines are the needed value of $\log(Y_{\tilde{d}_M} m_{\tilde{d}_M})$ to cause  $\Omega_{3/2}=0.23$ while the contour plot shows the calculated values of $\log(Y_{\tilde{d}_M} m_{\tilde{d}_M})$. Left: $T_{RH}=10^{12}$~GeV. Right: $T_{RH}=10^{6}$~GeV. The exclusion lines are the same as in Fig.~\ref{fig:Nu_TD}. We used here $f=1$ and $k=10^{-4}$.}
\label{fig:TRH_D1}
\end{figure}

So far, we have not taken the $SU(3)_C$-messengers into account. They can just decay in the lightest messenger due to interactions of the additional \(SU(5)\) gauge bosons \(X\) and \(\bar{X}\). However, those decays are suppressed by the mass of these particles which are highly constraint by bounds on proton decay. Hence, the strongly interacting messenger decay dominantly into SUSY particles because of the mixing induced by the superpotential term eq.~(\ref{eq:deltaW}).  As it can be seen from Fig.~\ref{fig:BrD}, the decay into gluino and SM fermion dominates and the width can be approximated by 
\begin{equation}
\Gamma_D \simeq \frac{g_3^2}{4 \pi} \left(\frac{f m_{3/2}}{m_{\tilde{d}_M}}{M^2}\right)^2 m_{\tilde{d}_M} \thickspace.
\end{equation}
Decays in \(W^-\) are just possible due to a tiny left-right mixing. Hence, there is no enhancement like in case of the decays in massive vector bosons and the width of $\tilde{d}_M$ is always smaller then the decay width of $\tilde{v}_M$ by a factor $\frac{m_{\tilde{\nu}_M}^2}{M_W^2}$. So, the tuning of $f$ to a value that the decay of the weakly interacting messengers happens to an appropriate time for dilution leads to another cosmological problem: in this region of parameter space the life-time of the $SU(3)_C$ messenger gets so large that it destroys the successful BBN predictions. They are decaying at temperature in the low keV range or below. At this time they are dominating the energy density of the universe, so the energy injection due to their decays would dissociate BBN products to an very high extent. In addition, the reheating temperature \(T_{RH}\) of this decay is according to eq.~(\ref{eq:entropyDecay}) related to the decay temperature \(T_D\) and the dilution factor \(\Delta^{\tilde{d}_M}\) by 
\begin{equation}
T^{\tilde{d}_M}_{RH} = T^{\tilde{d}_M}_D \sqrt[3]{\Delta^{\tilde{d}_M}} \thickspace .
\end{equation}
This reheating is too low to restart BBN again, especially since the number density of the  \(SU(3)_C\)-messengers gets diluted in the same way by the decay of the \(SU(2)_L\)-messengers. To get an rough impression of the reheating temperature, we can approximate \(Y_{\tilde{d}_M} \simeq Y_{\tilde{\nu}_M}\) and \(m_{\tilde{d}_M} \simeq m_{\tilde{\nu}_M}\). This leads to 
\begin{equation}
T^{\tilde{d}_M}_{RH} = T^{\tilde{d}_M}_{RH} \left(\frac{Y_{\tilde{d}_M}}{Y_{\tilde{\nu}_M}} \frac{m_{\tilde{d}_M}}{m_{\tilde{\nu}_M}} \frac{T_D^{\tilde{\nu}_M}}{T_D^{\tilde{d}_M}}\right)^{\frac{1}{3}} \simeq 
T^{\tilde{d}_M}_D \left(\frac{g}{g_3} \frac{{\tilde{\nu}_M}}{M_W}\right)^{\frac{1}{3}}, 
\end{equation}
i.e. the reheating temperature remains in the keV range.\\ 
The last possibility for a sufficient dilution in GMSB models with a gauge sector coming in $\bf 5$ and $\bf \bar{5}$ under $SU(5)$, which doesn't spoil BBN without assuming very low reheating temperatures, is to tune the decay width of the strong interacting messenger: maybe, it is possible to produce enough entropy just by the decays of \({\tilde{d}_M}\) for very low decay temperatures. We can do the same analysis as in case for the sneutrino-like messengers. Since the main annihilation channel is into gluons, the relic density of the strongly interacting messenger is decreased by a factor \(\left(\frac{\alpha}{\alpha_s}\right)^2\) in comparison to the weakly interacting messenger. So, the relic density of $SU(3)_C$ messenger is for a large region in the parameter space about two orders smaller then the relic density of the $SU(2)_L$ messengers. We do not expect a solution for those areas. On the other side, there are also regions, in which these messengers have a larger relic density: for the limit, in which only Goldstino interactions are dominant, the relic density of $\tilde{d}_M$ is exactly three times the relic density of $\tilde{\nu}_M$. This is  a direct consequence of the three colors. Therefore, we have done again a check for the complete $(M,r)$-plane by setting the decay temperature to 1~MeV through a tuning \(f\). We plotted the needed value of $Y_{\tilde{d}_M} m_{\tilde{d}_M}$ for a sufficient dilution in comparison with the calculated value by \micrOmegas. The result is that there exists no region, in which this scenario can work in principle. We show as example the results for \(k=0\) and \(k=10^{-4}\) in Fig.~\ref{fig:TD_D1}.\\
Also adjusting the reheating temperature \(T_{RH}\) to small values, so that the gravitinos were never in thermal equilibrium does not work for a reasonable choice of parameters. We show this at the example of Fig.~\ref{fig:TRH_D1}, where we set \(f=1\) and \(k = 10^{-4}\). The left plot shows the result for \(T_{RH} = 10^{12}\)~GeV, the right one for \(T_{RH} = 10^{6}\)~GeV. The change in the slope at which the gravitinos never reached thermal equilibrium is obvious, but that change is much too small. However, there might already be solutions without the need for a dilution if the gravitino is heavy and the reheating temperature is low enough. The relic density is given by eq.~(\ref{eq:Omega32_TR}).  We can use \(M_3 \approxeq \Lambda\) and \(m_{3/2} \approxeq \frac{\Lambda M}{M_P}\) to get 
\begin{equation}
\Omega_{3/2} h^2 \approxeq 10^7 \frac{T_{RH}}{1\, \TeV} \frac{\Lambda}{M} \thickspace ,
\end{equation}
i.e. the relic density depends for a fixed reheating temperature just on the ratio $r$. Therefore, it is possible to find solutions for given reheating temperatures by using fine-tuned values of \(r\). However, this would be often in conflict with Baryogenesis and is completely independent of the messenger sector. \\
Moreover, it is not possible to adjust the parameters in a way that a combined dilution resulting from the decays of \(\tilde{\nu}_M\) and \(\tilde{d}_M\) leads to a correct gravitino abundance and circumvents BBN bounds. The reason is that the ratios of the yields and of the decay widths are fixed by the gauge couplings, the messenger mass and \(W^-\) mass. It has been checked that this result holds for all possible messenger-matter-interacting terms presented in \cite{Jedamzik:2005ir}.\\
\vspace{0.5cm}\\
To summarize, the minimal model of GMSB suffers from a big cosmological gravitino problem which can not be solved by late time decays of the messengers. The reason is that the decay of the weakly interacting messengers is largely enhanced by the decay channel in massive vector bosons. Therefore, they decay for a natural choice of parameters too early to produce significant or sufficient entropy. A fine-tuning of their decay width could produce the needed entropy in a small area of parameter space, but this is ruled out by the strongly interacting messengers that, independently of the concrete realization of the mixing, decay at temperatures some order below. Therefore, the question is now if extensions of the messenger sector can cure that problem. Before that, we  discuss  also the effect of $R$-parity violation.

\section{Effect of bilinear \texorpdfstring{$R$}{R}-parity violation on entropy production}
\label{sec:GMSB_RpV}
We have seen that there is no solution to the cosmological gravitino problem in minimal GMSB by the decays of the messengers. In this section, we show that this doesn't change even if we take the effects of bilinear $R$-parity violation into account. 

\subsection{The Model including \texorpdfstring{$R$}{R}-parity and messenger number violation}
We consider the mixing terms between messenger and MSSM fields of eq.~(\ref{eq:deltaW}) and add the bilinear $R$-parity violating terms of eq.~(\ref{eq:RpV}). The complete superpotential reads
\begin{equation}
W = W_{MSSM} +  M \hat{D}_R \hat{\bar{D}}_R + M \hat{L} \hat{\bar{L}} + f_i m_{3/2} \left(\hat{D} \hat{d} + \hat{L} \hat{l} \right) + \epsilon_i \hat{l}_i \hat{H}_u \thickspace . 
\end{equation}
We have suppressed again the color and \(SU(2)_L\) indices. \(\epsilon_i\) are the bilinear $R$-parity violating parameters and we choose \(f_i = f\) as free parameter. The corresponding soft breaking terms are 
\begin{eqnarray}
\nonumber \La_{SB} &=& \La_{SB,MSSM} - B_{\epsilon_i} \epsilon_i \SL_i \SHZ - B_M \left(\tilde{D}_R \tilde{\bar{D}}^*_R + \tilde{L} \tilde{\bar{L}}^*\right) -  B_{f_i} m_{3/2} \left(\tilde{D}_R  \tilde{d}_R^* + \tilde{L} \tilde{l}^* \right)  \\
&& - m_{H l } \left(\SL \SHE^* +\SL^* \SHE \right)- m^2_{\tilde{D}_R} \tilde{D}_R \tilde{D}_R^*  - m^2_{\tilde{\bar{D}}_R} \tilde{\bar{D}}_R \tilde{\bar{D}}^*_R+ m^2_{\tilde{L}} \tilde{L} \tilde{L}^*  - m^2_{\tilde{\bar{L}}} \tilde{\bar{L}} \tilde{\bar{L}}^* \thickspace . 
\end{eqnarray}
The $R$-parity violating parameters \(\epsilon\) and \(m_{H l}\) are fixed by neutrino data as discussed in sec.~\ref{sec:R-parity}. We neglect again the soft breaking terms \(B_M\), since they aren't generated in GMSB.  In this scenario, the sneutrino-like messenger and the sneutrinos receive a VEV after EWSB like the neutral Higgs according to eq.~(\ref{higgsvev}) do
\begin{equation}
\tilde{\nu}_L^i =  \frac{1}{\sqrt{2}} \left(\phi^L_i + i \sigma^L_i + v_{L_i} \right)  \thickspace, \hspace{1cm} \tilde{\nu}_- =  \frac{1}{\sqrt{2}} \left(\phi_- + i \sigma_- + v_{M} \right) \thickspace .
\end{equation}
As we will see, \(v_{L_i}\) are of \(\Ord(10^{-2})\)~GeV and \(v_M\) is much smaller. The scalar and pseudo scalar components of these fields mix separately with the components of the neutral Higgs fields to give six CP even and odd mass eigenstates. Other eigenstates are build by mixings of the three neutrinos with the neutralinos. In addition, the 'charginos' are a mixture of charged SM leptons and MSSM charginos. In the squark sector, a mixing between the down squarks and the lightest, strongly interacting messenger takes place.  We have summarized the mixings in Table~\ref{tab:mix_GSMB_S}. For fermions, we have written the Weyl components for the electroweak eigenstates, but expressed the mass eigenstates already in 4-component Dirac spinors. Since the heaviest eigenstates of \(h_i\), \(A^0_i\), \(\tilde{e}_i\) and \(\tilde{d}_i\) consists almost completely out of their messenger components, we name them \(h_M\), \(A^0_M\), \(\tilde{e}_M\) and \(\tilde{d}_M\). \\ 
\begin{table}[t]
\begin{center}
\begin{center} \begin{tabular}{|c|c|c|}
\hline Electroweak Eigenstates & Mass Eigenstates \\
\hline \hline
 \(\phi_d, \phi_u, \phi^L_i, \phi_- \) & \(h_i\) \\
 \(\sigma_d, \sigma_u, \sigma^L_i, \sigma_-   \) & \(A^0_i\) \\
 \(\tilde{e}_{L,i}, \tilde{e}_{R,i},H_d^-, \left(H_u^+\right)^*, \tilde{E}_- \) & \(H^-_i\) \\
 \(\tilde{d}_{L,i}, \tilde{d}_{R,i}, \tilde{D}_- \) & \(\tilde{d}_i\)\\
\hline
 \(\nu_i, \tilde{H}_d^0, \tilde{H}_u^0,\tilde{B},\tilde{W}^0\) & \(\Neu_i=(\lambda^0_i,\bar{\lambda}^0_i)\) \\
 \(e_L,\tilde{W}^-, \tilde{H}_d^-\) / \(e^*_R,\tilde{W}^+, \tilde{H}_u^+\)  & \(\Cha_i=(\lambda^+_i,\bar{\lambda}^-_i)\) \\
\hline \hline
\end{tabular} \end{center}
\end{center}
\caption[Mass of a model with messenger number and $R$-parity violation]{Mass eigenstates after EWSB. The additional mixings in comparison to the MSSM are an effect off bilinear $R$-parity violation and messenger number violation. We name the heaviest eigenstates $h_M = h_6$, $A^0_M = A^0_6, \tilde{e}_M = \tilde{e}_9$ and $\tilde{d}_M = \tilde{d}_7$. } 
\label{tab:mix_GSMB_S}
\end{table} 
\subsection{Analysis}
We have again created an input file for \SARAH to calculate all mass matrices and vertices for this model. The model file is given in app.~\ref{Modelfile_GMSB_RpV}. In addition, we give in apps.~\ref{Masses_GMSB_RpV} and \ref{Vertices_GMSB_RpV} the mass matrices and vertices for the neutral messengers which are different to the case of conserved $R$-parity. Using this model file, also the tadpole equations were computed. The six equations are 
{\allowdisplaybreaks
\begin{align} 
\frac{\partial V}{\partial v_d} = \, & \frac{1}{8} \Big(v_d \Big(8 m_{H_d}^2  + \Big(g_1^2 + g_2^2\Big)\Big(- v_u^2  + v_d^2\Big)\Big)+8 v_d |\mu|^2 -8 v_u \mathrm{Re}\big\{B_{\mu}\big\}\nonumber \\ 
 &+8 \sum_{j=1}^{3} \mathrm{Re}\big\{m^{2}_{{H l},{j}} \big\} v_{L,{j}}  +\Big(g_1^2 + g_2^2\Big)v_d \sum_{j=1}^{3}{v^{2}_{L,{j}}} -8 \mathrm{Re}\big\{\mu\sum_{j=1}^{3}v_{L,{j}} \epsilon_{j}\big\}  \Big) \thickspace ,\\ 
\frac{\partial V}{\partial v_u} = \, & \frac{1}{8} \Big(8 v_u |\mu|^2 -8 v_d \mathrm{Re}\big\{B_{\mu}\big\}- \Big(g_1^2 + g_2^2\Big)v_u \sum_{j=1}^{3}{v^{2}_{L,{j}}} -4 \sqrt{2} v_M m_{3/2} \sum_{j=1}^{3} \mathrm{Re}\big\{f_{j} \epsilon_{j} \big\} \nonumber \\ 
 &+v_u \Big(8 m_{H_u}^2  + 8 \sum_{j=1}^{3}{|\epsilon|^{2}_{j}}  - \Big(g_1^2 + g_2^2\Big)\Big(- v^2_u  + v^2_d\Big)\Big)+8 \sum_{j=1}^{3}v_{L,{j}} \mathrm{Re}\big\{B_{\epsilon_{j}}\big\}  \Big) \thickspace ,\\ 
\frac{\partial V}{\partial v_{L_i}} = \, & \frac{1}{8} \Big(8 v_d \mathrm{Re}\big\{m^{2}_{{H l},{i}}\big\} +2 m_{3/2} \Big(2 m_{3/2} \mathrm{Re}\big\{f_{i} \sum_{j=1}^{3}f_{j} v_{L,{j}} \big\}  + 2 \sqrt{2} v_M \mathrm{Re}\big\{ M_L f_{i}- B_{f_{i}} \big\}\Big)\nonumber \\ 
 &+4 \sum_{j=1}^{3}\Big( m^2_{l,{i j}} +  m^2_{l,{j i}} \Big)v_{L,{j}} - \Big(g_1^2+g_2^2\Big) \Big(\Big(v_u^2 - v_d^2\Big) v_{L,{i}} - \sum_{j=1}^{3}{v_{L,{j}}}^{2} v_{L,{i}}\Big)\nonumber \\ 
 &-8 v_d \mathrm{Re}\big\{\mu \epsilon_{i} +8 \sum_{j=1}^{3}v_{L,{j}} \epsilon_{j}  \epsilon_{i}+8 v_u B_{\epsilon_{i}}\big\} \Big) \thickspace ,\\ 
\frac{\partial V}{\partial v_M} = \, & \frac{1}{4} \Big(2 \Big(- 2 F_L  + m_{\tilde{L}}^2 + m_{\tilde{\bar{L}}}^2\Big)v_M +4 v_M M_L^2 +2 \sqrt{2} M_L m_{3/2} \mathrm{Re}\big\{\sum_{j=1}^{3}f_{j} v_{L,{j}} \big\}  \nonumber \\ 
 &+2 v_M m_{3/2}^2 \sum_{j=1}^{3} |f_{j} |^2 - 2 \sqrt{2}  m_{3/2} \mathrm{Re}\big\{ \sum_{j=1}^{3}B_{f_{j}} v_{L,{j}} \big\} -2 \sqrt{2}  v_u m_{3/2} \mathrm{Re}\big\{\sum_{j=1}^{3}f_{j} \epsilon_{j} \big\}  \Big) \thickspace .
\label{eq:tadVM}
\end{align}
The $R$-parity violating parameters \(\epsilon_i\) and \(v_{L_i}\) were calculated with the routine {\tt RPTools} of \SPheno, which does a fit of those parameters to neutrino data. The used relations and experimental values were described in sec.~\ref{sec:R-parity}. Furthermore, we have used \(f_1=f_2=f_3=1\). The VEV of the neutral messenger is fixed by the tadpole equations: if complex phases and terms of \(\Ord(m_{3/2}^2)\) are neglected, an approximated solution of eq.~(\ref{eq:tadVM}) is given by
\begin{equation}
\label{eq:VM}
v_M = \frac{\sqrt{2} m_{3/2} \left(M \sum_i v_i f_i - v_u \sum_i f_i \epsilon_i - \sum_i v_i B_{f_i} \right) }{2 (F - M^2) - m^2_{\tilde{L}}  - m^2_{\tilde{\bar{L}}} } \simeq \frac{m_{3/2}}{\sqrt{2} M} \thickspace .
\end{equation}
Since we used equal values for \(f_i\), we also set \(B_{f_2}=B_{f_3} = B_{f_1} \equiv B_f\). In addition, we have chosen  \(B_{\epsilon_i} \equiv B_{\epsilon}\). Using these assumptions, the remaining set of tadpole equations was solved with respect to  \(B_f, \mu, B_\mu, B_{\epsilon}\) and \(m_{H l}\). }

\begin{figure}[htb]
\begin{center}
\includegraphics[scale=1.]{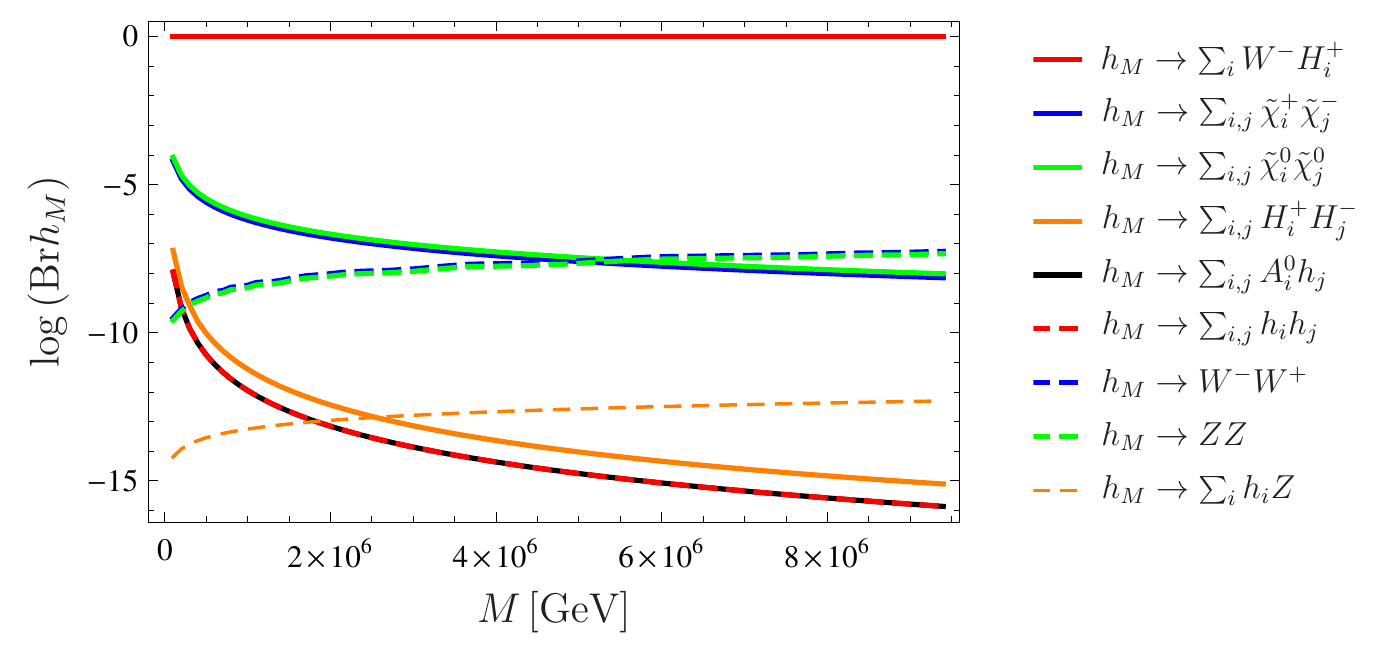} 
\end{center}
\caption[Branching ratios of sneutrino-like messengers]{Branching ratios for the decay of the neutral, scalar messenger. In this plot, $\Lambda$ is fixed by the assumption that the messenger has the lightest possible mass and $f=1$. For the messenger mass the one-loop corrections are take into account. All other parameters were chosen as described in the text.}
\label{fig:BrV_RpV}
\end{figure}

\begin{figure}[htb]
\begin{minipage}{16cm}
\includegraphics[scale=1.]{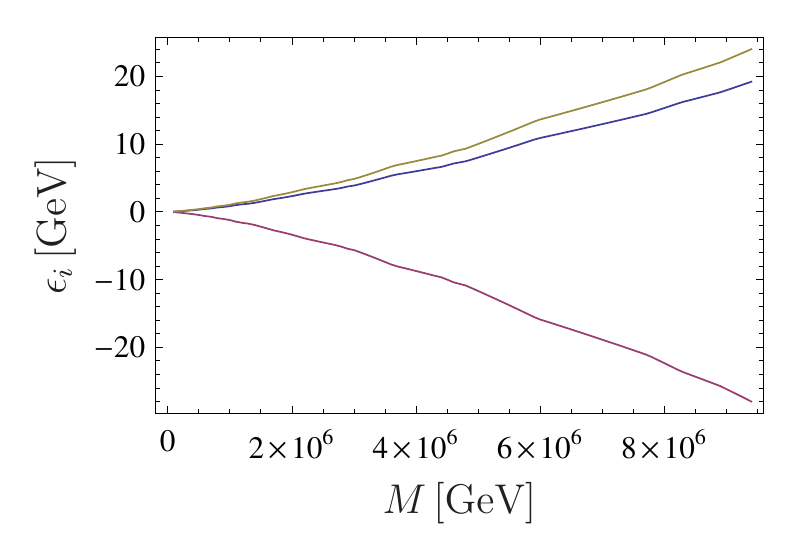} 
\hfill
\includegraphics[scale=1.]{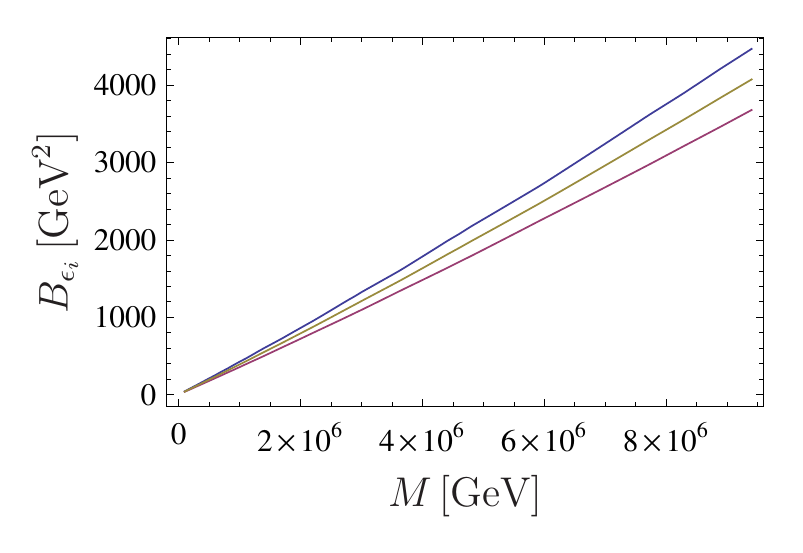} \\
\includegraphics[scale=1.]{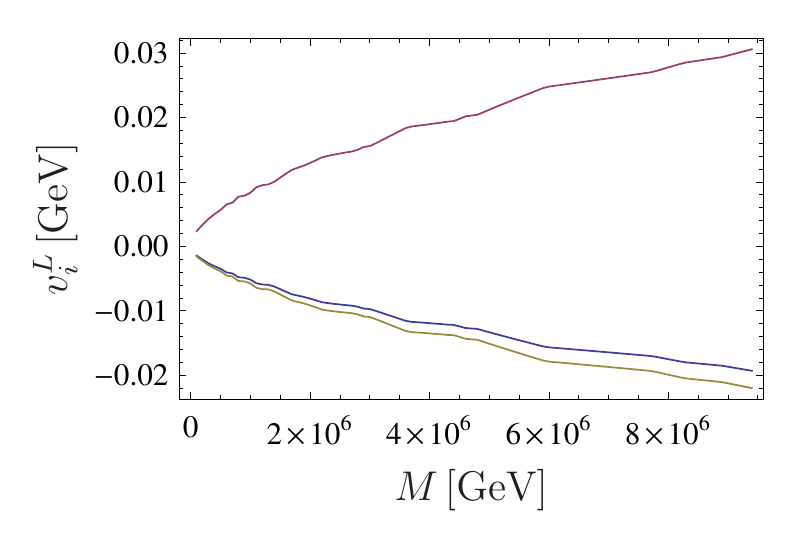} 
\hfill
\includegraphics[scale=1.]{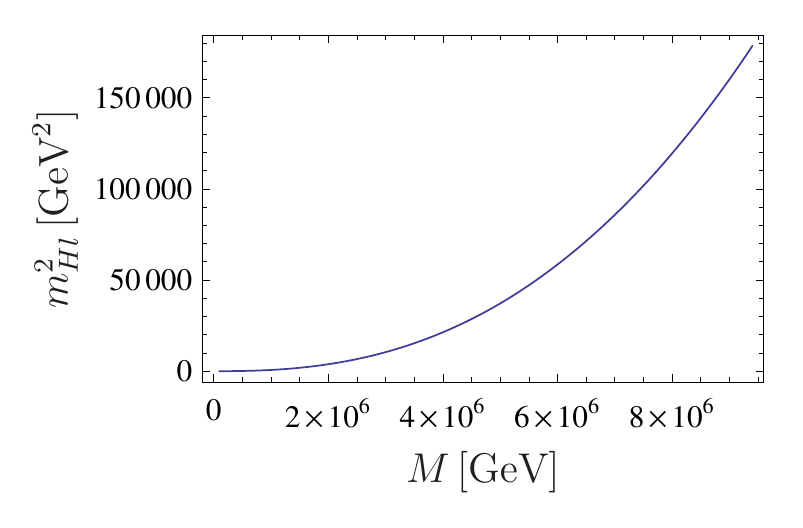} 
\end{minipage}%
\caption[$R$-parity violating parameters]{Values for the parameters associated with bilinear $R$-parity violation. Upper row: bilinear superpotential couplings $\epsilon_i$ and the corresponding soft breaking couplings  $B_{\epsilon_i}$. The second row shows the VEVs of the sneutrinos (left) and the soft breaking parameter $m_{Hl}^2$ (right).}%
\label{fig:Rparameters}%
\end{figure}%

\subsection{Results}
The inclusion of $R$-parity violating doesn't change the decay properties of the messenger particles significantly. The reason is that all $R$-parity violating interactions of the messenger fields are suppressed doubly: first, by the tiny mixing with MSSM fields, secondly, by the small $R$-parity violation. In Fig.~\ref{fig:BrV_RpV}, we plotted the different branching ratios of the neutral, scalar messenger. Obviously, there is no visible different for the branching ratios in comparison with their equivalents in  case of conserved $R$-parity shown in Fig.~\ref{fig:BrV}. Furthermore, also the new decays channels with two vector bosons in the final state are much smaller than the decay in a scalar and \(W^-\). The reason for that is the tiny VEV of the messengers according to eq.~(\ref{eq:VM}). Therefore, cosmologically relevant difference to the scenario with conserved $R$-parity might just come from late time decays of MSSM particles, especially of the lightest neutralino. \\
To give an impression about the size of the $R$-parity violating parameters, we have depicted in Fig.~\ref{fig:Rparameters} common values for those parameters, which are consistent with neutrino data. Obviously, we can neglect the decays of MSSM particles, in particular of the lightest neutralino, induced by those parameters of two reasons: the decays generated by the $R$-parity violating parameters are suppressed by factors proportional to \(\epsilon/M_\chi\). Here, \(M_\chi\) is an common neutralino mass of \(\Ord(100)\)~GeV. This ratio is several orders bigger than \(m_{3/2}/M\) what defines the order of the messenger decays. Hence, the decay temperature of MSSM particles is significantly larger. As we have also  seen when calculating the minimal possible mass of the messenger fields (see eq.~(\ref{eq:MinMass})), the mass of the lightest messenger is always some orders larger than the masses of MSSM fields, i.e. they always dominate the energy budget of the universe and can therefore produce a lot more of entropy. These two facts together with the dilution factor in eq.~(\ref{eq:dilutionfactor}) show that the entropy production by $R$-parity violating decays won't play any role in contrast to the messenger decays induced by eq.~(\ref{eq:deltaW}). Because of the huge difference in the masses of the messengers and SUSY particles this even holds if we drop the demand to explain neutrino data and set the $R$-parity violating parameters to much smaller values in order to suppress the decays.

\begin{figure}[t]
\begin{minipage}{16cm}
\includegraphics[scale=0.95]{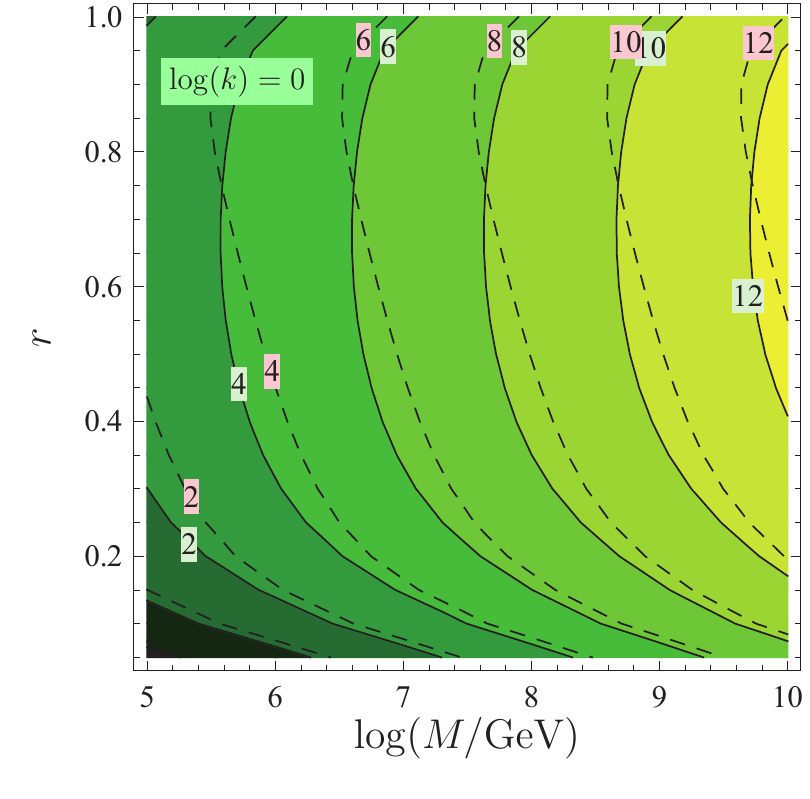} 
\hfill
\includegraphics[scale=0.95]{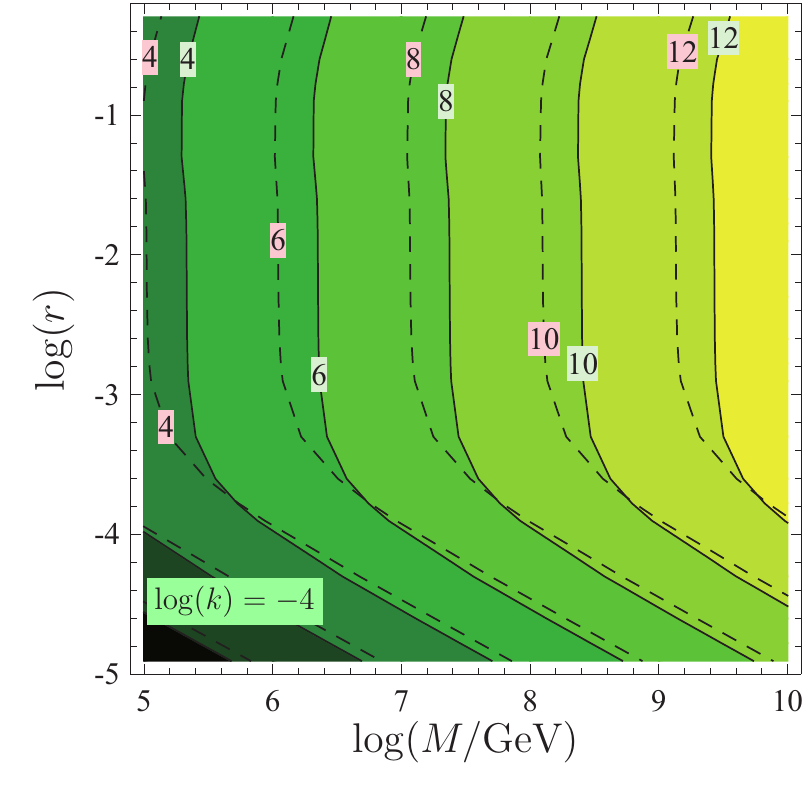} 
\end{minipage}
\caption[Comparison of the relic density of the $\tilde{v}_L$-like messengers with the $\tilde{e}_R$-like Messengers]{Comparison of the relic density of the sneutrino-like messengers (solid lines) with the selectron-like Messengers (dashed lines). The labels show $\log\Omega_{\tilde{\nu}_M}$ and $\log\Omega_{\tilde{e}^R_M}$, respectively.}
\label{fig:Omega_eR_vL}
\end{figure}

\begin{figure}[htb]
\begin{minipage}{16cm}
\includegraphics[scale=0.95]{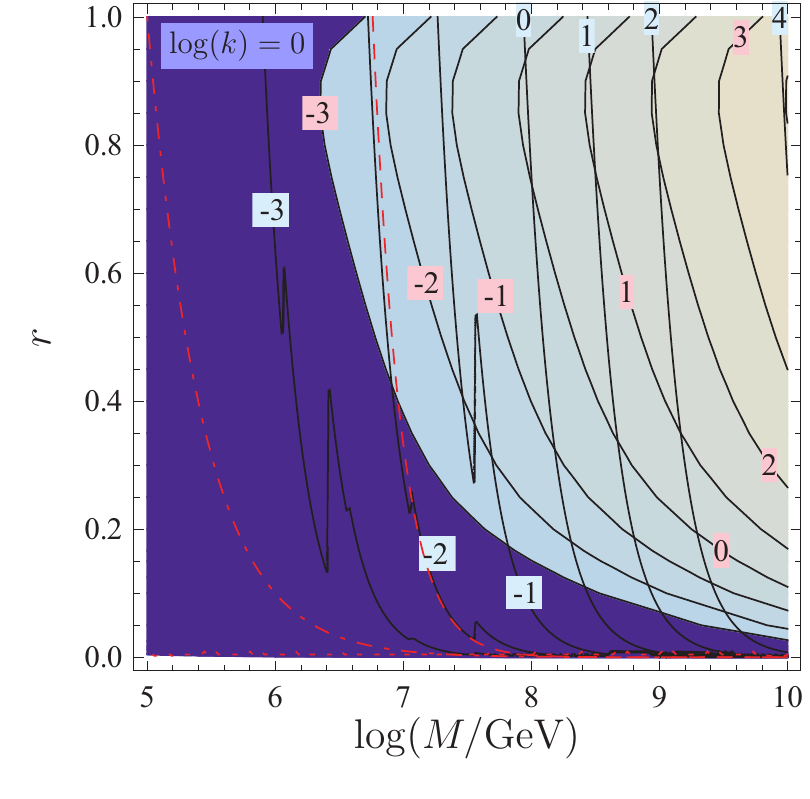} 
\hfill
\includegraphics[scale=0.95]{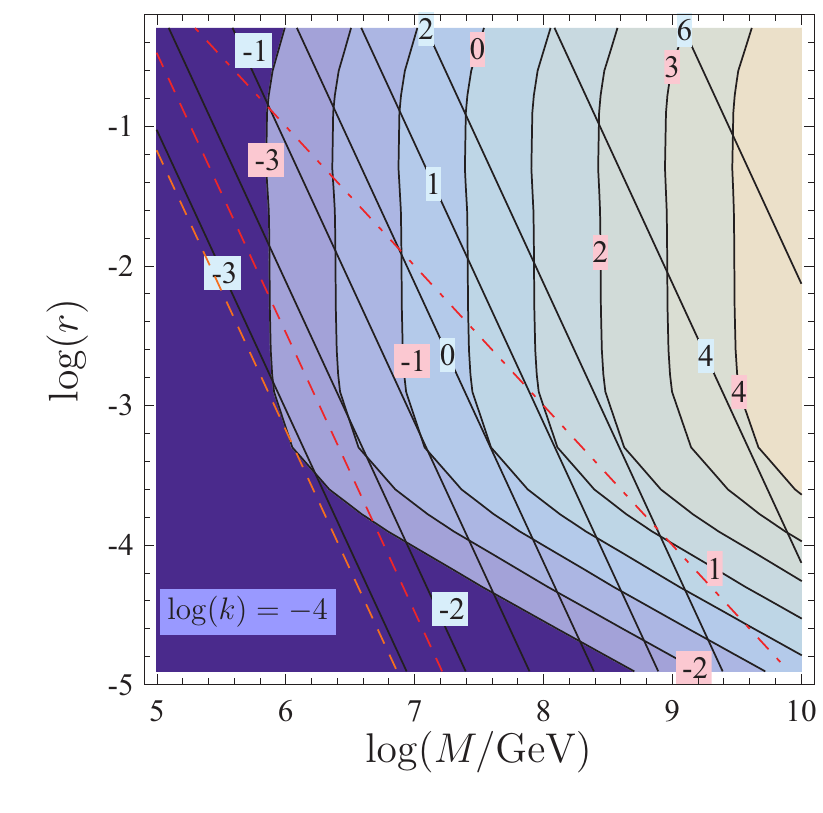} 
\end{minipage}
\caption[Yield and dilution for $U(1)_Y$-messengers]{Comparison of the needed value of $\log(Y_{\tilde{e}^R_M} m_{\tilde{e}^R_M})$ (contour plot) against the calculated one (black lines) in  case of $U(1)_Y$-messenger. The decay temperature is $T_D = 1$~MeV and the demanded, diluted relic density is $\Omega_{3/2}=0.23$. The exclusion lines are  $m_{3/2} < 1.6 $~keV (orange dashed line), $m_{3/2} < 8.0 $~keV (red dashed line), $\Lambda < 10^5 $~GeV (red dot dashed line).  The kinks are an effect of the different degrees of freedom depending on the freeze out temperature of the gravitino, see sec.~\ref{sec:entroy}. In this setup, there are tiny areas providing a solution to the gravitino problem.}
\label{fig:eR_TD}
\end{figure}

\section{Non-minimal messenger sectors}
The picture changes a little bit when the messenger multiplets contain $U(1)_Y$ charged particles which are \(SU(3)_C\times SU(2)_L\) singlets, i.e.\ \(e_R\)-like messengers. The light scalar components have a relic  density comparable to \(\tilde{v}_L\)-like messengers but their decay width is not enhanced by decays into vector-bosons.  Therefore, these messengers can have the smallest decay width of all messengers. Another possibility would be to analyze the case of a 16-plet of \(SO(10)\), which includes a gauge singlet. We discuss both cases in the following. 

\subsection{\texorpdfstring{$U(1)_Y$}{U1-Y}-Messenger}

\begin{figure}[!ht]
\begin{minipage}{16cm}
\includegraphics[scale=0.95]{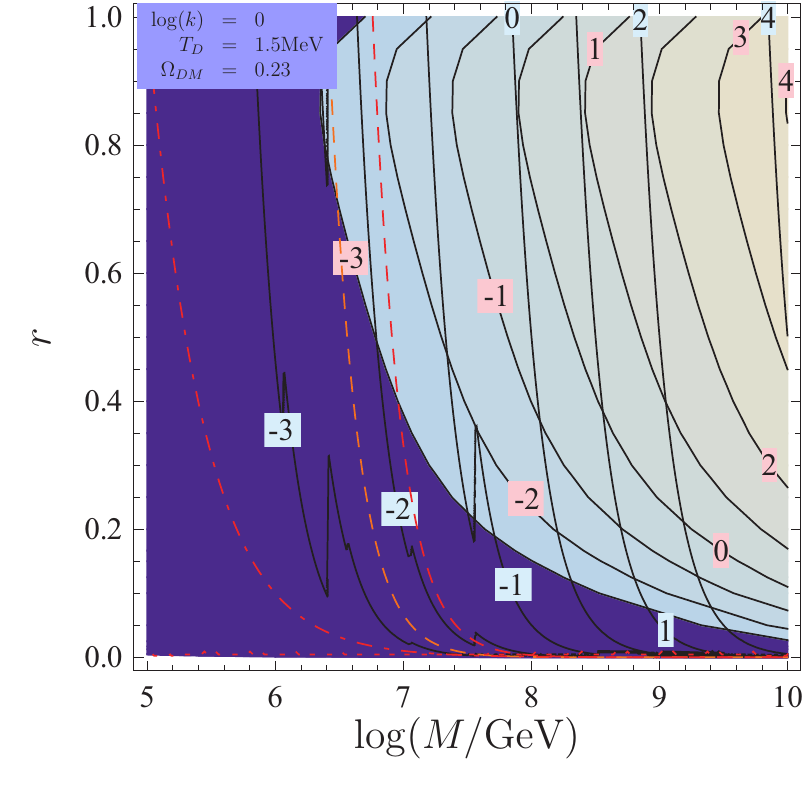} 
\hfill
\includegraphics[scale=0.95]{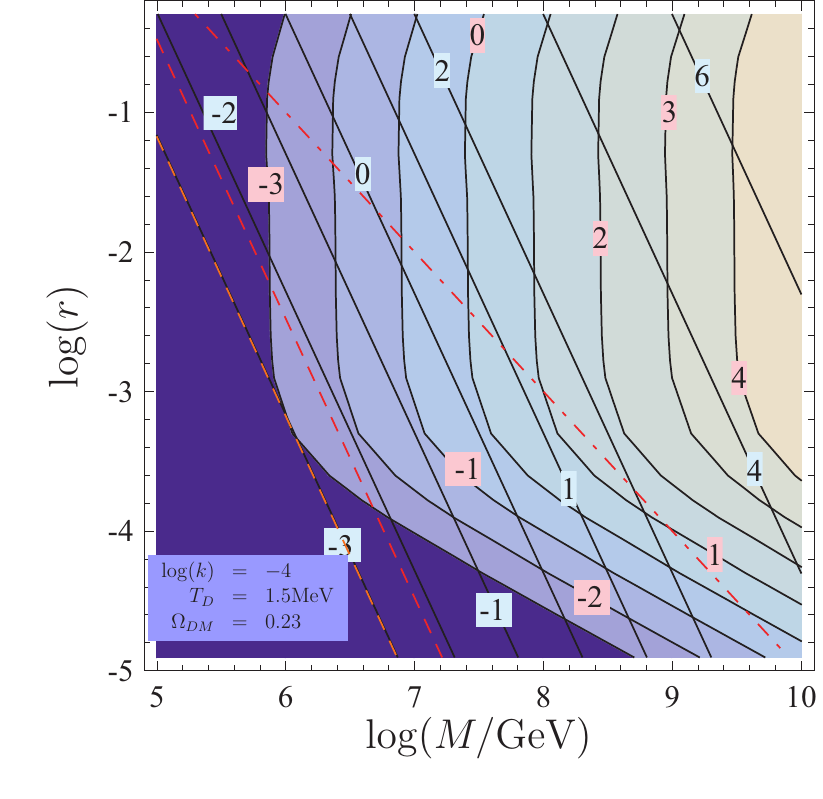} 
\end{minipage}
\caption[Yield and dilution for decay temperature of $T_D = 1.5\, \MeV$]{The same as Fig.~\ref{fig:eR_TD}, but the decay temperature was fixed to $T_D = 1.5\, \MeV$. All areas with correct gravitino relic density have disappeared.}
\label{fig:eR_ChangedTD}
\end{figure}
\begin{figure}[!ht]
\begin{minipage}{16cm}
\includegraphics[scale=0.95]{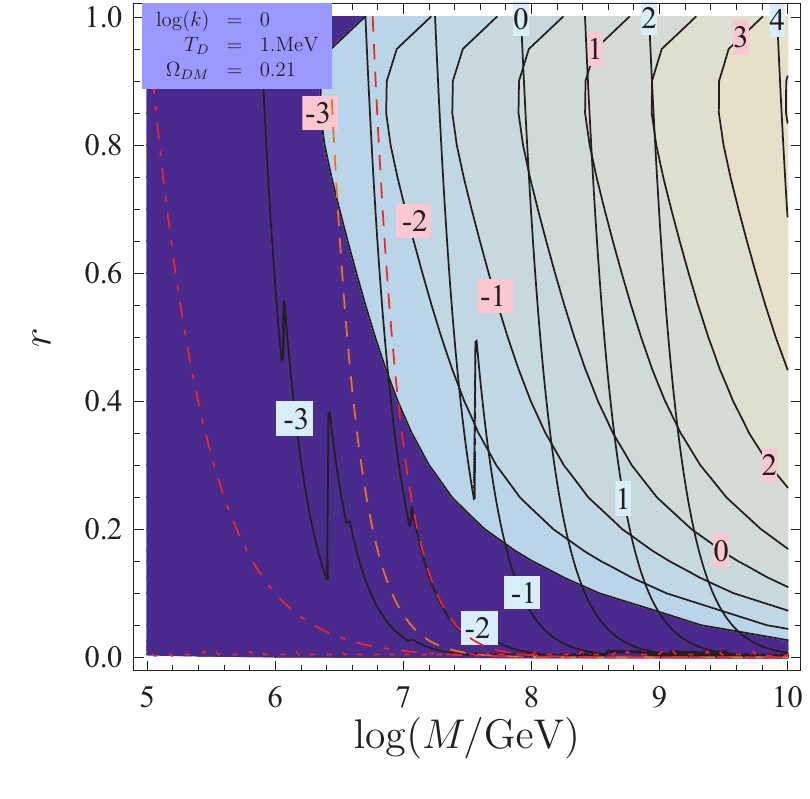} 
\hfill
\includegraphics[scale=0.95]{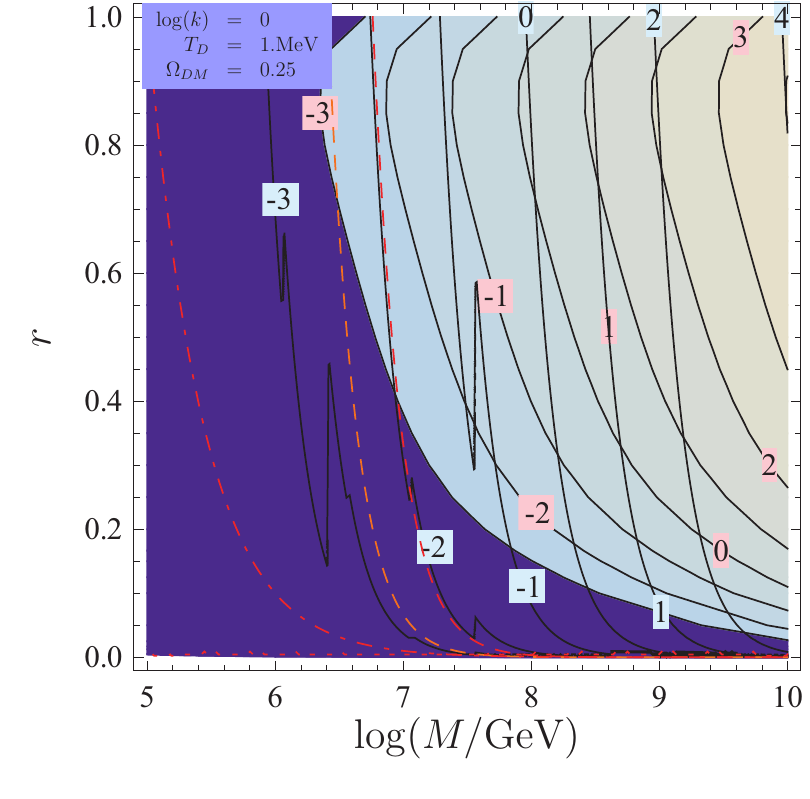} \\
\includegraphics[scale=0.95]{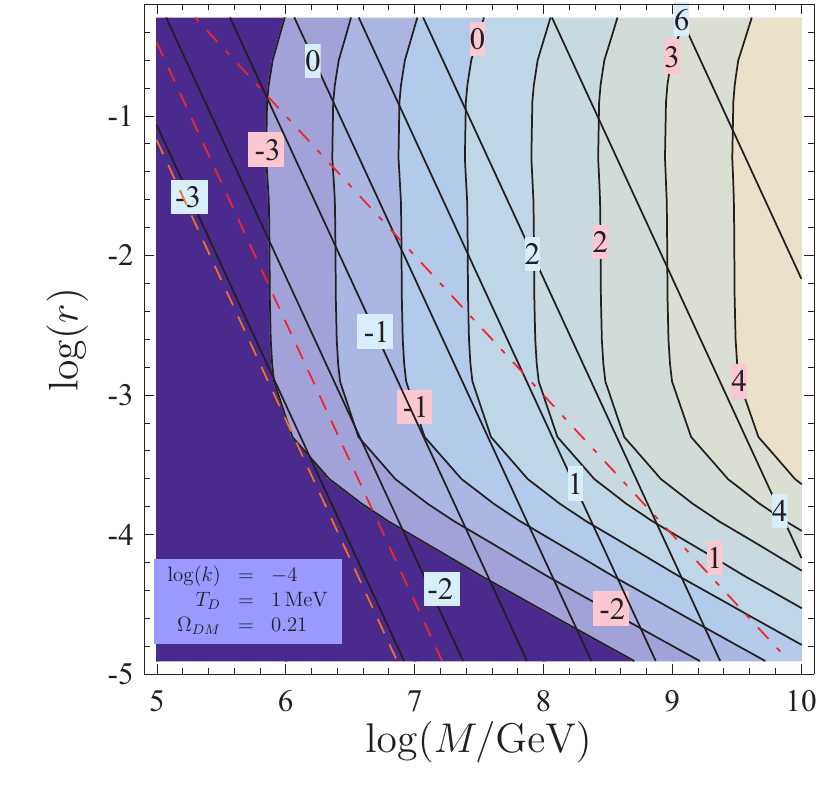} 
\hfill
\includegraphics[scale=0.95]{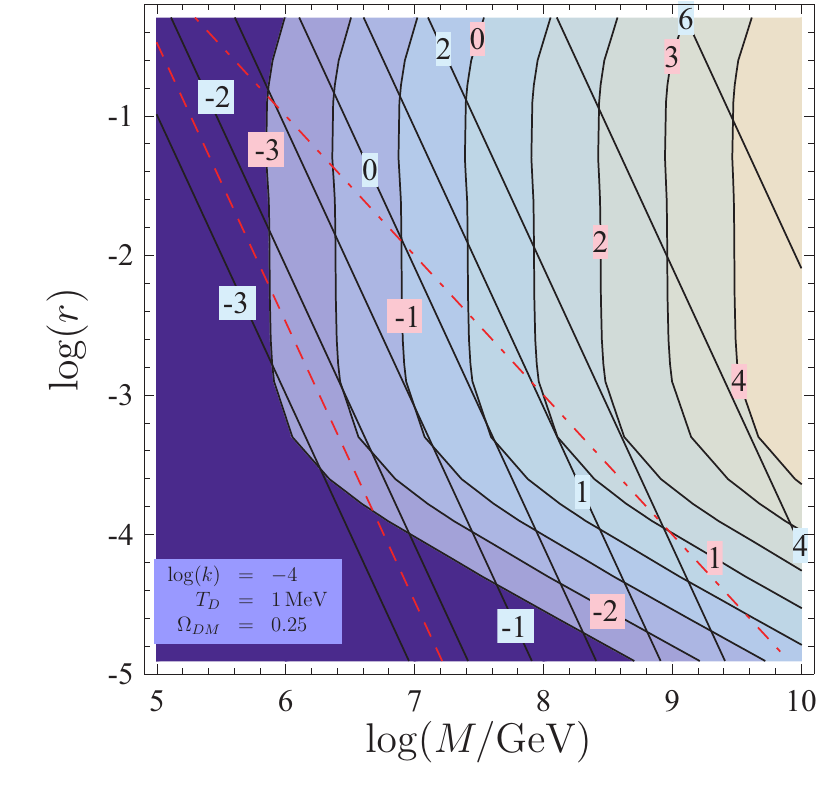} 
\end{minipage}
\caption[Yield and dilution for a variation of $\Omega_{3/2}$]{The same as Fig.~\ref{fig:eR_TD}, but the demanded relic density for the gravitino was varied between the 3$\sigma$ range of WMAP-7: $\Omega_{3/2} = 0.21,0.25$.}
\label{fig:eR_ChangedOm}
\end{figure}

The behavior of the messengers in the last sections can easily generalized to all messengers that are charged under $SU(3)_C$ or $SU(2)_L$. Therefore, there is no change in our results by adding $\tilde{d}_L$-, $\tilde{u}_L$- or $\tilde{u}_R$-like messengers. However, there will be a difference, by adding a messenger particle that is just charged under $U(1)_Y$. The relic density of this particle is expected to be similar to the relic density of the $SU(2)_L$ messengers. We call this field ${\tilde{e}^R_M}$ and checked that assumption numerically: it can be seen in Fig.~\ref{fig:Omega_eR_vL} that for large values of $r$ the relic density is even larger, i.e. the effect of the two times bigger hypercharge is more than compensated by the missing weak interactions. For smaller values of $r$, in regions where the Goldstino annihilations start to dominate, the relic densities of both types of messengers are nearly the same with only small differences due to the different one-loop corrected masses. \\
Even if the decay width is not enhanced by decays in \(W\) bosons, the choice $f=1$ doesn't lead to a solution. Only when we adjust \(f\) to small values in order to get the minimal allowed decay temperature of \(T_D = 1\)~MeV, there is sufficient dilution of gravitinos as can be seen in Fig.~\ref{fig:eR_TD}. However, this demands some fine-tuning. How big the fine-tuning must be in order to get a sufficient dilution, can be seen in the Fig.~\ref{fig:eR_ChangedTD}.  Here, we have changed the decay temperature to a slightly bigger value of \(T_D = 1.5\)~MeV to probe the sensitivity on the decay temperature. No allowed areas are left in this case. An increment of \(T_D\) from 1~MeV to 1.5~MeV has the same effect as reducing the relic density of the messengers by a factor \(\frac{3}{2}\). Therefore, the result is very susceptible to the calculation of the yield of the messengers. Here, we have to keep in mind that in \micrOmegas the RGE running of the electromagnetic coupling is not included, but the value at \(M_Z\) is used. However, the messengers freeze out at temperatures some orders above. The RGE running for the weak couplings is quite small in comparison to the running of the strong coupling. Nevertheless, only 2~\% running of the coupling leads to 16~\% difference in our result in areas of the phase space which are dominated by annihilation in electroweak gauge bosons.\\
In Fig.~\ref{fig:eR_ChangedOm} we have varied the demanded relic density after dilution according to the 3\(\sigma\) range of WMAP-7, i.e. \(0.21 < \Omega_{3/2} < 0.25\). For the smaller relic density demanding more dilution, the allowed regions are seriously reduced. The need for of such a fine-tuning makes also this scenario highly unattractive.
\subsection{\texorpdfstring{$SO(10)$}{SO10} messenger sector: gauge singlets as messengers}
\begin{figure}[!ht]
\begin{minipage}{16cm}
\includegraphics[scale=0.95]{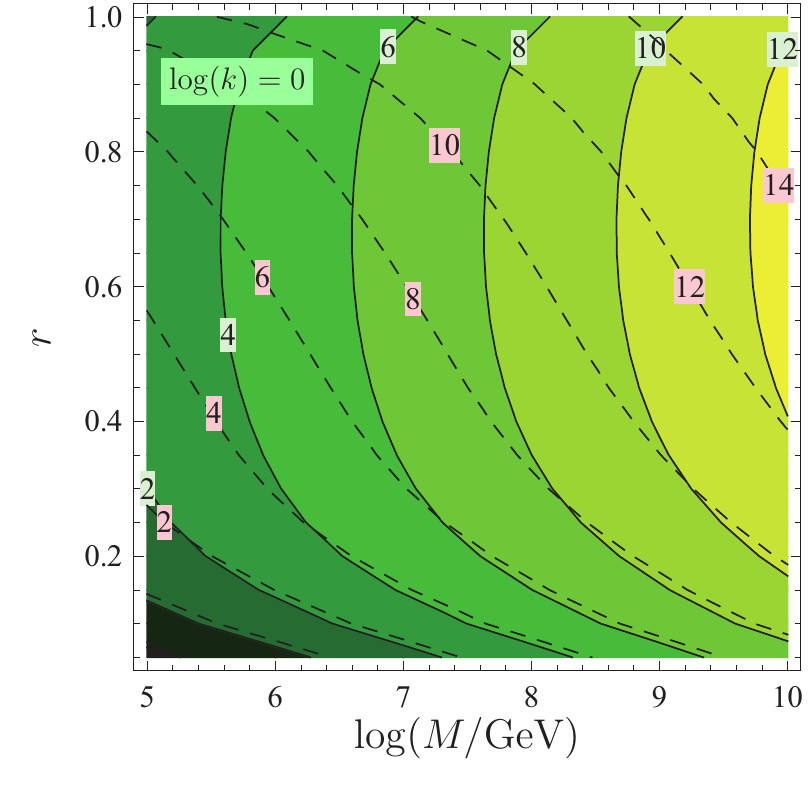} 
\hfill
\includegraphics[scale=0.95]{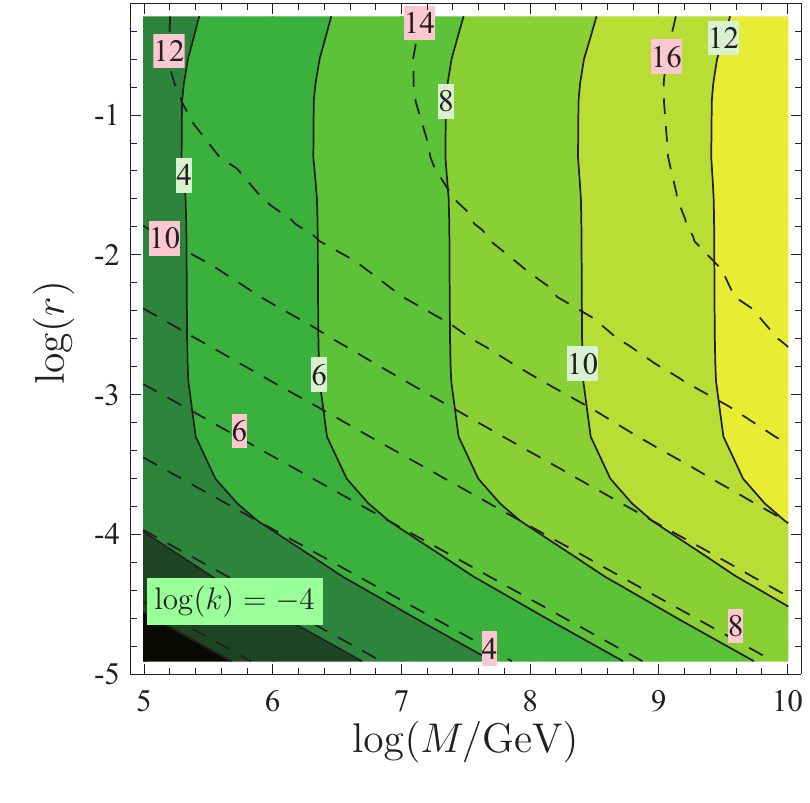} 
\end{minipage}
\caption[Comparison of the relic density of the $\tilde{v}_L$-like messengers and the $\tilde{v}_R$-like messengers.]{Comparison of the relic density of the $\tilde{v}_L$-like messengers (contour plot) and the $\tilde{v}_R$-like messengers (dashed lines). The labels give the value for $\log\Omega$.}
\label{fig:Omega_vR}
\end{figure}
The last possibility is to consider a messenger sector containing singlets under all gauge groups (\(\tilde{v}_R\)-like) as it appears for example in the 16-plet of $SO(10)$. There are different SUSY models which involve gauge singlets at the low scale. The best known is the NMSSM, which we will describe in more detail in chapter~\ref{chapter:NMSSM}. Another one is the \(\mu\nu\)SSM which combines the idea of the solution of the \(\mu\)-problem from the NMSSM with bilinear $R$-parity violation \cite{Bartl:2009an}. Since in those models gauge singlets are included, mixing terms like eq.~(\ref{eq:deltaW}) including messenger singlets  can be present in the superpotential. Nevertheless, we will focus on a model independent analysis of the singlet messengers and neglect possible contributions from additional trilinear superpotential interactions and other mixing terms. This is a good approximation as long as the superpotential interactions of the singlets are smaller than the interactions with the Goldstino component of the gravitino. We will see that this is indeed the case in the preferred regions of parameter space which we will find.  \\ 
\begin{figure}[!htb]
\begin{minipage}{16cm}
\includegraphics[scale=0.95]{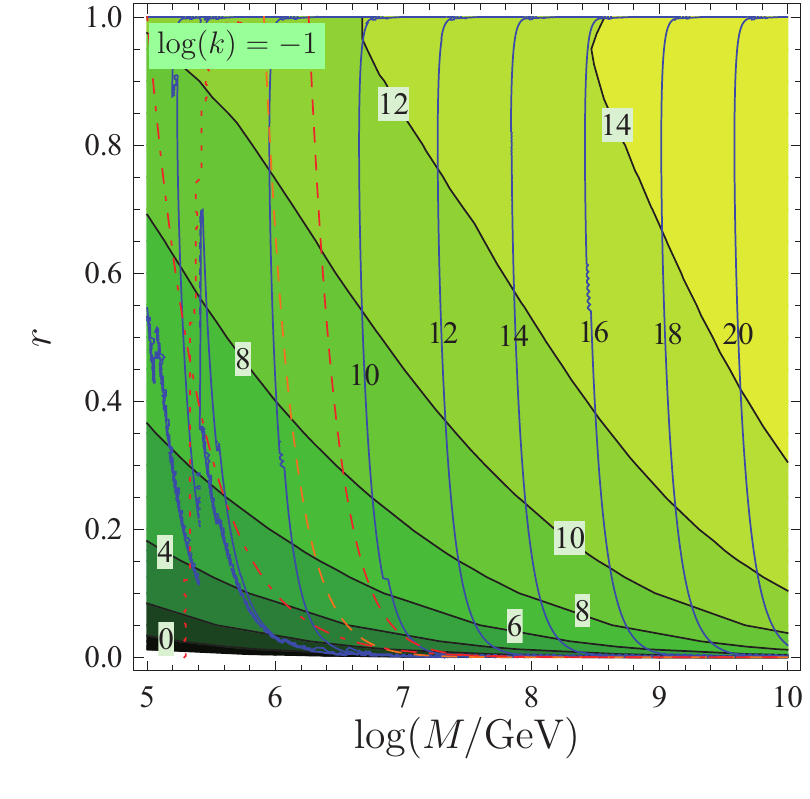} 
\hfill
\includegraphics[scale=0.95]{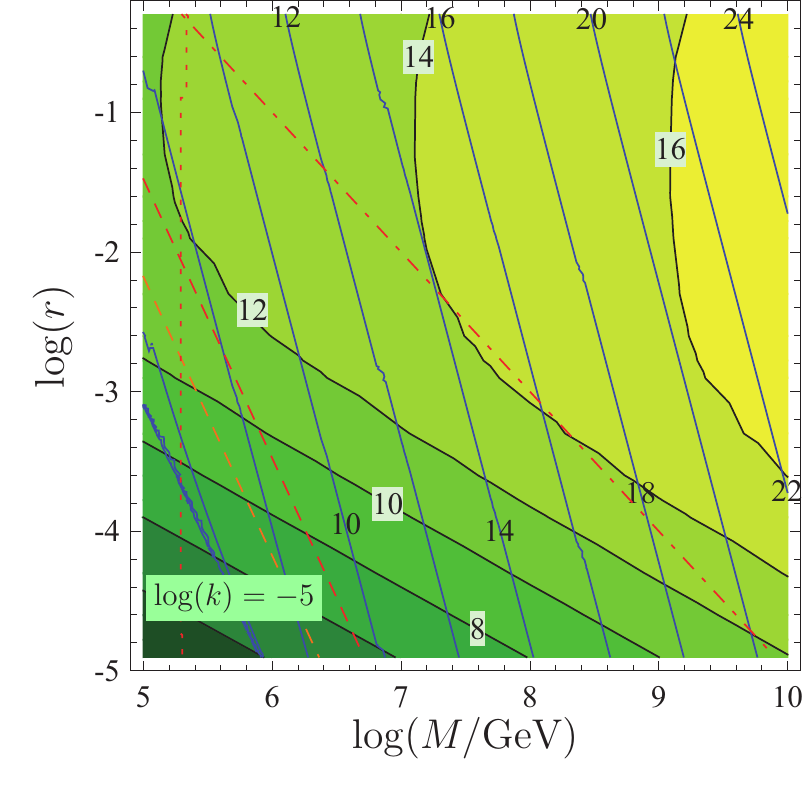} 
\end{minipage}
\caption[Calculated  against needed value of the relic density of the singlet messengers]{Calculated (plain lines) against needed (dashed lines) value of the relic density of the singlet messengers for a diluted relic density of the gravitinos of $\Omega_{3/2}=0.23$. The labels give $\log\Omega_{\tilde{\nu}^R_M}$ and the usual exclusion bounds were used.}
\label{fig:vR_omega}
\end{figure}

The 16-plet was already analyzed in \cite{Lemoine:2005hu} where it has been shown that this scenario works in principle.  Their analysis was based on annihilation due to loop vertices, which are dominant if the spurion mass is of the order of the messenger mass of \(M\simeq 10^6\)~GeV. The approximated yield was calculated to be
\begin{equation}
Y \simeq 10^{-5} \left(\frac{M}{10^6\, \GeV}\right)^2 \thickspace .
\end{equation}
This is five orders bigger then the yield for sneutrino-like messengers. Thus, even if the messenger masses are small, they can produce enough entropy and dilute the gravitinos to the needed amount.

To complement the analysis of \cite{Lemoine:2005hu}, we reconsider this scenario for the case that all particles of the hidden sector are heavier than the messengers. As a consequence, the dominant interaction of the messengers is always with the Goldstino component of the gravitino.  Their relic density and their decay width depend only on this interaction. Thus, the decay in MSSM particles is given by

\begin{equation}
\Gamma = \frac{1}{16\pi}\left(\frac{m_-^2}{m_{3/2} m_{Pl}}\right)^2 \delta^2 m_- .
\end{equation}
The relic density is shown in Fig.~\ref{fig:Omega_vR}. The result of our numerical studies is that for the case \(M^2 \gg F\) and \(f=1\), this scenario works for very small values of \(k\) of \(\Ord(10^{-5})\) and messenger masses of \(\Ord(10^7)\)~GeV as can be seen in Fig.~\ref{fig:vR_omega} on the right side. For larger values of \(k\) or \(M\), the annihilation is too effective and the additional entropy production too small. This can be partly compensated by reducing \(f\) as the dilution behaves like \(\Delta \sim k^{-4} f^{-1}\) and we reach the BBN bound very fast. However, if we assume \(\Lambda \simeq M\), one finds solutions for larger \(k\) as can be seen on the left side of Fig.~\ref{fig:vR_omega}. The reason is that the annihilation in Goldstinos, a t-channel interaction, is suppressed by the large mass splitting. The messengers are in the PeV-range and the gravitino mass is about 10~keV, i.e.\ it is still warm dark matter but not in conflict with the Lyman-\(\alpha\) observations. In contrast to charged messenger particles, a $\tilde{v}_R$ messenger does not receive large one-loop corrections to its mass due to gauge interactions.
\cleardoublepage
\chapter{Seesaw Scenarios and Neutralino Dark Matter}
\label{chapter:SU5}
Neutrino masses are zero in the MSSM. We have shown in the last chapter the possibility to create neutrino masses due to bilinear $R$-parity violation. In this chapter, we consider the possibility to produce neutrino masses via a dimension 5 operator by extending the particle content by heavy states. If neutrinos are Majorana particles, all models of neutrino mass reduce at low energy reduce to the unique Weinberg operator  \cite{Weinberg:1979sa,Weinberg:1980bf}
\begin{equation}
\label{eq:Dim5N}
 \kappa_{\alpha \beta} = \frac{f_{\alpha \beta}}{\Lambda} (H_u l) (H_u l) \thickspace. 
\end{equation}
While neutrino experiments determine only the ratio \(\frac{f_{\alpha\beta}}{\Lambda}\), they contain neither information about the origin of this operator nor about the absolute size of \(\Lambda\). If \(f\) is a coefficient of \(\Ord(1)\), current neutrino data indicates \(\Lambda \leq \Ord(10^{15})\)~GeV. Producing tiny neutrino masses by introducing very heavy fields is the basic ingredient of the so called 'seesaw mechanism'. In literature, three different types of seesaw mechanisms are distinguished \cite{Ma:1998dn} which are presented in sec~\ref{sec:Seesaw}. All seesaw scenarios have in common that new fields are introduced. In order not to spoil gauge unification, complete multiplets of \(SU(5)\) or of larger groups containing \(SU(5)\) as subgroup have to be added. This is explained in more detail in the sec.~\ref{sec:seesaw_models}. The new fields will not only generate neutrino masses but also influence the RGE running of the parameters of the model. Therefore, they change the properties of the LSP. Furthermore, the new interactions can be the source of dangerous flavor changing neutral currents. Hence, it must be checked that the considered scenarios are consistent with the bounds of precision data for rare events like \(\mu \rightarrow e \gamma\). We will discuss these topics in our numerical analysis at the end of this chapter. 
\section{Seesaw mechanism}
\label{sec:Seesaw}

\begin{figure}[t]
\begin{center}
\begin{tabular}{cccc}

\begin{picture}(20,10)(0,0)
\linethickness{1pt}
\put(0,50){(a)}
\end{picture}
&

\begin{fmffile}{Feynmangraphen/Wop2}
  \fmfframe(10,10)(10,10){
    \begin{fmfgraph*}(80,80)
    \fmftop{l1,l2}
    \fmfbottom{r1,r2}
    \fmf{plain}{l1,v1}
    \fmf{plain}{l2,v2}
    \fmf{plain, label=$M$}{v1,v2}
    \fmf{dashes}{r1,v1}
    \fmf{dashes}{v2,r2}
    \fmflabel{$l$}{l2}
    \fmflabel{$H_u$}{r2}
    \fmflabel{$l$}{l1}
    \fmflabel{$H_u$}{r1}
\end{fmfgraph*}}
\end{fmffile}

&

\begin{picture}(20,50)(0,0)
\linethickness{1pt}
\put(0,50){\(\rightarrow\)}
\end{picture}

&

\begin{fmffile}{Feynmangraphen/Wop1}
  \fmfframe(10,10)(10,10){
    \begin{fmfgraph*}(80,80)
    \fmftop{l1,r1}
    \fmfbottom{r2,l2}
    \fmfblob{.15w}{v1}
    \fmf{plain}{l1,v1,r1}
    \fmf{dashes}{l2,v1,r2}
    \fmflabel{$l$}{l1}
    \fmflabel{$H_u$}{l2}
    \fmflabel{$l$}{r1}
    \fmflabel{$H_u$}{r2}
\end{fmfgraph*}}
\end{fmffile}
 \\
\begin{picture}(20,10)(0,0)
\linethickness{1pt}
\put(0,50){(b)}
\end{picture}

&

\begin{fmffile}{Feynmangraphen/Wop3}
  \fmfframe(10,10)(10,10){
    \begin{fmfgraph*}(80,80)
    \fmftop{l1,l2}
    \fmfbottom{r1,r2}
    \fmf{plain}{l1,v1,l2}
    \fmf{dashes}{r1,v2,r2}
    \fmf{dashes, label=$M$}{v1,v2}
    \fmflabel{$l$}{l2}
    \fmflabel{$H_u$}{r2}
    \fmflabel{$l$}{l1}
    \fmflabel{$H_u$}{r1}
\end{fmfgraph*}}
\end{fmffile}

&

\begin{picture}(20,50)(0,0)
\linethickness{1pt}
\put(0,50){\(\rightarrow\)}
\end{picture}

&

\begin{fmffile}{Feynmangraphen/Wop1}
  \fmfframe(10,10)(10,10){
    \begin{fmfgraph*}(80,80)
    \fmftop{l1,r1}
    \fmfbottom{r2,l2}
    \fmfblob{.15w}{v1}
    \fmf{plain}{l1,v1,r1}
    \fmf{dashes}{l2,v1,r2}
    \fmflabel{$l$}{l1}
    \fmflabel{$H_u$}{l2}
    \fmflabel{$l$}{r1}
    \fmflabel{$H_u$}{r2}
\end{fmfgraph*}}
\end{fmffile}
\end{tabular}
\end{center}
\caption[Generic diagram for different seesaw mechanisms]{Generic diagram for different seesaw mechanisms. (a) Seesaw~I and III: the heavy fermion in the propagator can be a gauge singlet or a  $SU(2)_L$-triplet. (b) Seesaw~II: the heavy scalar in the propagator must be a $SU(2)_L$-triplet.}
\label{fig:Seesaw}
\end{figure}
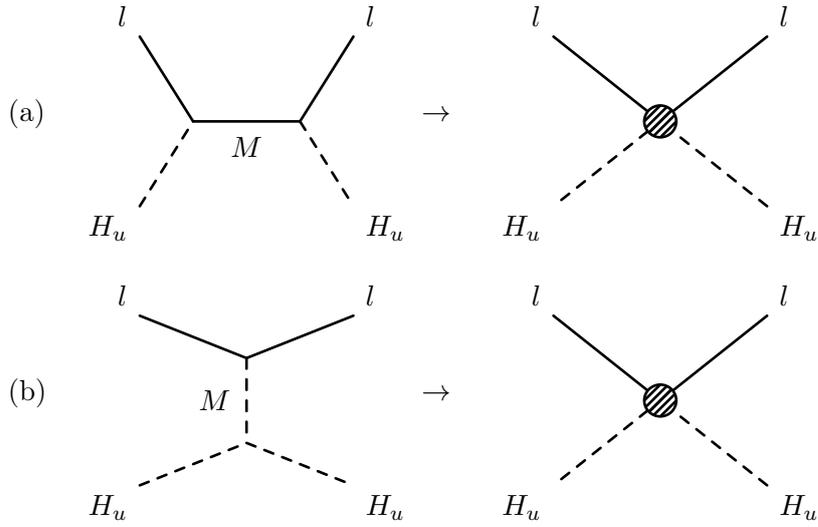

As already mentioned, there is an unique dimension 5 operator for neutrino masses. This effective operator is generated by integrating out heavy fields from the spectrum according to the decoupling theorem of Appelquist and Carazzone \cite{Appelquist:1974tg}
\begin{equation}
\label{eq:EffLag}
 \La_{\rm{eff}}(\phi) = \int d\Phi \La(\phi, \Phi) \thickspace.
\end{equation}
\(\Phi\) are heavy fields, while the \(\phi\) fields have masses of the energy scale or below. Removing the heavy field from the particle spectrum generates effective four-point interactions of light fields with an effective coupling \(C_{\rm{eff}}\) of the order
\begin{equation}
\label{eq:effC}
 C_{\rm{eff}} = \frac{1}{M^n} Y^2 \thickspace .
\end{equation}
Here, \(Y\) is the coupling between two light and one heavy state, while \(M\) is the mass of the heavy field. The power \(n\) depends on the nature of the particle in the propagator: it is \(n=2\) for a scalar or vector boson  and  \(n=1\) for a fermion. \\
There are three different possibilities to generate the operator of eq.~(\ref{eq:Dim5N}) stemming from the  two generically different Feynman diagrams of Fig.~\ref{fig:Seesaw}. The mass of the heavy particle is again called \(M\). The coupling between two fermions and one scalar is \(Y\), while the three-scalar interaction is named  \(\lambda\). We assume that \(\lambda\) is a complex number, while \(Y\) is a matrix. Referring to the cases (a) and (b) of Fig.~\ref{fig:Seesaw}, one obtains:
\begin{itemize}
 \item {\bf Seesaw I}: If the particle in case (a) is a gauge singlet fermion, the neutrino mass matrix is given by \cite{Mohapatra:1979ia,GellMann:1980vs,Minkowski:1977sc}
\begin{equation}
\label{eq:mvI}
m_\nu = \frac{1}{2} Y^T \frac{1}{M} Y v_u^2 \thickspace . 
\end{equation} 
 \item {\bf Seesaw II}: The case (b) is called seesaw~II. The diagram is only possible if the scalar is a \(SU(2)_L\) triplet which also carries hypercharge \cite{Schechter:1980gr,Cheng:1980qt}. The neutrino mass matrix is given by
\begin{equation}
\label{eq:mvII}
m_\nu = \frac{1}{2} Y^T \frac{1}{M^2} \lambda v_u^2
\end{equation}
 \item {\bf Seesaw III}: If the fermion in case (a) is a \(SU(2)_L\) triplet, this is called seesaw~III \cite{Foot:1988aq}. The corresponding mass matrix reads
\begin{equation}
\label{eq:mvIII}
m_\nu = \frac{1}{2} Y^T \frac{1}{M} Y v_u^2
\end{equation}
\end{itemize}
\paragraph*{Diagonalization of the mass matrix}
The neutrino mass term given in eqs.~(\ref{eq:mvI})-(\ref{eq:mvIII}) is a complex, symmetric matrix which can be diagonalized by an unitary $3\times 3$ matrix $U$~\cite{Schechter:1980gr}
\begin{equation}
\label{diagmeff}
{\hat m_{\nu}} = U^T  m_{\nu} U \thickspace .
\end{equation}
We can parametrizes \(U\) by using three angles and three phases. The standard form is 
\begin{eqnarray}
\label{def:unu}
U=
\left(
\begin{array}{ccc}
 c_{12}c_{13} & s_{12}c_{13}  & s_{13}e^{-i\delta}  \\
-s_{12}c_{23}-c_{12}s_{23}s_{13}e^{i\delta}  & 
c_{12}c_{23}-s_{12}s_{23}s_{13}e^{i\delta}  & s_{23}c_{13}  \\
s_{12}s_{23}-c_{12}c_{23}s_{13}e^{i\delta}  & 
-c_{12}s_{23}-s_{12}c_{23}s_{13}e^{i\delta}  & c_{23}c_{13}  
\end{array}
\right) 
 \times
 \left(
 \begin{array}{ccc}
 e^{i\alpha_1/2} & 0 & 0 \\
 0 & e^{i\alpha_2/2}  & 0 \\
 0 & 0 & 1
 \end{array}
 \right)
\end{eqnarray}
with $c_{ij} = \cos \Theta_{ij}$ and $s_{ij} = \sin \Theta_{ij}$. The angles $\Theta_{12}$,  $\Theta_{13}$ and  $\Theta_{23}$ are the solar neutrino angle, the reactor (or CHOOZ) angle and the atmospheric neutrino mixing angle, respectively. $\delta$ is the Dirac phase and $\alpha_i$ are Majorana phases. In the following, we will set the latter to 0 and consider for $\delta$ the cases $0$ and $\pi$.

\paragraph*{Running of the neutrino mass parameter}  The RGE for the effective operator of eq.~(\ref{eq:Dim5N}) can easily be calculated in SUSY. The reason is the so called non-renormalization theorem which states that the RGEs for superpotential parameters can be expressed by the anomalous dimensions of the external fields \cite{Wess:1973kz,Ferrara:1974fv,Grisaru:1979wc}. This leads to 
\begin{equation}
\label{eq:runWeinberg}
 \frac{d \kappa}{d t} = \gamma_{\hat{l}}^T \kappa + \kappa \gamma_{\hat{l}} + 2 \gamma_{\hat{H}_u} \kappa \thickspace .
\end{equation}
Here, we used as usual \(t=\ln Q\). \(\gamma_{\hat{l}}\) and \(\gamma_{\hat{H}_u}\) are the anomalous dimensions of the left-lepton and down-type Higgs superfields. In the MSSM, i.e. below the lowest seesaw scale, they are given at one-loop level by
\begin{eqnarray}
 \gamma^{(1)}_{\hat{l}} &=& Y_e^\dagger Y_e - \frac{3}{2} g_2^2 - \frac{3}{10} g_1^2 \thickspace ,\\
 \gamma^{(1)}_{\hat{H}_u} &=& 3 \mbox{Tr}\left(Y_u^\dagger Y_u\right) - \frac{3}{2} g_2^2 - \frac{3}{10} g_1^2 \thickspace .
\end{eqnarray}
\section{Seesaw models}
\label{sec:seesaw_models}
We have seen in the last section that it is necessary for the different seesaw scenarios  to add either a gauge singlet or a \(SU(2)_L\) triplet with or without hypercharge to the particle spectrum. However, just adding a superfield which is a triplet under \(SU(2)_L\) would spoil gauge unification. Therefore, it is necessary to embed all new fields in complete multiplets of a GUT group like \(SU(5)\) or larger. We present in the subsequent sections the particle content and superpotential for the three scenarios assuming an underlying \(SU(5)\)-theory. The complete superpotential for the different models is always defined as 
\begin{equation}
 W = W_{MSSM} + W^i \thickspace. 
\end{equation}
\(W^i\) with \(i=I,II,III\) is the superpotential involving the additional fields. \(W_{MSSM}\) is the \(SU(5)\) invariant form of the superpotential of the MSSM given in eq.~(\ref{SU5superpotential}). We use also the standard nomenclature of the \(SU(5)\) multiplets shown in eq.~(\ref{eq:SU5matter}) and eq.~(\ref{eq:SU5Higgs}). A detailed discussion about these models can also be found in \cite{Borzumati:2009hu}. \\
We will do a complete two-loop analysis and calculated therefore with \SARAH  the full set of RGEs at two-loop level: in app.~\ref{app:RGE_Seesaw}, the two-loop results for the anomalous dimensions and the \(\beta\)-functions for the gauge couplings are given. The RGEs for all other parameters can easily be calculated with \SARAH, or they can be derived by using the presented results and the relations given in  app.~\ref{app:RGE_con}. For the running of the Weinberg operator, we use eq.~(\ref{eq:runWeinberg}) and the two-loop results for the anomalous dimensions of \(\hat{l}\) and \(\hat{H}_u\). Each generation of heavy fields is separately integrated out at a certain threshold scale. Since we are dealing with two-loop RGEs, we need one-loop boundary conditions at these threshold scales. It is known that the main corrections take place in the gauge sector. The gauge couplings \(g_i\) and gaugino masses \(M_i\) receive at the boundaries  a shift due to a heavy field transforming as representation \(r\) under the corresponding gauge group in $\overline{\mbox{DR}}$-scheme of \cite{Hall:1980kf}
\begin{eqnarray}
\label{eq:shift1}
 g_i & \rightarrow & g_i \left( 1\pm \frac{1}{16 \pi^2} g_i^2 I^i_2(r) \log\left(\frac{M^2}{M_t^2}\right)\right)  \thickspace ,\\
\label{eq:shift2}
 M_i & \rightarrow & M_i \left( 1\pm \frac{1}{16 \pi^2} g_i^2 I^i_2(r) \log\left(\frac{M^2}{M_t^2}\right)\right) \thickspace .
\end{eqnarray}
Here, \(I_2^i(r)\) is the Dynkin index of a representation \(r\)  under the gauge group corresponding to the gauge coupling \(g_i\). \(M\) is the mass of that particle and \(M_t\) is the threshold scale. Obviously, the finite shifts vanish for those particle which are integrated out at their mass scale. The sign of the correction in eqs.~(\ref{eq:shift1}) and (\ref{eq:shift2}) for running up to the GUT scale is + and for running down from the GUT to the low scale is - .  The shifts for the other parameters like the soft breaking masses of the MSSM particles are of the order \(\frac{1}{16 \pi^2} \left(\frac{m}{M}\right)^2\) if only one heavy field is in each loop like in our cases. \(m\) is a usual soft breaking parameter of order \(\Ord(10^2 - 10^3)\)~GeV and \(M\) is a \(SU(5)\) mass parameter of the order \(10^{12}\)~GeV and above. Hence, these corrections are tiny and negligible.  After this general introduction, we present now the different realization of the seesaw types in more detail. 

\subsection{Seesaw I}
In the case of  seesaw~I, the particle content is extended by three generations of a gauge singlet \(\hat{N}\).  The new part of the superpotential with unbroken \(SU(5)\) reads
\begin{equation}
 W^I =  Y^{a b}_\nu\, \hat{N}_a \, \bar{5}_{M,b}\, 5_H + \frac{1}{2} \hat{N}_a M^{a b}_N \hat{N}_b \thickspace .
\end{equation}
Here, \(a,b=1,2,3\) are flavor indices. \(Y_\nu\) and \(M_N\) are complex \(3 \times 3\) matrices. After \(SU(5)\) breaking and integration over the colored Higgs fields,  the first term can  be written as
\begin{equation}
Y^{a b}_\nu \, \hat{N}_a \, \bar{5}_{M,b} \, 5_H \rightarrow  Y^{ab}_\nu \, \hat{N}_a \, \hat{l}_b \, \hat{H}_u  \thickspace .
\end{equation}
We have suppressed here the \(SU(2)_L\) indices. The generated neutrino mass term after EWSB is
\begin{equation}
\label{eq:mnuSeesawI}
 m_\nu = - \frac{v_u^2}{2} \, Y_\nu^T \, M_N^{-1} \, Y_\nu \thickspace . 
\end{equation}
This model is well known and was already studied in literature with respect to flavor changing neutral currents and LHC phenomenology, see e.g. \cite{Hirsch:2008dy} and references therein. We include these aspects for completeness in our analysis and will also discuss the dark matter properties of type~I. \\
The analysis of the seesaw~I is the easiest one of all three seesaw scenarios. As it can be seen in apps.~\ref{sec:AnaDimI} and \ref{sec:betaI}, the RGE do only change slightly. In addition, it is not necessary to calculate finite shifts to the gauge couplings or gaugino masses since the new heavy fields are gauge singlets. Therefore, we expect in this seesaw scenario the smallest impact  of all seesaw models on the masses of SUSY fields and, therefore, on properties of the lightest neutralino.  \\
We use as additional input parameters to the standard mSugra parameters the demanded values of the neutrino masses and the three masses of the gauge singlets. The Yukawa coupling \(Y_{\nu}\) is afterwards calculated in an iterative way to reproduce the input values of the neutrino masses by inverting eq.~(\ref{eq:mnuSeesawI}) \cite{Casas:2001sr}
\begin{equation}
\label{eq:inverseMassTerm}
Y_\nu = \sqrt{2} \frac{i}{v_u} \sqrt{\tilde{M}_N} R \sqrt{\tilde{m}_{\nu}} U^\dagger \thickspace,
\end{equation}
where \(\tilde{M}_N\) and \(\tilde{m}_{\nu}\) are diagonal matrices containing the corresponding eigenvalues. \(R\) is in general a complex, orthogonal matrix. \\
The effects on flavor violation can roughly be approximated by an one-step integration of the RGEs. The off-diagonal elements (\(i\neq j\)) in the slepton mass parameters and trilinear, scalar couplings are estimated to \cite{Hisano:1995cp}
\begin{eqnarray}
\label{eq:FCNCIa}
m^2_{\tilde{l},ij} &\simeq& -\frac{1}{8 \pi^2 }  \left( 3 m^2_0 + A^2_0 \right) \left(Y_\nu^\dagger L Y_\nu\right)_{ij} \thickspace , \\
\label{eq:FCNCIb}
A_{e,ij} &\simeq& -\frac{3}{16 \pi^2 }   A_0  \left(Y_e Y_\nu^\dagger L Y_\nu \right)_{ij} \thickspace .
\end{eqnarray}
Here, it was assumed that \(Y_e\) is diagonal and we defined \(L_{ij} = \log\left(\frac{M_{GUT}}{M_{N_i}}\right) \delta_{ij}\). While the off-diagonal elements in \(m^2_L\) and \(A_e\) depend strongly on the seesaw type as we will see, all models have in common 
\begin{equation}
m_{\tilde{e},ij}^2 \simeq 0 
\end{equation}
for \(i \neq j\). These approximation can be used to give a rough idea about the size of rare leptonic decays like $l_i \to l_j \gamma$ which scale like 
\begin{equation}
\mbox{Br}(l_i \to l_j \gamma) \propto \alpha^3 m_{l_i}^5
	\frac{| m^2_{\tilde L,ij}|^2}{\widetilde{m}^8}\tan^2\beta \thickspace .
\label{eq:LLGapprox}	
\end{equation}
Here,  $\widetilde{m}$ is the average of the SUSY masses involved in the loops and \(\beta\) is the common mixing angle in the MSSM Higgs sector.
\subsection{Seesaw II}
For the seesaw~II, it is necessary to add a scalar \(SU(2)_L\) triplet which also carries hypercharge \(Y_i\). Such a particle is part of the 15-plet of \(SU(5)\). Therefore, we add to the \(SU(5)\) invariant superpotential the interactions of a pair of \({\bf 15}\) and \({\bf \overline{15}}\) 
\begin{equation}
 W^{II} = \frac{1}{\sqrt{2}} Y_{15}^{ab}\, \overline{5}_a\,15\,\overline{5}_b + \frac{1}{\sqrt{2}} \lambda_1\, \overline{5}_H\,15\,\overline{5}_H + \frac{1}{\sqrt{2}} \lambda_2\, 5_H\,\overline{15}\, 5_H +  M_{15}\, 15\,\overline{15} \thickspace .
\end{equation}
Again, \(a\) and \(b\) are flavor indices. \(Y_{15}\) is a symmetric, complex \(3 \times 3\) matrix and \(\lambda_i\) as well as \(M_{15}\) are scalar, complex parameters. After \(SU(5)\) breaking, the {\bf 15} splits into irreducible representations of \(SU(3)_C\times SU(2)_L\times U(1)_Y\)  
\begin{equation}
\label{eq:split15}
 {\bf 15} = \hat{S} + \hat{T} + \hat{Z} \thickspace .
\end{equation}
The corresponding quantum numbers are 
\begin{equation}
 \hat{S}: \left(\bf{6}, \bf{1}\right)_{-2/3} \thickspace, \hspace{1cm} \hat{T}: \left(\bf{1},\bf{3}\right)_1 \thickspace, \hspace{1cm} \hat{Z}: \left(\bf{3},\bf{2}\right)_{1/6} \thickspace.
\end{equation}
Furthermore, also the coupling \(Y_{15}\) and mass term \(M_{15}\) split into different parameters below the \(SU(5)\) breaking scale because they belong to different representations of SM groups according to eq.~(\ref{eq:split15}). Again, after integrating out the colored Higgs fields, we end up with the superpotential terms 
\begin{eqnarray}
 \frac{1}{\sqrt{2}} Y_{15} \overline{5}\,15\,\overline{5} &\rightarrow& \frac{1}{\sqrt{2}} \left(Y_T\, \hat{l}\, \hat{T}\, \hat{l} + Y_S\, \hat{d}\, \hat{S}\, \hat{d}\right) + Y_Z\, \hat{d}\, \hat{Z}\, \hat{l} \thickspace ,\\
\frac{1}{\sqrt{2}} \lambda_1 \overline{5}_H\,15\,\overline{5}_H &\rightarrow& \frac{1}{\sqrt{2}} \lambda_1 \hat{H}_d \hat{T} \hat{H}_d  \thickspace ,\\
\frac{1}{\sqrt{2}} \lambda_2 5_H \overline{15}\,5_H &\rightarrow& \frac{1}{\sqrt{2}} \lambda_2 \hat{H}_u \hat{\bar{T}} \hat{H}_u \thickspace ,\\
 M_{15} 15\,\overline{15} &\rightarrow& M_T \hat{T} \hat{\bar{T}} + M_Z \hat{Z} \hat{\bar{Z}} + M_S \hat{S} \hat{\bar{S}} \thickspace .
\end{eqnarray}
When calculating the neutrino mass matrix using eq.~(\ref{eq:mvII}), one has to keep in mind that the three scalar coupling is stemming from the F-Terms and given by \(\lambda = \frac{1}{\sqrt{2}} \lambda_2 M_T\). This together with the normalization of the Yukawa coupling results for this model in the neutrino mass term
\begin{equation}
 m_\nu = \frac{1}{4} v_u^2 \lambda_2 M_T^{-1} Y_T  \thickspace . 
\end{equation}
\(Y_T\) is diagonalized by the same matrix as \(m_\nu\)
\begin{equation}
 Y_T^{dia} = U^T Y_T U \thickspace .
\end{equation}
If all eigenvalues, angles and phases of \(m_\nu\) were known, \(Y_T\) would be fixed up to an overall constant which can be estimated to
\begin{equation}
 \frac{M_T}{\lambda_2} \simeq 10^{15}\,\GeV\,\left(\frac{0.05 \eV}{m_\nu}\right) \thickspace . 
\end{equation}
Also the seesaw~II was already discussed in literature \cite{Esteves:2009vg,Esteves:2009qr}. However, these analyses were done without using the complete set of two-loop RGEs. In addition, as already pointed out in \cite{Borzumati:2009hu}, also some one-loop RGEs are wrong in literature. Therefore, we reconsidered this model performing a complete two-loop analysis. \\
We use as threshold scale the mass of the \(SU(2)_L\) triplet \(M_T\). This leads to the following boundary conditions:
\begin{eqnarray}
\label{eq:shiftI_1}
 g_1 & \rightarrow & g_1 \left(1 \pm \frac{g_1^2}{16 \pi^2} \left(\frac{8}{5} \log\frac{M_S}{M_T} + \frac{1}{6} \log\frac{M_Z}{M_T} \right) \right) \thickspace , \\
 g_2 & \rightarrow & g_2 \left(1 \pm \frac{g_2^2}{16 \pi^2} \frac{3}{2} \log\frac{M_Z}{M_T} \right) \thickspace , \\
 g_3 & \rightarrow & g_3 \left(1 \pm \frac{g_3^2}{16 \pi^2} \left(\frac{5}{2} \log\frac{M_S}{M_T} + \log\frac{M_Z}{M_T} \right)  \right) \thickspace . \label{eq:shiftI_3}
\end{eqnarray}
The shifts for the gauginos can easily derived from these results by comparing eq.~(\ref{eq:shift1}) and eq.~(\ref{eq:shift2}). For our numerically analysis, we use as input parameters \(M_T\), \(\lambda_1\), \(\lambda_2\) and \(Y_T\) at the threshold scale. We run with this parameters up to the GUT scale and set the following boundary conditions in addition to the standard mSugra boundary conditions 
\begin{align}
 Y_S = Y_Z \equiv Y_T \thickspace , \hspace{1cm}  M_S = M_Z \equiv M_T \thickspace .
\end{align}
Moreover, the boundary conditions for the soft breaking parameters read
\begin{align}
& T_{\lambda_1} \equiv A_0 \lambda_1 \thickspace , \hspace{1cm}  T_{\lambda_2} \equiv A_0 \lambda_2 \thickspace , \hspace{1cm}
 T_S = T_Z = T_T \equiv A_0 Y_T \thickspace ,\\
& m_S^2 = m_{\bar{S}}^2 = m_T^2 = m_{\bar{T}}^2 = m_Z^2 = m_{\bar{Z}}^2 \equiv m_0^2 \thickspace .
\end{align}
In principle, we have also to set \(B_S = B_Z = B_T \equiv B_0 M_T\). However, the \(\beta\)-function of those parameters decouple from the rest and they have no influence on our results. \\
The additional fields have not only a large impact on the running gauge couplings because of the finite shifts of eqs.~(\ref{eq:shiftI_1})-(\ref{eq:shiftI_3}) but also because of changed  RGEs.  The \(\beta\)-functions of the gauge couplings and gaugino masses depend on the Dynkin index \(I_2(R)\) summed over all chiral superfields. Generally, the contributions of additional  {\bf 15} and {\(\bf  \overline{15}\)} are
\begin{equation}
\label{eq:Ir_II}
\Delta I^i_2(R) = \frac{7}{2} (N_{15} +N_{\overline{15}}) 
\end{equation}
Finally, we can do again an rough approximation of the effects on flavor violation according to eqs.~(\ref{eq:FCNCIa}) and (\ref{eq:FCNCIb}). The result are
\begin{eqnarray}
\label{eq:LFVentriesII}
m^2_{\tilde l,ij} &\simeq& -\frac{1}{8 \pi^2 }  \left( 18 m^2_0 + 6 A^2_0 \right) 
 \left(Y^{\dagger}_T  Y_T\right)_{ij} \log(\frac{M_{GUT}}{M_T})  \thickspace ,\\
A_{e,ij} &\simeq& -\frac{18}{16 \pi^2 }   A_0  \left(Y_e Y^{\dagger}_T  Y_T\right)_{ij}\log(\frac{M_{GUT}}{M_T}) \thickspace .
\end{eqnarray}
As we will discuss in sec.~\ref{sec:Seesaw_LFV}, these approximations work only well for seesaw~I. However, they give at least a rough idea for the behavior of type~II and III. Hence, we expect a much larger effect in this case as for the seesaw~I. 
\subsection{Seesaw III}
The type~III seesaw is based on  additional fields belonging to the adjoint representation of \(SU(2)_L\). Hence, we add particles sitting in the 24-plet, the adjoint representation of \(SU(5)\), to the spectrum. It is not sufficient to add just one generation of 24-plets to explain all neutrino data if we assume \(SU(5)\) invariant boundary conditions at the GUT scale: the induced mass splitting between the different generations of neutrinos won't be large enough. Therefore, we will study the case with three generations of 24-plets. \\  
The superpotential involving the {\bf 24} is
\begin{equation}
\label{3sup}
W^{III} = Y^{a b}_{24} \, 5_H \, 24_{M,a} \, \bar{5}_{M,b} + \frac{1}{2} M^{a b}_{24} \, 24_{M,a} \, 24_{M,b}  \thickspace .
\end{equation}
We have added an index \(M\) to the {\bf 24} in order to clearly distinguish it from the Higgs 24-plet \(24_H\) common in \(SU(5)\) models (see eq.~(\ref{eq:SU5breaking})). The {\bf 24} has the same gauge quantum numbers as the gauge bosons of \(SU(5)\) and can be decomposed in SM representations by 
\begin{equation}
 24_M = \hat{G}_M + \hat{W}_M + \hat{B}_M + \hat{X}_M + \hat{\bar{X}}_M 
\end{equation}
with
\begin{equation}
  \hat{G}_M: \left(\bf{8},\bf{1}\right)_0, \thickspace \hat{W}_M: \left(\bf{1},\bf{3}\right)_0, \thickspace \hat{B}_M: \left(\bf{1},\bf{1}\right)_0, \hat{X}_M: \left(\bf{3},\bf{2} \right)_{-5/6}, \thickspace \hat{\bar{X}}_M: \left(\bf{\bar{3}},\bf{2} \right)_{5/6} \thickspace .
\end{equation}
For the decomposition of eq.~(\ref{3sup}) after \(SU(5)\) breaking we use 
\begin{equation}
24_M = \left(\begin{array}{cc} \left(\hat{G}_M\right)_{a b} & \left(\hat{X}_M\right)_{a \beta} \\
			      \left(\epsilon \hat{\bar{X}}_M \right)_{\alpha b} & \left(\epsilon \hat{W}_M \right)_{\alpha \beta} \end{array} \right) + \hat{B}_M \sqrt{\frac{6}{5}} \left( \begin{array}{cc} \frac{1}{3} \delta_{a b} & 0 \\ 0 & - \frac{1}{2} \delta_{\alpha \beta} \end{array} \right) \thickspace .
\end{equation}
Here, we have introduced the abbreviations
\begin{eqnarray}
\label{eq:IIIfields}
 \left(\hat{G}_M\right)_{a b} &=& \sqrt{2} \hat{G}_M^i T^i_{a b}\thickspace, \hspace{1cm} i=1 \dots 8 \thickspace , \\
\left(\hat{W}_M\right)_{\alpha \beta} &=& \sqrt{2} \hat{W}_M^i T^{i+20}_{\alpha \beta}\thickspace, \hspace{1cm} i=1,2,3 \thickspace .
\end{eqnarray}
\(T^i\) are the generators of \(SU(5)\), \(a,b\) range from 1 to 3 and \(\alpha,\beta\) from 4 to 5.  Using this definition, the superpotential eq.~(\ref{3sup}) reads in the \(SU(3)_C\times SU(2)_L \times U(1)_Y \) basis without the colored Higgs
\begin{eqnarray}
 5_H 24 Y_{24} \bar{5}_M  &\rightarrow& Y_W \hat{H}_u \hat{W}_M \hat{l} - \sqrt{\frac{3}{10}} Y_B \hat{H}_u \hat{B}_M \hat{l} + Y_X \hat{H}_u \hat{\bar{X}}_M \hat{d} \thickspace , \\
\frac{1}{2} M_{24} 24_M 24_M  &\rightarrow& \frac{1}{2} M_B \hat{B}_M \hat{B}_M + \frac{1}{2} M_G \hat{G}_M \hat{G}_M + \frac{1}{2} M_W \hat{W}_M \hat{W}_M + M_X \hat{X}_M \hat{\bar{X}}_M \thickspace .
\end{eqnarray}
The fermionic \(SU(2)_L\) triplet \(\tilde{W}_M\)  and the gauge singlet \(\tilde{B}_M\) contribute to the mass term of the neutrino via seesaw type~III and type~I, respectively. The normalization of the triplet in eq.~(\ref{eq:IIIfields}) causes an additional factor of \(\frac{1}{2}\) in comparison to eq.~(\ref{eq:mvIII}). The resulting mass term is
\begin{equation}
 m_\nu = - \frac{v_u^2}{2} \left(\frac{1}{2} Y^T_W M_W^{-1} Y_W +  \frac{3}{10} Y^T_B M_B^{-1} Y_B \right) \thickspace . 
\end{equation}
The analysis of type~III is slightly more complicated than of type~II. The reason is that we have added several generations of heavy fields and the masses of the different generations of {\bf 24} can be very different. Thus, we have to deal with separated threshold scales for each generation. Moreover, the matrices \(M_W\), \(M_G\), \(M_B\) and \(M_X\) can receive large off-diagonal elements due to the RGE running. Therefore, we have to go to the new mass eigenbasis before integrating out the heavy fields by
\begin{eqnarray}
 M_i^{Dia} = U_i M^\dagger_i M_i U_i \thickspace , \hspace{1cm} 
 M_i^{Dia} = V_i M_i M^\dagger_i V_i \thickspace , 
\end{eqnarray}
with \(i=W,G,B,X\). The masses of the fields which are no longer part of the spectrum are the largest eigenvalues of \(M_i^{Dia}\). Of course, all other parameters have also to be rotated to the new basis 
\begin{eqnarray*}
Y_i & \rightarrow &  V_i \, Y_i\thickspace, \hspace{1cm} i = W,B,X \thickspace ,\\
T_i & \rightarrow &  V_i \, T_i\thickspace, \hspace{1cm} i = W, B,X \thickspace , \\
M_i & \rightarrow & U_i^\dagger \, M_i \, V_i \thickspace,  \hspace{1cm} i = W,G,X,B \thickspace ,\\
m_i^2 & \rightarrow & V_i^\dagger \, m_i^2 \, V_i \thickspace, \hspace{1cm} i = W,G,X,B,\bar{X} \thickspace ,
\end{eqnarray*}
We use as threshold scales the three eigenvalues of \(M_W\) at the GUT scale. Hence, no particle is integrated out at its mass and we have to take also the shifts caused by the \(SU(2)_L\) triplet into account. This leads to the  following boundary conditions for gauge couplings at the \(i\)-th threshold scale
\begin{eqnarray}
 g_1 & \rightarrow & g_1 \left(1 \pm \frac{g_1^2}{16 \pi^2} \frac{5}{2} \log\frac{m^i_X}{M^{0,i}_W}  \right) \thickspace , \\
 g_2 & \rightarrow & g_2 \left(1 \pm \frac{g_2^2}{16 \pi^2} \left(\frac{3}{2} \log\frac{m^i_X}{M^{0,i}_W} + 2 \log\frac{m^i_W}{M^{0,i}_W}\right) \right) \thickspace , \\
 g_3 & \rightarrow & g_3 \left(1 \pm \frac{g_2^2}{16 \pi^2} \left(\log\frac{m^i_X}{M^{0,i}_W} + 3 \log\frac{m^i_G}{M^{0,i}_W}\right) \right) \thickspace .
\end{eqnarray}
\(m^i_{G,W,X}\) are the physical masses of the \(i.\) generation of \(G_M, W_M\) and \(X_M\). \(M^{0,i}_W\) is the \(i.\) eigenvalue of the matrix \(M_W\) at the GUT scale. The shifts for the gauginos can again easily extracted from these equations. \\
In presence of  24-plets, \(I_2(R)\) changes by
\begin{equation}
\label{eq:Ir_III}
\Delta I^i_2(R) = 5 N_{24} \thickspace ,
\end{equation}
and the estimation of the induced lepton flavor violation is in this case
\begin{eqnarray}
\label{eq:LFVentriesIII}
m^2_{\tilde l,ij} &\simeq& -\frac{1}{8 \pi^2 } 
 \left( \frac{27}{5} m^2_0 + \frac{9}{5} A^2_0 \right) 
 \left(Y_B^\dagger L Y_B\right)_{ij}  \thickspace ,\\
A_{e,ij} &\simeq& -\frac{27}{5} \frac{1}{ 16 \pi^2 }   A_0 
 \left(Y_e Y_B^\dagger L Y_B\right)_{ij} \thickspace .
\end{eqnarray}
Here, we have used the same conventions as in eqs.~(\ref{eq:FCNCIa}) and (\ref{eq:FCNCIb}). The expected effect is again larger as for type~I and seems to be smaller than for type~II. However, the two-loop RGEs are very important in this case and the spectrum is in general lighter as for seesaw~II. \\
Finally, some words about the boundary conditions. As in the other cases, we use mSugra GUT scale conditions. In addition, we take for the seesaw~III as input the values of  \(M_W\) and \(Y_B\) at the GUT scale. The additional terms should be \(SU(5)\) invariant at the GUT scale, thus, the boundary conditions have to be
\begin{align}
 & M_B = M_G = M_X \equiv M_W \thickspace ,\\
 & Y_W = Y_X \equiv Y_B \thickspace ,\\
 & m_{G_M}^2 = m_{W_M}^2 =  m_{B_M}^2 = m_{X_M}^2 =  m_{\bar{X}_M}^2 \equiv m_0^2 \thickspace ,\\
 & T_W = T_X = T_B \equiv A_0 Y_B \thickspace .
\end{align}

\section{Results}
We present in the following our numerical results \cite{Esteves:2010ff}. We show first the effect of the additional particles on the masses of the MSSM particles. Afterwards, we discuss the lepton flavor violating processes. In the third subsection, we come to the features of dark matter in the different seesaw scenarios.  
\subsection{Effect of the heavy particles on the MSSM spectrum}
\label{sec:Seesaw_Mass_Spectrum}
\begin{figure}[ht]
  \centering
\begin{minipage}{16cm}
\includegraphics[scale=0.95]{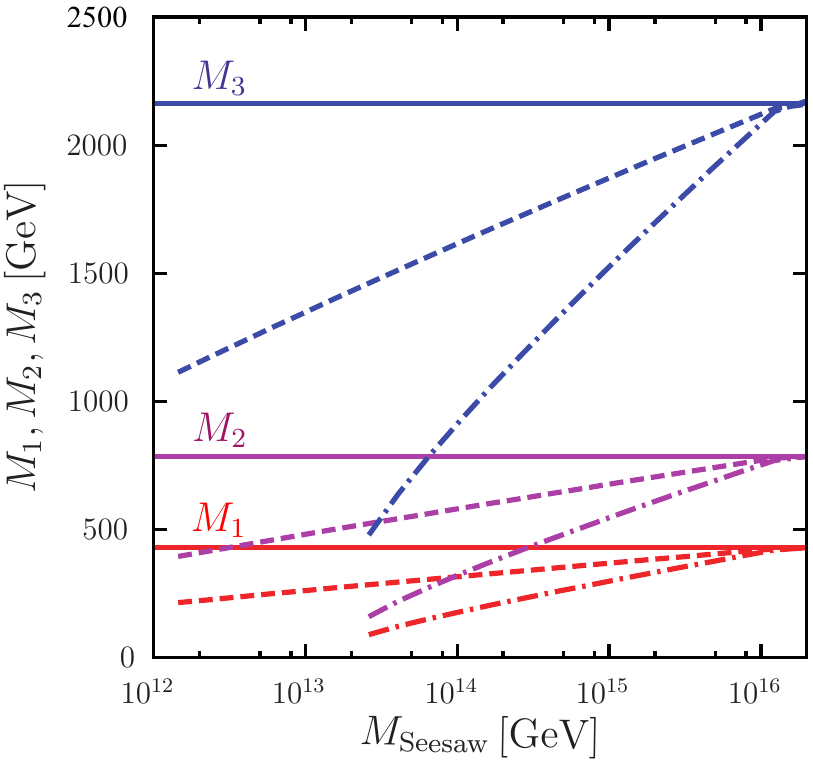} \hfill
\includegraphics[scale=0.95]{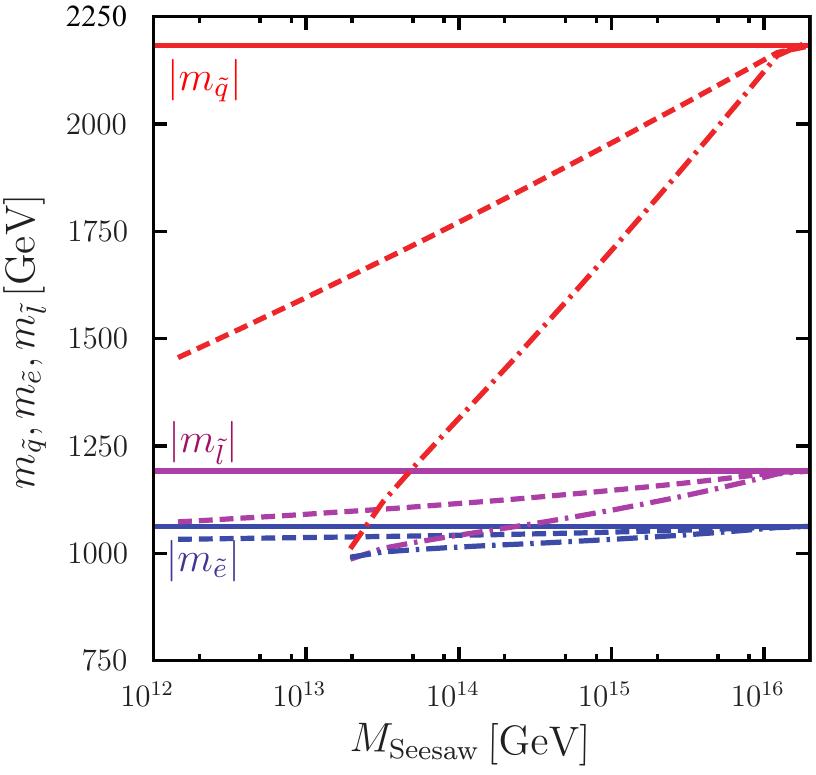}
\end{minipage}
\caption[Mass parameters versus seesaw scale]{Mass parameters at $Q=1$ TeV versus the seesaw scale for fixed
high scale parameters $m_0= M_{1/2} = 1$~TeV. The full lines correspond to seesaw type~I, the dashed ones to type~II and the
dash-dotted ones to type~III. In all cases, a degenerate spectrum of the seesaw particles has been assumed.}
  \label{fig:mass1000}
\end{figure}
\begin{figure}[!htb]
  \centering
\begin{minipage}{16cm}
\includegraphics[scale=0.95]{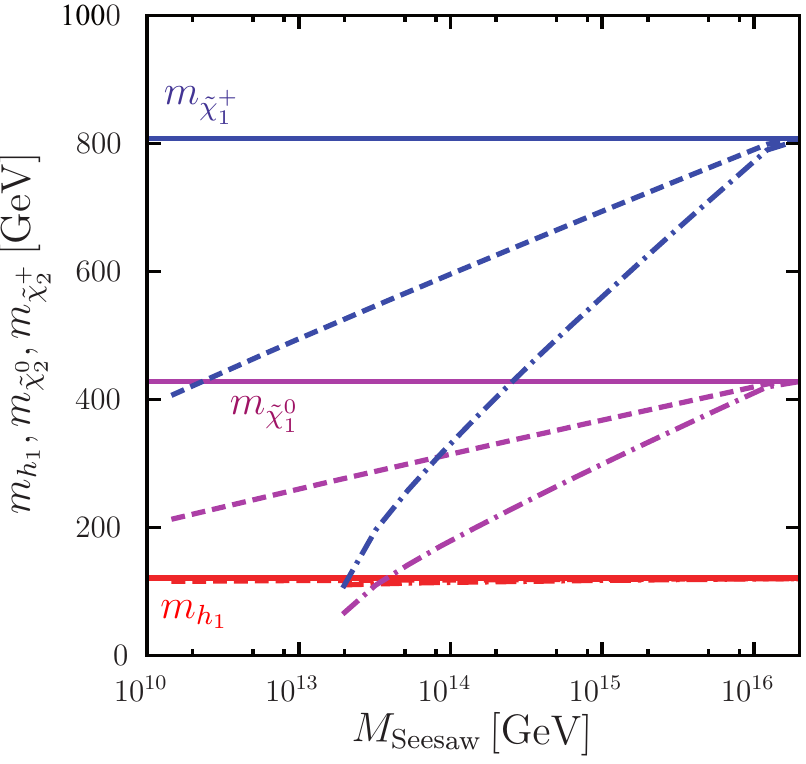} \hfill
\includegraphics[scale=0.95]{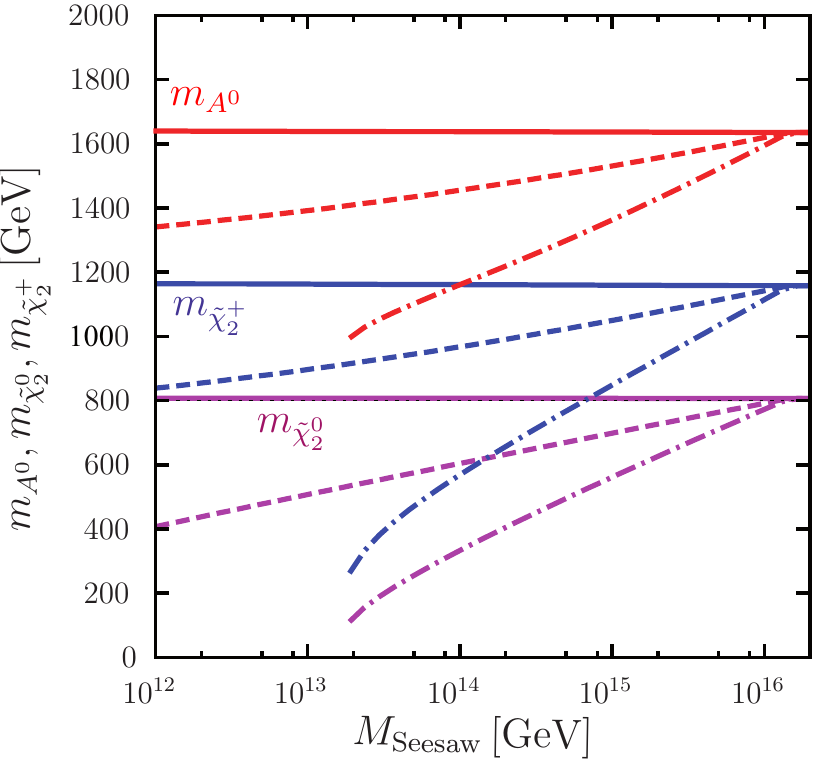}
\end{minipage}
\caption[Masses versus seesaw scale]{Example of different masses at $Q=1$ TeV versus the seesaw scale for fixed
high scale parameters $m_0= M_{1/2} = 1$~TeV, $\tan\beta=10$ and $\mu>0$. On left panel are $M_h$, $m_{\Neu_1}$ and $m_{\Cha_1}$ shown, while on the right panel we have  $M_A$, $m_{\Neu_2}$ and $m_{\Cha_2}$. The line codes are as in Fig.~\ref{fig:mass1000}.}
  \label{fig:spectra}
\end{figure}
The presence of charged particles at scales between the electroweak scale and the GUT scale leads to changes in the \(\beta\)-functions of the gauge couplings. In the MSSM, the one-loop \(\beta\)-functions of the gauge couplings are of the form 
\begin{equation}
\beta^{(1)}_{g_i} = \frac{1}{16 \pi^2} b_i g_i^3 \thickspace .
\end{equation}
The coefficients \(b_i\) are just the Dynkin index \(I_2(R)\) and the values for the MSSM particle content are $b_1=\frac{33}{5},b_2=1,b_3 = -3$. As already shown in eqs.~(\ref{eq:Ir_II}) and (\ref{eq:Ir_III}), these coefficients can receive large contributions if 15- or 24-plets are present.  In case of seesaw~II, this results in a total shift of  $\Delta b_i=7$ and in case of type~III assuming three generations of {\bf 24}, the total difference is  $\Delta b_i=15$. Furthermore, the heavy fields  change not only the evolution of the gauge couplings but also the evolution of the gaugino and scalar mass parameters with strong implications on the particle spectrum \cite{Buckley:2006nv,Hirsch:2008gh}. Additional effects on the spectrum of the scalars can be present if some of the Yukawa  couplings get large \cite{Hirsch:2008gh,Calibbi:2009wk,Biggio:2010me}. In Fig.~\ref{fig:mass1000}, we  show the dependence of several mass parameters at the scale $Q=1$~TeV on the seesaw scale \(M_{\rm{seesaw}}\). Here, we have chosen for seesaw~I and III a degenerated scale for all three generations of heavy fields.  Furthermore, we have fixed the 
GUT scale parameters to $m_0= M_{1/2} = 1$ TeV and we have set all additional Yukawa couplings to zero. As expected, the effects in case of seesaw type~II and type~III grow with decreasing seesaw scale. This implies that for 'low' values of the seesaw scale the spectrum depends strongly on the given seesaw scenario for a fixed set of GUT parameters.  For illustration, we give in Fig.~\ref{fig:spectra} some examples for the corresponding masses.  Here, we have fixed $\tan\beta=10$ and $A_0=0$~GeV. The ratio of the gaugino mass parameters is nearly the same in all three seesaw types  as in usual mSugra scenarios. As opposed to that, the ratios of the sfermion mass parameters change \cite{Buckley:2006nv,Hirsch:2008gh}. \\

\begin{figure}[ht]
  \begin{minipage}{16cm}
  \centering
  \includegraphics[scale=1.]{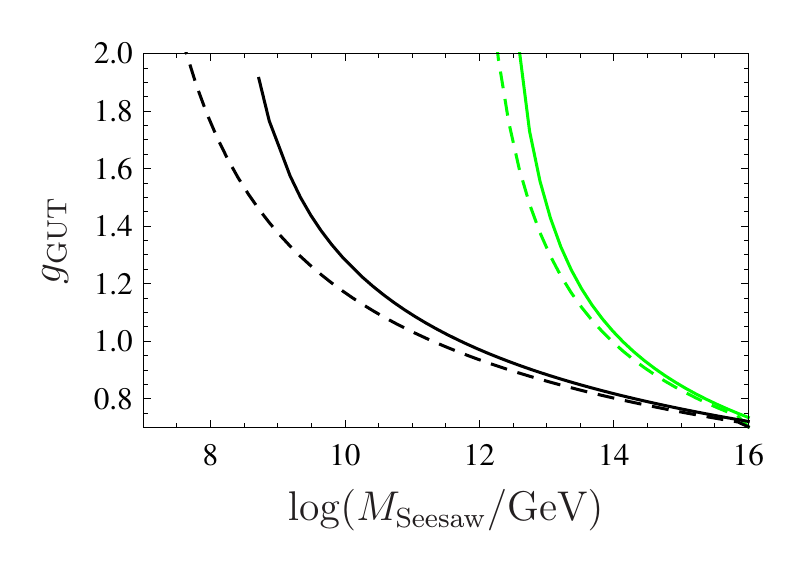} 
  \end{minipage}
 \caption[Landau poles for seesaw~II and III]{Values of the gauge coupling at $M_{GUT}=2 \cdot 10^{16}$~GeV as a function of the seesaw scale. The black lines are for seesaw II and green lines for type~III with three {\bf 24}-plets with degenerate mass
spectrum, the  full  lines correspond to  two-loop and the dashed ones to one-loop. For the calculation of the electroweak threshold the spectrum corresponds to $m_0=M_{1/2}=1$~TeV, $A_0=0$~GeV, $\tan\beta=10$ and $\mu>0$.}
  \label{fig:landau}
\end{figure}

To underline the importance of the two-loop RGEs, we note that the use of the two-loop RGEs for seesaw~III  leads to a shift of the GUT scale defined as $g_{U(1)_Y} = g_{SU(2)_L}$: using a degenerate seesaw scale of \(10^{13}\)~GeV, the one-loop value of about $1.0 \cdot 10^{16}$~GeV changes to roughly $4.2 \cdot 10^{16}$~GeV at two-loop level. This implies also that there is some difference for the strong coupling of  5-10\%. This can easily be accounted by threshold effects of the new GUT particles, e.g.~the missing members of the gauge fields and the Higgs fields responsible for the breaking of the GUT group. A detailed discussion about these threshold effects in \(SU(5)\) up to three-loop is given in \cite{Martens:2010nm}. A second effect of the heavy fields is that for one-loop and for two-loop RGEs the increase of the \(\beta\) coefficients cause larger values of the gauge couplings at the GUT scale. This implies that one reaches a Landau pole for sufficiently low values of the seesaw scale. As an example we show in
Fig.~\ref{fig:landau} the value of the gauge coupling at $M_{GUT}=2 \cdot 10^{16}$~GeV as a function of the seesaw scale for type~II with a pair of 15-plets and for type~III with three degenerate 24-plets. In both cases, the two-loop RGEs lead to a larger gauge coupling for a fixed seesaw scale. One can see that in case of seesaw~II, in principle, one could reach a seesaw scale of about $10^{8}$~GeV while for type~III a lower limit of $10^{13}$~GeV exists. However, we believe that we can no longer trust even the two-loop calculation for such large values of the $g_i$, as the neglected higher order terms become more and more important.

\subsection{Lepton flavor violation}
\label{sec:Seesaw_LFV}
\begin{figure}[ht]
  \centering
  \begin{minipage}{16cm}
  \includegraphics[scale=0.95]{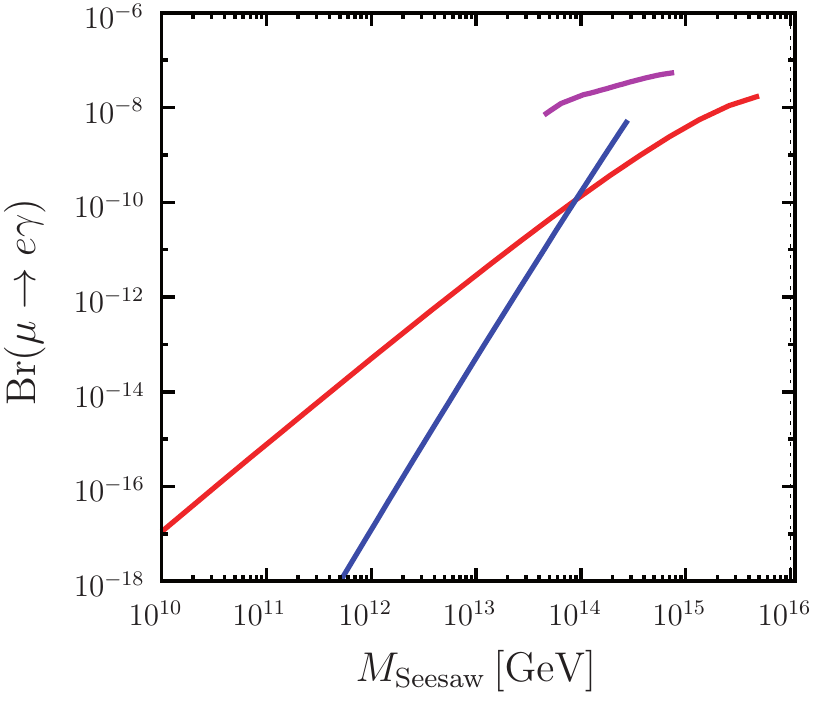}  \hfill
  \includegraphics[scale=0.95]{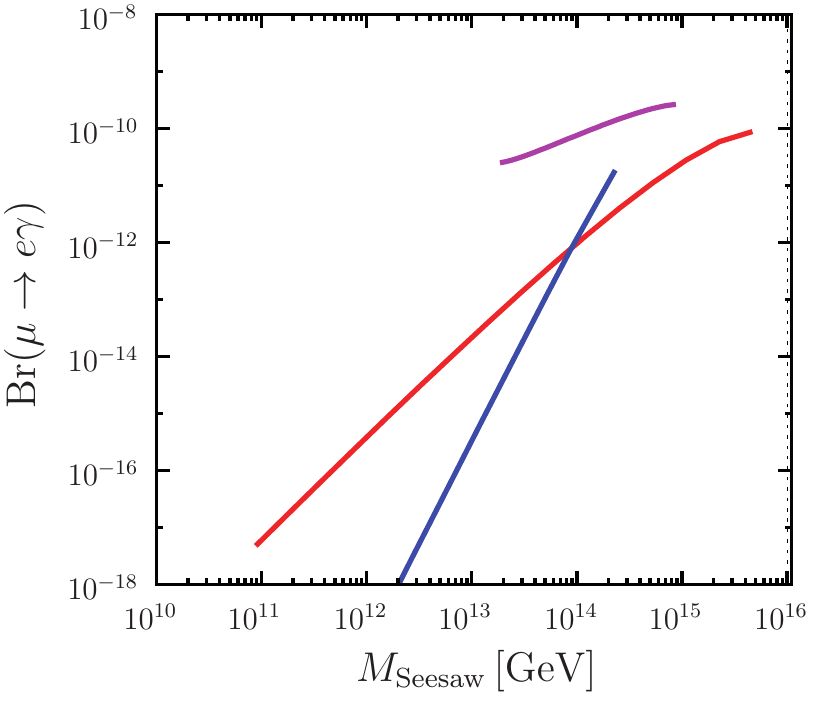}      
  \end{minipage}
  \caption[$\mu \rightarrow e \gamma$ for seesaw~I-III]{$\mu \rightarrow e \gamma$ for seesaw~I (red), seesaw~II (blue) and seesaw~III (pink) versus seesaw scale. On the left panel $m_0=m_{1/2}=300$~GeV, on the right panel $m_0=m_{1/2}=1000$~GeV. In both cases we take $\tan\beta=10$, $A_0$=0~GeV and $\mu>0$.}
  \label{fig:5a}
\end{figure}
\begin{figure}[!htb]
  \centering
\includegraphics[scale=0.95]{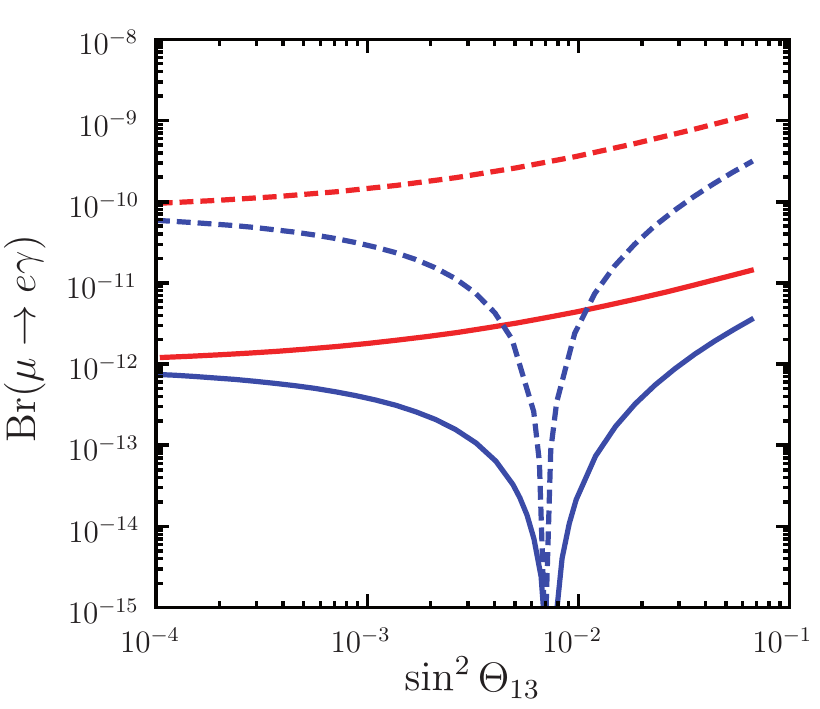}  
  \caption[$\mu \rightarrow e \gamma$ versus $\sin^2 \Theta_{13}$]{$\mu \rightarrow e \gamma$ versus $\sin^2 \Theta_{13}$ for $m_0=M_{1/2}=1000$~GeV, $\tan\beta=10$, $A_0=0$~GeV, $\mu>0$ and $M_{\rm seesaw} =10^{14}$~GeV. The color code is: seesaw type~I (solid lines) and seesaw type~III (dashed lines). The curves shown are for $\delta=0$ (red) and $\delta=\pi$ (blue) for normal hierarchy.}
  \label{fig:3a}
\end{figure}
We have already seen in eq.~(\ref{eq:LLGapprox})  that rates for the lepton flavor violating decays of $\mu$ and $\tau$ depend quadratically on the LFV entries in the slepton mass squared matrix. Furthermore, they are inverse proportional  to the overall SUSY mass to the power eight. Using this, one can immediately conclude that the rates for rare leptonic decays are in general larger for type~II and III than for type~I if the SUSY masses and seesaw scales are fixed. The only exception would be the presence of special cancellations. On the one hand, the LFV entries are larger in case of type~II compared to type~III as can be  seen in eqs.~(\ref{eq:LFVentriesIII}) and (\ref{eq:LFVentriesII}). On the other hand, the spectrum is lighter for fixed mSugra parameter in case of type~III compared to type~II. In sum, the largest rates for the rare lepton decays take place for seesaw~III. As example, we depict in Fig.~\ref{fig:5a} the case of $\mu\to e \gamma$. Here, we show  for all three seesaw models the branching ratio \(\mbox{Br}(\mu\to e \gamma)\) versus the seesaw scale. For this purpose, we used again degenerate seesaw spectra in case of type~I and III. The additional Yukawa couplings were fixed to obtain correct neutrino data. As we can see, it is only possible to show a relatively short interval  for the seesaw scale in case of type~III. There are mainly two reasons for that: first,  the gauge coupling get large at $M_{GUT}$ for small \(M_{\rm{seesaw}}\) as already discussed and, thus, perturbation  theory breaks down. Second, the squared masses of some scalars, in particular of the lighter stau and/or lighter sbottom, are getting negative. The second point can be circumvented to some extend by using larger soft breaking parameters at the GUT scale.

\begin{figure}[ht]
  \centering
  \begin{minipage}{16cm}
\includegraphics[scale=0.95]{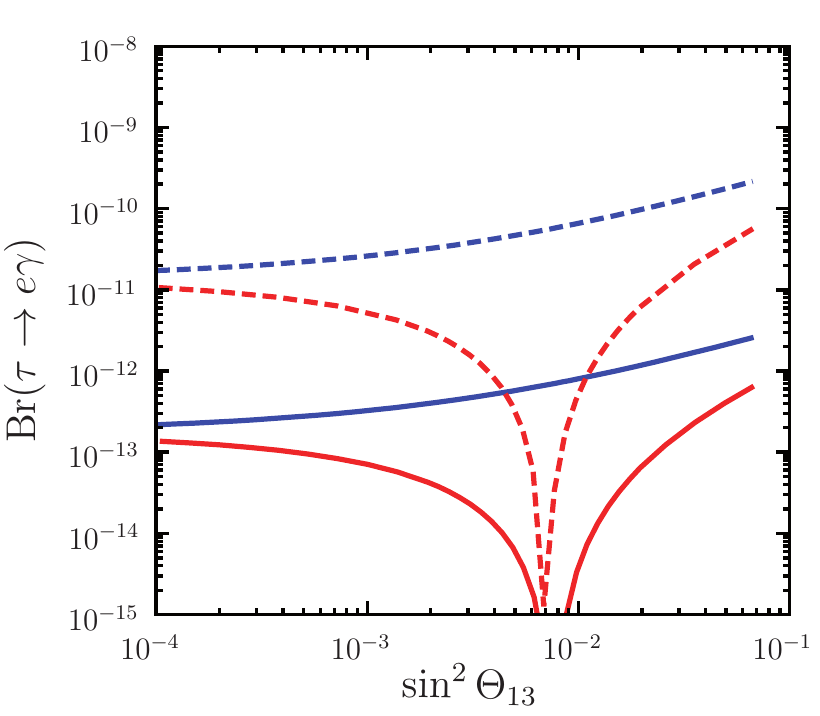} \hfill
\includegraphics[scale=0.95]{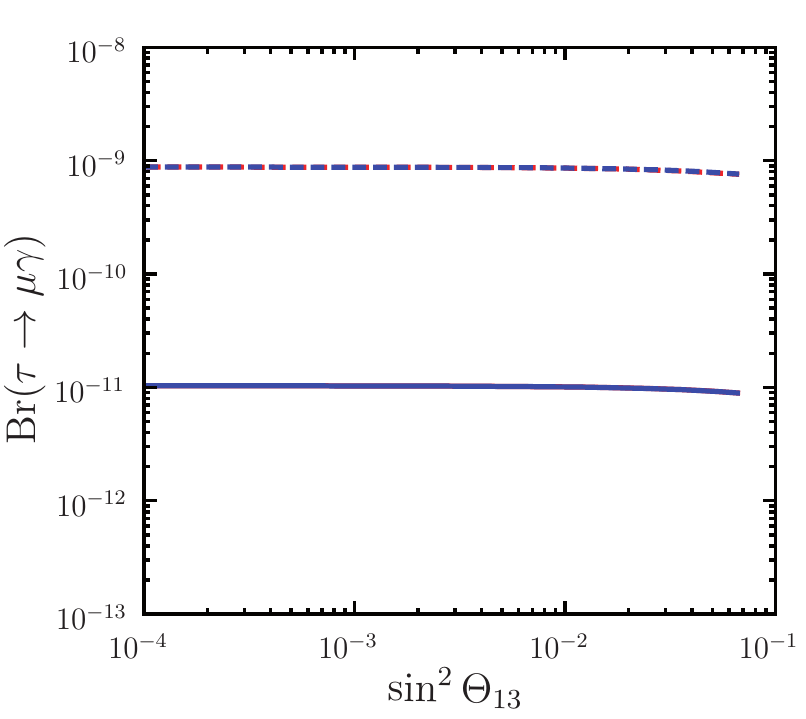}  
  \end{minipage}
  \caption[$\tau \rightarrow e \gamma$ versus $s_{13}^2$ and  $\tau \rightarrow \mu \gamma$ versus $\sin^2 \Theta_{13}$]{$\tau \rightarrow e \gamma$ versus $\sin^2 \Theta_{13}$ (left) and $\tau \rightarrow \mu \gamma$ versus $\sin^2 \Theta_{13}$ (right)  for $m_0=M_{1/2}=1000$~GeV,   $\tan\beta=10$, $A_0=0$~GeV and $\mu>0$, for seesaw type~I (solid lines) and seesaw type~III (dashed lines), for $M_{\rm seesaw} =10^{14}$~GeV. The curves shown are for $\delta=0$ (red) and $\delta=\pi$ (blue) for normal hierarchy.}
  \label{fig:3b}
\end{figure}

Especially for the seesaw~III, heavy SUSY masses are preferred: using generic Yukawa couplings which can explain neutrino data, always a heavy spectrum is needed to be consistent with experimental bounds on rare leptonic decays. Nevertheless, there might still an accidental  cancellation  take place between different contributions to these rare decays. As an example, we depict in the left plot of  Fig.~\ref{fig:3b}  again $\mbox{Br}(\mu\to e\gamma)$ for seesaw~I and III. This time, we show the dependence on the reactor angle  $\sin^2 \Theta_{13}$ for different values $\delta$. Obviously, there is an allowed range of $\sin^2 \Theta_{13}$ for  $\delta \simeq \pi$. Even if this looks like a fine-tuning of parameters, we can interpret it also from another point of view: if we assume that MEG collaboration has found a non-vanishing value for $\mbox{Br}(\mu\to e\gamma)$ and the measured spectrum at LHC is consistent with seesaw~III,  it would be possible to obtain for a fixed \(R\)-matrix of eq.~(\ref{eq:inverseMassTerm})  a relation between $\sin^2 \Theta_{13}$ and $M_{24}$ assuming a degenerate seesaw spectrum. This relation can be used to receive bounds on \(M_{24}\), or, if we are lucky, even to determine it.\\
In the right plot of Fig.~\ref{fig:3b} we show the corresponding rare \(\tau\)-decays. Also for $\tau\to e \gamma$, similar cancellation as just discussed might exist. However, this range in parameter space are already excluded by $\mu\to e\gamma$. In contrast, \(\tau \rightarrow \mu \gamma\) is insensitive to the reactor angle and should be measurable in the near future. 

\begin{figure}[!htb]
  \centering
\begin{minipage}{16cm}
  \includegraphics[scale=0.9]{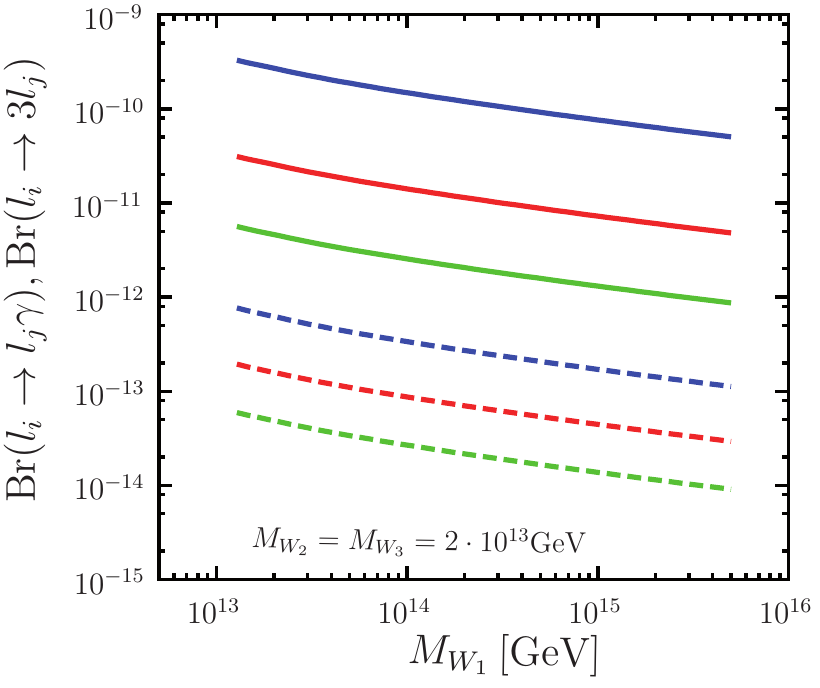} \hfill
  \includegraphics[scale=0.9]{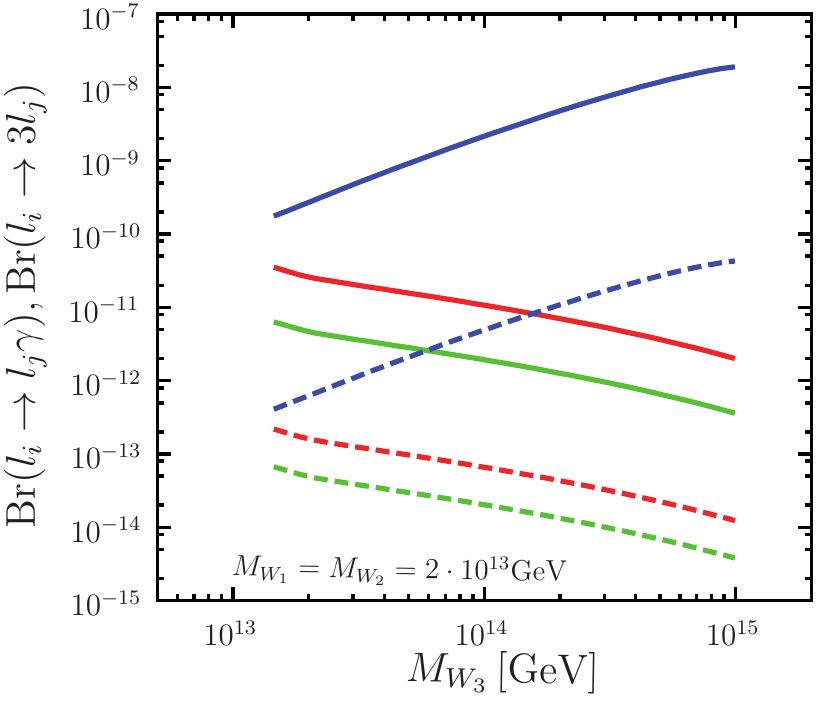}  
 \end{minipage}
  \caption[Branching ratios for $l_i \to l_j \gamma$ and $l_i \to
    3l_j$ versus the seesaw scale]{Branching ratios for $l_i \to l_j \gamma$ (solid lines) and $l_i \to 3l_j$ (dashed lines) versus the seesaw scale for $\tan\beta=10$, $\mu>0$, $A_0=0$~GeV, $M_{1/2}=m_0=1$~TeV. On the left panel, we varied $M_{W_1}$ with $M_{W_2}=M_{W_3}=2\cdot 10^{13}$~GeV, while on the right panel, we scan over $M_{W_3}$ with $M_{W_1}=M_{W_2}=2\cdot 10^{13}$~GeV. The color code is red for $\mu\to e \gamma$ or $\mu \to 3 e$, blue for $\tau\to \mu \gamma$ or $\tau \to 3 \mu$ and green for $\tau\to e \gamma$ or $\tau \to 3 e$.}
  \label{fig:NonDeg1}
\end{figure}

So far, we have assumed that the heavy fields are nearly degenerated in case of type~I and type~III. In Fig.~\ref{fig:NonDeg1}, we have softened this assumption for the seesaw~III and fixed two masses but varied the third one. We note that the indices of the heavy fields are generation indices and do not relate to a particular mass ordering: \(M_{W_2}\) corresponds to the 'solar neutrino scale' and \(M_{W_3}\) to the 'atmospheric neutrino scale'. In the left plot of  Fig.~\ref{fig:NonDeg1}, we varied the mass \(M_{W_1}\) of the first state. In that case, the branching ratios decrease with increasing  \(M_{W_1}\). The reason is that the SUSY spectrum also increases. In contrast, the implication on the neutrino physics is only small  and mainly caused by a light increase of the corresponding Yukawa couplings to obtain the correct neutrino masses. In the right plot of Fig.~\ref{fig:NonDeg1}, we varied the mass of the third heavy state. Here, a stronger correlation between the mass and the branching ratios is visible: the Yukawa couplings must increase to obtain the correct neutrino  mass difference squared for the atmospheric sector. This causes the increasing branching ratios for the \(\tau\) decays, while the ratios for \(e\) and \(\mu\) decrease and the \(\tau\) decays might be more important than \(\mu \rightarrow e \gamma\).

\subsection{Dark Matter}
\begin{figure}[!htb]
  \centering
  \begin{minipage}{16cm}
  \includegraphics[scale=0.95]{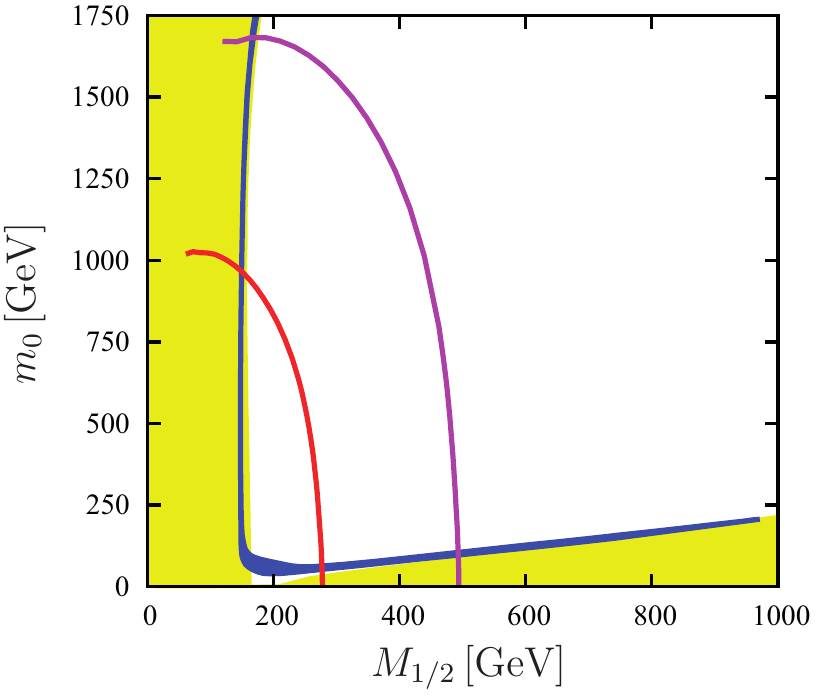} \hfill
  \includegraphics[scale=0.95]{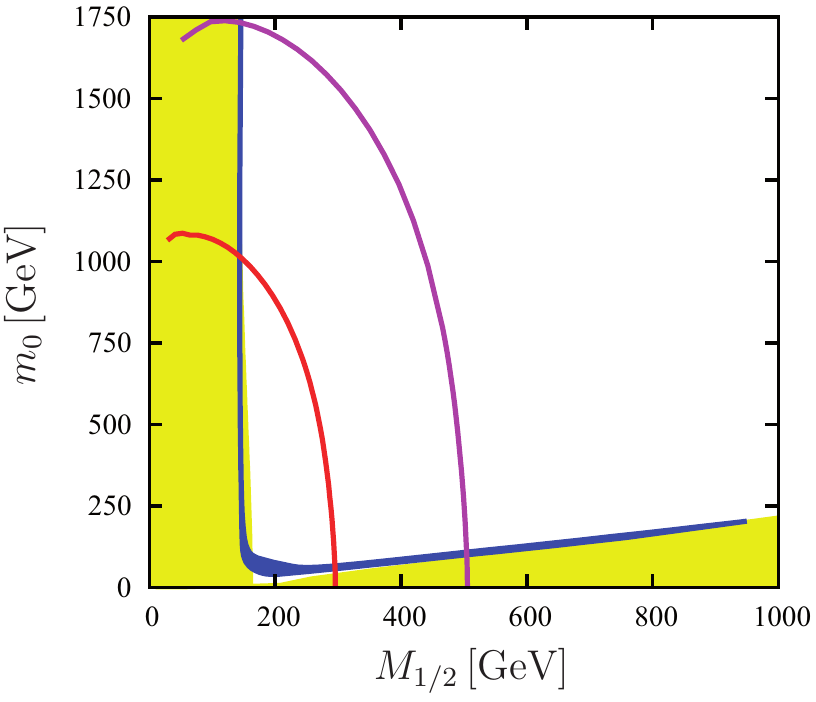} \\
  \includegraphics[scale=0.95]{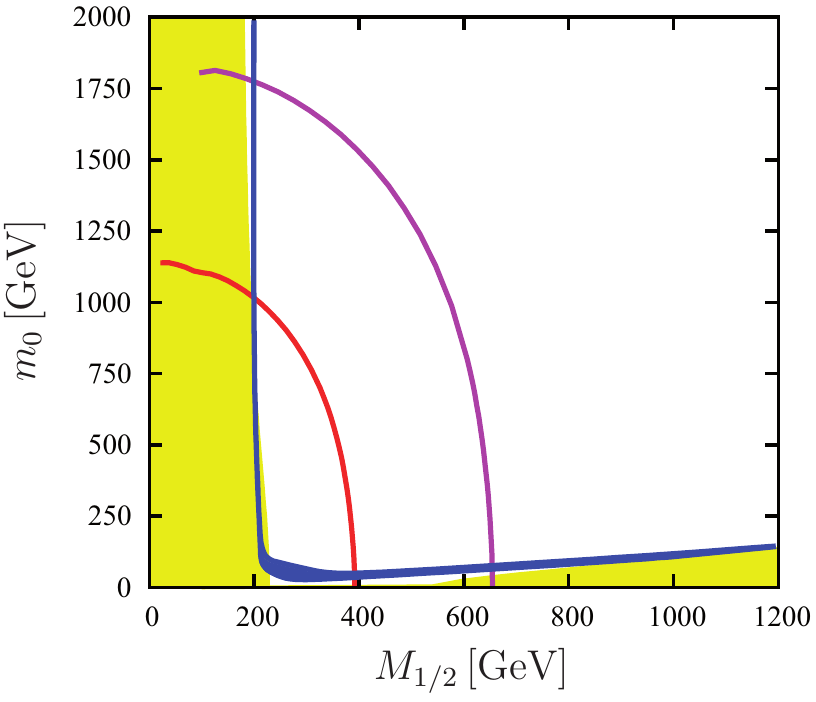} \hfill
  \includegraphics[scale=0.95]{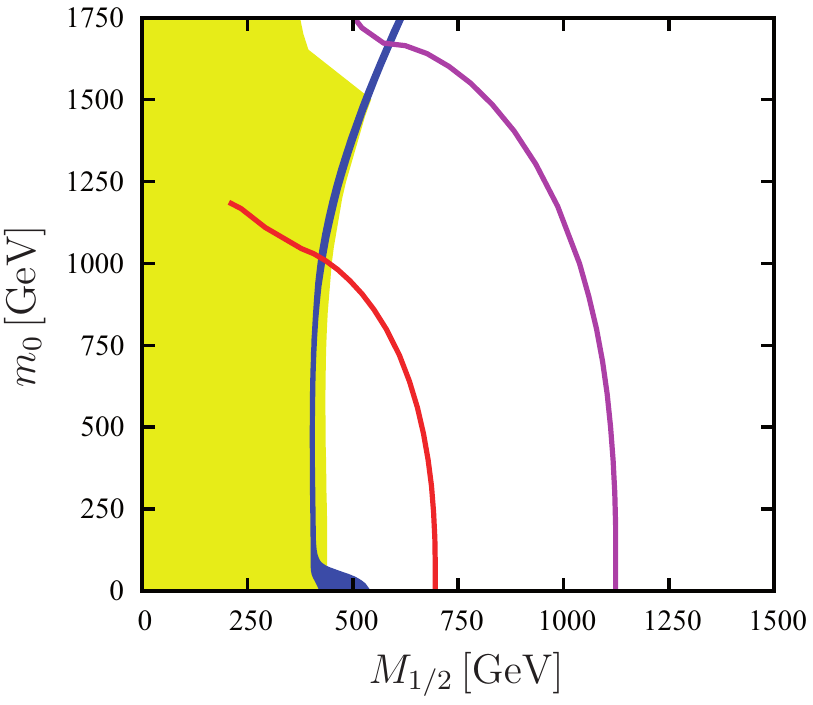}
\end{minipage}
   \caption[Dark matter in mSugra and seesaw~I-III]{Dark matter allowed region (in blue) for mSugra (left upper plot), for seesaw~I (right upper plot), seesaw~II (left lower plot) and seesaw~III (right lower plot). The parameters are $\tan\beta=10$, $A_0=0$, $\mu>0$ and  $M_T=10^{14}$ GeV for  $m_{top}=171.2$ GeV. The yellow regions are excluded by LEP (small values of $M_{1/2}$) or by a charged LSP (small values of $m_0$). Also shown are the Higgs mass curves for $M_H=110$ GeV (in red) and for $M_H=114.4$ GeV (in magenta).}
   \label{fig:2NewA}
\end{figure}
We turn now to the discussion of the dark matter properties in the different seesaw scenarios. For this purpose, we show in Fig.~\ref{fig:2NewA} a comparison between a standard mSugra scenario and the three seesaw models. The standard high scale parameters of mSugra parameters were fixed to
\begin{eqnarray}
 \tan\beta=10 \thickspace, \quad A_0=0\,\GeV \thickspace, \quad \mu>0 \thickspace . 
\end{eqnarray}
In addition, the seesaw scale was in all scenarios set to \(M_{N_i}=M_{W_i}=M_T = 10^{14}\)~GeV and we used a top mass of $m_{top}=171.2$~GeV if not stated different. We have calculated the relic density using {\tt micrOMEGAs} \cite{Belanger:2006is}. The blue bands show the 3\(\sigma\) range according to \cite{Amsler:2008zzb}
\begin{equation}
 \label{eq:Range3SWMAP5}
 0.0810 < \Omega h^2 < 0.129  \thickspace .
\end{equation}
The standard mSugra case is shown in the upper, left plot of Fig.~\ref{fig:2NewA}. Three distinct regions with the correct relic density are visible: the stau-coannihilation with a lighter stau mass close to the mass of the LSP for \(M_{1/2} > 300\,\GeV\), the bulk-region for moderate values of \(M_{1/2}\) and \(m_0\) resulting in small sfermion masses as well as the focus point region for \(M_{1/2} \simeq 170 \, \GeV\) and larger values of \(m_0\). In the  last region, the Higgsino component of the lightest neutralino is in the percent range. However, all regions but the coannihilation region are ruled out by experimental constraints: the mass of the lightest chargino is in the focus point region below the LEP limit. This is a direct consequence of the small value for \(M_{1/2}\). Furthermore, the focus point region and also the bulk region are ruled out by bounds from Higgs searches even if a moderate limit of \(110\)~GeV is applied. At first glance, this limit seems too low since LEP has already excluded Higgs masses up to 114.4~GeV \cite{Amsler:2008zzb}. However, this limit is only valid for a Higgs which couples to \(b \overline{b}\) with the same strength as the SM Higgs does, while it is known that this coupling is reduced in some areas of the MSSM parameter space. Moreover, there are theoretical uncertainties in the calculated Higgs mass. Although, we have used two-loop RGEs and two-loop self-energies for calculating the Higgs mass, the known uncertainty in \(\DR\)-scheme are 3-5~GeV as discussed in \cite{Allanach:2004rh,Heinemeyer:2004xw}. This leads to the shown, conservative exclusion line of 110~GeV when all these uncertainties are taken into account. As we will discuss for seesaw~II and III, it is possible to reduce the areas excluded by Higgs masses by choosing larger, negative values of \(A_0\).

\begin{figure}[t]
  \centering
  \begin{minipage}{16cm}
  \includegraphics[scale=1.]{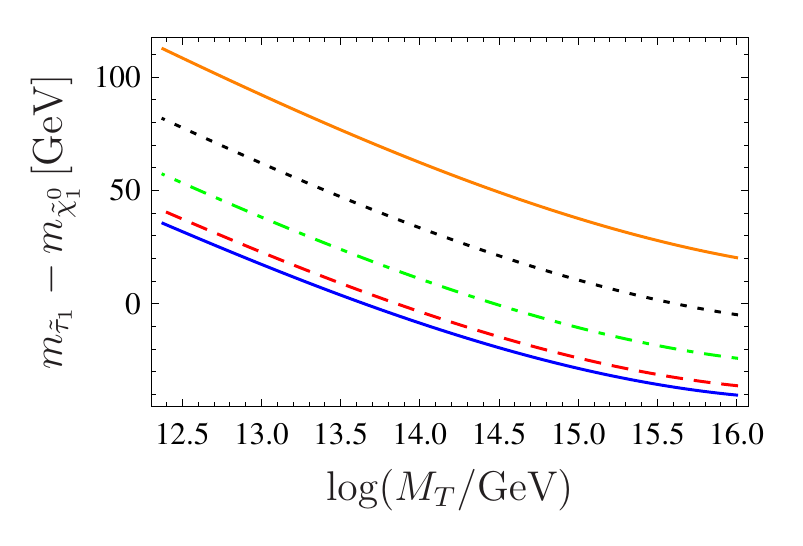} \hfill
  \includegraphics[scale=1.]{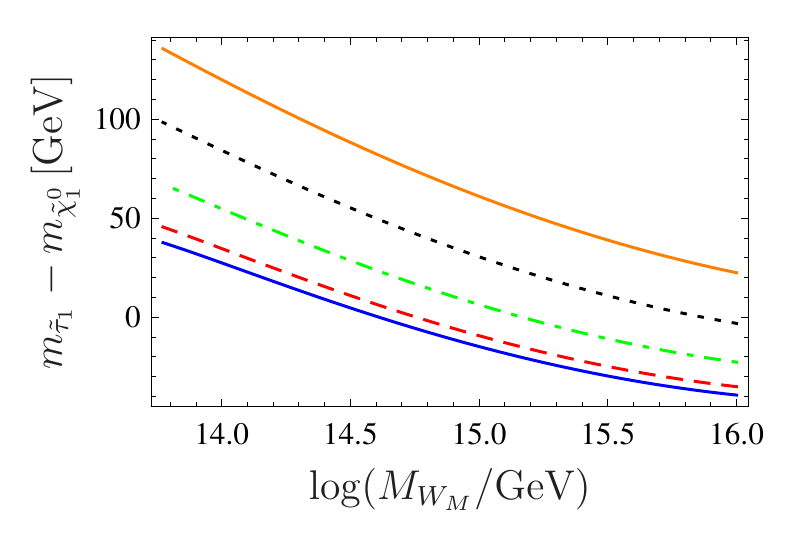} \\
  \includegraphics[scale=1.]{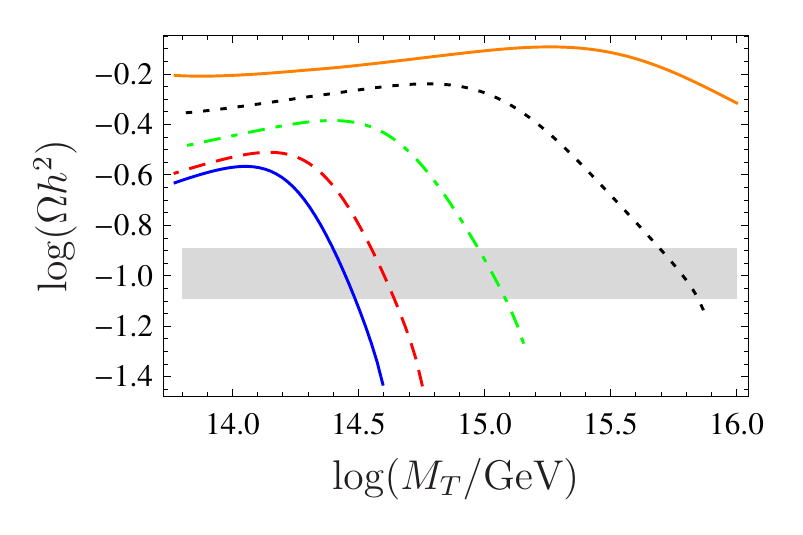} \hfill
  \includegraphics[scale=1.]{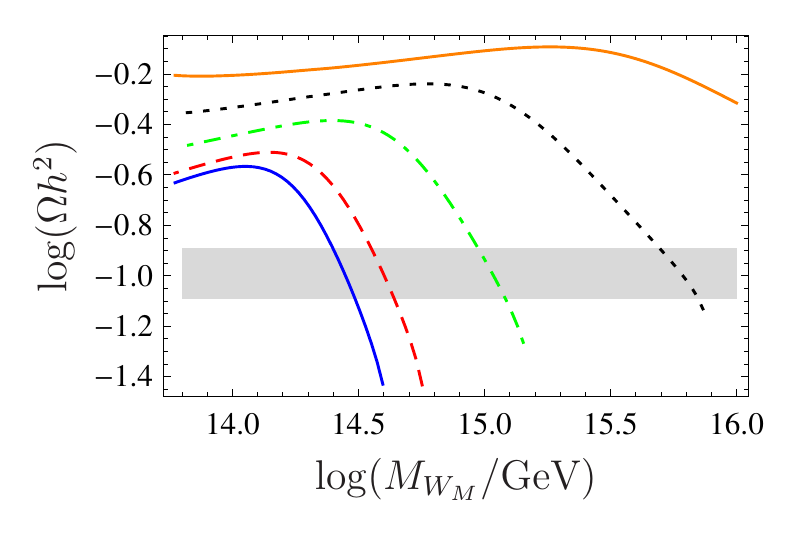} \\
\end{minipage}
   \caption[Dependence of mass difference between the lightest neutralino $\Neu_1$ and the lighter stau $\tilde{\tau}_1$ on the seesaw scale]{The upper row shows the dependence of the mass difference between the lightest neutralino $\Neu_1$ and the lighter stau $\tilde{\tau}_1$ on the seesaw scale for different values of $m_0$. The other parameters were fixed to $A_0=0$~GeV, $\mu>0$ and $M_{1/2} = 800$~GeV. The second row shows the corresponding relic density $\Omega h^2$. The left hand side is for seesaw~II, the right hand side for seesaw~III. The color code is as follows: 200~GeV (orange line), 150~GeV (dotted black line),  100~GeV (dotdashed green line), 100~GeV (dashed red line)  and 0~GeV (blue line). The gray band shows the preferred range according to eq.~(\ref{eq:Range3SWMAP5}). }
   \label{fig:Coann_II_III}
\end{figure}

\begin{figure}[!htb]
   \begin{minipage}{16cm}
  \includegraphics[scale=0.95]{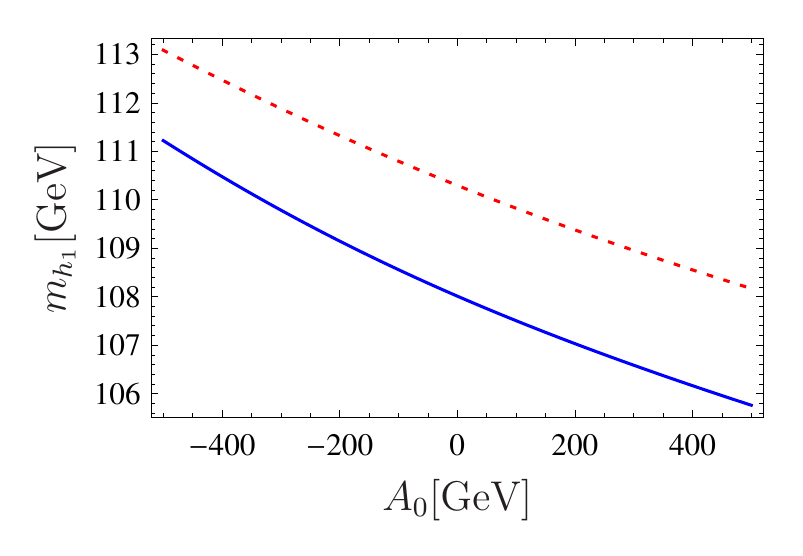}  \hfill
\includegraphics[scale=0.95]{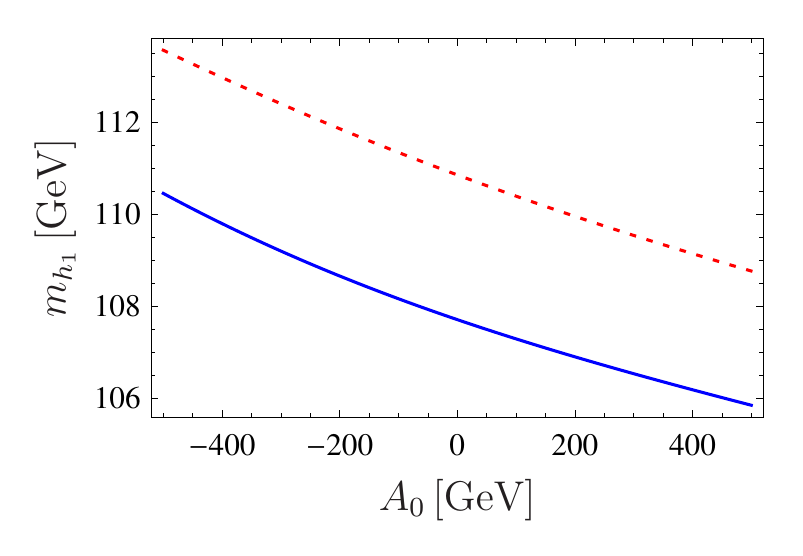} \\
\includegraphics[scale=0.95]{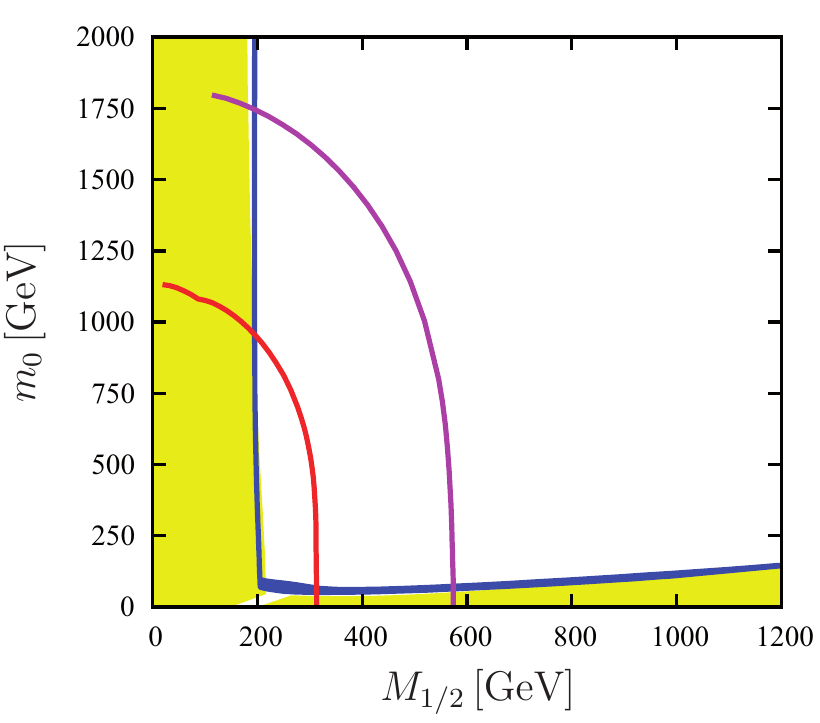}  \hfill
\includegraphics[scale=0.95]{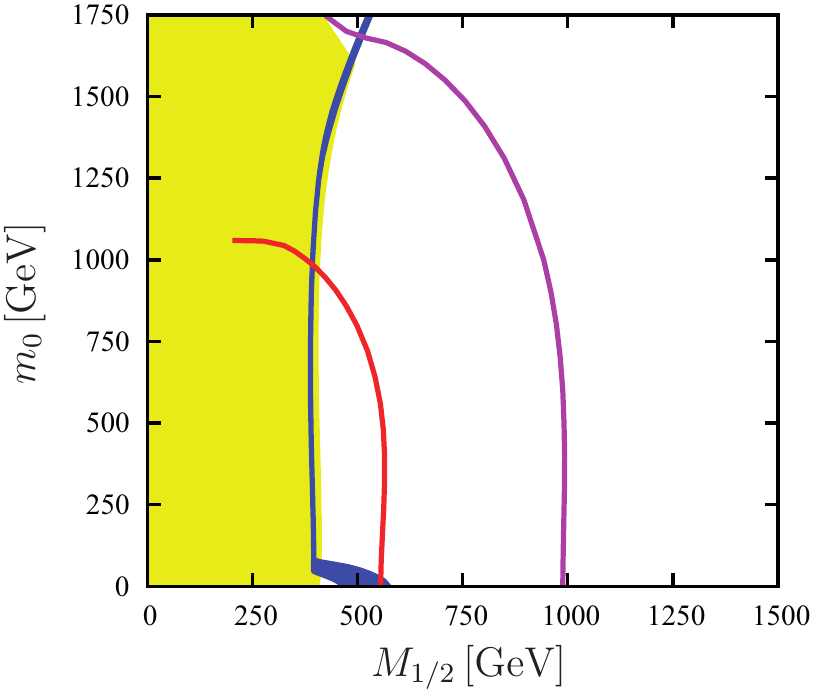} \\
\end{minipage}
     \caption[Dependence of the lightest Higgs mass $m_{h_1}$ on the GUT value of $A_0$]{The upper row shows the dependence of the lightest Higgs mass $m_{h_1}$ on the GUT value of $A_0$. We have chosen $\tan\beta=10$ and $\mu >0$. Left: seesaw~II with  $m_0 = M_{1/2} = 300\,\GeV$ (blue line) and $m_0 = 100\,\GeV,M_{1/2} = 400\,\GeV$ (dotted red line). Right: seesaw~III with  $m_0 = M_{1/2} = 500\,\GeV$ (blue line) and $m_0 = 100\,\GeV,M_{1/2} = 750\,\GeV$ (dotted red line). In the second row is the effect on the $(m_0,M_{1/2})$-plane shown: the parameters are the same as in in Fig.~\ref{fig:2NewA}, but $A_0=-300\,\GeV$ was used. The left plane is for type~II, the right one for type~III. }
   \label{fig:h_A0}
\end{figure}

The upper right plot of Fig.~\ref{fig:2NewA} shows the result for seesaw~I. As expected, there is no obvious difference in that case in comparison to the standard mSugra. This is an effect of the unchanged mass spectrum what we have already seen in sec.~\ref{sec:Seesaw_Mass_Spectrum}. We can immediately go to the discussion of type~II, depicted in the lower, left plot of Fig.~\ref{fig:2NewA}. Since the heavy particles have a significant impact on the SUSY masses in this case, the areas with correct relic density are also shifted. First, the coannihilation prefers regions with smaller values of \(m_0\): for a lower seesaw scale, the LSP as well as the NLSP are lighter. However, this effect is larger for the LSP and, thus, the mass ratio of NLSP to LSP increases. This has to be compensated, for sufficient coannihilation, by smaller values of \(m_0\). For \(\log(\frac{M_T}{\GeV}) < 13.2\), the coannihilation region disappears completely as shown in Fig.~\ref{fig:Coann_II_III}: the gap in masses between the lighter stau and the lightest neutralino is even for \(m_0=0\,\GeV\) too large to enable coannihilation. Second, the focus point region is shifted to a slightly larger value of \(M_{1/2}\) of 200~GeV. The reason is that neutralinos are generally lighter, as we have seen, in comparison to a pure mSugra scenario: for \(m_0 = 500\)~GeV and \(M_{1/2}=180\)~GeV the mass of the LSP is for standard mSugra 68.3~GeV, while it is for seesaw~II with 46.8~GeV notedly smaller. The Higgsino component is in both cases the same with 5.2~\%. Although for \(M_{1/2}=200\)~GeV, the neutralino is in the seesaw~II case still lighter with a mass of 53.0~GeV, but the Higgsino fraction is reduced to 4.3~\% resulting in the correct relic density. Even if the focus point region is shifted to larger values, it is still in conflict with LEP bounds because the chargino mass is reduced in the same way the neutralino mass is (see sec.~\ref{sec:Seesaw_Mass_Spectrum}). Moreover, also the exclusion regions stemming from Higgs searches cover larger regions of the parameter space.  However, these constraints can be reduced by changing the GUT value of \(A_0\), since the loop-corrections to the Higgs masses are sensitive to the mixing in the stop sector. We show the dependence of the lightest Higgs mass on \(A_0\) for two pairs of values of \(M_{1/2}\) and \(m_0\) in Fig.~\ref{fig:h_A0}. The upper, left plot of Fig.~\ref{fig:h_A0} shows the corresponding change in the \((m_0,M_{1/2})\)-plane for choosing \(A_0 = - 300\,\GeV\): the features of the regions with correct relic density are kept untouched but a larger range of the coannihilation region is no longer ruled out by Higgs search. Using larger, negative values of \(A_0\) would also enable parts for the bulk region. \\

\begin{figure}[!ht]
  \centering
  \begin{minipage}{16cm}
  \includegraphics[scale=1.]{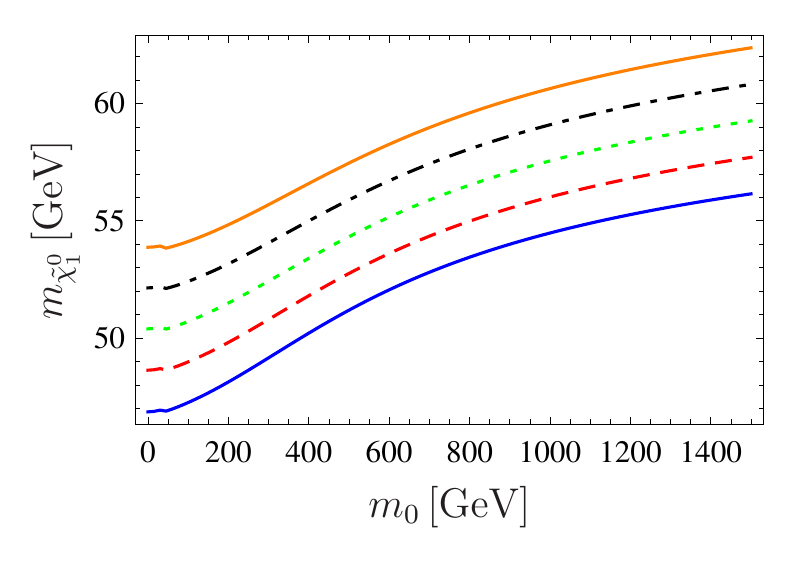} \hfill \includegraphics[scale=1.]{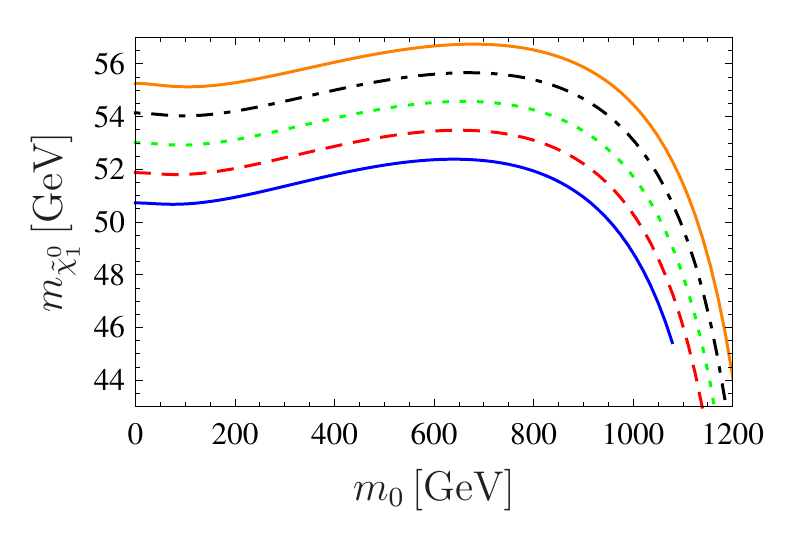} \\
  \includegraphics[scale=1.]{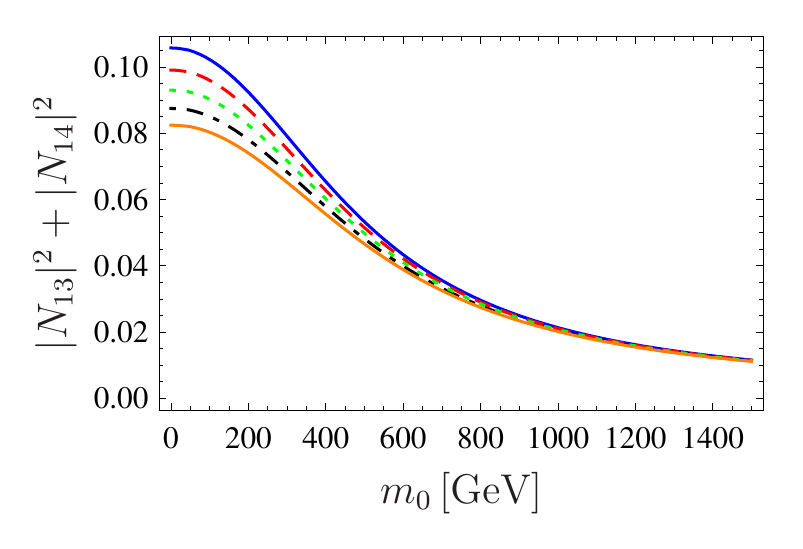} \hfill \includegraphics[scale=1.]{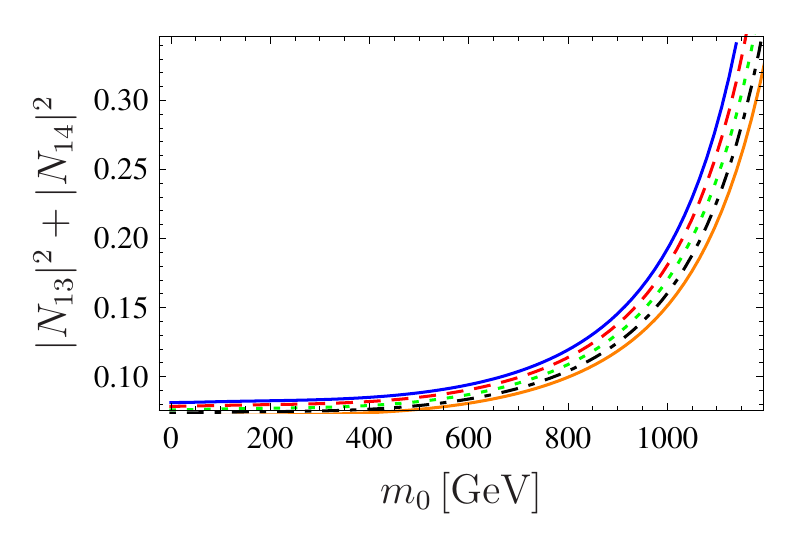} \\
  \includegraphics[scale=1.]{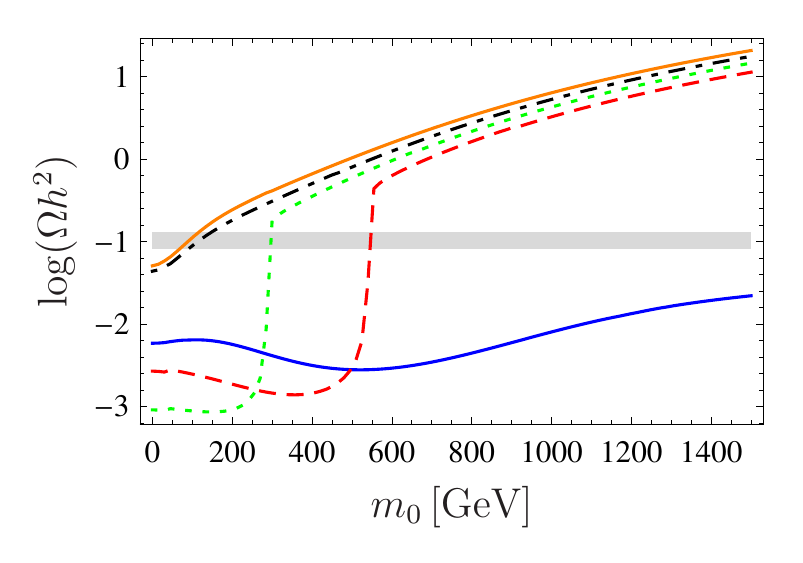} \hfill \includegraphics[scale=1.]{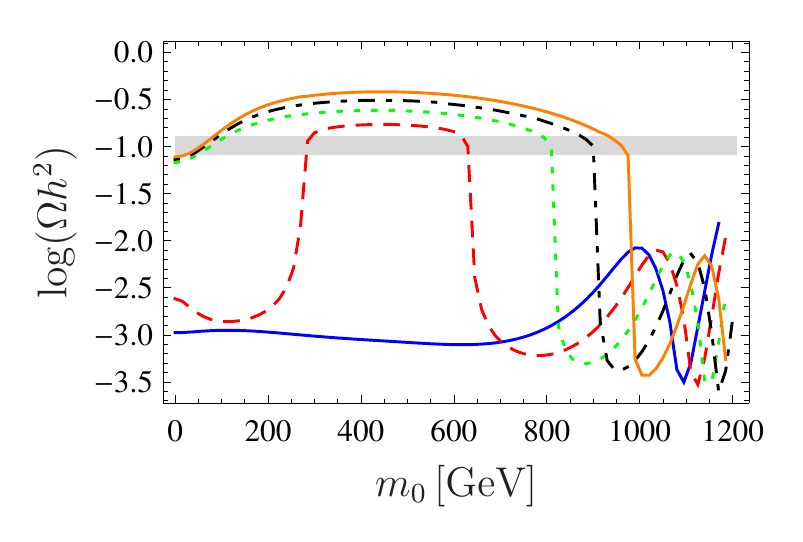}
\end{minipage}
   \caption[Neutralino mass $m_{\Neu_1}$, Higgsino fraction $|N_{13}|^2 + |N_{14}|^2$ and relic density for a variation of $m_0$ for seesaw~II and III]{Mass of the lightest neutralino (first row), its Higgsino content (second row) and the corresponding  $\Omega h^2$ (third row) as a function of $m_0$ for a seesaw type~II (left column) and type~III (right column) model with a degenerate seesaw scale $M_T = M_W=10^{14}$~GeV. The other parameters were set to  $A_0=0$~GeV, $\tan\beta=10$ and $\mu>0$. The lines correspond for seesaw~II to $M_{1/2}=195$~GeV (full blue line), $M_{1/2}=200$~GeV (red dashed line), $M_{1/2}=205$~GeV (green dashed dotted line), $M_{1/2}=210$~GeV (black dashed line) and  $M_{1/2}=215$~GeV (orange full line). For type~III, the color code is:  $M_{1/2}=400$~GeV (full blue line), $M_{1/2}=405$~GeV (red dashed line), $M_{1/2}=410$~GeV (green dashed dotted line), $M_{1/2}=415$~GeV (black dashed line ) and $M_{1/2}=420$~GeV (orange full line). The gray band shows the preferred range according to eq.~(\ref{eq:Range3SWMAP5}).}
   \label{fig:F0_II_III}
\end{figure}

Finally, we turn to the type~III seesaw, the lower, right plot of Fig.~\ref{fig:2NewA}. We have already seen that the mass spectrum changes the most. Hence, also the dark matter regions are affected at most. The coannihilation region has complete disappeared. The reason is the same as for the seesaw~II, namely, the neutralino mass tends faster to smaller values for lower seesaw scales as the stau mass does. This effect is stronger for type~III  as for type~II, therefore the coannihilation vanishes already for \(\log(\frac{M_{W_i}}{\GeV}) < 14.5\) as shown in the right plot of Fig.~\ref{fig:Coann_II_III}. The behavior of the focus point region of type~III is completely different to the other scenarios: while this distinct region of correct relic density is nearly a straight line for seesaw~I and II as well as for standard mSugra in the depicted range of \(m_0\), it  is bent for type~III. In the former cases, the Higgsino component and the mass of the LSP  vary only slightly with increasing \(m_0\) what can be compensated by a mildly smaller value of \(M_{1/2}\). In contrast, the LSP becomes Higgsino-like  at approximately \(m_0 = 1000\,\GeV\) in the seesaw~III scenario and the mass decreases rapidly. This is depicted in Fig.~\ref{fig:F0_II_III}. The origin of this behavior is the large value of the gauge couplings at the GUT scale up to 1.17 in comparison to usual mSugra values of about 0.72. That's why the gauge contributions to the running of the up-type Higgs compensate the large top Yukawa coupling and prevent large negative values of \(m_{H_u}^2\) and cause small values of \(\mu\). However, in this region of the parameter space, \(\mu\) grows faster with increasing \(M_{1/2}\) than \(M_1\) does because it is very sensitive to the GUT scale. Hence, to keep the Higgsino fraction at small values in order to circumvent too efficient annihilation, \(M_{1/2}\) has to be increased what explains the slope. The large constraints coming from Higgs bounds can also for the seesaw~III softened by larger, negative values of \(A_0\) as shown on the right column of in Fig.~\ref{fig:h_A0}. The upper plot gives the general dependence of the lightest Higgs mass on \(A_0\), while the lower plot shows the \((m_0,M_{1/2})\)-plane for $A_0=-300$~GeV. Obviously, the bulk regions starts to be consistent with Higgs bounds at this value for \(A_0\). Furthermore, also the kink for large values of \(m_0\) has disappeared in the focus point region since the Higgs masses are generally heavier and, hence, they are no longer crossed by the LSP mass. Finally, the focus point region with low values of \(m_0\) is also in that case forbidden by the small chargino mass. 

\begin{figure}[ht]
  \centering
  \begin{minipage}{16cm}
  \includegraphics[scale=0.95]{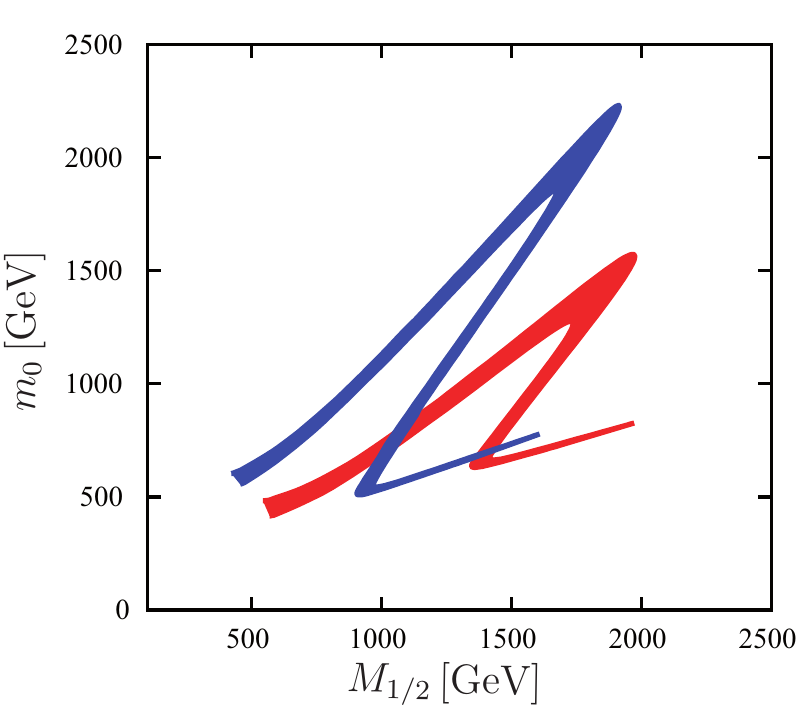} \hfill
  \includegraphics[scale=0.95]{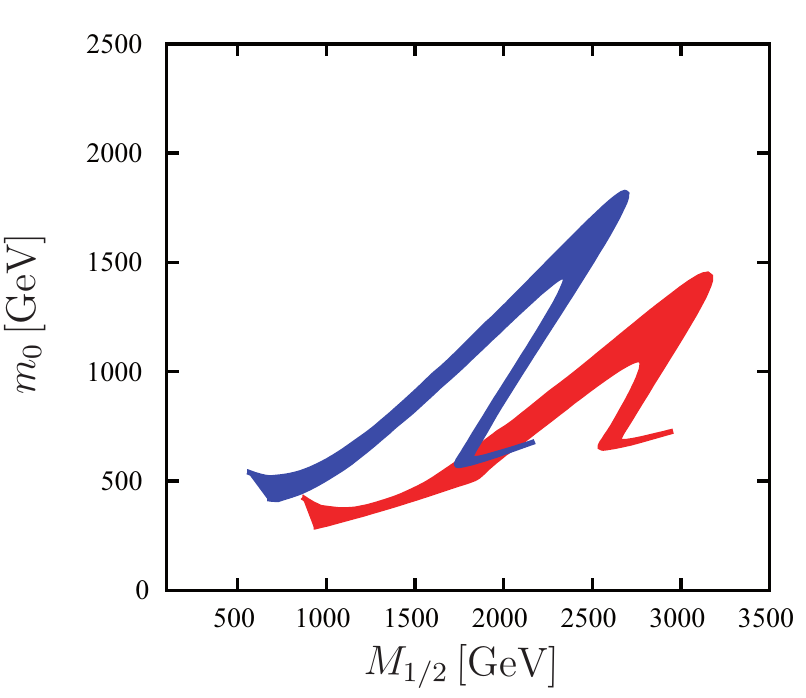}
\end{minipage}
   \caption[Comparison between using one-loop or two-loop
    RGEs on the dark matter allowed region for type-II  and type-III]{Comparison of a Higgs funnel using one-loop (blue) or two-loop
     (red) RGEs. Left plot shows type-II and the right one type-III. The parameters are: $A_0=0$~GeV, $\mu>0$ and $M_{\rm seesaw}=10^{14}$~GeV and $m_{top}=171.2$~GeV. In addition, we used $\tan\beta=52$ for type~II and $\tan\beta=49$ for type~III. }
   \label{fig:4}
\end{figure}

\begin{figure}[t]
\centering
  \begin{minipage}{16cm}
 \includegraphics[scale=0.95]{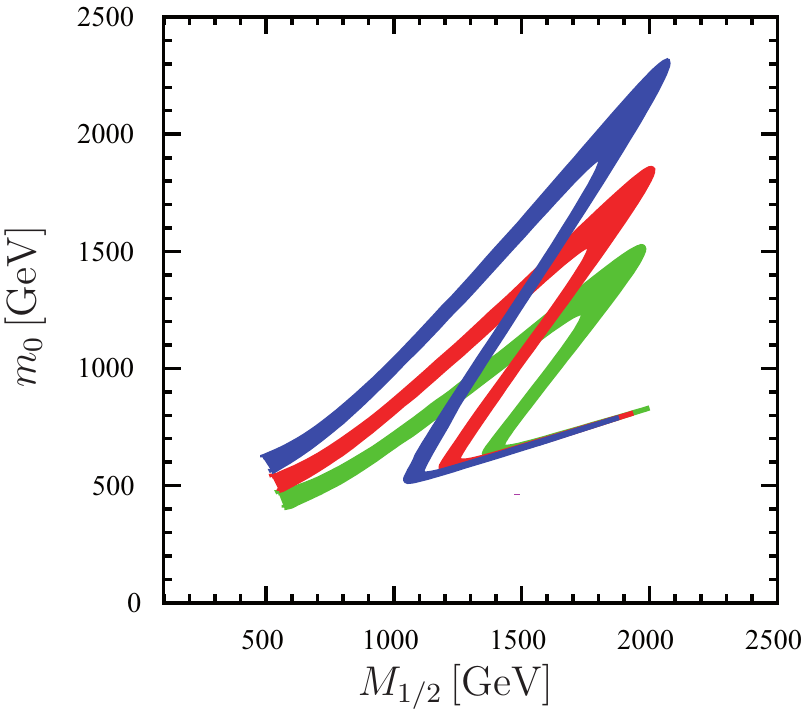} \hfill
\includegraphics[scale=0.95]{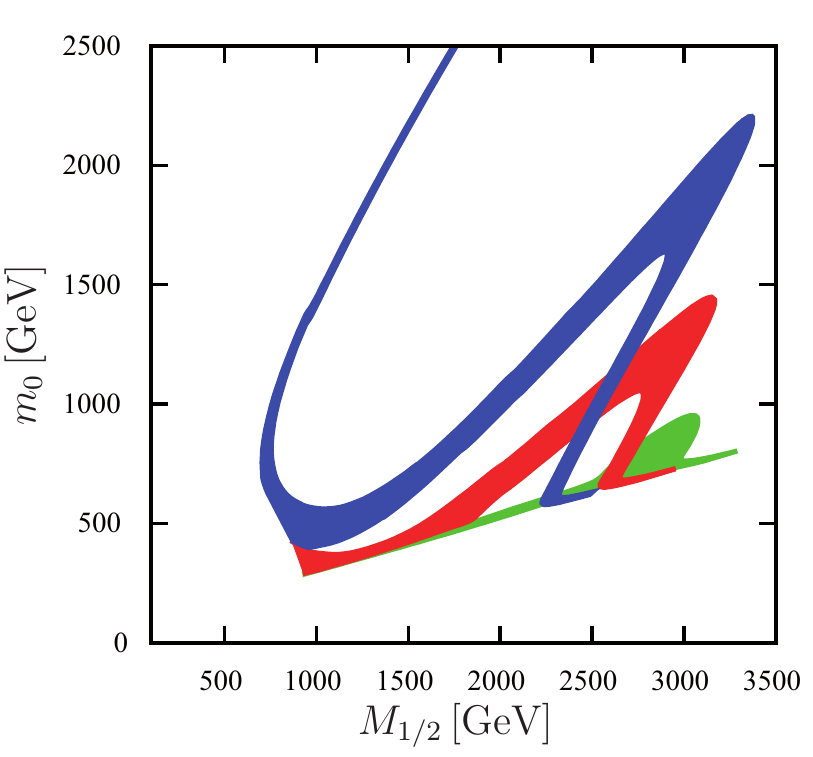}
\end{minipage}
\caption[Dark Matter iso curves for a variation of the top mass]{Dark Matter iso curves for a variation of the top mass $m_{top}=169.1$~GeV (blue), $m_{top}=171.2$~GeV (red) and $m_{top}=173.3$~GeV (green). Left plot: seesaw~II with $A_0=0,\mu>0$, $M_T=10^{14}$~GeV and $\tan\beta=52$. Right plot: seesaw~III with $A_0=0,\mu>0$, $M_{W_i}=10^{14}$~GeV and $\tan\beta=49$.}
\label{fig:HF_top}
\end{figure}

To show the importance of the two-loop RGEs, we have depicted in Fig.~\ref{fig:4} a comparison of a Higgs funnel calculated using the one- and two-loop RGEs. The Higgs funnel is characterized by a resonance effect, that means, the pseudo scalar mass is close to twice the neutralino mass. The band with correct relic density is in this case much broader than the other regions because of the large width of the Higgs boson. In the shown examples, we have fixed again \(A_0 = 0\,\GeV\) and \(\mu>0\). In addition, we used \(\tan\beta = 52\) for seesaw~II and \(\tan\beta = 49\) for seesaw~III because the resonances appear for different values of \(\tan\beta\). The seesaw scale was set to \(10^{14}\)~GeV and we used for type~III degenerated, heavy states. Obviously, there is a huge difference between both calculations: using two-loop RGEs pushes the Higgs funnel to much larger values of \(M_{1/2}\). The reason is that the two-loop contributions decrease the neutralino mass, while they increase the mass of the pseudo scalar Higgs. For example, in case of seesaw~II and for fixed values of \(m_0 = M_{1/2} = 1500\,\GeV\), the masses using one-loop RGEs are \(m_{\Neu_1} = 560.2\,\GeV, m_{A^0} = 1089.8\,\GeV\) and they are \(m_{\Neu_1} = 497.6\,\GeV, m_{A^0_1} = 1099.8\,\GeV\)  using two-loop RGEs. The two-loop contributions slightly increase the difference between the Higgs soft breaking masses squared \(m_{H_u}/m_{H_d}\) and lead therefore to a larger value of \(B_\mu\) what causes the increased pseudo scalar mass. In contrast, the bino soft breaking mass \(M_1\) is reduced by the large gauge contributions at two-loop level. Since this effect is for the type~III larger because of the three generations of {\bf 24}, the absolute difference between the curves at one- and two-loop level is also bigger. The Higgs funnel is not only sensitive to the used loop order of RGEs but also to the seesaw scale and the input value of the top mass. The last point is demonstrated at one example for seesaw~II and one example for seesaw~III in Fig.~\ref{fig:HF_top}, where the top mass was varied in the known 1\(\sigma\) range \cite{Amsler:2008zzb}
\begin{equation}
 169.1\,\GeV < m_{\top} < 173.3\,\GeV
\end{equation} 
A larger top mass leads to larger pseudo scalar masses because of the increased running of the up-type Higgs mass parameter but affects the neutralino mass much less. This has to be compensated by smaller values of \(m_0\) in order to stay near the resonance.  

\cleardoublepage
\chapter[Dark Matter in the NMSSM]{Dark Matter in the Next-to-minimal supersymmetric Standard Model}
\label{chapter:NMSSM}
The superpotential of the MSSM contains a parameter with  dimension mass, namely the so called $\mu$ parameter which gives mass to the Higgs bosons and Higgsinos. From a purely theoretical point of view, the value of this parameter is expected to be either of the order of the GUT/Planck scale or exactly zero, if it is protected by a symmetry. For phenomenological aspects, however, it is necessary that \(\mu\) is of the order of the scale of EWSB and it has to be non-zero to be consistent with experimental data. This discrepancy is the so called \(\mu\)-problem of the MSSM \cite{Kim:1983dt} already introduced in sec.~\ref{sec:SUSY_EWSB}. \\
The Next-to-Minimal Supersymmetric Standard Model (NMSSM) \cite{Fayet:1974pd,Fayet:1976et,Fayet:1977yc,Nilles:1982dy,Frere:1983ag,Derendinger:1983bz,Ellis:1988er,Drees:1988fc,Ellwanger:1993xa,Ellwanger:1995ru,Ellwanger:1996gw,Elliott:1994ht,King:1995vk} provides an elegant solution to this problem. The particle content of the MSSM is extended by an additional gauge singlet, which receives a VEV when SUSY gets broken. The corresponding term in the superpotential gives then rise to an effective $\mu$-term which is naturally of the order of the SUSY breaking scale. \\
Also in the NMSSM,  several regions  in parameter space exist with the correct relic density to explain the observed dark matter 
\cite{Belanger:2005kh,Hugonie:2007vd}. Similar to the MSSM, it turns out that in constraint NMSSM scenarios different regions are rather susceptible to mass differences of the various supersymmetric particles. In particular, the masses of the Higgs bosons, the neutralinos, the staus  and stops are important and require a precise calculation. Motivated by this observation, we calculated all SUSY masses of the NMSSM at one-loop level \cite{Staub:2010ty}. \\
\section{The Next-to-minimal supersymmetric Standard Model}
As already stated above, the solution to the \(\mu\)-problem of the MSSM is the replacement of the bilinear \(\mu\)-term by a coupling between the Higgs superfields and an additional gauge singlet $\hat{S}$ leading to the superpotential
\begin{equation}
 W_{\rm{NMSSM}} = -\hat{H}_u  \hat{q} Y_u \hat{u} 
                + \hat{H}_d\hat{q} Y_d  \hat{d} 
                + \hat{H}_d\hat{l} Y_e  \hat{e} 
                + \lambda \hat{H}_u \hat{H}_d \hat{S} 
                + \frac{1}{3} \kappa \hat{S} \hat{S} \hat{S} \thickspace,
\label{eq:superpotential}
\end{equation}
where the last term is introduced to forbid a Peccei-Quinn symmetry which would lead to an axion in contradiction to experimental results. Moreover, we have only taken into account dimensionless couplings to avoid the $\mu$-problem of the MSSM. For a recent review of the NMSSM see \cite{Ellwanger:2009dp} and references therein.  \\
The scalar component $S$ of \(\hat{S}\) receives after SUSY breaking a VEV, denoted $v_s$, which leads to 
\begin{equation}
 \mu_{\rm{eff}} = \frac{1}{\sqrt{2}} \lambda v_s \thickspace,
\end{equation}
where we have used the  decomposition
\begin{equation}
\label{VEVsinglet}
 S = \frac{1}{\sqrt{2}}  \left(\phi_s + i \sigma_s + v_s \right) \thickspace.
\end{equation}
Since \(\mu_{\rm{eff}}\) is a consequence of SUSY breaking, the natural value for this parameter is of the order of the SUSY breaking scale \cite{Ellwanger:1996gw}. The soft breaking terms are
\begin{eqnarray}
\nonumber
 V_{\rm SB,2} &=&  m_{H_u}^2 |H_u|^2 + m_{H_d}^2 |H_d|^2+ m_S^2 |S|^2 
    + \tilde{q}^\dagger m_{\tilde{q}}^2 \tilde{q} + \tilde{e}^\dagger m_{\tilde{e}}^2 \tilde{e} +\tilde{l}^\dagger m_{\tilde{l}}^2 \tilde{l}
          + \tilde{d}^\dagger m_{\tilde{d}}^2 \tilde{d}
          + \tilde{u}^\dagger m_{\tilde{u}}^2 \tilde{u} + \nonumber \\
 &&   + \frac{1}{2}\left(M_1 \, \tilde{B} \tilde{B} + M_2 \, \tilde{W}_a \tilde{W}^a  + M_3 \, \tilde{g}_\alpha \tilde{g}^\alpha + h.c.\right) \\
V_{\rm SB,3} &=& - H_u \tilde{q}  T_u\tilde{u}^\dagger
                      +H_d  \tilde{q} T_d \tilde{d}^\dagger 
                      + H_d \tilde{l} T_e \tilde{e}^\dagger 
                      + T_\lambda H_u H_d S  + \frac{1}{3} T_\kappa S S S \thickspace .
\end{eqnarray}

\section{Calculation of the one-loop mass spectrum}
\label{sec:spectrum}
We discuss in this section the $\overline{\rm DR}$ renormalization of the relevant masses, where we follow closely ref.~\cite{Pierce:1996zz}. We start with the tadpole equations at one-loop level. 

\subsection{One-loop tadpoles}
Once  electroweak symmetry gets broken, both Higgs doublets receive a VEV and we decompose the scalars similar to eq.~(\ref{VEVsinglet})
\begin{equation}
 H_{u,d} = \frac{1}{\sqrt{2}} \left( \phi_{u,d} + i \sigma_{u,d} + v_{u,d}\right).
\end{equation}
At tree level, the minimum conditions for the vacuum are calculated according to eq.~(\ref{EqTadPole}). The results read
{\allowdisplaybreaks
\begin{eqnarray} 
T_d = \frac{\partial V}{\partial v_d} &=& m_{H_d}^2 v_d + \frac{1}{8} v_d \big( v_d^2-v_u^2 \big) \big( g_1^{2} + g_2^2 \big) + \frac{1}{2} v_d \big( v_u^2 + v_s^{2} \big) |\lambda|^{2}
\label{eq:tadpoleD} \nonumber \\ 
 && -\frac{1}{2} v_s^{2} v_u \mathrm{Re}\big\{\kappa \lambda\big\} - \frac{1}{\sqrt{2}} v_s v_u \mathrm{Re}\big\{ T_{\lambda}\big\}, \\ 
T_u = \frac{\partial V}{\partial v_u} &=& m_{H_u}^2 v_d + \frac{1}{8} v_u \big( v_u^2-v_d^2 \big) \big( g_1^{2} + g_2^2 \big) + \frac{1}{2} v_u \big( v_d^2 + v_s^{2} \big) |\lambda|^{2}
\label{eq:tadpoleU} \nonumber  \\ 
 && -\frac{1}{2} v_s^{2} v_d \mathrm{Re}\big\{\kappa \lambda\big\} - \frac{1}{\sqrt{2}} v_s v_d \mathrm{Re}\big\{ T_{\lambda}\big\}, \\ 
T_s = \frac{\partial V}{\partial v_s} &=& m_S^2 v_s +v_s^{3} |\kappa|^{2} - v_d v_s v_u \mathrm{Re}\big\{ \kappa \lambda\big\} +\frac{1}{2} \big( v_d^{2} + v_u^2 \big) v_s |\lambda|^{2} \nonumber  \\ 
 &&+ \frac{1}{\sqrt{2}} v_s^{2} \mathrm{Re}\big\{ T_{\kappa}\big\}   - \frac{1}{\sqrt{2}} v_d v_u \mathrm{Re}\big\{ T_{\lambda}\big\} 
\label{eq:tadpoleS}.
\end{eqnarray}
Here, we have chosen a phase convention where all VEVs are real. For the later calculation of the one-loop corrections to the Higgs boson masses, one needs the evaluation of the tadpole equations at the one-loop level, leading to  corrections \(\delta t^{(1)}_i\). As renormalization condition we demand}
\begin{equation}
 T_i + \delta t^{(1)}_i = 0 \qquad{\rm for}\quad i=d,u,s \thickspace .
\label{eq:oneloop}
\end{equation} 
All calculations are performed in 't Hooft gauge using the diagrammatic approach. The explicit formulas for \(\delta t^{(1)}_i\) are given in app.~\ref{onelooptad}. In our subsequent analysis we will solve eqs.~(\ref{eq:oneloop}) for the soft SUSY breaking masses squared: $m^2_{H_d}$, $m^2_{H_u}$ and $m^2_{S}$ \thickspace.  \\
All parameters in eqs.~(\ref{eq:tadpoleD})-(\ref{eq:tadpoleS}) are understood as running parameters at a given renormalization scale $Q$. Note that the VEVs $v_d$ and $v_u$ are obtained from the running mass $M_Z(Q)$ of the $Z$ boson,  which is related to the pole mass $M_Z$ through
\begin{equation}
M^2_Z(Q) = \frac{g^2_1+g^2_2}{4} (v^2_u+v^2_d) = M^2_Z + \mathrm{Re}\big\{ \Pi^T_{ZZ}(M^2_Z) \big\} \thickspace.  
\end{equation}
The transverse self-energy $\Pi^T_{ZZ}$ is given in app.~\ref{app:Zself}. Details on the calculation of the running gauge couplings at $Q=M_Z$ can be found in ref.~\cite{Pierce:1996zz}. The ratio of these VEVs is denoted as in the MSSM by $\tan\beta=\frac{v_u}{v_d}$.
%
\subsection{Masses of the Higgs bosons}
The tree level mass matrices for the neutral scalar  Higgs bosons and pseudo scalar Higgs bosons can be calculated from the scalar potential according to
\begin{equation}
 	m^{2,h}_{T,i j} = \frac{\partial^2 V}{\partial \phi_i \partial \phi_j}
 	   \Bigg|_{\phi_k=0,\sigma_k=0}, \hspace{2cm}
 	m^{2,A^0}_{T,i j} = \frac{\partial^2 V}{\partial\sigma_i \partial\sigma_j}
 	\Bigg|_{\phi_k=0,\sigma_k=0} \thickspace,
\label{ScalarMass}
\end{equation}
with $i,j=1,2,3=u,d,s$. The matrices are symmetric and the entries in case of the scalar Higgs bosons are 
\begin{eqnarray} 
	m^{2,h}_{T,11} &=& m_{H_d}^2 + \frac{1}{2} \big( v_u^2 + v_s^{2} \big) |\lambda|^{2} + \frac{1}{8} \big( g_1^{2} + g_2^2 \big) \big(3 v_d^{2}  - v_u^{2} \big),\\ 
	m^{2,h}_{T,12} &=& -\frac{1}{\sqrt{2}} v_s \mathrm{Re}\big\{T_{\lambda}\big\} - \frac{1}{2} v_s^{2} \mathrm{Re}\big\{\kappa \lambda\big\} - \frac{1}{4} \big( g_1^{2}+g_2^2 -4 |\lambda|^2 \big) v_d v_u,\\ 
	m^{2,h}_{T,13} &=& - \frac{1}{\sqrt{2}} v_u \mathrm{Re}\big\{T_{\lambda}\big\}  + v_s v_d |\lambda|^2  - v_s v_u  \mathrm{Re}\big\{ \kappa \big\},\\ 
	m^{2,h}_{T,22} &=& m_{H_u}^2 + \frac{1}{2} \big( v_d^{2} + v_s^{2} \big) |\lambda|^{2} + \frac{1}{8} \big( g_1^{2} + g_2^2 \big) \big( 3 v_u^{2}-v_d^{2} \big),\\ 
	m^{2,h}_{T,23} &=& - \frac{1}{\sqrt{2}} v_d \mathrm{Re}\big\{T_{\lambda}\big\}  + v_s v_u |\lambda|^2 - v_s v_d \mathrm{Re}\big\{\lambda \kappa \big\},\\ 
	m^{2,h}_{T,33} &=& m_S^2 + 3 v_s^{2} |\kappa|^{2}  + \frac{1}{2} \big( v_d^{2} + v_u^{2} \big) |\lambda|^{2}  + \sqrt{2} v_s \mathrm{Re}\big\{T_{\kappa}\big\}  - v_d v_u \mathrm{Re}\big\{\kappa \lambda \big\} \thickspace,
\end{eqnarray} 
while those of the pseudo scalar ones are
\begin{eqnarray} 
m^{2,A^0}_{T,11} &=& m_{H_d}^2 + \frac{1}{2} \big( v_u^2 + v_s^{2} \big) |\lambda|^{2} + \frac{1}{8} \big( g_1^{2} + g_2^2 \big) \big( v_d^{2} - v_u^{2} \big),\\ 
m^{2,A^0}_{T,12} &=& \frac{1}{\sqrt{2}} v_s \mathrm{Re}\big\{T_{\lambda}\big\} + \frac{1}{2} v_s^2 \mathrm{Re}\big\{\kappa \lambda \big\},\\ 
m^{2,A^0}_{T,13} &=& \frac{1}{\sqrt{2}} v_u \mathrm{Re}\big\{T_{\lambda}\big\}  - v_s v_u \mathrm{Re}\big\{\kappa \lambda \big\}, \\ 
m^{2,A^0}_{T,22} &=& m_{H_u}^2 + \frac{1}{2} \big( v_d^2 + v_s^{2} \big) |\lambda|^{2} + \frac{1}{8} \big( g_1^{2} + g_2^2 \big) \big( v_u^{2} - v_d^{2} \big),\\ 
m^{2,A^0}_{T,23} &=& \frac{1}{\sqrt{2}} v_d \mathrm{Re}\big\{ T_{\lambda}\big\}  - v_d v_s \mathrm{Re}\big\{\kappa \lambda \big\}, \\ 
m^{2,A^0}_{T,33} &=& m_S^2 + v_s^{2} {\kappa}^{2} + \frac{1}{2} \big( v_d^{2} + v_u^{2} \big) |\lambda|^{2}  - \sqrt{2} v_s \mathrm{Re}\big\{T_{\kappa}\big\} + v_d v_u\mathrm{Re}\big\{\kappa \lambda \big\} \thickspace, 
\end{eqnarray} 
where $m_{H_d}^2$, $m_{H_u}^2$ and $m_{S}^2$ satisfy the tadpole equations. \\
The diagonalization of the mass matrices \(m^{2,h}_T\) and \(m^{2,A^0}_T\) leads in total to five physical mass eigenstates and one neutral  Goldstone boson which becomes the longitudinal component of the $Z$ boson. The five physical degrees of freedom are: three CP-even Higgs bosons denoted \(h_{1,2,3}\) and two CP-odd bosons denoted \(A^0_{1,2}\). The corresponding rotation  matrices $Z^H$ und $Z^A$ are defined through
\begin{equation}
Z^{X} m^{2,x} Z^{X,T} = m^{2,x}_{\mathrm{diag}} ,\qquad x=h,A^0,\qquad X=H,A \thickspace.
\end{equation}
Moreover, we note that we order all masses by $m_i \le m_j$ if $i<j$. \\
The one-loop scalar Higgs masses are then calculated by taking the real part of the poles of the corresponding propagator matrices
\begin{equation}
\mathrm{Det}\left[ p^2_i \mathbf{1} - m^{2,h}_{1L}(p^2) \right] = 0 \thickspace,
\label{eq:propagator}
\end{equation}
where
\begin{equation}
 m^{2,h}_{1L}(p^2) = \tilde{m}^{2,h}_T -  \Pi_{hh}(p^2) \thickspace.
\end{equation}
Here, \(\tilde{m}_h\) is the tree level mass matrix from eq.~(\ref{ScalarMass}) where to solution of eq.~(\ref{eq:oneloop}) was used to express the soft breaking squared mass parameters. Eq.~(\ref{eq:propagator}) has to be solved for each eigenvalue $p^2=m^2_i$. The same procedure is applied for the pseudo scalar Higgs bosons, too. \\
The complete one-loop expressions for the self energy of the CP-odd and even Higgs bosons are given in apps.~\ref{app:H0self} and \ref{app:A0self}. 
The charged Higgs sector consists of \(H_d^-\) and \(H_u^+\). The mass matrix in the basis \(\left(H_d^-,H_u^{+,*}\right)\) is diagonalized by an unitary matrix \(Z^+\)
\begin{equation}
Z^+ m^{2,H^+} Z^{+,\dagger} = m^{2,H^+}_{\mathrm{diag}} \thickspace.
\end{equation}
The eigenstates yield as in the MSSM the longitudinal component of the $W$ boson and a charged Higgs boson $H^+$ with mass
\begin{equation}
m^{2,H^+}_T = \frac{(v_d^2 + v_u^2)(2 v_s \mathrm{Re}\big\{\kappa \lambda\big\} + v_d v_u (g_2^2 - 2 |\lambda|^2) + 2 \sqrt{2} v_s \mathrm{Re}\big\{T_\lambda\big\})}{4 v_d v_u} \thickspace .
\end{equation}
The one-loop mass is
\begin{equation}
m^{2,H^+}_{1L} =  m^{2,H^+}_T
 - \mathrm{Re}\big\{ \Pi_{H^+H^+}(m^{2,H^+}_{1L})\big\} \thickspace ,
\end{equation}
where the self-energy \(\Pi_{H^+H^+}\) can be found in app.~\ref{app:Hpself}
%
\subsection{Chargino and neutralino masses}
As for the Higgs bosons discussed in the previous section, one has to find the real parts of the poles of the  propagator matrix to obtain the masses of charginos and neutralinos. At the tree level, the chargino mass matrix in the basis $\tilde \psi^-= ({\tilde{W}^-}, {\tilde{H}_d^-})^T$, 
$\tilde \psi^+= (\tilde{W}^+, {\tilde{H}_u^+})$ is given by
\begin{equation}
\mathscr{L}_{\Cha} = - {\tilde \psi^-}{}^T M^{\tilde\chi^+}_T \tilde \psi^+
+ \mathrm{h.c.}
\end{equation} 
with
\begin{equation} 
M^{\tilde\chi^+}_T = \left( 
\begin{array}{cc}
M_2 &\frac{1}{\sqrt{2}} g_2 v_u \\ 
\frac{1}{\sqrt{2}} g_2 v_d  &\frac{1}{\sqrt{2}} v_s \lambda \end{array}  
\right) \thickspace.
\end{equation} 
This mass matrix is diagonalized by a biunitary transformation such that $U^* M^{\tilde\chi^+}_T V^\dagger$ is diagonal. The matrices $U$ and $V$
are obtained by diagonalizing $ M^{\tilde\chi^+}_T ( M^{\tilde\chi^+}_T)^\dagger$ and $(M^{\tilde\chi^+}_T)^* ( M^{\tilde\chi^+}_T)^T$, respectively. At the one-loop level, the mass matrix is given by  
\begin{eqnarray}
\label{eq:cha1loop}
M^{\tilde\chi^+}_{1L}(p^2_i) =  M^{\tilde\chi^+}_T - \Sigma^+_S(p^2_i)
 - \Sigma^+_R(p^2_i) M^{\tilde \chi^+}_T - M^{\tilde \chi^+}_T \Sigma^+_L(p^2_i) \thickspace.
\end{eqnarray}
In case of the neutralinos, one has a complex symmetric $5\times 5$ mass matrix which in the basis  $\tilde\psi^0 = ({\tilde{B}}, {\tilde{W}^0}, {\tilde{H}_d^0}, {\tilde{H}_u^0}, \tilde{S})^T$ is at the tree level given by
\begin{equation} 
M^{\tilde\chi^0}_T =
\left( 
\begin{array}{ccccc}
M_1 &0 &-\frac{1}{2} g_1 v_d  &\frac{1}{2} g_1 v_u  &0\\ 
0 &M_2 &\frac{1}{2} g_2 v_d  &-\frac{1}{2} g_2 v_u  &0\\ 
-\frac{1}{2} g_1 v_d  &\frac{1}{2} g_2 v_d  &0 &- \frac{1}{\sqrt{2}} v_s \lambda  &- \frac{1}{\sqrt{2}} v_u \lambda \\ 
\frac{1}{2} g_1 v_u  &-\frac{1}{2} g_2 v_u  &- \frac{1}{\sqrt{2}} v_s \lambda  &0 &- \frac{1}{\sqrt{2}} v_d \lambda \\ 
0 &0 &- \frac{1}{\sqrt{2}} v_u \lambda  &- \frac{1}{\sqrt{2}} v_d \lambda  &\sqrt{2} v_s \kappa \end{array} 
\right) \thickspace.
\end{equation} 
One can show that for real parameters the matrix $M^{\tilde\chi^0}_T$ can be diagonalized directly by a $5\times 5$ mixing matrix $N$ such that
$N^* M^{\tilde\chi^0}_T N^\dagger$ is diagonal. In the complex case, one has to diagonalize $M^{\tilde\chi^0}_T (M^{\tilde\chi^0}_T)^\dagger$. At the one-loop level, we obtain
\begin{eqnarray}
M^{\tilde\chi^0}_{1L} (p^2_i) &=& M^{\tilde\chi^0}_T - 
\frac{1}{2} \bigg[ \Sigma^0_S(p^2_i) + \Sigma^{0,T}_S(p^2_i)
 + \left(\Sigma^{0,T}_L(p^2_i)+   \Sigma^0_R(p^2_i)\right) M^{\tilde\chi^0}_T
 \nonumber \\
&& \hspace{16mm}
+ M^{\tilde\chi^0}_T \left(\Sigma^{0,T}_R(p^2_i) +  \Sigma^0_L(p^2_i) \right) \bigg] \thickspace .
\label{eq:OneloopN}
\end{eqnarray}
The complete self-energies for neutralinos and charginos are given in apps.~\ref{sec:OneLoopNeu} and \ref{sec:OneLoopCha}.
%
\subsection{Masses of sleptons}
In the basis \( \left(\tilde{e}_{L}, \tilde{\mu}_{L}, \tilde{\tau}_{L},
 \tilde{e}_{R}, \tilde{\mu}_{R}, \tilde{\tau}_{R}\right) \), the mass matrix of the charged sleptons at the tree level is given by 
\begin{equation} 
m^{2,\tilde l}_{T} = 
\left( 
\begin{array}{cc}
m^2_{LL} &-\frac{1}{2} v_s v_u \lambda^* Y_e^T  + \frac{1}{\sqrt{2}} v_d T_e^T \\ 
-\frac{1}{2} v_s v_u \lambda Y_e^* + \frac{1}{\sqrt{2}} v_d T_e^* 
 &m^2_{RR}\end{array} 
\right) \thickspace ,
\end{equation} 
with the diagonal entries
\begin{eqnarray} 
m^2_{LL} &=& m^2_{\tilde l} + \frac{v^2_d}{2} (Y_{e})^* (Y_{e})^T
 + \frac{1}{8}
 \Big( g_1^2- g_2^2 \Big)\Big(v_d^2 - v_u^2 \Big) {\bf 1}_3, \\ 
m^2_{RR} &=& m^2_{\tilde e} + \frac{v^2_d}{2} (Y_{e})^T (Y_{e})^*
 + \frac{g_1^2}{4}  \Big(v_u^2- v_d^2  \Big) {\bf 1}_3 \thickspace.
\end{eqnarray}
${\bf 1}_3$ is the $3\times 3$ unit matrix. This matrix is diagonalized by an unitary mixing matrix $Z^E$:
\begin{equation}
Z^E m^{2,\tilde l}_{T} Z^{E \dagger} = m^{2,\tilde l}_{\mathrm{diag}} \thickspace.
\end{equation}
The corresponding mass matrix at the one-loop level is again obtained by taking into account the self-energy according to
\begin{equation}
m^{2,\tilde l}_{1L}(p^2_i) = m^{2,\tilde l}_{T} - \Pi_{\tilde l\tilde l}(p^2_i) \thickspace,
\end{equation}
and the one-loop masses are obtained by calculating the real parts of the poles of the propagator matrix. The expression for $\Pi_{\tilde l\tilde l}(p^2_i) $ can be  found in app.~\ref{app:SleptonsSelf}. \\
Finally, the tree level sneutrino mass matrix is given in the basis $\left(\tilde{\nu}_{e}, \tilde{\nu}_{\mu}, \tilde{\nu}_{\tau} \right)$  by
\begin{equation} 
m^{2,\tilde \nu}_T = m^2_{\tilde l}
 + \frac{1}{8} \Big( g_1^2+ g_2^2 \Big)\Big(v_d^2 - v_u^2 \Big) {\bf 1}_3 \thickspace .
\end{equation} 
This matrix is diagonalized by an unitary mixing matrix $Z^E$:
\begin{equation}
Z^\nu m^{2,\tilde \nu}_{T} Z^{\nu \dagger} = m^{2,\tilde \nu}_{\mathrm{diag}} \thickspace.
\end{equation}
Similarly as above, the one-loop mass matrix is given by
\begin{equation}
	m^{2,\tilde \nu}_{1L}(p^2_i) = m^{2,\tilde \nu}_{T} - \Pi_{\tilde \nu\tilde \nu}(p^2_i) \thickspace.
\end{equation}
The one-loop masses are obtained by calculating the real parts of the poles of the propagator matrix. The expression for $\Pi_{\tilde \nu\tilde \nu}(p^2_i) $ can be found in app.~\ref{app:SneutrinoSelf}.
\subsection{Masses of squarks}
In the basis $\big(\tilde{d}_{L},\tilde{s}_{L},\tilde{b}_{L},
                   \tilde{d}_{R},\tilde{s}_{R},\tilde{b}_{R}\big)$ \thickspace, 
the mass matrix for the down-type squarks is given by
\begin{equation} 
m^{2,\tilde d}_{T} = \left( 
\begin{array}{cc}
m^{2,d}_{LL} &
\frac{1}{2}  \big(\sqrt{2} v_d T^{T}_d  - v_s v_u \lambda^* Y^{T}_d \big)\\ 
\frac{1}{2} \big(\sqrt{2} v_d T^*_{d} - v_s v_u \lambda Y^{*}_d \big) 
& m^{2,d}_{RR} \end{array} 
\right) \thickspace,
\end{equation} 
where the diagonal entries read 
\begin{eqnarray} 
\nonumber
m^{2,d}_{LL} &=&  m^2_{\tilde q}  + \frac{v_d^2}{2}  Y^*_d Y^T_d
 - \frac{3 g_2^{2} + g_1^{2}}{24} \big(v_d^{2} - v_u^{2}\big)  {\bf 1}_3 \thickspace,\\ 
m^{2,d}_{RR} &=& m^2_{\tilde d} +  \frac{v_d^2}{2}  Y^T_d Y^*_d 
 + \frac{g_1^{2}}{12} \Big( v_u^{2} - v_d^{2}\Big)  {\bf 1}_3 \thickspace.
\end{eqnarray} 
The corresponding expressions for the up-type squarks in the basis 
$\big(\tilde{u}_{L}, \tilde{c}_{L}, \tilde{t}_{L},
 \tilde{u}_{R}, \tilde{c}_{R}, \tilde{t}_{R}\big)$ are
\begin{equation} 
m^{2,\tilde u}_{T} = \left( 
\begin{array}{cc}
m^{2,u}_{LL}  &\frac{1}{2} \big(\sqrt{2} v_u T^{T}_u  - v_d v_s \lambda^* Y^{T}_u \big)\\ 
\frac{1}{2} \big(\sqrt{2} v_u T^{*}_u - v_d v_s \lambda Y^{*}_u \big)
& m^{2,u}_{RR}  \end{array} 
\right) 
\end{equation} 
with
\begin{eqnarray} 
\nonumber
m^{2,u}_{LL}  &=&  m^2_{\tilde q}  + \frac{v_u^2}{2}  Y^*_u Y^T_u
 + \frac{3 g_2^{2} - g_1^{2}}{24} \big(v_d^{2} - v_u^{2}\big){\bf 1}_3 \thickspace, \\ 
 m^{2,u}_{RR} &=&  m^2_{\tilde u} +  \frac{v_u^2}{2}  Y^T_u Y^*_u 
 + \frac{g_1^{2}}{6} \Big(v_d^{2}- v_u^{2} \Big)  {\bf 1}_3 \thickspace .
\end{eqnarray} 
These matrix are diagonalized by
 unitary mixing matrices $Z^Q$:
\begin{equation}
Z^Q m^{2,\tilde q}_{T} Z^{Q \dagger} = m^{2,\tilde q}_{\mathrm{diag}} \,\,,\,\, q=d,u \, Q=D,U \thickspace .
\end{equation}
\subsection{Masses of quarks, leptons and the gluino}
The masses of the SM fermions are exactly the same as in the MSSM and the tree level relations are given in app.~\ref{chapter:MSSM}. Furthermore, the formulas for the one-loop self-energies of the SM fermions are very similar to the MSSM results. The MSSM results can be found e.g. in \cite{Pierce:1996zz} and we list our NMSSM results in apps.~\ref{app:SEup}~-~\ref{app:SElep}: only the upper limit of the sums over internal Higgs fields and neutralinos has to be adjusted in comparison to the MSSM. Since all SM fermions are Dirac spinors, the connection between the tree level and one-loop mass matrix is the same as for the chargino given by eq.~(\ref{eq:cha1loop}). \\
In the case of the gluino there is no difference in comparison to the MSSM. The gluino mass is given by the absolute value of the gaugino parameter \(|M_3|\) and the one-loop self energies has exactly the same form. For completeness, we added the one-loop self energy of the gluino to app.~\ref{app:SEglu}. Since the gluino is a Majorana fermion like the neutralinos, the one-loop mass can by calculated analog to eq.~(\ref{eq:OneloopN}).
%
%
%
\section{The constrained NMSSM}
\label{sec:msugra}
\subsection{The model and its free parameters}
In the subsequent numerical analysis, we are mainly interested in precision calculation of the SUSY masses and potential effects in the calculation of  the relic density. To reduce the number of free parameters, we   focus on a scenario motivated by minimal supergravity (mSugra) \cite{Chamseddine:1982jx}. More precisely, we study a variant of the constrained NMSSM \cite{Djouadi:2008uw,Djouadi:2008uj}  where we allow for non-universal Higgs mass parameters squared at the GUT scale. In our setup, these parameters are determined with the help of the tadpole equations eq.~(\ref{eq:oneloop}) at the electroweak scale. \\
We define the GUT scale as the scale where the $U_Y(1)$ and $SU(2)_L$ couplings fulfill $\sqrt{\frac{5}{3}} g_1=g_2$. First, we apply the same boundary conditions for the gaugino masses \(M_1,M_2, M_3\) and the soft breaking masses of the squarks and sleptons \(m_i^2\) at the GUT scale as in the standard MSSM
\begin{eqnarray}
 	M_1 = M_2 = M_3 &\equiv& M_{1/2}, \\
 	m_{\tilde{d}}^2 = m_{\tilde{u}}^2 = m_{\tilde{q}}^2 = m_{\tilde{e}}^2 = m_{\tilde{l}}^2 &\equiv& m_0^2 \, {\bf 1}_3 \thickspace.
\end{eqnarray}
Second, the trilinear scalar couplings \(T_i\) are given by
\begin{equation}
 	T_u = A_0 Y_u, \quad T_d = A_0 Y_d, \quad T_e = A_0 Y_e, \quad
 	T_\lambda = A_\lambda \lambda, \quad {\rm and} \quad T_\kappa = A_\kappa \kappa \thickspace.
\end{equation}
Here, \(A_0\) is defined at the GUT scale, while \(\lambda\), \(\kappa\), \(T_\lambda\) and \(T_\kappa\) can be defined either at the GUT or at the SUSY scale in our numerical analysis.  Together with the values for \(\tan\beta = \frac{v_u}{v_d}\) and \(v_s\), the spectrum is fixed. To summarize, we have nine input parameters,
\begin{equation}
 	M_{1/2}, \quad m_0, \quad A_0, \quad \lambda, \quad \kappa, \quad A_\lambda, \quad A_\kappa, \quad v_s, \quad \rm {and}\quad \tan\beta.
	\label{eq:params}
\end{equation}
We allow for non-universalities in the trilinear parameters for an easier comparison with the existing literature but in principal we could take all $A$-parameters equal at the GUT scale. We choose in the following \(v_s > 0\) and \(\lambda,\kappa \in [-1,1]\). 
\subsection{Procedure to evaluate the SUSY parameters at the electroweak scale}
\label{sec:procedure}
In order to connect the parameters at various scales, we use the RGEs, which are calculated with \SARAH in the most general form. We have compared the obtained expressions for the RGEs with those given in ref.~\cite{Ellwanger:2009dp} in the limit where only the third generation Yukawa couplings contributes. There has been a slight difference in the two-loop \(\beta\)-function of \(A_\lambda = T_\lambda/\lambda\), but it was confirmed by the authors of ref.~\cite{Ellwanger:2009dp} that our result is correct. We list the RGEs different to the MSSM in app.~\ref{gamma}. The complete set of RGEs can easily be calculated by the \verb"CalcRGEs" command of {\tt SARAH} as explained in app.~\ref{examplesRGEs}.\\
In the calculation of the gauge and Yukawa couplings, we follow closely the procedure described in ref.\ \cite{Porod:2003um}: the values for the Yukawa couplings giving mass to the SM fermions  and the gauge couplings are determined at the scale \(M_Z\) based on the measured values for the quark, lepton and vector boson masses as well as for the gauge couplings.  Here, we have included the one-loop corrections to the masses of $W$ and $Z$ boson as well as the SUSY contributions to \(\delta_{VB}\) for calculating the gauge couplings. Similarly, we have included the complete one-loop corrections to the self-energies of SM fermions extending the formulas of \cite{Pierce:1996zz} to include the additional neutralino and Higgs bosons. Moreover, we have resummed the $\tan\beta$ enhanced terms for the calculation of the Yukawa couplings of the $b$-quark and the $\tau$-lepton as in \cite{Porod:2003um}. \\
The vacuum expectation values \(v_d\) and \(v_u\) are calculated with respect to the given value of \(\tan\beta\) at \(M_Z\). Furthermore, we solve the tadpole equations to get initial values for \(m_{H_d}^2\), \(m_{H_u}^2\) and \(m_S^2\). Afterwards the RGEs are used to obtain the values at the GUT scale and all boundary conditions including $\lambda$ and $\kappa$ are set as described above. Then, an RGE running to the SUSY scale is performed and the SUSY masses are calculated  at the one-loop level. For the neutral and pseudo scalar Higgs bosons we  include beside the one-loop contributions presented here in additionally the known two-loop contributions \cite{Degrassi:2009yq}. For that purpose, also the numerical the values for the VEVs at \(M_{SUSY}\) are needed. These are derived using the two-loop RGEs 
\begin{equation}
	\beta_{v_i} = - v_i \left(\gamma^{(1)}_i + \gamma^{(2)}_i \right) \thickspace ,
\end{equation}
with \(i=u,d\). Here, $\gamma^{(1)}_i$ and $\gamma^{(2)}_i$ are the anomalous dimensions for the two Higgs-doublets at the one- and two-loop level, respectively. The corresponding expressions are given in app.~\ref{gamma}. Let us recall that the input value for $v_s$ is  already given at \(M_{\rm SUSY}\). These steps are iterated until the  masses converge with a relative precision of \(10^{-5}\). The complete procedure has been implemented in {\tt SPheno} as discussed in sec.~\ref{sec:SARAH_SPheno} \cite{Porod:2003um}.
\subsection{An example spectrum}
\begin{figure}[!ht]
\begin{minipage}{16cm}
\includegraphics[scale=0.92]{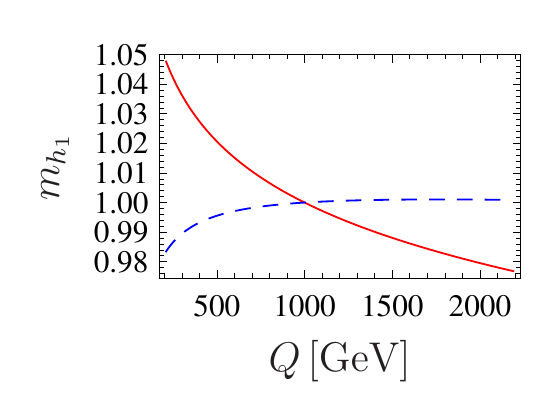}
\hfill
\includegraphics[scale=0.92]{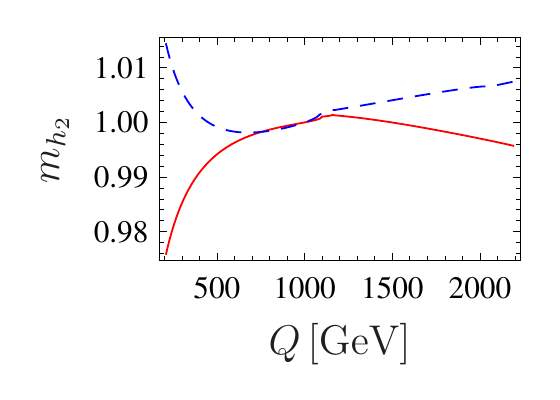} 
\hfill
\includegraphics[scale=0.92]{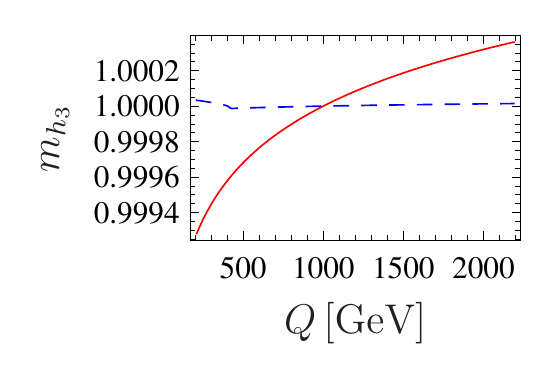}
\end{minipage}
\caption[Dependence of CP even Higgs masses on the renormalization
  scale $Q$]{Dependence of CP even Higgs masses on the renormalization
  scale $Q$ at one-loop (red) and two-loop level (dashed blue) normalized to the value at $Q=1$~TeV.
From left to right: $m_{h_1}$, $m_{h_2}$ and $m_{h_3}$.}
\label{fig:Scalar_Q}
\end{figure}

\begin{table}[!t]
\begin{center}
\begin{tabular}{|c|c|c|c|c|c|}
\hline
Particle & \(m_T\) [GeV] & \(m_{1L}\) [GeV] & \(\Delta\) [\%] & \(m_{2L}\) [GeV] & \(\Delta\) [\%]\\
\hline
\(h_1\) & 86.7 & 113.3 & 23.5 & 119.6 & 5.2 \\
\(h_2\) & 863.1 & 934.2 & 7.6 & 937.3 &  0.3 \\
\(h_3\) & 2073.9 & 2073.9 & \(<\) 0.1  & 2073.9 & \(<\) 0.1 \\
\(A^0_1\) & 76.4 & 69.3 & 10.2 & 69.5 &  0.3 \\
\(A^0_2\) & 865.2 & 937.2 & 7.7 & 940.4 & 0.3  \\
\hline
\(\tilde\chi^0_1\) & 211.6 & 207.6 & 1.9 & - & - \\
\(\tilde\chi^0_2\) & 388.2 & 398.4 & 2.6& - & -\\
\(\tilde\chi^0_3\) & 987.9 & 980.5 & 0.7& - & -\\
\(\tilde\chi^0_4\) & 993.0 & 985.1 & 0.8& - & -\\
\(\tilde\chi^0_5\) & 2074.8 & 2074.9 & \(<\) 0.1& - & -\\
\(\tilde\chi_1^+\) & 388.2 & 398.6 & 2.6& - & -\\
\(\tilde\chi_2^+\) & 993.3 & 985.9 & 0.7& - & -\\
\hline 
\(\tilde{\tau}_1\) & 191.1 & 193.3 & 1.2 & - & -\\
\(\tilde{\tau}_2\) & 388.1 & 393.1 & 1.1& - & -\\
\(\tilde{t}_1\) & 506.9 & 541.8 & 6.4 & - & -\\
\(\tilde{t}_2\) & 914.4 & 949.3 & 3.7 & - & -\\
\(\tilde{b}_1\) & 845.3 & 880.4 & 3.9 & - & -\\
\(\tilde{b}_2\) & 961.9 & 1008.5 & 4.6 & - & -\\
\hline
\(\tilde{g}\) & 1107.2 & 1154.2 & 4.1 & - & - \\
\hline
\end{tabular}
\end{center}
\caption[Comparison of the tree level $m_T$ and loop masses at one-loop ($m_{1L}$) and two-loop ($m_{2L}$)]{Comparison of the tree level $m_T$ and loop masses at one-loop ($m_{1L}$) and two-loop ($m_{2L}$). $\Delta$ is the relative difference $|1-\frac{m_T}{m_{1L}}|$ respectively $|1-\frac{m_{1L}}{m_{2L}}|$. }
\label{tab:spectrum1}
\end{table}
In Table~\ref{tab:spectrum1}, we give as an example the masses of the Higgs bosons, charginos, neutralinos and third generation sfermions at tree level as well as at the one- and two- loop level for the parameter set
\begin{eqnarray}
\nonumber &m_0 = 180\,\GeV \thickspace,\quad M_{1/2} = 500\,\GeV\thickspace, \quad 
A_0 = A_\lambda^{\rm GUT} =- 1500\,\GeV \thickspace, \quad A_\kappa^{\rm GUT} = -36\,\GeV\thickspace,& \\
&\tan\beta = 10\thickspace, \quad \kappa^{\rm GUT} = 0.11\thickspace, 
\quad \lambda^{\rm GUT} = 0.1\thickspace, \quad v_s = 13689\,\GeV \thickspace .&
\end{eqnarray}
which is close to the benchmark scenario 1 of ref.~\cite{Djouadi:2008uw}. As can be seen in Table~\ref{tab:spectrum1}, the corrections are sizable ranging from 0.1~\% to 23.6~\% in case of the lightest Higgs boson. This large correction is well known and the main reason for including the two-loop corrections. The corresponding two-loop Higgs masses as well as the relative correction with respect to the one-loop results are displayed in Table~\ref{tab:spectrum1}. Again, the largest correction with 5.2~\% is in case of the lightest Higgs boson mass. \\
As an estimate of the remaining theoretical uncertainty, we have varied the renormalization scale  in {\tt SPheno}.  We show in  Fig.~\ref{fig:Scalar_Q} the scale dependence  for masses of neutral scalar Higgs bosons at the one- and two-loop masses  normalized  to their values at $Q=1$~TeV and vary  the renormalization scale $Q$  between 200~GeV and 2.2~TeV. As can be seen, the large variation of 8~\%  at one-loop for the lightest Higgs, which is mainly the lighter $SU(2)_L$ doublet Higgs in this case, is reduced at two-loop to less than 2~\%. In case of the heavier Higgs bosons, the scale dependence  is significantly smaller and similarly showing a significant improvement when going from the one-loop level to the two-loop level. However, we remark that the values of  $\lambda$ and $\kappa$ are small in this scenario and we expect a stronger dependence in case of larger couplings.

\begin{figure}[!ht]
\begin{minipage}{16cm}
\includegraphics[scale=1.]{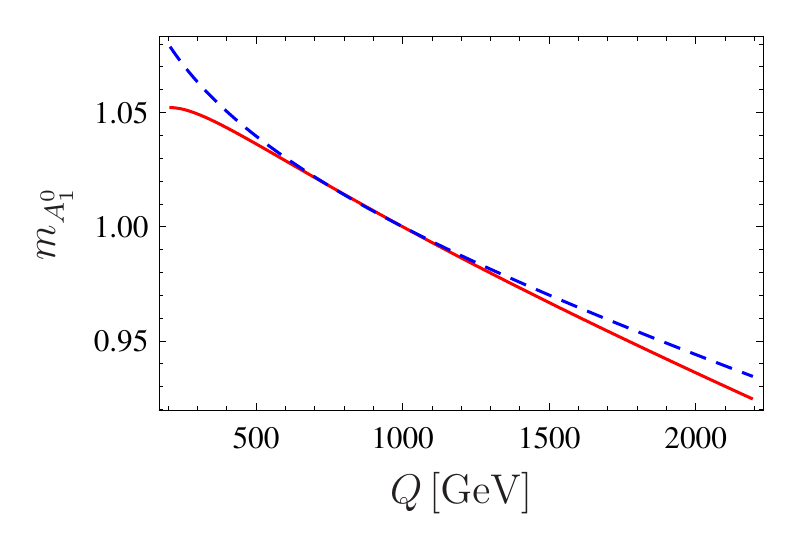}
\hfill
\includegraphics[scale=1.]{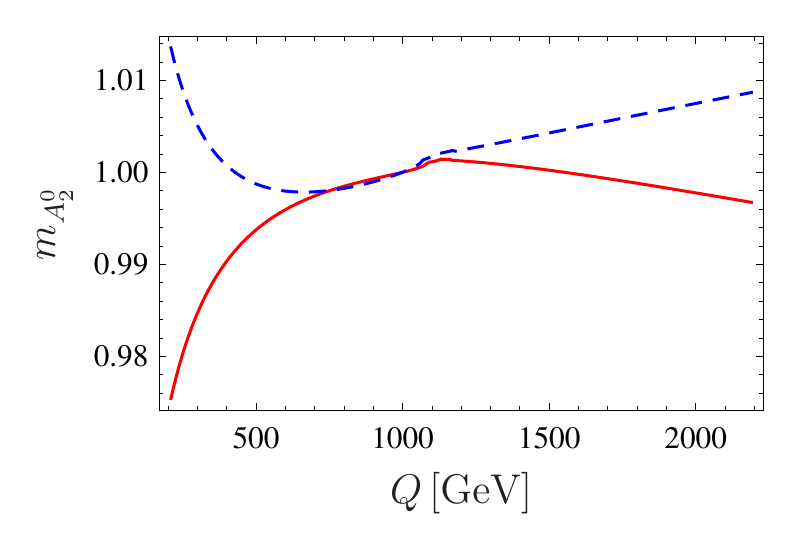}
\caption[Dependence of CP odd Higgs masses on the renormalization
scale $Q$]{Dependence of CP odd Higgs masses on the renormalization
scale $Q$ at one-loop (red) and two-loop level (dashed blue) normalized to the value at 
$Q=1$~TeV.  Left:   $m_{A_1^0}$. Right: $m_{A_2^0}$.}
\label{fig:PseudoScalar_Q}
\end{minipage}
\end{figure}

The picture changes slightly in case of the pseudo scalar bosons as can be seen in  Fig.~\ref{fig:PseudoScalar_Q}. While the heavier pseudo scalar behaves exactly as the  second scalar field since both originate  to 99.5~\% from \(H_d\), the scale dependence for the lighter pseudo scalar is smaller compared to the lightest scalar field, but hardly improves at the two-loop level. The reason is that in the two-loop part contain 'only' the strong contributions of the third generation squarks  whereas this state is mainly a singlet state and, thus, the contributions due to the NMSSM specific couplings would be needed for a further improvement.\\

\begin{figure}[!ht]
\begin{minipage}{16cm}
\includegraphics[scale=1.]{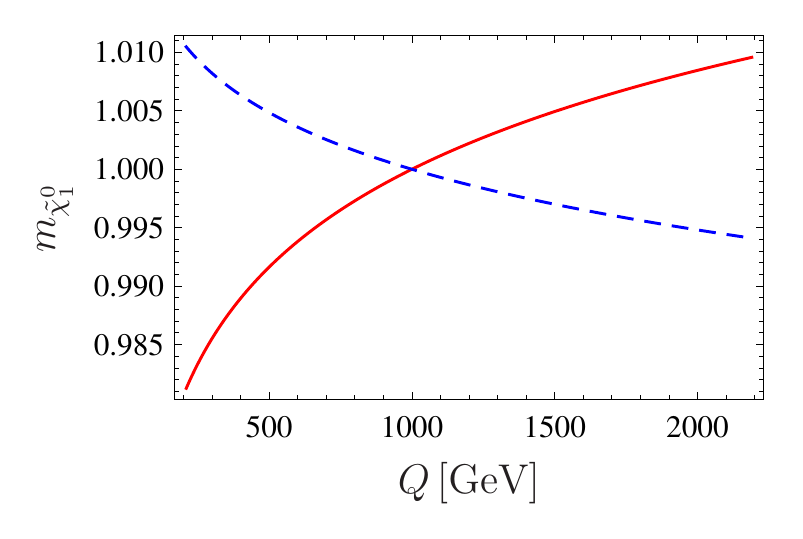}
\hfill
\includegraphics[scale=1.]{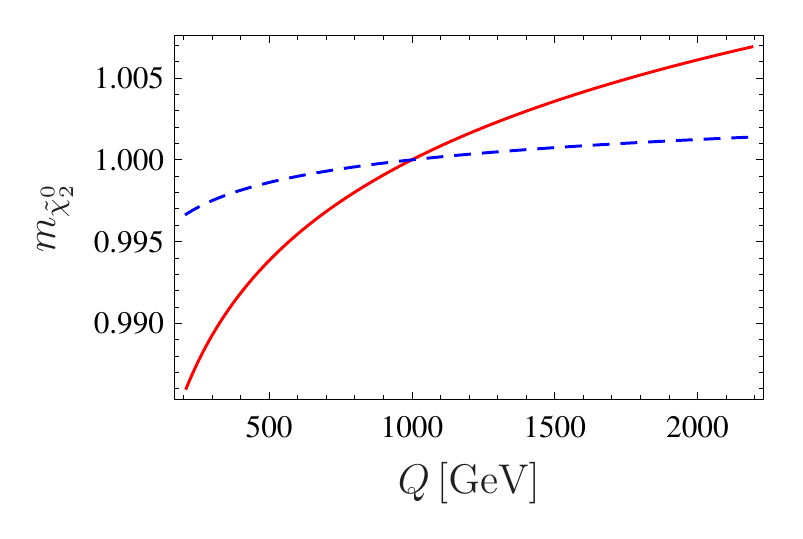} \\
\includegraphics[scale=1.]{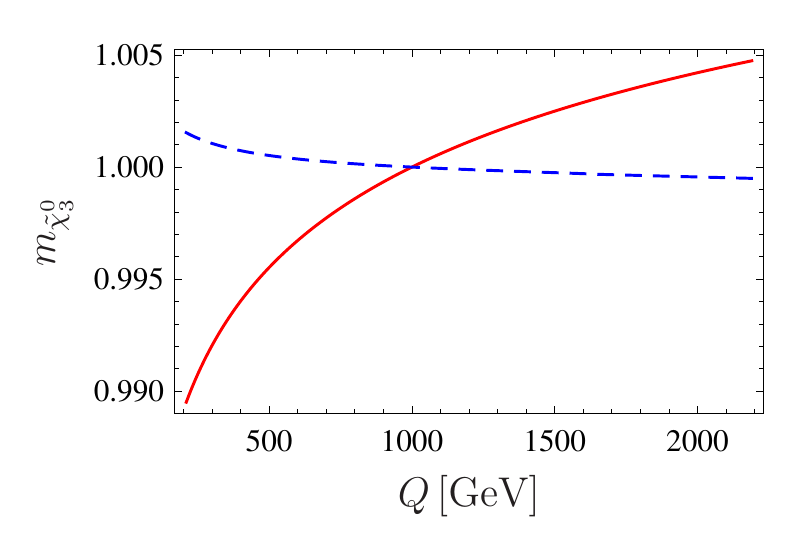}
\hfill
\includegraphics[scale=1.]{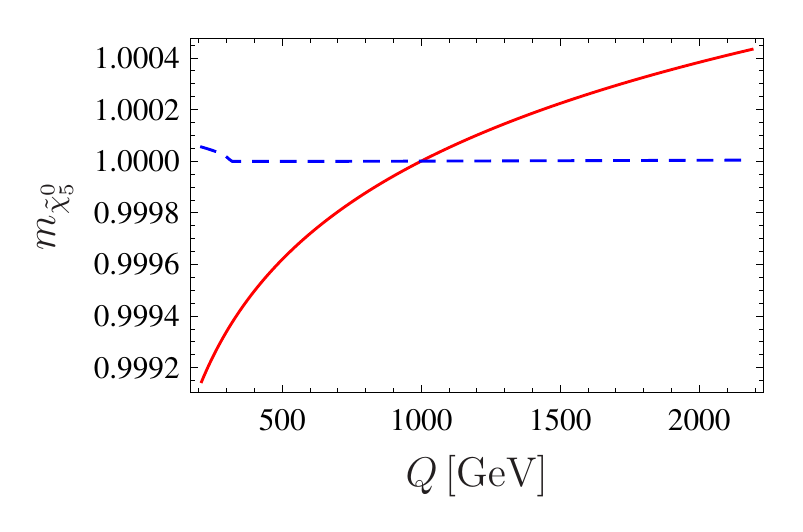}
\caption[Dependence of the masses of the light neutralinos on the renormalization
scale $Q$]{Dependence of the masses of the light neutralinos on the renormalization
scale $Q$ at tree (red) and one-loop level (dashed blue) normalized to the value at $Q=1$~TeV.
From left to right and from above to below: $m_{\tilde \chi^0_1}$, $m_{\tilde \chi^0_2}$,
$m_{\tilde \chi^0_3}$ and  $m_{\tilde \chi^0_5}$.
}
\label{fig:Neutralino_Q}
\end{minipage}
\end{figure}

\begin{figure}[!ht]
\begin{minipage}{16cm}
\includegraphics[scale=1.]{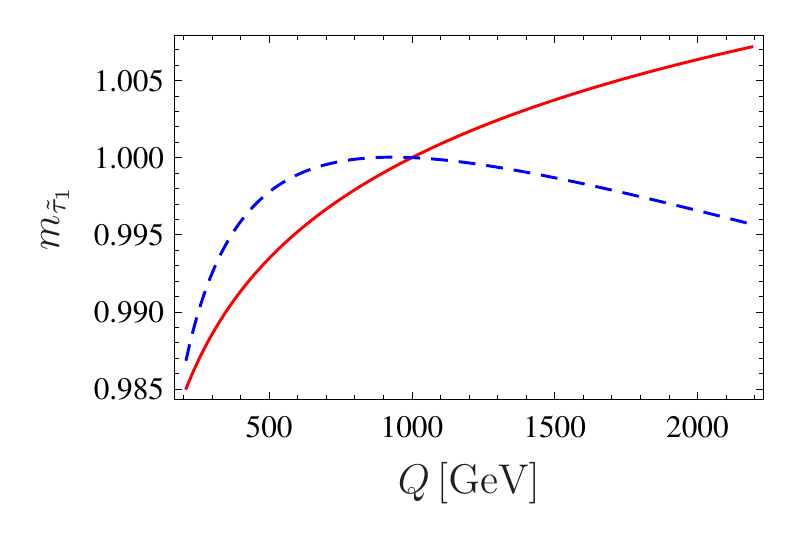}
\hfill
\includegraphics[scale=1.]{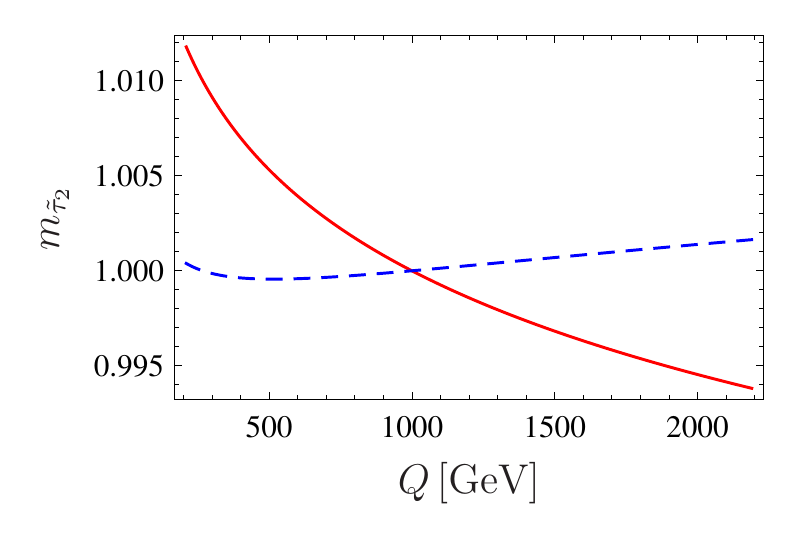} 
\caption[Dependence of the stau masses of the sleptons on the renormalization
  scale $Q$]{Dependence of the stau masses of the sleptons on the renormalization
  scale $Q$ at tree (red) and one-loop level (dashed blue) normalized to the value at $Q=1$~TeV.
}
\label{fig:Sleptons_Q}
\end{minipage}
\end{figure}

In Fig.~\ref{fig:Neutralino_Q}, the scale dependence for  different neutralinos is shown.  As can be seen, in case of the three lighter states the scale dependence is reduced from the tree level of about 1.5~\% to 3-5 per-mill. In case of the singlet state $\tilde \chi_5$,  the scale dependence is already small due to the small values of $\lambda$ and $\kappa$. We note that the scale dependence of $\tilde \chi^+_1$ ($\tilde \chi^+_2$ and $\tilde \chi^0_4$) is nearly the same as that of $\tilde \chi^0_2$ ($\tilde \chi^0_3$) as these state have their main origin in the same electroweak multiplet.
Finally we show in  Fig.~\ref{fig:Sleptons_Q} the scale dependence of the staus. The scale dependence at tree level amounts to about 2-2.5~\% and is  reduced at one-loop level to about 1~\% and less where the $\tilde \tau_1$ shows still the larger dependence. The sleptons of the first two generations show a somewhat smaller scale dependence because in their cases the Yukawa couplings do not play any role.
\section{Comparison with the literature}
\label{sec:comparison}
To date, the program package {\tt NMSSM-Tools} \cite{Ellwanger:2006rn,Ellwanger:2005dv,Ellwanger:2004xm} has been the only complete spectrum calculator for the NMSSM. {\tt NMSSM-Tools}
uses for the constrained NMSSM the parameters \(m_0\), \(M_{1/2}\), \(A_0\) and \(A_\kappa\) at the GUT scale whereas   \(\tan\beta\) and \(\lambda\) are given at the electroweak scale. Moreover, in {\tt NMSSM-Tools}  the tadpole equations are solved with respect to \(|v_s|\), \(\kappa\), and \(m_S^2\). We have performed a detailed numerical comparison of our implementation in \SPheno with the version {\tt NMSSM-Tools} 2.3.1 and present here a few typical examples.
\subsection{Differences between the programs}
Since both programs use different methods to calculate the spectrum,  we have done a comparison where we modified the codes such that  both codes use equivalent methods except for small details. First, the implementation of {\tt NMSSM-Tools} involves two different scales, namely, the SUSY scale defined as
\begin{equation}
 Q^2_{\rm SUSY} = M^2_{\rm SUSY} = \frac{1}{4} \left(2 m_{\tilde{q}}^2 + m_{\tilde{u}}^2 + m_{\tilde{d}}^2\right),
\end{equation}
and the scale at which the masses are calculated,
\begin{equation}
 Q^2_{\rm STSB} = m_{\tilde{q}_3} m_{\tilde{u}_3}.   
\end{equation}
In \SPheno, all masses are evaluated at the SUSY scale, so that we had to set \(Q_{\rm STSB} = Q_{\rm SUSY}\) in the relevant routines of {\tt NMSSM-Tools}. Second, as already stated in sec.~\ref{sec:procedure}, the two-loop \(\beta\) function of \(A_\lambda\) has been  corrected in the
public version of {\tt NMSSM-Tools}.  However, in general the numerical effect on the spectrum is rather small. \\
In the Higgs sector, the loop contributions are taken into account differently in both codes. While {\tt SPheno} takes the complete one-loop correction including the dependence of the external momenta, {\tt NMSSM-Tools} uses the effective potential approach, i.e. setting the external momenta to zero. Also the included contributions differ: in {\tt SPheno}, the complete one-loop corrections to both, scalar and pseudo scalar Higgs bosons,  and the two-loop contributions as given in \cite{Degrassi:2009yq} are included. In  {\tt NMSSM-Tools} beside the dominant contributions due to third generation sfermions also electroweak corrections and some leading two-loop corrections for the scalars are calculated: for the pseudo scalars only the dominant one-loop corrections due to tops, stops, bottoms, and sbottoms are included. To account for these differences, we have switched off the two-loop parts in both codes. Furthermore, we have set the external momenta of the loop-diagrams of scalars in  {\tt SPheno} to zero. Finally, we have kept only those corrections to the pseudo scalar masses in {\tt SPheno} which are also included in {\tt NMSSM-Tools}. In the following, we refer to these modified versions by {\tt SPheno mod} and  {\tt NMSSM-Tools mod}, respectively. \\
Also in the chargino and neutralino sector the implementations are different: in {\tt SPheno}, the complete one-loop corrections are implemented whereas in  {\tt NMSSM-Tools} the corrections to the parameters $M_1$, $M_2$ and $\mu_{\rm eff}$  are taken into account. In the slepton sector, the differences are larger: {\tt SPheno} contains the complete one-loop corrections whereas in {\tt NMSSM-Tools} the calculation is done at tree level. For completeness, we note that the data transfer has been done using the SLHA2 conventions \cite{Allanach:2008qq}. 
\begin{figure}[t]
\begin{center}
\begin{minipage}{16cm}
\includegraphics[scale=1.]{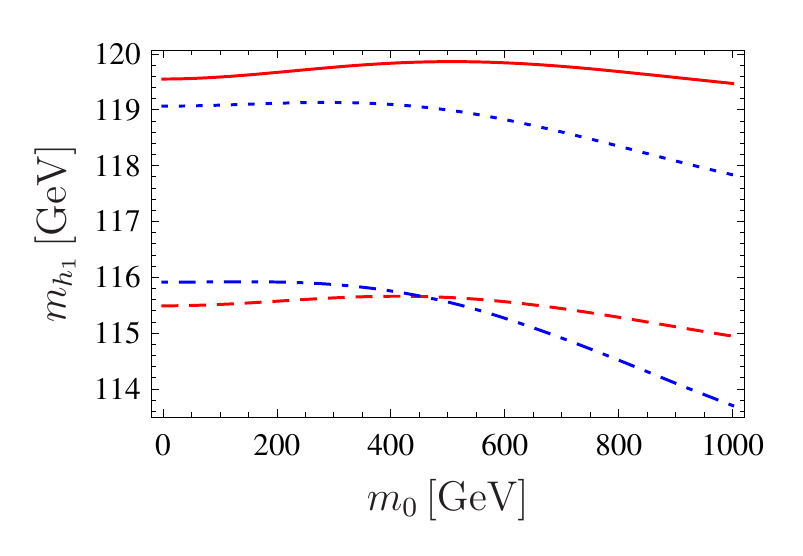} 
\hfill
\includegraphics[scale=1.]{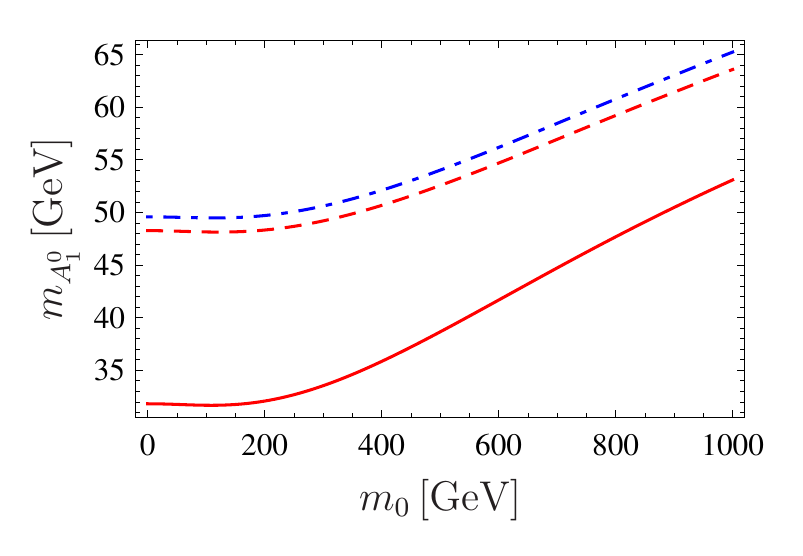} \\
\includegraphics[scale=1.]{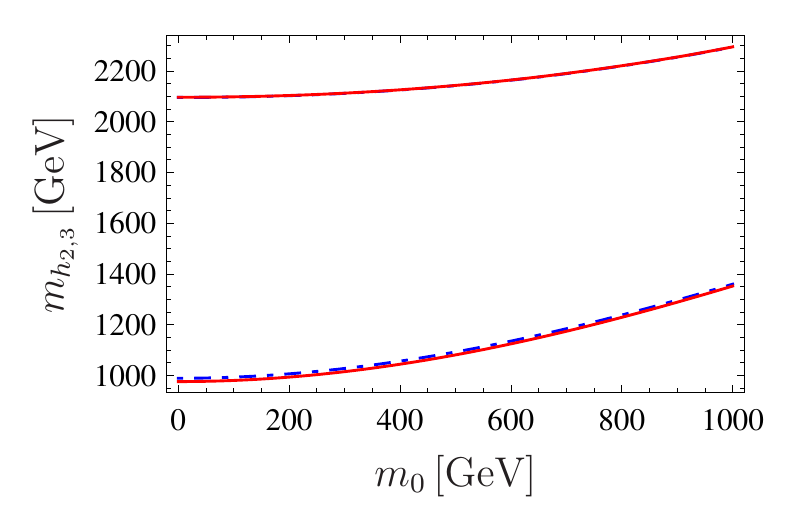}
\hfill
\includegraphics[scale=1.]{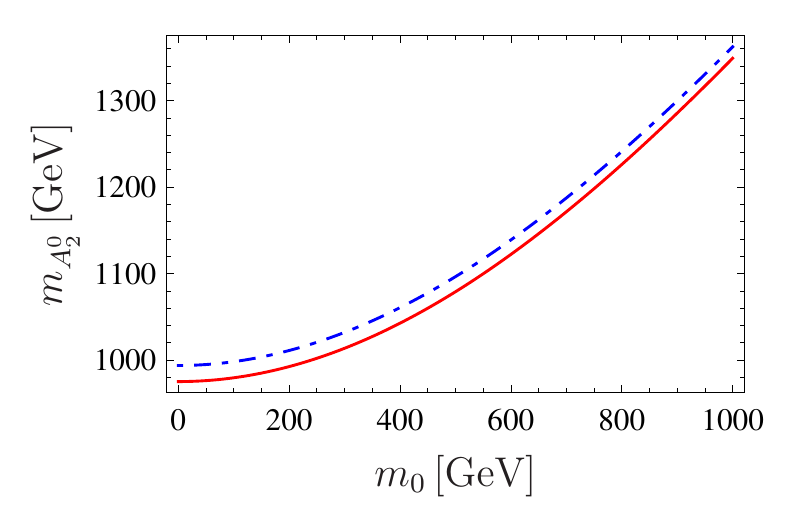}
\end{minipage}
\caption[Comparison of Higgs masses between  {\tt SPheno}  and  {\tt NMSSM-Tools} ]{Comparison of the masses in GeV  of the lightest scalar (upper left), the lightest pseudo scalar (upper right), heavier scalar masses (lower left) and heavier pseudo scalar mass (lower right)  as a function of $m_0$ (in GeV). All other parameters are fixed as in eq.~(\ref{eq:defP1}). The lines are for unmodified version of {\tt SPheno} (full red), {\tt NMSSM-Tools} (dotted blue),  {\tt SPheno mod} (dashed red) and {\tt NMSSM-Tools mod}  (dot-dashed blue).  }
\label{fig_scalar}
\end{center}
\end{figure}
\begin{figure}[t]
\begin{minipage}{16cm}
\includegraphics[scale=1.0]{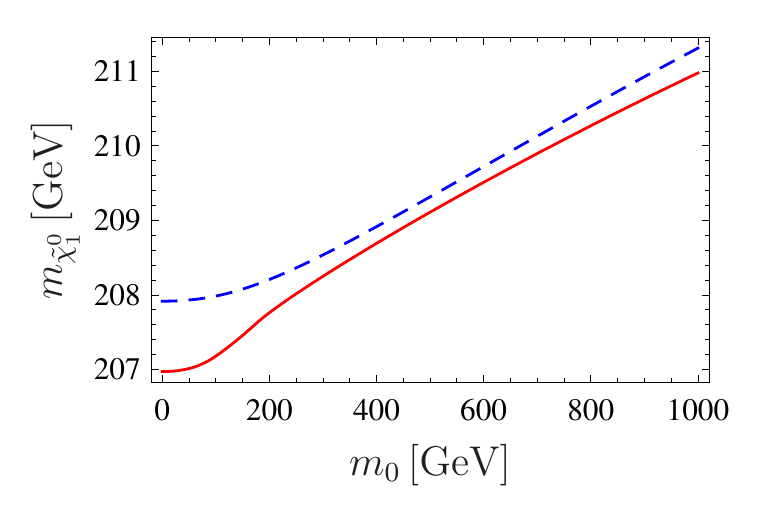}
\hfill
\includegraphics[scale=1.0]{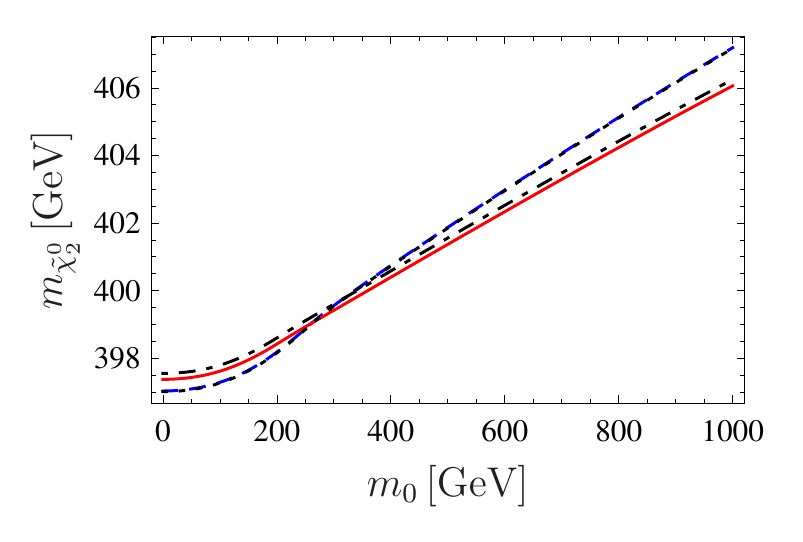}
\\
\includegraphics[scale=1.0]{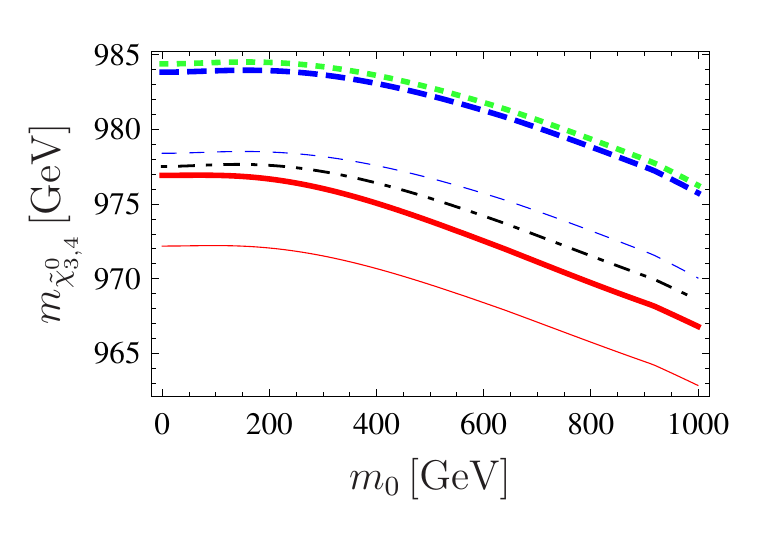}
\hfill
\includegraphics[scale=1.0]{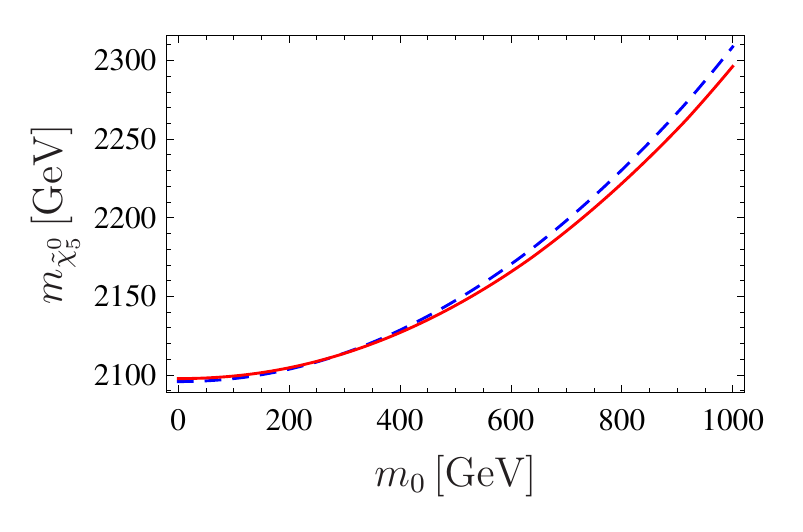}
\caption[Comparison of chargino and neutralino  masses as a function
  of $m_0$ ]{Comparison of chargino and neutralino  masses (in GeV) as a function
  of $m_0$ (in GeV). 
  All other parameters are fixed as in eq.~(\ref{eq:defP1}). The lines correspond to the unmodified versions of
  {\tt SPheno} (full red) and {\tt NMSSM-Tools} (dashed blue).
 Up left: light neutralinos $\tilde \chi^0_1$. Up right: 
 neutralino $\tilde \chi_2$ and chargino $\tilde \chi_1^+$ ({\tt SPheno}: black dotdashed,
  {\tt NMSSM-Tools}: black dotted).  
 Down left: neutralinos $\tilde \chi_3$ (thin lines), $\tilde \chi_4$  
 (thick lines) and chargino  $\tilde \chi_2^+$  ({\tt SPheno}: black dotdashed, {\tt NMSSM-Tools}: green dotted). Down right: $\tilde \chi_5$. 
}
\label{fig_others}
\end{minipage}
\end{figure}
\begin{figure}[t]
\begin{minipage}{16cm}
\includegraphics[scale=1.]{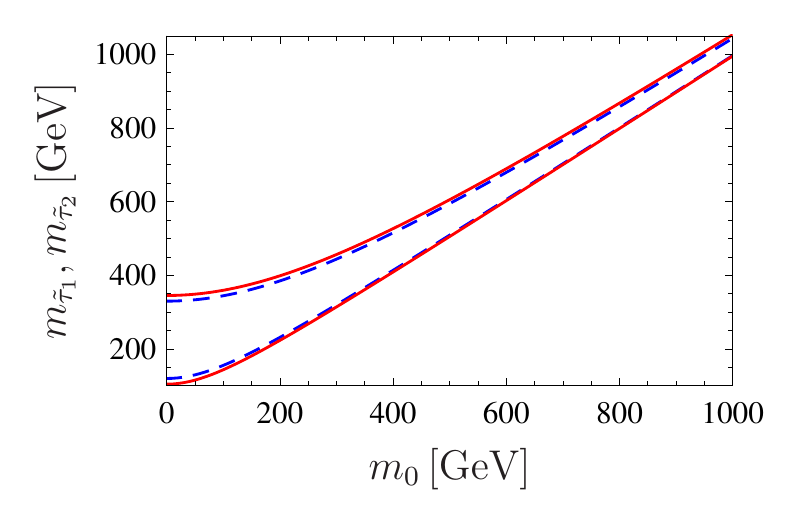}
\hfill
\includegraphics[scale=1.]{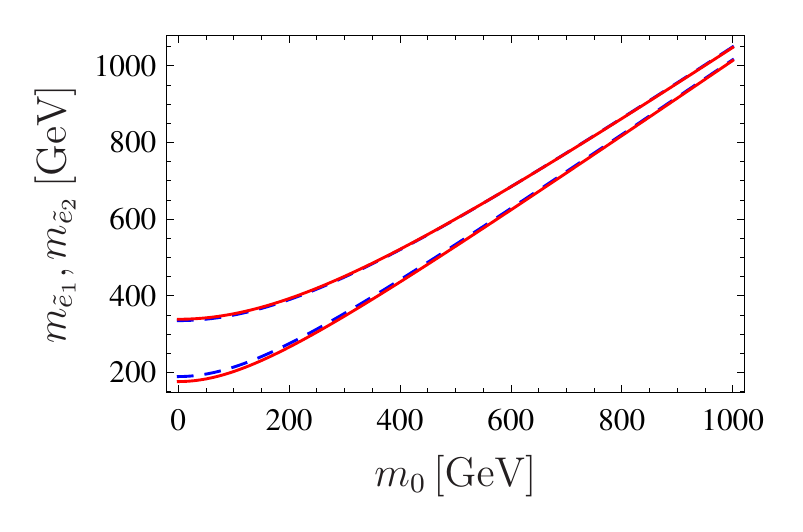}
\caption[Comparison of selectron and stau masses  as a function
  of $m_0$]{Comparison of selectron and stau masses (in GeV) as a function
  of $m_0$. All other parameters are fixed as
  in eq.~(\ref{eq:defP1}). The lines correspond to the unmodified versions of
  {\tt SPheno} (full red) and {\tt NMSSM-Tools} (dashed blue).}
\label{fig:stau_m0}
\end{minipage}
\end{figure}
\subsection{Results of the comparison}
As a first reference scenario, we take the benchmark point 1 proposed in \cite{Djouadi:2008uw}. The corresponding input parameters for
{\tt NMSSM-Tools} are
\begin{eqnarray}
&m_0=180\,{\rm GeV}\thickspace,\quad M_{1/2}=500\,\rm GeV\thickspace,\quad A_0=-1500\,{\rm GeV}\thickspace, \quad\tan\beta=10 \thickspace,&\nonumber\\
&\lambda^{\rm SUSY}=0.1\thickspace,\quad A_{\kappa}^{\rm GUT}=-33.45 \thickspace, \quad\mu_{\rm eff}>0 \thickspace.& 
\label{eq:defP1}
\end{eqnarray}
In the following, we will vary $m_0$ and keep the other parameters to the values shown here. In the left graph of Fig.~\ref{fig_scalar}, we show the mass of the lightest scalar $h_1$ as a function of $m_0$. The largest discrepancies
arise for the lighter scalar and pseudo scalar boson, where the relative  differences between the complete calculation of both  programs amount up to 2.5 and 35~\%, respectively. In case of $h^0_1$, this is a combination of the $p^2$ terms in the loop-functions and the additional two-loop contributions. The differences in case of $A^0_1$ can easily be understood by noting that in {\tt NMSSM-Tools}  only the contribution of third-generation sfermions are taken into account whereas we include the complete one-loop corrections plus the known two-loop contributions. In case of the modified program codes, these differences reduce to at most 2~\% which is meanly due to two differences: first, the way the top Yukawa coupling is calculated  and, secondly, the way the tadpole equations are solved. There is no visible difference between  {\tt NMSSM-Tools}  and  {\tt NMSSM-Tools mod} for the pseudo scalar and the heavy scalars. The reason is that in case of the pseudo scalar no two-loop corrections are calculated in {\tt NMSSM-Tools} and in case of the heavy scalars they are very small. \\
Finally, we have also  cross-checked our results in the Higgs sector with ref.~\cite{Degrassi:2009yq} and we have found agreement better than one per-mill when using the set of soft SUSY parameters at the scale $Q_{\rm STSB}$.  This small difference is an effect of the Yukawa and scalar-trilinear couplings of the first two generations which we take also into account. If we restrict ourself to third generation couplings, there is an exact agreement between both calculations. \\
Concerning the chargino and neutralino masses, the agreement between the two spectrum calculators is rather good as can be seen in  Fig.~\ref{fig_others}. The relative differences are  at most 1~\% and in general slightly below 0.5~\%. In case of the sleptons, the differences are more pronounced as can be seen in  Fig.~\ref{fig:stau_m0} which is due to the differences between tree level and one-loop calculation and amounts to 3~\% and 0.6~\% for the light and heavy stau, respectively.  Although, one expects similar experimental uncertainties for LHC physics. However, the precision which is necessary for a future linear collider or dark matter calculation require the inclusion of the radiative corrections to the slepton masses.
\section{Effects of one-loop corrections on the relic density of dark matter}
\label{sec:DM}
It is well known that the prediction of the dark matter relic density $\Omega h^2$ is very susceptible to the exact mass configuration of the scenario under consideration \cite{Belanger:2005jk}. For a neutralino LSP, this is for instance the case for the annihilation through Higgs-resonances, but also in case of neutralino-sfermion coannihilation. For the latter, the mass difference between the two particles plays a key role in the calculation of the resulting relic density. For that reason, it is necessary to calculate the complete spectrum as precisely as possible to get viable results of allowed regions of parameter space with respect to the constraints imposed by the presence of dark matter. Let us recall that recent measurements by the WMAP satellite in combination with further cosmological data lead to the favored interval 
\begin{equation}
	0.1018 < \Omega h^2 < 0.1228
	\label{eq:WMAP} 
\end{equation}
at 3$\sigma$ confidence level \cite{Komatsu:2010fb}. \\
We compute the relic density of the lightest neutralino using the program  {\tt micrOMEGAs\,2.4.O} \cite{Belanger:2006is}. To this end, we have implemented the NMSSM particle content and corresponding interactions into a model file for {\tt CalcHEP}  \cite{Pukhov:2004ca}, which is used by {\tt micrOMEGAs} to evaluate the (co)annihilation cross-section. The relevant interactions have again been calculated and written into the  model files by {\tt SARAH}. Let us note, that we take into account  important QCD effects, such as the running strong coupling constant and the running  quark masses \cite{Herrmann:2007ku,Herrmann:2009wk,Herrmann:2009mp}. \\
\begin{figure}[t]
\begin{minipage}{16cm}
	\includegraphics[scale=1.]{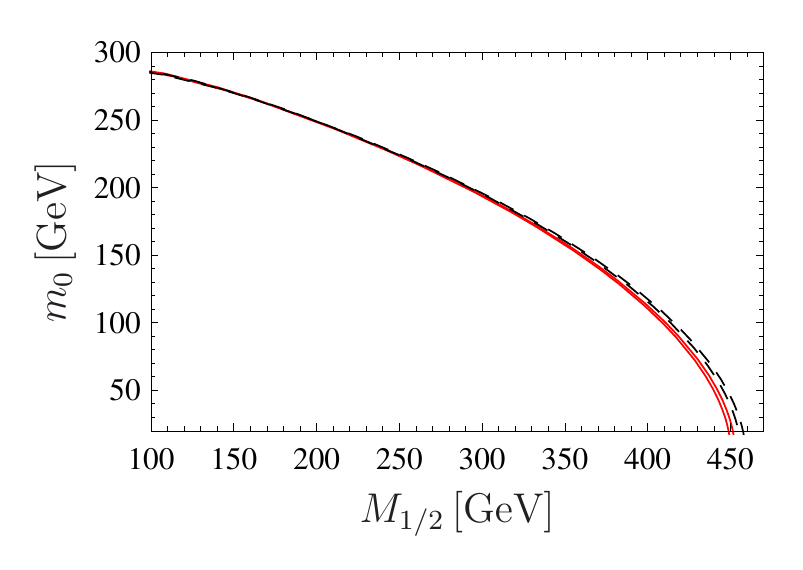}
	\hfill
	\includegraphics[scale=1.]{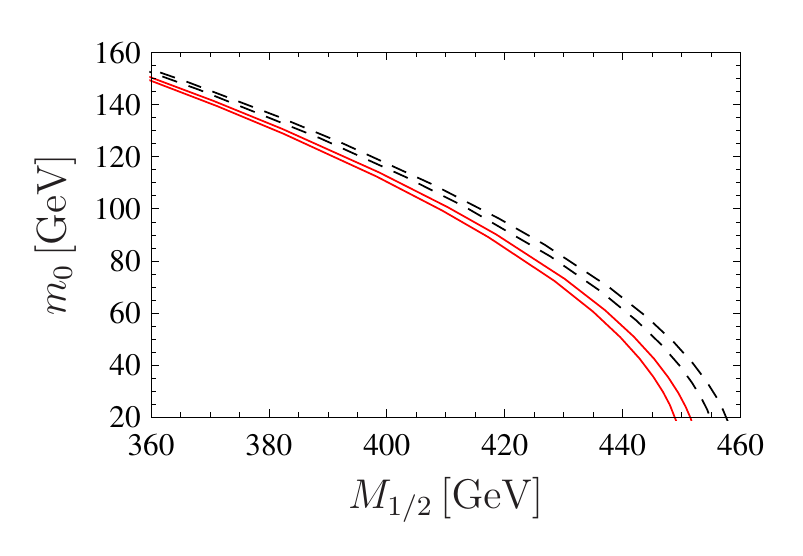} \\
	\includegraphics[scale=1.]{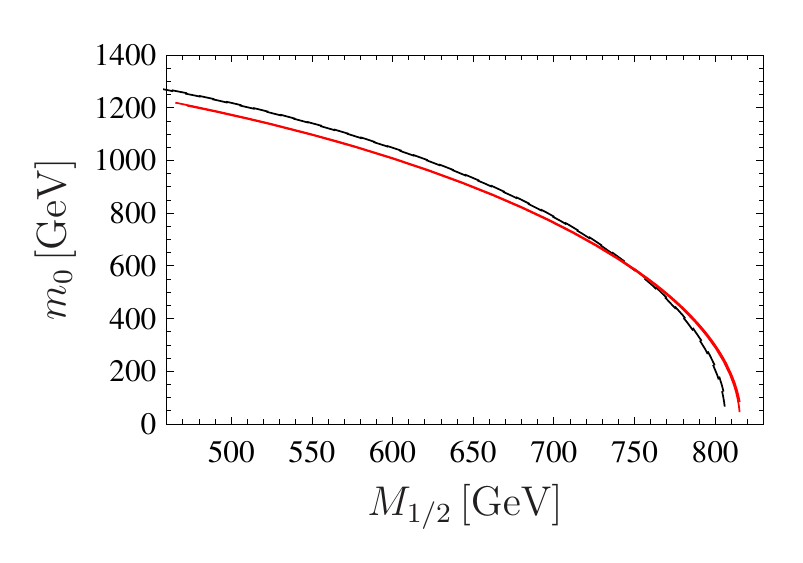}
	\hfill
	\includegraphics[scale=1.]{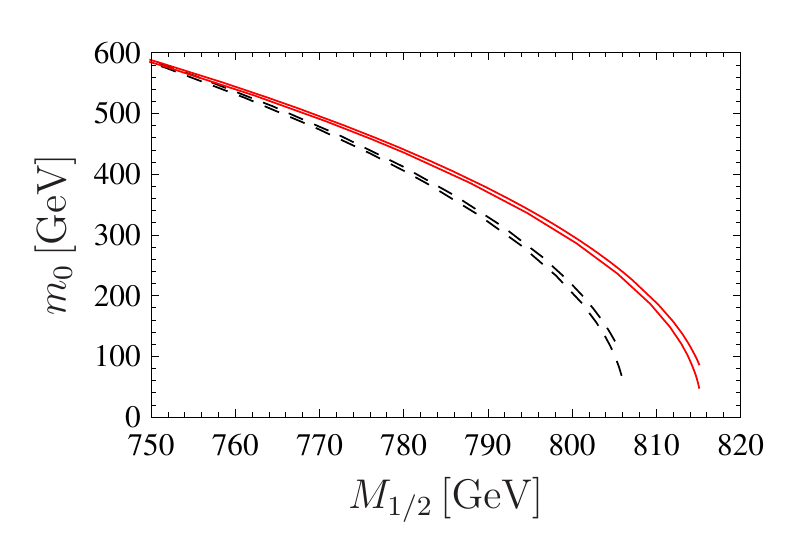}
	\caption[($m_{0}$,$M_{1/2}$)-plane for dominant neutralino-stau and neutralino-stop coannihilations]{The isolines corresponding to $\Omega h^2 = 0.1018$ and $\Omega h^2 = 0.1228$ in the ($m_{0}$,$M_{1/2}$)-plane for dominant neutralino-stau (up) and neutralino-stop (down) coannihilations. All other parameters are fixed as in eq.~(\ref{eq:paramsDM}) (stau-coannihilation) respectively as in eq.~(\ref{eq:paramStop}) (stop-coannihilation). The red solid lines have been obtained for the complete mass spectrum at the one-loop level, while for the black dashed line the loop corrections to the slepton (up) or up-squark (down) masses have been disabled. The right graphs correspond to zooms into the left ones.}
\label{fig:stau_coann}
\end{minipage}
\end{figure}
\paragraph*{Coannihilation} As first example, we illustrate the effect of the one-loop correction to the slepton masses on the dark matter relic density in a region of dominant neutralino-stau coannihilation. In the two upper plots of Fig.~\ref{fig:stau_coann}, we show the isolines corresponding to the upper and lower limit of eq.~(\ref{eq:WMAP}) in the ($m_0,M_{1/2}$)-plane. All remaining parameters of eq.~(\ref{eq:params}) are fixed to
\begin{eqnarray}
 &\tan\beta = 15\thickspace, \quad \kappa^{\rm SUSY} = -0.05\thickspace, \quad \lambda^{\rm SUSY} = -0.1\thickspace,&\nonumber \\ &\thinspace A_\kappa^{\rm GUT} = 30\,{\rm GeV}\thickspace, \quad A_0 = A_\lambda^{\rm GUT} = 1000\,{\rm GeV} \thickspace,\quad v_s = 2\cdot 10^4\,{\rm GeV}\thickspace.& 
\label{eq:paramsDM}
\end{eqnarray}
One clearly sees that the allowed parameter range gets shifted depending on the precision with which the spectrum is calculated. Moreover, the two regions shown do not overlap as can be seen clearly  in the figure to the right.\\
For a point with $\Omega h^2 = 0.112$ at $M_{1/2} \simeq 451.2$~GeV, the resulting one-loop corrected masses of the lightest neutralino and the lighter stau are $m_{\tilde{\chi}_1^0} = 186.0$~GeV and $m_{\tilde{\tau}_1} = 196.8$~GeV, respectively. As consequence, coannihilations account for about 60~\% of the total annihilation cross-section, where the most important final states are $\tau h_1$ (27\%) and $\tau Z^0$ (15~\%). In addition, a sizable contribution of about 14~\% (5~\%) comes  from stau-antistau (stau-stau) annihilation. The remaining contributions are mainly from neutralino pair annihilation. For lower values of $M_{1/2} \lesssim 200$~GeV, the coannihilations become less important within the WMAP-favored region, the dominant mechanism is then neutralino pair annihilation into $\tau^+\tau^-$ pairs through stau-exchange.\\
Even if we have focused in our discussion so far on the one-loop correction in the electroweak sector, we have also calculated the squark masses at one-loop level. Large values of \(A_0\) can cause a big mass splitting in the stop sector. The lighter stop can in those cases be the NLSP. As example, we consider the spectrum based on the input parameters
\begin{eqnarray}
 &\tan\beta = 10\thickspace, \quad \kappa^{\rm GUT} = 0.5\thickspace, \quad \lambda^{\rm GUT} = 0.4\thickspace,\quad A_\kappa^{\rm GUT} = -2510\,{\rm GeV}\thickspace, &\nonumber \\ 
&A_0 = -2200\,{\rm GeV},\quad  A_\lambda^{\rm GUT} = -770\,{\rm GeV}\thickspace,\quad v_s = 2763\,{\rm GeV} \thickspace.& 
\label{eq:paramStop}
\end{eqnarray}
In large areas of the ($m_{0}$,$M_{1/2}$)-plane, the mass of the lighter stop is close to the mass of the LSP. This leads to a sufficient coannihilation between both. We show in the second row of Fig.~\ref{fig:stau_coann} the result for the range of eq.~(\ref{eq:WMAP}) with and without the one-loop corrections to the up-squark masses. It is well known that the loop corrections in the stop sector are more important than for staus. Hence, the impact on the relic density of the neutralino is much larger in this case. In addition, even in areas of the (\(m_0,M_{1/2}\)) where the coannihilation is absent, the loop corrections to the stops have a significant impact on the relic density. This is the case for smaller values of \(M_{1/2}\): the annihilation into \(t\bar{t}\) due to a squark in the t-channel is important. The crossing between the two bands in the lower, left plot in Fig.~\ref{fig:stau_coann} happens because the loop correction to the up-squarks by fermions become more important at small \(M_{1/2}\). This changes the sign of the mass difference between tree and one-loop mass. 
\paragraph*{Higgs funnel}
\begin{figure}[!ht]
\begin{minipage}{16cm}
\begin{center}
\includegraphics[scale=1.]{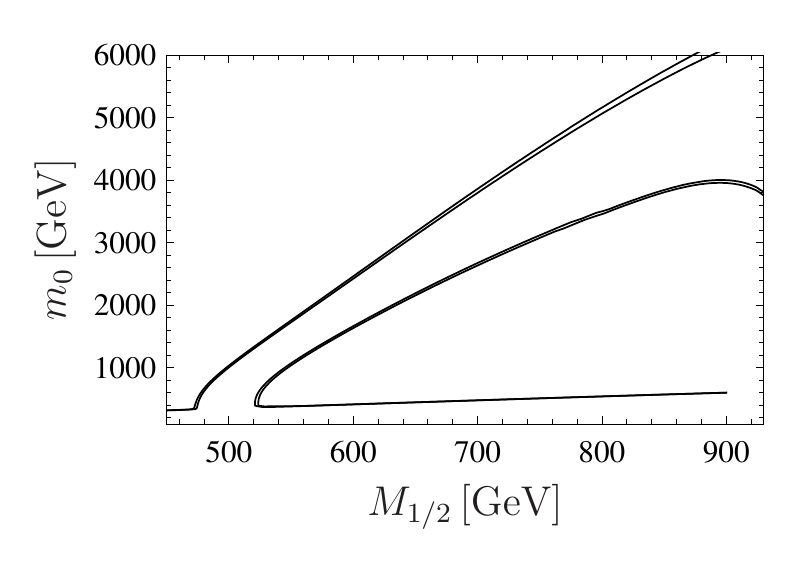}
\end{center}
\caption[Higgs funnel in the NMSSM]{The isolines corresponding to $\Omega h^2 = 0.1018,0.1228$ in the ($m_{0},M_{1/2}$)-plane for a dominant annihilation due to a Higgs resonance using two-loop corrections.  All other parameters are fixed as in eq.~(\ref{eq:HF}).  }
\label{fig:HF}
\end{minipage}
\end{figure}

\begin{figure}[!ht]
\begin{minipage}{16cm}
\includegraphics[scale=1.]{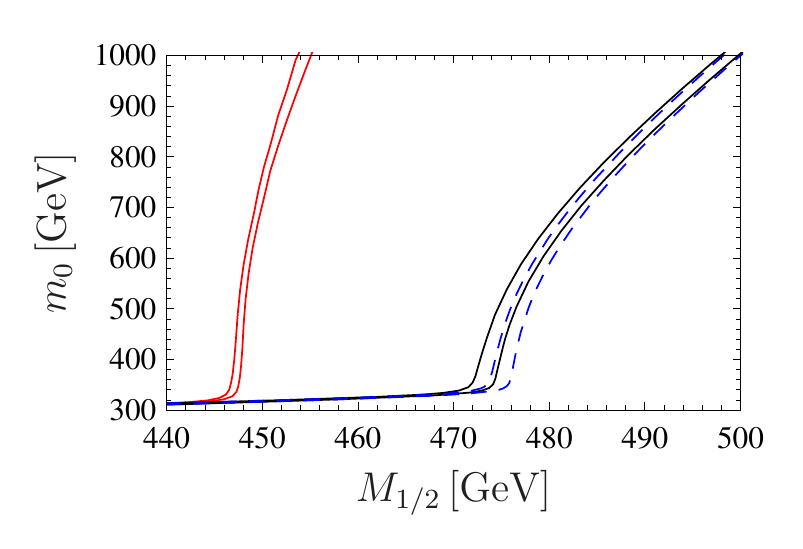}
\hfill
\includegraphics[scale=1.]{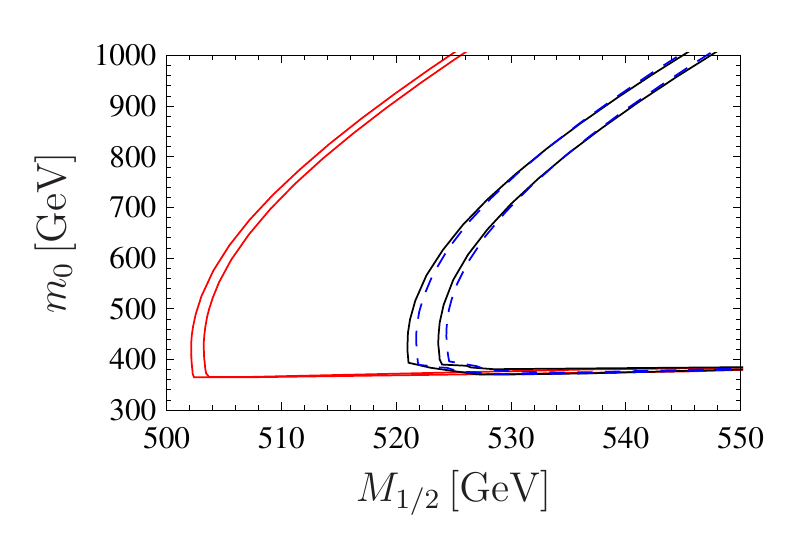}
\caption[Higgs funnel in the NMSSM]{Zooms into areas of Fig.~\ref{fig:HF} with pseudo scalar masses very close to twice the mass of the LSP. The color code is: complete calculation of the mass spectrum (black), pseudo scalar masses without two-loop corrections (blue dashed), and  pseudo scalar masses only at tree level (red).}
\label{fig:HF_zoom}
\end{minipage}
\end{figure}

A second scenario in which the loop corrections of other particles have large influence on the neutralino relic density is the Higgs funnel. As already stated, the mass of the pseudo scalar is close to twice the mass of the lightest neutralino in that case. This resonance causes a strong annihilation and prevents the overclosure of the universe by neutralinos in those areas of the parameter space.  The Higgs funnel in the MSSM  and in high scale extension like the seesaw scenarios in chapter~\ref{chapter:SU5} with unified Higgs masses at the GUT scale has a cuspy shape (see e.g. Fig.~\ref{fig:4}). However, it consists in our considered NMSSM scenario of two separated lines. This is shown in Fig.~\ref{fig:HF}. The line show again the 3\(\sigma\) interval of WMAP-7 given in eq.~(\ref{eq:WMAP}).  We have used as input
\begin{eqnarray}
 &\tan\beta = 38 \thickspace, \quad \kappa^{\rm GUT} = 0.15 \thickspace, \quad \lambda^{\rm GUT} = -0.3 \thickspace, \quad A_\kappa^{\rm GUT} = 310\,{\rm GeV} \thickspace, &\nonumber \\ 
&A_0 = 1000\,{\rm GeV} \thickspace,\quad A_\lambda^{\rm GUT} = 1550\,{\rm GeV} \thickspace, \quad v_s = 5900\,{\rm GeV} \thickspace.& 
\label{eq:HF}
\end{eqnarray}
The reason for this feature is the behavior of the Higgsino component of the LSP: in the MSSM, \(\mu\) is usually calculated by the tadpole equations. That's why it is significantly affected by the difference
\begin{equation}
m_{H_d}^2 \cos\beta - m_{H_u}^2 \sin\beta \thickspace ,
\end{equation}
i.e. it depends strongly on \(m_0\) and due to RGE running also on \(M_{1/2}\). In contrast, \(\mu_{\rm eff}\) does not vary much by changing \(m_0\) or \(M_{1/2}\) in the NMSSM. The Higgsino mass is therefore relatively constant in the (\(m_0,M_{1/2}\))-plane what has large impact on the composition of the LSP. For small values of \(M_{1/2}=500\,\GeV\), \(|N_{13}| = 0.10\) and \(|N_{14}| = 0.04\) holds, hence, the pseudo scalar mass must be close to the resonance in order enable a sufficient annihilation. With increasing \(M_{1/2}\) the Higgsino fraction gets larger, e.g for \(M_{1/2}=1000\,\GeV\) we have \(|N_{13}| = 0.25\) and \(|N_{14}| = 0.21\). Thus, the difference to the resonance must be bigger. Otherwise, the relic density would be much too small. This explains why the distance between the two bands with correct relic density is getting larger with increasing \(M_{1/2}\) and they don't meet. Furthermore, the pseudo scalar mass in the NMSSM is not as strong correlated to \(m_0\) as it is in standard mSugra scenarios. This behavior is more similar to a constrained MSSM with free Higgs mass parameters \cite{Ellis:2002iu}. Therefore, the slope with respect to \(m_0\) is much bigger than for \(M_{1/2}\): a change of a few GeV in \(M_{1/2}\) demands a change of some hundred GeV in \(m_0\). Hence, it can be expected that missing loop corrections of the pseudo scalars have to be compensated by a large shift in \(m_0\) and \(M_{1/2}\) in order to have the same relic density. 
Fig.~\ref{fig:HF_zoom} shows zooms into the areas of Fig.~\ref{fig:HF} where the pseudo scalar mass is near the resonance. The result for the complete two-loop calculation of the pseudo scalar masses (black) as well as for tree level (red) and one-loop  (dashed blue) are shown. The difference between one- and two-loop calculation is obvious, but the bands still overlap. In contrast, the band with the tree level calculation is shifted by 20-30~GeV in comparison to the calculations including loop corrections. That corresponds to more than 10\(\sigma\) difference.

\section{Singlino dark matter}
\begin{figure}[htb]
\begin{minipage}{16cm}
	\includegraphics[scale=1.]{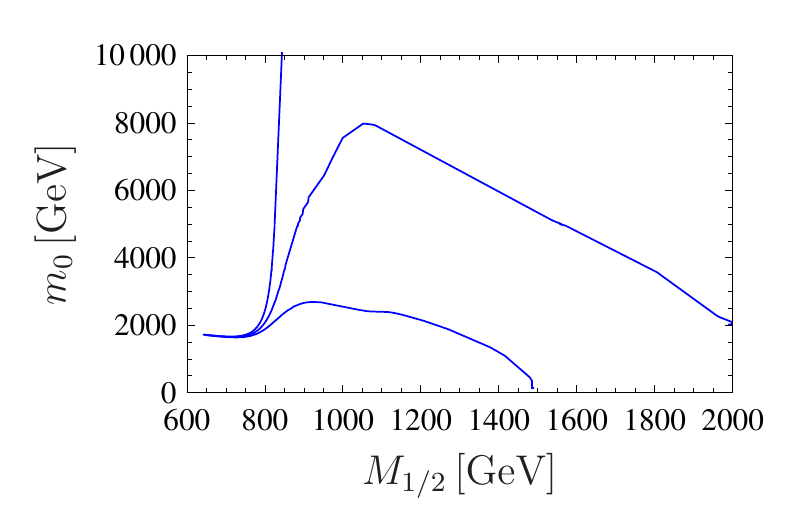}
	\hfill
	\includegraphics[scale=1.]{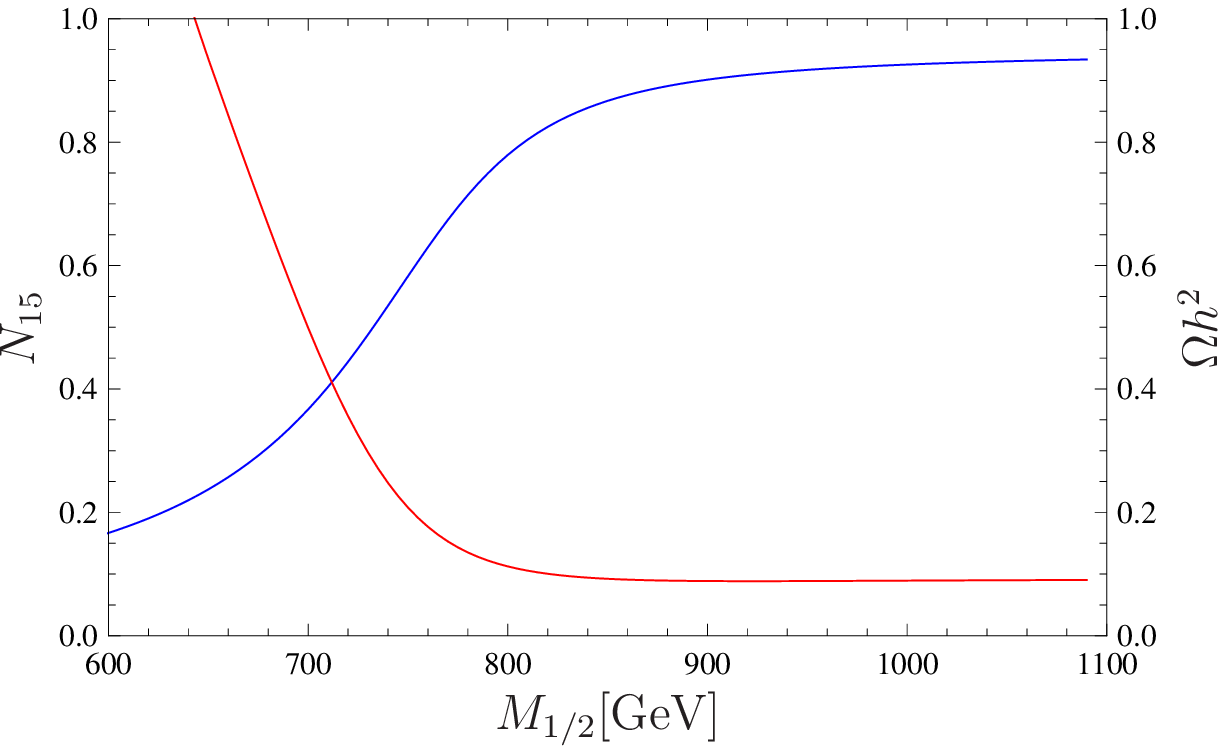}
	\caption[Singlino Dark Matter]{Singlino dark matter. The left plot shows the ($m_0,M_{1/2}$)-plane for the parameters of eq.~(\ref{par:SinglinoDM}). The lines correspond to $\Omega h^2 = 0.1018,0.1123,0.1228$. The plot on the right gives the singlino fraction $|N_{15}|$ (blue) and relic density $\Omega h^2$ (red) of the lightest neutralino for a variation of $M_{1/2}$. $m_0$ was set to 2250~GeV.}
\label{fig:Singlino_DM}
\end{minipage}
\end{figure}

\begin{figure}[htb]
\begin{minipage}{16cm}
\begin{center}
\includegraphics[scale=1.]{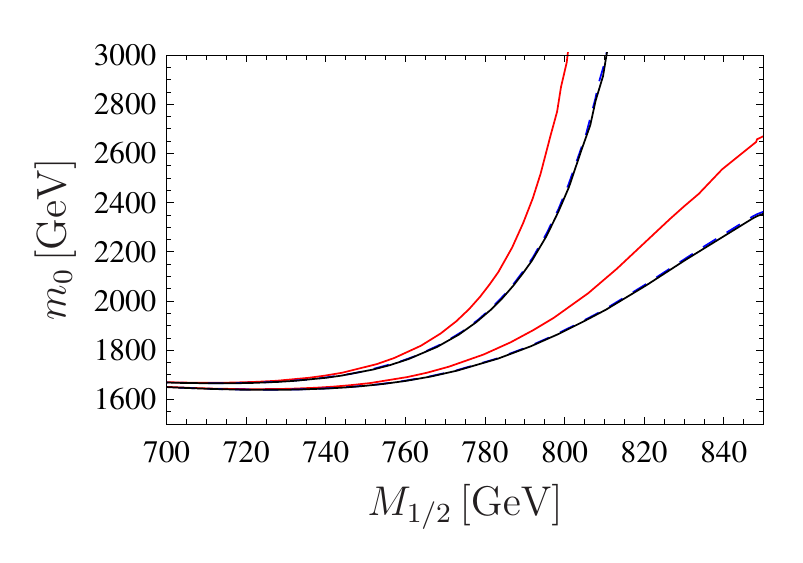}
\end{center}
\caption[Loop corrections to Higgs funnel]{The isolines corresponding to $\Omega h^2 = 0.1018$ and $\Omega h^2 = 0.1228$ in the ($m_{0}$-$M_{1/2}$) plane for a dominant annihilation due to a Higgs Funnel.  All other parameters are fixed as in eq.~(\ref{par:SinglinoDM}). Black: complete calculation of the mass spectrum. Blue dashed: pseudo scalar masses without two-loop corrections. Red: pseudo scalar masses only at tree level.}
\label{fig:Higgs_Funnel}
\end{minipage}
\end{figure}

As already discussed,  only tiny regions with the correct amount of dark matter exist in the mSugra MSSM. This is an effect of the nature of the lightest neutralino which is gaugino- or Higgsino-like. As opposed to that, the lightest neutralino can also be singlet-like in the NMSSM. The mass of a singlino LSP is quite insensitive to the GUT value of \(M_{1/2}\). Hence, it is expected that also the relic density is constant for a large range of \(M_{1/2}\). In the left plot of Fig.~\ref{fig:Singlino_DM}, the relic density is shown in the (\(m_0,M_{1/2}\))-plane. The other parameters are chosen as
\begin{eqnarray}
 &\tan\beta = 9 \thickspace, \quad \kappa^{\rm GUT} = 0.22 \thickspace, \quad \lambda^{\rm GUT} = 0.54 \thickspace, \quad A_\kappa^{\rm GUT} = 1979\,{\rm GeV} \thickspace, &\nonumber \\ 
 &A_0 = -1948\,{\rm GeV} \thickspace,\quad A_\lambda^{\rm GUT} = 4435\,{\rm GeV} \thickspace, \quad v_s = 1491\,{\rm GeV}.& 
\label{par:SinglinoDM}
\end{eqnarray}
We emphasize that the lines show really \(\Omega h^2 = 0.1018,0.1123,0.1228\). It can be seen that for large values for \(M_{1/2}\) the 3\(\sigma\) range of WMAP-7 covers the complete range of \(m_0\) from 0~GeV to more than 10~TeV. If such a scenario is realized in nature, there is not any constraint on the GUT parameters of \(m_0\) and \(M_{1/2}\) coming from relic density observations. The plot in the middle of Fig.~\ref{fig:Singlino_DM} shows the singlino component of the LSP for \(m_0 = 2250\,\GeV\) for a variation of \(M_{1/2}\). For values of \(M_{1/2}\) above 800~GeV, the singlino fraction of the lightest neutralino is more than 70~\%. This is the range at which the relic density starts to vary much less in the (\(m_0,M_{1/2}\))-plane. \\
A singlino LSP annihilates mostly through a Higgs in the propagator. The most important final states for \(M_{1/2}\simeq 800\)~GeV are \(t \bar{t}\) (53~\%) and \(h_1 A^0_1\) (32~\%). The second one is more important as the \(\tau \bar{\tau}\) and \(b \bar{b}\) channel because of the large value of \(\lambda\) and the small \(H_d\)-component of the lightest pseudo scalar. Therefore, the correct shape of the isocurves is sensitive to the loop corrections in the scalar and pseudo scalar sector. We show in Fig.~\ref{fig:Higgs_Funnel} a zoom in the (\(m_0,M_{1/2}\))-plane of Fig.~\ref{fig:Singlino_DM}. In this figure, the isocurves according to eq.~(\ref{eq:WMAP}) are shown for a calculation of the pseudo scalar masses at tree, one-loop or two-loop level. Even if the Higgs masses are not near a resonance at this example, there is a large difference between the curves at tree and one-loop level. \\

\begin{figure}[htb]
\begin{minipage}{16cm}
\begin{center}
\includegraphics[scale=1]{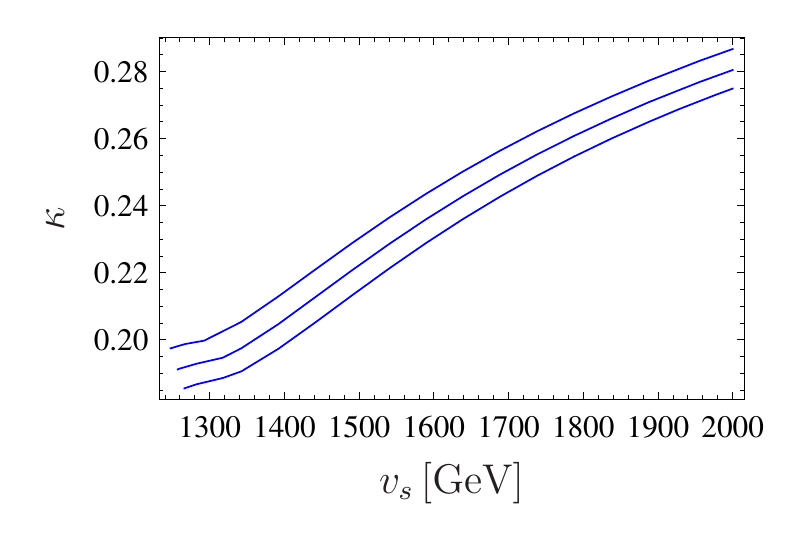}
\end{center}
\caption[($\kappa,v_s$)-plane for singlino dark matter]{ ($\kappa,v_s$)-plane for the parameters of eq.~(\ref{par:SinglinoDM}) and $m_0=2200$~GeV, $M_{1/2}=2000$~GeV. The isolines corresponding to $\Omega h^2 = 0.1018$ and $\Omega h^2 = 0.1228$.}
\label{fig:KappaVS}
\end{minipage}
\end{figure}

The relic density of a singlino-like neutralino has its largest dependence on \(\kappa\) and \(v_s\). The reason is that the product \(\frac{1}{\sqrt{2}}\kappa v_s\) defines the mass of the singlino. Furthermore, \(\kappa\) fixes together with \(\lambda\) the interactions of the singlino. The resulting isocurves in the (\(\kappa,v_s\))-plane are shown in Fig.~\ref{fig:KappaVS}. \\

\cleardoublepage
\chapter{Summary}
\label{chapter:summary}
We have seen that there are interesting dark matter scenarios in supersymmetric models beside the standard case of a neutralino LSP in the MSSM. These scenarios are often appealing because they can solve problems which still exist in the MSSM. We studied models which can explain neutrino masses via the seesaw mechanism and bilinear $R$-parity violation, respectively. In addition, we looked at the  NMSSM which solves the \(\mu\)-problem of the MSSM.\\
We have reconsidered the question of dark matter in gauge mediated SUSY breaking. Even if $R$-parity is broken, the gravitino in a standard GMSB scenario has a sufficient life time to be still a valid dark matter candidate. Assuming that the reheat temperature is large enough so that the gravitino gets into thermal equilibrium, we  find that in the simplest models it is not possible to obtain the correct amount of dark matter once all constraints are taken into account: observation of the Lyman-\(\alpha\) forest implies that a gravitino with a mass above approximately 8~keV is needed which yields too much dark matter if only the standard history of the universe is considered.  The mechanisms to produce additional entropy via messenger number violating terms, proposed so far in literature, do not work in scenarios where the messengers come in complete $SU(5)$ representations. This is a consequence of two facts which have been overlooked so far: first, the decay of the messenger particle has to occur after electroweak phase transition and thus decays into $W$ bosons have to be considered. Second, for each representation of the SM gauge group one has a lightest messenger scalar which would be stable if messenger number was a conserved quantum number. Roughly spoken, one finds two main possibilities: the $SU(2)_L$ messengers decay too fast and the $SU(3)_C$ messengers do not produce sufficient entropy to dilute the gravitinos, or the $SU(2)_L$ messengers decay sufficiently late and the $SU(3)_C$ messengers destroy the predictions for BBN. However, by sufficiently lowering the reheat temperature, one can avoid this problem once the gravitino mass is in the range of 1~MeV and above. This conclusion does also hold when we add bilinear $R$-parity violation to explain neutrino data. Although, we have mainly focused on terms of the form $f m_{3/2} \Phi_{MSSM} \bar{\Phi}_M$ in detail, one can easily extend this discussion to other scenarios. We have checked this for all cases proposed in ref.~\cite{Jedamzik:2005ir}, which contains an exhaustive list of possibilities, the same conclusions hold but for a tiny region in parameter space where a $\tilde e_R$ like messenger is the lightest messenger particle. Moreover, we have seen that scenarios with a SM gauge singlet lightest messenger, occurring for example in 16-plet of $SO(10)$ models, are still viable. \\
The seesaw mechanism is an attractive possibility to generate neutrino masses in the MSSM by an unique dimension 5 operator. We have discussed the three different seesaw types: seesaw~I is a result of a fermionic gauge singlet, seesaw~II stems from an additional scalar which transforms as triplet under \(SU(2)_L\) and carries hypercharge and seesaw~III is based on additional fermions in the adjoint representation of \(SU(2)_L\). Since additional \(SU(2)_L\) triplets alone would spoil gauge unification, we have embedded all additional fields in complete representations of \(SU(5)\). Furthermore, we used GUT scale conditions for the arising couplings and mass parameters which are invariant under \(SU(5)\) and performed an analysis of all three models using two-loop RGEs. We have seen that especially the seesaw types~II and III have a large impact on the mass spectrum of the SUSY fields. This effect increases with decreasing seesaw scale. In addition, large contributions to flavor violating processes are present  in these scenarios  in the parameter range consistent with neutrino data: branching ratios like \(\mbox{Br}(\mu \rightarrow e \gamma)\) are in general largest for type~III followed by type~II and type~I. This is a consequence of the fact that for a given set of mSugra parameters the mass spectrum in seesaw type~III is lighter than in the other scenarios. In general, the lepton flavor violating decays can be reduced by decreasing the seesaw scale or increasing the SUSY masses. However, the first possibility works hardly for type~III because a Landau pole is reached already for high values of the seesaw scale, therefore, always a heavy spectrum is preferred in type~III. The effects on the mass spectrum are also reflected by the regions consistent with the measured relic density in the (\(m_0,M_{1/2}\))-plane. While there is no visible difference between the seesaw~I and a standard mSugra scenario with the same GUT scale parameters, the regions are clearly shifted or have disappeared for seesaw~II and III: the coannihilation region prefers smaller values for \(m_0\) with decreasing seesaw scale and therefore vanishes, when the seesaw scale has crossed a distinct value. This particular scale is for seesaw~III roughly \(\Ord(10^{14})~\GeV\) and one order higher than for type~II. In contrast, the focus point region gets shifted to larger values of \(M_{1/2}\) because of the reduced SUSY masses. We have also shown that the two-loop RGEs have a significant influence on the areas with correct relic density. In particular, the Higgs funnel regions for seesaw~II and III differ extremely between one- and two-loop calculations. \\
The NMSSM is an attractive extension of the MSSM, in particular, as it solves  the $\mu$-problem of the MSSM and as it leads to new phenomenology at present and future collider experiments. It can also explain the observed amount of dark matter in the universe. However, improved theoretical predictions are necessary  for comparison of the WMAP data. We therefore have performed the complete one-loop calculation of all masses: Higgs bosons, charginos, neutralinos, squarks and sleptons. While in case of the Higgs bosons we have reproduced known results, the corrections to the other particles have not yet been discussed in the literature. We have shown that the corrections amount to the order of a few percent. While the corrections are most likely below the precision of the coming LHC data, they  are clearly important for comparison with WMAP data and also with a future international linear collider, and thus crucial for precision investigations of the NMSSM parameter space. Especially the so far unknown, radiative corrections to the stau and stop masses are important for the coannihilation regions.  Furthermore, we have shown that the properties of a singlet-like dark matter state, which is  possible in the NMSSM, can have completely different properties as gaugino- or Higgsino-like dark matter in the MSSM. In this case, it can happen that ranges over several TeV in both directions in the  \((m_0,M_{1/2})\)-plane  are consistent with WMAP-7 data. \\
Now, in the light of the first LHC results and astronomical observations coming from the satellite experiments PLANCK and PAMELA, it will most likely be necessary to test many different supersymmetric models to explain this data. We are sure that we have done with the development of the Mathematica package \SARAH a big step to simplify and speed up the way from SUSY model building to numerical results. \SARAH calculates for a given model all necessary analytical expressions for the analysis at tree and one-loop level. This contains, for instance, all vertices, all masses, the one-loop self energies and the one- and two-loop RGEs. The models can easily be changed and new models can be added in an intuitive way. Nevertheless, a large variety of models can be handled with \SARAH: there is no restriction to the number of chiral superfields, \(SU(N)\) gauge groups, number of field rotations or symmetry breakings.  \SARAH can use the derived information about a new SUSY model to write model files for \FeynArts/\FormCalc or \CalcHep/\CompHep. Furthermore, it is possible to generate source code for \SPheno. Using this source code, a fully functional version of \SPheno is generated for a new SUSY model without any need to change the source code by hand: it calculates the complete mass spectrum based on a high scale model containing only some free parameters, calculates two- and three-body decays of the SUSY particles and checks the low energy constraints. Since the model file written by \SARAH can also be used with \micrOmegas, this provides a very fast way to precise, comprehensive numerical results and phenomenology.

 \begin{appendix}
\cleardoublepage
  \chapter{Calculation of the Lagrangian of supersymmetric models}
\label{sec:SARAH_Lag}
We describe in this section the calculation of the complete Lagrangian for a supersymmetric model based on the superpotential and the gauge structure. 
\paragraph*{Interactions of chiral superfields}
If we call the superpotential for a given theory \(W\) and use \(\phi_i\) for the scalar and \(\psi_i\) for the fermionic component of a chiral supermultiplet, the matter interactions can by derived by
\begin{equation}
\label{eq:MatterInteractions}
\La_Y = - \frac{1}{2} W^{ij} \psi_i \psi_j + \mbox{h.c.} \thickspace, \hspace{1cm} \La_F = F^{* i} F_i + \mbox{h.c.}
\end{equation}
with 
\begin{equation}
\label{eq:FTerms}
W^{ij} = \frac{\delta^2}{\delta \phi_i \delta \phi_j} W \hspace{1cm} \mbox{and} \hspace{1cm} F^i = - W^{* i} = \frac{\delta W}{\delta \phi_i} \thickspace .
\end{equation}
The first term of eq. (\ref{eq:MatterInteractions}) describes  the interaction of two fermions with one scalar, while the second term forms the so called F-terms which describe  four-scalar interactions.  

\paragraph*{Interactions of vector superfields}
We name  the spin-\(\frac{1}{2}\) component of a vector supermultiplet \(\lambda\) and  the spin-1 component \(A^\mu\). The most general Lagrangian only involving these fields is
\begin{equation}
\label{eq:LagVS}
\La = - \frac{1}{4} F^a_{\mu\nu} F^{\mu\nu a} - i \lambda^{\dagger a } \bar{\sigma}^\mu D_\mu \lambda^a
\end{equation}
with the field strength
\begin{equation}
\label{eq:FieldStrength}
F_{\mu\nu}^a = \partial_\mu A_\nu^a - \partial_\nu A_\mu^a + g f^{abc} A_\mu^b A_\nu^c \thickspace ,
\end{equation}
and the covariant derivative
\begin{equation}
D_\mu \lambda^a = \partial_\mu \lambda^a + g f^{abc} A_\mu^b \lambda^c \thickspace .
\end{equation}
Here, \(f^{abc}\) is the structure constant of the gauge group. Plugging eq.~(\ref{eq:FieldStrength}) in the first term of eq.~(\ref{eq:LagVS}) leads to self-interactions of three and four gauge bosons
\begin{equation}
\La_{V} = - \frac{1}{4} (\partial_\mu A_\nu^a - \partial_\nu A_\mu^a)  g f^{abc} A^{\mu,b} A^{\nu,c}  - \frac{1}{4} g^2 (f_{abc} A_\mu^b A_\nu^c) (f^{ade} A^{\mu,e} A^{\nu,e})  \thickspace .
\end{equation}
The second term of eq.~(\ref{eq:LagVS})  describes the  interactions between vector bosons and gauginos.
\paragraph*{Supersymmetric gauge interactions}
The parts of the Lagrangian with both chiral and vector superfields are the kinetic terms for the fermions and scalars
\begin{equation}
\La_{kin} = - D^\mu \phi^{*i} D_\mu \phi_i  - i \psi^{\dagger i } \bar{\sigma}^\mu D_\mu \psi_i 
\end{equation}
as well as the interaction between a gaugino and a matter fermion and scalar
\begin{equation}
\La_{GFS} = - \sqrt{2} g (\phi^* T^a \psi) \lambda^a + \mbox{h.c.} \thickspace .
\end{equation} 
Here, \(T^a\) are the fundamental generators of the gauge group. Furthermore, the covariant derivatives are 
\begin{eqnarray}
D_\mu \phi_i &=& \partial_\mu \phi_i - ig A^a_\mu (T^a \phi)_i  \thickspace ,\\
D_\mu \phi^{*i} &=& \partial_\mu \phi^{*i} + ig A^a_\mu (\phi^* T^a)^i \thickspace , \\
D_\mu \psi_i &=& \partial_\mu \psi_i - i g A^a_\mu (T^a \psi)_i \thickspace,
\end{eqnarray}
In addition, the D-Terms are defined by
\begin{equation}
\La_D =  \frac{1}{2} D^a D^a \thickspace .
\end{equation}
The solution of the equations of motion for the auxiliary fields leads to 
\begin{equation}
\label{eq:DTerms}
D^a = - g (\phi^* T^a \phi) \thickspace .
\end{equation}
\paragraph*{Soft-breaking terms} SUSY must be a broken. This can be parametrized by adding soft-breaking terms to the Lagrangian. The possible terms are the mass terms for all scalar matter fields and gauginos
\begin{equation}
\La_{SB} = - m_{\phi_i}^2 \phi_i \phi_i^* - \frac{1}{2} M_{\lambda_i} \lambda_i \lambda_i
\end{equation}
as well as soft-breaking interaction corresponding to the superpotential terms
\begin{equation}
\La_{Soft,W} = T \phi_i \phi_j \phi_k + B \phi_i \phi_j + S \phi_i \thickspace .
\end{equation}
\paragraph*{Gauge fixing terms and ghost interactions}
\label{GaugeFixing}
The Lagrangian of a theory without further restrictions is invariant under a general gauge transformation. This invariance leads to severe problems in the quantization of the theory as can be seen in the divergence of functional integrals. Therefore, it is in necessary to add gauge fixing terms to break this gauge invariance. \\
The general form of the gauge fixing Lagrangian is
\begin{equation}
\La_{GF} = - \frac{1}{2} \sum_a|f(x)^a|^2 \thickspace .
\end{equation} 
\(f_a\) can be a function of partial derivatives of a gauge boson and a Goldstone boson. The corresponding ghost terms of the ghost fields \(\bar{\eta}\) and \(\eta\)  are
\begin{equation}
\La_{Ghost} = - \bar{\eta}_a (\delta f^a) \thickspace. 
\end{equation}
Here, \(\delta\) assigns the operator for a BRST transformation. For an unbroken gauge symmetry, the gauge fixing terms in the often chosen \(R_\xi\)-gauge  are
\begin{equation}
\La_{GF} = - \frac{1}{2 R_\xi} \sum_a \left(\partial^\mu V_\mu^a \right)^2  \thickspace.
\end{equation}
Here, \(V_\mu\) are the gauge boson of the unbroken gauge group. It is often common to choose a distinct value for \(R_\xi\). The most popular gauges are the unitary gauge \(R_\xi \rightarrow \infty\) and the Feynman-'t Hooft-gauge \(R_\xi = 1\). For broken symmetries, the gauge fixings terms are chosen in a way that the mixing terms between vector bosons and scalars disappear from the Lagrangian. Therefore, the common choice for the gauge fixing Lagrangian for theories with the standard model gauge sector after EWSB is  
\begin{equation}
\label{GFewsb}
\La_{GF, R_\xi} =  - \frac{1}{2 \xi_\gamma} \left( \partial^\mu \gamma_\mu\right)^2 - \frac{1}{2 \xi_Z} \left( \partial^\mu Z_\mu + \xi_Z M_Z G^0 \right)^2 + - \frac{1}{\xi_{W^+}} \left( \partial^\mu W^+_\mu + \xi_{W^+} M_W G^+\right)^2 \thickspace . 
\end{equation}
Here, \(G^0\) and \(G^+\) are the Goldstone bosons, which build the longitudinal component of the massive vector bosons.

\chapter{Mass eigenstates of the minimal supersymmetric standard model}
\label{chapter:MSSM}
\section*{Gauge bosons}
\label{sec:EWSB_G}
After EWSB, the photon and \(Z\) boson are mixtures of the the \(B\) and \(W_3\) boson
\begin{eqnarray}
 Z = -\sin\Theta_W B + \cos\Theta_W W_3 \thickspace,\hspace{1cm}
 A = \sin\Theta_W W_3 + \cos\Theta_W B \thickspace,
\end{eqnarray}
while the superposition of \(W_1\) and \(W_2\) form the charged eigenstates \(W^+\) and \(W^-=\left(W^+\right)^*\), with
\begin{equation}
 W^\pm = \frac{1}{\sqrt{2}} ( W_1 \mp i W_2) \thickspace.
\end{equation}
\section*{Neutralinos and charginos}
\label{sec:NeutralinoMix}
In analogy to the charged bosons, the first and second wino \(\tilde{W}_{1,2}\) mix after EWSB to charged fermions
\begin{equation}
 \tilde{W}^\pm = \frac{1}{\sqrt{2}} ( \tilde{W}_1 \mp i \tilde{W}_2 ) \thickspace.
\end{equation}
Furthermore, the charged components of the Higgsinos, \(H_u^+,H_d^-\), and the charged winos \(\tilde{W}^\pm\) form new mass eigenstates, called charginos \(\tilde{\chi}^\pm\). The mass matrix with respect to the basis \((\tilde{W}^-,H_d^-)^T\)/\((\tilde{W}^+,H_u^+)\) is
\begin{equation} 
M^{\tilde\chi^+} = \left( 
\begin{array}{cc}
M_2 &\frac{1}{\sqrt{2}} g_2 v_u \\ 
\frac{1}{\sqrt{2}} g_2 v_d  &\mu\end{array} 
\right) \thickspace .
\end{equation} 
This matrix is diagonalized by two unitary matrices \(U\) and \(V\)
\begin{equation}
U^* M^{\tilde\chi^+} V^\dagger = M^{\tilde\chi^+}_{dia} \thickspace .
\end{equation}
The neutral components of the Higgsinos, the bino and the third wino mix to four Majorana fermions, the neutralinos \(\chi^0_i\). The mass matrix in the basis \((\tilde{B}, \tilde{W}, \tilde{H}_d^0, \tilde{H}_u^0)\) reads
\begin{equation} 
M^{\tilde\chi^0} = \left( 
\begin{array}{cccc}
M_1 &0 &-\frac{1}{2} g_1 v_d  &\frac{1}{2} g_1 v_u \\ 
0 &M_2 &\frac{1}{2} g_2 v_d  &-\frac{1}{2} g_2 v_u \\ 
-\frac{1}{2} g_1 v_d  &\frac{1}{2} g_2 v_d  &0 &- \mu \\ 
\frac{1}{2} g_1 v_u  &-\frac{1}{2} g_2 v_u  &- \mu  &0\end{array} 
\right)  \thickspace .
\end{equation} 
This matrix is diagonalized by an unitary matrix \(N\)
\begin{equation}
 N^* M^{\tilde\chi^0} N^\dagger = M^{\tilde\chi^0}_{\mathrm{diag}} \thickspace .
\end{equation}

\section*{Sleptons}
The three generations of charged sleptons \(\tilde{e}_{L,i}\) and \(\tilde{e}_{R,i}\) mix to six charged eigenstates
 \(\tilde{e}_{1} \dots \tilde{e}_{6}\) with
\begin{eqnarray}
\tilde{e}_{L,i} = Z^{E,*}_{j i} \tilde{e}_j \thickspace,\hspace{1cm}
\tilde{e}_{R,i} = Z^{E,*}_{j i+3} \tilde{e}_j \thickspace .
\end{eqnarray}
\(Z^E\) diagonalizes the corresponding mass matrix
\begin{equation}
\label{eq:DiaSf}
Z^E m^{2,\tilde l}_{T} Z^{E \dagger} = m^{2,\tilde l}_{\mathrm{diag}} \thickspace .
\end{equation}
In the basis \( \left(\tilde{e}_{L,i}, \tilde{e}_{R,i}\right) \), the mass matrix of the charged sleptons at the tree level is given by 
\begin{equation} 
m^{2,\tilde l} = 
\left( 
\begin{array}{cc}
m^2_{LL} &-\frac{1}{2} v_u \mu^* Y_e^T  + \frac{1}{\sqrt{2}} v_d T_e^T \\ 
-\frac{1}{2}  v_u \mu Y_e^* + \frac{1}{\sqrt{2}} v_d T_e^* 
 &m^2_{RR}\end{array} 
\right) 
\end{equation} 
with the diagonal entries
\begin{eqnarray} 
m^2_{LL} &=& m^2_{\tilde l} + \frac{v^2_d}{2} (Y_{e})^* (Y_{e})^T
 + \frac{1}{8}
 \Big( g_1^2- g_2^2 \Big)\Big(v_d^2 - v_u^2 \Big) {\bf 1}_3 \thickspace, \\ 
m^2_{RR} &=& m^2_{\tilde e} + \frac{v^2_d}{2} (Y_{e})^T (Y_{e})^*
 + \frac{g_1^2}{4}  \Big(v_u^2- v_d^2  \Big) {\bf 1}_3 \thickspace.
\end{eqnarray}
where ${\bf 1}_3$ is the $3\times 3$ unit matrix.

\section*{Squarks}
\label{mixingDS}
The three generations of up-type squarks \(\tilde{u}_{L,i}\) and \(\tilde{u}_{R,i}\) mix to
\(\tilde{u}_{i}\)
\begin{eqnarray}
\tilde{u}_{L,i}= Z^{U,*}_{j i} \tilde{u}_{j} \thickspace,\hspace{1cm}
\tilde{u}_{R,i} = Z^{U,*}_{j i+3} \tilde{u}_{j} \thickspace ,
\end{eqnarray}
while the down-type squarks  \(\tilde{d}_{L,i}\) and \(\tilde{d}_{R,i}\) rotate  to 
 \(\tilde{d}_{i}\)
\begin{eqnarray}
\tilde{d}_{L,i} = Z^{D,*}_{j i} \tilde{d}_{j} \thickspace,\hspace{1cm}
\tilde{d}_{R,i} = Z^{D,*}_{j i+3} \tilde{d}_{j} \thickspace.
\end{eqnarray}
The mass matrix for the down-type squarks  in the basis \( \left(\tilde{d}_{L,i}, \tilde{d}_{R,i}\right) \) is given by
\begin{equation} 
m^{2,\tilde d} = \left( 
\begin{array}{cc}
m^{2,d}_{LL} &
\frac{1}{2}  \big(\sqrt{2} v_d T^{T}_d  -  v_u \mu^* Y^{T}_d \big)\\ 
\frac{1}{2} \big(\sqrt{2} v_d T^*_{d} - v_u \mu Y^{*}_d \big) 
& m^{2,d}_{RR} \end{array} 
\right) \thickspace,
\end{equation} 
where the diagonal entries read 
\begin{eqnarray} 
\nonumber
m^{2,d}_{LL} &=&  m^2_{\tilde q}  + \frac{v_d^2}{2}  Y^*_d Y^T_d
 - \frac{3 g_2^{2} + g_1^{2}}{24} \big(v_d^{2} - v_u^{2}\big)  {\bf 1}_3 \thickspace,\\ 
m^{2,d}_{RR} &=& m^2_{\tilde d} +  \frac{v_d^2}{2}  Y^T_d Y^*_d 
 + \frac{g_1^{2}}{12} \Big( v_u^{2} - v_d^{2}\Big)  {\bf 1}_3 \thickspace.
\end{eqnarray} 
The corresponding expressions for the up-type squarks in the basis \( \left(\tilde{u}_{L,i}, \tilde{u}_{R,i}\right) \) are
\begin{equation} 
m^{2,\tilde u} = \left( 
\begin{array}{cc}
m^{2,u}_{LL}  &\frac{1}{2} \big(\sqrt{2} v_u T^{T}_u  - v_d  \mu^* Y^{T}_u \big)\\ 
\frac{1}{2} \big(\sqrt{2} v_u T^{*}_u - v_d \mu Y^{*}_u \big)
& m^{2,u}_{RR}  \end{array} 
\right) 
\end{equation} 
with
\begin{eqnarray} 
\nonumber
m^{2,u}_{LL}  &=&  m^2_{\tilde q}  + \frac{v_u^2}{2}  Y^*_u Y^T_u
 + \frac{3 g_2^{2} - g_1^{2}}{24} \big(v_d^{2} - v_u^{2}\big){\bf 1}_3 \thickspace, \\ 
 m^{2,u}_{RR} &=&  m^2_{\tilde u} +  \frac{v_u^2}{2}  Y^T_u Y^*_u 
 + \frac{g_1^{2}}{6} \Big(v_d^{2}- v_u^{2} \Big)  {\bf 1}_3 \thickspace .
\end{eqnarray} 
\(m^{2,\tilde{u}}\) and \(m^{2,\tilde{d}}\) are diagonalized in the same way as the charged sleptons in eq.~(\ref{eq:DiaSf}) by the matrices \(Z^U\) and \(Z^D\), respectively.
%
\section*{Sneutrinos}
The three sneutrinos mix to the mass eigenstates \(\tilde{\nu}\)
\begin{equation}
 \tilde{\nu}_{L,i} = Z^{\nu,*}_{j i} \tilde{\nu}_{i} \thickspace ,
\end{equation}
and the mass matrix is
\begin{equation}
M_{\tilde{\nu}}^2 = \frac{e^2 (v_d^2 - v_u^2)}{8 s_w^2 c_w^2} {\bf 1}_3 + m_{\tilde l}^2 \thickspace.
\end{equation}
This matrix is diagonalized by the unitary matrix \(Z^{\nu}\).

\section*{Charged leptons, neutrinos and quarks}
If the Yukawa couplings are not diagonal, also the SM leptons and quarks rotate. The lepton mass matrix is diagonalized by the two matrices \(Z^{E,L}\) and \(Z^{E,R}\), while in the quark sector the matrices \(Z^{U,L}\), \(Z^{U,R}\), \(Z^{D,L}\) and \(Z^{D,R}\) are used:
\begin{equation}
Z^{x,R,*} M^X Z^{x,L,\dagger} = M^X_{dia}, \hspace{1cm} X = E, D, L \thickspace .
\end{equation}
The mass matrices are given by the product of Yukawa couplings and VEVs
\begin{eqnarray}
(M^E)_{ij} = \frac{v_d}{\sqrt{2}} Y_{e,ij} \thickspace,\hspace{1cm}
(M^D)_{ij} = \frac{v_d}{\sqrt{2}} Y_{d,ij} \thickspace,\hspace{1cm}
(M^U)_{ij} = \frac{v_u}{\sqrt{2}} Y_{u,ij} \thickspace .
\end{eqnarray}
The symmetric neutrino mass matrix is diagonalized by the unitary matrix \(Z^V\)  
\begin{equation}
 m_{\nu_i} = ((Z^V)^T m_\nu Z^V)_{ii} \thickspace .
\end{equation}
The CKM and PMNS matrix are defined as
\begin{equation}
 (Z^{U,L})^\dagger Z^{D,L} = V_{\rm CKM} \thickspace, \hspace{1cm}   (Z^{E,L})^\dagger Z^V   =  V_{\rm PMNS} \thickspace .
\end{equation}

\section*{Higgs}
The Higgs fields are parameterized after EWSB by
\begin{eqnarray}
\label{eq:Higgs1}
   H_d^0 = \frac{1}{\sqrt{2}}( v_d + i \sigma_d + \phi_d) \thickspace,\hspace{1cm}
\label{eq:Higgs2}
   H_u^0 =  \frac{1}{\sqrt{2}}(v_u + i \sigma_u +  \phi_u) \thickspace.
\end{eqnarray}
Here \(v_u\) and \(v_d\) are the VEVs of the Higgs fields and they are connected to the mass of the \(W\) boson and $Z$ boson via
\begin{eqnarray}
M_W = \frac{e}{2 \sin\Theta_W} \sqrt{v_d^2 + v_u^2} \thickspace,\hspace{1cm}
M_Z = \frac{e}{2 \sin\Theta_W \cos\Theta_W} \sqrt{v_d^2 + v_u^2} \thickspace .
\end{eqnarray}
\(\Theta_W\) is the Weinberg angle. The ratio of the VEVs defines the mixing angle \(\beta\)
 \begin{equation}
  \tan\beta = \frac{v_d}{v_u} \thickspace.
 \end{equation}
\(\phi_d\) and \(\phi_u\) are the scalar components, while \(\sigma_d\) and \(\sigma_u\) are the pseudo scalar ones. The different mass eigenstates are 
\begin{itemize}
 \item {\bf neutral, scalar Higgs \(h\)}: The mass matrix in the basis \((\phi_d, \phi_u)\) reads
\begin{equation} 
m^{2,h} = \left( 
\begin{array}{cc}
 m_{H_d}^2  +  |\mu|^2  + \frac{1}{8}\Big({g_1}^{2} + {g_2}^{2}\Big)\Big(3 {v_d}^{2}  - {v_u}^{2} \Big) & - \mathrm{Re}\big\{ B_{\mu}\big\}  - \frac{1}{4} \Big({g_1}^{2} + {g_2}^{2}\Big)v_d v_u \\ 
- \mathrm{Re}\big\{B_{\mu}\big\}   - \frac{1}{4} \Big({g_1}^{2} + {g_2}^{2}\Big)v_d v_u & m_{H_u}^2  +  |\mu|^2  - \frac{1}{8} \Big({g_1}^{2} + {g_2}^{2}\Big)\Big({v_d}^{2}-3 {v_u}^{2} \Big)\end{array} 
\right) 
\end{equation} 
with \(Z^H m^{2,h} Z^{H,T} = m^{2,h}_{\mathrm{diag}} \).
\item {\bf charged, scalar Higgs \(H^+\)}: The mass matrix in the basis \((H_d^-, (H_u^+)^*)\)  is
\begin{equation} 
m^{2,H^+} = \left( 
\begin{array}{cc}
 m_{H_d}^2  + |\mu|^2  + \frac{1}{8} \Big({g_1}^{2} + g_2^2 \Big)\Big( {v_d}^{2}  - {v_u}^{2} \Big) &\frac{1}{4} {g_2}^{2} v_d v_u  + B_{\mu}\\ 
\frac{1}{4} {g_2}^{2} v_d v_u  + B_{\mu}{}^* &  m_{H_u}^2  + |\mu|^2   - \frac{1}{8} \Big({g_1}^{2} + g_2^2 \Big)\Big( {v_d}^{2}  - {v_u}^{2} \Big) \Big)\end{array} 
\right) 
\end{equation} 
with \(Z^+ m^{2,H^+} Z^{+,\dagger} = m^{2,H^+}_{\mathrm{diag}} \). The mass matrix has rank 1, i.e. one eigenvalue is 0. This eigenvalue \(G^\pm\) is the Goldstone boson belonging to \(W^\pm\).  
\item {\bf CP-odd Higgs \(A^0\)}: The mass matrix in the basis \((\sigma_d, \sigma_u)\)  is
\begin{equation} 
m^{2,A^0} = \left( 
\begin{array}{cc}
 m_{H_d}^2  + |\mu|^2  + \frac{1}{8}  \Big({g_1}^{2} + {g_2}^{2}\Big)\Big( {v_d}^{2}- {v_u}^{2}\Big) & \mathrm{Re}\big\{B_{\mu}\big\}\\ 
 \mathrm{Re}\big\{B_{\mu}\big\} & m_{H_u}^2  + |\mu|^2  - \frac{1}{8}  \Big({g_1}^{2} + {g_2}^{2}\Big)\Big({v_d}^{2}- {v_u}^{2} \Big)\end{array} 
\right) 
\end{equation} 
with \(Z^H m^{2,A^0} Z^{H,T} = m^{2,A^0}_{\mathrm{diag}}\). The first eigenvalue is massless and  build the longitudinal component of the \(Z\) boson. 

\end{itemize}

\section*{Gluons and gluinos}
The gluons and gluinos don't mix. The mass of the gluino is
\begin{equation}
 m_{\tilde{g}} = |M_3| \thickspace ,
\end{equation}
and the gluons remain massless.

  \chapter{Basics of group theory}
\label{chapter:group}
We  give in this chapter a very brief introduction to the basics of group and representation theory. For more information about this topic and its application in particle physics, see for example \cite{Georgi:1982jb} or \cite{Slansky:1981yr} and references therein.

\section{Roots and weights}
\paragraph*{Cartan subalgebra}: If \(T^i\) are the generators of a Lie group, the Cartan subalgebra is defined as the largest subset of commuting, hermitian generators \(H_i\):
\begin{equation}
 H_i = H_i^\dagger, \hspace{1cm} [H_i,H_j] = 0, \hspace{1cm} \{H_i\} \subseteq \{T_i\} \thickspace .
\end{equation}
\(E_i\) are the generators \(T_i\) which do not belong to the Cartan subalgebra. Cartan generators can be diagonalized simultaneously and they have the following properties:
\begin{enumerate}
 \item The action on a state \(|\mu,r\ket\) of some representation \(r\) is
\begin{equation}
 H_i |\mu, r\ket = \mu_i |\mu,r\ket \thickspace .
\end{equation}
The eigenvalues \(\mu_i\) are called 'weights'. 
\item The action on the adjoint representation is
\begin{equation}
 H_i |H_j\ket = 0, \hspace{1cm} H_i |E_\alpha\ket = \alpha_i |E_\alpha\ket \thickspace .
\end{equation}
The eigenvalues \(\alpha_i\) are called 'roots'.
\end{enumerate}
We define
\begin{enumerate}
 \item {\bf Positive roots and weights}: Roots and weights are called 'positive', if the sign of the first, non-zero component is positive.
 \item {\bf Dual basis}: The basis formed by the positive roots is called 'Dual Basis'.
\item {\bf Simple roots}:  'Simple roots' are positive roots which can't be written as sum of other roots. 
\end{enumerate}
We can  now introduce an ordering of two weights \(\mu\) and \(\nu\):
\begin{equation}
\mu > \nu \hspace{1cm} \mbox{if} \hspace{1cm} \mu - \nu \mbox{is positive}
\end{equation}
Using this definition, it is possible to assign the highest weight of an irreducible representation.

\paragraph*{Dynkin labels}: The Dynkin coefficients \(l^j\) for the irreducible representation with highest weight \(\mu\) are
\begin{equation}
 l^j = \frac{ 2 \alpha^j \mu}{(\alpha^j)^2} \thickspace ,
\end{equation}
where the simple roots \(\alpha^j\) are used. The 'fundamental weights' \(\mu^k\) are defined as
\begin{equation}
 \frac{2 \alpha^j \mu^k}{(\alpha^j)^2} = \delta_{j k} \thickspace .
\end{equation}
Using these definitions, every weight can be written as 
\begin{equation}
 \mu = \sum_{j=1}^m l^j \mu^j \thickspace .
\end{equation}
The vector \([l^j]\) defines the representation in the so called 'Dynkin basis' in an unique way. 
\paragraph*{Weyl formula}: We have now all needed definitions to present formulas for the calculation of the dimension \(N(r)\) and the quadratic Casimir \(C_2(r)\) of any irreducible representation \(r\) with highest weight \(\mu\). The dimension of an irreducible representation with highest weight \(\mu\) is
\begin{equation}
 N(\mu) = \Pi_\alpha \frac{\bra \mu + \delta, \alpha \ket}{\bra \delta, \alpha \ket} \thickspace .
\end{equation}
The product runs over all positive roots \(\alpha\), and \(\delta\) is the Weyl vector defined as half the sum of all positive roots
\begin{equation}
 \delta = \frac{1}{2} \sum_\alpha \alpha \thickspace . 
\end{equation}
The quadratic Casimir can be calculated by the Weyl formula
\begin{equation}
 C_2(\mu) = \bra \mu, \mu +  \delta \ket \thickspace . 
\end{equation}
\paragraph*{Realization for $SU(N)$}: For any \(SU(N)\), the positive roots and fundamental weights can be written as
\begin{itemize}
 \item positive roots
\begin{equation}
 \alpha^i = \nu^i - \nu^{i+1} \thickspace , \hspace{1cm} i=1,\dots N
\end{equation}
\item fundamental weights
\begin{equation}
 \mu^j = \sum_{k=1}^j \nu^k \quad \mbox{with} \quad  [\nu^j]_m = \frac{1}{\sqrt{2 m (m+1)}} \left( \sum_{k=1}^m \delta_{j k} - m \delta_{j,m+1} \right) \thickspace .
\end{equation}
\end{itemize}
The Dynkin labels for any irreducible representation of \(SU(N)\) can be extracted from the corresponding Young Tableaux. This we will show now. 
\section{Young Tableaux}
\label{chapter:young}
The transformation properties of a field under \(SU(N) \otimes \dots \otimes SU(N)\) are given by
\begin{equation}
 \tilde{\Psi}^{i_1 \dots i_n}_{j_1 \dots j_m} = U^{i_1}_{a_1} \dots U^{i_n}_{a_n} U^{b_1,*}_{j_1} \dots U^{b_m,*}_{j_m} \Psi^{a_1 \dots a_n}_{b_1 \dots b_m} \thickspace .
\end{equation}
\(U\) are the fundamental generators of \(SU(N)\) and \(U^*\) are the conjugated generators. A handy tool for representation theory in \(SU(N)\) groups are the so called Young Tableauxs.  The basic idea of Young Tableauxs is to draw for each index of an irreducible representation a \parbox{0.5cm}{\yng(1)}:
\begin{equation}
\parbox{0.5cm}{\yng(1)}: \hspace{1.5cm}  \Psi^j
\end{equation}
While two boxes below each other correspond to symmetrized indices, 
\begin{equation}
\parbox{0.5cm}{\yng(1,1)}: \hspace{1.5cm} \Psi^{i j} = \frac{1}{\sqrt{2}} \left( u^i v^j + u^j v^i \right) \thickspace,
\end{equation}
boxes besides each other define antisymmetrized indices 
\begin{equation}
\parbox{1.cm}{\yng(2)}: \hspace{1.cm}  \Psi^i_j = \frac{1}{\sqrt{2}} \left( u^i v^j - u^j v^i \right) \thickspace.
\end{equation}
The combination of several boxes is called tableaux. A tableaux is called 'regular', if the number of boxed per row doesn't increase from top to down.\\
The dimension of a tableaux can be either counted by the number of possible boxer per topology, or by using the hook formula
\begin{equation}
\label{eq:hook}
D = \Pi_i \frac{N + d_i}{h_i} \thickspace .
\end{equation}
Here, \(h_i\) is the so called hook of the \(i.\) box: it's the number of all boxes below and right of the considered box plus 1. \(N\) is the dimension of the gauge group and \(d\) is the distance to the upper left corner: going right counts +1 and going down -1. Let us clarify this by two examples of \(SU(3)\):
\begin{enumerate}
 \item The fundamental representation consists just of one box, i.e. \(d=0\) and \(h=1\). Hence, the dimension is equal to the dimension of the gauge group, \(D=3\). 
 \item The adjoint representation for \(N\) is given by a tableaux with \(N-1\) rows in the first columns and one row in the second column. We will check that by applying the hook formula:
\begin{center}
\begin{tabular}{l c l c}
Distances: & \parbox{1cm}{\young(01,\minusOne)} & \hspace{1cm} Hooks: & \parbox{1cm}{\young(31,1)}
\end{tabular}
\end{center}
This leads to \( D = \frac{3}{3} \frac{4}{1} \frac{2}{1} = 8\), exactly as expected.
\end{enumerate}

\paragraph*{Quantum numbers of subgroups} The decomposition of a GUT scale multiplet transforming under \(SU(N+M)\) into the representation corresponding to \(SU(N) \times SU(M)\) can be done in the following way:
\begin{enumerate}
 \item Draw the Young Tableaux for the representation with respect to the unbroken gauge group
 \item Decompose the tableaux into all possible combinations with respect to the product group.  
\end{enumerate}
Let us demonstrate this at the example of 
\begin{equation}
SU(N+M) \rightarrow SU(N) \times SU(M) \times U(1) \thickspace .
\end{equation}
\(SU(N)\) acts on the first \(N\) indices and \(SU(M)\) acts on last \(M\) indices. \(SU(N) \times SU(M)\) commutes with \(U(1)\), which is \(M\) on the first \(N\) indices and \(-N\) on the last \(M\) indices. We can therefore define an \(U(1)\) charge \(Y\) for all multiplets by
\begin{equation}
 Y = N_N M - N_M N
\end{equation}
\(N_N\) and \(N_M\) are the numbers of boxes corresponding to the different gauge groups. This leads for the fundamental fundamental representation to
\begin{equation}
\Yvcentermath1
\yng(1)  \hspace{1.5cm}  \rightarrow \hspace{1.5cm} \Big(\thickspace \yng(1) \hspace{0.3cm} \bullet \thickspace \Big)_M \hspace{0.5cm} \oplus \hspace{0.5cm} \Big( \thickspace \bullet \hspace{0.3cm} \yng(1) \thickspace \Big)_{-N}
\end{equation}
The $\bullet$ assigns a singlet. This means, the fundamental representation of \(SU(N+M)\) decomposes into two multiplets with the quantum numbers \((N,1)_M\) and \((1,M)_{-N}\) with respect to \(SU(N) \times SU(M) \times U(1)\). If we use this result and set \(N+M=5\), \(N=3\) and \(M=2\), we have the decomposition of a {\bf 5} of \(SU(5)\) in SM gauge groups. The quantum numbers are 
\begin{equation}
 {\bf 5} \rightarrow  \left(3,1\right)_{2} \oplus\left(1,2\right)_{-3}\thickspace,
\end{equation}
i.e. up to a normalization of the \(U(1)\)-charge exactly the quantum numbers of \(d^c\) and \(l\). We can now do the same with the {\bf 10} of \(SU(5)\) and end up with:
\begin{equation}
\Yvcentermath1
\yng(1,1) \hspace{1.5cm}  \rightarrow \hspace{1.5cm}  \left(\thickspace \yng(1,1) \hspace{0.3cm} \bullet \thickspace \right)_4 \hspace{0.3cm} \oplus \hspace{0.3cm}  \left(\thickspace  \bullet \hspace{0.3cm} \yng(1,1) \thickspace \right)_{-6} \hspace{0.3cm} \oplus \hspace{0.3cm} \Big(\thickspace \yng(1) \hspace{0.3cm} \yng(1) \thickspace  \Big)_{1} \hspace{1cm} .
\end{equation}
We can identify the boxes at first position belonging to \(SU(3)\) as
\begin{center}
 {\bf 3}: \parbox{0.5cm}{\yng(1)} \thickspace, \hspace{1cm}
 \({\bf \bar{3}}\): \parbox{0.5cm}{ \yng(1,1)} \thickspace ,
\end{center}
and the boxes at second position belonging \(SU(2)\) as
\begin{center}
 {\bf 2}: \parbox{0.5cm}{ \yng(1)} \thickspace, \hspace{1cm}
 {\bf 1}:  \parbox{0.5cm}{ \yng(1,1) } \thickspace .
\end{center}
This reproduces the quantum numbers for the other SM fermions \(u^c\), \(e^c\) and \(q\)
\begin{equation}
 {\bf 10} \rightarrow \left(\bar{3},1\right)_4 \oplus \left(1,1\right)_{-6} \oplus \left(3,2\right)_1 \thickspace .
\end{equation}

  \chapter[Using \SARAH]{Examples for the work with \SARAH}
\label{section:appendix_SARAH}
\section{Defining models in \SARAH: The MSSM}
\label{sec:SARAH_Modelfiles}
\subsection{The model file}
We begin this section about details of the use of \SARAH with a discussion of the different parts of the model file for the MSSM. 
\begin{enumerate}
\item The gauge sector is \(U(1)_Y\times SU(2)_L\times SU(3)_C\) and is just defined by adding the corresponding vector superfields. 
\begin{verbatim}
Gauge[[1]]={B,   U[1], hypercharge, g1, False};
Gauge[[2]]={WB, SU[2], left,        g2, True};
Gauge[[3]]={G,  SU[3], color,       g3, False};
\end{verbatim}
\item The doublet superfields are \(\hat{q}\), \(\hat{l}\), \(\hat{H}_d\) and \(\hat{H}_u\) are added by 
\begin{verbatim}
Fields[[1]] = {{uL,  dL},  3, q,   1/6, 2, 3};  
Fields[[2]] = {{vL,  eL},  3, l,  -1/2, 2, 1};
Fields[[3]] = {{Hd0, Hdm}, 1, Hd, -1/2, 2, 1};
Fields[[4]] = {{Hup, Hu0}, 1, Hu,  1/2, 2, 1};
\end{verbatim}
The different parts are: the name of the up- and down-component, the number of generations, the name of the superfield and the transformation properties under the gauge groups. 
\item The singlet superfields  \(\hat{d}\), \(\hat{u}\) and \(\hat{e}\) are added by
\begin{verbatim}
Fields[[5]] = {conj[dR], 3, d,  1/3, 1, -3};
Fields[[6]] = {conj[uR], 3, u, -2/3, 1, -3};
Fields[[7]] = {conj[eR], 3, e,    1, 1,  1};
\end{verbatim}
The definition is analog to the definition of the doublets. 
\item The  superpotential of the MSSM is
\begin{equation}
\label{superpotential_MSSM}
W =  \hat{q} Y_u \hat{u} \hat{H}_u -  \hat{q} Y_d \hat{d} \hat{H}_d  - \hat{l} Y_e \hat{e} \hat{H}_d  +\mu \hat{H}_u \hat{H}_d
\end{equation}
and given in \SARAH by
\begin{verbatim}
SuperPotential = { {{1, Yu},{u,q,Hu}}, {{-1,Yd},{d,q,Hd}},
                   {{-1,Ye},{e,l,Hd}}, {{1,\[Mu]},{Hu,Hd}}  };
\end{verbatim}
\item There are two different sets of eigenstates: the gauge eigenstates before EWSB and the mass eigenstates after EWSB. The internal names are
\begin{verbatim}
NameOfStates={GaugeES, EWSB};
\end{verbatim}
\item The gauge fixing terms for the unbroken gauge groups are
\begin{verbatim}
DEFINITION[GaugeES][GaugeFixing]=
  { {Der[VWB],  -1/(2 RXi[W])},
    {Der[VG],   -1/(2 RXi[G]) }};	
\end{verbatim}
This has the same meaning as
\begin{equation}
\La_{GF} = -\frac{1}{2 \xi_W}|\partial_\mu W^{\mu,i}|^2  -\frac{1}{2 \xi_g}|\partial_\mu g^{\mu,i}|^2
\end{equation}
\item The vector bosons and gauginos rotate after EWSB as follows
\begin{verbatim}
DEFINITION[EWSB][GaugeSector]= 
{ {VWB, {1,{VWm,             1/Sqrt[2]}, {conj[VWm],            1/Sqrt[2]}},
        {2,{VWm,-\[ImaginaryI]/Sqrt[2]}, {conj[VWm],\[ImaginaryI]/Sqrt[2]}},
        {3,{VP,            Sin[ThetaW]}, {VZ,                Cos[ThetaW]}}},
  {VB,  {1,{VP,            Cos[ThetaW]}, {VZ,               -Sin[ThetaW]}}},
  {fWB, {1,{fWm,             1/Sqrt[2]}, {fWp,                  1/Sqrt[2]}}, 
        {2,{fWm,-\[ImaginaryI]/Sqrt[2]}, {fWp,      \[ImaginaryI]/Sqrt[2]}},
        {3,{fW0,                    1}}}                                                         
      };   
\end{verbatim}
This is the common mixing of vector bosons and gauginos after EWSB, see app.~\ref{chapter:MSSM}.
\item  The neutral components of the scalar Higgs receive VEVs \(v_d\)/\(v_u\) and split in scalar and pseudo scalar components according to eqs.~(\ref{eq:Higgs1}) and (\ref{eq:Higgs2}). This is added to \SARAH by 
\begin{verbatim}
DEFINITION[EWSB][VEVs]= 
{{SHd0, {vd, 1/Sqrt[2]}, {sigmad, \[ImaginaryI]/Sqrt[2]},{phid, 1/Sqrt[2]}},
 {SHu0, {vu, 1/Sqrt[2]}, {sigmau, \[ImaginaryI]/Sqrt[2]},{phiu, 1/Sqrt[2]}}};
\end{verbatim}
\item The particles mix after EWSB to new mass eigenstates
\begin{verbatim}
DEFINITION[EWSB][MatterSector]= 
{{{SdL, SdR           }, {Sd, ZD}},
 {{SuL, SuR           }, {Su, ZU}},
 {{SeL, SeR           }, {Se, ZE}},
 {{SvL                }, {Sv, ZV}},
 {{phid, phiu         }, {hh, ZH}},
 {{sigmad, sigmau     }, {Ah, ZA}},
 {{SHdm, conj[SHup]   }, {Hpm,ZP}},
 {{fB, fW0, FHd0, FHu0}, {L0, ZN}}, 
 {{{fWm, FHdm}, {fWp, FHup}}, {{Lm,Um},  {Lp,Up}}},
 {{{FeL},       {conj[FeR]}}, {{FEL,ZEL},{FER,ZER}}},
 {{{FdL},       {conj[FdR]}}, {{FDL,ZDL},{FDR,ZDR}}},
 {{{FuL},       {conj[FuR]}}, {{FUL,ZUL},{FUR,ZUR}}} }; 
\end{verbatim}
This defines the mixings to the mass eigenstates described in app.~\ref{chapter:MSSM}. 
\item The new gauge fixing terms according to eq.~(\ref{GFewsb}) are
\begin{verbatim}
DEFINITION[EWSB][GaugeFixing]= 
{{Der[VP],                                            - 1/(2 RXi[P])},	
 {Der[VWm]+\[ImaginaryI] Mass[VWm] RXi[W] Hpm[{1}],   - 1/(RXi[W])},
 {Der[VZ] + Mass[VZ] RXi[Z] Ah[{1}],                  - 1/(2 RXi[Z])},
 {Der[VG],                                            - 1/(2 RXi[G])}};
\end{verbatim}
Because of this definition, \(A^0_1\) and \(H^\pm_1\) are recognized in all calculations as Goldstone bosons.
\item No particles should be integrated out or deleted
\begin{verbatim}
IntegrateOut={};
DeleteParticles={};
\end{verbatim}
\item The Dirac spinors for the mass eigenstates are the following
\begin{verbatim}
dirac[[1]] = {Fd,  FdL, FdR};
dirac[[2]] = {Fe,  FeL, FeR};
dirac[[3]] = {Fu,  FuL, FuR};
dirac[[4]] = {Fv,  FvL, 0};
dirac[[5]] = {Chi, L0, conj[L0]};
dirac[[6]] = {Cha, Lm, conj[Lp]};
dirac[[7]] = {Glu, fG, conj[fG]};
\end{verbatim}
\end{enumerate}
\subsection{Parameter and particle files}
\paragraph*{Parameter file}
Additional properties and information about the parameters and particles of a model are saved in the files \verb"parameters.m" and \verb"particles.m". An entry in the parameter file looks like
\begin{verbatim}
{Yu, { LaTeX -> "Y^u",
       Real -> True,
       Form -> Diagonal,
       Dependence ->  None, 
       Value -> None, 
       LesHouches -> Yu
             }}
\end{verbatim}
and contains information about the numerical value ({\tt Value} \(\rightarrow\) number), the position in a LesHouches accord file ({\tt LesHouches} \(\rightarrow\) position) or the dependence on other parameters \linebreak ({\tt Dependence} \(\rightarrow\) equation). Also simplifying assumptions can be made: it can be defined that parameters contain only real entries ({\tt Real \(\rightarrow\) True}) or that the parameter is diagonal ({\tt Form \(\rightarrow\) Diagonal}). Also a \LaTeX{} name can be given ({\tt LaTeX} \(\rightarrow\) name). Furthermore,   the GUT normalization can be assigned ({\tt GUTnormalization} \(\rightarrow\) value) for the gauge couplings of an  \(U(1)\) gauge groups. 
\paragraph*{Particle file}
\label{sec:ParametersFile}
The particles file contains entries like 
\begin{verbatim}
{Su ,  {  RParity -> -1,
          PDG ->  {1000002,2000002,1000004,2000004,1000006,2000006},
          Width -> Automatic,
          Mass -> Automatic,
          FeynArtsNr -> 13,   
          LaTeX -> "\\tilde{u}",
          OutputName -> "um" }},   
\end{verbatim}
and defines properties of all particles such as the $R$-parity ({\tt RParity} \(\rightarrow\) number) or the mass ({\tt Mass} \(\rightarrow\) value or {\tt Automatic}). {\tt Automatic} means that for the output for \FeynArts or \CalcHep not a fixed numerical value is used, but that the masses are calculated using tree level relations. In addition, the PDG code is given ({\tt PDG} \(\rightarrow\) number), the number for the particle class used in the \FeynArts model file can be fixed ({\tt FeynArts} \(\rightarrow\) number) and the name in \LaTeX{} form is given ({\tt LaTeX} \(\rightarrow\) name). If a \CalcHep or \CompHep model file should be written, it is also helpful to define an appropriate name in this context ({\tt OutputName} \(\rightarrow\) name). 
\paragraph*{Global definitions}
It is also possible to define global properties for parameters or particles which are present in more than one model file. These properties are afterwards used for all models. The global definitions are saved in the files \verb"particles.m" and \verb"parametes.m" directly in the main model directory. For each parameter or particle, an entry like 
\begin{verbatim}
{{        Descriptions -> "Down Squark", 
          RParity -> -1,
          PDG ->  {1000002,2000002,1000004,2000004,1000006,2000006},
          Width -> Automatic,
          Mass -> Automatic,
          FeynArtsNr -> 13,   
          LaTeX -> "\\tilde{u}",
          OutputName -> "um" }},   
\end{verbatim}
can be added. In particular, the entry \verb"Description" is important. This should be an unique identifier for each particle or parameter. This identifier can later on be used in the different files of the different models, e.g.     
\begin{verbatim}
{Su ,  {  Descriptions -> "Down Squark"}},   
\end{verbatim}
Of course, it is also possible to overwrite some of the global definitions by defining them locally, too. For instance, to use \verb"u" instead of \verb"um" as output name in a specific model, the entry should be changed to  
\begin{verbatim}
{Su ,  {  Descriptions -> "Down Squark",
          OutputName -> "u" }},   
\end{verbatim}
in the corresponding particle file of the model.
\section{Working with \SARAH}
\subsection{Starting \SARAH}
\SARAH is loaded and a model initialized and evaluated by the Mathematica commands
\begin{verbatim}
 <<SARAH.m
 Start["Modelname"];
\end{verbatim}
Our examples in the following are based on the MSSM, therefore we chose \verb"MSSM" as \verb"Modelname".
\subsection{Mass matrices and tadpole equations} 
\label{sec:SARAHmass}
\begin{enumerate}
\item {\bf Higgs mass matrix} The \((1,2)\)-entry of the mass matrix of the scalar Higgs in the MSSM is saved in
{\tt MassMatrix[hh][[1,2]]}. This returns
\begin{verbatim}
-(g1^2*vd*vu)/4 - (g2^2*vd*vu)/4 - B[\[Mu]]/2 -conj[B[\[Mu]]]/2 
\end{verbatim}
\item {\bf Squark mass matrix} In the same way, the (1,1)-entry of the \(6 \times 6\) down squark mass matrix is obtained by
 {\tt MassMatrix[Sd][[1,1]]}. The output is
\begin{verbatim}
(-3*g2^2*(vd^2 + vu^2) + g1^2*(vu^2 - vd^2) + 24*mq2[1,1] +
    12*vd^2*sum[j1, 1, 3,conj[Yd[j1, 1]]*Yd[j1, 1]])/24
\end{verbatim}
\item {\bf Squark mass matrix with generation indices as variable} To get the result for the \(2 \times 2\) down squark matrix without the explicit insertion of generation indices, 
\begin{verbatim}
 MassMatrixUnexpanded[Sd][[1,1]]
\end{verbatim} 
is used. The output is
\begin{verbatim}
(Delta[cm1,cn1]*(-((g1^2+3*g2^2)*(vd^2-vu^2)*Delta[gm1,gn1]) 
  + vd^2*sum[j1,1,3,conj[Yd[j1,gn1]]*Yd[j1,gm1]] + 12*(2*mq2[gm1,gn1] )))/24
\end{verbatim}
\item {\bf Tadpole equation} The tadpole equation corresponding to  \(\frac{\partial V}{\partial v_d}=0\) is obtained by 
\begin{verbatim}
TadpoleEquation[vd]  
\end{verbatim}
 and reads
\begin{verbatim}
(8*mHd2*vd+g1^2*vd^3+g2^2*vd^3-g1^2*vd*vu^2-g2^2*vd*vu^2-4*vu*B[\[Mu]] +
   (8*vd*\[Mu]*conj[\[Mu]]-4*vu*conj[B[\[Mu]]])/8   ==   0
\end{verbatim}
\end{enumerate}
\subsection{Calculating vertices}
One main function of \SARAH is to calculate the vertices for a model. In contrast to the most other calculations, vertices normally are not calculated automatically: it can last several minutes to calculate all vertices of a model and sometimes these are not needed. Calculating vertices is done via
\begin{verbatim}
Vertex[ParticleList,Options]
\end{verbatim}
\verb"ParticleList" is a list containing the involved fields. This list can include up to 6 particles. The following \verb"Options" are supported by the \verb"Vertex" command:
\begin{itemize}
\item \verb"Eigenstates", value: \verb" Name of Eigenstates", default: Last entry of \verb"NameOfStates". \\ 
Fixes the considered eigenstates 
\item \verb"UseDependences", value: \verb"True" or \verb"False", default: \verb"False". \\
Optional relations between the parameters will be used, if \verb"UseDependences" is set to \verb"True".   
\end{itemize} 
The Output of \verb"Vertex" is an array: 
\begin{verbatim}
{{ParticleList},{{Coefficient 1, Lorentz 1},{Coefficient 2, Lorentz 2},...} 
\end{verbatim}
First, the list of the involved particles is given and the indices are inserted. The second part consists of the value of the vertex and can be also a list, if different Lorentz structures are present. 
\paragraph*{Examples}
Some examples to clarify the usage and output of \verb"Vertex":
\begin{enumerate}
\item {\bf One possible Lorentz structure}: {\tt Vertex[\{hh,Ah,Z\}] } leads to the vertex of scalar and a pseudo scalar Higgs with a \(Z\) boson
\begin{verbatim}
{{hh[{gt1}], Ah[{gt2}], VZ[{lt3}]}, 
{((ZA[gt2,1]*ZH[gt1,1]-ZA[gt2,2]*ZH[gt1,2])*(g2*Cos[ThetaW]+
  g1*Sin[ThetaW]))/2, Mom[Ah[{gt2}], lt3] - Mom[hh[{gt1}],lt3]}}
\end{verbatim}
The output is divided in two parts. First, the involved particles are given, second, the value of the vertex is given. This second part is again split in two parts: one is the Lorentz independent part and the second part defines the transformation under the Lorentz group.  
\item {\bf Several possible Lorentz structures}: {\tt Vertex[\{bar[Fd],Fd,hh\}]} is the interaction between two  d-quarks and a Higgs:
\begin{verbatim}
{{bar[Fd[{gt1, ct1}]], Fd[{gt2, ct2}], hh[{gt3}]}, 
 {((-I)*Delta[ct1,ct2]*Delta[gt1,gt2]*ZH[gt3,2]*Yd[gt2,gt1])/Sqrt[2],PL}, 
 {((-I)*Delta[ct1,ct2]*Delta[gt1,gt2]*ZH[gt3,2]*Yd[gt1,gt2])/Sqrt[2],PR}}
\end{verbatim}
Obviously, there are three parts: one for the involved particles and two for the different Lorentz structures. \verb"PL" and \verb"PR" are the polarization projectors \(P_L = \frac{1}{2} (1 - \gamma_5), P_R = \frac{1}{2} (1 + \gamma_5)\).
\item {\bf Changing the considered eigenstates and using Weyl fermions}: It is also possible to calculate the vertices for Weyl fermions and/or to consider the gauge eigenstates. For instance,
\begin{verbatim}
Vertex[{fB, FdL, conj[SdL]}, Eigenstates -> GaugeES]
\end{verbatim}
gives
\begin{verbatim}
{{fB, FdL[{gt2, ct2}], conj[SdL[{gt3, ct3}]]}, 
 {((-I/3)*g1*Delta[ct2, ct3]*Delta[gt2, gt3])/Sqrt[2],1}}
\end{verbatim}
\item {\bf Using dependences}: With {\tt Vertex[\{conj[Se], Se, VP\}, UseDependences -> True]} \(g_1\) and \(g_2\) are replaced by the electric charge \(e\). This and similar relations can be defined in the parameters file (see sec.~\ref{sec:ParametersFile}). 
\begin{verbatim}
{{conj[Se[{gt1}]], Se[{gt2}], VP[{lt3}]}, 
{(-I)*e*Delta[gt1,gt2],-Mom[conj[Se[{gt1}]],lt3]+Mom[Se[{gt2}],lt3]}}
\end{verbatim}
\item {\bf Fixing the generations}: It is possible to give the indices of the particles already as input
\begin{verbatim}
Vertex[{hh[{1}], hh[{1}], Ah[{2}], Ah[{2}]}]
\end{verbatim} 
leads to
\begin{verbatim}
{{hh[{1}], hh[{1}], Ah[{2}], Ah[{2}]}, 
 {(-I/4)*(g1^2 + g2^2)*Cos[2*\[Alpha]]*Cos[2*\[Beta]], 1}}
\end{verbatim}
Obviously, the given definition of the mixing matrices for the Higgs fields were automatically inserted.  
\item {\bf Effective operators}: In effective theories, also interactions between two fermions and two scalars are possible. For example, an effective vertex for a model in which the gluino was integrated out:
\begin{verbatim}
Vertex[{Fd, Fd, conj[Sd], conj[Sd]}]
\end{verbatim}
Returns
\begin{verbatim}
{{Fd[{gt1,ct1}],Fd[{gt2,ct2}],conj[Sd[{gt3,ct3}]],conj[Sd[{gt4,ct4}]]}, 
{-(g3^2*(sum[j1, 1, 8, (Lam[j1, ct3, ct2]*Lam[j1, ct4, ct1])/Mass[fG][j1]]*
           ZD[gt3, gt2]*ZD[gt4, gt1] + 
         sum[j1, 1, 8, (Lam[j1, ct3, ct1]*Lam[j1, ct4, ct2])/Mass[fG][j1]]*
           ZD[gt3, gt1]*ZD[gt4, gt2])),
    LorentzProduct[PL, PL]}, {0, LorentzProduct[PR, PL]}, 
{g3^2*(sum[j1, 1, 8, (Lam[j1, ct2, ct3]*Lam[j1, ct4, ct1])/Mass[fG][j1]]*
       ZD[gt3, 3 + gt2]*ZD[gt4, gt1] + 
       sum[j1, 1, 8, (Lam[j1, ct2, ct4]*Lam[j1, ct3, ct1])/Mass[fG][j1]]*
       ZD[gt3, gt1]*ZD[gt4, 3 + gt2]),
    LorentzProduct[PL, PR]}, {0, LorentzProduct[PR, PR]}, 
{0, LorentzProduct[gamma,PL, PL]}, {0, LorentzProduct[gamma, PR, PL]}, 
{0, LorentzProduct[gamma, PL, PR]}, {0, LorentzProduct[gamma,PR, PR]}}
\end{verbatim}
Obviously, \SARAH checks the eight possible operators (4 different combination of polarization operators with and without a \(\gamma\) matrix) and returns the result for each operator. 
\end{enumerate}
In addition, all vertices can be calculated at once using 
\begin{verbatim}
 MakeVertexList[Eigenstates]
\end{verbatim}
This searches for all possible interactions present in the Lagrangian and creates lists for the generic subclasses of interactions, e.g. {\tt VertexList[FFS]} or {\tt VertexList[SSVV]} for all two-fermion-one-scalar interactions and all two-scalar-two-vector-boson interactions, respectively. 
\subsection{Renormalization group equations}
\label{examplesRGEs}
The calculation of the RGEs can be started after the initialization of a model via
\begin{verbatim}
CalcRGEs[Options];
\end{verbatim}
\paragraph*{Options}
The options offer a possibility to disable the calculation of the two-loop RGEs (\verb"TwoLoop" \(\rightarrow\) \verb"False"). Another option is to handle the number of generations of specific chiral superfields as variable ({\tt VariableGenerations} \(\rightarrow\) list of fields). This might be used for models which include chiral superfields much heavier than the SUSY scale to make the dependence on these fields obvious. Normally, the \(\beta\)-function are written in a compact form using matrix multiplication, as explained below. This can be switched off by the option \verb"NoMatrixMultiplication"  \(\rightarrow\) \verb"True". The last option offers the possibility to read the results of former calculations (\verb"ReadLists"  \(\rightarrow\) \verb"True")\\
\paragraph*{GUT normalization}
The gauge couplings of abelian  gauge groups are often normalized at the GUT scale. Therefore, it is possible to define for each  \(U(1)\) gauge coupling the GUT-normalization by the corresponding entry in the parameters file. See app.~\ref{sec:ParametersFile} for more information. 
\paragraph*{Results}
The RGEs are saved in different arrays in Mathematica whose names are given in brackets: anomalous dimensions of all superfields (\verb"Gij"), trilinear (\verb"BetaYijk"), bilinear (\verb"BetaMuij") and  linear (\verb"BetaLi") superpotential parameters, trilinear  (\verb"BetaTijk"), bilinear   (\verb"BetaBij") and  linear (\verb"BetaLSi") soft breaking parameters, scalar soft breaking masses (\verb"Betam2ij"), gaugino soft breaking masses (\verb"BetaMi"), gauge couplings (\verb"BetaGauge") and VEVs (\verb"BetaVEVs").\\
All entries of these arrays are three dimensional: the first entry gives the name of the parameter, the second one the one-loop \(\beta\)-function and the third one the two-loop \(\beta\)-function. Furthermore, the results for the RGEs of the scalar masses are simplified by using abbreviations for often appearing traces (see also \cite{Martin:1993zk}). The definition of the traces are saved in the array \verb"TraceAbbr". In \verb"TraceAbbr[[1]]" all abbreviations needed for the one-loop RGEs are given, and in \verb"TraceAbbr[[2]]" those are for the two-loop part.\\
The results are also saved as text files in the directory
\begin{verbatim}
PackageDirectory/Output/Modelname/RGEs/ 
\end{verbatim}

\paragraph*{Matrix Multiplication}
Generally, the results contain sums over the generation indices of the particles in the loop. \SARAH always tries to write them as matrix multiplications, in order to shorten the expressions. Therefore, new symbols are introduced:
\begin{itemize}
\item \verb"MatMul[A,B,C,...][i,j]": \((A B C \dots)_{i,j}\). Matrix multiplication, also used for vector-matrix and vector-vector multiplication.
\item \verb"trace[A,B,C,...]": \(\mbox{Tr}(A B C \dots)\). Trace of a matrix or product of matrices.
\item \verb"Adj[M]": \(M^\dagger\). Adjoint of a matrix
\item \verb"Tp[M]": \(M^T\). Transposed of a matrix  
\end{itemize} 
As already mentioned, the usage of matrix multiplication can be switched off with the corresponding option. In addition, it is automatically switched off, if the model contains a parameter with three generation indices. 
\paragraph*{Examples}
\begin{enumerate}
\item {\bf \(\beta\)-function of Yukawa coupling}: The Yukawa couplings of the MSSM are saved in \verb"BetaYijk". The first entry consists of
\begin{verbatim}
BetaYijk[[1,1]]:  Ye[i1,i2] ,
\end{verbatim}
i.e. this entry contains the \(\beta\)-functions for the electron Yukawa coupling. The corresponding one-loop \(\beta\)-function is
\begin{verbatim}
BetaYijk[[1,2]]:
(-9*g1^2*Ye[i1,i2])/5-3*g2^2*Ye[i1,i2]+3*trace[Yd,Adj[Yd]]*Ye[i1,i2]+ 
  trace[Ye,Adj[Ye]]*Ye[i1, i2]+3*MatMul[Ye,Adj[Ye],Ye][i1, i2]
\end{verbatim}
The two-loop \(\beta\)-function is saved in \verb"BetaYijk[[1,3]]" but we skip it here because of its length. 
\item {\bf \(\beta\)-function of soft breaking masses and abbreviations for traces}:  The first entry of \verb"Betam2ij" corresponds to the soft breaking mass of the selectron
\begin{verbatim}
 Betam2ij[[1,1]]:           me2[i1,i2]
\end{verbatim}
 and the one-loop \(\beta\)-function is saved in  \verb"Betam2ij[[1,2]]":
\begin{verbatim}
(-24*g1^2*MassB*conj[MassB]+10*g1^2*Tr1[1])*Kronecker[i1,i2]/5 + 
 4*mHd2*MatMul[Ye,Adj[Ye]][i1,i2]+4*MatMul[T[Ye],Adj[T[Ye]]][i1,i2] + 
  2*MatMul[me2,Ye,Adj[Ye]][i1,i2]+4*MatMul[Ye, ml2, Adj[Ye]][i1,i2] + 
  2*MatMul[Ye,Adj[Ye],me2][i1,i2]
\end{verbatim}
The definition of the element \verb"Tr1[1]" is saved in \verb"TraceAbbr[[1,1]]":
\begin{verbatim}
{Tr1[1], -mHd2 + mHu2 + trace[md2] + trace[me2] - trace[ml2] +
          trace[mq2] - 2*trace[mu2]}
\end{verbatim}
\item {\bf Number of generations as variable}: With
\begin{verbatim}
CalcRGEs[VariableGenerations -> {q}]
\end{verbatim}
the number of generations of the left-quark superfield is handled as variable. Therefore, the one-loop \(\beta\)-function of the hypercharge couplings reads 
\begin{verbatim}
 (63*g1^3)/10 + (g1^3*NumberGenerations[q])/10
\end{verbatim}
\item{\bf No matrix multiplication}: Using matrix multiplication can be switched off by
\begin{verbatim}
CalcRGEs[NoMatrixMultiplication -> True]
\end{verbatim}
The one-loop \(\beta\)-function for the electron Yukawa coupling is now written as
\begin{verbatim}
  sum[j2,1,3,sum[j1,1,3,conj[Yd[j2,j1]]*Yu[i1,j1]]*Yd[j2,i2]] + 
2*sum[j2,1,3,sum[j1,1,3,conj[Yu[j1,j2]]*Yu[j1,i2]]*Yu[i1,j2]] + 
  sum[j2,1,3,sum[j1,1,3,conj[Yu[j2,j1]]*Yu[i1,j1]]*Yu[j2,i2]] + 
(3*sum[j2,1,3,sum[j1,1,3,conj[Yu[j1,j2]]*Yu[j1,j2]]]*Yu[i1,i2])/2 + 
(3*sum[j2,1,3,sum[j1,1,3,conj[Yu[j2,j1]]*Yu[j2,j1]]]*Yu[i1,i2])/2 - 
 (13*g1^2*Yu[i1,i2])/15-3*g2^2*Yu[i1,i2]-(16*g3^2*Yu[i1,i2])/3
\end{verbatim}
\end{enumerate}

\subsection{Loop calculations}
\label{resultsLoop}
The command to start the loop calculation is
\begin{verbatim}
CalcLoopCorrections[Eigenstates];
\end{verbatim}
As usual, \verb"Eigenstates" can in case of the MSSM either be \verb"GaugeES" for the gauge eigenstates or \verb"EWSB" for the eigenstates after EWSB. If the vertices for the given set of eigenstates were not calculated before, this is done before the calculation of the loop contributions begins. \\
\paragraph*{Conventions} Using the conventions of \cite{Pierce:1996zz}, the results will contain the  Passarino Veltman integrals listed in app.~\ref{sec:Integrals}. The involved couplings are abbreviated by
\begin{itemize}
\item \verb"Cp[p1,p2,p3]" and \verb"Cp[p1,p2,p3,p4]" for non-chiral, three- and four-point interactions involving the particles \verb"p1" - \verb"p4".
\item \verb"Cp[p1,p2,p3][PL]" and \verb"Cp[p1,p2,p3][PR]" for chiral, three-point interactions involving the fields \verb"p1" - \verb"p3". 
\end{itemize}
The self energies can be used for calculating the radiative corrections to masses and mass matrices, respectively. For calculating the loop corrections to a mass matrix, it is convenient to use unrotated, external fields, while the fields in the loop are rotated. Therefore, \SARAH adds to the symbols of the external particle in the interaction an \verb"U" for 'unrotated', e.g. \verb"Sd" \(\rightarrow\) \verb"USd". The mixing matrix associated to this field in the vertex has to be replaced. 
\paragraph*{Results} The results for the loop corrections are saved in two different ways. First as list containing the different loop contributions for each particle. Every entry includes the following information: internal particles, generic type of the diagram, numerical factors coming from symmetry considerations and possible charges in the loop. The second output is a sum of all contributions, where the generic results of app.~\ref{sec:Integrals} have already be inserted. This can afterwards written as pdf file using the \LaTeX{} output of \SARAH.\\
The results for the self energies are saved in \verb"SelfEnergy1LoopList" as list of the contributions and in \verb"SelfEnergy1LoopSum" written as sums. The last one is a two-dimensional array. The first column gives the name of the external particle, the entry in the second column depends on the type of the field. For scalars, the one-loop self energy \(\Pi(p^2)\)  is given,  for fermions, the one-loop self energies for the different polarizations (\(\Sigma^L(p^2)\),\(\Sigma^R(p^2)\), \(\Sigma^S(p^2)\)) are written in a 3 dimensional array, while for vector bosons, the transversal part of the self energy \(\Pi^T(p^2)\) is shown. \\
Also the corrections to the tadpoles are saved twice:  in \verb"Tadpoles1LoopSum[Eigenstates]" explicitly written as sum and secondly in \verb"Tadpoles1LoopList[Eigenstates]"  as list of the different contributions.   
\paragraph*{Examples}
\begin{enumerate}
\item {\bf One-loop tadpoles} The radiative correction of the tadpoles due to a chargino loop is saved in 
\begin{verbatim}
Tadpoles1LoopList[EWSB][[1]]; 
\end{verbatim}
and reads
\begin{verbatim}
 {bar[Cha],Cp[Uhh[{gO1}],bar[Cha[{gI1}]],Cha[{gI1}]],FFS,1,1/2}
\end{verbatim}
The meaning of the different entries is: (i) a chargino (\verb"Cha") is in the loop, (ii) the vertex with an external, unrotated Higgs (\verb"Uhh") with generation index \verb"gO1" and two charginos with index \verb"gI1" is needed, (iii) the generic type of the diagram is \verb"FFS", (iv) the charge factor is 1, (v) the diagram is weighted by a factor \(\frac{1}{2}\) with respect to the generic expression (see app.~\ref{sec:Integrals}).\\
The corresponding term in \verb"Tadpoles1LoopSum[EWSB]" is
\begin{verbatim}
4*sum[gI1,1,2, A0[Mass[bar[Cha[{gI1}]]]^2]*
   Cp[phid,bar[Cha[{gI1}]],Cha[{gI1}]]*Mass[Cha[{gI1}]]] 
\end{verbatim}
\item {\bf One-loop self-energies}
\begin{enumerate}
\item The correction to the down squark matrix due to  a four-point interaction with a pseudo scalar Higgs is saved in
{\tt  SelfEnergy1LoopList[EWSB][[1, 12]]} and reads
\begin{verbatim}
{Ah,Cp[conj[USd[{gO1}]],USd[{gO2}],Ah[{gI1}],Ah[{gI1}]],SSSS,1,1/2}
\end{verbatim}
This has the same meaning as the term
\begin{verbatim}
-sum[gI1,1,2,A0[Mass[Ah[{gI1}]]^2]*
   Cp[conj[USd[{gO1}]],USd[{gO2}],Ah[{gI1}],Ah[{gI1}]]]/2
\end{verbatim}
in \verb"SelfEnergy1LoopSum[EWSB]". 
\item Corrections to the \(Z\) boson are saved in {\tt SelfEnergy1LoopList[EWSB][[15]]}. An arbitrary entry looks like
\begin{verbatim}
{bar[Fd], Fd, Cp[VZ, bar[Fd[{gI1}]], Fd[{gI2}]], FFV, 3, 1/2}
\end{verbatim}
and corresponds to
\begin{verbatim}
(3*sum[gI1, 1, 3, sum[gI2, 1, 3, 
     H0[p^2, Mass[bar[Fd[{gI1}]]]^2, Mass[Fd[{gI2}]]^2]*
 (conj[Cp[VZ,bar[Fd[{gI1}]],Fd[{gI2}]][PL]]*
     Cp[VZ,bar[Fd[{gI1}]],Fd[{gI2}]][PL] + 
  conj[Cp[VZ,bar[Fd[{gI1}]],Fd[{gI2}]][PR]]*
     Cp[VZ,bar[Fd[{gI1}]],Fd[{gI2}]][PR]) + 
 2*B0[p^2,Mass[bar[Fd[{gI1}]]]^2,Mass[Fd[{gI2}]]^2]*
     Mass[bar[Fd[{gI1}]]]*Mass[Fd[{gI2}]]*
  Re[Cp[VZ,bar[Fd[{gI1}]],Fd[{gI2}]][PL]*
     Cp[VZ,bar[Fd[{gI1}]],Fd[{gI2}]][PR])]])/2 
\end{verbatim}
in \verb"SelfEnergy1LoopListSum[EWSB]". 
\end{enumerate}
\end{enumerate}
\subsection{Calculations for \texorpdfstring{$SU(N)$}{SU(N)} gauge groups}
\label{calcIrrep}
The user can use the functions to obtain information about different representations of \(SU(N)\) independently from the model using the new command
\begin{verbatim}
CheckIrrepSUN[Dim,N]
\end{verbatim}
\verb"Dim" is the dimension of the irreducible representation and \verb"N" is the dimension of the \(SU(N)\) gauge group. 
The result is a vector containing the following information: (i) repeating the dimension of the field, (ii) number of covariant indices, (iii) number of contravariant indices, (iv) value of the quadratic Casimir \(C_2(r)\), (v) value of the Dynkin index \(I_2(r)\), (vi) Dynkin labels for the highest weight.
\paragraph*{Examples}
\begin{enumerate}
\item {\bf Fundamental representation}: The properties of a particle, transforming under the fundamental representation of \(SU(3)\) are calculated via {\tt CheckIrrepSUN[3,3]}. The output is the well known result
\begin{verbatim}
{3, 1, 0, 4/3, 1/2, {1, 0}}
\end{verbatim}
\item {\bf Adjoint representation}: The properties of a field transforming as {\bf 24} of \(SU(5)\) are calculated by
{\tt  CheckIrrepSUN[24,5] }. The output will be
\begin{verbatim}
{24, 1, 1, 5, 5, {1, 0, 0, 1}}
\end{verbatim}
\item {\bf  Different representations with the same dimension}:
The {\bf{70}} under \(SU(5)\) is not unique. Therefore, {\tt CheckIrrepSUN[\{70, \{0, 0, 0, 4\}\}, 5] }  returns
\begin{verbatim}
{70, 0, 4, 72/5, 42, {0, 0, 0, 4}} 
\end{verbatim}
while {\tt CheckIrrepSUN[\{70, \{2, 0, 0, 1\}\}, 5] } leads to
\begin{verbatim}
{70, 2, 1, 42/5, 49/2, {2, 0, 0, 1}} 
\end{verbatim}
\end{enumerate}

  \chapter{Conventions and generic expressions}
\section{Vertices}
\label{app:ConVertices}
We present in the following appendices important vertices for the different models. Our conventions are the following: \(\Gamma\) stands for a vertex without the Lorentz structure. The involved fields are given as lower indices. Chiral interactions between two fermions \(F_1,F_2\) and one scalar \(S\) are parametrized by
\begin{equation}
 \Gamma_{F_1 F_2 S}^L P_L + \Gamma_{F_1 F_2 S}^R P_R
\end{equation}
with the projection operators \(P_L = \halb \left(1-\gamma_5\right)\) and \(P_R = \halb \left(1+\gamma_5\right)\). Interactions between two fermions  \(F_1,F_2\) and one vector boson \(V_\mu\) are parametrized by
\begin{equation}
 \Gamma_{F_1 F_2 V_\mu}^L \gamma_\mu P_L + \Gamma_{F_1 F_2 V_\mu}^R \gamma_\mu P_R \thickspace . 
\end{equation}
The Lorentz structure of interactions between one or two scalars \(S_1,S_2\) and  two vector bosons \(V_\mu,V_\nu\) is \(g_{\mu \nu}\). The vertex \(\Gamma_{S_1 S_2 V^\mu}\) involving two scalars \(S_1,S_2\) with momenta \(p_1,p_2\) and one vector boson \(V_\mu\) has the momentum flow \begin{align}
p^\mu_1  - p^\mu_2 \thickspace.  
\end{align}
Furthermore, we use the abbreviations \(s_{\Theta_W} = \sin\left(\Theta_W\right)\) and \(c_{\Theta_W} = \cos\left(\Theta_W\right)\). %
\section{Renormalization group equations}
\label{app:RGE_con}
We present also in the following appendices our results for the RGEs of the different models. We will only show the \(\beta\)-functions for the gauge couplings and the anomalous dimensions of all chiral superfields. We discuss in this section briefly how these results were calculated. Furthermore, we show how they can be used to calculate the other \(\beta\)-functions of the models. \\
For a  general $N=1$ supersymmetric gauge theory with superpotential  
\begin{equation}
 W (\phi) = \frac{1}{2}{\mu}^{ij}\phi_i\phi_j + \frac{1}{6}Y^{ijk}
\phi_i\phi_j\phi_k
\end{equation}
the  soft SUSY-breaking scalar terms are given by
\begin{equation}
V_{\hbox{soft}} = \left(\frac{1}{2}b^{ij}\phi_i\phi_j
+ \frac{1}{6}h^{ijk}\phi_i\phi_j\phi_k +\hbox{c.c.}\right)
+(m^2)^i{}_j\phi_i\phi_j^* \thickspace.
\end{equation}
The anomalous dimensions are given by \cite{Martin:1993zk}
\begin{align}
 \gamma_i^{(1)j} = & \frac{1}{2} Y_{ipq} Y^{jpq} - 2 \delta_i^j g^2 C_2(i) \thickspace, \\
 \gamma_i^{(2)j}  = &  -\frac{1}{2} Y_{imn} Y^{npq} Y_{pqr} Y^{mrj} + g^2 Y_{ipq} Y^{jpq} [2C_2(p)- C_2(i)] \nonumber \\
 & \; \;  + 2 \delta_i^j g^4 [ C_2(i) S(R)+ 2 C_2(i)^2 - 3 C_2(G) C_2(i)] \thickspace,
\end{align}
and the \(\beta\)-functions for the gauge couplings are given by
\begin{align}
 \beta_g^{(1)}  =  & g^3 \left[S(R) - 3 C_2(G) \right] \thickspace,\\
 \beta_g^{(2)}  =  & g^5 \left\{ - 6[C_2(G)]^2 + 2 C_2(G) S(R) + 4 S(R) C_2(R) \right\}
    - g^3 Y^{ijk} Y_{ijk}C_2(k)/d(G) \thickspace .
\end{align}
Here, \(C_2(i)\) is the quadratic Casimir for a specific superfield and $C_2(R),C_2(G)$ are the quadratic Casimirs for the matter and adjoint  representations, respectively. \(d(G)\) is the dimension of the adjoint representation.  \\
The $\beta$-functions for the superpotential parameters can be obtained by using superfield technique. The obtained expressions are \cite{West:1984dg,Jones:1984cx}. 
\begin{eqnarray}
 \beta_Y^{ijk} &= & Y^{p(ij} {\gamma_p}^{k)} \thickspace, \\
 \beta_{\mu}^{ij} &= & \mu^{p(i} {\gamma_p}^{j)} \thickspace .
\end{eqnarray}
The most \(\beta\)-functions of the models can derived from these results using the procedure given in \cite{Jack:1997eh} based on the spurion formalism \cite{Yamada:1994id}. In the following, we briefly summarize the basic ideas of this calculation for completeness. \\

The  exact results for the soft $\beta$-functions are given by \cite{Jack:1997eh}:
\begin{eqnarray}
\label{eq:betaM}
\beta_M &=& 2{\cal O} \left[\frac{\beta_g}{g}\right] \thickspace, \\
\beta_{h}^{ijk} &=& h{}^{l(jk}\gamma^{i)}{}_l -
2Y^{l(jk}\gamma_1{}^{i)}{}_l \thickspace, \cr
\beta_{b}^{ij} &=&   
b{}^{l(i}\gamma^{j)}{}_l-2\mu{}^{l(i}\gamma_1{}^{j)}{}_l \thickspace,\cr
\left(\beta_{m^2}\right){}^i{}_j &=& \Delta\gamma^i{}_j \thickspace.
\label{eq:betam2}
\end{eqnarray}
The $(..)$ in the superscripts denote symmetrization and we defined
\begin{eqnarray}
{\cal O}  &=& Mg^2\frac{\partial}{\partial g^2}-h^{lmn}
\frac{\partial}{\partial Y^{lmn}} \thickspace, \\
(\gamma_1)^i{}_j  &=& {\cal O}\gamma^i{}_j \thickspace, \\
\Delta &=& 2{\cal O} {\cal O}^* +2MM^* g^2 \frac{\partial}{\partial g^2}
+\left[{\tilde Y}^{lmn} \frac{\partial}{\partial Y^{lmn}} + \hbox{c.c.}\right]
+X \frac{\partial}{\partial g} \thickspace.
\end{eqnarray}
Here, $M$ is the gaugino mass and  
${\tilde Y}^{ijk} = (m^2)^i{}_lY^{jkl} +  (m^2)^j{}_lY^{ikl} + (m^2)^k{}_lY^{ijl}.$
Eqs.~(\ref{eq:betaM})-(\ref{eq:betam2}) hold  in a class of renormalization schemes that includes the DRED$'$-one \cite{Jack:1994rk}. We take the known contributions of $X$ from \cite{Jack:1998iy}:
\begin{eqnarray}
X^{\mathrm{DRED}'(1)}&=&-2g^3S \thickspace, \\
X^{\mathrm{DRED}'(2)}&=& (2r)^{-1}g^3 \mathrm{tr} [ W C_2(R)]
-4g^5C_2(G)S-2g^5C_2(G)QMM^* \thickspace,\end{eqnarray}
where
\begin{eqnarray}
S &=&  r^{-1} \mathrm{tr}[m^2C_2(R)] -MM^* C_2(G) \thickspace,  \\
W^j{}_i&=&\frac{1}{2}Y_{ipq}Y^{pqn}(m^2)^j{}_n+ \frac{1}{2}Y^{jpq}Y_{pqn}(m^2)^n{}_i
+2Y_{ipq}Y^{jpr}(m^2)^q{}_r 
+h_{ipq}h^{jpq}-8g^2MM^*C_2(R)^j{}_i \thickspace .
\nonumber \\
\end{eqnarray}
With $Q = T(R) - 3C_2(G)$, and $T(R) = \mathrm{tr} \left[C_2(R)\right]$, $r$ being  the number of group generators. \\

\section{One-loop amplitudes for one- and two-point functions}
\label{sec:Integrals}
We used for the calculation of the one-loop self energies and the one-loop corrections to the tadpoles in \(\DR\)-scheme the scalar functions defined in \cite{Pierce:1996zz}. The basis integrals are
\begin{eqnarray}
A_0(m) &=& 16\pi^2Q^{4-n}\int{\frac{d^nq}{ i\,(2\pi)^n}}{\frac{1}{ q^2-m^2+i\varepsilon}} \thickspace ,\\
B_0(p, m_1, m_2) &=& 16\pi^2Q^{4-n}\int{\frac{d^nq}{ i\,(2\pi)^n}} {\frac{1}{\biggl[q^2-m^2_1+i\varepsilon\biggr]\biggl[
(q-p)^2-m_2^2+i\varepsilon\biggr]}} \thickspace ,
\label{B0 def}
\end{eqnarray}
with the renormalization scale \(Q\). The integrals are regularized by integrating in $n=4-2\epsilon$ dimensions. The result for \(A_0\) is
\begin{equation}
A_0(m)\ =\ m^2\left({\frac{1}{\hat\epsilon}} + 1 - \ln{\frac{m^2}{Q^2}}\right)~,\label{A}
\end{equation}
where $1/\hat\epsilon =1/\epsilon-\gamma_E+\ln 4\pi$. The function $B_0$  has the analytic expression
\begin{equation}
 B_0(p, m_1, m_2) \ =\ {\frac{1}{\hat\epsilon}} - \ln\left(\frac{p^2}{Q^2}\right) - f_B(x_+) - f_B(x_-)~,
\end{equation}
with
\begin{equation}
 x_{\pm}\ =\ \frac{s \pm \sqrt{s^2 - 4p^2(m_1^2-i\varepsilon)}}{2p^2}~,
\qquad f_B(x) \ =\ \ln(1-x) - x\ln(1-x^{-1})-1~,
\end{equation}
and $s=p^2-m_2^2+m_1^2$. All the other functions can be expressed by $A_0$ and $B_0$. For instance,
\begin{equation}
 B_1(p, m_1,m_2) \ =\ {\frac{1}{2p^2}}\biggl[ A_0(m_2) - A_0(m_1) + (p^2
+m_1^2 -m_2^2) B_0(p, m_1, m_2)\biggr]~,
\end{equation}
and
\begin{eqnarray}
B_{22}(p, m_1,m_2) &=& \frac{1}{6}\ \Bigg\{\,
\frac{1}{2}\biggl(A_0(m_1)+A_0(m_2)\biggr)
+\left(m_1^2+m_2^2-\frac{1}{2}p^2\right)B_0(p,m_1,m_2)\nonumber \\ &&+ \frac{m_2^2-m_1^2}{2p^2}\ \biggl[\,A_0(m_2)-A_0(m_1)-(m_2^2-m_1^2)
B_0(p,m_1,m_2)\,\biggr] \nonumber\\ && +  m_1^2 + m_2^2
-\frac{1}{3}p^2\,\Bigg\}~.
\end{eqnarray}
Furthermore, for the scalar self-energies it is useful to define
\begin{eqnarray}
F(p,m_1,m_2) &=& A_0(m_1)-2A_0(m_2)- (2p^2+2m^2_1-m^2_2)B_0(p,m_1,m_2)
\ ,\\[2mm] G(p,m_1,m_2) &=&
(p^2-m_1^2-m_2^2)B_0(p,m_1,m_2)-A_0(m_1)-A_0(m_2)\ ,\\[2mm] H(p,m_1,m_2)
&=& 4B_{22}(p,m_1,m_2) + G(p,m_1,m_2)\ ,\\[1mm] \tilde
B_{22}(p,m_1,m_2) &=& B_{22}(p,m_1,m_2) - \frac{1}{4}A_0(m_1) -
\frac{1}{4}A_0(m_2)~.\label{B22}
\end{eqnarray}
We will use for long expressions often the abbreviations
\begin{align}
  B_i(m_1^2,m_2^2) & =  B_i(p^2,m_1^2, m_2^2) \thickspace , \\
  F_0(m_1^2,m_2^2)  & =  F_0(p^2,m_1^2, m_2^2) \thickspace , \\
 G_0(m_1^2,m_2^2) & =  G_0(p^2,m_1^2, m_2^2) \thickspace . 
\end{align}
We list in the following the results for the generic different diagrams contributing to the one- and two-point functions. Several coefficient are involved:
\begin{itemize}
 \item \(c_S\) is a symmetry factor: if the particles in the loop are indistinguishable, the weight of the contribution is only half of the weight in the case of distinguishable particles. If two different charge flows are possible in the loop, the weight of the diagram is doubled.
 \item \(c_C\) is a charge factor: for corrections due to vector bosons in the adjoint representation, this is the Casimir \(C_2\) of the corresponding group. For corrections due to matter fields, this can be for example be a color factor for quarks/squarks. For corrections of vector bosons in the adjoint representation this is the Dynkin index \(I_2\) of the gauge group.
 \item  \(c_R\) is 2 for real fields and Majorana fermions in the loop and 1 otherwise. \\
\end{itemize}
\subsection{One-loop tadpoles}
\begin{enumerate}
\item Fermion loop (generic name in \SARAH: \verb"FFS"):
\begin{equation}
T = 8 c_S c_C m_F \Gamma A_0(m_F^2) 
\end{equation}
\item Scalar loop (generic name in \SARAH: \verb"SSS"):
\begin{equation}
T = - 2 c_S c_C \Gamma A_0(m_S^2) 
\end{equation}
\item Vector boson loop (generic name in \SARAH: \verb"SVV"):
\begin{equation}
T = 6 c_S c_C \Gamma A_0(m_V^2) 
\end{equation}
\end{enumerate}

\subsection{One-loop self-energies}
\paragraph*{Corrections to fermion}
\begin{enumerate}
\item Fermion-scalar loop (generic name in \SARAH: \verb"FFS"):
\begin{eqnarray*}
\Sigma^S(p^2) &=& m_F c_S c_C c_R \Gamma^1_R \Gamma^{2,*}_L B_0(p^2,m_F^2,m_S^2) \\
\Sigma^R(p^2) &=& - c_S c_C c_R \frac{1}{2} \Gamma^1_R \Gamma^{2,*}_R B_1(p^2,m_F^2,m_S^2) \\
\Sigma^L(p^2) &=& - c_S c_C c_R \frac{1}{2} \Gamma^1_L \Gamma^{2,*}_L B_1(p^2,m_F^2,m_S^2) 
\end{eqnarray*}
\item Fermion-vector boson loop (generic name in \SARAH: \verb"FFV"):
\begin{eqnarray*}
\Sigma^S(p^2) &=& - 4 c_S c_C c_R m_F \Gamma^1_L \Gamma^{2,*}_R B_0(p^2,m_F^2,m_S^2) \\
\Sigma^R(p^2) &=& - c_S c_C c_R \Gamma^1_L \Gamma^{2,*}_L B_1(p^2,m_F^2,m_S^2) \\
\Sigma^L(p^2) &=& - c_S c_C c_R \Gamma^1_R \Gamma^{2,*}_R B_1(p^2,m_F^2,m_S^2) 
\end{eqnarray*}
\end{enumerate}

\paragraph*{Corrections to scalar}
\begin{enumerate}
\item Fermion loop (generic name in \SARAH: \verb"FFS"): 
\begin{equation}
\label{eq:SE_FFS}
\Pi(p^2) = c_S c_C c_R \left((\Gamma^1_L \Gamma^{2,*}_L + \Gamma^1_R \Gamma^{2,*}_R) G_0(p^2,m_F^2,m_S^2) +  (\Gamma^1_L \Gamma^{2,*}_R + \Gamma^1_R \Gamma^{2,*}_L) B_0(p^2,m_F^2,m_S^2) \right)
\end{equation}
\item Scalar loop (two 3-point interactions, generic name in \SARAH: \verb"SSS"):
\begin{equation}
\Pi(p^2) = c_S c_C c_R \Gamma^1 \Gamma^{2,*} B_0(p^2,m_F^2,m_S^2) 
\end{equation}
\item Scalar loop (4-point interaction, generic name in \SARAH: \verb"SSSS"):
\begin{equation}
\label{eq:SE_SSSS}
\Pi(p^2) =  - c_S c_C \Gamma A_0(m_S^2) 
\end{equation}
\item Vector boson-scalar loop (generic name in \SARAH: \verb"SSV"):
\begin{equation}
\label{eq:SE_SSV}
\Pi(p^2) = c_S c_C c_R \Gamma^1 \Gamma^{2,*} F_0(p^2,m_F^2,m_S^2) 
\end{equation}
\item Vector boson loop (two 3-point interactions, generic name in \SARAH: \verb"SVV"):
\begin{equation}
\Pi(p^2) =  c_S c_C c_R \frac{7}{2} \Gamma^1 \Gamma^{2,*} B_0(p^2,m_F^2,m_S^2) 
\end{equation}
\item Vector boson loop (4-point interaction, generic name in \SARAH: \verb"SSVV"):
\begin{equation}
\Pi(p^2) =   c_S c_C \Gamma A_0(m_V^2) 
\end{equation}
\end{enumerate}

\paragraph*{Corrections to vector boson}
\begin{enumerate}
\item Fermion loop (generic name in \SARAH: \verb"FFV"):
\begin{equation}
\Pi^T(p^2) = c_S c_C c_R \left((|\Gamma^1_L|^2+|\Gamma^1_R|^2) H_0(p^2,m_V^2,m_F^2)+ 4 \Re(\Gamma^1_L \Gamma^2_R)B_0(p^2,m_V^2,m_F^2) \right)
\end{equation}
\item Scalar loop (generic name in \SARAH: \verb"SSV"):
\begin{equation}
\Pi^T(p^2) = -4 c_S c_C c_R |\Gamma|^2 B_{22}(p^2,m_{S_1}^2,m_{S_2}^2)
\end{equation}
\item Vector boson loop (generic name in \SARAH: \verb"VVV"):
\begin{equation}
\Pi^T(p^2) =  |\Gamma|^2  c_S c_C c_R \left(-(4 p^2 + m_{V_1}^2 + m_{V_2}^2 ) B_0(p^2,m_{V_1}^2,m_{V_1}^2) - 8 B_{22}(p^2,m_{S_1}^2,m_{S_2}^2) \right)
\end{equation}
\item Vector-Scalar-Loop (generic name in \SARAH: \verb"SVV"):
\begin{equation}
\Pi^T(p^2) =  |\Gamma|^2 c_S c_C c_R B_0(p^2,m_V^2,m_S^2)
\end{equation}
\end{enumerate}
We need here only the diagrams involving three-point interactions because the four-point interactions are related to them due to gauge invariance.

  \chapter{Mass matrices and interactions of messenger particles}
\section{Model file for SARAH without \texorpdfstring{$R$}{R}-parity violation}
\label{Modelfile_GMSB}
\begin{verbatim}
Off[General::spell]
Print["Model file for the MSSM with SU(5)-GMSB Messengers"];
ModelNameLaTeX ="GMSB_eff";

(*-------------------------------------------*)
(*   Particle Content*)
(*-------------------------------------------*)

(* Gauge Superfields *)

Gauge[[1]]={B,   U[1], hypercharge, g1,False};
Gauge[[2]]={WB, SU[2], left,        g2,True};
Gauge[[3]]={G,  SU[3], color,       g3,False};


(* Chiral Superfields *)

Fields[[1]] = {{uL,  dL},  3, q,   1/6, 2, 3};  
Fields[[2]] = {{vL,  eL},  3, l,  -1/2, 2, 1};
Fields[[3]] = {{Hd0, Hdm}, 1, Hd, -1/2, 2, 1};
Fields[[4]] = {{Hup, Hu0}, 1, Hu,  1/2, 2, 1};

Fields[[5]] = {conj[dR], 3, d,  1/3, 1, -3};
Fields[[6]] = {conj[uR], 3, u, -2/3, 1, -3};
Fields[[7]] = {conj[eR], 3, e,    1, 1,  1};

Fields[[8]] = {conj[dRM1], 1, dM1, 1/3, 1, -3};
Fields[[9]] = {conj[dRM2], 1, dM2, -1/3, 1, 3};
Fields[[10]] = {{vLM1,eLM1}, 1, lM1, -1/2, 2, 1};
Fields[[11]] = {{eLM2,vLM2}, 1, lM2, 1/2, 2, 1};

(*------------------------------------------------------*)
(* Superpotential *)
(*------------------------------------------------------*)

         
SuperPotential = { {{1, Yu},{u,q,Hu}}, {{-1,Yd},{d,q,Hd}},
                   {{-1,Ye},{e,l,Hd}}, {{1,\[Mu]},{Hu,Hd}},
                   {{1,MessD},{dM1,dM2}}, {{1,MessL},{lM1,lM2}},
                   {{m32,f}, {d,dM2}},     {{m32,f},{l,lM2}}  };
  

(*-------------------------------------------*)
(* Integrate Out or Delete Particles         *)
(*-------------------------------------------*)

IntegrateOut={};
DeleteParticles={SvMh,SdMh,SeMh,FvLM1,FvLM2,FeLM1,FeLM2,FdRM1,FdRM2};

(*-------------------------------------------*)
(* Before EWSB                               *)
(*-------------------------------------------*)

NameOfStates={GaugeES, TEMP,TEMP2, EWSB};


DEFINITION[TEMP][MatterSector]= 
{    {{SvLM1,conj[SvLM2]}, {SvM,ZMV}}, {{SeLM1,conj[SeLM2]}, {SeM,ZME}}, 
     {{SdRM1,conj[SdRM2]}, {SdM,ZMD}}    }; 

DEFINITION[TEMP2][Flavors]= 
{{SvM,{SvMl,SvMh}},  {SeM,{SeMl,SeMh}},
 {SdM,{SdMl,SdMh}} };

DEFINITION[GaugeES][Additional]={
{{SdRM1,SdRM2},{1,FD}},
{{SeLM1,SeLM2},{1,FL}},
{{SvLM1,SvLM2},{1,FL}},
{{conj[SvL], SHd0}, {1, mHL2}},
{{conj[SeL], SHdm}, {1, mHL2}} 
};

DEFINITION[GaugeES][GaugeFixing]=
{ {Der[VWB],  -1/(2 RXi[VWB])},
  {Der[VG],   -1/(2 RXi[VG]) } };

(*-------------------------------------------*)
(* After EWSB                                *)
(*-------------------------------------------*)

(* Gauge Sector *)

DEFINITION[EWSB][GaugeSector]= 
{{VWB,{1,{VWm,1/Sqrt[2]},{conj[VWm],1/Sqrt[2]}},
      {2,{VWm,-\[ImaginaryI]/Sqrt[2]},{conj[VWm],\[ImaginaryI]/Sqrt[2]}},
      {3,{VP,Sin[ThetaW]},{VZ,Cos[ThetaW]}}},
 {VB, {1,{VP,Cos[ThetaW]},{VZ,-Sin[ThetaW]}}},
 {fWB,{1,{fWm,1/Sqrt[2]},{fWp,1/Sqrt[2]}},
      {2,{fWm,-\[ImaginaryI]/Sqrt[2]},{fWp,\[ImaginaryI]/Sqrt[2]}},
      {3,{fW0,1}}}};
        

(* ----- VEVs ---- *)

DEFINITION[EWSB][VEVs]= 
  {{SHd0, {vd, 1/Sqrt[2]}, {sigmad, \[ImaginaryI]/Sqrt[2]},{phid,1/Sqrt[2]}},
   {SHu0, {vu, 1/Sqrt[2]}, {sigmau, \[ImaginaryI]/Sqrt[2]},{phiu,1/Sqrt[2]}}};
 
(* ---- Mixings ---- *)

DEFINITION[EWSB][MatterSector]= 
{    {{SdL, SdR,SdMl}, {Sd, ZD}},
     {{SvL, SvMl}, {Sv, {Z^V}}},
     {{SuL, SuR}, {Su, ZU}},
     {{SeL, SeR,SeMl}, {Se, ZE}},
     {{phid, phiu}, {hh, ZH}},
     {{sigmad, sigmau}, {Ah, ZA}},
     {{SHdm,conj[SHup]},{Hpm,ZP}},
     {{fB, fW0, FHd0, FHu0}, {L0, ZN}}, 
     {{{fWm, FHdm}, {fWp, FHup}}, {{Lm,UM}, {Lp,UP}}},
     {{{FeL},{conj[FeR]}},{{FEL,ZEL},{FER,ZER}}},
     {{{FdL},{conj[FdR]}},{{FDL,ZDL},{FDR,ZDR}}},
     {{{FuL},{conj[FuR]}},{{FUL,ZUL},{FUR,ZUR}}}  
       };

(*--- Gauge Fixing ---- *)

DEFINITION[EWSB][GaugeFixing]=
  {	{Der[VP],                                            - 1/(2 RXi[VP])},	
	{Der[VWm]-\[ImaginaryI] Mass[VWm] RXi[VWm] Hpm[{1}], - 1/(RXi[VWm])},
	{Der[VZ] + Mass[VZ] RXi[VZ] Ah[{1}],                 - 1/(2 RXi[VZ])},
	{Der[VG],                                            - 1/(2 RXi[VG])}
																	};

(*------------------------------------------------------*)
(* Dirac-Spinors *)
(*------------------------------------------------------*)

dirac[[1]] = {Fd,  FDL, conj[FDR]};
dirac[[2]] = {Fe,  FEL, conj[FER]};
dirac[[3]] = {Fu,  FUL, conj[FUR]};
dirac[[4]] = {Fv,  FvL, 0};
dirac[[5]] = {Chi, L0, conj[L0]}; 
dirac[[6]] = {Cha, Lm, conj[Lp]};
dirac[[7]] = {Glu, fG, conj[fG]}; 
\end{verbatim}

\section{Mass matrices of light messengers}
\label{Masses_GMSB}
We show here just the mass matrices involving light scalar messenger fields. 
\begin{itemize} 
\item {\bf Mass matrix for Down-Squarks}: \(\left(\tilde{d}_{L,{{o_1} {\alpha_1}}}, \tilde{d}_{R,{{o_2} {\alpha_2}}}, \tilde{D}_{{{-,\alpha_3}}} \right)\) \\
\begin{align} 
m_{11} &= \frac{1}{24} \delta_{{\alpha_1},{\beta_1}} \Big(24 m^2_{\tilde{q},{{o_1} {p_1}}} - \Big(3 g_2^2  + g_1^2\Big)\delta_{{o_1},{p_1}} \Big(- v_u^2  + v_d^2 + \sum_{{a}=1}^{3}{v_{L,{{a}}}}^{2}\Big)\nonumber \\ 
 &+12 v_d^2 \sum_{{a}=1}^{3}Y^{*}_{d,{a} {p_1}} Y_{d,{a} {o_1}}  \Big)\\ 
m_{12} &= \frac{1}{\sqrt{2}} \delta_{{\alpha_1},{\beta_2}} \Big(v_d T^d_{{p_2} {o_1}}  - v_u \mu^* Y_{d,{p_2} {o_1}} \Big) \\
m_{13} &= -\frac{1}{2} v_d m_{3/2} \delta_{{\alpha_1},{\beta_3}} \sum_{{a}=1}^{3}f^*_{{a}} Y_{d,{a} {o_1}} \\
m_{22} &= \frac{1}{12} \delta_{{\alpha_2},{\beta_2}} \Big(12 \Big(|m_{3/2}|^2 f^*_{{o_2}} f_{{p_2}}  + m^2_{\tilde{d},{{p_2} {o_2}}}\Big)- g_1^2 \delta_{{o_2},{p_2}} \Big(- v_u^2  + v_d^2 + \sum_{{a}=1}^{3}{v_{L,{{a}}}}^{2}\Big)\nonumber \\ 
 &+6 v_d^2 \sum_{{a}=1}^{3}Y^{*}_{d,{o_2} {a}} Y_{d,{p_2} {a}}  \Big)\\ 
m_{23} &= \frac{1}{\sqrt{2}} m_{3/2} \Big(- B^*_{f,{{o_2}}}  + M_D f^*_{{o_2}} \Big)\delta_{{\alpha_2},{\beta_3}} \\
m_{33} &= \frac{1}{2} \Big(\delta_{{\alpha_3},{\beta_3}} \Big(2 |M_D|^2  + |m_{3/2}|^2 \sum_{{a}=1}^{3}f^*_{{a}} f_{{a}}   + m_{D_1}^2 + m_{D_2}^2\Big) - F_D  - F_D^* \Big)
\end{align} 

\item {\bf Mass matrix for Sneutrinos} \(\left(\tilde{\nu}_{L,{{o_1}}}, \tilde{\nu}_-\right)\) \\
\begin{equation} 
m^2_{\tilde{\nu}} = \left( 
\begin{array}{cc}
m_{11} &\frac{1}{\sqrt{2}} m_{3/2} \Big(- B_{f,{{o_1}}}  + M_L^* f_{{{o_1}}} \Big)\\ 
\frac{1}{\sqrt{2}} m_{3/2}^* \Big(- B^*_{f,{{p_1}}}  + M_L f^*_{{p_1}} \Big) &m_{22}\end{array} 
\right) 
\end{equation} 
\begin{align} 
m_{11} &= \frac{1}{8} \Big(8 \Big(|m_{3/2}|^2 f^*_{{p_1}} f_{{{o_1}}}  + m_{\tilde{l},{{o_1} {p_1}}}^{2}\Big) + \Big(g_{1}^{2} + g_{2}^{2}\Big)\Big( v_{d}^{2}- v_{u}^{2}\Big)\delta_{{o_1}{p_1}} \Big)\\ 
m_{22} &= \frac{1}{2} \Big(2 |M_L|^2  -2 {\Re\Big(F_L\Big)}  + |m_{3/2}|^2 \sum_{a=1}^{3}f^*_{a} f_{{a}}   + m_{L_1}^2 + m_{L_2}^2\Big)
\end{align} 

\item {\bf Mass matrix for charged Sleptons} \(\left(\tilde{e}_{L,{{o_1}}}, \tilde{e}_{R,{{o_2}}}, \tilde{E}_-\right)\) \\
\begin{equation} 
m^2_{\tilde{e}} = \left( 
\begin{array}{ccc}
m_{11} &\frac{1}{\sqrt{2}} \Big(v_d T_{{{p_2} {o_1}}}^{e}  - v_u \mu^* Y_{{{p_2} {o_1}}}^{e} \Big) &\frac{1}{\sqrt{2}} m_{3/2} \Big(M_L^* f_{{{o_1}}}  + B_{f,{{o_1}}}\Big)\\ 
\frac{1}{\sqrt{2}} \Big(v_d T^{e,*}_{{o_2} {p_1}}  - v_u \mu Y^{e,*}_{{o_2} {p_1}} \Big) &m_{22} &\frac{1}{2} m_{3/2} v_d \sum_{a=1}^{3}f_{{a}} \sum_{b=1}^{3}Y^{e,*}_{{o_2} b} \\ 
\frac{1}{\sqrt{2}} m_{3/2}^* \Big(M_L f^*_{{p_1}}  + B^*_{f,{{p_1}}}\Big) &\frac{1}{2} v_d m_{3/2}^* \sum_{a=1}^{3}f^*_{a} \sum_{b=1}^{3}Y_{{{p_2} b}}^{e}  &m_{33}\end{array} 
\right) 
\end{equation} 
\begin{align} 
m_{11} &= \frac{1}{8} \Big(4 v_{d}^{2} \sum_{a=1}^{3}Y^{e,*}_{a {p_1}} Y_{{a {o_1}}}^{e}   + 8 |m_{3/2}|^2 f^*_{{p_1}} f_{{{o_1}}}  + 8 m_{\tilde{l},{{o_1} {p_1}}}^{2}  + \Big(g_{1}^{2}- g_{2}^{2}\Big)\Big( v_{d}^{2}- v_{u}^{2}\Big)\delta_{{o_1}{p_1}} \Big)\\ 
m_{22} &= \frac{1}{4} \Big(2 v_{d}^{2} \sum_{a=1}^{3}Y^{e,*}_{{o_2} a} Y_{{{p_2} a}}^{e}   + 4 m_{\tilde{e},{{p_2} {o_2}}}^{2}  + g_{1}^{2} \Big(- v_{d}^{2}  + v_{u}^{2}\Big)\delta_{{o_2}{p_2}} \Big)\\ 
m_{33} &= \frac{1}{2} \Big(2 |M_L|^2  -2 {\Re\Big(F_L\Big)}  + |m_{3/2}|^2 \sum_{a=1}^{3}f^*_{a} f_{{a}}   + m_{L_1}^2 + m_{L_2}^2\Big)
\end{align} 
\end{itemize}

{\allowdisplaybreaks
\section{Interactions}
\label{Vertices_GMSB}
We list here only the interactions involving gauge bosons and all interactions of the neutral and strongly interacting messengers. The pure scalar interactions are not so important for our purposes and rather lengthy. Also four-point interactions are not present, since we have considered only two-body decays in our studies. Our conventions are explained in \ref{app:ConVertices}.
\subsection*{Interactions between two fermions and one scalar}
\begin{align} 
\Gamma^L_{\tilde{g}_{{i}}d_{{j \beta}}\tilde{d}^*_{{k \gamma}}}  =  & \,
-i \frac{1}{\sqrt{2}} g_3 \lambda^{i}_{\gamma,\beta} \sum_{a=1}^{3}U^{d,*}_{L,{j a}} Z_{{k a}}^{D}  \\ 
 \Gamma^R_{\tilde{g}_{{i}}d_{{j \beta}}\tilde{d}^*_{{k \gamma}}}  =  & \,i \frac{1}{\sqrt{2}} g_3 \lambda^{i}_{\gamma,\beta} \sum_{a=1}^{3}Z_{{k 3 + a}}^{D} U_{R,{j a}}^{d}   
\\ 
\Gamma^L_{\tilde{\chi}_{{i}}\nu_{{j}}\tilde{\nu}^*_{{k}}}  =  & \,
i \frac{1}{\sqrt{2}} \Big(g_1 N^*_{i 1}  - g_2 N^*_{i 2} \Big)\Theta_{j,3} {Z^V}_{{k j}} \\ 
\\ 
\Gamma^L_{\tilde{\chi}_{{i}}d_{{j \beta}}\tilde{d}^*_{{k \gamma}}}  =  & \,
-\frac{i}{6} \delta_{\beta\gamma} \Big(-3 \sqrt{2} g_2 N^*_{i 2} \sum_{a=1}^{3}U^{d,*}_{L,{j a}} Z_{{k a}}^{D}   + 6 N^*_{i 3} \sum_{b=1}^{3}U^{d,*}_{L,{j b}} \sum_{a=1}^{3}Y_{{a b}}^{d} Z_{{k 3 + a}}^{D}    \nonumber \\
 & + \sqrt{2} g_1 N^*_{i 1} \sum_{a=1}^{3}U^{d,*}_{L,{j a}} Z_{{k a}}^{D}  \Big)\\ 
 \Gamma^R_{\tilde{\chi}_{{i}}d_{{j \beta}}\tilde{d}^*_{{k \gamma}}}  =  & \,-\frac{i}{3} \delta_{\beta\gamma} \Big(3 \sum_{b=1}^{3}\sum_{a=1}^{3}Y^{d,*}_{a b} U_{R,{j a}}^{d}  Z_{{k b}}^{D}  N_{{i 3}}  + \sqrt{2} g_1 \sum_{a=1}^{3}Z_{{k 3 + a}}^{D} U_{R,{j a}}^{d}  N_{{i 1}} \Big) 
\\ 
\Gamma^L_{\tilde{\chi}^-_{{i}}e_{{j}}\tilde{\nu}^*_{{k}}}  =  & \,
-i g_2 U^*_{i 1} \sum_{a=1}^{3}U^{e,*}_{L,{j a}} {Z^V}_{{k a}}  \\ 
 \Gamma^R_{\tilde{\chi}^-_{{i}}e_{{j}}\tilde{\nu}^*_{{k}}}  =  & \,i \sum_{b=1}^{3}\sum_{a=1}^{3}Y^{e,*}_{a b} U_{R,{j a}}^{e}  {Z^V}_{{k b}}  V_{{i 2}}  
\\ 
\Gamma^L_{\tilde{\chi}^-_{{i}}\bar{u}_{{j \beta}}\tilde{d}_{{k \gamma}}}  =  & \,
i U^*_{i 2} \delta_{\beta\gamma} \sum_{b=1}^{3}Z^{D,*}_{k b} \sum_{a=1}^{3}U^{u,*}_{R,{j a}} Y_{{a b}}^{u}   \\ 
 \Gamma^R_{\tilde{\chi}^-_{{i}}\bar{u}_{{j \beta}}\tilde{d}_{{k \gamma}}}  =  & \,i \delta_{\beta\gamma} \Big(- g_2 \sum_{a=1}^{3}Z^{D,*}_{k a} U_{L,{j a}}^{u}  V_{{i 1}}  + \sum_{b=1}^{3}\sum_{a=1}^{3}Y^{d,*}_{a b} Z^{D,*}_{k 3 + a}  U_{L,{j b}}^{u}  V_{{i 2}} \Big) 
\\ 
\Gamma^L_{\bar{e}_{{i}}\tilde{\chi}^+_{{j}}\tilde{\nu}_{{k}}}  =  & \,
i V^*_{j 2} \sum_{b=1}^{3}{Z^V}^*_{k b} \sum_{a=1}^{3}U^{e,*}_{R,{i a}} Y_{{a b}}^{e}   \\ 
 \Gamma^R_{\bar{e}_{{i}}\tilde{\chi}^+_{{j}}\tilde{\nu}_{{k}}}  =  & \,-i g_2 \sum_{a=1}^{3}{Z^V}^*_{k a} U_{L,{i a}}^{e}  U_{{j 1}}  
\\ 
\Gamma^L_{\bar{d}_{{i \alpha}}\tilde{\chi}_{{j}}\tilde{d}_{{k \gamma}}}  =  & \,
-\frac{i}{3} \delta_{\alpha\gamma} \Big(3 N^*_{j 3} \sum_{b=1}^{3}Z^{D,*}_{k b} \sum_{a=1}^{3}U^{d,*}_{R,{i a}} Y_{{a b}}^{d}    + \sqrt{2} g_1 N^*_{j 1} \sum_{a=1}^{3}Z^{D,*}_{k 3 + a} U^{d,*}_{R,{i a}}  \Big)\\ 
 \Gamma^R_{\bar{d}_{{i \alpha}}\tilde{\chi}_{{j}}\tilde{d}_{{k \gamma}}}  =  & \,-\frac{i}{6} \delta_{\alpha\gamma} \Big(6 \sum_{b=1}^{3}\sum_{a=1}^{3}Y^{d,*}_{a b} Z^{D,*}_{k 3 + a}  U_{L,{i b}}^{d}  N_{{j 3}}  + \sqrt{2} \sum_{a=1}^{3}Z^{D,*}_{k a} U_{L,{i a}}^{d}  \Big(-3 g_2 N_{{j 2}}  + g_1 N_{{j 1}} \Big)\Big) 
\\ 
\end{align}

\subsection*{Interactions between two scalars and one vector boson}
\begin{align} 
\Gamma_{\tilde{d}_{{i \alpha}}\tilde{d}^*_{{j \beta}}Z_{{\mu}}}  = & \, 
\frac{i}{6} \delta_{\alpha\beta} \Big(-2 g_1 {s_{\Theta_W}}  \Big(Z^{D,*}_{i 7} Z_{{j 7}}^{D}  + \sum_{a=1}^{3}Z^{D,*}_{i 3 + a} Z_{{j 3 + a}}^{D} \Big) + \Big(3 g_2 {c_{\Theta_W}} \nonumber \\
 & \hspace{1cm}    + g_1 {s_{\Theta_W}}  \Big)\sum_{a=1}^{3}Z^{D,*}_{i a} Z_{{j a}}^{D}  \Big) 
\\ 
\Gamma_{\tilde{\nu}_{{i}}\tilde{e}^*_{{j}}W^-_{{\mu}}}  = & \, 
-i \frac{1}{\sqrt{2}} g_2 \Big({Z^V}^*_{i 4} Z_{{j 7}}^{E}  + \sum_{a=1}^{3}{Z^V}^*_{i a} Z_{{j a}}^{E} \Big) 
\\ 
\Gamma_{\tilde{\nu}_{{i}}\tilde{\nu}^*_{{j}}Z_{{\mu}}}  = & \, 
-\frac{i}{2} \Big(g_1 {s_{\Theta_W}}   + g_2 {c_{\Theta_W}}  \Big)\Big({Z^V}^*_{i 4} {Z^V}_{{j 4}}  + \sum_{a=1}^{3}{Z^V}^*_{i a} {Z^V}_{{j a}} \Big) 
\\ 
\Gamma_{\tilde{u}_{{i \alpha}}\tilde{d}^*_{{j \beta}}W^-_{{\mu}}}  = & \, 
-i \frac{1}{\sqrt{2}} g_2 \delta_{\alpha\beta} \sum_{a=1}^{3}Z^{U,*}_{i a} Z_{{j a}}^{D}   
\end{align} 
}

\section{Model file for SARAH including \texorpdfstring{$R$}{R}-parity violation}
\label{Modelfile_GMSB_RpV}
We show here only the differences to the model file without $R$-parity violation
\begin{verbatim}
Off[General::spell]
Print["Model file for the MSSM with SU(5)-GMSB Messengers 
          and R-Parity Violation loaded"];

ModelNameLaTeX ="GMSB_RpV";

...

(*------------------------------------------------------*)
(* Superpotential *)
(*------------------------------------------------------*)

         
SuperPotential = { {{1, Yu},{u,q,Hu}}, {{-1,Yd},{d,q,Hd}},
                   {{-1,Ye},{e,l,Hd}}, {{1,\[Mu]},{Hu,Hd}},
                   {{1,MessD},{dM1,dM2}}, {{1,MessL},{lM1,lM2}},
                   {{m32,f},{d,dM2}},     {{m32,f},{l,lM2}}, 
                   {{1,\[Epsilon]},{l,Hu}}  };
  
...

(*----------------------------------------------*)
(*   DEFINITION                                 *)
(*----------------------------------------------*)

...

DEFINITION[GaugeES][Additional]={
{{SdRM1,SdRM2},{1,FD}},
{{SeLM1,SeLM2},{1,FL}},
{{SvLM1,SvLM2},{1,FL}},
{{conj[SvL], SHd0}, {1, mHL2}},
{{conj[SeL], SHdm}, {1, mHL2}} 
};
        
(* ----- VEVs ---- *)

DEFINITION[EWSB][VEVs]= 
  {{SHd0, {vd, 1/Sqrt[2]}, {sigmad, \[ImaginaryI]/Sqrt[2]},{phid,1/Sqrt[2]}},
   {SHu0, {vu, 1/Sqrt[2]}, {sigmau, \[ImaginaryI]/Sqrt[2]},{phiu,1/Sqrt[2]}},
   {SvL, {vL, 1/Sqrt[2]}, {sigmaL, \[ImaginaryI]/Sqrt[2]},{phiL,1/Sqrt[2]}},
   {SvMl, {vM, 1/Sqrt[2]}, {sigmaM, \[ImaginaryI]/Sqrt[2]},{phiM,1/Sqrt[2]}}
};
 
(* ---- Mixings ---- *)

DEFINITION[EWSB][MatterSector]= 
{    {{SdL, SdR,SdMl}, {Sd, ZD}},
     {{SuL, SuR}, {Su, ZU}},
     {{phid, phiu,phiL,phiM}, {hh, ZH}},
     {{sigmad, sigmau,sigmaL,sigmaM}, {Ah, ZA}},
     {{SHdm,conj[SHup],SeL,SeR,SeMl},{Hpm,ZP}},
     {{FvL,fB, fW0, FHd0, FHu0}, {L0, ZN}}, 
     {{{FeL,fWm, FHdm}, {conj[FeR],fWp, FHup}},  {{Lm,UM}, {Lp,UP}}},
     {{{FdL},{conj[FdR]}},{{FDL,ZDL},{FDR,ZDR}}},
     {{{FuL},{conj[FuR]}},{{FUL,ZUL},{FUR,ZUR}}}      \
       }; 

\end{verbatim}

{\allowdisplaybreaks 
\section{Mass matrices of light messengers including \texorpdfstring{$R$}{R}-parity violation}
\label{Masses_GMSB_RpV}
\begin{itemize}

\item {\bf Mass matrix for Higgs} \(\left(\phi_{d}, \phi_{u}, \phi_{{l},{{o_3}}}, \phi_-\right)\) \\
\begin{align} 
m_{11} &= \frac{1}{8} \Big(8 m_{H_d}^2  + 8 |\mu|^2  + \Big(g_1^2 + g_2^2\Big)\Big(3 v_d^2  - v_u^2 \Big) + \Big(g_1^2 + g_2^2\Big)\sum_{{a}=1}^{3}{v_{L,{{a}}}}^{2} \Big)\\ 
m_{21} &= \frac{1}{4} \Big(-2 B_{\mu}  -2 B_{\mu}^*  - \Big(g_1^2 + g_2^2\Big)v_d v_u \Big)\\ 
m_{22} &= m_{H_u}^2  + |\mu|^2  + \sum_{{a}=1}^{3}{\epsilon_{{a}}}^{2} + \frac{1}{8} \Big( - \Big(g_1^2 + g_2^2\Big)\Big(v^2_d-3 v_u^2\Big) - \Big(g_1^2 + g_2^2\Big)\sum_{{a}=1}^{3}{v_{L,{{a}}}}^{2} \Big)\\ 
m_{31} &= \frac{1}{4} \Big(2 m^{2,*}_{{H l},{{o_3}}}  + 2 m^2_{{H l},{{o_3}}}  -2 \Big(\mu + \mu^*\Big)\epsilon_{{o_3}}  + \Big(g_1^2 + g_2^2\Big)v_d \sum_{{a}=1}^{3}v_{L,{{a}}} \Big)\\ 
m_{32} &= -\frac{1}{4} \Big(g_1^2 + g_2^2\Big)v_u \sum_{{a}=1}^{3}v_{L,{{a}}}  + B_{\epsilon,{{o_3}}}\\ 
m_{33} &= \frac{1}{8} \Big(\Big(g_1^2 + g_2^2\Big)\Big(2 \sum_{{a}=1}^{3}v_{L,{{a}}} \sum_{{b}=1}^{3}v_{L,{{b}}}  + \delta_{{o_3},{p_3}} \Big(- v_u^2  + v_d^2 + \sum_{{a}=1}^{3}{v_{L,{{a}}}}^{2}\Big)\Big)\nonumber \\ 
 &+4 \Big(2 \epsilon_{{o_3}} \epsilon_{{p_3}}  + |m_{3/2}|^2 \Big(f^*_{{o_3}} f_{{p_3}}  + f^*_{{p_3}} f_{{o_3}} \Big) + m^2_{\tilde{l},{{o_3} {p_3}}} + m^2_{\tilde{l},{{p_3} {o_3}}}\Big)\Big)\\ 
m_{42} &= -\frac{1}{2} \frac{1}{\sqrt{2}} \Big(m_{3/2} \sum_{{a}=1}^{3}f^*_{{a}} \epsilon_{{a}}   + m_{3/2} \sum_{{a}=1}^{3}f_{{a}} \epsilon_{{a}}  \Big)\\ 
m_{43} &= \frac{1}{2} \frac{1}{\sqrt{2}} \Big(- m_{3/2} B^*_{f,{{p_3}}}  - m_{3/2} B_{f,{{p_3}}}  + m_{3/2} M_L^* f_{{p_3}}  + M_L m_{3/2} f^*_{{p_3}} \Big)\\ 
m_{44} &= \frac{1}{2} \Big(2 |M_L|^2  - 2  \mathrm{Re}\big\{  F_L \big\}  + |m_{3/2}|^2 \sum_{{a}=1}^{3}f^*_{{a}} f_{{a}}   + m_{L_1}^2 + m_{L_2}^2\Big)
\end{align} 
\item {\bf Mass matrix for Pseudo-Scalar Higgs} \(\left(\sigma_{d}, \sigma_{u}, \sigma_{{l},{{o_3}}}, \sigma_-\right) \)\\

\begin{align} 
m_{11} &= \frac{1}{8} \Big(8 m_{H_d}^2  + 8 |\mu|^2  + \Big(g_1^2 + g_2^2\Big)\sum_{{a}=1}^{3}{v_{L,{{a}}}}^{2}  + \Big(g_1^2 + g_2^2\Big)\Big(- v_u^2  + v_d^2\Big)\Big)\\ 
m_{12} & = \mathrm{Re}\big\{B_{\mu}\big\} \\
m_{22} &= \frac{1}{8} \Big(8 m_{H_u}^2  + 8 |\mu|^2  + 8 \sum_{{a}=1}^{3}{\epsilon_{{a}}}^{2}  - \Big(g_1^2 + g_2^2\Big)\sum_{{a}=1}^{3}{v_{L,{{a}}}}^{2} - \Big(g_1^2 + g_2^2\Big)\Big(v_d^2 - v_u^2\Big)\Big)\\ 
m_{31} &= \frac{1}{2} \Big(- \Big(\mu + \mu^*\Big)\epsilon_{{o_3}}  + m^{2,*}_{{H l},{{o_3}}} + m^2_{{H l},{{o_3}}}\Big)\\ 
m_{32} &= - B_{\epsilon,{{p_3}}} \\
m_{33} &= \frac{1}{8} \Big(\Big(g_1^2 + g_2^2\Big)\delta_{{o_3},{p_3}} \Big(- v_u^2  + v_d^2 + \sum_{{a}=1}^{3}{v_{L,{{a}}}}^{2}\Big)\nonumber \\ 
 &+4 \Big(2 \epsilon_{{o_3}} \epsilon_{{p_3}}  + |m_{3/2}|^2 \Big(f^*_{{o_3}} f_{{p_3}}  + f^*_{{p_3}} f_{{o_3}} \Big) + m^2_{\tilde{l},{{o_3} {p_3}}} + m^2_{\tilde{l},{{p_3} {o_3}}}\Big)\Big)\\ 
m_{42} &= \frac{1}{2} \frac{1}{\sqrt{2}} \Big(m_{3/2} \sum_{{a}=1}^{3}f^*_{{a}} \epsilon_{{a}}   + m_{3/2} \sum_{{a}=1}^{3}f_{{a}} \epsilon_{{a}}  \Big)\\ 
m_{43} &= \frac{1}{2} \frac{1}{\sqrt{2}} \Big(- m_{3/2} B^*_{f,{{p_3}}}  - m_{3/2} B_{f,{{p_3}}}  + m_{3/2} M_L^* f_{{p_3}}  + M_L m_{3/2} f^*_{{p_3}} \Big)\\ 
m_{44} &= \frac{1}{2} \Big(2 |M_L|^2  - 2  \mathrm{Re}\big\{  F_L \big\}  + |m_{3/2}|^2 \sum_{{a}=1}^{3}f^*_{{a}} f_{{a}}   + m_{L_1}^2 + m_{L_2}^2\Big)
\end{align} 
\item {\bf Mass matrix for Charged Higgs} \(\left(H_d^-, H_u^{+,*}, \tilde{e}_{L,{{o_3}}}, \tilde{e}_{R,{{o_4}}}, \tilde{E}_l \right)\)\\

\begin{align} 
m_{11} &= \frac{1}{8} \Big(8 m_{H_d}^2 +g_1^2 \Big(- v_u  + v_d\Big)\Big(v_d + v_u\Big)+g_2^2 \Big(v_d^2 + v_u^2\Big)+8 |\mu|^2 \nonumber \\ 
 &+\Big(- g_2  + g_1\Big)\Big(g_1 + g_2\Big)\sum_{{a}=1}^{3}{v_{L,{{a}}}}^{2} +4 \sum_{{c}=1}^{3}\sum_{{b}=1}^{3}\sum_{{a}=1}^{3}Y^{*}_{e,{a} {c}} Y_{e,{a} {b}}  v_{L,{{b}}}  v_{L,{{c}}}  \Big)\\ 
m_{14} &= \frac{1}{4} g_2^2 v_d v_M \\
m_{21} &= \frac{1}{4} g_2^2 v_d v_u  + B_{\mu}^*\\ 
m_{22} &=  m_{H_u}^2  +  |\mu|^2  + \sum_{{a}=1}^{3}{\epsilon_{{a}}}^{2}  +\frac{1}{8} \Big( \Big(g_2^2- g_1^2 \Big)\sum_{{a}=1}^{3}{v_{L,{{a}}}}^{2} + g_1^2 \Big(v_u^2- v_d^2 \Big) + g_2^2 \Big(v_d^2 + v_u^2\Big)\Big)\\ 
m_{31} &= +m^{2,*}_{{H l},{{o_3}}}+\frac{1}{4} \Big(-2 v_d \sum_{{b}=1}^{3}\sum_{{a}=1}^{3}Y^{*}_{e,{a} {b}} Y_{e,{a} {o_3}}  v_{L,{{b}}}   -4 \mu^* \epsilon_{{o_3}}  + g_2^2 v_d \sum_{{a}=1}^{3}v_{L,{{a}}} \Big)\\ 
m_{32} &= - B_{\epsilon,{{o_3}}}  + \frac{1}{4} g_2^2 v_u \sum_{{a}=1}^{3}v_{L,{{a}}} \\ 
m_{33} &= \frac{1}{8} \Big(\Big(g_1- g_2 \Big)\Big(g_1 + g_2\Big)\delta_{{o_3},{p_3}} \Big(- v_u^2  + v_d^2 + \sum_{{a}=1}^{3}{v_{L,{{a}}}}^{2}\Big)\nonumber \\ 
 &+4 v_d^2 \sum_{{a}=1}^{3}Y^{*}_{e,{a} {p_3}} Y_{e,{a} {o_3}}  +2 g_2^2 \sum_{{a}=1}^{3}v_{L,{{a}}} \sum_{{b}=1}^{3}v_{L,{{b}}} +8 \Big(\epsilon_{{o_3}} \epsilon_{{p_3}}  + |m_{3/2}|^2 f^*_{{p_3}} f_{{o_3}}  + m^2_{\tilde{l},{{o_3} {p_3}}}\Big)\Big)\\ 
m_{41} &= \frac{1}{2} \Big(\Big(m_{3/2} v_M \sum_{{a}=1}^{3}f_{{a}}  - \sqrt{2} v_u \sum_{{a}=1}^{3}\epsilon_{{a}} \Big)\sum_{{b}=1}^{3}Y^{*}_{e,{o_4} {b}}  - \sqrt{2} \sum_{{a}=1}^{3}T^{e,*}_{{o_4} {a}} v_{L,{{a}}}  \Big)\\ 
m_{42} &= - \frac{1}{\sqrt{2}} \Big(\mu \sum_{{a}=1}^{3}Y^{*}_{e,{o_4} {a}} v_{L,{{a}}}   + v_d \sum_{{a}=1}^{3}\epsilon_{{a}} \sum_{{b}=1}^{3}Y^{*}_{e,{o_4} {b}} \Big)\\ 
m_{43} &= \frac{1}{\sqrt{2}} \Big(v_d T^{e,*}_{{o_4} {p_3}}  - v_u \mu Y^{*}_{e,{o_4} {p_3}} \Big)\\ 
m_{44} &= \frac{1}{4} \Big(4 m^2_{\tilde{e},{{p_4} {o_4}}} - g_1^2 \delta_{{o_4},{p_4}} \Big(- v_u^2  + v_d^2 + \sum_{{a}=1}^{3}{v_{L,{{a}}}}^{2}\Big)\nonumber \\ 
 &+2 v_d^2 \sum_{{a}=1}^{3}Y^{*}_{e,{o_4} {a}} Y_{e,{p_4} {a}}  +2 \sum_{{a}=1}^{3}v_{L,{{a}}} Y_{e,{p_4} {a}}  \sum_{{b}=1}^{3}Y^{*}_{e,{o_4} {b}} v_{L,{{b}}}  \Big)\\ 
m_{45} &= \frac{1}{2} m_{3/2} v_d \sum_{{a}=1}^{3}f_{{a}} \sum_{{b}=1}^{3}Y^{*}_{e,{o_4} {b}} \\
m_{52} &= \frac{1}{4} \Big(-2 \sqrt{2} m_{3/2} \sum_{{a}=1}^{3}f^*_{{a}} \epsilon_{{a}}   + g_2^2 v_M v_u \Big)\\ 
m_{53} &= \frac{1}{4} g_2^2 v_M \sum_{{a}=1}^{3}v_{L,{{a}}}  + \frac{1}{\sqrt{2}} m_{3/2} \Big(M_L f^*_{{p_3}}  + B^*_{f,{{p_3}}}\Big)\\ 
m_{55} &= \frac{1}{4} \Big( -2  \mathrm{Re}\big\{  F_L \big\}  + 2 |m_{3/2}|^2 \sum_{{a}=1}^{3}f^*_{{a}} f_{{a}}+ 2 \Big(m_{L_1}^2   + m_{L_2}^2\Big) + 4 |M_L|^2  + g_2^2 {v_M}^{2} \Big)
\end{align} 
\end{itemize} 

\section{Interactions}
We concentrate on the  interactions between two scalars and one vector boson responsible for the dominant decay channels. In addition, we list the interactions between one scalar and two vector boson, since they are new in comparison to the case with conserved $R$-parity. 
\label{Vertices_GMSB_RpV}

\subsection*{Interactions between two vector bosons and one scalar}
\begin{align} 
\Gamma_{h_{{i}}W^+_{{\sigma}}W^-_{{\mu}}}  = & \, 
\frac{i}{2} g_{2}^{2} \Big(v_d Z^{H,*}_{i 1}  + v_M Z^{H,*}_{i 6}  + v_u Z^{H,*}_{i 2}  + \sum_{a=1}^{3}Z^{H,*}_{i 2 + a} v_{L,{a}} \Big) 
\\ 
\Gamma_{h_{{i}}Z_{{\sigma}}Z_{{\mu}}}  = & \, 
\frac{i}{2} \Big(g_1 {s_{\Theta_W}}   + g_2 {c_{\Theta_W}}  \Big)^{2} \Big(v_d Z^{H,*}_{i 1}  + v_M Z^{H,*}_{i 6}  + v_u Z^{H,*}_{i 2}  + \sum_{a=1}^{3}Z^{H,*}_{i 2 + a} v_{L,{a}} \Big) 
\end{align} 

\subsection*{Interactions between two scalars and one vector boson}
\begin{align} 
\Gamma_{h_{{i}}H^+_{{j}}W^-_{{\mu}}}  = & \, 
-\frac{i}{2} g_2 \Big(Z^{H,*}_{i 1} Z_{{j 1}}^{+}  - Z^{H,*}_{i 2} Z_{{j 2}}^{+}  + Z^{H,*}_{i 6} Z_{{j 9}}^{+}  + \sum_{a=1}^{3}Z^{H,*}_{i 2 + a} Z_{{j 2 + a}}^{+} \Big) 
\\ 
\Gamma_{h_{{i}}A_{h,{j}}Z_{{\mu}}}  = & \, 
\frac{1}{2} \Big(- g_1 {s_{\Theta_W}}   - g_2 {c_{\Theta_W}}  \Big)\Big(Z^{A,*}_{j 1} Z^{H,*}_{i 1}  - Z^{A,*}_{j 2} Z^{H,*}_{i 2}  + Z^{A,*}_{j 6} Z^{H,*}_{i 6}  + \sum_{a=1}^{3}Z^{A,*}_{j 2 + a} Z^{H,*}_{i 2 + a} \Big) 
\\ 
\Gamma_{A_{h,{i}}H^+_{{j}}W^-_{{\mu}}}  = & \, 
\frac{1}{2} g_2 \Big(Z^{A,*}_{i 1} Z_{{j 1}}^{+}  + Z^{A,*}_{i 2} Z_{{j 2}}^{+}  + Z^{A,*}_{i 6} Z_{{j 9}}^{+}  + \sum_{a=1}^{3}Z^{A,*}_{i 2 + a} Z_{{j 2 + a}}^{+} \Big) 
\end{align} 
}
  \chapter{Renormalization group equations for seesaw scenarios}
\label{app:RGE_Seesaw}
The complete expressions for all parameters can also be produced with \verb"SARAH" using the model files \verb"Seesaw2" and \verb"Seesaw3". In addition, we show for completeness the main results for the Seesaw I which were already known in literature. Here and in the subsequent sections ${\bf 1}$ denotes the 3$\times$3 unit matrix. \(N_X\) is the number of generations of heavy field \(X\). Furthermore, we define
\begin{equation}
 \tilde{N}_X = N_X + N_{\bar{X}}
\end{equation}  
\section{Seesaw I}
\subsection{Anomalous dimensions}
\label{sec:AnaDimI}
{\allowdisplaybreaks \begin{align} 
\gamma_{\hat{q}}^{(1)} = \, &
-\frac{1}{30} \Big(45 g_{2}^{2}  + 80 g_{3}^{2}  + g_{1}^{2}\Big){\bf 1}  + {Y_{d}^{\dagger}  Y_d} + {Y_{u}^{\dagger}  Y_u}\\ 
\gamma_{\hat{q}}^{(2)} = \, &
+\Big[8 g_{2}^{2} g_{3}^{2}  + \frac{15}{4} g_{2}^{4}  + \frac{1}{90} g_{1}^{2} \Big(16 g_{3}^{2}  + 9 g_{2}^{2} \Big) + \frac{199}{900} g_{1}^{4}  -\frac{8}{9} g_{3}^{4} \Big]{\bf 1} +\frac{4}{5} g_{1}^{2} {Y_{u}^{\dagger}  Y_u} -2 {Y_{d}^{\dagger}  Y_d  Y_{d}^{\dagger}  Y_d} \nonumber \\ 
 &-2 {Y_{u}^{\dagger}  Y_u  Y_{u}^{\dagger}  Y_u} +{Y_{d}^{\dagger}  Y_d} \Big[-3 \mbox{Tr}\Big({Y_d  Y_{d}^{\dagger}}\Big)  + \frac{2}{5} g_{1}^{2}  - \mbox{Tr}\Big({Y_e  Y_{e}^{\dagger}}\Big) \Big]-3 {Y_{u}^{\dagger}  Y_u} \mbox{Tr}\Big({Y_u  Y_{u}^{\dagger}}\Big) \nonumber \\ 
 &- {Y_{u}^{\dagger}  Y_u} \mbox{Tr}\Big({Y_v  Y_{v}^{\dagger}}\Big) \\ 
\gamma_{\hat{l}}^{(1)} = \, &
-\frac{3}{10} \Big(5 g_{2}^{2}  + g_{1}^{2}\Big){\bf 1}  + {Y_{e}^{\dagger}  Y_e} + {Y_{v}^{\dagger}  Y_v}\\ 
\gamma_{\hat{l}}^{(2)} = \, &
+\frac{3}{100} \Big(125 g_{2}^{4}  + 30 g_{1}^{2} g_{2}^{2}  + 69 g_{1}^{4} \Big){\bf 1} -2 {Y_{e}^{\dagger}  Y_e  Y_{e}^{\dagger}  Y_e} -2 {Y_{v}^{\dagger}  Y_v  Y_{v}^{\dagger}  Y_v} \nonumber \\ 
 &+{Y_{e}^{\dagger}  Y_e} \Big[-3 \mbox{Tr}\Big({Y_d  Y_{d}^{\dagger}}\Big)  + \frac{6}{5} g_{1}^{2}  - \mbox{Tr}\Big({Y_e  Y_{e}^{\dagger}}\Big) \Big]-3 {Y_{v}^{\dagger}  Y_v} \mbox{Tr}\Big({Y_u  Y_{u}^{\dagger}}\Big) \nonumber \\ 
 &- {Y_{v}^{\dagger}  Y_v} \mbox{Tr}\Big({Y_v  Y_{v}^{\dagger}}\Big) \\ 
\gamma_{\hat{H}_d}^{(1)} = \, &
3 \mbox{Tr}\Big({Y_d  Y_{d}^{\dagger}}\Big)  -\frac{3}{10} \Big(5 g_{2}^{2}  + g_{1}^{2}\Big) + \mbox{Tr}\Big({Y_e  Y_{e}^{\dagger}}\Big)\\ 
\gamma_{\hat{H}_d}^{(2)} = \, &
+\frac{207}{100} g_{1}^{4} +\frac{9}{10} g_{1}^{2} g_{2}^{2} +\frac{15}{4} g_{2}^{4} -\frac{2}{5} \Big(-40 g_{3}^{2}  + g_{1}^{2}\Big)\mbox{Tr}\Big({Y_d  Y_{d}^{\dagger}}\Big) +\frac{6}{5} g_{1}^{2} \mbox{Tr}\Big({Y_e  Y_{e}^{\dagger}}\Big) \nonumber \\ 
 &-9 \mbox{Tr}\Big({Y_d  Y_{d}^{\dagger}  Y_d  Y_{d}^{\dagger}}\Big) -3 \mbox{Tr}\Big({Y_d  Y_{u}^{\dagger}  Y_u  Y_{d}^{\dagger}}\Big) -3 \mbox{Tr}\Big({Y_e  Y_{e}^{\dagger}  Y_e  Y_{e}^{\dagger}}\Big) - \mbox{Tr}\Big({Y_e  Y_{v}^{\dagger}  Y_v  Y_{e}^{\dagger}}\Big) \\ 
\gamma_{\hat{H}_u}^{(1)} = \, &
3 \mbox{Tr}\Big({Y_u  Y_{u}^{\dagger}}\Big)  -\frac{3}{10} \Big(5 g_{2}^{2}  + g_{1}^{2}\Big) + \mbox{Tr}\Big({Y_v  Y_{v}^{\dagger}}\Big)\\ 
\gamma_{\hat{H}_u}^{(2)} = \, &
+\frac{207}{100} g_{1}^{4} +\frac{9}{10} g_{1}^{2} g_{2}^{2} +\frac{15}{4} g_{2}^{4} +\frac{4}{5} \Big(20 g_{3}^{2}  + g_{1}^{2}\Big)\mbox{Tr}\Big({Y_u  Y_{u}^{\dagger}}\Big) -3 \mbox{Tr}\Big({Y_d  Y_{u}^{\dagger}  Y_u  Y_{d}^{\dagger}}\Big) \nonumber \\ 
 &- \mbox{Tr}\Big({Y_e  Y_{v}^{\dagger}  Y_v  Y_{e}^{\dagger}}\Big) -9 \mbox{Tr}\Big({Y_u  Y_{u}^{\dagger}  Y_u  Y_{u}^{\dagger}}\Big) -3 \mbox{Tr}\Big({Y_v  Y_{v}^{\dagger}  Y_v  Y_{v}^{\dagger}}\Big) \\ 
\gamma_{\hat{d}}^{(1)} = \, &
2 {Y_d^*  Y_{d}^{T}}  -\frac{2}{15} \Big(20 g_{3}^{2}  + g_{1}^{2}\Big){\bf 1} \\ 
\gamma_{\hat{d}}^{(2)} = \, &
+\frac{2}{225} \Big(-100 g_{3}^{4}  + 101 g_{1}^{4}  + 80 g_{1}^{2} g_{3}^{2} \Big){\bf 1} -2 \Big({Y_d^*  Y_{d}^{T}  Y_d^*  Y_{d}^{T}} + {Y_d^*  Y_{u}^{T}  Y_u^*  Y_{d}^{T}}\Big)\nonumber \\ 
 &+{Y_d^*  Y_{d}^{T}} \Big(-2 \mbox{Tr}\Big({Y_e  Y_{e}^{\dagger}}\Big)  + 6 g_{2}^{2}  -6 \mbox{Tr}\Big({Y_d  Y_{d}^{\dagger}}\Big)  + \frac{2}{5} g_{1}^{2} \Big)\\ 
\gamma_{\hat{u}}^{(1)} = \, &
2 {Y_u^*  Y_{u}^{T}}  -\frac{8}{15} \Big(5 g_{3}^{2}  + g_{1}^{2}\Big){\bf 1} \\ 
\gamma_{\hat{u}}^{(2)} = \, &
+\frac{8}{225} \Big(107 g_{1}^{4}  -25 g_{3}^{4}  + 80 g_{1}^{2} g_{3}^{2} \Big){\bf 1} -\frac{2}{5} \Big(5 \Big({Y_u^*  Y_{d}^{T}  Y_d^*  Y_{u}^{T}} + {Y_u^*  Y_{u}^{T}  Y_u^*  Y_{u}^{T}}\Big) \nonumber \\ 
 &+ {Y_u^*  Y_{u}^{T}} \Big(-15 g_{2}^{2}  + 15 \mbox{Tr}\Big({Y_u  Y_{u}^{\dagger}}\Big)  + 5 \mbox{Tr}\Big({Y_v  Y_{v}^{\dagger}}\Big)  + g_{1}^{2}\Big)\Big)\\ 
\gamma_{\hat{e}}^{(1)} = \, &
2 {Y_e^*  Y_{e}^{T}}  -\frac{6}{5} g_{1}^{2} {\bf 1} \\ 
\gamma_{\hat{e}}^{(2)} = \, &
+\frac{234}{25} g_{1}^{4} {\bf 1} -2 \Big({Y_e^*  Y_{e}^{T}  Y_e^*  Y_{e}^{T}} + {Y_e^*  Y_{v}^{T}  Y_v^*  Y_{e}^{T}}\Big)\nonumber \\ 
 &+{Y_e^*  Y_{e}^{T}} \Big(-2 \mbox{Tr}\Big({Y_e  Y_{e}^{\dagger}}\Big)  + 6 g_{2}^{2}  -6 \mbox{Tr}\Big({Y_d  Y_{d}^{\dagger}}\Big)  -\frac{6}{5} g_{1}^{2} \Big)\\ 
\gamma_{\hat{\nu}}^{(1)} = \, &
2 {Y_v^*  Y_{v}^{T}} \\ 
\gamma_{\hat{\nu}}^{(2)} = \, &
-2 \Big({Y_v^*  Y_{e}^{T}  Y_e^*  Y_{v}^{T}} + {Y_v^*  Y_{v}^{T}  Y_v^*  Y_{v}^{T}}\Big) \nonumber \\ 
 &+ {Y_v^*  Y_{v}^{T}} \Big(-2 \mbox{Tr}\Big({Y_v  Y_{v}^{\dagger}}\Big)  + 6 g_{2}^{2}  -6 \mbox{Tr}\Big({Y_u  Y_{u}^{\dagger}}\Big)  + \frac{6}{5} g_{1}^{2} \Big)
\end{align} } 
\subsection{Gauge couplings}
\label{sec:betaI}
{\allowdisplaybreaks  \begin{align} 
\beta_{g_1}^{(1)} = \, &
\frac{33}{5} g_{1}^{3} \\ 
\beta_{g_1}^{(2)} = \, &
\frac{1}{25} g_{1}^{3} \Big[-130 \mbox{Tr}\Big({Y_u  Y_{u}^{\dagger}}\Big)  + 135 g_{2}^{2}  + 199 g_{1}^{2}  -30 \mbox{Tr}\Big({Y_v  Y_{v}^{\dagger}}\Big)  \nonumber \\ 
 & \hspace{1cm}+ 440 g_{3}^{2}  -70 \mbox{Tr}\Big({Y_d  Y_{d}^{\dagger}}\Big)  -90 \mbox{Tr}\Big({Y_e  Y_{e}^{\dagger}}\Big) \Big]\\ 
\beta_{g_2}^{(1)} = \, &
g_{2}^{3}\\ 
\beta_{g_2}^{(2)} = \, &
\frac{1}{5} g_{2}^{3} \Big[-10 \mbox{Tr}\Big({Y_e  Y_{e}^{\dagger}}\Big)  -10 \mbox{Tr}\Big({Y_v  Y_{v}^{\dagger}}\Big)  + 120 g_{3}^{2}  + 125 g_{2}^{2}  \nonumber \\ 
 &\hspace{1cm}-30 \mbox{Tr}\Big({Y_d  Y_{d}^{\dagger}}\Big)  -30 \mbox{Tr}\Big({Y_u  Y_{u}^{\dagger}}\Big)  + 9 g_{1}^{2} \Big]\\ 
\beta_{g_3}^{(1)} = \, &
-3 g_{3}^{3} \\ 
\beta_{g_3}^{(2)} = \, &
\frac{1}{5} g_{3}^{3} \Big[11 g_{1}^{2}  -20 \mbox{Tr}\Big({Y_d  Y_{d}^{\dagger}}\Big)  -20 \mbox{Tr}\Big({Y_u  Y_{u}^{\dagger}}\Big)  + 45 g_{2}^{2}  + 70 g_{3}^{2} \Big]
\end{align}} 

\section{Seesaw II}
\label{RGEgaugeSeesaw2}
\subsection{Anomalous dimensions}
\label{AnaDimSeesaw2}
{\allowdisplaybreaks \begin{align} 
\gamma_{\hat{q}}^{(1)}  = \, &
-\frac{1}{30} \Big(45 g_{2}^{2}  + 80 g_{3}^{2}  + g_{1}^{2}\Big){\bf 1}  + {Y_{d}^{\dagger}  Y_d} + {Y_{u}^{\dagger}  Y_u}\\ 
\gamma_{\hat{q}}^{(2)}  = \, &
+\frac{4}{5} g_{1}^{2} {Y_{u}^{\dagger}  Y_u} -3 |\lambda_2|^2 {Y_{u}^{\dagger}  Y_u} -2 {Y_{d}^{\dagger}  Y_d  Y_{d}^{\dagger}  Y_d} -4 {Y_{d}^{\dagger}  Y_s  Y_s^*  Y_d} -2 {Y_{d}^{\dagger}  Y_z  Y_{z}^{\dagger}  Y_d} -2 {Y_{u}^{\dagger}  Y_u  Y_{u}^{\dagger}  Y_u} \nonumber \\ 
 &+{\bf 1} \Big[199 g_{1}^{4} +90 g_{1}^{2} g_{2}^{2} +3375 g_{2}^{4} +160(g_{1}^{2} g_{3}^{2} +5(4 g_{2}^{2} g_{3}^{2} - g_{3}^{4})) +48 \Big(125 g_{3}^{4}  + g_{1}^{4}\Big)\tilde{N}_S  \nonumber \\ 
 &+\left(54 g_{1}^{4}+2700 g_{2}^{4}\right) \tilde{N}_{T}+ \left(3 g_{1}^{4}+2025 g_{2}^{4}+2400 g_3^4 \right) \tilde{N}_{Z} \Big] \frac{1}{900} \nonumber \\ 
& +{Y_{d}^{\dagger}  Y_d} \Big[-3 |\lambda_1|^2  -3 \mbox{Tr}\Big({Y_d  Y_{d}^{\dagger}}\Big)  + \frac{2}{5} g_{1}^{2} - \mbox{Tr}\Big({Y_e  Y_{e}^{\dagger}}\Big) \Big]-3 {Y_{u}^{\dagger}  Y_u} \mbox{Tr}\Big({Y_u  Y_{u}^{\dagger}}\Big) \\ 
\gamma_{\hat{l}}^{(1)}  = \, &
3 \Big({Y_{z}^{\dagger}  Y_z} + {Y_t^*  Y_t}\Big) -\frac{3}{10} \Big(5 g_{2}^{2}  + g_{1}^{2}\Big){\bf 1}  + {Y_{e}^{\dagger}  Y_e}\\ 
\gamma_{\hat{l}}^{(2)}  = \, &
-\frac{2}{5} g_{1}^{2} {Y_{z}^{\dagger}  Y_z} +16 g_{3}^{2} {Y_{z}^{\dagger}  Y_z} +\frac{18}{5} g_{1}^{2} {Y_t^*  Y_t} +12 g_{2}^{2} {Y_t^*  Y_t} -3 |\lambda_1|^2 {Y_t^*  Y_t} -2 {Y_{e}^{\dagger}  Y_e  Y_{e}^{\dagger}  Y_e} \nonumber \\ 
 &-6 {Y_{z}^{\dagger}  Y_d  Y_{d}^{\dagger}  Y_z} -12 {Y_{z}^{\dagger}  Y_s  Y_s^*  Y_z} -6 {Y_{z}^{\dagger}  Y_z  Y_{z}^{\dagger}  Y_z} -9 {Y_t^*  Y_t  Y_t^*  Y_t} -3 {Y_t^*  Y_{e}^{T}  Y_e^*  Y_t} \nonumber \\ 
 & -9 {Y_t^*  Y_{z}^{T}  Y_z^*  Y_t} +\frac{3}{100} {\bf 1} \Big[69 g_{1}^{4} +30 g_{1}^{2} g_{2}^{2} +125 g_{2}^{4} +16 g_{1}^{4} \tilde{N}_S +\left(18 g_{1}^{4}  +100 g_{2}^{4}  \right)\tilde{N}_T \nonumber \\ 
 &+\left(g_{1}^{4}  +75 g_{2}^{4} \right)\tilde{N}_Z \Big]+{Y_{e}^{\dagger}  Y_e} \Big[-3 |\lambda_1|^2  -3 \mbox{Tr}\Big({Y_d  Y_{d}^{\dagger}}\Big)  + \frac{6}{5} g_{1}^{2}  - \mbox{Tr}\Big({Y_e  Y_{e}^{\dagger}}\Big) \Big]\nonumber \\ 
 &-3 {Y_t^*  Y_t} \mbox{Tr}\Big({Y_t  Y_t^*}\Big) -3 {Y_{z}^{\dagger}  Y_z} \mbox{Tr}\Big({Y_z  Y_{z}^{\dagger}}\Big) \\ 
\gamma_{\hat{H}_d}^{(1)}  = \, &
3 |\lambda_1|^2  + 3 \mbox{Tr}\Big({Y_d  Y_{d}^{\dagger}}\Big)  -\frac{3}{10} g_{1}^{2}  -\frac{3}{2} g_{2}^{2}  + \mbox{Tr}\Big({Y_e  Y_{e}^{\dagger}}\Big)\\ 
\gamma_{\hat{H}_d}^{(2)}  = \, &
-12 |\lambda_{1}^{2}|^4+\frac{3}{5} |\lambda_1|^2 \Big[-15 \mbox{Tr}\Big({Y_d  Y_{d}^{\dagger}}\Big)  + 20 g_{2}^{2}  -5 \mbox{Tr}\Big({Y_e  Y_{e}^{\dagger}}\Big)  -5 \mbox{Tr}\Big({Y_t  Y_t^*}\Big)  + 6 g_{1}^{2} \Big]\nonumber \\ 
 &+\frac{1}{100} \Big[207 g_{1}^{4} +90 g_{1}^{2} g_{2}^{2} +375 g_{2}^{4} +48 g_{1}^{4}\tilde{N}_S +\left(54 g_{1}^{4} +300 g_{2}^{4} \right)\tilde{N}_T \nonumber \\ 
 &+ (3 g_{1}^{4} +225 g_{2}^{4})\tilde{N}_Z -40 g_{1}^{2} \mbox{Tr}\Big({Y_d  Y_{d}^{\dagger}}\Big)\Big] +16 g_{3}^{2} \mbox{Tr}\Big({Y_d  Y_{d}^{\dagger}}\Big)  -9 \mbox{Tr}\Big({Y_d  Y_{d}^{\dagger}  Y_d  Y_{d}^{\dagger}}\Big) \nonumber \\ 
 &-12 \mbox{Tr}\Big({Y_d  Y_{d}^{\dagger}  Y_s  Y_s^*}\Big) -6 \mbox{Tr}\Big({Y_d  Y_{d}^{\dagger}  Y_z  Y_{z}^{\dagger}}\Big) -3 \mbox{Tr}\Big({Y_d  Y_{u}^{\dagger}  Y_u  Y_{d}^{\dagger}}\Big) -3 \mbox{Tr}\Big({Y_e  Y_{e}^{\dagger}  Y_e  Y_{e}^{\dagger}} \nonumber \\ 
 &-3 \mbox{Tr}\Big({Y_e  Y_{z}^{\dagger}  Y_z  Y_{e}^{\dagger}}\Big) -3 \mbox{Tr}\Big({Y_e  Y_t^*  Y_t  Y_{e}^{\dagger}}\Big)+1.2 g_{1}^{2} \mbox{Tr}\Big({Y_e  Y_{e}^{\dagger}}\Big) \Big)\\ 
\gamma_{\hat{H}_u}^{(1)}  = \, &
3 |\lambda_2|^2  -\frac{3}{10} \Big(-10 \mbox{Tr}\Big({Y_u  Y_{u}^{\dagger}}\Big)  + 5 g_{2}^{2}  + g_{1}^{2}\Big)\\ 
\gamma_{\hat{H}_u}^{(2)}  = \, &
\frac{1}{100} \Big[207 g_{1}^{4} +90 g_{1}^{2} g_{2}^{2} +375 g_{2}^{4} -1200 |\lambda_{2}|^4 +48 g_{1}^{4}\tilde{N}_S + \left(54 g_{1}^{4}+300 g_{2}^{4}\right)\tilde{N}_T  \nonumber \\ 
 &+\left(3 g_{1}^{4} +225 g_{2}^{4}\right)\tilde{N}_Z +60 |\lambda_2|^2 \Big(-15 \mbox{Tr}\Big({Y_u  Y_{u}^{\dagger}}\Big)  + 20 g_{2}^{2}  + 6 g_{1}^{2} \Big)+80 g_{1}^{2} \mbox{Tr}\Big({Y_u  Y_{u}^{\dagger}}\Big) \Big] \nonumber \\ 
 &+16 g_{3}^{2} \mbox{Tr}\Big({Y_u  Y_{u}^{\dagger}}\Big) -3 \mbox{Tr}\Big({Y_d  Y_{u}^{\dagger}  Y_u  Y_{d}^{\dagger}}\Big) -9\mbox{Tr}\Big({Y_u  Y_{u}^{\dagger}  Y_u  Y_{u}^{\dagger}}\Big) \\ 
\gamma_{\hat{d}}^{(1)}  = \, &
2 \Big(2 {Y_s^*  Y_s}  + {Y_d^*  Y_{d}^{T}} + {Y_z^*  Y_{z}^{T}}\Big) -\frac{2}{15} \Big(20 g_{3}^{2}  + g_{1}^{2}\Big){\bf 1} \\ 
\gamma_{\hat{d}}^{(2)}  = \, &
+\frac{32}{15} g_{1}^{2} {Y_s^*  Y_s} +\frac{80}{3} g_{3}^{2} {Y_s^*  Y_s} +\frac{2}{5} g_{1}^{2} {Y_z^*  Y_{z}^{T}} +6 g_{2}^{2} {Y_z^*  Y_{z}^{T}} -2 {Y_d^*  Y_{d}^{T}  Y_d^*  Y_{d}^{T}} -2 {Y_d^*  Y_{u}^{T}  Y_u^*  Y_{d}^{T}} \nonumber \\ 
 &-8 {Y_s^*  Y_d  Y_{d}^{\dagger}  Y_s} -16 {Y_s^*  Y_s  Y_s^*  Y_s} -8 {Y_s^*  Y_z  Y_{z}^{\dagger}  Y_s} -6 {Y_z^*  Y_t  Y_t^*  Y_{z}^{T}} -2 {Y_z^*  Y_{e}^{T}  Y_e^*  Y_{z}^{T}} \nonumber \\ 
 &-6 {Y_z^*  Y_{z}^{T}  Y_z^*  Y_{z}^{T}} +\frac{1}{225} {\bf 1} \Big[202 g_{1}^{4} +160 g_{1}^{2} g_{3}^{2} -200 g_{3}^{4} +12 \Big(125 g_{3}^{4}  + 4 g_{1}^{4} \Big)\tilde{N}_S \nonumber \\ 
 &+54 g_{1}^{4}\tilde{N}_T +\left(3 g_{1}^{4}  +600 g_{3}^{4}\right)\tilde{N}_Z \Big]-2 {Y_z^*  Y_{z}^{T}} \mbox{Tr}\Big({Y_z  Y_{z}^{\dagger}}\Big)-4 {Y_s^*  Y_s} \mbox{Tr}\Big({Y_s  Y_s^*}\Big) \nonumber \\ &+{Y_d^*  Y_{d}^{T}} \Big[-2 \mbox{Tr}\Big({Y_e  Y_{e}^{\dagger}}\Big)  + 6 g_{2}^{2}  -6 |\lambda_1|^2 -6 \mbox{Tr}\Big({Y_d  Y_{d}^{\dagger}}\Big)   + \frac{2}{5} g_{1}^{2} \Big]  \\ 
\gamma_{\hat{u}}^{(1)}  = \, &
2 {Y_u^*  Y_{u}^{T}}  -\frac{8}{15} \Big(5 g_{3}^{2}  + g_{1}^{2}\Big){\bf 1} \\ 
\gamma_{\hat{u}}^{(2)}  = \, &
\frac{2}{225} \Big[2 {\bf 1} \Big(214 g_{1}^{4} +160 g_{1}^{2} g_{3}^{2} -50 g_{3}^{4} +\Big(375 g_{3}^{4}  + 48 g_{1}^{4} \Big) \tilde{N}_S +54 g_{1}^{4} \tilde{N}_T\nonumber \\ 
 & + \left(3 g_{1}^{4} +150 g_{3}^{4}\right)\tilde{N}_Z \Big)-45 \Big\{5 \Big({Y_u^*  Y_{d}^{T}  Y_d^*  Y_{u}^{T}} + {Y_u^*  Y_{u}^{T}  Y_u^*  Y_{u}^{T}}\Big) \nonumber \\ 
 &+ {Y_u^*  Y_{u}^{T}} \Big(-15 g_{2}^{2}  + 15 |\lambda_2|^2  + 15 \mbox{Tr}\Big({Y_u  Y_{u}^{\dagger}}\Big)  + g_{1}^{2}\Big)\Big\}\Big]\\ 
\gamma_{\hat{e}}^{(1)}  = \, &
2 {Y_e^*  Y_{e}^{T}}  -\frac{6}{5} g_{1}^{2} {\bf 1} \\ 
\gamma_{\hat{e}}^{(2)}  = \, &
\frac{1}{25} \Big[3 g_{1}^{4} {\bf 1} \Big(16\tilde{N}_S  + 18\tilde{N}_T  + 78 + \tilde{N}_Z\Big)-10 \Big\{5 \Big(3 {Y_e^*  Y_t  Y_t^*  Y_{e}^{T}}  + 3 {Y_e^*  Y_{z}^{T}  Y_z^*  Y_{e}^{T}} \nonumber \\ 
 & + {Y_e^*  Y_{e}^{T}  Y_e^*  Y_{e}^{T}}\Big)+{Y_e^*  Y_{e}^{T}} \Big(-15 g_{2}^{2}  + 15 |\lambda_1|^2  + 15 \mbox{Tr}\Big({Y_d  Y_{d}^{\dagger}}\Big)  + 3 g_{1}^{2}  + 5 \mbox{Tr}\Big({Y_e  Y_{e}^{\dagger}}\Big) \Big)\Big\}\Big]\\ 
\gamma_{\hat{T}}^{(1)}  = \, &
-4 g_{2}^{2}  -\frac{6}{5} g_{1}^{2}  + |\lambda_1|^2 + \mbox{Tr}\Big({Y_t  Y_t^*}\Big)\\ 
\gamma_{\hat{T}}^{(2)}  = \, &
\frac{1}{25} \Big[234 g_{1}^{4} +240 g_{1}^{2} g_{2}^{2} +500 g_{2}^{4} -150 |\lambda_{1}|^4  +48 g_{1}^{4} \tilde{N}_S  +\left(54 g_{1}^{4}+200 g_{2}^{4}\right)\tilde{N}_T \nonumber \\ 
 &+ \left(3 g_{1}^{4}  +150 g_{2}^{4}\right)\tilde{N}_Z -5 |\lambda_1|^2 \Big(10 \mbox{Tr}\Big({Y_e  Y_{e}^{\dagger}}\Big)  + 30 \mbox{Tr}\Big({Y_d  Y_{d}^{\dagger}}\Big)  + 3 g_{1}^{2}  + 5 g_{2}^{2} \Big)\nonumber \\ 
 &-15 g_{1}^{2} \mbox{Tr}\Big({Y_t  Y_t^*}\Big) -25 g_{2}^{2} \mbox{Tr}\Big({Y_t  Y_t^*}\Big) -50 \mbox{Tr}\Big({Y_e  Y_t^*  Y_t  Y_{e}^{\dagger}}\Big) -150 \mbox{Tr}\Big({Y_t  Y_{z}^{\dagger}  Y_z  Y_t^*}\Big) \nonumber \\
& -150 \mbox{Tr}\Big({Y_t  Y_t^*  Y_t  Y_t^*}\Big) \Big]\\ 
\gamma_{\hat{\bar{T}}}^{(1)}  = \, &
-4 g_{2}^{2}  -\frac{6}{5} g_{1}^{2}  + |\lambda_2|^2\\ 
\gamma_{\hat{\bar{T}}}^{(2)}  = \, &
\frac{1}{25} \Big[234 g_{1}^{4} +240 g_{1}^{2} g_{2}^{2} +500 g_{2}^{4} -150 |\lambda_{2}|^4 +48 g_{1}^{4} \tilde{N}_S +\left(54 g_{1}^{4} +200 g_{2}^{4}\right)\tilde{N}_T \nonumber \\ 
 &+\left(3 g_{1}^{4} +150 g_{2}^{4}\right)\tilde{N}_Z  -5 |\lambda_2|^2 \Big(30 \mbox{Tr}\Big({Y_u  Y_{u}^{\dagger}}\Big)  + 3 g_{1}^{2}  + 5 g_{2}^{2} \Big)\Big]\\ 
\gamma_{\hat{S}}^{(1)}  = \, &
-\frac{4}{15} \Big(25 g_{3}^{2}  + 2 g_{1}^{2} \Big) + \mbox{Tr}\Big({Y_s  Y_s^*}\Big)\\ 
\gamma_{\hat{S}}^{(2)}  = \, &
\frac{2}{225} \Big[3 \Big(32 g_{1}^{4}  + 625 g_{3}^{4} \Big)\tilde{N}_S +2 \Big\{214 g_{1}^{4} +400 g_{1}^{2} g_{3}^{2} +1375 g_{3}^{4} +54 g_{1}^{4} \tilde{N}_T \nonumber \\ 
 & + \left(3 g_{1}^{4} +375 g_{3}^{4}\right)\tilde{N}_Z -15 g_{1}^{2} \mbox{Tr}\Big({Y_s  Y_s^*}\Big) -75 g_{3}^{2} \mbox{Tr}\Big({Y_s  Y_s^*}\Big) -225 \mbox{Tr}\Big({Y_d  Y_{d}^{\dagger}  Y_s  Y_s^*}\Big) \nonumber \\ 
 &-450 \mbox{Tr}\Big({Y_s  Y_s^*  Y_s  Y_s^*}\Big) -225 \mbox{Tr}\Big({Y_s  Y_s^*  Y_z  Y_{z}^{\dagger}}\Big) \Big\}\Big]\\ 
\gamma_{\hat{\bar{S}}}^{(1)}  = \, &
-\frac{4}{15} \Big(25 g_{3}^{2}  + 2 g_{1}^{2} \Big)\\ 
\gamma_{\hat{\bar{S}}}^{(2)}  = \, &
\frac{2}{225} \Big[428 g_{1}^{4} +800 g_{1}^{2} g_{3}^{2} +2750 g_{3}^{4} +3 \Big(32 g_{1}^{4}  + 625 g_{3}^{4} \Big)\tilde{N}_S\nonumber \\ 
 & \hspace{1cm}+108 g_{1}^{4}\tilde{N}_T + \left(6 g_{1}^{4}+750 g_{3}^{4}\right)\tilde{N}_Z \Big]\\ 
\gamma_{\hat{Z}}^{(1)}  = \, &
\frac{1}{30} \Big(30 \mbox{Tr}\Big({Y_z  Y_{z}^{\dagger}}\Big)  -45 g_{2}^{2}  -80 g_{3}^{2}  - g_{1}^{2} \Big)\\ 
\gamma_{\hat{Z}}^{(2)}  = \, &
+\frac{199}{900} g_{1}^{4} +\frac{1}{10} g_{1}^{2} g_{2}^{2} +\frac{15}{4} g_{2}^{4} +\frac{8}{45} g_{1}^{2} g_{3}^{2} +8 g_{2}^{2} g_{3}^{2} -\frac{8}{9} g_{3}^{4} +\frac{4}{75} \Big(125 g_{3}^{4}  + g_{1}^{4}\Big)\tilde{N}_S \nonumber \\ 
 & +\left(\frac{3}{50} g_{1}^{4} +3 g_{2}^{4}\right)\tilde{N}_T +\left(\frac{1}{300} g_{1}^{4} +\frac{9}{4} g_{2}^{4}+\frac{8}{3} g_{3}^{4}\right)\tilde{N}_Z +\frac{2}{5} g_{1}^{2} \mbox{Tr}\Big({Y_z  Y_{z}^{\dagger}}\Big) \nonumber \\ 
 &-2 \mbox{Tr}\Big({Y_d  Y_{d}^{\dagger}  Y_z  Y_{z}^{\dagger}}\Big) - \mbox{Tr}\Big({Y_e  Y_{z}^{\dagger}  Y_z  Y_{e}^{\dagger}}\Big) -4 \mbox{Tr}\Big({Y_s  Y_s^*  Y_z  Y_{z}^{\dagger}}\Big) -3 \mbox{Tr}\Big({Y_t  Y_{z}^{\dagger}  Y_z  Y_t^*}\Big)\nonumber \\ 
 &-5 \mbox{Tr}\Big({Y_z  Y_{z}^{\dagger}  Y_z  Y_{z}^{\dagger}}\Big) \\ 
\gamma_{\hat{\bar{Z}}}^{(1)}  = \, &
\frac{1}{30} \Big(-45 g_{2}^{2}  -80 g_{3}^{2}  - g_{1}^{2} \Big)\\ 
\gamma_{\hat{\bar{Z}}}^{(2)}  = \, &
\frac{1}{900} \Big[199 g_{1}^{4} +90 g_{1}^{2} g_{2}^{2} +3375 g_{2}^{4} +160(g_{1}^{2} g_{3}^{2} +20 g_{2}^{2} g_{3}^{2} -5 g_{3}^{4}) +48 \Big(125 g_{3}^{4}  + g_{1}^{4}\Big)\tilde{N}_S \nonumber \\ 
 &+\left(54 g_{1}^{4} +2700 g_{2}^{4}\right)\tilde{N}_T +\left(3 g_{1}^{4}+2025 g_{2}^{4} +2400 g_{3}^{4}\right)\tilde{N}_Z \Big]
\end{align} } 
\subsection{Gauge couplings}
\label{sec:betaII}
{\allowdisplaybreaks  \begin{align} 
\beta_{g_1}^{(1)}  = \, &
\frac{1}{10} g_{1}^{3} \Big(16 \tilde{N}_S  + 18 \tilde{N}_T  + 66 + \tilde{N}_Z\Big)\\ 
\beta_{g_1}^{(2)}  = \, &
\frac{1}{150} g_{1}^{3} \Big[1194 g_{1}^{2} +810 g_{2}^{2} +2640 g_{3}^{2} -810(|\lambda_1|^2 +|\lambda_2|^2) +\left(256 g_{1}^{2} +3200 g_{3}^{2}\right)\tilde{N}_S  \nonumber \\ 
 &+\left(648 g_{1}^{2} +2160 g_{2}^{2}\right)\tilde{N}_T +\left(g_{1}^{2}+45 g_{2}^{2} +80 g_{3}^{2}\right)\tilde{N}_Z -420 \mbox{Tr}\Big({Y_d  Y_{d}^{\dagger}}\Big) \nonumber \\ 
 &-540 \mbox{Tr}\Big({Y_e  Y_{e}^{\dagger}}\Big) -720 \mbox{Tr}\Big({Y_s  Y_s^*}\Big) -810 \mbox{Tr}\Big({Y_t  Y_t^*}\Big) -780 \mbox{Tr}\Big({Y_u  Y_{u}^{\dagger}}\Big) -420 \mbox{Tr}\Big({Y_z  Y_{z}^{\dagger}}\Big) \Big]\\ 
\beta_{g_2}^{(1)}  = \, &
\frac{1}{2} g_{2}^{3} \Big(3 \tilde{N}_Z  + 4 \tilde{N}_T  + 2\Big)\\ 
\beta_{g_2}^{(2)}  = \, &
\frac{1}{10} g_{2}^{3} \Big[18 g_{1}^{2} +250 g_{2}^{2} +240 g_{3}^{2} -70 |\lambda_1|^2 -70 |\lambda_2|^2 +\left(48 g_{1}^{2} +240 g_{2}^{2} \right)\tilde{N}_T \nonumber \\ 
 &+\left(g_{1}^{2}  +105 g_{2}^{2} +80 g_{3}^{2}\right)\tilde{N}_Z -60 \mbox{Tr}\Big({Y_d  Y_{d}^{\dagger}}\Big) -20 \mbox{Tr}\Big({Y_e  Y_{e}^{\dagger}}\Big) -70 \mbox{Tr}\Big({Y_t  Y_t^*}\Big)\nonumber \\ 
 & -60 \mbox{Tr}\Big({Y_u  Y_{u}^{\dagger}}\Big) -60 \mbox{Tr}\Big({Y_z  Y_{z}^{\dagger}}\Big) \Big]\\ 
\beta_{g_3}^{(1)}  = \, &
\frac{1}{2} g_{3}^{3} \Big(2 \Big(-3 + \tilde{N}_Z\Big) + 5\tilde{N}_S \Big)\\ 
\beta_{g_3}^{(2)}  = \, &
\frac{1}{15} g_{3}^{3} \Big[33 g_{1}^{2} +135 g_{2}^{2} +210 g_{3}^{2} +5 \Big(145 g_{3}^{2}  + 8 g_{1}^{2} \Big)\tilde{N}_S-135 \mbox{Tr}\Big({Y_s  Y_s^*}\Big) \nonumber \\ & +\left(g_{1}^{2} +45 g_{2}^{2} +170 g_{3}^{2}\right)\tilde{N}_Z -60 \mbox{Tr}\Big({Y_d  Y_{d}^{\dagger}}\Big) -60 \mbox{Tr}\Big({Y_u  Y_{u}^{\dagger}}\Big) -60 \mbox{Tr}\Big({Y_z  Y_{z}^{\dagger}}\Big) \Big]
\end{align}} 
\section{Seesaw III}
\label{RGEgaugeSeesaw3}
\subsection{Anomalous dimensions}
\label{AnaDimSeesaw3}
{\allowdisplaybreaks \begin{align} 
\gamma_{\hat{q}}^{(1)}  = \, &
-\frac{1}{30} \Big(45 g_{2}^{2}  + 80 g_{3}^{2}  + g_{1}^{2}\Big){\bf 1}  + {Y_{d}^{\dagger}  Y_d} + {Y_{u}^{\dagger}  Y_u}\\ 
\gamma_{\hat{q}}^{(2)}  = \, &
+\frac{4}{5} g_{1}^{2} {Y_{u}^{\dagger}  Y_u} -2 {Y_{d}^{\dagger}  Y_d  Y_{d}^{\dagger}  Y_d} -2 {Y_{d}^{\dagger}  Y_{x}^{T}  Y_x^*  Y_d} -2 {Y_{u}^{\dagger}  Y_u  Y_{u}^{\dagger}  Y_u} +{\bf 1} \Big[\frac{199}{900} g_{1}^{4} +\frac{1}{10} g_{1}^{2} g_{2}^{2} \nonumber \\ 
 &+\frac{15}{4} g_{2}^{4}  +\frac{8}{45} g_{1}^{2} g_{3}^{2} +8 g_{2}^{2} g_{3}^{2} -\frac{8}{9} g_{3}^{4} +8 g_{3}^{4} N_{G_M} +3 g_{2}^{4} N_{W_M} \nonumber \\ 
 &+ \left(\frac{1}{12} g_{1}^{4} +\frac{9}{4} g_{2}^{4} +\frac{8}{3} g_{3}^{4}\right)\tilde{N}_{X_M} \Big]   -\frac{3}{10} {Y_{u}^{\dagger}  Y_u} \,\mbox{Tr}\Big({Y_b  Y_{b}^{\dagger}}\Big) +{Y_{d}^{\dagger}  Y_d} \Big(-3 \,\mbox{Tr}\Big({Y_d  Y_{d}^{\dagger}}\Big)  \nonumber \\ 
 &+ \frac{2}{5} g_{1}^{2}  - \,\mbox{Tr}\Big({Y_e  Y_{e}^{\dagger}}\Big) \Big)-3 {Y_{u}^{\dagger}  Y_u} \,\mbox{Tr}\Big({Y_u  Y_{u}^{\dagger}}\Big)-\frac{3}{2} {Y_{u}^{\dagger}  Y_u} \,\mbox{Tr}\Big({Y_w  Y_{w}^{\dagger}}\Big) -3 {Y_{u}^{\dagger}  Y_u} \,\mbox{Tr}\Big({Y_x  Y_{x}^{\dagger}}\Big) \\ 
\gamma_{\hat{l}}^{(1)}  = \, &
\frac{1}{10} \Big(10 {Y_{e}^{\dagger}  Y_e}  + 15 {Y_{w}^{\dagger}  Y_w}  -3 \Big(5 g_{2}^{2}  + g_{1}^{2}\Big){\bf 1}  + 3 {Y_{b}^{\dagger}  Y_b} \Big)\\ 
\gamma_{\hat{l}}^{(2)}  = \, &
\frac{1}{200} \Big[240 g_{1}^{2} {Y_{e}^{\dagger}  Y_e} +1200 g_{2}^{2} {Y_{w}^{\dagger}  Y_w} -36 {Y_{b}^{\dagger}  Y_b  Y_{b}^{\dagger}  Y_b} -60 {Y_{b}^{\dagger}  Y_b  Y_{w}^{\dagger}  Y_w} -400 {Y_{e}^{\dagger}  Y_e  Y_{e}^{\dagger}  Y_e} \nonumber \\ 
 &-45 {Y_{w}^{\dagger}  Y_w  Y_{b}^{\dagger}  Y_b} -300 {Y_{w}^{\dagger}  Y_w  Y_{w}^{\dagger}  Y_w} +6 {\bf 1} \Big(100 g_{2}^{4} N_{W_M}  + 125 g_{2}^{4}  + 25 \Big(3 g_{2}^{4}  + g_{1}^{4}\Big)\tilde{N}_{X_M}  \nonumber \\ 
 &  + 30 g_{1}^{2} g_{2}^{2}  + 69 g_{1}^{4}  \Big)-18 {Y_{b}^{\dagger}  Y_b} \,\mbox{Tr}\Big({Y_b  Y_{b}^{\dagger}}\Big) -90 {Y_{w}^{\dagger}  Y_w} \,\mbox{Tr}\Big({Y_b  Y_{b}^{\dagger}}\Big) -600 {Y_{e}^{\dagger}  Y_e} \,\mbox{Tr}\Big({Y_d  Y_{d}^{\dagger}}\Big) \nonumber \\ 
 &-200 {Y_{e}^{\dagger}  Y_e} \,\mbox{Tr}\Big({Y_e  Y_{e}^{\dagger}}\Big) -180 {Y_{b}^{\dagger}  Y_b} \,\mbox{Tr}\Big({Y_u  Y_{u}^{\dagger}}\Big) -900 {Y_{w}^{\dagger}  Y_w} \,\mbox{Tr}\Big({Y_u  Y_{u}^{\dagger}}\Big)\nonumber \\ 
 &-90 {Y_{b}^{\dagger}  Y_b} \,\mbox{Tr}\Big({Y_w  Y_{w}^{\dagger}}\Big)  -450 {Y_{w}^{\dagger}  Y_w} \,\mbox{Tr}\Big({Y_w  Y_{w}^{\dagger}}\Big) -180 {Y_{b}^{\dagger}  Y_b} \,\mbox{Tr}\Big({Y_x  Y_{x}^{\dagger}}\Big) \nonumber \\ 
 &-900 {Y_{w}^{\dagger}  Y_w} \,\mbox{Tr}\Big({Y_x  Y_{x}^{\dagger}}\Big) \Big]\\ 
\gamma_{\hat{H}_d}^{(1)}  = \, &
3 \,\mbox{Tr}\Big({Y_d  Y_{d}^{\dagger}}\Big)  -\frac{3}{10} \Big(5 g_{2}^{2}  + g_{1}^{2}\Big) + \,\mbox{Tr}\Big({Y_e  Y_{e}^{\dagger}}\Big)\\ 
\gamma_{\hat{H}_d}^{(2)}  = \, &
+\frac{207}{100} g_{1}^{4} +\frac{9}{10} g_{1}^{2} g_{2}^{2} +\frac{15}{4} g_{2}^{4} +3 g_{2}^{4} N_{W_M} +\frac{3}{4} \Big(3 g_{2}^{4}  + g_{1}^{4}\Big)\tilde{N}_{X_M}-9 \,\mbox{Tr}\Big({Y_d  Y_{d}^{\dagger}  Y_d  Y_{d}^{\dagger}}\Big) \nonumber \\ 
 &-\frac{2}{5} g_{1}^{2} \,\mbox{Tr}\Big({Y_d  Y_{d}^{\dagger}}\Big) +16 g_{3}^{2} \,\mbox{Tr}\Big({Y_d  Y_{d}^{\dagger}}\Big) +\frac{6}{5} g_{1}^{2} \,\mbox{Tr}\Big({Y_e  Y_{e}^{\dagger}}\Big) -\frac{3}{10} \,\mbox{Tr}\Big({Y_b  Y_{e}^{\dagger}  Y_e  Y_{b}^{\dagger}}\Big)  \nonumber \\ 
 &-6 \,\mbox{Tr}\Big({Y_d  Y_{d}^{\dagger}  Y_{x}^{T}  Y_x^*}\Big) -3 \,\mbox{Tr}\Big({Y_d  Y_{u}^{\dagger}  Y_u  Y_{d}^{\dagger}}\Big) -3 \,\mbox{Tr}\Big({Y_e  Y_{e}^{\dagger}  Y_e  Y_{e}^{\dagger}}\Big) -\frac{3}{2} \,\mbox{Tr}\Big({Y_e  Y_{w}^{\dagger}  Y_w  Y_{e}^{\dagger}}\Big)   \\ 
\gamma_{\hat{H}_u}^{(1)}  = \, &
-\frac{3}{10} \Big(-10 \,\mbox{Tr}\Big({Y_u  Y_{u}^{\dagger}}\Big)  -10 \,\mbox{Tr}\Big({Y_x  Y_{x}^{\dagger}}\Big)  + 5 g_{2}^{2}  -5 \,\mbox{Tr}\Big({Y_w  Y_{w}^{\dagger}}\Big)  - \,\mbox{Tr}\Big({Y_b  Y_{b}^{\dagger}}\Big)  + g_{1}^{2}\Big)\\ 
\gamma_{\hat{H}_u}^{(2)}  = \, &
+\frac{207}{100} g_{1}^{4} +\frac{9}{10} g_{1}^{2} g_{2}^{2} +\frac{15}{4} g_{2}^{4} +3 g_{2}^{4} N_{W_M} +\frac{3}{4} \Big(3 g_{2}^{4}  + g_{1}^{4}\Big)\tilde{N}_{X_M} +\frac{4}{5} g_{1}^{2} \,\mbox{Tr}\Big({Y_u  Y_{u}^{\dagger}}\Big) \nonumber \\ 
 &+16 g_{3}^{2} \,\mbox{Tr}\Big({Y_u  Y_{u}^{\dagger}}\Big) +6 g_{2}^{2} \,\mbox{Tr}\Big({Y_w  Y_{w}^{\dagger}}\Big) +2 g_{1}^{2} \,\mbox{Tr}\Big({Y_x  Y_{x}^{\dagger}}\Big) +16 g_{3}^{2} \,\mbox{Tr}\Big({Y_x  Y_{x}^{\dagger}}\Big)  \nonumber \\ 
 &-\frac{3}{10} \,\mbox{Tr}\Big({Y_b  Y_{e}^{\dagger}  Y_e  Y_{b}^{\dagger}}\Big) -\frac{57}{40} \,\mbox{Tr}\Big({Y_b  Y_{w}^{\dagger}  Y_w  Y_{b}^{\dagger}}\Big) -6 \,\mbox{Tr}\Big({Y_d  Y_{d}^{\dagger}  Y_{x}^{T}  Y_x^*}\Big) -3 \,\mbox{Tr}\Big({Y_d  Y_{u}^{\dagger}  Y_u  Y_{d}^{\dagger}}\Big) \nonumber \\ 
 &-\frac{3}{2} \,\mbox{Tr}\Big({Y_e  Y_{w}^{\dagger}  Y_w  Y_{e}^{\dagger}}\Big) -9 \,\mbox{Tr}\Big({Y_u  Y_{u}^{\dagger}  Y_u  Y_{u}^{\dagger}}\Big) -\frac{15}{4} \,\mbox{Tr}\Big({Y_w  Y_{w}^{\dagger}  Y_w  Y_{w}^{\dagger}}\Big) -9 \,\mbox{Tr}\Big({Y_x  Y_{x}^{\dagger}  Y_x  Y_{x}^{\dagger}}\Big) \nonumber \\ &  -\frac{27}{100} \,\mbox{Tr}\Big({Y_b  Y_{b}^{\dagger}  Y_b  Y_{b}^{\dagger}}\Big) \\
\gamma_{\hat{d}}^{(1)}  = \, &
2 \Big({Y_{x}^{\dagger}  Y_x} + {Y_d^*  Y_{d}^{T}}\Big) -\frac{2}{15} \Big(20 g_{3}^{2}  + g_{1}^{2}\Big){\bf 1} \\ 
\gamma_{\hat{d}}^{(2)}  = \, &
+\frac{2}{5} g_{1}^{2} {Y_d^*  Y_{d}^{T}} +6 g_{2}^{2} {Y_d^*  Y_{d}^{T}} -2 {Y_{x}^{\dagger}  Y_x  Y_{x}^{\dagger}  Y_x} -2 {Y_d^*  Y_{d}^{T}  Y_d^*  Y_{d}^{T}} -2 {Y_d^*  Y_{u}^{T}  Y_u^*  Y_{d}^{T}} \nonumber \\ 
 &+\frac{1}{225} {\bf 1} \Big[160 g_{1}^{2} g_{3}^{2}  + 1800 g_{3}^{4} N_{G_M}  -200 g_{3}^{4}  + 202 g_{1}^{4}   + 75 \Big(8 g_{3}^{4}  + g_{1}^{4}\Big)\tilde{N}_{X_M}  \Big] \nonumber \\ 
 &-6 {Y_d^*  Y_{d}^{T}} \,\mbox{Tr}\Big({Y_d  Y_{d}^{\dagger}}\Big) -2 {Y_d^*  Y_{d}^{T}} \,\mbox{Tr}\Big({Y_e  Y_{e}^{\dagger}}\Big) +{Y_{x}^{\dagger}  Y_x} \Big(2 g_{1}^{2}  -3 \,\mbox{Tr}\Big({Y_w  Y_{w}^{\dagger}}\Big)  + 6 g_{2}^{2}  \nonumber \\  & -6 \,\mbox{Tr}\Big({Y_u  Y_{u}^{\dagger}}\Big)  -6 \,\mbox{Tr}\Big({Y_x  Y_{x}^{\dagger}}\Big)  -\frac{3}{5} \,\mbox{Tr}\Big({Y_b  Y_{b}^{\dagger}}\Big) \Big)\\ 
\gamma_{\hat{u}}^{(1)}  = \, &
2 {Y_u^*  Y_{u}^{T}}  -\frac{8}{15} \Big(5 g_{3}^{2}  + g_{1}^{2}\Big){\bf 1} \\ 
\gamma_{\hat{u}}^{(2)}  = \, &
-2 \Big({Y_u^*  Y_{d}^{T}  Y_d^*  Y_{u}^{T}} + {Y_u^*  Y_{u}^{T}  Y_u^*  Y_{u}^{T}}\Big)+\frac{4}{225} {\bf 1} \Big[160 g_{1}^{2} g_{3}^{2}  + 214 g_{1}^{4}  + 450 g_{3}^{4} N_{G_M} -50 g_{3}^{4}  \nonumber \\ 
 & + 75 \Big(2 g_{3}^{4}  + g_{1}^{4}\Big)\tilde{N}_{X_M} \Big]-\frac{1}{5} {Y_u^*  Y_{u}^{T}} \Big[15 \,\mbox{Tr}\Big({Y_w  Y_{w}^{\dagger}}\Big)  + 2 g_{1}^{2}  -30 g_{2}^{2}  + 30 \,\mbox{Tr}\Big({Y_u  Y_{u}^{\dagger}}\Big) \nonumber \\ 
 & + 30 \,\mbox{Tr}\Big({Y_x  Y_{x}^{\dagger}}\Big)  + 3 \,\mbox{Tr}\Big({Y_b  Y_{b}^{\dagger}}\Big) \Big]\\ 
\gamma_{\hat{e}}^{(1)}  = \, &
2 {Y_e^*  Y_{e}^{T}}  -\frac{6}{5} g_{1}^{2} {\bf 1} \\ 
\gamma_{\hat{e}}^{(2)}  = \, &
-\frac{3}{5} {Y_e^*  Y_{b}^{T}  Y_b^*  Y_{e}^{T}} -2 {Y_e^*  Y_{e}^{T}  Y_e^*  Y_{e}^{T}} -3 {Y_e^*  Y_{w}^{T}  Y_w^*  Y_{e}^{T}} +\frac{3}{25} g_{1}^{4} {\bf 1} \Big(25 \tilde{N}_{X_M}  + 78\Big)\nonumber \\ 
 &+{Y_e^*  Y_{e}^{T}} \Big[-2 \,\mbox{Tr}\Big({Y_e  Y_{e}^{\dagger}}\Big)  + 6 g_{2}^{2}  -6 \,\mbox{Tr}\Big({Y_d  Y_{d}^{\dagger}}\Big)  -\frac{6}{5} g_{1}^{2} \Big]\\ 
\gamma_{\hat{W}_M}^{(1)}  = \, &
-4 g_{2}^{2} {\bf 1}  + {Y_w^*  Y_{w}^{T}}\\ 
\gamma_{\hat{W}_M}^{(2)}  = \, &
+2 g_{2}^{4} {\bf 1} \Big(3 \tilde{N}_{X_M}  + 4 N_{W_M}  + 10\Big)+\frac{1}{10} \Big[-3 {Y_w^*  Y_{b}^{T}  Y_b^*  Y_{w}^{T}} -10 {Y_w^*  Y_{e}^{T}  Y_e^*  Y_{w}^{T}} \nonumber \\ 
 &-15 {Y_w^*  Y_{w}^{T}  Y_w^*  Y_{w}^{T}} +{Y_w^*  Y_{w}^{T}} \Big\{-10 g_{2}^{2}  -15 \,\mbox{Tr}\Big({Y_w  Y_{w}^{\dagger}}\Big)  -30 \,\mbox{Tr}\Big({Y_u  Y_{u}^{\dagger}}\Big)  -30 \,\mbox{Tr}\Big({Y_x  Y_{x}^{\dagger}}\Big) \nonumber \\ 
 & -3 \,\mbox{Tr}\Big({Y_b  Y_{b}^{\dagger}}\Big)  + 6 g_{1}^{2} \Big\}\Big]\\ 
\gamma_{\hat{G}_M}^{(1)}  = \, &
-6 g_{3}^{2} {\bf 1} \\ 
\gamma_{\hat{G}_M}^{(2)}  = \, &
6 g_{3}^{4} {\bf 1} \Big(3 N_{G_M}  + 3 + \tilde{N}_{X_M} \Big)\\ 
\gamma_{\hat{B}_M}^{(1)}  = \, &
\frac{3}{5} {Y_b^*  Y_{b}^{T}} \\ 
\gamma_{\hat{B}_M}^{(2)}  = \, &
\frac{3}{50} \Big[-3 {Y_b^*  Y_{b}^{T}  Y_b^*  Y_{b}^{T}} -5 \Big(2 {Y_b^*  Y_{e}^{T}  Y_e^*  Y_{b}^{T}}  + 3 {Y_b^*  Y_{w}^{T}  Y_w^*  Y_{b}^{T}} \Big)+3 {Y_b^*  Y_{b}^{T}} \Big\{10 g_{2}^{2}  \nonumber \\ 
 &-10 \,\mbox{Tr}\Big({Y_u  Y_{u}^{\dagger}}\Big)  -10 \,\mbox{Tr}\Big({Y_x  Y_{x}^{\dagger}}\Big)  + 2 g_{1}^{2}  -5 \,\mbox{Tr}\Big({Y_w  Y_{w}^{\dagger}}\Big)  - \,\mbox{Tr}\Big({Y_b  Y_{b}^{\dagger}}\Big) \Big\}\Big]\\ 
\gamma_{\hat{X}_M}^{(1)}  = \, &
-\frac{1}{6} \Big(16 g_{3}^{2}  + 5 g_{1}^{2}  + 9 g_{2}^{2} \Big){\bf 1} \\ 
\gamma_{\hat{X}_M}^{(2)}  = \, &
\frac{1}{36} {\bf 1} \Big[223 g_{1}^{4} +90 g_{1}^{2} g_{2}^{2} +135 g_{2}^{4} +160 g_{1}^{2} g_{3}^{2} +288 (g_{2}^{2} g_{3}^{2}+g_{3}^{4} N_{G_M}) -32 g_{3}^{4}\nonumber \\ 
 &+108 g_{2}^{4} N_{W_M} +\left(75 g_{1}^{4}  +81 g_{2}^{4}  +96 g_{3}^{4} \right)\tilde{N}_{X_M} \Big]\\ 
\gamma_{\hat{\bar{X}}_M}^{(1)}  = \, &
\frac{1}{6} \Big(- \Big(16 g_{3}^{2}  + 5 g_{1}^{2}  + 9 g_{2}^{2} \Big){\bf 1}  + 6 {Y_x^*  Y_{x}^{T}} \Big)\\ 
\gamma_{\hat{\bar{X}}_M}^{(2)}  = \, &
+\frac{1}{36} {\bf 1} \Big[223 g_{1}^{4} +90 g_{1}^{2} g_{2}^{2} +135 g_{2}^{4} +160 g_{1}^{2} g_{3}^{2} +288 (g_{2}^{2} g_{3}^{2}+ g_{3}^{4} N_{G_M}) -32 g_{3}^{4}  \nonumber \\ 
 &+108 g_{2}^{4} N_{W_M}+\left(75 g_{1}^{4} +81 g_{2}^{4} +96 g_{3}^{4}\right)\tilde{N}_{X_M} \Big]+\nonumber \\ 
 & + \frac{1}{10} \Big[-20 \Big({Y_x^*  Y_d  Y_{d}^{\dagger}  Y_{x}^{T}} + {Y_x^*  Y_{x}^{T}  Y_x^*  Y_{x}^{T}}\Big)- {Y_x^*  Y_{x}^{T}} \Big\{15 \,\mbox{Tr}\Big({Y_w  Y_{w}^{\dagger}}\Big)  \nonumber \\ 
 & + 30 \,\mbox{Tr}\Big({Y_u  Y_{u}^{\dagger}}\Big)+ 30 \,\mbox{Tr}\Big({Y_x  Y_{x}^{\dagger}}\Big)  + 3 \,\mbox{Tr}\Big({Y_b  Y_{b}^{\dagger}}\Big)  + 4 g_{1}^{2} \Big\}\Big]
\end{align} } 
\subsection{Gauge couplings}
\label{sec:betaIII}
{\allowdisplaybreaks  
\begin{align} 
\beta_{g_1}^{(1)}  = \, &
\frac{1}{10} g_{1}^{3} \Big(25 \tilde{N}_{X_M}  + 66\Big)\\ 
\beta_{g_1}^{(2)}  = \, &
\frac{1}{150} g_{1}^{3} \Big[125 \Big(16 g_{3}^{2}  + 5 g_{1}^{2}  + 9 g_{2}^{2} \Big)\tilde{N}_{X_M} +6 \Big\{199 g_{1}^{2} +135 g_{2}^{2} +440 g_{3}^{2} -9 \,\mbox{Tr}\Big({Y_b  Y_{b}^{\dagger}}\Big) \nonumber \\ 
 &-70 \,\mbox{Tr}\Big({Y_d  Y_{d}^{\dagger}}\Big) -90 \,\mbox{Tr}\Big({Y_e  Y_{e}^{\dagger}}\Big)-130 \,\mbox{Tr}\Big({Y_u  Y_{u}^{\dagger}}\Big) -60 \,\mbox{Tr}\Big({Y_w  Y_{w}^{\dagger}}\Big) -190 \,\mbox{Tr}\Big({Y_x  Y_{x}^{\dagger}}\Big) \Big\}\Big]\\ 
\beta_{g_2}^{(1)}  = \, &
\frac{1}{2} g_{2}^{3} \Big(3 \tilde{N}_{X_M} + 4 N_{W_M}  + 2\Big)\\ 
\beta_{g_2}^{(2)}  = \, &
\frac{1}{30} g_{2}^{3} \Big[54 g_{1}^{2} +750 g_{2}^{2} +720(g_{3}^{2} +g_{2}^{2} N_{W_M}) +15 \Big(16 g_{3}^{2}  + 21 g_{2}^{2}  + 5 g_{1}^{2} \Big)\tilde{N}_{X_M} \nonumber \\ 
 &-18 \,\mbox{Tr}\Big({Y_b  Y_{b}^{\dagger}}\Big)-280 \,\mbox{Tr}\Big({Y_w  Y_{w}^{\dagger}}\Big)  -60 \,\mbox{Tr}\Big({Y_e  Y_{e}^{\dagger}}\Big) -180 \Big(\mbox{Tr}\Big({Y_d  Y_{d}^{\dagger}}\Big) +\mbox{Tr}\Big({Y_u  Y_{u}^{\dagger}}\Big) \nonumber \\
& +\mbox{Tr}\Big({Y_x  Y_{x}^{\dagger}}\Big)\Big) \Big]\\ 
\beta_{g_3}^{(1)}  = \, &
g_{3}^{3} \Big(3 N_{G_M}  -3 + \tilde{N}_{X_M}\Big)\\ 
\beta_{g_3}^{(2)}  = \, &
\frac{1}{15} g_{3}^{3} \Big[33 g_{1}^{2} +135 g_{2}^{2} +210 g_{3}^{2} +810 g_{3}^{2} N_{G_M} +5 \Big(34 g_{3}^{2}  + 5 g_{1}^{2}  + 9 g_{2}^{2} \Big)\tilde{N}_{X_M} \nonumber \\ 
 &+170 g_{3}^{2} N_{\bar{X}_M} -60 \,\mbox{Tr}\Big({Y_d  Y_{d}^{\dagger}}\Big) -60 \,\mbox{Tr}\Big({Y_u  Y_{u}^{\dagger}}\Big) -60 \,\mbox{Tr}\Big({Y_x  Y_{x}^{\dagger}}\Big) \Big]
\end{align}}

  \chapter{Analytical expressions for the NMSSM}
\section{Anomalous dimensions and beta functions for gauge couplings}
\label{gamma}
In this app., we give the detailed expressions of the anomalous dimension of the Higgs-fields, which are needed for the RGE evaluation of the VEVs.
\allowdisplaybreaks
\begin{align}
\gamma_{\hat{H}_d}^{(1)} =  \, &  
3 \mbox{Tr}\Big({Y_d  Y_d^\dagger}\Big)  -\frac{3}{10} g_1^2  -\frac{3}{2} g_2^2  + |\lambda|^2 + \mbox{Tr}\Big({Y_e  Y_e^\dagger}\Big)\\ 
\gamma_{\hat{H}_d}^{(2)} =  \, & 
\frac{207}{100} g_1^4 +\frac{9}{10} g_1^2 g_2^2 +\frac{15}{4} g_2^4 -2 |\lambda|^2 |\kappa|^2  -3 |\lambda|^4 -\frac{2}{5} g_1^2 \mbox{Tr}\Big({Y_d  Y_d^\dagger}\Big) \nonumber \\ 
 &+16 {g_3}^{2} \mbox{Tr}\Big({Y_d  Y_d^\dagger}\Big) +\frac{6}{5} g_1^2 \mbox{Tr}\Big({Y_e  Y_e^\dagger}\Big) -3 |\lambda|^2 \mbox{Tr}\Big({Y_u  Y_u^\dagger}\Big) -9 \mbox{Tr}\Big({Y_d  Y_d^\dagger  Y_d  Y_d^\dagger}\Big) \nonumber \\ 
 &-3 \mbox{Tr}\Big({Y_d  Y_d^\dagger  Y_u  Y_u^\dagger}\Big) -3 \mbox{Tr}\Big({Y_e  Y_e^\dagger  Y_e  Y_e^\dagger}\Big) \\ 
\gamma_{\hat{H}_u}^{(1)} =  \, &  
3 \mbox{Tr}\Big({Y_u  Y_u^\dagger}\Big)  -\frac{3}{10} g_1^2  -\frac{3}{2} g_2^2  + |\lambda|^2\\ 
\gamma_{\hat{H}_u}^{(2)} =  \, & 
\frac{207}{100} g_1^4 +\frac{9}{10} g_1^2 g_2^2 +\frac{15}{4} g_2^4 -2 |\lambda|^2 |\kappa|^2  -3 |\lambda|^4  -3 |\lambda|^2 \mbox{Tr}\Big({Y_d  Y_d^\dagger}\Big) \nonumber \\ 
 &- |\lambda|^2 \mbox{Tr}\Big({Y_e  Y_e^\dagger}\Big) +\frac{4}{5} g_1^2 \mbox{Tr}\Big({Y_u  Y_u^\dagger}\Big) +16 {g_3}^{2} \mbox{Tr}\Big({Y_u  Y_u^\dagger}\Big) -3 \mbox{Tr}\Big({Y_d  Y_d^\dagger  Y_u  Y_u^\dagger}\Big) \nonumber \\
& -9 \mbox{Tr}\Big({Y_u  Y_u^\dagger  Y_u  Y_u^\dagger}\Big) \\ 
\gamma_{\hat{S}}^{(1)} =  \, &  
2 |\kappa|^2  + 2 |\lambda|^2 \\ 
\gamma_{\hat{S}}^{(2)} =  \, &
\frac{6}{5} g_1^2 |\lambda|^2 +6 g_2^2 |\lambda|^2 -8 |\kappa|^4 -8 |\lambda|^2 |\kappa|^2 -4 |\lambda|^4 -6 |\lambda|^2 \mbox{Tr}\Big({Y_d  Y_d^\dagger}\Big) \nonumber \\ 
 &-2 |\lambda|^2 \mbox{Tr}\Big({Y_e  Y_e^\dagger}\Big) -6 |\lambda|^2 \mbox{Tr}\Big({Y_u  Y_u^\dagger}\Big) 
\end{align} 
The other anomalous dimensions  differ just at two-loop level from the MSSM expressions.
{\allowdisplaybreaks 
\begin{align} 
\gamma_{\hat{q}}^{(2)} =  \, & 
+\Big(8 g_{2}^{2} g_{3}^{2}  + \frac{15}{4} g_{2}^{4}  + \frac{1}{90} g_{1}^{2} \Big(16 g_{3}^{2}  + 9 g_{2}^{2} \Big) + \frac{199}{900} g_{1}^{4}  -\frac{8}{9} g_{3}^{4} \Big){\bf 1} +\frac{4}{5} g_{1}^{2} {Y_u^*  Y_{u}^{T}} - |\lambda|^2 {Y_u^*  Y_{u}^{T}} \nonumber \\ 
 &-2 {Y_d^*  Y_{d}^{T}  Y_d^*  Y_{d}^{T}} -2 {Y_u^*  Y_{u}^{T}  Y_u^*  Y_{u}^{T}} +{Y_d^*  Y_{d}^{T}} \Big(-3 \mbox{Tr}\Big({Y_d  Y_{d}^{\dagger}}\Big)  + \frac{2}{5} g_{1}^{2}  - |\lambda|^2  - \mbox{Tr}\Big({Y_e  Y_{e}^{\dagger}}\Big) \Big)\nonumber \\ 
 &-3 {Y_u^*  Y_{u}^{T}} \mbox{Tr}\Big({Y_u  Y_{u}^{\dagger}}\Big) \\ 
\gamma_{\hat{l}}^{(2)} =  \, & 
+\frac{3}{100} \Big(125 g_{2}^{4}  + 30 g_{1}^{2} g_{2}^{2}  + 69 g_{1}^{4} \Big){\bf 1} -2 {Y_e^*  Y_{e}^{T}  Y_e^*  Y_{e}^{T}} \nonumber \\ 
 &+{Y_e^*  Y_{e}^{T}} \Big(-3 \mbox{Tr}\Big({Y_d  Y_{d}^{\dagger}}\Big)  + \frac{6}{5} g_{1}^{2}  - |\lambda|^2  - \mbox{Tr}\Big({Y_e  Y_{e}^{\dagger}}\Big) \Big)\\ 
\gamma_{\hat{d}}^{(2)} =  \, &  
+\frac{2}{225} \Big(-100 g_{3}^{4}  + 101 g_{1}^{4}  + 80 g_{1}^{2} g_{3}^{2} \Big){\bf 1} -2 \Big({Y_{d}^{\dagger}  Y_d  Y_{d}^{\dagger}  Y_d} + {Y_{d}^{\dagger}  Y_u  Y_{u}^{\dagger}  Y_d}\Big)\nonumber \\ 
 &+{Y_{d}^{\dagger}  Y_d} \Big(-2 |\lambda|^2  -2 \mbox{Tr}\Big({Y_e  Y_{e}^{\dagger}}\Big)  + 6 g_{2}^{2}  -6 \mbox{Tr}\Big({Y_d  Y_{d}^{\dagger}}\Big)  + \frac{2}{5} g_{1}^{2} \Big)\\ 
\gamma_{\hat{u}}^{(2)} =  \, &  
+\frac{8}{225} \Big(107 g_{1}^{4}  -25 g_{3}^{4}  + 80 g_{1}^{2} g_{3}^{2} \Big){\bf 1} \nonumber \\ 
 &-\frac{2}{5} \Big(5 \Big({Y_{u}^{\dagger}  Y_d  Y_{d}^{\dagger}  Y_u} + {Y_{u}^{\dagger}  Y_u  Y_{u}^{\dagger}  Y_u}\Big) + {Y_{u}^{\dagger}  Y_u} \Big(-15 g_{2}^{2}  + 15 \mbox{Tr}\Big({Y_u  Y_{u}^{\dagger}}\Big)  + 5 |\lambda|^2  + g_{1}^{2}\Big)\Big)\\ 
\gamma_{\hat{e}}^{(2)} =  \, &  
-2 {Y_{e}^{\dagger}  Y_e  Y_{e}^{\dagger}  Y_e}  + \frac{234}{25} g_{1}^{4} {\bf 1}  + {Y_{e}^{\dagger}  Y_e} \Big(-2 |\lambda|^2  -2 \mbox{Tr}\Big({Y_e  Y_{e}^{\dagger}}\Big)  + 6 g_{2}^{2}  -6 \mbox{Tr}\Big({Y_d  Y_{d}^{\dagger}}\Big)  -\frac{6}{5} g_{1}^{2} \Big)\\ 
\end{align} } 
Also the \(\beta\)-functions of the gauge couplings are the same at one-loop level but receive at two-loop contributions involving \(\lambda\) and \(\kappa\).
{\allowdisplaybreaks  
\begin{align} 
\beta_{g_1}^{(2)}  =  \, &
\frac{1}{25} g_{1}^{3} \Big(-130 \mbox{Tr}\Big({Y_u  Y_{u}^{\dagger}}\Big)  + 135 g_{2}^{2}  + 199 g_{1}^{2}  -30 |\lambda|^2  + 440 g_{3}^{2} \nonumber \\
 & \hspace{3cm}  -70 \mbox{Tr}\Big({Y_d  Y_{d}^{\dagger}}\Big)  -90 \mbox{Tr}\Big({Y_e  Y_{e}^{\dagger}}\Big) \Big)\\ 
\beta_{g_2}^{(2)}  = \, & 
\frac{1}{5} g_{2}^{3} \Big(-10 |\lambda|^2  -10 \mbox{Tr}\Big({Y_e  Y_{e}^{\dagger}}\Big)  + 120 g_{3}^{2}  + 125 g_{2}^{2}  -30 \mbox{Tr}\Big({Y_d  Y_{d}^{\dagger}}\Big)  -30 \mbox{Tr}\Big({Y_u  Y_{u}^{\dagger}}\Big)  + 9 g_{1}^{2} \Big)\\ 
\beta_{g_3}^{(2)}  =  \, &
\frac{1}{5} g_{3}^{3} \Big(11 g_{1}^{2}  -20 \mbox{Tr}\Big({Y_d  Y_{d}^{\dagger}}\Big)  -20 \mbox{Tr}\Big({Y_u  Y_{u}^{\dagger}}\Big)  + 45 g_{2}^{2}  + 70 g_{3}^{2} \Big)
\end{align}} 
These expressions can be used according to \ref{app:RGE_con} to calculate all RGEs of the NMSSM. 

\section{Couplings}
We list in the following all couplings of the NMSSM which contribute to the electroweak self-energies or influence the annihilation or coannihilation of the neutralino. These and all other couplings of the NMSSM can be derived with the command {\tt MakeVertexList[EWSB]} of {\tt SARAH}.Our conventions are given in \ref{app:ConVertices}. In addition, we define the following abbreviations:
\begin{eqnarray}
\tilde{\lambda} &\equiv&  + g_1^2 + g_2^2  -4 {\lambda}^{2} \\
  \bar{\lambda}  &\equiv& g_2^2  -2 {\lambda}^{2}  \\
g_-^2   &\equiv&  g_2^2 - g_1^2 \\
  g_+^2 &\equiv& g_1^2 + g_2^2  \\
\Lambda_1 &\equiv& 2 v_s \kappa \lambda^*  + 2 v_s \lambda \kappa^*  + \sqrt{2}\,2\, \mathrm{Re}\big\{T_\lambda\big\} \\
\Lambda_2 &\equiv& -2 v_u \lambda  + v_d \kappa \\
\Lambda_3 &\equiv& -2 v_d \lambda  + v_u \kappa
\end{eqnarray}
Furthermore, \(c_\Theta\) is \(\cos(\Theta_W)\) and \(s_\Theta\) is \(\sin(\Theta_W)\).
\setlength{\mathindent}{0cm}
\subsection*{Interactions between two fermions and one scalar}
\begin{align} 
%
%
 &\Gamma_{\tilde{\chi}^0_{{{i}}}\tilde{\chi}^0_{{{j}}}h_{{{k}}}}^L \, =  \,
\frac{i}{2} \Big(- g_2 N^*_{{i} 2} N^*_{{j} 3} Z^H_{{k} 1} +\sqrt{2} \lambda N^*_{{i} 5} N^*_{{j} 4} Z^H_{{k} 1} +\sqrt{2} \lambda N^*_{{i} 4} N^*_{{j} 5} Z^H_{{k} 1}- g_1 N^*_{{i} 4} N^*_{{j} 1} Z^H_{{k} 2} +g_2 N^*_{{i} 4} N^*_{{j} 2} Z^H_{{k} 2}  \nonumber \\ 
 &\; \;+\sqrt{2} \lambda N^*_{{i} 5} N^*_{{j} 3} Z^H_{{k} 2} +g_2 N^*_{{i} 2} N^*_{{j} 4} Z^H_{{k} 2} +g_1 N^*_{{i} 1} \Big(N^*_{{j} 3} Z^H_{{k} 1}  - N^*_{{j} 4} Z^H_{{k} 2} \Big)+\sqrt{2} \lambda N^*_{{i} 4} N^*_{{j} 3} Z^H_{{k} 3} \nonumber \\ 
 &-2 \sqrt{2} \kappa N^*_{{i} 5} N^*_{{j} 5} Z^H_{{k} 3} +N^*_{{i} 3} \Big(g_1 N^*_{{j} 1} Z^H_{{k} 1}  - g_2 N^*_{{j} 2} Z^H_{{k} 1}  + \sqrt{2} \lambda \Big(N^*_{{j} 4} Z^H_{{k} 3}  + N^*_{{j} 5} Z^H_{{k} 2} \Big)\Big)\Big)\\
 &\Gamma_{\tilde{\chi}^0_{{{i}}}\tilde{\chi}^0_{{{j}}}h_{{{k}}}}^R \, =  \,
\frac{i}{2} \Big(Z^H_{{k} 1} \Big(N_{{i} 3} \Big(g_1 N_{{j} 1}  - g_2 N_{{j} 2} \Big)+g_1 N_{{i} 1} N_{{j} 3} - g_2 N_{{i} 2} N_{{j} 3} +\sqrt{2} \lambda^* N_{{i} 5} N_{{j} 4} +\sqrt{2} \lambda^* N_{{i} 4} N_{{j} 5} \Big)\nonumber \\ 
 &\; \;+\sqrt{2} Z^H_{{k} 3} \Big(-2 \kappa^* N_{{i} 5} N_{{j} 5}  + \lambda^* \Big(N_{{i} 3} N_{{j} 4}  + N_{{i} 4} N_{{j} 3} \Big)\Big)+Z^H_{{k} 2} \Big(N_{{i} 4} \Big(- g_1 N_{{j} 1}  + g_2 N_{{j} 2} \Big)\nonumber \\ 
 &\; \;+\Big(- g_1 N_{{i} 1}  + g_2 N_{{i} 2} \Big)N_{{j} 4}  +\sqrt{2} \lambda^* \Big(N_{{i} 3} N_{{j} 5}  + N_{{i} 5} N_{{j} 3} \Big)\Big)\Big) \\
%
%
&\Gamma_{\tilde{\chi}^0_{{{i}}}\tilde{\chi}^0_{{{j}}}A^0_{{{k}}}}^L\, =  \,
\frac{1}{2} \Big(- g_2 N^*_{{i} 2} N^*_{{j} 3} Z^A_{{k} 1} - \sqrt{2} \lambda N^*_{{i} 5} N^*_{{j} 4} Z^A_{{k} 1} - \sqrt{2} \lambda N^*_{{i} 4} N^*_{{j} 5} Z^A_{{k} 1} - g_1 N^*_{{i} 4} N^*_{{j} 1} Z^A_{{k} 2} +g_2 N^*_{{i} 4} N^*_{{j} 2} Z^A_{{k} 2} \nonumber \\ 
 & \; \;- \sqrt{2} \lambda N^*_{{i} 5} N^*_{{j} 3} Z^A_{{k} 2} +g_2 N^*_{{i} 2} N^*_{{j} 4} Z^A_{{k} 2} - N^*_{{i} 1} \Big(- g_1 N^*_{{j} 3} Z^A_{{k} 1}  + g_1 N^*_{{j} 4} Z^A_{{k} 2} \Big)- \sqrt{2} \lambda N^*_{{i} 4} N^*_{{j} 3} Z^A_{{k} 3} \nonumber \\ 
 &\;\;+2 \sqrt{2} \kappa N^*_{{i} 5} N^*_{{j} 5} Z^A_{{k} 3} - N^*_{{i} 3} \Big(- g_1 N^*_{{j} 1} Z^A_{{k} 1}  + g_2 N^*_{{j} 2} Z^A_{{k} 1} + \sqrt{2} \lambda \Big(N^*_{{j} 4} Z^A_{{k} 3}  + N^*_{{j} 5} Z^A_{{k} 2} \Big)\Big)\Big) \\
&\Gamma_{\tilde{\chi}^0_{{{i}}}\tilde{\chi}^0_{{{j}}}A^0_{{{k}}}}^R \, =  \,
\frac{1}{2} \Big(- Z^A_{{k} 1} \Big(N_{{i} 3} \Big(g_1 N_{{j} 1}  - g_2 N_{{j} 2} \Big)+g_1 N_{{i} 1} N_{{j} 3} - g_2 N_{{i} 2} N_{{j} 3} - \sqrt{2} \lambda^*(N_{{i} 5} N_{{j} 4} +N_{{i} 4} N_{{j} 5}) \Big)\nonumber \\ 
 &\; \;+\sqrt{2} Z^A_{{k} 3} \Big(-2 \kappa^* N_{{i} 5} N_{{j} 5}  + \lambda^* \Big(N_{{i} 3} N_{{j} 4}  + N_{{i} 4} N_{{j} 3} \Big)\Big)+Z^A_{{k} 2} \Big(N_{{i} 4} \Big(g_1 N_{{j} 1}  - g_2 N_{{j} 2} \Big) \nonumber \\ 
 &\; \;+\Big(g_1 N_{{i} 1}  - g_2 N_{{i} 2} \Big)N_{{j} 4} +\sqrt{2} \lambda^* \Big(N_{{i} 3} N_{{j} 5}  + N_{{i} 5} N_{{j} 3} \Big)\Big)\Big) \\
%
%
&\Gamma_{\tilde{\chi}^0_{{{i}}}\tilde{\chi}^-_{{{j}}}H^+_{{{k}}}}^L \, =  \, %
i \Big(- g_2 V^*_{{j} 1} N^*_{{i} 3} Z^+_{{k} 1}  + V^*_{{j} 2} \Big(\frac{1}{\sqrt{2}} g_1 N^*_{{i} 1} Z^+_{{k} 1}  + \frac{1}{\sqrt{2}} g_2 N^*_{{i} 2} Z^+_{{k} 1}  - \lambda N^*_{{i} 5} Z^+_{{k} 2} \Big)\Big) \\
&\Gamma_{\chi_{{{i}}}\tilde{\chi}^-_{{{j}}}H^+_{{{k}}}}^R \, =  \,
i \Big(-\frac{1}{2} \Big(2 g_2 U_{{j} 1} N_{{i} 4}  + \sqrt{2} U_{{j} 2} \Big(g_1 N_{{i} 1}  + g_2 N_{{i} 2} \Big)\Big)Z^+_{{k} 2}  - \lambda^* U_{{j} 2} N_{{i} 5} Z^+_{{k} 1} \Big) \\
%
%
&\Gamma_{\tilde{\chi}^-_{{{i}}}\tilde{\chi}^+_{{{j}}}h_{{{k}}}}^L \, =  \,
-i \frac{1}{\sqrt{2}} \Big(g_2 V^*_{{j} 1} U^*_{{i} 2} Z^H_{{k} 2}  + V^*_{{j} 2} \Big(g_2 U^*_{{i} 1} Z^H_{{k} 1}  + \lambda U^*_{{i} 2} Z^H_{{k} 3} \Big)\Big) \\
&\Gamma_{\tilde{\chi}^-_{{{i}}}\tilde{\chi}^+_{{{j}}}h_{{{k}}}}^R  \, =  \,
-i \frac{1}{\sqrt{2}} \Big(g_2 V_{{i} 1} U_{{j} 2} Z^H_{{k} 2}  + V_{{i} 2} \Big(g_2 U_{{j} 1} Z^H_{{k} 1}  + \lambda^* U_{{j} 2} Z^H_{{k} 3} \Big)\Big) \\
%
%
&\Gamma_{\tilde{\chi}^-_{{{i}}}\tilde{\chi}^+_{{{j}}}A^0_{{{k}}}}^L \, =  \,
- \frac{1}{\sqrt{2}} \Big(g_2 V^*_{{j} 1} U^*_{{i} 2} Z^A_{{k} 2}  + V^*_{{j} 2} \Big(g_2 U^*_{{i} 1} Z^A_{{k} 1}  - \lambda U^*_{{i} 2} Z^A_{{k} 3} \Big)\Big) \\
&\Gamma_{\tilde{\chi}^-_{{{i}}}\tilde{\chi}^+_{{{j}}}A^0_{{{k}}}}^R \, =  \,
 \frac{1}{\sqrt{2}} \Big(g_2 V_{{i} 1} U_{{j} 2} Z^A_{{k} 2}  + V_{{i} 2} \Big(g_2 U_{{j} 1} Z^A_{{k} 1}  - \lambda^* U_{{j} 2} Z^A_{{k} 3} \Big)\Big) \\
%
%
&\Gamma^R_{\bar{\nu}_{{{i}}}\tilde{\chi}^0_{{{j}}}\tilde{\nu}_{{{k}}}} \, =  \,i \frac{1}{\sqrt{2}} Z^{\nu,*}_{{k} {i}}  \Big(g_1 N_{{j} 1}  - g_2 N_{{j} 2} \Big) \\
%
%
& \Gamma^R_{\bar{\nu}_{{{i}}}\tilde{\chi}^+_{{{j}}}\tilde{e}_{{{k}}}}  \, =  \,i \Big(- g_2 Z^{E,*}_{{k} {i}} V_{{j} 1}  + \sum_{{a}=1}^{3}Y^*_{e,{{i} {a}}} Z^{E,*}_{{k} 3 + {a}}  V_{{j} 2} \Big) \\
%
%
& \Gamma^R_{\bar{\nu}_{{{i}}}e_{{{j}}}H^+_{{{k}}}}  \, =  \,i \sum_{{a}=1}^{3}Y^*_{e,{{i} {a}}} U^e_{R,{{j} {a}}}  Z^+_{{k} 1}  \\
%
%
&\Gamma^L_{\tilde{\chi}^0_{{{i}}}e_{{{j}}}\tilde{e}^*_{{{k}}}} \, =  \,
i \Big(\frac{1}{\sqrt{2}} g_1 N^*_{{i} 1} \sum_{{a}=1}^{3}U^{e,*}_{L,{{j} {a}}} Z^E_{{k} {a}}  +\frac{1}{\sqrt{2}} g_2 N^*_{{i} 2} \sum_{{a}=1}^{3}U^{e,*}_{L,{{j} {a}}} Z^E_{{k} {a}}  - N^*_{{i} 3} \sum_{{a,b}=1}^{3}U^{e,*}_{L,{{j} {a}}} Y_{e,{{a} {b}}}  Z^E_{{k} 3 + {b}}  \Big)\\ 
& \Gamma^R_{\tilde{\chi}^0_{{{i}}}e_{{{j}}}\tilde{e}^*_{{{k}}}}  \, =  \,i \Big(- \sqrt{2} g_1 \sum_{{a}=1}^{3}Z^E_{{k} 3 + {a}} U^e_{R,{{j} {a}}}  N_{{i} 1} - \sum_{{a,b}=1}^{3}Y^*_{e,{{a} {b}}} Z^E_{{k} {a}}  U^e_{R,{{j} {b}}}  N_{{i} 3} \Big) \\
%
%
&\Gamma^L_{\tilde{\chi}^0_{{{i}}}d_{{{j} {\beta}}}\tilde{d}^*_{{{k} {\gamma}}}}  \, =  \,
-\frac{i}{6} \delta_{{\beta},{\gamma}} \Big(\sqrt{2} g_1 N^*_{{i} 1} \sum_{{a}=1}^{3}U^{d,*}_{L,{{j} {a}}} Z^D_{{k} {a}}  -3 \sqrt{2} g_2 N^*_{{i} 2} \sum_{{a}=1}^{3}U^{d,*}_{L,{{j} {a}}} Z^D_{{k} {a}} +6 N^*_{{i} 3} \sum_{{a,b}=1}^{3}U^{d,*}_{L,{{j} {a}}} Y_{d,{{a} {b}}}  Z^D_{{k} 3 + {b}}  \Big)\\ 
& \Gamma^R_{\tilde{\chi}^0_{{{i}}}d_{{{j} {\beta}}}\tilde{d}^*_{{{k} {\gamma}}}} \, =  \,-\frac{i}{3} \delta_{{\beta},{\gamma}} \Big(\sqrt{2} g_1 \sum_{{a}=1}^{3}Z^D_{{k} 3 + {a}} U^d_{R,{{j} {a}}}  N_{{i} 1} +3 \sum_{{a,b}=1}^{3}Y^*_{d,{{a} {b}}} Z^D_{{k} {a}}  U^d_{R,{{j} {b}}}  N_{{i} 3} \Big) \\
%
%
&\Gamma^L_{\tilde{\chi}^0_{{{i}}}u_{{{j} {\beta}}}\tilde{u}^*_{{{k} {\gamma}}}}  \, =  \,
-\frac{i}{6} \delta_{{\beta},{\gamma}} \Big(\sqrt{2} g_1 N^*_{{i} 1} \sum_{{a}=1}^{3}U^{u,*}_{L,{{j} {a}}} Z^U_{{k} {a}}  +3 \sqrt{2} g_2 N^*_{{i} 2} \sum_{{a}=1}^{3}U^{u,*}_{L,{{j} {a}}} Z^U_{{k} {a}} +6 N^*_{{i} 4} \sum_{{a,b}=1}^{3}U^{u,*}_{L,{{j} {a}}} Y_{u,{{a} {b}}}  Z^U_{{k} 3 + {b}}  \Big)\\ 
& \Gamma^R_{\tilde{\chi}^0_{{{i}}}u_{{{j} {\beta}}}\tilde{u}^*_{{{k} {\gamma}}}}   \, =  \,\frac{i}{3} \delta_{{\beta},{\gamma}} \Big(2 \sqrt{2} g_1 \sum_{{a}=1}^{3}Z^U_{{k} 3 + {a}} U^u_{R,{{j} {a}}}  N_{{i} 1} -3 \sum_{{a,b}=1}^{3}Y^*_{u,{{a} {b}}} Z^U_{{k} {a}}  U^u_{R,{{j} {b}}}  N_{{i} 4} \Big) \\
%
%
&\Gamma^L_{\tilde{\chi}^+_{{{i}}}e_{{{j}}}\tilde{\nu}^*_{{{k}}}}  \, =  \,
-i g_2 U^*_{{i} 1} \sum_{{a}=1}^{3}U^{e,*}_{L,{{j} {a}}} Z^{\nu}_{{k} {a}}  \\ 
& \Gamma^R_{\tilde{\chi}^+_{{{i}}}e_{{{j}}}\tilde{\nu}^*_{{{k}}}}  \, =  \,i \sum_{{a,b}=1}^{3}Y^*_{e,{{a} {b}}} Z^{\nu}_{{k} {a}}  U^e_{R,{{j} {b}}}  V_{{i} 2} \\
%
%
&\Gamma^L_{\tilde{\chi}^+_{{{i}}}d_{{{j} {\beta}}}\tilde{u}^*_{{{k} {\gamma}}}}  \, =  \,
i \delta_{{\beta},{\gamma}} \Big(- g_2 U^*_{{i} 1} \sum_{{a}=1}^{3}U^{d,*}_{L,{{j} {a}}} Z^U_{{k} {a}}  +U^*_{{i} 2} \sum_{{a,b}=1}^{3}U^{d,*}_{L,{{j} {a}}} Y_{u,{{a} {b}}}  Z^U_{{k} 3 + {b}}  \Big)\\ 
& \Gamma^R_{\tilde{\chi}^+_{{{i}}}d_{{{j} {\beta}}}\tilde{u}^*_{{{k} {\gamma}}}}  \, =  \,i \delta_{{\beta},{\gamma}} \sum_{{a,b}=1}^{3}Y^*_{d,{{a} {b}}} Z^U_{{k} {a}}  U^d_{R,{{j} {b}}}  V_{{i} 2} \\
%
%
&\Gamma^L_{\bar{e}_{{{i}}}e_{{{j}}}h_{{{k}}}} \, =  \,
-i \frac{1}{\sqrt{2}} \sum_{{b}=1}^{3}U^{e,*}_{R,{{i} {b}}} \sum_{{a}=1}^{3}U^{e,*}_{L,{{j} {a}}} Y_{e,{{a} {b}}}   Z^H_{{k} 1} \\ 
& \Gamma^R_{\bar{e}_{{{i}}}e_{{{j}}}h_{{{k}}}}  \, =  \,-i \frac{1}{\sqrt{2}} \sum_{{a,b}=1}^{3}Y^*_{e,{{a} {b}}} U^e_{L,{{i} {a}}}  U^e_{R,{{j} {b}}}  Z^H_{{k} 1} \\
%
%
&\Gamma^L_{\bar{e}_{{{i}}}e_{{{j}}}A^0_{{{k}}}} \, =  \,
\frac{1}{\sqrt{2}} \sum_{{b}=1}^{3}U^{e,*}_{R,{{i} {b}}} \sum_{{a}=1}^{3}U^{e,*}_{L,{{j} {a}}} Y_{e,{{a} {b}}}   Z^A_{{k} 1} \\ 
& \Gamma^R_{\bar{e}_{{{i}}}e_{{{j}}}A^0_{{{k}}}}  \, =  \,- \frac{1}{\sqrt{2}} \sum_{{a,b}=1}^{3}Y^*_{e,{{a} {b}}} U^e_{L,{{i} {a}}}  U^e_{R,{{j} {b}}}  Z^A_{{k} 1} \\
%
%
&\Gamma^L_{\bar{d}_{{{i} {\alpha}}}d_{{{j} {\beta}}}h_{{{k}}}}  \, =  \,
-i \frac{1}{\sqrt{2}} \delta_{{\alpha},{\beta}} \sum_{{b}=1}^{3}U^{d,*}_{R,{{i} {b}}} \sum_{{a}=1}^{3}U^{d,*}_{L,{{j} {a}}} Y_{d,{{a} {b}}}   Z^H_{{k} 1} \\ 
& \Gamma^R_{\bar{d}_{{{i} {\alpha}}}d_{{{j} {\beta}}}h_{{{k}}}} \, =  \,-i \frac{1}{\sqrt{2}} \delta_{{\alpha},{\beta}} \sum_{{a,b}=1}^{3}Y^*_{d,{{a} {b}}} U^d_{L,{{i} {a}}}  U^d_{R,{{j} {b}}}  Z^H_{{k} 1}\\
%
%
&\Gamma^L_{\bar{d}_{{{i} {\alpha}}}d_{{{j} {\beta}}}A^0_{{{k}}}} \, =  \,
\frac{1}{\sqrt{2}} \delta_{{\alpha},{\beta}} \sum_{{b}=1}^{3}U^{d,*}_{R,{{i} {b}}} \sum_{{a}=1}^{3}U^{d,*}_{L,{{j} {a}}} Y_{d,{{a} {b}}}   Z^A_{{k} 1} \\ 
 &\Gamma^R_{\bar{d}_{{{i} {\alpha}}}d_{{{j} {\beta}}}A^0_{{{k}}}}  \, =  \,- \frac{1}{\sqrt{2}} \delta_{{\alpha},{\beta}} \sum_{{a,b}=1}^{3}Y^*_{d,{{a} {b}}} U^d_{L,{{i} {a}}}  U^d_{R,{{j} {b}}}  Z^A_{{k} 1} \\
%
%
&\Gamma^L_{\bar{d}_{{{i} {\alpha}}}u_{{{j} {\beta}}}H^-_{{{k}}}} \, =  \,
i Z^{+,*}_{{k} 1} \delta_{{\alpha},{\beta}} \sum_{{b}=1}^{3}U^{d,*}_{R,{{i} {b}}} \sum_{{a}=1}^{3}U^{u,*}_{L,{{j} {a}}} Y_{d,{{a} {b}}}   \\ 
& \Gamma^R_{\bar{d}_{{{i} {\alpha}}}u_{{{j} {\beta}}}H^-_{{{k}}}}  \, =  \,i Z^{+,*}_{{k} 2} \delta_{{\alpha},{\beta}} \sum_{{a,b}=1}^{3}Y^*_{u,{{a} {b}}} U^d_{L,{{i} {a}}}  U^u_{R,{{j} {b}}} \\
%
%
&\Gamma^L_{\bar{u}_{{{i} {\alpha}}}u_{{{j} {\beta}}}h_{{{k}}}}  \, =  \,
-i \frac{1}{\sqrt{2}} \delta_{{\alpha},{\beta}} \sum_{{b}=1}^{3}U^{u,*}_{R,{{i} {b}}} \sum_{{a}=1}^{3}U^{u,*}_{L,{{j} {a}}} Y_{u,{{a} {b}}}   Z^H_{{k} 2} \\ 
& \Gamma^R_{\bar{u}_{{{i} {\alpha}}}u_{{{j} {\beta}}}h_{{{k}}}}  \, =  \,-i \frac{1}{\sqrt{2}} \delta_{{\alpha},{\beta}} \sum_{{a,b}=1}^{3}Y^*_{u,{{a} {b}}} U^u_{L,{{i} {a}}}  U^u_{R,{{j} {b}}}  Z^H_{{k} 2} \\
%
%
&\Gamma^L_{\bar{u}_{{{i} {\alpha}}}u_{{{j} {\beta}}}A^0_{{{k}}}}\, =  \,
\frac{1}{\sqrt{2}} \delta_{{\alpha},{\beta}} \sum_{{b}=1}^{3}U^{u,*}_{R,{{i} {b}}} \sum_{{a}=1}^{3}U^{u,*}_{L,{{j} {a}}} Y_{u,{{a} {b}}}   Z^A_{{k} 2} \\ 
& \Gamma^R_{\bar{u}_{{{i} {\alpha}}}u_{{{j} {\beta}}}A^0_{{{k}}}} \, =  \,- \frac{1}{\sqrt{2}} \delta_{{\alpha},{\beta}} \sum_{{a,b}=1}^{3}Y^*_{u,{{a} {b}}} U^u_{L,{{i} {a}}}  U^u_{R,{{j} {b}}}  Z^A_{{k} 2}
\end{align} 
\setlength{\mathindent}{1cm}
\subsection*{Interactions between two fermions and one vector boson}
\begin{align} 
\Gamma^L_{\bar{\nu}_{{{i}}}\nu_{{{j}}}Z_{{\mu}}}  =  & \,
-\frac{i}{2} \delta_{{i},{j}} \Big(g_1 s_{\Theta}  + g_2 c_{\Theta} \Big)\\ 
 \Gamma^L_{\bar{\nu}_{{{i}}}e_{{{j}}}W^+_{{\mu}}}  =  & \,
-i \frac{1}{\sqrt{2}} g_2 U^{e,*}_{L,{{j} {i}}}  \\ 
\Gamma^L_{\tilde{\chi}^0_{{{i}}}\tilde{\chi}^0_{{{j}}}\gamma_{{\mu}}}  =  & \,
\frac{i}{2} \Big(g_1 c_{\Theta}  - g_2 s_{\Theta} \Big)\Big(N^*_{{j} 3} N_{{i} 3}  - N^*_{{j} 4} N_{{i} 4} \Big)\\ 
 \Gamma^R_{\tilde{\chi}^0_{{{i}}}\tilde{\chi}^0_{{{j}}}\gamma_{{\mu}}}  =  & \,-\frac{i}{2} \Big(g_1 c_{\Theta}  - g_2 s_{\Theta} \Big)\Big(N^*_{{i} 3} N_{{j} 3}  - N^*_{{i} 4} N_{{j} 4} \Big) \\
\Gamma^L_{\tilde{\chi}^0_{{{i}}}\tilde{\chi}^0_{{{j}}}Z_{{\mu}}}  =  & \,
-\frac{i}{2} \Big(g_1 s_{\Theta}  + g_2 c_{\Theta} \Big)\Big(N^*_{{j} 3} N_{{i} 3}  - N^*_{{j} 4} N_{{i} 4} \Big)\\ 
 \Gamma^R_{\tilde{\chi}^0_{{{i}}}\tilde{\chi}^0_{{{j}}}Z_{{\mu}}}  =  & \,\frac{i}{2} \Big(g_1 s_{\Theta}  + g_2 c_{\Theta} \Big)\Big(N^*_{{i} 3} N_{{j} 3}  - N^*_{{i} 4} N_{{j} 4} \Big) \\
\Gamma^L_{\tilde{\chi}^0_{{{i}}}\tilde{\chi}^-_{{{j}}}W^+_{{\mu}}}  =  & \,
-\frac{i}{2} g_2 \Big(2 V^*_{{j} 1} N_{{i} 2}  + \sqrt{2} V^*_{{j} 2} N_{{i} 3} \Big)\\ 
 \Gamma^R_{\tilde{\chi}^0_{{{i}}}\tilde{\chi}^-_{{{j}}}W^+_{{\mu}}}  =  & \,-\frac{i}{2} g_2 \Big(2 N^*_{{i} 2} U_{{j} 1}  - \sqrt{2} N^*_{{i} 4} U_{{j} 2} \Big) \\
\Gamma^L_{\tilde{\chi}^+_{{{i}}}\tilde{\chi}^-_{{{j}}}\gamma_{{\mu}}}  =  & \,
\frac{i}{2} \Big(2 g_2 V^*_{{j} 1} s_{\Theta} V_{{i} 1}  + V^*_{{j} 2} \Big(g_1 c_{\Theta}  + g_2 s_{\Theta} \Big)V_{{i} 2} \Big)\\ 
 \Gamma^R_{\tilde{\chi}^+_{{{i}}}\tilde{\chi}^-_{{{j}}}\gamma_{{\mu}}}  =  & \,\frac{i}{2} \Big(2 g_2 U^*_{{i} 1} s_{\Theta} U_{{j} 1}  + U^*_{{i} 2} \Big(g_1 c_{\Theta}  + g_2 s_{\Theta} \Big)U_{{j} 2} \Big) \\
\Gamma^L_{\tilde{\chi}^+_{{{i}}}\tilde{\chi}^-_{{{j}}}Z_{{\mu}}}  =  & \,
\frac{i}{2} \Big(2 g_2 V^*_{{j} 1} c_{\Theta} V_{{i} 1}  + V^*_{{j} 2} \Big(- g_1 s_{\Theta}  + g_2 c_{\Theta} \Big)V_{{i} 2} \Big)\\ 
 \Gamma^R_{\tilde{\chi}^+_{{{i}}}\tilde{\chi}^-_{{{j}}}Z_{{\mu}}}  =  & \,\frac{i}{2} \Big(2 g_2 U^*_{{i} 1} c_{\Theta} U_{{j} 1}  + U^*_{{i} 2} \Big(- g_1 s_{\Theta}  + g_2 c_{\Theta} \Big)U_{{j} 2} \Big) \\
\Gamma_{\bar{e}_{{{i}}}e_{{{j}}}\gamma_{{\mu}}}  =  & \,
\frac{i}{2} \delta_{{i},{j}} \Big(g_1 c_{\Theta}  + g_2 s_{\Theta} \Big) \gamma_\mu P_L + 
i g_1 c_{\Theta} \delta_{{i},{j}}  \gamma_\mu P_R \\
\Gamma_{\bar{e}_{{{i}}}e_{{{j}}}Z_{{\mu}}}  =  & \,
\frac{i}{2} \delta_{{i},{j}} \Big(- g_1 s_{\Theta}  + g_2 c_{\Theta} \Big) \gamma_\mu P_L + 
-i g_1 \delta_{{i},{j}} s_{\Theta} \gamma_\mu P_R \\
\Gamma_{\bar{d}_{{{i} {\alpha}}}d_{{{j} {\beta}}}\gamma_{{\mu}}}  =  & \,
-\frac{i}{6} \delta_{{\alpha},{\beta}} \delta_{{i},{j}} \Big(-3 g_2 s_{\Theta}  + g_1 c_{\Theta} \Big)\gamma_\mu P_L + 
 \frac{i}{3} g_1 c_{\Theta} \delta_{{\alpha},{\beta}} \delta_{{i},{j}} \gamma_\mu P_R \\
\Gamma_{\bar{d}_{{{i} {\alpha}}}d_{{{j} {\beta}}}Z_{{\mu}}}  =  & \,
\frac{i}{6} \delta_{{\alpha},{\beta}} \delta_{{i},{j}} \Big(3 g_2 c_{\Theta}  + g_1 s_{\Theta} \Big) \gamma_\mu P_L +  
 -\frac{i}{3} g_1 \delta_{{\alpha},{\beta}} \delta_{{i},{j}} s_{\Theta} \gamma_\mu P_R \\
\Gamma^L_{\bar{d}_{{{i} {\alpha}}}u_{{{j} {\beta}}}W^-_{{\mu}}}  =  & \,
-i \frac{1}{\sqrt{2}} g_2 \delta_{{\alpha},{\beta}} \sum_{{a}=1}^{3}U^{u,*}_{L,{{j} {a}}} U^d_{L,{{i} {a}}}  \\ 
\Gamma_{\bar{u}_{{{i} {\alpha}}}u_{{{j} {\beta}}}\gamma_{{\mu}}}  =  & \,
-\frac{i}{6} \delta_{{\alpha},{\beta}} \delta_{{i},{j}} \Big(3 g_2 s_{\Theta}  + g_1 c_{\Theta} \Big) \gamma_\mu P_L + 
 -\frac{2 i}{3} g_1 c_{\Theta} \delta_{{\alpha},{\beta}} \delta_{{i},{j}}  \gamma_\mu P_R \\
\Gamma_{\bar{u}_{{{i} {\alpha}}}u_{{{j} {\beta}}}Z_{{\mu}}}  =  & \,
-\frac{i}{6} \delta_{{\alpha},{\beta}} \delta_{{i},{j}} \Big(3 g_2 c_{\Theta}  - g_1 s_{\Theta} \Big) \gamma_\mu P_L + 
 \frac{2 i}{3} g_1 \delta_{{\alpha},{\beta}} \delta_{{i},{j}} s_{\Theta} \gamma_\mu P_R
\end{align} 

\subsection*{Interactions between  two scalars and one vector boson}
\begin{align} 
\Gamma_{\tilde{d}_{{{i} {\alpha}}}\tilde{d}^*_{{{j} {\beta}}}\gamma_{{\mu}}}  = & \, 
-\frac{i}{6} \delta_{{\alpha},{\beta}} \Big(\Big(-3 g_2 s_{\Theta}  + g_1 c_{\Theta} \Big)\sum_{{a}=1}^{3}Z^{D,*}_{{i} {a}} Z^D_{{j} {a}}  -2 g_1 c_{\Theta} \sum_{{a}=1}^{3}Z^{D,*}_{{i} 3 + {a}} Z^D_{{j} 3 + {a}}  \Big) \\
\Gamma_{\tilde{d}_{{{i} {\alpha}}}\tilde{d}^*_{{{j} {\beta}}}Z_{{\mu}}}  = & \, 
\frac{i}{6} \delta_{{\alpha},{\beta}} \Big(\Big(3 g_2 c_{\Theta}  + g_1 s_{\Theta} \Big)\sum_{{a}=1}^{3}Z^{D,*}_{{i} {a}} Z^D_{{j} {a}}  -2 g_1 s_{\Theta} \sum_{{a}=1}^{3}Z^{D,*}_{{i} 3 + {a}} Z^D_{{j} 3 + {a}}  \Big) \\
\Gamma_{\tilde{\nu}_{{{i}}}\tilde{e}^*_{{{j}}}W^-_{{\mu}}}  = & \, 
-i \frac{1}{\sqrt{2}} g_2 \sum_{{a}=1}^{3}Z^{\nu,*}_{{i} {a}} Z^E_{{j} {a}} \\
\Gamma_{\tilde{\nu}_{{{i}}}\tilde{\nu}^*_{{{j}}}Z_{{\mu}}}  = & \, 
-\frac{i}{2} \sum_{{a}=1}^{3}Z^{\nu,*}_{{i} {a}} Z^\nu_{{j} {a}} \Big(g_1 s_{\Theta}  + g_2 c_{\Theta} \Big) \\
\Gamma_{\tilde{u}_{{{i} {\alpha}}}\tilde{d}^*_{{{j} {\beta}}}W^-_{{\mu}}}  = & \, 
-i \frac{1}{\sqrt{2}} g_2 \delta_{{\alpha},{\beta}} \sum_{{a}=1}^{3}Z^{U,*}_{{i} {a}} Z^D_{{j} {a}} \\
\Gamma_{\tilde{u}_{{{i} {\alpha}}}\tilde{u}^*_{{{j} {\beta}}}\gamma_{{\mu}}}  = & \, 
-\frac{i}{6} \delta_{{\alpha},{\beta}} \Big(\Big(3 g_2 s_{\Theta}  + g_1 c_{\Theta} \Big)\sum_{{a}=1}^{3}Z^{U,*}_{{i} {a}} Z^U_{{j} {a}}  +4 g_1 c_{\Theta} \sum_{{a}=1}^{3}Z^{U,*}_{{i} 3 + {a}} Z^U_{{j} 3 + {a}}  \Big) \\
\Gamma_{\tilde{u}_{{{i} {\alpha}}}\tilde{u}^*_{{{j} {\beta}}}Z_{{\mu}}}  = & \, 
-\frac{i}{6} \delta_{{\alpha},{\beta}} \Big(\Big(3 g_2 c_{\Theta}  - g_1 s_{\Theta} \Big)\sum_{{a}=1}^{3}Z^{U,*}_{{i} {a}} Z^U_{{j} {a}}  -4 g_1 s_{\Theta} \sum_{{a}=1}^{3}Z^{U,*}_{{i} 3 + {a}} Z^U_{{j} 3 + {a}}  \Big) \\
\Gamma_{\tilde{e}_{{{i}}}\tilde{e}^*_{{{j}}}\gamma_{{\mu}}}  = & \, 
\frac{i}{2} \Big(\Big(g_1 c_{\Theta}  + g_2 s_{\Theta} \Big)\sum_{{a}=1}^{3}Z^{E,*}_{{i} {a}} Z^E_{{j} {a}}  +2 g_1 c_{\Theta} \sum_{{a}=1}^{3}Z^{E,*}_{{i} 3 + {a}} Z^E_{{j} 3 + {a}}  \Big) \\
\Gamma_{\tilde{e}_{{{i}}}\tilde{e}^*_{{{j}}}Z_{{\mu}}}  = & \, 
\frac{i}{2} \Big(\Big(- g_1 s_{\Theta}  + g_2 c_{\Theta} \Big)\sum_{{a}=1}^{3}Z^{E,*}_{{i} {a}} Z^E_{{j} {a}}  -2 g_1 s_{\Theta} \sum_{{a}=1}^{3}Z^{E,*}_{{i} 3 + {a}} Z^E_{{j} 3 + {a}}  \Big) \\
\Gamma_{h_{{{i}}}H^+_{{{j}}}W^-_{{\mu}}}  = & \, 
-\frac{i}{2} g_2 \Big(Z^H_{{i} 1} Z^+_{{j} 1}  - Z^H_{{i} 2} Z^+_{{j} 2} \Big) \\
\Gamma_{h_{{{i}}}A^0_{{{j}}}Z_{{\mu}}}  = & \, 
\frac{1}{2} \Big(- g_1 s_{\Theta}  - g_2 c_{\Theta} \Big)\Big(Z^A_{{j} 1} Z^H_{{i} 1}  - Z^A_{{j} 2} Z^H_{{i} 2} \Big) \\
\Gamma_{A^0_{{{i}}}H^+_{{{j}}}W^-_{{\mu}}}  = & \, 
\frac{1}{2} g_2 \Big(Z^A_{{i} 1} Z^+_{{j} 1}  + Z^A_{{i} 2} Z^+_{{j} 2} \Big) \\
\Gamma_{H^-_{{{i}}}H^+_{{{j}}}\gamma_{{\mu}}}  = & \, 
\frac{i}{2} \Big(g_1 c_{\Theta}  + g_2 s_{\Theta} \Big)\Big(Z^{+,*}_{{i} 1} Z^+_{{j} 1}  + Z^{+,*}_{{i} 2} Z^+_{{j} 2} \Big) \\
\Gamma_{H^-_{{{i}}}H^+_{{{j}}}Z_{{\mu}}}  = & \, 
\frac{i}{2} \Big(- g_1 s_{\Theta}  + g_2 c_{\Theta} \Big)\Big(Z^{+,*}_{{i} 1} Z^+_{{j} 1}  + Z^{+,*}_{{i} 2} Z^+_{{j} 2} \Big)
\end{align}

\subsection*{Interactions between  one scalar and two vector bosons}
\begin{align} 
\Gamma_{h_{{{i}}}W^+_{{\sigma}}W^-_{{\mu}}}  = & \, 
\frac{i}{2} g_2^2 \Big(v_d Z^H_{{i} 1}  + v_u Z^H_{{i} 2} \Big) \\
\Gamma_{h_{{{i}}}Z_{{\sigma}}Z_{{\mu}}}  = & \, 
\frac{i}{2} \left(g_1 s_{\Theta}  + g_2 c_{\Theta} \right)^{2} \Big(v_d Z^H_{{i} 1}  + v_u Z^H_{{i} 2} \Big) \\
\Gamma_{H^-_{{{i}}}W^+_{{\sigma}}\gamma_{{\mu}}}  = & \, 
-\frac{i}{2} g_1 g_2 \Big(v_d Z^{+,*}_{{i} 1}  - v_u Z^{+,*}_{{i} 2} \Big)c_{\Theta}  \\
\Gamma_{H^-_{{{i}}}W^+_{{\sigma}}Z_{{\mu}}}  = & \, 
\frac{i}{2} g_1 g_2 \Big(v_d Z^{+,*}_{{i} 1}  - v_u Z^{+,*}_{{i} 2} \Big)s_{\Theta} 
\end{align}

\subsection*{Interactions between  two scalars and two vector bosons}
\begin{align} 
\Gamma_{\tilde{\nu}_{{{i}}}W^-_{{\sigma}}\tilde{\nu}^*_{{{k}}}W^+_{{\nu}}}  = & \, 
\frac{i}{2} g_2^2 \delta_{{i},{k}} \\
\Gamma_{\tilde{\nu}_{{{i}}}Z_{{\sigma}}\tilde{\nu}^*_{{{k}}}Z_{{\nu}}}  = & \, 
\frac{i}{2} \delta_{{i},{k}} \left(g_1 s_{\Theta}  + g_2 c_{\Theta} \right)^{2} \\
\Gamma_{\tilde{e}_{{{i}}}W^-_{{\sigma}}\tilde{e}^*_{{{k}}}W^+_{{\nu}}}  = & \, 
\frac{i}{2} g_2^2 \sum_{{b}=1}^{3}Z^{E,*}_{{i} {b}} Z^E_{{k} {b}} \\
\Gamma_{\tilde{e}_{{{i}}}Z_{{\sigma}}\tilde{e}^*_{{{k}}}Z_{{\nu}}}  = & \, 
-i \Big(-\frac{1}{2} \left(- g_1 s_{\Theta}  + g_2 c_{\Theta} \right)^{2} \sum_{{b}=1}^{3}Z^{E,*}_{{i} {b}} Z^E_{{k} {b}}  -2 g_1^2 {s_{\Theta}}^{2} \sum_{{b}=1}^{3}Z^{E,*}_{{i} 3 + {b}} Z^E_{{k} 3 + {b}}  \Big) \\
\Gamma_{h_{{{i}}}W^-_{{\sigma}}h_{{{k}}}W^+_{{\nu}}}  = & \, 
\frac{i}{2} g_2^2 \Big(Z^H_{{i} 1} Z^H_{{k} 1}  + Z^H_{{i} 2} Z^H_{{k} 2} \Big) \\
\Gamma_{h_{{{i}}}Z_{{\sigma}}h_{{{k}}}Z_{{\nu}}}  = & \, 
\frac{i}{2} \left(g_1 s_{\Theta}  + g_2 c_{\Theta} \right)^{2} \Big(Z^H_{{i} 1} Z^H_{{k} 1}  + Z^H_{{i} 2} Z^H_{{k} 2} \Big) \\
\Gamma_{A^0_{{{i}}}W^-_{{\sigma}}A^0_{{{k}}}W^+_{{\nu}}}  = & \, 
\frac{i}{2} g_2^2 \Big(Z^A_{{i} 1} Z^A_{{k} 1}  + Z^A_{{i} 2} Z^A_{{k} 2} \Big) \\
\Gamma_{A^0_{{{i}}}Z_{{\sigma}}A^0_{{{k}}}Z_{{\nu}}}  = & \, 
\frac{i}{2} \left(g_1 s_{\Theta}  + g_2 c_{\Theta} \right)^{2} \Big(Z^A_{{i} 1} Z^A_{{k} 1}  + Z^A_{{i} 2} Z^A_{{k} 2} \Big) \\
\Gamma_{H^-_{{{i}}}W^-_{{\sigma}}H^+_{{{k}}}W^+_{{\nu}}}  = & \, 
\frac{i}{2} g_2^2 \Big(Z^{+,*}_{{i} 1} Z^+_{{k} 1}  + Z^{+,*}_{{i} 2} Z^+_{{k} 2} \Big) \\
\Gamma_{H^-_{{{i}}}Z_{{\sigma}}H^+_{{{k}}}Z_{{\nu}}}  = & \, 
\frac{i}{2} \left(- g_1 s_{\Theta}  + g_2 c_{\Theta} \right)^{2} \Big(Z^{+,*}_{{i} 1} Z^+_{{k} 1}  + Z^{+,*}_{{i} 2} Z^+_{{k} 2} \Big)
\end{align}

\subsection*{Interactions between  four scalars}
We define 
\begin{align}
A_1 = & \frac{i}{12} \Big(\Big(C_L^1 g_1^2 +3 C_L^2 g_2^2\Big) \sum_{{a}=1}^{3}Z^{F,*}_{{k} {a}} 
Z^F_{{i} {a}}  +2 C_R^1 g_1^2 \sum_{{a}=1}^{3}Z^{F,*}_{{k} 3 + {a}} 
Z^F_{{i} 3 + {a}}  \Big) \\
A_2 = & -i \Big(\sum_{{c}=1}^{3}Z^{F,*}_{{k} 3 + {c}}
\sum_{{a,b}=1}^{3}Y^*_{f,{{a} {c}}} Y_{f,{{a} {b}}}
 Z^F_{{i} 3 + {b}} +\sum_{{c}=1}^{3}\sum_{{b}=1}^{3}Z^{F,*}_{{k} {b}}
\sum_{{a}=1}^{3}Y^*_{f,{{c} {a}}} Y_{f,{{b} {a}}}   Z^F_{{i} {c}}  \Big) \\
A_3 = & \frac{i}{2} \Big( \lambda \sum_{{b}=1}^{3}Z^{F,*}_{{k} 3 + {b}}
\sum_{{a}=1}^{3}Y^*_{f,{{a} {b}}} Z^F_{{i} {a}}   + \lambda^* 
\sum_{{a,b}=1}^{3}Z^{F,*}_{{k} {a}} Y_{f,{{a} {b}}} 
 Z^F_{{i} 3 + {b}}  \Big) \\
A_4 = & -\frac{i}{ \sqrt{2} } \Big(\sum_{{b}=1}^{3}Z^{F,*}_{{i} 3 + {b}} 
\sum_{{a}=1}^{3}T^*_{f,{{a} {b}}} Z^F_{{j} {a}}  +
\sum_{{a,b}=1}^{3}Z^{F,*}_{{i} {a}} T_{f,{{a} {b}}} 
 Z^F_{{j} 3 + {b}}  \Big) \\
A_5 = & \frac{i}{12} \Big(\Big( C_R^1 g_1^2  +3 C_L^1 g_2^2 \Big) \sum_{{a}=1}^{3}Z^{F,*}_{{i} {a}} 
Z^F_{{j} {a}}  +2 g_1^2 \sum_{{a}=1}^{3}Z^{F,*}_{{i} 3 + {a}} 
Z^F_{{j} 3 + {a}}  \Big) \\
A_6 = & \frac{i}{2} \Big(\lambda \sum_{{b}=1}^{3}Z^{F,*}_{{i} 3 + {b}} 
\sum_{{a}=1}^{3}Y^*_{f,{{a} {b}}} Z^F_{{j} {a}}   +\lambda^* 
\sum_{{a,b}=1}^{3}Z^{F,*}_{{i} {a}} Y_{f,{{a} {b}}} 
 Z^F_{{j} 3 + {b}}  \Big) \\
A_7 = & -i \Big( \sum_{{c}=1}^{3}Z^{F,*}_{{i} 3 + {c}} \
\sum_{{a,b}=1}^{3}Y^*_{f,{{a} {c}}} Y_{f,{{a} {b}}} \
 Z^F_{{j} 3 + {b}}  +
\sum_{{c}=1}^{3}\sum_{{b}=1}^{3}Z^{F,*}_{{i} {b}} \
\sum_{{a}=1}^{3}Y^*_{f,{{c} {a}}} Y_{f,{{b} {a}}}   Z^F_{{j} \
{c}}  \Big) \\
A_8 = & \frac{1}{\sqrt{2} } \Big(- \sum_{{b}=1}^{3}Z^{F,*}_{{i} 3 + {b}} 
\sum_{{a}=1}^{3}T^*_{f,{{a} {b}}} Z^F_{{j} {a}}   +
\sum_{{a,b}=1}^{3}Z^{F,*}_{{i} {a}} T_{f,{{a} {b}}} \ Z^F_{{j} 3 + {b}}  \Big) \\
A_9 = & \frac{1}{2} \Big(- \lambda \sum_{{b}=1}^{3}Z^{F,*}_{{i} 3 + {b}} 
\sum_{{a}=1}^{3}Y^*_{f,{{a} {b}}} Z^F_{{j} {a}}   +\lambda^* 
\sum_{{a,b}=1}^{3}Z^{F,*}_{{i} {a}} Y_{f,{{a} {b}}} 
 Z^F_{{j} 3 + {b}}  \Big)
\end{align}
With this definitions often appearing terms in the vertices involving squarks and sleptons are given by
\begin{align}
 D_i = & A_i \hspace{0.2cm}  \mbox{with} \hspace{0.2cm}  Y_f \rightarrow Y_d, T_f \rightarrow T_d, Z^F \rightarrow Z^D, C^1_L \rightarrow 1, C^1_R \rightarrow 1, C^2_L \rightarrow 1 \\
U_i = & A_i \hspace{0.2cm}  \mbox{with} \hspace{0.2cm}  Y_f \rightarrow Y_u, T_f \rightarrow T_u, Z^F \rightarrow Z^U, C^1_L \rightarrow 1, C^1_R \rightarrow -2, C^2_L \rightarrow -1 \\
E_i = & A_i \hspace{0.2cm} \mbox{with} \hspace{0.2cm}  Y_f \rightarrow Y_e, T_f \rightarrow T_e, Z^F \rightarrow Z^E, C^1_L \rightarrow 3, C^1_R \rightarrow -3, C^2_L \rightarrow -1 
\end{align}

\setlength{\mathindent}{0cm}
\begin{align} 
%
%
&  \Gamma_{\tilde{d}_{{{i} {\alpha}}}h_{{{j}}}\tilde{d}^*_{{{k} {\gamma}}}h_{{{l}}}}  \, =  \,
\delta_{{\alpha},{\beta}} \Big(D_1 \Big(Z^H_{{j} 1} Z^H_{{l} 1}  - Z^H_{{j} 2} Z^H_{{l} 2} 
\Big) + D_2 Z^H_{{j} 1} Z^H_{{l} 1}  + D_3 \Big(Z^H_{{j} 2} 
Z^H_{{l} 3}  + Z^H_{{j} 3} Z^H_{{l} 2} \Big) \Big)\\
%
%
&  \Gamma_{\tilde{d}_{{{i} {\alpha}}}A^0_{{{j}}}\tilde{d}^*_{{{k} {\gamma}}}A^0_{{{l}}}} \, =  \,
\delta_{{\alpha},{\beta}} \Big(D_1 \Big(Z^A_{{j} 1} Z^A_{{l} 1}  - Z^A_{{j} 2} Z^A_{{l} 2}\Big) + D_2 Z^A_{{j} 1} Z^A_{{l} 1}  + D_3 \Big(- Z^A_{{j} 
2} Z^A_{{l} 3}  - Z^A_{{j} 3} Z^A_{{l} 2} \Big)\Big)\\
%
%
&  \Gamma_{\tilde{u}_{{{i} {\alpha}}}h_{{{j}}}\tilde{u}^*_{{{k} {\gamma}}}h_{{{l}}}} \, =  \,
\delta_{{\alpha},{\beta}} \Big(U_1 \Big(Z^H_{{j} 1} Z^H_{{l} 1}  - Z^H_{{j} 2} Z^H_{{l} 2} \Big) + U_2 Z^H_{{j} 2} Z^H_{{l} 2} + U_3 \Big(Z^H_{{j} 1}Z^H_{{l} 3}  + Z^H_{{j} 3} Z^H_{{l} 1} \Big) \Big)\\
%
%
&  \Gamma_{\tilde{u}_{{{i} {\alpha}}}A^0_{{{j}}}\tilde{u}^*_{{{k} {\gamma}}}A^0_{{{l}}}} \, =  \,
\delta_{{\alpha},{\beta}} \Big(U_1 \Big(Z^A_{{j} 1} Z^A_{{l} 1}  - Z^A_{{j} 2} Z^A_{{l} 2} 
\Big) + U_2 Z^A_{{j} 2} Z^A_{{l} 2} + U_3 \Big(- Z^A_{{j} 1} Z^A_{{l} 3}  - Z^A_{{j} 3} Z^A_{{l} 1} \Big) \Big)\\
%
%
&  \Gamma_{\tilde{e}_{{{i}}}h_{{{j}}}\tilde{e}^*_{{{k}}}h_{{{l}}}} \, =  \,
E_1 \Big(- Z^H_{{j} 1} Z^H_{{l} 1}  + Z^H_{{j} 2} Z^H_{{l} 2} \Big) + E_2 Z^H_{{j} 1} Z^H_{{l} 1}  + E_3 \Big(Z^H_{{j} 2} Z^H_{{l} 3}  + Z^H_{{j} 3} Z^H_{{l} 2} \Big) \\
%
%
&  \Gamma_{\tilde{e}_{{{i}}}A_{{{h,j}}}\tilde{e}^*_{{{k}}}A_{{{h,l}}}}  \, =  \,
E_1 \Big(- Z^A_{{j} 1} Z^A_{{l} 1}  + Z^A_{{j} 2} Z^A_{{l} 2} \Big) + E_2 Z^A_{{j} 1} Z^A_{{l} 1}  + E_3 \Big(- Z^A_{{j} 2} Z^A_{{l} 3}  - Z^A_{{j} 3} Z^A_{{l} 2} \Big) \\
%
%
&  \Gamma_{h_{{{i}}}h_{{{j}}}h_{{{k}}}h_{{{l}}}}\, =  \,
\frac{i}{4} \Big(Z^H_{{i} 1} \Big(Z^H_{{j} 1} \Big(-3 g_+^2Z^H_{{k} 1} Z^H_{{l} 1}  + \tilde{\lambda}Z^H_{{k} 2} Z^H_{{l} 2}  -4 |\lambda|^2 Z^H_{{k} 3} Z^H_{{l} 3} \Big)\nonumber \\ 
 & \; \;+Z^H_{{j} 2} \Big(\tilde{\lambda}Z^H_{{k} 2} Z^H_{{l} 1} +\tilde{\lambda}Z^H_{{k} 1} Z^H_{{l} 2} +4 \mathrm{Re}\big\{\kappa \lambda\big\} Z^H_{{k} 3} Z^H_{{l} 3} \Big)+2 Z^H_{{j} 3} \Big(\lambda \kappa^* \Big(Z^H_{{k} 2} Z^H_{{l} 3}  + Z^H_{{k} 3} Z^H_{{l} 2} \Big)\nonumber \\ 
 & \; \;+\lambda^* \Big(\Big(\kappa Z^H_{{k} 2} -2 \lambda Z^H_{{k} 1}  \Big)Z^H_{{l} 3}  + Z^H_{{k} 3} \Big(  \kappa Z^H_{{l} 2} -2 \lambda Z^H_{{l} 1} \Big)\Big)\Big)\Big)+Z^H_{{i} 2} \Big(Z^H_{{j} 2} \Big(-3 g_+^2Z^H_{{k} 2} Z^H_{{l} 2} \nonumber \\
& \; \; + \tilde{\lambda}Z^H_{{k} 1} Z^H_{{l} 1}   -4 |\lambda|^2 Z^H_{{k} 3} Z^H_{{l} 3} \Big)+Z^H_{{j} 1} \Big(\tilde{\lambda}Z^H_{{k} 2} Z^H_{{l} 1} +\tilde{\lambda}Z^H_{{k} 1} Z^H_{{l} 2} + 4 \mathrm{Re}\big\{\kappa \lambda\big\}Z^H_{{k} 3} Z^H_{{l} 3} \Big)\nonumber \\ 
 & \; \;+2 Z^H_{{j} 3} \Big(\lambda \kappa^* \Big(Z^H_{{k} 1} Z^H_{{l} 3}  + Z^H_{{k} 3} Z^H_{{l} 1} \Big)+\lambda^* \Big(\Big( \kappa Z^H_{{k} 1}-2 \lambda Z^H_{{k} 2}   \Big)Z^H_{{l} 3}  + Z^H_{{k} 3} \Big(  \kappa Z^H_{{l} 1}-2 \lambda Z^H_{{l} 2}  \Big)\Big)\Big)\Big)\nonumber \\ 
 & \; \;+2 Z^H_{{i} 3} \Big(\lambda^* \Big(Z^H_{{j} 3} \Big(Z^H_{{k} 1} \Big( \kappa Z^H_{{l} 2}-2 \lambda Z^H_{{l} 1} \Big) + Z^H_{{k} 2} \Big(\kappa Z^H_{{l} 1}-2 \lambda Z^H_{{l} 2}   \Big)\Big)+Z^H_{{j} 1} \Big(\Big( \kappa Z^H_{{k} 2} -2 \lambda Z^H_{{k} 1}  \Big)Z^H_{{l} 3}  \nonumber \\
& \; \; + Z^H_{{k} 3} \Big(  \kappa Z^H_{{l} 2} -2 \lambda Z^H_{{l} 1}\Big)\Big)+Z^H_{{j} 2} \Big(\Big(\kappa Z^H_{{k} 1} -2 \lambda Z^H_{{k} 2}  \Big)Z^H_{{l} 3}  + Z^H_{{k} 3} \Big(\kappa Z^H_{{l} 1} -2 \lambda Z^H_{{l} 2}  \Big)\Big)\Big)\nonumber \\ 
 & \; \;+\kappa^* \Big(Z^H_{{j} 3} \Big(-12 \kappa Z^H_{{k} 3} Z^H_{{l} 3}  + \lambda Z^H_{{k} 1} Z^H_{{l} 2}  + \lambda Z^H_{{k} 2} Z^H_{{l} 1} \Big)\nonumber \\ 
 & \; \;+\lambda \Big(Z^H_{{j} 1} \Big(Z^H_{{k} 2} Z^H_{{l} 3}  + Z^H_{{k} 3} Z^H_{{l} 2} \Big) + Z^H_{{j} 2} \Big(Z^H_{{k} 1} Z^H_{{l} 3}  + Z^H_{{k} 3} Z^H_{{l} 1} \Big)\Big)\Big)\Big)\Big)\\
%
%
&  \Gamma_{h_{{{i}}}h_{{{j}}}A^0_{{{k}}}A^0_{{{l}}}} \, =  \, 
\frac{i}{4} \Big(Z^A_{{k} 2} \Big(Z^A_{{l} 2} \Big(\tilde{\lambda}Z^H_{{i} 1} Z^H_{{j} 1}  -4 |\lambda|^2 Z^H_{{i} 3} Z^H_{{j} 3}  - g_+^2Z^H_{{i} 2} Z^H_{{j} 2} \Big)\nonumber \\ 
 & \; \;+2 \Big(\kappa \lambda^*  + \lambda \kappa^* \Big)\Big(- Z^A_{{l} 1} Z^H_{{i} 3} Z^H_{{j} 3}  + Z^A_{{l} 3} \Big(Z^H_{{i} 1} Z^H_{{j} 3}  + Z^H_{{i} 3} Z^H_{{j} 1} \Big)\Big)\Big)+Z^A_{{k} 1} \Big(- Z^A_{{l} 1} \Big(- \tilde{\lambda}Z^H_{{i} 2} Z^H_{{j} 2} \nonumber \\ 
 & \; \; + 4 |\lambda|^2 Z^H_{{i} 3} Z^H_{{j} 3}  + g_+^2Z^H_{{i} 1} Z^H_{{j} 1} \Big)+2 \Big(\kappa \lambda^*  + \lambda \kappa^* \Big)\Big(- Z^A_{{l} 2} Z^H_{{i} 3} Z^H_{{j} 3}  + Z^A_{{l} 3} \Big(Z^H_{{i} 2} Z^H_{{j} 3}  + Z^H_{{i} 3} Z^H_{{j} 2} \Big)\Big)\Big)\nonumber \\ 
 & \; \;+2 Z^A_{{k} 3} \Big(\lambda^* \Big(- Z^A_{{l} 3} \Big(Z^H_{{i} 1} \Big(2 \lambda Z^H_{{j} 1}  + \kappa Z^H_{{j} 2} \Big) + Z^H_{{i} 2} \Big(2 \lambda Z^H_{{j} 2}  + \kappa Z^H_{{j} 1} \Big)\Big)+\kappa \Big(Z^A_{{l} 1} \Big(Z^H_{{i} 2} Z^H_{{j} 3}  + Z^H_{{i} 3} Z^H_{{j} 2} \Big) \nonumber \\ 
 & \; \;+ Z^A_{{l} 2} \Big(Z^H_{{i} 1} Z^H_{{j} 3}  + Z^H_{{i} 3} Z^H_{{j} 1} \Big)\Big)\Big)+\kappa^* \Big(- Z^A_{{l} 3} \Big(4 \kappa Z^H_{{i} 3} Z^H_{{j} 3}  + \lambda Z^H_{{i} 1} Z^H_{{j} 2}  + \lambda Z^H_{{i} 2} Z^H_{{j} 1} \Big)\nonumber \\ 
 & \; \;+\lambda \Big(Z^A_{{l} 1} \Big(Z^H_{{i} 2} Z^H_{{j} 3}  + Z^H_{{i} 3} Z^H_{{j} 2} \Big) + Z^A_{{l} 2} \Big(Z^H_{{i} 1} Z^H_{{j} 3}  + Z^H_{{i} 3} Z^H_{{j} 1} \Big)\Big)\Big)\Big)\Big)\\
%
%
&  \Gamma_{h_{{{i}}} H^+_{{{j}}}h_{{{k}}} H^+_{{{l}}}} \, =  \, 
\frac{i}{4} \Big(- Z^{+,*}_{{j} 2} \Big(4 \lambda^* Z^H_{{i} 3} Z^H_{{k} 3} \Big(\kappa Z^+_{{l} 1}  + \lambda Z^+_{{l} 2} \Big)+Z^H_{{i} 1} \Big(\bar{\lambda}Z^H_{{k} 2} Z^+_{{l} 1}  + g_-^2Z^H_{{k} 1} Z^+_{{l} 2} \Big)\nonumber \\ 
 & \; \;+Z^H_{{i} 2} \Big(\bar{\lambda}Z^H_{{k} 1} Z^+_{{l} 1}  + g_+^2Z^H_{{k} 2} Z^+_{{l} 2} \Big)\Big)- Z^{+,*}_{{j} 1} \Big(4 \lambda Z^H_{{i} 3} Z^H_{{k} 3} \Big(\kappa^* Z^+_{{l} 2}  + \lambda^* Z^+_{{l} 1} \Big)\nonumber \\ 
 & \; \;+Z^H_{{i} 2} \Big(\bar{\lambda}Z^H_{{k} 1} Z^+_{{l} 2}  + g_-^2Z^H_{{k} 2} Z^+_{{l} 1} \Big)+Z^H_{{i} 1} \Big(\bar{\lambda}Z^H_{{k} 2} Z^+_{{l} 2}  + g_+^2Z^H_{{k} 1} Z^+_{{l} 1} \Big)\Big)\Big) \\
%
%
&  \Gamma_{A^0_{{{i}}}A^0_{{{j}}}A^0_{{{k}}}A^0_{{{l}}}} \, =  \, 
\frac{i}{4} \Big(Z^A_{{i} 1} \Big(Z^A_{{j} 1} \Big(-3 g_+^2Z^A_{{k} 1} Z^A_{{l} 1}  + \tilde{\lambda}Z^A_{{k} 2} Z^A_{{l} 2}  -4 |\lambda|^2 Z^A_{{k} 3} Z^A_{{l} 3} \Big)\nonumber \\ 
 & \; \;+Z^A_{{j} 2} \Big(\tilde{\lambda}Z^A_{{k} 2} Z^A_{{l} 1} +\tilde{\lambda}Z^A_{{k} 1} Z^A_{{l} 2} +4 \mathrm{Re}\big\{\kappa \lambda \big\}Z^A_{{k} 3} Z^A_{{l} 3} \Big)+2 Z^A_{{j} 3} \Big(\lambda \kappa^* \Big(Z^A_{{k} 2} Z^A_{{l} 3}  + Z^A_{{k} 3} Z^A_{{l} 2} \Big)\nonumber \\ 
 & \; \;+\lambda^* \Big(\Big(-2 \lambda Z^A_{{k} 1}  + \kappa Z^A_{{k} 2} \Big)Z^A_{{l} 3} + Z^A_{{k} 3} \Big(-2 \lambda Z^A_{{l} 1}  + \kappa Z^A_{{l} 2} \Big)\Big)\Big)\Big)\nonumber \\ 
 & \; \;+Z^A_{{i} 2} \Big(Z^A_{{j} 2} \Big(-3 g_+^2Z^A_{{k} 2} Z^A_{{l} 2}  + \tilde{\lambda}Z^A_{{k} 1} Z^A_{{l} 1}  -4 |\lambda|^2 Z^A_{{k} 3} Z^A_{{l} 3} \Big)\nonumber \\ 
 & \; \;+Z^A_{{j} 1} \Big(\tilde{\lambda}Z^A_{{k} 2} Z^A_{{l} 1} +\tilde{\lambda}Z^A_{{k} 1} Z^A_{{l} 2} +4 \mathrm{Re}\big\{\kappa \lambda\big\}Z^A_{{k} 3} Z^A_{{l} 3} \Big)2 Z^A_{{j} 3} \Big(\lambda \kappa^* \Big(Z^A_{{k} 1} Z^A_{{l} 3}  + Z^A_{{k} 3} Z^A_{{l} 1} \Big)\nonumber \\ 
 & \; \;+\lambda^* \Big(\Big(-2 \lambda Z^A_{{k} 2}  + \kappa Z^A_{{k} 1} \Big)Z^A_{{l} 3}  + Z^A_{{k} 3} \Big(\kappa Z^A_{{l} 1}-2 \lambda Z^A_{{l} 2}  \Big)\Big)\Big)\Big)\nonumber \\ 
 & \; \;+2 Z^A_{{i} 3} \Big(\lambda^* \Big(Z^A_{{j} 3} \Big(Z^A_{{k} 1} \Big(-2 \lambda Z^A_{{l} 1}  + \kappa Z^A_{{l} 2} \Big) + Z^A_{{k} 2} \Big(\kappa Z^A_{{l} 1} -2 \lambda Z^A_{{l} 2} \Big)\Big)\nonumber \\ 
 & \; \;+Z^A_{{j} 1} \Big(\Big(-2 \lambda Z^A_{{k} 1}  + \kappa Z^A_{{k} 2} \Big)Z^A_{{l} 3}  + Z^A_{{k} 3} \Big(-2 \lambda Z^A_{{l} 1}  + \kappa Z^A_{{l} 2} \Big)\Big)\nonumber \\ 
 & \; \;+Z^A_{{j} 2} \Big(\Big(-2 \lambda Z^A_{{k} 2}  + \kappa Z^A_{{k} 1} \Big)Z^A_{{l} 3}  + Z^A_{{k} 3} \Big(-2 \lambda Z^A_{{l} 2}  + \kappa Z^A_{{l} 1} \Big)\Big)\Big)\nonumber \\ 
 & \; \;+\kappa^* \Big(Z^A_{{j} 3} \Big(-12 \kappa Z^A_{{k} 3} Z^A_{{l} 3}  + \lambda Z^A_{{k} 1} Z^A_{{l} 2}  + \lambda Z^A_{{k} 2} Z^A_{{l} 1} \Big)\nonumber \\ 
 & \; \;+\lambda \Big(Z^A_{{j} 1} \Big(Z^A_{{k} 2} Z^A_{{l} 3}  + Z^A_{{k} 3} Z^A_{{l} 2} \Big) + Z^A_{{j} 2} \Big(Z^A_{{k} 1} Z^A_{{l} 3}  + Z^A_{{k} 3} Z^A_{{l} 1} \Big)\Big)\Big)\Big)\Big) \\
%
%
&  \Gamma_{A^0_{{{i}}}H^-_{{{j}}}A^0_{{{k}}}H^+_{{{l}}}}\, =  \,
\frac{i}{4} \Big(- Z^{+,*}_{{j} 2} \Big(4 \lambda^* Z^A_{{i} 3} Z^A_{{k} 3} \Big(\lambda Z^+_{{l} 2}- \kappa Z^+_{{l} 1}  \Big)+Z^A_{{i} 1} \Big(g_-^2Z^A_{{k} 1} Z^+_{{l} 2}- \bar{\lambda}Z^A_{{k} 2} Z^+_{{l} 1}  \Big)\nonumber \\ 
 & \; \;+Z^A_{{i} 2} \Big(g_+^2Z^A_{{k} 2} Z^+_{{l} 2} - \bar{\lambda}Z^A_{{k} 1} Z^+_{{l} 1}  \Big)\Big)- Z^{+,*}_{{j} 1} \Big(4 \lambda Z^A_{{i} 3} Z^A_{{k} 3} \Big( \lambda^* Z^+_{{l} 1}- \kappa^* Z^+_{{l} 2} \Big)\nonumber \\ 
 & \; \;+Z^A_{{i} 2} \Big(g_-^2Z^A_{{k} 2} Z^+_{{l} 1}- \bar{\lambda}Z^A_{{k} 1} Z^+_{{l} 2} \Big)+Z^A_{{i} 1} \Big( g_+^2Z^A_{{k} 1} Z^+_{{l} 1} -\bar{\lambda}Z^A_{{k} 2} Z^+_{{l} 2}   \Big)\Big)\Big) \\
%
%
&  \Gamma_{H^-_{{{i}}}H^-_{{{j}}}H^+_{{{k}}}H^+_{{{l}}}} \, =  \, 
\frac{i}{4} \Big(Z^{+,*}_{{i} 2} \Big(-2 g_+^2Z^{+,*}_{{j} 2} Z^+_{{k} 2} Z^+_{{l} 2}  + \tilde{\lambda}Z^{+,*}_{{j} 1} \Big(Z^+_{{k} 1} Z^+_{{l} 2}  + Z^+_{{k} 2} Z^+_{{l} 1} \Big)\Big)\nonumber \\ 
 & \; \;+Z^{+,*}_{{i} 1} \Big(-2 g_+^2Z^{+,*}_{{j} 1} Z^+_{{k} 1} Z^+_{{l} 1}  + \tilde{\lambda}Z^{+,*}_{{j} 2} \Big(Z^+_{{k} 1} Z^+_{{l} 2}  + Z^+_{{k} 2} Z^+_{{l} 1} \Big)\Big)\Big) \\
%
%
&  \Gamma_{\tilde{d}_{{{i} {\alpha}}}\tilde{\nu}_{{{j}}}\tilde{d}^*_{{{k} {\gamma}}}\tilde{\nu}^*_{{{l}}}} \, =  \, 
\frac{i}{24} \delta_{{\alpha},{\gamma}} \Big(\delta_{{j},{l}} \Big(2 g_1^2 \sum_{{a}=1}^{3}Z^{D,*}_{{i} 3 + {a}} Z^D_{{k} 3 + {a}}   + \Big(3 g_2^2  + g_1^2\Big)\sum_{{a}=1}^{3}Z^{D,*}_{{i} {a}} Z^D_{{k} {a}}  \Big)\nonumber \\ 
 & \; \;+\delta_{{j},{l}} \Big(2 g_1^2 \sum_{{b}=1}^{3}Z^{D,*}_{{i} 3 + {b}} Z^D_{{k} 3 + {b}}   + \Big(3 g_2^2  + g_1^2\Big)\sum_{{b}=1}^{3}Z^{D,*}_{{i} {b}} Z^D_{{k} {b}}  \Big)\Big) \\
%
%
&  \Gamma_{\tilde{d}_{{{i} {\alpha}}}\tilde{e}_{{{j}}}\tilde{d}^*_{{{k} {\gamma}}}\tilde{e}^*_{{{l}}}}\, =  \, 
-\frac{i}{24} \delta_{{\alpha},{\gamma}} \Big(2 g_1^2 
\sum_{{a}=1}^{3}Z^{E,*}_{{j} 3 + {a}} Z^E_{{l} 3 + {a}}  \Big(2 
\sum_{{b}=1}^{3}Z^{D,*}_{{i} 3 + {b}} Z^D_{{k} 3 + {b}}   + 
\sum_{{b}=1}^{3}Z^{D,*}_{{i} {b}} Z^D_{{k} {b}} \Big)\nonumber \\ 
 & \; \;- \sum_{{a}=1}^{3}Z^{E,*}_{{j} {a}} Z^E_{{l} {a}}  \Big(2 
g_1^2 \sum_{{b}=1}^{3}Z^{D,*}_{{i} 3 + {b}} Z^D_{{k} 3 + {b}}   
+ \Big( g_1^2-3 g_2^2\Big)\sum_{{b}=1}^{3}Z^{D,*}_{{i} {b}} 
Z^D_{{k} {b}}  \Big)- \Big(2 g_1^2 \sum_{{a}=1}^{3}Z^{D,*}_{{i} 3 + {a}} Z^D_{{k} 3 
+ {a}}  \nonumber \\ 
 & \; \; + \Big(g_1^2-3 g_2^2\Big)\sum_{{a}=1}^{3}Z^{D,*}_{{i} {a}} Z^D_{{k} {a}}  
\Big)\sum_{{b}=1}^{3}Z^{E,*}_{{j} {b}} Z^E_{{l} {b}}  +2 g_1^2 \Big(2 \sum_{{a}=1}^{3}Z^{D,*}_{{i} 3 + {a}} Z^D_{{k} 
3 + {a}}   + \sum_{{a}=1}^{3}Z^{D,*}_{{i} {a}} Z^D_{{k} {a}} 
\Big)\sum_{{b}=1}^{3}Z^{E,*}_{{j} 3 + {b}} Z^E_{{l} 3 + {b}}  \nonumber \\ 
 & \; \;+24 \Big(\sum_{{a,b}=1}^{3}Z^{E,*}_{{j} {a}} 
Y_{e,{{a} {b}}}  Z^E_{{l} 3 + {b}}  \sum_{{d}=1}^{3}Z^{D,*}_{{i} 
3 + {d}} \sum_{{c}=1}^{3}Y^*_{d,{{c} {d}}} Z^D_{{k} {c}}   +\sum_{{a,b}=1}^{3}Z^{D,*}_{{i} {a}} Y_{d,{{a} 
{b}}}  Z^D_{{k} 3 + {b}}  \sum_{{d}=1}^{3}Z^{E,*}_{{j} 3 + {d}} 
\sum_{{c}=1}^{3}Y^*_{e,{{c} {d}}} Z^E_{{l} {c}}   \Big)\Big) \\
%
%
&  \Gamma_{\tilde{d}_{{{i} {\alpha}}}H^-_{{{j}}}\tilde{d}^*_{{{k} {\gamma}}}H^+_{{{l}}}}\, =  \, 
\frac{i}{12} \delta_{{\alpha},{\gamma}} \Big(Z^{+,*}_{{j} 1} 
\Big(\Big( g_1^2-3 g_2^2\Big)\sum_{{a}=1}^{3}Z^{D,*}_{{i} 
{a}} Z^D_{{k} {a}}  +2 g_1^2 \sum_{{a}=1}^{3}Z^{D,*}_{{i} 3 + 
{a}} Z^D_{{k} 3 + {a}}  \nonumber \\ 
 & \; \;-12 \sum_{{c}=1}^{3}Z^{D,*}_{{i} 3 + {c}} 
\sum_{{a,b}=1}^{3}Y^*_{d,{{a} {c}}} Y_{d,{{a} {b}}} 
 Z^D_{{k} 3 + {b}}   \Big)Z^+_{{l} 1} - Z^{+,*}_{{j} 2} \Big(\Big(g_1^2-3 g_2^2\Big)\sum_{{a}=1}^{3}Z^{D,*}_{{i} {a}} Z^D_{{k} {a}}  +2 
g_1^2 \sum_{{a}=1}^{3}Z^{D,*}_{{i} 3 + {a}} Z^D_{{k} 3 + {a}}  
\nonumber \\ 
 & \; \;+12 \sum_{{c}=1}^{3}\sum_{{b}=1}^{3}Z^{D,*}_{{i} {b}} 
\sum_{{a}=1}^{3}Y^*_{u,{{c} {a}}} Y_{u,{{b} {a}}}   Z^D_{{k} 
{c}}  \Big)Z^+_{{l} 2} \Big)\\
%
%
&  \Gamma_{\tilde{\nu}_{{{i}}}\tilde{\nu}_{{{j}}}\tilde{\nu}^*_{{{k}}}\tilde{\nu}^*_{{{l}}}}  \, =  \, 
-\frac{i}{8} \Big(g_1^2 + g_2^2\Big)\Big(2 \delta_{{i},{k}} \delta_{{j},{l}}  + 2 \delta_{{i},{l}} \delta_{{j},{k}} \Big) \\
%
%
&  \Gamma_{\tilde{\nu}_{{{i}}}\tilde{u}_{{{j} {\beta}}}\tilde{\nu}^*_{{{k}}}\tilde{u}^*_{{{l} {\delta}}}}  \, =  \, 
\frac{i}{24} \delta_{{\beta},{\delta}} \Big(\delta_{{i},{k}} \Big(\Big(-3 g_2^2  + g_1^2\Big)\sum_{{a}=1}^{3}Z^{U,*}_{{j} {a}} Z^U_{{l} {a}}   -4 g_1^2 \sum_{{a}=1}^{3}Z^{U,*}_{{j} 3 + {a}} Z^U_{{l} 3 + {a}}  \Big)\nonumber \\ 
 & \; \;+\delta_{{i},{k}} \Big(\Big(-3 g_2^2  + g_1^2\Big)\sum_{{b}=1}^{3}Z^{U,*}_{{j} {b}} Z^U_{{l} {b}}   -4 g_1^2 \sum_{{b}=1}^{3}Z^{U,*}_{{j} 3 + {b}} Z^U_{{l} 3 + {b}}  \Big)\Big) \\
%
%
&  \Gamma_{\tilde{\nu}_{{{i}}}\tilde{e}_{{{j}}}\tilde{\nu}^*_{{{k}}}\tilde{e}^*_{{{l}}}} \, =  \, 
-\frac{i}{4} \Big(\delta_{{i},{k}} \Big(\Big(g_1^2- g_2^2 \Big)\sum_{{a}=1}^{3}Z^{E,*}_{{j} {a}} Z^E_{{l} {a}}  -2 
g_1^2 \sum_{{a}=1}^{3}Z^{E,*}_{{j} 3 + {a}} Z^E_{{l} 3 + {a}}  
\Big)\nonumber \\ 
 & \; \;+g_2^2 \Big(\sum_{{a,b}=1}^{3}\Big[Z^{E,*}_{{j} {a}} Z^{\nu}_{{k} {a}}  
Z^{\nu,*}_{{i} {b}} Z^E_{{l} {b}}  
+Z^{\nu,*}_{{i} {a}} Z^E_{{l} {a}}  
Z^{E,*}_{{j} {b}} Z^{\nu}_{{k} {b}}\Big]  \Big)+4 \sum_{{a,b,c,d}=1}^{3}Z^{\nu,*}_{{i} {a}} Y_{e,{{a} 
{b}}}  Z^E_{{l} 3 + {b}}  Z^{E,*}_{{j} 3 + {d}} 
Y^*_{e,{{c} {d}}} Z^{\nu}_{{k} {c}}   \Big) \\
%
%
&  \Gamma_{\tilde{\nu}_{{{i}}}h_{{{j}}}\tilde{\nu}^*_{{{k}}}h_{{{l}}}}  \, =  \, 
-\frac{i}{4} \Big(g_1^2 + g_2^2\Big)\delta_{{i},{k}} \Big(Z^H_{{j} 1} Z^H_{{l} 1}  - Z^H_{{j} 2} Z^H_{{l} 2} \Big) \\
%
%
&  \Gamma_{\tilde{\nu}_{{{i}}}A^0_{{{j}}}\tilde{\nu}^*_{{{k}}}A^0_{{{l}}}}\, =  \, 
-\frac{i}{4} \Big(g_1^2 + g_2^2\Big)\delta_{{i},{k}} \Big(Z^A_{{j} 1} Z^A_{{l} 1}  - Z^A_{{j} 2} Z^A_{{l} 2} \Big) \\
%
%
&  \Gamma_{\tilde{\nu}_{{{i}}}H^-_{{{j}}}\tilde{\nu}^*_{{{k}}}H^+_{{{l}}}}  \, =  \, 
\frac{i}{4} \Big(Z^{+,*}_{{j} 1} \Big(\Big( g_2^2- g_1^2  \Big)\delta_{{i},{k}} -4 \sum_{{a,b,c}=1}^{3}Z^{\nu,*}_{{i} {b}} Y^*_{e,{{c} {a}}} Y_{e,{{b} {a}}}   Z^{\nu}_{{k} {c}}  \Big)Z^+_{{l} 1}+\Big(g_1^2- g_2^2 \Big)Z^{+,*}_{{j} 2} \delta_{{i},{k}} Z^+_{{l} 2} \Big) \\
%
%
&  \Gamma_{\tilde{u}_{{{i} {\alpha}}}\tilde{e}_{{{j}}}\tilde{u}^*_{{{k} {\gamma}}}\tilde{e}^*_{{{l}}}} \, =  \,
\frac{i}{24} \delta_{{\alpha},{\gamma}} \Big(-4 g_1^2 
\sum_{{a}=1}^{3}Z^{U,*}_{{i} 3 + {a}} Z^U_{{k} 3 + {a}}  \Big(\sum_{{b}=1}^{3}Z^{E,*}_{{j} {b}} Z^E_{{l} {b}} -2 
\sum_{{b}=1}^{3}Z^{E,*}_{{j} 3 + {b}} Z^E_{{l} 3 + {b}}  \Big)\nonumber \\ 
 & \; \;+\sum_{{a}=1}^{3}Z^{U,*}_{{i} {a}} Z^U_{{k} {a}}  \Big(\Big(3 g_2^2  + g_1^2\Big)\sum_{{b}=1}^{3}Z^{E,*}_{{j} {b}} 
Z^E_{{l} {b}}-2 
g_1^2 \sum_{{b}=1}^{3}Z^{E,*}_{{j} 3 + {b}} Z^E_{{l} 3 + {b}}   
  \Big)+\Big(\Big(3 g_2^2  + 
g_1^2\Big)\sum_{{a}=1}^{3}Z^{E,*}_{{j} {a}} Z^E_{{l} {a}} \nonumber \\ 
 & \; \; -2 g_1^2 \sum_{{a}=1}^{3}Z^{E,*}_{{j} 3 + {a}} Z^E_{{l} 3 
+ {a}}  
\Big)\sum_{{b}=1}^{3}Z^{U,*}_{{i} {b}} Z^U_{{k} {b}}  -4 g_1^2 \Big(\sum_{{a}=1}^{3}Z^{E,*}_{{j} {a}} Z^E_{{l} {a}} -2 \sum_{{a}=1}^{3}Z^{E,*}_{{j} 3 + {a}} Z^E_{{l} 
3 + {a}}  
\Big)\sum_{{b}=1}^{3}Z^{U,*}_{{i} 3 + {b}} Z^U_{{k} 3 + {b}}  \Big)\\
%
%
&  \Gamma_{\tilde{u}_{{{i} {\alpha}}}H^-_{{{j}}}\tilde{u}^*_{{{k} {\gamma}}}H^+_{{{l}}}} \, =  \,
\frac{i}{12} \delta_{{\alpha},{\gamma}} \Big(Z^{+,*}_{{j} 1} \Big(\Big(3 g_2^2  + g_1^2\Big)\sum_{{a}=1}^{3}Z^{U,*}_{{i} {a}} Z^U_{{k} {a}}  -4 \Big(g_1^2 \sum_{{a}=1}^{3}Z^{U,*}_{{i} 3 + {a}} Z^U_{{k} 3 + {a}} \nonumber \\ 
 & \; \; +3 \sum_{{c}=1}^{3}\sum_{{b}=1}^{3}Z^{U,*}_{{i} {b}} \sum_{{a}=1}^{3}Y^*_{d,{{c} {a}}} Y_{d,{{b} {a}}}   Z^U_{{k} {c}}  \Big)\Big)Z^+_{{l} 1} - Z^{+,*}_{{j} 2} \Big(\Big(3 g_2^2  + g_1^2\Big)\sum_{{a}=1}^{3}Z^{U,*}_{{i} {a}} Z^U_{{k} {a}}  -4 g_1^2 \sum_{{a}=1}^{3}Z^{U,*}_{{i} 3 + {a}} Z^U_{{k} 3 + {a}}  \nonumber \\ 
 & \; \;+12 \sum_{{c}=1}^{3}Z^{U,*}_{{i} 3 + {c}} \sum_{{a,b}=1}^{3}Y^*_{u,{{a} {c}}} Y_{u,{{a} {b}}}  Z^U_{{k} 3 + {b}}   \Big)Z^+_{{l} 2} \Big) \\
%
%
&  \Gamma_{\tilde{e}_{{{i}}}\tilde{e}_{{{j}}}\tilde{e}^*_{{{k}}}\tilde{e}^*_{{{l}}}}  \, =  \, 
-\frac{i}{8} \Big(g_1^2 \sum_{{a}=1}^{3}Z^{E,*}_{{i} {a}} 
Z^E_{{l} {a}}  \sum_{{b}=1}^{3}Z^{E,*}_{{j} {b}} Z^E_{{k} {b}}  
+g_2^2 \sum_{{a}=1}^{3}Z^{E,*}_{{i} {a}} Z^E_{{l} {a}}  
\sum_{{b}=1}^{3}Z^{E,*}_{{j} {b}} Z^E_{{k} {b}}  \nonumber \\ 
 & \; \;-2 g_1^2 \sum_{{a}=1}^{3}Z^{E,*}_{{i} 3 + {a}} Z^E_{{l} 3 + 
{a}}  \sum_{{b}=1}^{3}Z^{E,*}_{{j} {b}} Z^E_{{k} {b}}  -2 g_1^2 \sum_{{a}=1}^{3}Z^{E,*}_{{j} 3 + {a}} Z^E_{{l} 3 + 
{a}}  \Big(\sum_{{b}=1}^{3}Z^{E,*}_{{i} {b}} Z^E_{{k} {b}}-2 \sum_{{b}=1}^{3}Z^{E,*}_{{i} 3 + {b}} Z^E_{{k} 3 + 
{b}}    \Big)\nonumber \\ 
 & \; \;+\sum_{{a}=1}^{3}Z^{E,*}_{{j} {a}} Z^E_{{l} {a}}  \Big(\Big(g_1^2 + g_2^2\Big)\sum_{{b}=1}^{3}Z^{E,*}_{{i} {b}} 
Z^E_{{k} {b}} -2 
g_1^2 \sum_{{b}=1}^{3}Z^{E,*}_{{i} 3 + {b}} Z^E_{{k} 3 + {b}}    \Big)-2 g_1^2 \Big( \sum_{{a,b}=1}^{3}Z^{E,*}_{{i} {a}} Z^E_{{l} {a}}  Z^{E,*}_{{j} 3 + {b}} Z^E_{{k} 3 + {b}} \nonumber \\ 
 & \; \;+ \sum_{{a,b}=1}^{3}Z^{E,*}_{{j} {a}} Z^E_{{k} {a}}  
Z^{E,*}_{{i} 3 + {b}} Z^E_{{l} 3 + {b}}   \Big)  +4 
g_1^2 \Big(\sum_{{a,b}=1}^{3}\Big[Z^{E,*}_{{i} 3 + {a}} Z^E_{{l} 3 + {a}}  
Z^{E,*}_{{j} 3 + {b}} Z^E_{{k} 3 + {b}}  +Z^{E,*}_{{j} 3 + {a}} Z^E_{{k} 3 + {a}}  
Z^{E,*}_{{i} 3 + {b}} Z^E_{{l} 3 + {b}}\Big] \Big) \nonumber 
\\ 
 & \; \;+\sum_{{a}=1}^{3} \sum_{{b}=1}^{3} \Big[ g_1^2 Z^{E,*}_{{j} {a}} Z^E_{{k} {a}}  
Z^{E,*}_{{i} {b}} Z^E_{{l} {b}}  +g_2^2 
Z^{E,*}_{{j} {a}} Z^E_{{k} {a}}  
Z^{E,*}_{{i} {b}} Z^E_{{l} {b}}  -2 g_1^2 Z^{E,*}_{{j} 3 + {a}} Z^E_{{k} 3 + 
{a}}  Z^{E,*}_{{i} {b}} Z^E_{{l} {b}}   \nonumber \\ 
 & \; \;+g_1^2 
Z^{E,*}_{{i} {a}} Z^E_{{k} {a}}  
Z^{E,*}_{{j} {b}} Z^E_{{l} {b}} +g_2^2 Z^{E,*}_{{i} {a}} Z^E_{{k} {a}}  
Z^{E,*}_{{j} {b}} Z^E_{{l} {b}}  -2 g_1^2 
Z^{E,*}_{{i} 3 + {a}} Z^E_{{k} 3 + {a}}  
Z^{E,*}_{{j} {b}} Z^E_{{l} {b}} \Big] \nonumber \\ 
& \; \; + 2 g_1^2 \Big(\sum_{{a}=1}^{3}Z^{E,*}_{{i} {a}} Z^E_{{k} {a}} -2 \sum_{{a}=1}^{3}Z^{E,*}_{{i} 3 + {a}} Z^E_{{k} 
3 + {a}} 
\Big)\sum_{{b}=1}^{3}Z^{E,*}_{{j} 3 + {b}} Z^E_{{l} 3 + {b}}  \nonumber \\ 
 & \; \;+8 \Big(\sum_{{a,b}=1}^{3}Z^{E,*}_{{j} {a}} 
Y_{e,{{a} {b}}}  Z^E_{{l} 3 + {b}}  \sum_{{d}=1}^{3}Z^{E,*}_{{i} 
3 + {d}} \sum_{{c}=1}^{3}Y^*_{e,{{c} {d}}} Z^E_{{k} {c}}   +\sum_{{a,b}=1}^{3}Z^{E,*}_{{i} {a}} Y_{e,{{a} 
{b}}}  Z^E_{{l} 3 + {b}}  \sum_{{d}=1}^{3}Z^{E,*}_{{j} 3 + {d}} 
\sum_{{c}=1}^{3}Y^*_{e,{{c} {d}}} Z^E_{{k} {c}}   \nonumber \\ 
 & \; \;+\sum_{{a,b}=1}^{3}Z^{E,*}_{{j} {a}} Y_{e,{{a} 
{b}}}  Z^E_{{k} 3 + {b}}  \sum_{{d}=1}^{3}Z^{E,*}_{{i} 3 + {d}} 
\sum_{{c}=1}^{3}Y^*_{e,{{c} {d}}} Z^E_{{l} {c}}   +\sum_{{a,b}=1}^{3}Z^{E,*}_{{i} {a}} Y_{e,{{a} 
{b}}}  Z^E_{{k} 3 + {b}}  \sum_{{d}=1}^{3}Z^{E,*}_{{j} 3 + {d}} 
\sum_{{c}=1}^{3}Y^*_{e,{{c} {d}}} Z^E_{{l} {c}}   \Big)\Big) \\
%
%
&  \Gamma_{\tilde{e}_{{{i}}}H^-_{{{j}}}\tilde{e}^*_{{{k}}}H^+_{{{l}}}}  \, =  \, 
\frac{i}{4} \Big(- Z^{+,*}_{{j} 1} \Big(\Big(g_1^2 + g_2^2\Big)\sum_{{a}=1}^{3}Z^{E,*}_{{i} {a}} Z^E_{{k} {a}}  -2 g_1^2 \sum_{{a}=1}^{3}Z^{E,*}_{{i} 3 + {a}} Z^E_{{k} 3 + {a}}  \nonumber \\ 
 & \; \;+4 \sum_{{a,b,c}=1}^{3}Z^{E,*}_{{i} 3 + {c}} Y^*_{e,{{a} {c}}} Y_{e,{{a} {b}}}  Z^E_{{k} 3 + {b}}   \Big)Z^+_{{l} 1} +Z^{+,*}_{{j} 2} \Big(-2 g_1^2 \sum_{{a}=1}^{3}\Big[Z^{E,*}_{{i} 3 + {a}} Z^E_{{k} 3 + {a}}   + \Big(g_1^2 + g_2^2\Big)Z^{E,*}_{{i} {a}} Z^E_{{k} {a}} \Big] \Big)Z^+_{{l} 2} \Big)
\end{align} 

\subsection*{Interactions between  three scalars}
\begin{align} 
%
%
&  \Gamma_{\tilde{d}_{{{i} {\alpha}}}\tilde{d}^*_{{{j} {\beta}}}h_{{{k}}}}  \, =  \, 
\delta_{{\alpha},{\beta}} \Big(D_4 Z^H_{{k} 1} +D_7 v_d Z^H_{{k} 1} +D_5 \Big(v_d Z^H_{{k} 1}  - v_u Z^H_{{k} 2} \Big)+D_6 \Big(v_s Z^H_{{k} 2}  + v_u Z^H_{{k} 3} \Big) \Big)\\
%
%
&  \Gamma_{\tilde{d}_{{{i} {\alpha}}}\tilde{d}^*_{{{j} {\beta}}}A_{h^*,{{k}}}}  \, =  \, 
\delta_{{\alpha},{\beta}} \Big(D_8 Z^A_{{k} 1}  + D_9 \Big(v_s Z^A_{{k} 2}  + v_u Z^A_{{k}3} \Big)\Big) \\
%
& \; \;  \Gamma_{\tilde{u}_{{{i} {\alpha}}}\tilde{u}^*_{{{j} {\beta}}}h_{{{k}}}}  \, =  \, 
\delta_{{\alpha},{\beta}} \Big(U_4 Z^H_{{k} 2} +U_7 v_u Z^H_{{k} 2} +U_5 \Big(v_d Z^H_{{k} 1}  - v_u Z^H_{{k} 2} \Big)+U_6 \Big(v_d Z^H_{{k} 3}  + v_s Z^H_{{k} 1} \Big)\Big) \\
%
%
&  \Gamma_{\tilde{u}_{{{i} {\alpha}}}\tilde{u}^*_{{{j} {\beta}}}A^0_{{{k}}}}  \, =  \, 
\delta_{{\alpha},{\beta}} \Big(U_8 Z^A_{{k} 2}  + U_9 \Big(v_d Z^A_{{k} 3}  + v_s Z^A_{{k} 1} \Big)\Big) \\
%
%
&  \Gamma_{\tilde{e}_{{{i}}}\tilde{e}^*_{{{j}}}h_{{{k}}}} \, =  \, 
E_4 Z^H_{{k} 1} +E_7 v_d Z^H_{{k} 1} +E_5 \Big(- v_d Z^H_{{k} 1}  + v_u Z^H_{{k} 2} \Big)+E_6 \Big(v_s Z^H_{{k} 2}  + v_u Z^H_{{k} 3} \Big) \\
%
%
&  \Gamma_{\tilde{e}_{{{i}}}\tilde{e}^*_{{{j}}}A^0_{{{k}}}}  \, =  \, 
E_8 Z^A_{{k} 1}  + E_9 \Big(v_s Z^A_{{k} 2}  + v_u Z^A_{{k}3} \Big) \\
%
%
&  \Gamma_{\tilde{u}_{{{i} {\alpha}}}\tilde{d}^*_{{{j} {\beta}}} H^-_{{{k}}}}  \, =  \, 
-\frac{i}{4} \delta_{{\alpha},{\beta}} \Big(Z^{+,*}_{{k} 1} \Big(\sqrt{2} g_2^2 v_d \sum_{{a}=1}^{3}Z^{U,*}_{{i} {a}} Z^D_{{j} {a}} -2 \Big(\sqrt{2} v_s \lambda \sum_{{b}=1}^{3}Z^{U,*}_{{i} 3 + {b}} \sum_{{a}=1}^{3}Y^*_{u,{{a} {b}}} Z^D_{{j} {a}}  \nonumber \\ 
 & \; \;  +2 \sum_{{a,b}=1}^{3}Z^{U,*}_{{i} {a}} T_{d,{{a} {b}}}  Z^D_{{j} 3 + {b}} +\sqrt{2} \Big(v_u \sum_{{c}=1}^{3}Z^{U,*}_{{i} 3 + {c}} \sum_{{a,b}=1}^{3}Y^*_{u,{{a} {c}}} Y_{d,{{a} {b}}}  Z^D_{{j} 3 + {b}}   \nonumber \\ 
 & \; \;+v_d \sum_{{c}=1}^{3}\sum_{{b}=1}^{3}Z^{U,*}_{{i} {b}} \sum_{{a}=1}^{3}Y^*_{d,{{c} {a}}} Y_{d,{{b} {a}}}   Z^D_{{j} {c}}  \Big)\Big)\Big)+Z^{+,*}_{{k} 2} \Big(\sqrt{2} g_2^2 v_u \sum_{{a}=1}^{3}Z^{U,*}_{{i} {a}} Z^D_{{j} {a}}  \nonumber \\ 
 & \; \;-2 \Big(2 \sum_{{b}=1}^{3}Z^{U,*}_{{i} 3 + {b}} \sum_{{a}=1}^{3}T^*_{u,{{a} {b}}} Z^D_{{j} {a}}  +\sqrt{2} \Big(v_s \lambda^* \sum_{{a,b}=1}^{3}Z^{U,*}_{{i} {a}} Y_{d,{{a} {b}}}  Z^D_{{j} 3 + {b}}  \nonumber \\ 
 & \; \; +v_d \sum_{{c}=1}^{3}Z^{U,*}_{{i} 3 + {c}} \sum_{{a,b}=1}^{3}Y^*_{u,{{a} {c}}} Y_{d,{{a} {b}}}  Z^D_{{j} 3 + {b}} +v_u \sum_{{c}=1}^{3}\sum_{{b}=1}^{3}Z^{U,*}_{{i} {b}} \sum_{{a}=1}^{3}Y^*_{u,{{c} {a}}} Y_{u,{{b} {a}}}   Z^D_{{j} {c}}  \Big)\Big)\Big)\Big) \\
%
%
&  \Gamma_{\tilde{\nu}_{{{i}}}\tilde{\nu}^*_{{{j}}}h_{{{k}}}}  \, =  \,  
-\frac{i}{4} \Big(g_1^2 + g_2^2\Big)\delta_{{i},{j}} \Big(v_d Z^H_{{k} 1}  - v_u Z^H_{{k} 2} \Big) \\
%
%
&  \Gamma_{\tilde{\nu}_{{{i}}}\tilde{e}^*_{{{j}}}H^-_{{{k}}}}   \, =  \, 
\frac{i}{4} \Big(\sqrt{2} Z^{+,*}_{{k} 2} \Big(- g_2^2 v_u \sum_{{a}=1}^{3}Z^{\nu,*}_{{i} {a}} Z^E_{{j} {a}}  +2 v_s \lambda^* \sum_{{a,b}=1}^{3}Z^{\nu,*}_{{i} {a}} Y_{e,{{a} {b}}}  Z^E_{{j} 3 + {b}}  \Big)\nonumber \\ 
 & \; \;+Z^{+,*}_{{k} 1} \Big(- \sqrt{2} g_2^2 v_d \sum_{{a}=1}^{3}Z^{\nu,*}_{{i} {a}} Z^E_{{j} {a}}  +4 \sum_{{a,b}=1}^{3}Z^{\nu,*}_{{i} {a}} T_{e,{{a} {b}}}  Z^E_{{j} 3 + {b}}  +2 \sqrt{2} v_d \sum_{{c}=1}^{3}\sum_{{b}=1}^{3}Z^{\nu,*}_{{i} {b}} \sum_{{a}=1}^{3}Y^*_{e,{{c} {a}}} Y_{e,{{b} {a}}}   Z^E_{{j} {c}}  \Big)\Big) \\
%
%
&  \Gamma_{h_{{{i}}}h_{{{j}}}h_{{{k}}}}  \, =  \, 
\frac{i}{4} \Big(Z^H_{{i} 1} \Big(Z^H_{{j} 1} \Big(-3 g_+^2v_d Z^H_{{k} 1}  -4 v_s |\lambda|^2 Z^H_{{k} 3}  + v_u \tilde{\lambda}Z^H_{{k} 2} \Big)\nonumber \\ 
 & \; \;+Z^H_{{j} 2} \Big(v_u \tilde{\lambda}Z^H_{{k} 1} +v_d \tilde{\lambda}Z^H_{{k} 2} +\Lambda_1 Z^H_{{k} 3} \Big)+Z^H_{{j} 3} \Big(\sqrt{2}\, 2 \,\mathrm{Re}\big\{T_\lambda\big\}Z^H_{{k} 2} +2 \lambda \kappa^* \Big(v_s Z^H_{{k} 2}  + v_u Z^H_{{k} 3} \Big) \nonumber \\
 & \; \;+2 \lambda^* \Big(\Lambda_3Z^H_{{k} 3}  -2 v_s \lambda Z^H_{{k} 1}  + v_s \kappa Z^H_{{k} 2} \Big)\Big)\Big)+Z^H_{{i} 2} \Big(Z^H_{{j} 2} \Big(-3 g_+^2v_u Z^H_{{k} 2}  -4 v_s |\lambda|^2 Z^H_{{k} 3}  + v_d \tilde{\lambda}Z^H_{{k} 1} \Big)\nonumber \\ 
 & \; \;+Z^H_{{j} 1} \Big(v_u \tilde{\lambda}Z^H_{{k} 1} +v_d \tilde{\lambda}Z^H_{{k} 2} +\Lambda_1 Z^H_{{k} 3} \Big)+Z^H_{{j} 3} \Big(\sqrt{2} \mathrm{Re}\big\{T_\lambda\big\}Z^H_{{k} 1} +2 \lambda \kappa^* \Big(v_d Z^H_{{k} 3}  + v_s Z^H_{{k} 1} \Big) \nonumber \\
 & \; \;+2 \lambda^* \Big(-2 v_s \lambda Z^H_{{k} 2}  + \Lambda_2Z^H_{{k} 3}  + v_s \kappa Z^H_{{k} 1} \Big)\Big)\Big)+Z^H_{{i} 3} \Big(\sqrt{2} \Big(-4 \mathrm{Re}\big\{T_\kappa\big\}Z^H_{{j} 3} Z^H_{{k} 3} \nonumber \\ 
 & \; \; + T_{\lambda}^* \Big(Z^H_{{j} 1} Z^H_{{k} 2}  + Z^H_{{j} 2} Z^H_{{k} 1} \Big) + T_{\lambda} \Big(Z^H_{{j} 1} Z^H_{{k} 2}  + Z^H_{{j} 2} Z^H_{{k} 1} \Big)\Big)+2 \kappa^* \Big(\lambda Z^H_{{j} 2} \Big(v_d Z^H_{{k} 3}  + v_s Z^H_{{k} 1} \Big)\nonumber \\ 
 & \; \;+\lambda Z^H_{{j} 1} \Big(v_s Z^H_{{k} 2}  + v_u Z^H_{{k} 3} \Big)+Z^H_{{j} 3} \Big(-12 v_s \kappa Z^H_{{k} 3}  + v_d \lambda Z^H_{{k} 2}  + v_u \lambda Z^H_{{k} 1} \Big)\Big)\nonumber \\ 
 & \; \;+2 \lambda^* \Big(Z^H_{{j} 3} \Big(\Lambda_3Z^H_{{k} 1}  + \Lambda_2Z^H_{{k} 2} \Big)+Z^H_{{j} 1} \Big(\Lambda_3Z^H_{{k} 3}  -2 v_s \lambda Z^H_{{k} 1}  + v_s \kappa Z^H_{{k} 2} \Big)\nonumber \\ 
 & \; \;+Z^H_{{j} 2} \Big(-2 v_s \lambda Z^H_{{k} 2}  + \Lambda_2Z^H_{{k} 3}  + v_s \kappa Z^H_{{k} 1} \Big)\Big)\Big)\Big) \\
%
%
&  \Gamma_{h_{{{i}}}A^0_{{{j}}}A^0_{{{k}}}} \, =  \, 
\frac{i}{4} \Big(- Z^A_{{j} 1} \Big(-4 v_s \mathrm{Re}\big\{\lambda \kappa\big\} Z^A_{{k} 3} Z^H_{{i} 2} +\sqrt{2}\,2\,\mathrm{Re}\big\{T_{\lambda}\big\} Z^A_{{k} 3} Z^H_{{i} 2} \nonumber \\
& \; \;  +4 v_s \mathrm{Re}\big\{\lambda \kappa\} Z^A_{{k} 2} Z^H_{{i} 3}   +\sqrt{2}\,2\,\mathrm{Re}\big\{T_{\lambda}\big\} Z^A_{{k} 2} Z^H_{{i} 3} -4 v_u \mathrm{Re}\big\{\lambda \kappa\big\} Z^A_{{k} 3} Z^H_{{i} 3}  \nonumber \\ 
 & \; \;+Z^A_{{k} 1} \Big(4 v_s |\lambda|^2 Z^H_{{i} 3}  + g_+^2v_d Z^H_{{i} 1}  - v_u \tilde{\lambda}Z^H_{{i} 2} \Big)\Big)    
+Z^A_{{j} 2} \Big(2 v_s \kappa \lambda^* Z^A_{{k} 3} Z^H_{{i} 1} - \sqrt{2} T_{\lambda}^* Z^A_{{k} 3} Z^H_{{i} 1} \nonumber \\
& \; \;- \sqrt{2} T_{\lambda} Z^A_{{k} 3} Z^H_{{i} 1} -2 v_s \kappa \lambda^* Z^A_{{k} 1} Z^H_{{i} 3} - \sqrt{2} T_{\lambda}^* Z^A_{{k} 1} Z^H_{{i} 3}  - \sqrt{2} T_{\lambda} Z^A_{{k} 1} Z^H_{{i} 3} +2 v_d \kappa \lambda^* Z^A_{{k} 3} Z^H_{{i} 3} \nonumber \\
& \; \;  +Z^A_{{k} 2} \Big(-4 v_s |\lambda|^2 Z^H_{{i} 3}  - g_+^2v_u Z^H_{{i} 2}  + v_d \tilde{\lambda}Z^H_{{i} 1} \Big)+2 \lambda \kappa^* \Big(- v_s Z^A_{{k} 1} Z^H_{{i} 3}  + Z^A_{{k} 3} \Big(v_d Z^H_{{i} 3}  + v_s Z^H_{{i} 1} \Big)\Big)\Big)\nonumber \\ 
 & \; \;  +Z^A_{{j} 3} \Big(- \sqrt{2} \Big(-2 \mathrm{Re}\big\{T_\kappa\big\}Z^A_{{k} 3} Z^H_{{i} 3}  + T_{\lambda}^* \Big(Z^A_{{k} 1} Z^H_{{i} 2}  + Z^A_{{k} 2} Z^H_{{i} 1} \Big)  + T_{\lambda} \Big(Z^A_{{k} 1} Z^H_{{i} 2}  + Z^A_{{k} 2} Z^H_{{i} 1} \Big)\Big)\nonumber \\
& \; \; +2 \lambda^* \Big(- Z^A_{{k} 3} \Big(\Big(2 v_d \lambda  + v_u \kappa \Big)Z^H_{{i} 1}  + \Big(2 v_u \lambda  + v_d \kappa \Big)Z^H_{{i} 2} \Big)+\kappa Z^A_{{k} 2} \Big(v_d Z^H_{{i} 3}  + v_s Z^H_{{i} 1} \Big)\nonumber \\ 
 & \; \;+\kappa Z^A_{{k} 1} \Big(v_s Z^H_{{i} 2}  + v_u Z^H_{{i} 3} \Big)\Big) +2 \kappa^* \Big(\lambda Z^A_{{k} 2} \Big(v_d Z^H_{{i} 3}  + v_s Z^H_{{i} 1} \Big)+\lambda Z^A_{{k} 1} \Big(v_s Z^H_{{i} 2}  + v_u Z^H_{{i} 3} \Big) \nonumber \\ 
 & \; \; - Z^A_{{k} 3} \Big(4 v_s \kappa Z^H_{{i} 3}  + v_d \lambda Z^H_{{i} 2}  + v_u \lambda Z^H_{{i} 1} \Big)\Big)\Big)\Big) \\
%
%
&  \Gamma_{h_{{{i}}}H^+_{{{j}}}H^-_{{{k}}}}  \, =  \, 
\frac{i}{4} \Big(- Z^{+,*}_{{k} 1} \Big(Z^H_{{i} 2} \Big(g_-^2v_u Z^+_{{j} 1}  + v_d \bar{\lambda}Z^+_{{j} 2} \Big)+Z^H_{{i} 1} \Big(g_+^2v_d Z^+_{{j} 1}  + v_u \bar{\lambda}Z^+_{{j} 2} \Big)\nonumber \\ 
 & \; \; +2 Z^H_{{i} 3} \Big(2 v_s |\lambda|^2 Z^+_{{j} 1}  + \Big(2 v_s \lambda \kappa^*  + \sqrt{2} T_{\lambda} \Big)Z^+_{{j} 2} \Big)\Big)- Z^{+,*}_{{k} 2} \Big(Z^H_{{i} 1} \Big(g_-^2v_d Z^+_{{j} 2}  + v_u \bar{\lambda}Z^+_{{j} 1} \Big) \nonumber \\ 
 & \; \;  +Z^H_{{i} 2} \Big(g_+^2v_u Z^+_{{j} 2}  + v_d \bar{\lambda}Z^+_{{j} 1} \Big)+2 Z^H_{{i} 3} \Big(2 v_s \lambda^* \Big(\kappa Z^+_{{j} 1}  + \lambda Z^+_{{j} 2} \Big) + \sqrt{2} T_{\lambda}^* Z^+_{{j} 1} \Big)\Big)\Big) \\
&  \Gamma_{A^0_{{{i}}}H^+_{{{j}}}H^-_{{{k}}}}  \, =  \, 
\frac{1}{4} \Big(Z^{+,*}_{{k} 2} \Big(2 \Big(2 v_s \kappa \lambda^*  - \sqrt{2} T_{\lambda}^* \Big)Z^A_{{i} 3}  + v_d \bar{\lambda}Z^A_{{i} 2}  + v_u \bar{\lambda}Z^A_{{i} 1} \Big)Z^+_{{j} 1} \nonumber \\ 
 & \; \;  - Z^{+,*}_{{k} 1} \Big(2 \Big(2 v_s \lambda \kappa^*  - \sqrt{2} T_{\lambda} \Big)Z^A_{{i} 3}  + v_d \bar{\lambda}Z^A_{{i} 2}  + v_u \bar{\lambda}Z^A_{{i} 1} \Big)Z^+_{{j} 2} \Big)
\end{align}

\section{One-loop tadpoles}

In this and the subsequent Apps., particles that are denoted with a hat, e.g.
\(\hat{h}_i\), are the unrotated external states. In the corresponding
vertices the associated mixing matrix has to be replaced by the identity matrix.
Moreover, we have summed her and in the subsequent section
in all the vertices implicitly over the color
indices of quarks and squarks.

At the one-loop level the expressions for the tadpoles of eq.\ (\ref{eq:oneloop}) are given by
\label{onelooptad}
\begin{align} 
 \delta t_i &= \frac{3}{2} {A_0\big(m^2_{Z}\big)} {\Gamma_{\hat{h}_{{i}}ZZ}} +3 {A_0\big(m^2_W\big)} {\Gamma_{\hat{h}_{{i}}W^+W^-}} - \sum_{a=1}^{2}{A_0\big(m^2_{H^+_{{a}}}\big)} {\Gamma_{\hat{h}_{{i}}H^+_{{a}}H^-_{{a}}}}  \nonumber \\ 
 &+4 \sum_{a=1}^{2}{A_0\big(m^2_{{\tilde{\chi}}^+_{{a}}}\big)} {\Gamma_{\hat{h}_{{i}}\tilde{\chi}^+_{{a}}\tilde{\chi}^-_{{a}}}} m^2_{{\tilde{\chi}}^+_{{a}}}  -\frac{1}{2} \sum_{a=1}^{3}{A_0\big(m^2_{A^0_{{a}}}\big)} {\Gamma_{\hat{h}_{{i}}A^0_{{a}}A^0_{{a}}}}  -\frac{1}{2} \sum_{a=1}^{3}{A_0\big(m^2_{h_{{a}}}\big)} {\Gamma_{\hat{h}_{{i}}h_{{a}}h_{{a}}}}  \nonumber \\ 
 & +12 \sum_{a=1}^{3}{A_0\big(m^2_{{d}_{{a}}}\big)} {\Gamma_{\hat{h}_{{i}}\bar{d}_{{a}}d_{{a}}}} m^2_{{d}_{{a}}} +4 \sum_{a=1}^{3}{A_0\big(m^2_{{e}_{{a}}}\big)} {\Gamma_{\hat{h}_{{i}}\bar{e}_{{a}}e_{{a}}}} m^2_{{e}_{{a}}}  +12 \sum_{a=1}^{3}{A_0\big(m^2_{{u}_{{a}}}\big)} {\Gamma_{\hat{h}_{{i}}\bar{u}_{{a}}u_{{a}}}} m^2_{{u}_{{a}}}  \nonumber \\ 
 &+2 \sum_{a=1}^{5}{A_0\big(m^2_{\tilde{\chi}^0_{{a}}}\big)} {\Gamma_{\hat{h}_{{i}}\tilde{\chi}^0_{{a}}\tilde{\chi}^0_{{a}}}} m^2_{\tilde{\chi}^0_{{a}}}  -3 \sum_{a=1}^{6}{A_0\big(m^2_{\tilde{d}_{{a}}}\big)} {\Gamma_{\hat{h}_{{i}}\tilde{d}^*_{{a}}\tilde{d}_{{a}}}}    -3 \sum_{a=1}^{6}{A_0\big(m^2_{\tilde{u}_{{a}}}\big)} {\Gamma_{\hat{h}_{{i}}\tilde{u}^*_{{a}}\tilde{u}_{{a}}}}\nonumber \\ 
 & - \sum_{a=1}^{3}{A_0\big(m^2_{\tilde{\nu}_{{a}}}\big)} {\Gamma_{\hat{h}_{{i}}\tilde{\nu}^*_{{a}}\tilde{\nu}_{{a}}}}- \sum_{a=1}^{6}{A_0\big(m^2_{\tilde{e}_{{a}}}\big)} {\Gamma_{\hat{h}_{{i}}\tilde{e}^*_{{a}}\tilde{e}_{{a}}}}
\end{align} 

\section{One-loop self-energies} 
\label{oneloopselfenergy}
The definitions of the scalar one-loop functions and their explicit analytic
expressions can be found in app.~\ref{sec:Integrals}. 
\setlength{\mathindent}{0cm}
\subsection{Self-energy of \texorpdfstring{$Z$}{Z} boson}
\label{app:Zself}
In agreement with ref.\ \cite{Degrassi:2009yq} we obtain for the 
transverse self-energy of the $Z$ boson
\begin{align} 
& \Pi^T_{ZZ}(p^2) \, = \, \frac{1}{2} g_2^2 c_\Theta^2 \Big(-8 {B_{22}\Big(m^2_W,m^2_W\Big)}  - {B_0\Big(m^2_W,m^2_W\Big)} \Big(2 m^2_W  + 4 {p}^{2} \Big)\Big)\nonumber \\ 
 & \; \;-4 \sum_{a,b=1}^{2}|{\Gamma_{ZH^+_{{a}}H^-_{{b}}}}|^2 {B_{22}\Big(m^2_{H^+_{{a}}},m^2_{H^+_{{b}}}\Big)}  +\frac{1}{2} \sum_{a,b=1}^{2} \Big[\Big(|{\Gamma^L_{Z\tilde{\chi}^+_{{a}}\tilde{\chi}^-_{{b}}}}|^2 + |{\Gamma^R_{Z\tilde{\chi}^+_{{a}}\tilde{\chi}^-_{{b}}}}|^2\Big){H_0\Big(m^2_{{\tilde{\chi}}^+_{{a}}},m^2_{\tilde{\chi}^+_{{b}}}\Big)} \nonumber \\ 
& \; \; +4 {B_0\Big(m^2_{{\tilde{\chi}}^+_{{a}}},m^2_{\tilde{\chi}^+_{{b}}}\Big)} m_{\tilde{\chi}^+_{{a}}} m_{\tilde{\chi}^-_{{b}}} {\mathrm{Re}\Big\{{\Gamma^{L*}_{Z\tilde{\chi}^+_{{a}}\tilde{\chi}^-_{{b}}}} {\Gamma^R_{Z\tilde{\chi}^+_{{a}}\tilde{\chi}^-_{{b}}}} \Big\}} \Big] -4 \sum_{a,b=1}^{3}|{\Gamma_{ZA^0_{{a}}h_{{b}}}}|^2 {B_{22}\Big(m^2_{h_{{a}}},m^2_{A^0_{{b}}}\Big)} \nonumber \\ 
 & \; \; -4 \sum_{a,b=1}^{3}|{\Gamma_{Z\tilde{\nu}^*_{{a}}\tilde{\nu}_{{b}}}}|^2 {B_{22}\Big(m^2_{\tilde{\nu}_{{a}}},m^2_{\tilde{\nu}_{{b}}}\Big)} +\frac{3}{2} \sum_{a,b=1}^{3} \Big[\Big(|{\Gamma^L_{Z\bar{d}_{{a}}d_{{b}}}}|^2 + |{\Gamma^R_{Z\bar{d}_{{a}}d_{{b}}}}|^2\Big){H_0\Big(m^2_{{d}_{{a}}},m^2_{d_{{b}}}\Big)} \nonumber \\ 
 & \; \; \ +4 {B_0\Big(m^2_{{d}_{{a}}},m^2_{d_{{b}}}\Big)} m_{\bar{d}_{{a}}} m_{d_{{b}}} {\mathrm{Re}\Big\{{\Gamma^{L*}_{Z\bar{d}_{{a}}d_{{b}}}} {\Gamma^R_{Z\bar{d}_{{a}}d_{{b}}}} \Big\}} \Big] +\frac{1}{2} \sum_{a,b=1}^{3} \Big[\Big(|{\Gamma^L_{Z\bar{e}_{{a}}e_{{b}}}}|^2 + |{\Gamma^R_{Z\bar{e}_{{a}}e_{{b}}}}|^2\Big){H_0\Big(m^2_{{e}_{{a}}},m^2_{e_{{b}}}\Big)} \nonumber \\ 
 & \; \; \ +4 {B_0\Big(m^2_{{e}_{{a}}},m^2_{e_{{b}}}\Big)} m_{\bar{e}_{{a}}} m_{e_{{b}}} {\mathrm{Re}\Big\{{\Gamma^{L*}_{Z\bar{e}_{{a}}e_{{b}}}} {\Gamma^R_{Z\bar{e}_{{a}}e_{{b}}}} \Big\}} \Big]+\frac{3}{2} \sum_{a,b=1}^{3} \Big[\Big(|{\Gamma^L_{Z\bar{u}_{{a}}u_{{b}}}}|^2 + |{\Gamma^R_{Z\bar{u}_{{a}}u_{{b}}}}|^2\Big){H_0\Big(m^2_{{u}_{{a}}},m^2_{u_{{b}}}\Big)} \nonumber \\ 
 & \; \; \ +4 {B_0\Big(m^2_{{u}_{{a}}},m^2_{u_{{b}}}\Big)} m_{\bar{u}_{{a}}} m_{u_{{b}}} {\mathrm{Re}\Big\{{\Gamma^{L*}_{Z\bar{u}_{{a}}u_{{b}}}} {\Gamma^R_{Z\bar{u}_{{a}}u_{{b}}}} \Big\}} \Big] +\frac{1}{2} \sum_{a,b=1}^{3} \Big[\Big(|{\Gamma^L_{Z\bar{\nu}_{{a}}\nu_{{b}}}}|^2 + |{\Gamma^R_{Z\bar{\nu}_{{a}}\nu_{{b}}}}|^2\Big){H_0\Big(0,0\Big)} \nonumber \\ 
 & \; \;+\frac{1}{4} \sum_{a,b=1}^{5}\Big[\Big(|{\Gamma^L_{Z\tilde{\chi}^0_{{a}}\tilde{\chi}^0_{{b}}}}|^2 + |{\Gamma^R_{Z\tilde{\chi}^0_{{a}}\tilde{\chi}^0_{{b}}}}|^2\Big){H_0\Big(m^2_{\tilde{\chi}^0_{{a}}},m^2_{\tilde{\chi}^0_{{b}}}\Big)}  +4 {B_0\Big(m^2_{\tilde{\chi}^0_{{a}}},m^2_{\tilde{\chi}^0_{{b}}}\Big)} m_{\tilde{\chi}^0_{{a}}} m_{\tilde{\chi}^0_{{b}}} {\mathrm{Re}\Big\{{\Gamma^{L*}_{Z\tilde{\chi}^0_{{a}}\tilde{\chi}^0_{{b}}}} {\Gamma^R_{Z\tilde{\chi}^0_{{a}}\tilde{\chi}^0_{{b}}}} \Big\}} \Big] \nonumber \\ 
 & \; \;-12 \sum_{a,b=1}^{6}|{\Gamma_{Z\tilde{d}^*_{{a}}\tilde{d}_{{b}}}}|^2 {B_{22}\Big(m^2_{\tilde{d}_{{a}}},m^2_{\tilde{d}_{{b}}}\Big)} -4 \sum_{a,b=1}^{6}|{\Gamma_{Z\tilde{e}^*_{{a}}\tilde{e}_{{b}}}}|^2 {B_{22}\Big(m^2_{\tilde{e}_{{a}}},m^2_{\tilde{e}_{{b}}}\Big)}  \nonumber \\ 
 & \; \;-12 \sum_{a,b=1}^{6}|{\Gamma_{Z\tilde{u}^*_{{a}}\tilde{u}_{{b}}}}|^2 {B_{22}\Big(m^2_{\tilde{u}_{{a}}},m^2_{\tilde{u}_{{b}}}\Big)}  +\frac{1}{2} \sum_{b=1}^{3}|{\Gamma_{ZZh_{{b}}}}|^2 {B_0\Big(m^2_{Z},m^2_{h_{{b}}}\Big)}   
\end{align} 

\subsection{Self-energy of CP-even Higgs bosons}
\label{app:H0self}
\begin{align} 
& \Pi_{h_ih_j}(p^2) \, = \, \frac{7}{4} {B_0\Big(m^2_{Z},m^2_{Z}\Big)} {\Gamma^*_{\hat{h}_{{j}}ZZ}} {\Gamma_{\hat{h}_{{i}}ZZ}} +2 {A_0\Big(m^2_{Z}\Big)} {\Gamma_{\hat{h}_{{i}}\hat{h}_{{j}}ZZ}}-\frac{1}{2} \sum_{a=1}^{3}{A_0\Big(m^2_{A^0_{{a}}}\Big)} {\Gamma_{\hat{h}_{{i}}\hat{h}_{{j}}A^0_{{a}}A^0_{{a}}}}  \nonumber \\ 
 & \; \;+\frac{7}{2} {B_0\Big(m^2_W,m^2_W\Big)} {\Gamma^*_{\hat{h}_{{j}}W^+W^-}} {\Gamma_{\hat{h}_{{i}}W^+W^-}}  +4 {A_0\Big(m^2_W\Big)} {\Gamma_{\hat{h}_{{i}}\hat{h}_{{j}}W^+W^-}} - \sum_{a=1}^{3}{A_0\Big(m^2_{\tilde{\nu}_{{a}}}\Big)} {\Gamma_{\hat{h}_{{i}}\hat{h}_{{j}}\tilde{\nu}^*_{{a}}\tilde{\nu}_{{a}}}}\nonumber \\ 
 & \; \;- \sum_{a=1}^{2}{A_0\Big(m^2_{H^+_{{a}}}\Big)} {\Gamma_{\hat{h}_{{i}}\hat{h}_{{j}}H^+_{{a}}H^-_{{a}}}}  +\sum_{a,b=1}^{2}{B_0\Big(m^2_{H^+_{{a}}},m^2_{H^+_{{b}}}\Big)} {\Gamma^*_{\hat{h}_{{j}}H^+_{{a}}H^-_{{b}}}} {\Gamma_{\hat{h}_{{i}}H^+_{{a}}H^-_{{b}}}} \nonumber \\ 
 & \; \;-2 \sum_{a=1}^{2}m_{\tilde{\chi}^+_{{a}}} \sum_{b=1}^{2} \Big[ {B_0\Big(m^2_{{\tilde{\chi}}^+_{{a}}},m^2_{\tilde{\chi}^+_{{b}}}\Big)} m_{\tilde{\chi}^-_{{b}}} \Big({\Gamma^{L*}_{\hat{h}_{{j}}\tilde{\chi}^+_{{a}}\tilde{\chi}^-_{{b}}}} {\Gamma^R_{\hat{h}_{{i}}\tilde{\chi}^+_{{a}}\tilde{\chi}^-_{{b}}}} + {\Gamma^{R*}_{\hat{h}_{{j}}\tilde{\chi}^+_{{a}}\tilde{\chi}^-_{{b}}}} {\Gamma^L_{\hat{h}_{{i}}\tilde{\chi}^+_{{a}}\tilde{\chi}^-_{{b}}}} \Big) \Big]  \nonumber \\ 
 & \; \;+\sum_{a,b=1}^{2} \Big[{G_0\Big(m^2_{{\tilde{\chi}}^+_{{a}}},m^2_{\tilde{\chi}^+_{{b}}}\Big)} \Big({\Gamma^{L*}_{\hat{h}_{{j}}\tilde{\chi}^+_{{a}}\tilde{\chi}^-_{{b}}}} {\Gamma^L_{\hat{h}_{{i}}\tilde{\chi}^+_{{a}}\tilde{\chi}^-_{{b}}}}  + {\Gamma^{R*}_{\hat{h}_{{j}}\tilde{\chi}^+_{{a}}\tilde{\chi}^-_{{b}}}} {\Gamma^R_{\hat{h}_{{i}}\tilde{\chi}^+_{{a}}\tilde{\chi}^-_{{b}}}} \Big) \Big]   -\frac{1}{2} \sum_{a=1}^{3}{A_0\Big(m^2_{h_{{a}}}\Big)} {\Gamma_{\hat{h}_{{i}}\hat{h}_{{j}}h_{{a}}h_{{a}}}} \nonumber \\ 
 & \; \; +\frac{1}{2} \sum_{a,b=1}^{3}{B_0\Big(m^2_{A^0_{{a}}},m^2_{A^0_{{b}}}\Big)} {\Gamma^*_{\hat{h}_{{j}}A^0_{{a}}A^0_{{b}}}} {\Gamma_{\hat{h}_{{i}}A^0_{{a}}A^0_{{b}}}}  +\sum_{a,b=1}^{3}{B_0\Big(m^2_{A^0_{{a}}},m^2_{h_{{b}}}\Big)} {\Gamma^*_{\hat{h}_{{j}}A^0_{{a}}h_{{b}}}} {\Gamma_{\hat{h}_{{i}}A^0_{{a}}h_{{b}}}} \nonumber \\ 
 & \; \;+\sum_{a,b=1}^{3}{B_0\Big(m^2_{\tilde{\nu}_{{a}}},m^2_{\tilde{\nu}_{{b}}}\Big)} {\Gamma^*_{\hat{h}_{{j}}\tilde{\nu}^*_{{a}}\tilde{\nu}_{{b}}}} {\Gamma_{\hat{h}_{{i}}\tilde{\nu}^*_{{a}}\tilde{\nu}_{{b}}}} +\frac{1}{2} \sum_{a,b=1}^{3}{B_0\Big(m^2_{h_{{a}}},m^2_{h_{{b}}}\Big)} {\Gamma^*_{\hat{h}_{{j}}h_{{a}}h_{{b}}}} {\Gamma_{\hat{h}_{{i}}h_{{a}}h_{{b}}}}  \nonumber \\ 
 & \; \;-6 \sum_{a=1}^{3}m_{\bar{d}_{{a}}} \sum_{b=1}^{3}\Big[{B_0\Big(m^2_{{d}_{{a}}},m^2_{d_{{b}}}\Big)} m_{d_{{b}}} \Big({\Gamma^{L*}_{\hat{h}_{{j}}\bar{d}_{{a}}d_{{b}}}} {\Gamma^R_{\hat{h}_{{i}}\bar{d}_{{a}}d_{{b}}}}   + {\Gamma^{R*}_{\hat{h}_{{j}}\bar{d}_{{a}}d_{{b}}}} {\Gamma^L_{\hat{h}_{{i}}\bar{d}_{{a}}d_{{b}}}} \Big)\Big]  \nonumber \\ 
 & \; \;+3 \sum_{a,b=1}^{3} \Big[{G_0\Big(m^2_{{d}_{{a}}},m^2_{d_{{b}}}\Big)} \Big({\Gamma^{L*}_{\hat{h}_{{j}}\bar{d}_{{a}}d_{{b}}}} {\Gamma^L_{\hat{h}_{{i}}\bar{d}_{{a}}d_{{b}}}} + {\Gamma^{R*}_{\hat{h}_{{j}}\bar{d}_{{a}}d_{{b}}}} {\Gamma^R_{\hat{h}_{{i}}\bar{d}_{{a}}d_{{b}}}} \Big) \Big] -3 \sum_{a=1}^{6}{A_0\Big(m^2_{\tilde{d}_{{a}}}\Big)} {\Gamma_{\hat{h}_{{i}}\hat{h}_{{j}}\tilde{d}^*_{{a}}\tilde{d}_{{a}}}} \nonumber \\ 
 & \; \;-2 \sum_{a=1}^{3}m_{\bar{e}_{{a}}} \sum_{b=1}^{3} \Big[{B_0\Big(m^2_{{e}_{{a}}},m^2_{e_{{b}}}\Big)} m_{e_{{b}}} \Big({\Gamma^{L*}_{\hat{h}_{{j}}\bar{e}_{{a}}e_{{b}}}} {\Gamma^R_{\hat{h}_{{i}}\bar{e}_{{a}}e_{{b}}}}  + {\Gamma^{R*}_{\hat{h}_{{j}}\bar{e}_{{a}}e_{{b}}}} {\Gamma^L_{\hat{h}_{{i}}\bar{e}_{{a}}e_{{b}}}} \Big) \Big]  \nonumber \\ 
 & \; \;+\sum_{a,b=1}^{3} \Big[{G_0\Big(m^2_{{e}_{{a}}},m^2_{e_{{b}}}\Big)} \Big({\Gamma^{L*}_{\hat{h}_{{j}}\bar{e}_{{a}}e_{{b}}}} {\Gamma^L_{\hat{h}_{{i}}\bar{e}_{{a}}e_{{b}}}}  + {\Gamma^{R*}_{\hat{h}_{{j}}\bar{e}_{{a}}e_{{b}}}} {\Gamma^R_{\hat{h}_{{i}}\bar{e}_{{a}}e_{{b}}}} \Big) \Big] +\sum_{b=1}^{3}{\Gamma^*_{\hat{h}_{{j}}ZA^0_{{b}}}} {\Gamma_{\hat{h}_{{i}}ZA^0_{{b}}}} {F_0\Big(m^2_{A^0_{{b}}},m^2_{Z}\Big)}  \nonumber \\ 
 & \; \;-6 \sum_{a=1}^{3}m_{\bar{u}_{{a}}} \sum_{b=1}^{3} \Big[{B_0\Big(m^2_{{u}_{{a}}},m^2_{u_{{b}}}\Big)} m_{u_{{b}}} \Big({\Gamma^{L*}_{\hat{h}_{{j}}\bar{u}_{{a}}u_{{b}}}} {\Gamma^R_{\hat{h}_{{i}}\bar{u}_{{a}}u_{{b}}}}   + {\Gamma^{R*}_{\hat{h}_{{j}}\bar{u}_{{a}}u_{{b}}}} {\Gamma^L_{\hat{h}_{{i}}\bar{u}_{{a}}u_{{b}}}} \Big) \Big]  \nonumber \\ 
 & \; \;+3 \sum_{a,b=1}^{3} \Big[{G_0\Big(m^2_{{u}_{{a}}},m^2_{u_{{b}}}\Big)} \Big({\Gamma^{L*}_{\hat{h}_{{j}}\bar{u}_{{a}}u_{{b}}}} {\Gamma^L_{\hat{h}_{{i}}\bar{u}_{{a}}u_{{b}}}}  + {\Gamma^{R*}_{\hat{h}_{{j}}\bar{u}_{{a}}u_{{b}}}} {\Gamma^R_{\hat{h}_{{i}}\bar{u}_{{a}}u_{{b}}}} \Big) \Big] -3 \sum_{a=1}^{6}{A_0\Big(m^2_{\tilde{u}_{{a}}}\Big)} {\Gamma_{\hat{h}_{{i}}\hat{h}_{{j}}\tilde{u}^*_{{a}}\tilde{u}_{{a}}}} \nonumber \\ 
 & \; \;- \sum_{a=1}^{5}m_{\tilde{\chi}^0_{{a}}} \sum_{b=1}^{5} \Big[{B_0\Big(m^2_{\tilde{\chi}^0_{{a}}},m^2_{\tilde{\chi}^0_{{b}}}\Big)} m_{\tilde{\chi}^0_{{b}}} \Big({\Gamma^{L*}_{\hat{h}_{{j}}\tilde{\chi}^0_{{a}}\tilde{\chi}^0_{{b}}}} {\Gamma^R_{\hat{h}_{{i}}\tilde{\chi}^0_{{a}}\tilde{\chi}^0_{{b}}}}  + {\Gamma^{R*}_{\hat{h}_{{j}}\tilde{\chi}^0_{{a}}\tilde{\chi}^0_{{b}}}} {\Gamma^L_{\hat{h}_{{i}}\tilde{\chi}^0_{{a}}\tilde{\chi}^0_{{b}}}} \Big) \Big] \nonumber \\ 
 & \; \;+\frac{1}{2} \sum_{a,b=1}^{5}\Big[{G_0\Big(m^2_{\tilde{\chi}^0_{{a}}},m^2_{\tilde{\chi}^0_{{b}}}\Big)} \Big({\Gamma^{L*}_{\hat{h}_{{j}}\tilde{\chi}^0_{{a}}\tilde{\chi}^0_{{b}}}} {\Gamma^L_{\hat{h}_{{i}}\tilde{\chi}^0_{{a}}\tilde{\chi}^0_{{b}}}}  + {\Gamma^{R*}_{\hat{h}_{{j}}\tilde{\chi}^0_{{a}}\tilde{\chi}^0_{{b}}}} {\Gamma^R_{\hat{h}_{{i}}\tilde{\chi}^0_{{a}}\tilde{\chi}^0_{{b}}}} \Big) \Big] - \sum_{a=1}^{6}{A_0\Big(m^2_{\tilde{e}_{{a}}}\Big)} {\Gamma_{\hat{h}_{{i}}\hat{h}_{{j}}\tilde{e}^*_{{a}}\tilde{e}_{{a}}}}  \nonumber \\ 
  & \; \;+3 \sum_{a,b=1}^{6}{B_0\Big(m^2_{\tilde{d}_{{a}}},m^2_{\tilde{d}_{{b}}}\Big)} {\Gamma^*_{\hat{h}_{{j}}\tilde{d}^*_{{a}}\tilde{d}_{{b}}}} {\Gamma_{\hat{h}_{{i}}\tilde{d}^*_{{a}}\tilde{d}_{{b}}}}  +\sum_{a,b=1}^{6}{B_0\Big(m^2_{\tilde{e}_{{a}}},m^2_{\tilde{e}_{{b}}}\Big)} {\Gamma^*_{\hat{h}_{{j}}\tilde{e}^*_{{a}}\tilde{e}_{{b}}}} {\Gamma_{\hat{h}_{{i}}\tilde{e}^*_{{a}}\tilde{e}_{{b}}}} \nonumber \\ 
 & \; \;+3 \sum_{a,b=1}^{6}{B_0\Big(m^2_{\tilde{u}_{{a}}},m^2_{\tilde{u}_{{b}}}\Big)} {\Gamma^*_{\hat{h}_{{j}}\tilde{u}^*_{{a}}\tilde{u}_{{b}}}} {\Gamma_{\hat{h}_{{i}}\tilde{u}^*_{{a}}\tilde{u}_{{b}}}}  +2 \sum_{b=1}^{2}{\Gamma^*_{\hat{h}_{{j}}W^+H^-_{{b}}}} {\Gamma_{\hat{h}_{{i}}W^+H^-_{{b}}}} {F_0\Big(m^2_{H^+_{{b}}},m^2_W\Big)}  
\end{align} 

\subsection{Self-energy of CP-odd Higgs bosons}
\label{app:A0self}
\begin{align} 
& \Pi_{A^0_{i}A^0_{j}}(p^2) \ =  \ 2 {A_0\Big(m^2_{Z}\Big)} {\Gamma_{\hat{A}_{h{i}}\hat{A}_{h{j}}ZZ}} + 4 {A_0\Big(m^2_W\Big)} {\Gamma_{\hat{A}_{h{i}}\hat{A}_{h{j}}W^+W^-}} -\frac{1}{2} \sum_{a=1}^{3}{A_0\Big(m^2_{A^0_{{a}}}\Big)} {\Gamma_{\hat{A}_{h{i}}\hat{A}_{h{j}}A^0_{{a}}A^0_{{a}}}}   \nonumber \\ 
 & \; \;- \sum_{a=1}^{2}{A_0\Big(m^2_{H^+_{{a}}}\Big)} {\Gamma_{\hat{A}_{h{i}}\hat{A}_{h{j}}H^+_{{a}}H^-_{{a}}}}  +\sum_{a,b=1}^{2}{B_0\Big(m^2_{H^+_{{a}}},m^2_{H^+_{{b}}}\Big)} {\Gamma^*_{\hat{A}_{h{j}}H^+_{{a}}H^-_{{b}}}} {\Gamma_{\hat{A}_{h{i}}H^+_{{a}}H^-_{{b}}}} \nonumber \\ 
 & \; \;-2 \sum_{a=1}^{2}m_{\tilde{\chi}^+_{{a}}} \sum_{b=1}^{2}\Big[ {B_0\Big(m^2_{{\tilde{\chi}}^+_{{a}}},m^2_{\tilde{\chi}^+_{{b}}}\Big)} m_{\tilde{\chi}^-_{{b}}} \Big({\Gamma^{L*}_{\hat{A}_{h{j}}\tilde{\chi}^+_{{a}}\tilde{\chi}^-_{{b}}}} {\Gamma^R_{\hat{A}_{h{i}}\tilde{\chi}^+_{{a}}\tilde{\chi}^-_{{b}}}}   + {\Gamma^{R*}_{\hat{A}_{h{j}}\tilde{\chi}^+_{{a}}\tilde{\chi}^-_{{b}}}} {\Gamma^L_{\hat{A}_{h{i}}\tilde{\chi}^+_{{a}}\tilde{\chi}^-_{{b}}}} \Big) \Big] \nonumber \\ 
 & \; \;+\sum_{a,b=1}^{2} \Big[{G_0\Big(m^2_{{\tilde{\chi}}^+_{{a}}},m^2_{\tilde{\chi}^+_{{b}}}\Big)} \Big({\Gamma^{L*}_{\hat{A}_{h{j}}\tilde{\chi}^+_{{a}}\tilde{\chi}^-_{{b}}}} {\Gamma^L_{\hat{A}_{h{i}}\tilde{\chi}^+_{{a}}\tilde{\chi}^-_{{b}}}}  + {\Gamma^{R*}_{\hat{A}_{h{j}}\tilde{\chi}^+_{{a}}\tilde{\chi}^-_{{b}}}} {\Gamma^R_{\hat{A}_{h{i}}\tilde{\chi}^+_{{a}}\tilde{\chi}^-_{{b}}}} \Big) \Big]- \sum_{a=1}^{3}{A_0\Big(m^2_{\tilde{\nu}_{{a}}}\Big)} {\Gamma_{\hat{A}_{h{i}}\hat{A}_{h{j}}\tilde{\nu}^*_{{a}}\tilde{\nu}_{{a}}}}\nonumber \\ 
 & \; \;+\sum_{a,b=1}^{3}{B_0\Big(m^2_{A^0_{{a}}},m^2_{h_{{b}}}\Big)} {\Gamma^*_{\hat{A}_{h{j}}A^0_{{a}}h_{{b}}}} {\Gamma_{\hat{A}_{h{i}}A^0_{{a}}h_{{b}}}} +\frac{1}{2} \sum_{a,b=1}^{3}{B_0\Big(m^2_{h_{{a}}},m^2_{h_{{b}}}\Big)} {\Gamma^*_{\hat{A}_{h{j}}h_{{a}}h_{{b}}}} {\Gamma_{\hat{A}_{h{i}}h_{{a}}h_{{b}}}}  \nonumber \\ 
 & \; \;-6 \sum_{a=1}^{3}m_{\bar{d}_{{a}}} \sum_{b=1}^{3} \Big[{B_0\Big(m^2_{{d}_{{a}}},m^2_{d_{{b}}}\Big)} m_{d_{{b}}} \Big({\Gamma^{L*}_{\hat{A}_{h{j}}\bar{d}_{{a}}d_{{b}}}} {\Gamma^R_{\hat{A}_{h{i}}\bar{d}_{{a}}d_{{b}}}}  + {\Gamma^{R*}_{\hat{A}_{h{j}}\bar{d}_{{a}}d_{{b}}}} {\Gamma^L_{\hat{A}_{h{i}}\bar{d}_{{a}}d_{{b}}}} \Big)  \Big] \nonumber \\ 
 & \; \;+3 \sum_{a,b=1}^{3} \Big[{G_0\Big(m^2_{{d}_{{a}}},m^2_{d_{{b}}}\Big)} \Big({\Gamma^{L*}_{\hat{A}_{h{j}}\bar{d}_{{a}}d_{{b}}}} {\Gamma^L_{\hat{A}_{h{i}}\bar{d}_{{a}}d_{{b}}}}  + {\Gamma^{R*}_{\hat{A}_{h{j}}\bar{d}_{{a}}d_{{b}}}} {\Gamma^R_{\hat{A}_{h{i}}\bar{d}_{{a}}d_{{b}}}} \Big) \Big] -\frac{1}{2} \sum_{a=1}^{3}{A_0\Big(m^2_{h_{{a}}}\Big)} {\Gamma_{\hat{A}_{h{i}}\hat{A}_{h{j}}h_{{a}}h_{{a}}}}   \nonumber \\ 
 & \; \;-2 \sum_{a=1}^{3}m_{\bar{e}_{{a}}} \sum_{b=1}^{3} \Big[{B_0\Big(m^2_{{e}_{{a}}},m^2_{e_{{b}}}\Big)} m_{e_{{b}}} \Big({\Gamma^{L*}_{\hat{A}_{h{j}}\bar{e}_{{a}}e_{{b}}}} {\Gamma^R_{\hat{A}_{h{i}}\bar{e}_{{a}}e_{{b}}}}   + {\Gamma^{R*}_{\hat{A}_{h{j}}\bar{e}_{{a}}e_{{b}}}} {\Gamma^L_{\hat{A}_{h{i}}\bar{e}_{{a}}e_{{b}}}} \Big) \Big] \nonumber \\ 
 & \; \;+\sum_{a,b=1}^{3} \Big[{G_0\Big(m^2_{{e}_{{a}}},m^2_{e_{{b}}}\Big)} \Big({\Gamma^{L*}_{\hat{A}_{h{j}}\bar{e}_{{a}}e_{{b}}}} {\Gamma^L_{\hat{A}_{h{i}}\bar{e}_{{a}}e_{{b}}}}  + {\Gamma^{R*}_{\hat{A}_{h{j}}\bar{e}_{{a}}e_{{b}}}} {\Gamma^R_{\hat{A}_{h{i}}\bar{e}_{{a}}e_{{b}}}} \Big) \Big]-3 \sum_{a=1}^{6}{A_0\Big(m^2_{\tilde{d}_{{a}}}\Big)} {\Gamma_{\hat{A}_{h{i}}\hat{A}_{h{j}}\tilde{d}^*_{{a}}\tilde{d}_{{a}}}} \nonumber \\ 
 & \; \;-6 \sum_{a=1}^{3}m_{\bar{u}_{{a}}} \sum_{b=1}^{3} \Big[{B_0\Big(m^2_{{u}_{{a}}},m^2_{u_{{b}}}\Big)} m_{u_{{b}}} \Big({\Gamma^{L*}_{\hat{A}_{h{j}}\bar{u}_{{a}}u_{{b}}}} {\Gamma^R_{\hat{A}_{h{i}}\bar{u}_{{a}}u_{{b}}}}  + {\Gamma^{R*}_{\hat{A}_{h{j}}\bar{u}_{{a}}u_{{b}}}} {\Gamma^L_{\hat{A}_{h{i}}\bar{u}_{{a}}u_{{b}}}} \Big) \Big] \nonumber \\ 
 & \; \;+3 \sum_{a,b=1}^{3} \Big[{G_0\Big(m^2_{{u}_{{a}}},m^2_{u_{{b}}}\Big)} \Big({\Gamma^{L*}_{\hat{A}_{h{j}}\bar{u}_{{a}}u_{{b}}}} {\Gamma^L_{\hat{A}_{h{i}}\bar{u}_{{a}}u_{{b}}}}   + {\Gamma^{R*}_{\hat{A}_{h{j}}\bar{u}_{{a}}u_{{b}}}} {\Gamma^R_{\hat{A}_{h{i}}\bar{u}_{{a}}u_{{b}}}} \Big) \Big] -3 \sum_{a=1}^{6}{A_0\Big(m^2_{\tilde{u}_{{a}}}\Big)} {\Gamma_{\hat{A}_{h{i}}\hat{A}_{h{j}}\tilde{u}^*_{{a}}\tilde{u}_{{a}}}} \nonumber \\ 
 & \; \;- \sum_{a=1}^{5}m_{\tilde{\chi}^0_{{a}}} \sum_{b=1}^{5} \Big[{B_0\Big(m^2_{\tilde{\chi}^0_{{a}}},m^2_{\tilde{\chi}^0_{{b}}}\Big)} m_{\tilde{\chi}^0_{{b}}} \Big({\Gamma^{L*}_{\hat{A}_{h{j}}\tilde{\chi}^0_{{a}}\tilde{\chi}^0_{{b}}}} {\Gamma^R_{\hat{A}_{h{i}}\tilde{\chi}^0_{{a}}\tilde{\chi}^0_{{b}}}}   + {\Gamma^{R*}_{\hat{A}_{h{j}}\tilde{\chi}^0_{{a}}\tilde{\chi}^0_{{b}}}} {\Gamma^L_{\hat{A}_{h{i}}\tilde{\chi}^0_{{a}}\tilde{\chi}^0_{{b}}}} \Big) \Big]  \nonumber \\ 
 & \; \;+\frac{1}{2} \sum_{a,b=1}^{5} \Big[{G_0\Big(m^2_{\tilde{\chi}^0_{{a}}},m^2_{\tilde{\chi}^0_{{b}}}\Big)} \Big({\Gamma^{L*}_{\hat{A}_{h{j}}\tilde{\chi}^0_{{a}}\tilde{\chi}^0_{{b}}}} {\Gamma^L_{\hat{A}_{h{i}}\tilde{\chi}^0_{{a}}\tilde{\chi}^0_{{b}}}}  + {\Gamma^{R*}_{\hat{A}_{h{j}}\tilde{\chi}^0_{{a}}\tilde{\chi}^0_{{b}}}} {\Gamma^R_{\hat{A}_{h{i}}\tilde{\chi}^0_{{a}}\tilde{\chi}^0_{{b}}}} \Big) \Big]  - \sum_{a=1}^{6}{A_0\Big(m^2_{\tilde{e}_{{a}}}\Big)} {\Gamma_{\hat{A}_{h{i}}\hat{A}_{h{j}}\tilde{e}^*_{{a}}\tilde{e}_{{a}}}}  \nonumber \\ 
 & \; \; +3 \sum_{a,b=1}^{6}{B_0\Big(m^2_{\tilde{d}_{{a}}},m^2_{\tilde{d}_{{b}}}\Big)} {\Gamma^*_{\hat{A}_{h{j}}\tilde{d}^*_{{a}}\tilde{d}_{{b}}}} {\Gamma_{\hat{A}_{h{i}}\tilde{d}^*_{{a}}\tilde{d}_{{b}}}} +\frac{1}{2} \sum_{a,b=1}^{3}{B_0\Big(m^2_{A^0_{{a}}},m^2_{A^0_{{b}}}\Big)} {\Gamma^*_{\hat{A}_{h{j}}A^0_{{a}}A^0_{{b}}}} {\Gamma_{\hat{A}_{h{i}}A^0_{{a}}A^0_{{b}}}}   \nonumber \\ 
 & \; \;+\sum_{a,b=1}^{6}{B_0\Big(m^2_{\tilde{e}_{{a}}},m^2_{\tilde{e}_{{b}}}\Big)} {\Gamma^*_{\hat{A}_{h{j}}\tilde{e}^*_{{a}}\tilde{e}_{{b}}}} {\Gamma_{\hat{A}_{h{i}}\tilde{e}^*_{{a}}\tilde{e}_{{b}}}} +3 \sum_{a,b=1}^{6}{B_0\Big(m^2_{\tilde{u}_{{a}}},m^2_{\tilde{u}_{{b}}}\Big)} {\Gamma^*_{\hat{A}_{h{j}}\tilde{u}^*_{{a}}\tilde{u}_{{b}}}} {\Gamma_{\hat{A}_{h{i}}\tilde{u}^*_{{a}}\tilde{u}_{{b}}}}  \nonumber \\ 
 & \; \;+2 \sum_{b=1}^{2}{\Gamma^*_{\hat{A}_{h{j}}W^+H^-_{{b}}}} {\Gamma_{\hat{A}_{h{i}}W^+H^-_{{b}}}} {F_0\Big(m^2_{H^+_{{b}}},m^2_W\Big)} +\sum_{b=1}^{3}{\Gamma^*_{\hat{A}_{h{j}}Zh_{{b}}}} {\Gamma_{\hat{A}_{h{i}}Zh_{{b}}}} {F_0\Big(m^2_{h_{{b}}},m^2_{Z}\Big)}  
\end{align} 

\subsection{Self-energy of the charged Higgs boson}
\label{app:Hpself}
\begin{align} 
& \Pi_{H^-_iH^-_j}(p^2) \, = \, \frac{7}{2} {B_0\Big(m^2_{Z},m^2_W\Big)} {\Gamma^*_{\hat{H}^+_{{j}}W^-Z}} {\Gamma_{\hat{H}^+_{{i}}W^-Z}} +2 {A_0\Big(m^2_{Z}\Big)} {\Gamma_{\hat{H}^+_{{i}}\hat{H}^-_{{j}}ZZ}} \nonumber \\ 
 & \; \;+4 {A_0\Big(m^2_W\Big)} {\Gamma_{\hat{H}^+_{{i}}\hat{H}^-_{{j}}W^+W^-}} - \sum_{a=1}^{2}{A_0\Big(m^2_{H^+_{{a}}}\Big)} {\Gamma_{\hat{H}^+_{{i}}\hat{H}^-_{{j}}H^+_{{a}}H^-_{{a}}}}  -3 \sum_{a=1}^{6}{A_0\Big(m^2_{\tilde{u}_{{a}}}\Big)} {\Gamma_{\hat{H}^+_{{i}}\hat{H}^-_{{j}}\tilde{u}^*_{{a}}\tilde{u}_{{a}}}} \nonumber \\ 
 & \; \;-2 \sum_{a=1}^{2}m_{\tilde{\chi}^-_{{a}}} \sum_{b=1}^{5}\Big[{B_0\Big(m^2_{\tilde{\chi}^+_{{a}}},m^2_{\tilde{\chi}^0_{{b}}}\Big)} m_{\tilde{\chi}^0_{{b}}} \Big({\Gamma^{L*}_{\hat{H}^+_{{j}}\tilde{\chi}^-_{{a}}\tilde{\chi}^0_{{b}}}} {\Gamma^R_{\hat{H}^+_{{i}}\tilde{\chi}^-_{{a}}\tilde{\chi}^0_{{b}}}}   + {\Gamma^{R*}_{\hat{H}^+_{{j}}\tilde{\chi}^-_{{a}}\tilde{\chi}^0_{{b}}}} {\Gamma^L_{\hat{H}^+_{{i}}\tilde{\chi}^-_{{a}}\tilde{\chi}^0_{{b}}}} \Big)\Big]  \nonumber \\ 
 & \; \;+\sum_{a=1}^{2}\sum_{b=1}^{5} \Big[{G_0\Big(m^2_{\tilde{\chi}^+_{{a}}},m^2_{\tilde{\chi}^0_{{b}}}\Big)} \Big({\Gamma^{L*}_{\hat{H}^+_{{j}}\tilde{\chi}^-_{{a}}\tilde{\chi}^0_{{b}}}} {\Gamma^L_{\hat{H}^+_{{i}}\tilde{\chi}^-_{{a}}\tilde{\chi}^0_{{b}}}}  + {\Gamma^{R*}_{\hat{H}^+_{{j}}\tilde{\chi}^-_{{a}}\tilde{\chi}^0_{{b}}}} {\Gamma^R_{\hat{H}^+_{{i}}\tilde{\chi}^-_{{a}}\tilde{\chi}^0_{{b}}}} \Big) \Big]\nonumber \\ 
 & \; \;-\frac{1}{2} \sum_{a=1}^{3}{A_0\Big(m^2_{A^0_{{a}}}\Big)} {\Gamma_{\hat{H}^+_{{i}}\hat{H}^-_{{j}}A^0_{{a}}A^0_{{a}}}}  - \sum_{a=1}^{3}{A_0\Big(m^2_{\tilde{\nu}_{{a}}}\Big)} {\Gamma_{\hat{H}^+_{{i}}\hat{H}^-_{{j}}\tilde{\nu}^*_{{a}}\tilde{\nu}_{{a}}}} -\frac{1}{2} \sum_{a=1}^{3}{A_0\Big(m^2_{h_{{a}}}\Big)} {\Gamma_{\hat{H}^+_{{i}}\hat{H}^-_{{j}}h_{{a}}h_{{a}}}}  \nonumber \\ 
 & \; \;+\sum_{a=1}^{3}\sum_{b=1}^{2}{B_0\Big(m^2_{A^0_{{a}}},m^2_{H^+_{{b}}}\Big)} {\Gamma^*_{\hat{H}^+_{{j}}A^0_{{a}}H^-_{{b}}}} {\Gamma_{\hat{H}^+_{{i}}A^0_{{a}}H^-_{{b}}}} +\sum_{a=1}^{3}\sum_{b=1}^{2}{B_0\Big(m^2_{h_{{a}}},m^2_{H^+_{{b}}}\Big)} {\Gamma^*_{\hat{H}^+_{{j}}h_{{a}}H^-_{{b}}}} {\Gamma_{\hat{H}^+_{{i}}h_{{a}}H^-_{{b}}}} \nonumber \\ 
 & \; \;-6 \sum_{a=1}^{3}m_{d_{{a}}} \sum_{b=1}^{3}\Big[{B_0\Big(m^2_{d_{{a}}},m^2_{{u}_{{b}}}\Big)} m_{\bar{u}_{{b}}} \Big({\Gamma^{L*}_{\hat{H}^+_{{j}}d_{{a}}\bar{u}_{{b}}}} {\Gamma^R_{\hat{H}^+_{{i}}d_{{a}}\bar{u}_{{b}}}}   + {\Gamma^{R*}_{\hat{H}^+_{{j}}d_{{a}}\bar{u}_{{b}}}} {\Gamma^L_{\hat{H}^+_{{i}}d_{{a}}\bar{u}_{{b}}}} \Big)  \Big]\nonumber \\ 
 & \; \;+3 \sum_{a,b=1}^{3}\Big[{G_0\Big(m^2_{d_{{a}}},m^2_{{u}_{{b}}}\Big)} \Big({\Gamma^{L*}_{\hat{H}^+_{{j}}d_{{a}}\bar{u}_{{b}}}} {\Gamma^L_{\hat{H}^+_{{i}}d_{{a}}\bar{u}_{{b}}}}   + {\Gamma^{R*}_{\hat{H}^+_{{j}}d_{{a}}\bar{u}_{{b}}}} {\Gamma^R_{\hat{H}^+_{{i}}d_{{a}}\bar{u}_{{b}}}} \Big) \Big]  - \sum_{a=1}^{6}{A_0\Big(m^2_{\tilde{e}_{{a}}}\Big)} {\Gamma_{\hat{H}^+_{{i}}\hat{H}^-_{{j}}\tilde{e}^*_{{a}}\tilde{e}_{{a}}}}  \nonumber \\ 
 & \; \;-2 \sum_{a=1}^{3}m_{e_{{a}}} \sum_{b=1}^{3}\Big[{B_0\Big(m^2_{e_{{a}}},m^2_{{\nu}_{{b}}}\Big)} m_{\bar{\nu}_{{b}}} \Big({\Gamma^{L*}_{\hat{H}^+_{{j}}e_{{a}}\bar{\nu}_{{b}}}} {\Gamma^R_{\hat{H}^+_{{i}}e_{{a}}\bar{\nu}_{{b}}}}  + {\Gamma^{R*}_{\hat{H}^+_{{j}}e_{{a}}\bar{\nu}_{{b}}}} {\Gamma^L_{\hat{H}^+_{{i}}e_{{a}}\bar{\nu}_{{b}}}} \Big)\Big]  \nonumber \\ 
 & \; \;+\sum_{a,b=1}^{3}\Big[{G_0\Big(m^2_{e_{{a}}},m^2_{{\nu}_{{b}}}\Big)} \Big({\Gamma^{L*}_{\hat{H}^+_{{j}}e_{{a}}\bar{\nu}_{{b}}}} {\Gamma^L_{\hat{H}^+_{{i}}e_{{a}}\bar{\nu}_{{b}}}} + {\Gamma^{R*}_{\hat{H}^+_{{j}}e_{{a}}\bar{\nu}_{{b}}}} {\Gamma^R_{\hat{H}^+_{{i}}e_{{a}}\bar{\nu}_{{b}}}} \Big) \Big] -3 \sum_{a=1}^{6}{A_0\Big(m^2_{\tilde{d}_{{a}}}\Big)} {\Gamma_{\hat{H}^+_{{i}}\hat{H}^-_{{j}}\tilde{d}^*_{{a}}\tilde{d}_{{a}}}} \nonumber \\ 
 & \; \;+\sum_{a=1}^{6}\sum_{b=1}^{3}{B_0\Big(m^2_{\tilde{e}_{{a}}},m^2_{\tilde{\nu}_{{b}}}\Big)} {\Gamma^*_{\hat{H}^+_{{j}}\tilde{e}_{{a}}\tilde{\nu}^*_{{b}}}} {\Gamma_{\hat{H}^+_{{i}}\tilde{e}_{{a}}\tilde{\nu}^*_{{b}}}} +3 \sum_{a,b=1}^{6}{B_0\Big(m^2_{\tilde{d}_{{a}}},m^2_{\tilde{u}_{{b}}}\Big)} {\Gamma^*_{\hat{H}^+_{{j}}\tilde{d}_{{a}}\tilde{u}^*_{{b}}}} {\Gamma_{\hat{H}^+_{{i}}\tilde{d}_{{a}}\tilde{u}^*_{{b}}}}  \nonumber \\ 
 & \; \;+\sum_{b=1}^{2}\Big[{\Gamma^*_{\hat{H}^+_{{j}}\gamma H^-_{{b}}}} {\Gamma_{\hat{H}^+_{{i}}\gamma H^-_{{b}}}} {F_0\Big(m^2_{H^+_{{b}}},0\Big)} +{\Gamma^*_{\hat{H}^+_{{j}}ZH^-_{{b}}}} {\Gamma_{\hat{H}^+_{{i}}ZH^-_{{b}}}} {F_0\Big(m^2_{H^+_{{b}}},m^2_{Z}\Big)}\Big] \nonumber \\ 
 & \; \;+\sum_{b=1}^{3}\Big[{\Gamma^*_{\hat{H}^+_{{j}}W^-A^0_{{b}}}} {\Gamma_{\hat{H}^+_{{i}}W^-A^0_{{b}}}} {F_0\Big(m^2_{A^0_{{b}}},m^2_W\Big)}+{\Gamma^*_{\hat{H}^+_{{j}}W^-h_{{b}}}} {\Gamma_{\hat{H}^+_{{i}}W^-h_{{b}}}} {F_0\Big(m^2_{h_{{b}}},m^2_W\Big)}  \Big]
\end{align} 

\subsection{Self-energy of neutralinos}
\label{sec:OneLoopNeu}

\begin{align} 
&  \Sigma^S_{\tilde{\chi}^0_i\tilde{\chi}^0_j}(p^2) \ =  \ \sum_{a,b=1}^{2}{B_0\Big(m^2_{\tilde{\chi}^+_{{b}}},m^2_{H^+_{{a}}}\Big)} {\Gamma^{L*}_{\hat{\tilde{\chi}}^0_{{j}}H^+_{{a}}\tilde{\chi}^-_{{b}}}} m_{\tilde{\chi}^-_{{b}}} {\Gamma^R_{\hat{\tilde{\chi}}^0_{{i}}H^+_{{a}}\tilde{\chi}^-_{{b}}}} \nonumber \\ 
 & \; \;+\frac{1}{2} \sum_{a=1}^{3}\sum_{b=1}^{5}m_{\tilde{\chi}^0_{{b}}}\Big[{B_0\Big(m^2_{\tilde{\chi}^0_{{b}}},m^2_{A^0_{{a}}}\Big)} {\Gamma^{L*}_{\hat{\tilde{\chi}}^0_{{j}}A^0_{{a}}\tilde{\chi}^0_{{b}}}}  {\Gamma^R_{\hat{\tilde{\chi}}^0_{{i}}A^0_{{a}}\tilde{\chi}^0_{{b}}}} +{B_0\Big(m^2_{\tilde{\chi}^0_{{b}}},m^2_{h_{{a}}}\Big)} {\Gamma^{L*}_{\hat{\tilde{\chi}}^0_{{j}}h_{{a}}\tilde{\chi}^0_{{b}}}}  {\Gamma^R_{\hat{\tilde{\chi}}^0_{{i}}h_{{a}}\tilde{\chi}^0_{{b}}}} \Big]  \nonumber \\ 
 & \; \;+\sum_{a=1}^{6}\sum_{b=1}^{3}\Big[3{B_0\Big(m^2_{d_{{b}}},m^2_{\tilde{d}_{{a}}}\Big)} {\Gamma^{L*}_{\hat{\tilde{\chi}}^0_{{j}}\tilde{d}^*_{{a}}d_{{b}}}} m_{d_{{b}}} {\Gamma^R_{\hat{\tilde{\chi}}^0_{{i}}\tilde{d}^*_{{a}}d_{{b}}}} +{B_0\Big(m^2_{e_{{b}}},m^2_{\tilde{e}_{{a}}}\Big)} {\Gamma^{L*}_{\hat{\tilde{\chi}}^0_{{j}}\tilde{e}^*_{{a}}e_{{b}}}} m_{e_{{b}}} {\Gamma^R_{\hat{\tilde{\chi}}^0_{{i}}\tilde{e}^*_{{a}}e_{{b}}}} \nonumber \\ 
 & \; \;+3 {B_0\Big(m^2_{u_{{b}}},m^2_{\tilde{u}_{{a}}}\Big)} {\Gamma^{L*}_{\hat{\tilde{\chi}}^0_{{j}}\tilde{u}^*_{{a}}u_{{b}}}} m_{u_{{b}}} {\Gamma^R_{\hat{\tilde{\chi}}^0_{{i}}\tilde{u}^*_{{a}}u_{{b}}}}\Big] -4 \sum_{b=1}^{2}{B_0\Big(m^2_{\tilde{\chi}^+_{{b}}},m^2_W\Big)} {\Gamma^{R*}_{\hat{\tilde{\chi}}^0_{{j}}W^+\tilde{\chi}^-_{{b}}}} m_{\tilde{\chi}^-_{{b}}} {\Gamma^L_{\hat{\tilde{\chi}}^0_{{i}}W^+\tilde{\chi}^-_{{b}}}}  \nonumber \\ 
 & \; \;-2 \sum_{b=1}^{5}{B_0\Big(m^2_{\tilde{\chi}^0_{{b}}},m^2_{Z}\Big)} {\Gamma^{R*}_{\hat{\tilde{\chi}}^0_{{j}}Z\tilde{\chi}^0_{{b}}}} m_{\tilde{\chi}^0_{{b}}} {\Gamma^L_{\hat{\tilde{\chi}}^0_{{i}}Z\tilde{\chi}^0_{{b}}}} +\sum_{a,b=1}^{3}{B_0\Big(0,0\Big)} {\Gamma^{L*}_{\hat{\tilde{\chi}}^0_{{j}}\tilde{\nu}^*_{{a}}\nu_{{b}}}} m_{\nu_{{b}}} {\Gamma^R_{\hat{\tilde{\chi}}^0_{{i}}\tilde{\nu}^*_{{a}}\nu_{{b}}}}   \\ 
&  \Sigma^R_{\tilde{\chi}^0_i\tilde{\chi}^0_j}(p^2) \ =  \ \sum_{a,b=1}^{2} {B_0\Big(m^2_{\tilde{\chi}^+_{{b}}},m^2_{H^+_{{a}}}\Big)} {\Gamma^{L*}_{\hat{\tilde{\chi}}^0_{{j}}H^+_{{a}}\tilde{\chi}^-_{{b}}}} m_{\tilde{\chi}^-_{{b}}} {\Gamma^R_{\hat{\tilde{\chi}}^0_{{i}}H^+_{{a}}\tilde{\chi}^-_{{b}}}} \nonumber \\ 
 & \; \;+\frac{1}{2} \sum_{a=1}^{3}\sum_{b=1}^{5}m_{\tilde{\chi}^0_{{b}}}\Big[{B_0\Big(m^2_{\tilde{\chi}^0_{{b}}},m^2_{A^0_{{a}}}\Big)} {\Gamma^{L*}_{\hat{\tilde{\chi}}^0_{{j}}A^0_{{a}}\tilde{\chi}^0_{{b}}}}  {\Gamma^R_{\hat{\tilde{\chi}}^0_{{i}}A^0_{{a}}\tilde{\chi}^0_{{b}}}} +{B_0\Big(m^2_{\tilde{\chi}^0_{{b}}},m^2_{h_{{a}}}\Big)} {\Gamma^{L*}_{\hat{\tilde{\chi}}^0_{{j}}h_{{a}}\tilde{\chi}^0_{{b}}}} {\Gamma^R_{\hat{\tilde{\chi}}^0_{{i}}h_{{a}}\tilde{\chi}^0_{{b}}}} \Big] \nonumber \\ 
 & \; \;+\sum_{a=1}^{6}\sum_{b=1}^{3}\Big[3{B_0\Big(m^2_{d_{{b}}},m^2_{\tilde{d}_{{a}}}\Big)} {\Gamma^{L*}_{\hat{\tilde{\chi}}^0_{{j}}\tilde{d}^*_{{a}}d_{{b}}}} m_{d_{{b}}} {\Gamma^R_{\hat{\tilde{\chi}}^0_{{i}}\tilde{d}^*_{{a}}d_{{b}}}}  +{B_0\Big(m^2_{e_{{b}}},m^2_{\tilde{e}_{{a}}}\Big)} {\Gamma^{L*}_{\hat{\tilde{\chi}}^0_{{j}}\tilde{e}^*_{{a}}e_{{b}}}} m_{e_{{b}}} {\Gamma^R_{\hat{\tilde{\chi}}^0_{{i}}\tilde{e}^*_{{a}}e_{{b}}}} \nonumber \\ 
 & \; \;+3 {B_0\Big(m^2_{u_{{b}}},m^2_{\tilde{u}_{{a}}}\Big)} {\Gamma^{L*}_{\hat{\tilde{\chi}}^0_{{j}}\tilde{u}^*_{{a}}u_{{b}}}} m_{u_{{b}}} {\Gamma^R_{\hat{\tilde{\chi}}^0_{{i}}\tilde{u}^*_{{a}}u_{{b}}}}\Big]  -4 \sum_{b=1}^{2}{B_0\Big(m^2_{\tilde{\chi}^+_{{b}}},m^2_W\Big)} {\Gamma^{R*}_{\hat{\tilde{\chi}}^0_{{j}}W^+\tilde{\chi}^-_{{b}}}} m_{\tilde{\chi}^-_{{b}}} {\Gamma^L_{\hat{\tilde{\chi}}^0_{{i}}W^+\tilde{\chi}^-_{{b}}}}  \nonumber \\ 
 & \; \;-2 \sum_{b=1}^{5}{B_0\Big(m^2_{\tilde{\chi}^0_{{b}}},m^2_{Z}\Big)} {\Gamma^{R*}_{\hat{\tilde{\chi}}^0_{{j}}Z\tilde{\chi}^0_{{b}}}} m_{\tilde{\chi}^0_{{b}}} {\Gamma^L_{\hat{\tilde{\chi}}^0_{{i}}Z\tilde{\chi}^0_{{b}}}} +\sum_{a,b=1}^{3}{B_0\Big(0,0\Big)} {\Gamma^{L*}_{\hat{\tilde{\chi}}^0_{{j}}\tilde{\nu}^*_{{a}}\nu_{{b}}}} m_{\nu_{{b}}} {\Gamma^R_{\hat{\tilde{\chi}}^0_{{i}}\tilde{\nu}^*_{{a}}\nu_{{b}}}}  \\ 
&  \Sigma^L_{\tilde{\chi}^0_i\tilde{\chi}^0_j}(p^2) \, = \, \sum_{a,b=1}^{2}{B_0\Big(m^2_{\tilde{\chi}^+_{{b}}},m^2_{H^+_{{a}}}\Big)} {\Gamma^{L*}_{\hat{\tilde{\chi}}^0_{{j}}H^+_{{a}}\tilde{\chi}^-_{{b}}}} m_{\tilde{\chi}^-_{{b}}} {\Gamma^R_{\hat{\tilde{\chi}}^0_{{i}}H^+_{{a}}\tilde{\chi}^-_{{b}}}} \nonumber \\ 
 & \; \;+\frac{1}{2} \sum_{a=1}^{3}\sum_{b=1}^{5} m_{\tilde{\chi}^0_{{b}}}\Big[{B_0\Big(m^2_{\tilde{\chi}^0_{{b}}},m^2_{A^0_{{a}}}\Big)} {\Gamma^{L*}_{\hat{\tilde{\chi}}^0_{{j}}A^0_{{a}}\tilde{\chi}^0_{{b}}}} {\Gamma^R_{\hat{\tilde{\chi}}^0_{{i}}A^0_{{a}}\tilde{\chi}^0_{{b}}}} +{B_0\Big(m^2_{\tilde{\chi}^0_{{b}}},m^2_{h_{{a}}}\Big)} {\Gamma^{L*}_{\hat{\tilde{\chi}}^0_{{j}}h_{{a}}\tilde{\chi}^0_{{b}}}} {\Gamma^R_{\hat{\tilde{\chi}}^0_{{i}}h_{{a}}\tilde{\chi}^0_{{b}}}} \Big]  \nonumber \\ 
 & \; \;+ \sum_{a=1}^{6}\sum_{b=1}^{3}\Big[3{B_0\Big(m^2_{d_{{b}}},m^2_{\tilde{d}_{{a}}}\Big)} {\Gamma^{L*}_{\hat{\tilde{\chi}}^0_{{j}}\tilde{d}^*_{{a}}d_{{b}}}} m_{d_{{b}}} {\Gamma^R_{\hat{\tilde{\chi}}^0_{{i}}\tilde{d}^*_{{a}}d_{{b}}}}  +{B_0\Big(m^2_{e_{{b}}},m^2_{\tilde{e}_{{a}}}\Big)} {\Gamma^{L*}_{\hat{\tilde{\chi}}^0_{{j}}\tilde{e}^*_{{a}}e_{{b}}}} m_{e_{{b}}} {\Gamma^R_{\hat{\tilde{\chi}}^0_{{i}}\tilde{e}^*_{{a}}e_{{b}}}} \nonumber \\ 
 & \; \;+3 {B_0\Big(m^2_{u_{{b}}},m^2_{\tilde{u}_{{a}}}\Big)} {\Gamma^{L*}_{\hat{\tilde{\chi}}^0_{{j}}\tilde{u}^*_{{a}}u_{{b}}}} m_{u_{{b}}} {\Gamma^R_{\hat{\tilde{\chi}}^0_{{i}}\tilde{u}^*_{{a}}u_{{b}}}}\Big]  -4 \sum_{b=1}^{2}{B_0\Big(m^2_{\tilde{\chi}^+_{{b}}},m^2_W\Big)} {\Gamma^{R*}_{\hat{\tilde{\chi}}^0_{{j}}W^+\tilde{\chi}^-_{{b}}}} m_{\tilde{\chi}^-_{{b}}} {\Gamma^L_{\hat{\tilde{\chi}}^0_{{i}}W^+\tilde{\chi}^-_{{b}}}}  \nonumber \\ 
 & \; \;-2 \sum_{b=1}^{5}{B_0\Big(m^2_{\tilde{\chi}^0_{{b}}},m^2_{Z}\Big)} {\Gamma^{R*}_{\hat{\tilde{\chi}}^0_{{j}}Z\tilde{\chi}^0_{{b}}}} m_{\tilde{\chi}^0_{{b}}} {\Gamma^L_{\hat{\tilde{\chi}}^0_{{i}}Z\tilde{\chi}^0_{{b}}}} +\sum_{a,b=1}^{3}{B_0\Big(0,0\Big)} {\Gamma^{L*}_{\hat{\tilde{\chi}}^0_{{j}}\tilde{\nu}^*_{{a}}\nu_{{b}}}} m_{\nu_{{b}}} {\Gamma^R_{\hat{\tilde{\chi}}^0_{{i}}\tilde{\nu}^*_{{a}}\nu_{{b}}}}   
\end{align} 

\subsection{Self-energy of charginos}
\label{sec:OneLoopCha}
\begin{align} 
&  \Sigma^S_{\tilde{\chi}^+_i\tilde{\chi}^+_j}(p^2) \ =  \ \sum_{a=1}^{2}\sum_{b=1}^{5}{B_0\Big(m^2_{\tilde{\chi}^0_{{b}}},m^2_{H^+_{{a}}}\Big)} {\Gamma^{L*}_{\hat{\tilde{\chi}}^+_{{j}}H^-_{{a}}\tilde{\chi}^0_{{b}}}} m_{\tilde{\chi}^0_{{b}}} {\Gamma^R_{\hat{\tilde{\chi}}^+_{{i}}H^-_{{a}}\tilde{\chi}^0_{{b}}}} \nonumber \\ 
 & \; \;+\sum_{a=1}^{3}\sum_{b=1}^{2}m_{\tilde{\chi}^-_{{b}}}\Big[{B_0\Big(m^2_{\tilde{\chi}^+_{{b}}},m^2_{A^0_{{a}}}\Big)} {\Gamma^{L*}_{\hat{\tilde{\chi}}^+_{{j}}A^0_{{a}}\tilde{\chi}^-_{{b}}}}  {\Gamma^R_{\hat{\tilde{\chi}}^+_{{i}}A^0_{{a}}\tilde{\chi}^-_{{b}}}} +{B_0\Big(m^2_{\tilde{\chi}^+_{{b}}},m^2_{h_{{a}}}\Big)} {\Gamma^{L*}_{\hat{\tilde{\chi}}^+_{{j}}h_{{a}}\tilde{\chi}^-_{{b}}}} {\Gamma^R_{\hat{\tilde{\chi}}^+_{{i}}h_{{a}}\tilde{\chi}^-_{{b}}}} \Big] \nonumber \\ 
 & \; \;+\sum_{a=1}^{6}\sum_{b=1}^{3}\Big[3{B_0\Big(m^2_{d_{{b}}},m^2_{\tilde{u}_{{a}}}\Big)} {\Gamma^{L*}_{\hat{\tilde{\chi}}^+_{{j}}\tilde{u}^*_{{a}}d_{{b}}}} m_{d_{{b}}} {\Gamma^R_{\hat{\tilde{\chi}}^+_{{i}}\tilde{u}^*_{{a}}d_{{b}}}}  +3 {B_0\Big(m^2_{{u}_{{b}}},m^2_{\tilde{d}_{{a}}}\Big)} {\Gamma^{L*}_{\hat{\tilde{\chi}}^+_{{j}}\tilde{d}_{{a}}\bar{u}_{{b}}}} m_{\bar{u}_{{b}}} {\Gamma^R_{\hat{\tilde{\chi}}^+_{{i}}\tilde{d}_{{a}}\bar{u}_{{b}}}} \nonumber \\ 
 & \; \;+{B_0\Big(m^2_{{\nu}_{{b}}},m^2_{\tilde{e}_{{a}}}\Big)} {\Gamma^{L*}_{\hat{\tilde{\chi}}^+_{{j}}\tilde{e}_{{a}}\bar{\nu}_{{b}}}} m_{\bar{\nu}_{{b}}} {\Gamma^R_{\hat{\tilde{\chi}}^+_{{i}}\tilde{e}_{{a}}\bar{\nu}_{{b}}}}\Big]+\sum_{a,b=1}^{3}{B_0\Big(m^2_{e_{{b}}},m^2_{\tilde{\nu}_{{a}}}\Big)} {\Gamma^{L*}_{\hat{\tilde{\chi}}^+_{{j}}\tilde{\nu}^*_{{a}}e_{{b}}}} m_{e_{{b}}} {\Gamma^R_{\hat{\tilde{\chi}}^+_{{i}}\tilde{\nu}^*_{{a}}e_{{b}}}} \nonumber \\ 
 & \; \;-4 \sum_{b=1}^{2}{B_0\Big(m^2_{\tilde{\chi}^+_{{b}}},0\Big)} {\Gamma^{R*}_{\hat{\tilde{\chi}}^+_{{j}}\gamma\tilde{\chi}^-_{{b}}}} m_{\tilde{\chi}^-_{{b}}} {\Gamma^L_{\hat{\tilde{\chi}}^+_{{i}}\gamma\tilde{\chi}^-_{{b}}}}-4 \sum_{b=1}^{2}{B_0\Big(m^2_{\tilde{\chi}^+_{{b}}},m^2_{Z}\Big)} {\Gamma^{R*}_{\hat{\tilde{\chi}}^+_{{j}}Z\tilde{\chi}^-_{{b}}}} m_{\tilde{\chi}^-_{{b}}} {\Gamma^L_{\hat{\tilde{\chi}}^+_{{i}}Z\tilde{\chi}^-_{{b}}}}  \nonumber \\ 
 & \; \;-4 \sum_{b=1}^{5}{B_0\Big(m^2_{\tilde{\chi}^0_{{b}}},m^2_W\Big)} {\Gamma^{R*}_{\hat{\tilde{\chi}}^+_{{j}}W^-\tilde{\chi}^0_{{b}}}} m_{\tilde{\chi}^0_{{b}}} {\Gamma^L_{\hat{\tilde{\chi}}^+_{{i}}W^-\tilde{\chi}^0_{{b}}}}  \\
&  \Sigma^R_{\tilde{\chi}^+_i\tilde{\chi}^+_j}(p^2) \ =  \ \sum_{a=1}^{2}\sum_{b=1}^{5}{B_0\Big(m^2_{\tilde{\chi}^0_{{b}}},m^2_{H^+_{{a}}}\Big)} {\Gamma^{L*}_{\hat{\tilde{\chi}}^+_{{j}}H^-_{{a}}\tilde{\chi}^0_{{b}}}} m_{\tilde{\chi}^0_{{b}}} {\Gamma^R_{\hat{\tilde{\chi}}^+_{{i}}H^-_{{a}}\tilde{\chi}^0_{{b}}}} \nonumber \\ 
 & \; \;+\sum_{a=1}^{3}\sum_{b=1}^{2}m_{\tilde{\chi}^-_{{b}}}\Big[{B_0\Big(m^2_{\tilde{\chi}^+_{{b}}},m^2_{A^0_{{a}}}\Big)} {\Gamma^{L*}_{\hat{\tilde{\chi}}^+_{{j}}A^0_{{a}}\tilde{\chi}^-_{{b}}}}  {\Gamma^R_{\hat{\tilde{\chi}}^+_{{i}}A^0_{{a}}\tilde{\chi}^-_{{b}}}} +{B_0\Big(m^2_{\tilde{\chi}^+_{{b}}},m^2_{h_{{a}}}\Big)} {\Gamma^{L*}_{\hat{\tilde{\chi}}^+_{{j}}h_{{a}}\tilde{\chi}^-_{{b}}}}  {\Gamma^R_{\hat{\tilde{\chi}}^+_{{i}}h_{{a}}\tilde{\chi}^-_{{b}}}} \Big] \nonumber \\ 
 & \; \;+\sum_{a=1}^{6}\sum_{b=1}^{3}\Big[3 {B_0\Big(m^2_{d_{{b}}},m^2_{\tilde{u}_{{a}}}\Big)} {\Gamma^{L*}_{\hat{\tilde{\chi}}^+_{{j}}\tilde{u}^*_{{a}}d_{{b}}}} m_{d_{{b}}} {\Gamma^R_{\hat{\tilde{\chi}}^+_{{i}}\tilde{u}^*_{{a}}d_{{b}}}}  +3{B_0\Big(m^2_{{u}_{{b}}},m^2_{\tilde{d}_{{a}}}\Big)} {\Gamma^{L*}_{\hat{\tilde{\chi}}^+_{{j}}\tilde{d}_{{a}}\bar{u}_{{b}}}} m_{\bar{u}_{{b}}} {\Gamma^R_{\hat{\tilde{\chi}}^+_{{i}}\tilde{d}_{{a}}\bar{u}_{{b}}}} \nonumber \\ 
 & \; \;+{B_0\Big(m^2_{{\nu}_{{b}}},m^2_{\tilde{e}_{{a}}}\Big)} {\Gamma^{L*}_{\hat{\tilde{\chi}}^+_{{j}}\tilde{e}_{{a}}\bar{\nu}_{{b}}}} m_{\bar{\nu}_{{b}}} {\Gamma^R_{\hat{\tilde{\chi}}^+_{{i}}\tilde{e}_{{a}}\bar{\nu}_{{b}}}}\Big]+\sum_{a,b=1}^{3}{B_0\Big(m^2_{e_{{b}}},m^2_{\tilde{\nu}_{{a}}}\Big)} {\Gamma^{L*}_{\hat{\tilde{\chi}}^+_{{j}}\tilde{\nu}^*_{{a}}e_{{b}}}} m_{e_{{b}}} {\Gamma^R_{\hat{\tilde{\chi}}^+_{{i}}\tilde{\nu}^*_{{a}}e_{{b}}}} \nonumber \\ 
 & \; \;-4 \sum_{b=1}^{2}{B_0\Big(m^2_{\tilde{\chi}^+_{{b}}},0\Big)} {\Gamma^{R*}_{\hat{\tilde{\chi}}^+_{{j}}\gamma\tilde{\chi}^-_{{b}}}} m_{\tilde{\chi}^-_{{b}}} {\Gamma^L_{\hat{\tilde{\chi}}^+_{{i}}\gamma\tilde{\chi}^-_{{b}}}}  -4 \sum_{b=1}^{2}{B_0\Big(m^2_{\tilde{\chi}^+_{{b}}},m^2_{Z}\Big)} {\Gamma^{R*}_{\hat{\tilde{\chi}}^+_{{j}}Z\tilde{\chi}^-_{{b}}}} m_{\tilde{\chi}^-_{{b}}} {\Gamma^L_{\hat{\tilde{\chi}}^+_{{i}}Z\tilde{\chi}^-_{{b}}}}  \nonumber \\ 
 & \; \;-4 \sum_{b=1}^{5}{B_0\Big(m^2_{\tilde{\chi}^0_{{b}}},m^2_W\Big)} {\Gamma^{R*}_{\hat{\tilde{\chi}}^+_{{j}}W^-\tilde{\chi}^0_{{b}}}} m_{\tilde{\chi}^0_{{b}}} {\Gamma^L_{\hat{\tilde{\chi}}^+_{{i}}W^-\tilde{\chi}^0_{{b}}}}  \\ 
&  \Sigma^L_{\tilde{\chi}^+_i\tilde{\chi}^+_j}(p^2) \ =  \ \sum_{a=1}^{2}\sum_{b=1}^{5}{B_0\Big(m^2_{\tilde{\chi}^0_{{b}}},m^2_{H^+_{{a}}}\Big)} {\Gamma^{L*}_{\hat{\tilde{\chi}}^+_{{j}}H^-_{{a}}\tilde{\chi}^0_{{b}}}} m_{\tilde{\chi}^0_{{b}}} {\Gamma^R_{\hat{\tilde{\chi}}^+_{{i}}H^-_{{a}}\tilde{\chi}^0_{{b}}}} \nonumber \\ 
 & \; \;+\sum_{a=1}^{3}\sum_{b=1}^{2}m_{\tilde{\chi}^-_{{b}}}\Big[{B_0\Big(m^2_{\tilde{\chi}^+_{{b}}},m^2_{A^0_{{a}}}\Big)} {\Gamma^{L*}_{\hat{\tilde{\chi}}^+_{{j}}A^0_{{a}}\tilde{\chi}^-_{{b}}}}  {\Gamma^R_{\hat{\tilde{\chi}}^+_{{i}}A^0_{{a}}\tilde{\chi}^-_{{b}}}} +{B_0\Big(m^2_{\tilde{\chi}^+_{{b}}},m^2_{h_{{a}}}\Big)} {\Gamma^{L*}_{\hat{\tilde{\chi}}^+_{{j}}h_{{a}}\tilde{\chi}^-_{{b}}}} {\Gamma^R_{\hat{\tilde{\chi}}^+_{{i}}h_{{a}}\tilde{\chi}^-_{{b}}}}\Big] \nonumber \\ 
 & \; \;+ \sum_{a=1}^{6}\sum_{b=1}^{3}\Big[3 {B_0\Big(m^2_{d_{{b}}},m^2_{\tilde{u}_{{a}}}\Big)} {\Gamma^{L*}_{\hat{\tilde{\chi}}^+_{{j}}\tilde{u}^*_{{a}}d_{{b}}}} m_{d_{{b}}} {\Gamma^R_{\hat{\tilde{\chi}}^+_{{i}}\tilde{u}^*_{{a}}d_{{b}}}} +3 {B_0\Big(m^2_{{u}_{{b}}},m^2_{\tilde{d}_{{a}}}\Big)} {\Gamma^{L*}_{\hat{\tilde{\chi}}^+_{{j}}\tilde{d}_{{a}}\bar{u}_{{b}}}} m_{\bar{u}_{{b}}} {\Gamma^R_{\hat{\tilde{\chi}}^+_{{i}}\tilde{d}_{{a}}\bar{u}_{{b}}}}  \nonumber \\ 
 & \; \; +{B_0\Big(m^2_{{\nu}_{{b}}},m^2_{\tilde{e}_{{a}}}\Big)} {\Gamma^{L*}_{\hat{\tilde{\chi}}^+_{{j}}\tilde{e}_{{a}}\bar{\nu}_{{b}}}} m_{\bar{\nu}_{{b}}} {\Gamma^R_{\hat{\tilde{\chi}}^+_{{i}}\tilde{e}_{{a}}\bar{\nu}_{{b}}}}\Big] +\sum_{a,b=1}^{3}{B_0\Big(m^2_{e_{{b}}},m^2_{\tilde{\nu}_{{a}}}\Big)} {\Gamma^{L*}_{\hat{\tilde{\chi}}^+_{{j}}\tilde{\nu}^*_{{a}}e_{{b}}}} m_{e_{{b}}} {\Gamma^R_{\hat{\tilde{\chi}}^+_{{i}}\tilde{\nu}^*_{{a}}e_{{b}}}} \nonumber \\ 
 & \; \;-4 \sum_{b=1}^{2}{B_0\Big(m^2_{\tilde{\chi}^+_{{b}}},0\Big)} {\Gamma^{R*}_{\hat{\tilde{\chi}}^+_{{j}}\gamma\tilde{\chi}^-_{{b}}}} m_{\tilde{\chi}^-_{{b}}} {\Gamma^L_{\hat{\tilde{\chi}}^+_{{i}}\gamma\tilde{\chi}^-_{{b}}}}  -4 \sum_{b=1}^{2}{B_0\Big(m^2_{\tilde{\chi}^+_{{b}}},m^2_{Z}\Big)} {\Gamma^{R*}_{\hat{\tilde{\chi}}^+_{{j}}Z\tilde{\chi}^-_{{b}}}} m_{\tilde{\chi}^-_{{b}}} {\Gamma^L_{\hat{\tilde{\chi}}^+_{{i}}Z\tilde{\chi}^-_{{b}}}}  \nonumber \\ 
 & \; \;-4 \sum_{b=1}^{5}{B_0\Big(m^2_{\tilde{\chi}^0_{{b}}},m^2_W\Big)} {\Gamma^{R*}_{\hat{\tilde{\chi}}^+_{{j}}W^-\tilde{\chi}^0_{{b}}}} m_{\tilde{\chi}^0_{{b}}} {\Gamma^L_{\hat{\tilde{\chi}}^+_{{i}}W^-\tilde{\chi}^0_{{b}}}}   
\end{align} 

\subsection{Self-energy of sleptons}
\label{app:SleptonsSelf}

\begin{align} 
& \Pi_{\tilde{e}_i\tilde{e}_j}(p^2) \ =  \ 2 {A_0\Big(m^2_{Z}\Big)} {\Gamma_{\hat{\tilde{e}}^*_{{i}}\hat{\tilde{e}}_{{j}}ZZ}} +4 {A_0\Big(m^2_W\Big)} {\Gamma_{\hat{\tilde{e}}^*_{{i}}\hat{\tilde{e}}_{{j}}W^+W^-}}  - \sum_{a=1}^{2}{A_0\Big(m^2_{H^+_{{a}}}\Big)} {\Gamma_{\hat{\tilde{e}}^*_{{i}}\hat{\tilde{e}}_{{j}}H^+_{{a}}H^-_{{a}}}}  \nonumber \\ 
 & \; \;+\sum_{a=1}^{2}\sum_{b=1}^{3}{B_0\Big(m^2_{H^+_{{a}}},m^2_{\tilde{\nu}_{{b}}}\Big)} {\Gamma^*_{\hat{\tilde{e}}^*_{{j}}H^-_{{a}}\tilde{\nu}_{{b}}}} {\Gamma_{\hat{\tilde{e}}^*_{{i}}H^-_{{a}}\tilde{\nu}_{{b}}}}  - \sum_{a=1}^{6}{A_0\Big(m^2_{\tilde{e}_{{a}}}\Big)} {\Gamma_{\hat{\tilde{e}}^*_{{i}}\hat{\tilde{e}}_{{j}}\tilde{e}^*_{{a}}\tilde{e}_{{a}}}}-3 \sum_{a=1}^{6}{A_0\Big(m^2_{\tilde{u}_{{a}}}\Big)} {\Gamma_{\hat{\tilde{e}}^*_{{i}}\hat{\tilde{e}}_{{j}}\tilde{u}^*_{{a}}\tilde{u}_{{a}}}}   \nonumber \\ 
 & \; \;-2 \sum_{a=1}^{2}m_{\tilde{\chi}^-_{{a}}} \sum_{b=1}^{3}\Big[{B_0\Big(m^2_{\tilde{\chi}^+_{{a}}},0\Big)} m_{\nu_{{b}}} \Big({\Gamma^{L*}_{\hat{\tilde{e}}^*_{{j}}\tilde{\chi}^-_{{a}}\nu_{{b}}}} {\Gamma^R_{\hat{\tilde{e}}^*_{{i}}\tilde{\chi}^-_{{a}}\nu_{{b}}}}   + {\Gamma^{R*}_{\hat{\tilde{e}}^*_{{j}}\tilde{\chi}^-_{{a}}\nu_{{b}}}} {\Gamma^L_{\hat{\tilde{e}}^*_{{i}}\tilde{\chi}^-_{{a}}\nu_{{b}}}} \Big) \Big]  \nonumber \\ 
 & \; \;+\sum_{b=1}^{3}\Big[\sum_{a=1}^{2}{G_0\Big(m^2_{\tilde{\chi}^+_{{a}}},0\Big)} \Big({\Gamma^{L*}_{\hat{\tilde{e}}^*_{{j}}\tilde{\chi}^-_{{a}}\nu_{{b}}}} {\Gamma^L_{\hat{\tilde{e}}^*_{{i}}\tilde{\chi}^-_{{a}}\nu_{{b}}}}  + {\Gamma^{R*}_{\hat{\tilde{e}}^*_{{j}}\tilde{\chi}^-_{{a}}\nu_{{b}}}} {\Gamma^R_{\hat{\tilde{e}}^*_{{i}}\tilde{\chi}^-_{{a}}\nu_{{b}}}} \Big)   +{\Gamma^*_{\hat{\tilde{e}}^*_{{j}}W^-\tilde{\nu}_{{b}}}} {\Gamma_{\hat{\tilde{e}}^*_{{i}}W^-\tilde{\nu}_{{b}}}} {F_0\Big(m^2_{\tilde{\nu}_{{b}}},m^2_W\Big)} \Big] \nonumber \\ 
 & \; \;-\frac{1}{2} \sum_{a=1}^{3}{A_0\Big(m^2_{A^0_{{a}}}\Big)} {\Gamma_{\hat{\tilde{e}}^*_{{i}}\hat{\tilde{e}}_{{j}}A^0_{{a}}A^0_{{a}}}}  - \sum_{a=1}^{3}{A_0\Big(m^2_{\tilde{\nu}_{{a}}}\Big)} {\Gamma_{\hat{\tilde{e}}^*_{{i}}\hat{\tilde{e}}_{{j}}\tilde{\nu}^*_{{a}}\tilde{\nu}_{{a}}}}  -\frac{1}{2} \sum_{a=1}^{3}{A_0\Big(m^2_{h_{{a}}}\Big)} {\Gamma_{\hat{\tilde{e}}^*_{{i}}\hat{\tilde{e}}_{{j}}h_{{a}}h_{{a}}}}  \nonumber \\ 
 & \; \;+\sum_{a=1}^{3}\sum_{b=1}^{6}{B_0\Big(m^2_{A^0_{{a}}},m^2_{\tilde{e}_{{b}}}\Big)} {\Gamma^*_{\hat{\tilde{e}}^*_{{j}}A^0_{{a}}\tilde{e}_{{b}}}} {\Gamma_{\hat{\tilde{e}}^*_{{i}}A^0_{{a}}\tilde{e}_{{b}}}} +\sum_{a=1}^{3}\sum_{b=1}^{6}{B_0\Big(m^2_{h_{{a}}},m^2_{\tilde{e}_{{b}}}\Big)} {\Gamma^*_{\hat{\tilde{e}}^*_{{j}}h_{{a}}\tilde{e}_{{b}}}} {\Gamma_{\hat{\tilde{e}}^*_{{i}}h_{{a}}\tilde{e}_{{b}}}} \nonumber \\ 
 & \; \;-2 \sum_{a=1}^{5}m_{\tilde{\chi}^0_{{a}}} \sum_{b=1}^{3} \Big[{B_0\Big(m^2_{\tilde{\chi}^0_{{a}}},m^2_{e_{{b}}}\Big)} m_{e_{{b}}} \Big({\Gamma^{L*}_{\hat{\tilde{e}}^*_{{j}}\tilde{\chi}^0_{{a}}e_{{b}}}} {\Gamma^R_{\hat{\tilde{e}}^*_{{i}}\tilde{\chi}^0_{{a}}e_{{b}}}} + {\Gamma^{R*}_{\hat{\tilde{e}}^*_{{j}}\tilde{\chi}^0_{{a}}e_{{b}}}} {\Gamma^L_{\hat{\tilde{e}}^*_{{i}}\tilde{\chi}^0_{{a}}e_{{b}}}} \Big)  \Big] \nonumber \\ 
 & \; \;+\sum_{a=1}^{5}\sum_{b=1}^{3} \Big[{G_0\Big(m^2_{\tilde{\chi}^0_{{a}}},m^2_{e_{{b}}}\Big)} \Big({\Gamma^{L*}_{\hat{\tilde{e}}^*_{{j}}\tilde{\chi}^0_{{a}}e_{{b}}}} {\Gamma^L_{\hat{\tilde{e}}^*_{{i}}\tilde{\chi}^0_{{a}}e_{{b}}}} + {\Gamma^{R*}_{\hat{\tilde{e}}^*_{{j}}\tilde{\chi}^0_{{a}}e_{{b}}}} {\Gamma^R_{\hat{\tilde{e}}^*_{{i}}\tilde{\chi}^0_{{a}}e_{{b}}}} \Big) \Big]  -3 \sum_{a=1}^{6}{A_0\Big(m^2_{\tilde{d}_{{a}}}\Big)} {\Gamma_{\hat{\tilde{e}}^*_{{i}}\hat{\tilde{e}}_{{j}}\tilde{d}^*_{{a}}\tilde{d}_{{a}}}} \nonumber \\ 
 & \; \;+\sum_{b=1}^{6}\Big[{\Gamma^*_{\hat{\tilde{e}}^*_{{j}}\gamma\tilde{e}_{{b}}}} {\Gamma_{\hat{\tilde{e}}^*_{{i}}\gamma\tilde{e}_{{b}}}} {F_0\Big(m^2_{\tilde{e}_{{b}}},0\Big)} +{\Gamma^*_{\hat{\tilde{e}}^*_{{j}}Z\tilde{e}_{{b}}}} {\Gamma_{\hat{\tilde{e}}^*_{{i}}Z\tilde{e}_{{b}}}} {F_0\Big(m^2_{\tilde{e}_{{b}}},m^2_{Z}\Big)\Big] }  
\end{align} 

\subsection{Self-energy of sneutrinos}
\label{app:SneutrinoSelf}
\begin{align} 
& \Pi_{\tilde{\nu}_i\tilde{\nu}_j}(p^2) \, = \, 2 {A_0\Big(m^2_{Z}\Big)} {\Gamma_{\hat{\tilde{\nu}}^*_{{i}}\hat{\tilde{\nu}}_{{j}}ZZ}} +4 {A_0\Big(m^2_W\Big)} {\Gamma_{\hat{\tilde{\nu}}^*_{{i}}\hat{\tilde{\nu}}_{{j}}W^+W^-}} -\frac{1}{2} \sum_{a=1}^{3}{A_0\Big(m^2_{A^0_{{a}}}\Big)} {\Gamma_{\hat{\tilde{\nu}}^*_{{i}}\hat{\tilde{\nu}}_{{j}}A^0_{{a}}A^0_{{a}}}}  \nonumber \\
& \; \; -2 \sum_{a=1}^{2}m_{\tilde{\chi}^+_{{a}}} \sum_{b=1}^{3} \Big[{B_0\Big(m^2_{{\tilde{\chi}}^+_{{a}}},m^2_{e_{{b}}}\Big)} m_{e_{{b}}} \Big({\Gamma^{L*}_{\hat{\tilde{\nu}}^*_{{j}}\tilde{\chi}^+_{{a}}e_{{b}}}} {\Gamma^R_{\hat{\tilde{\nu}}^*_{{i}}\tilde{\chi}^+_{{a}}e_{{b}}}} + {\Gamma^{R*}_{\hat{\tilde{\nu}}^*_{{j}}\tilde{\chi}^+_{{a}}e_{{b}}}} {\Gamma^L_{\hat{\tilde{\nu}}^*_{{i}}\tilde{\chi}^+_{{a}}e_{{b}}}} \Big) \Big] \nonumber \\ 
 & \; \;+\sum_{a=1}^{2}\sum_{b=1}^{3} \Big[{G_0\Big(m^2_{{\tilde{\chi}}^+_{{a}}},m^2_{e_{{b}}}\Big)} \Big({\Gamma^{L*}_{\hat{\tilde{\nu}}^*_{{j}}\tilde{\chi}^+_{{a}}e_{{b}}}} {\Gamma^L_{\hat{\tilde{\nu}}^*_{{i}}\tilde{\chi}^+_{{a}}e_{{b}}}}   + {\Gamma^{R*}_{\hat{\tilde{\nu}}^*_{{j}}\tilde{\chi}^+_{{a}}e_{{b}}}} {\Gamma^R_{\hat{\tilde{\nu}}^*_{{i}}\tilde{\chi}^+_{{a}}e_{{b}}}} \Big) \Big] - \sum_{a=1}^{6}{A_0\Big(m^2_{\tilde{e}_{{a}}}\Big)} {\Gamma_{\hat{\tilde{\nu}}^*_{{i}}\hat{\tilde{\nu}}_{{j}}\tilde{e}^*_{{a}}\tilde{e}_{{a}}}} \nonumber \\ 
 & \; \;+\sum_{a=1}^{2}\sum_{b=1}^{6}{B_0\Big(m^2_{H^+_{{a}}},m^2_{\tilde{e}_{{b}}}\Big)} {\Gamma^*_{\hat{\tilde{\nu}}^*_{{j}}H^+_{{a}}\tilde{e}_{{b}}}} {\Gamma_{\hat{\tilde{\nu}}^*_{{i}}H^+_{{a}}\tilde{e}_{{b}}}} - \sum_{a=1}^{3}{A_0\Big(m^2_{\tilde{\nu}_{{a}}}\Big)} {\Gamma_{\hat{\tilde{\nu}}^*_{{i}}\hat{\tilde{\nu}}_{{j}}\tilde{\nu}^*_{{a}}\tilde{\nu}_{{a}}}}- \sum_{a=1}^{2}{A_0\Big(m^2_{H^+_{{a}}}\Big)} {\Gamma_{\hat{\tilde{\nu}}^*_{{i}}\hat{\tilde{\nu}}_{{j}}H^+_{{a}}H^-_{{a}}}}  \nonumber \\ 
 & \; \;-2 \sum_{a=1}^{5}m_{\tilde{\chi}^0_{{a}}} \sum_{b=1}^{3} \Big[{B_0\Big(m^2_{\tilde{\chi}^0_{{a}}},0\Big)} m_{\nu_{{b}}} \Big({\Gamma^{L*}_{\hat{\tilde{\nu}}^*_{{j}}\tilde{\chi}^0_{{a}}\nu_{{b}}}} {\Gamma^R_{\hat{\tilde{\nu}}^*_{{i}}\tilde{\chi}^0_{{a}}\nu_{{b}}}}  + {\Gamma^{R*}_{\hat{\tilde{\nu}}^*_{{j}}\tilde{\chi}^0_{{a}}\nu_{{b}}}} {\Gamma^L_{\hat{\tilde{\nu}}^*_{{i}}\tilde{\chi}^0_{{a}}\nu_{{b}}}} \Big) \Big] \nonumber \\ 
 & \; \;+\sum_{a=1}^{5}\sum_{b=1}^{3} \Big[{G_0\Big(m^2_{\tilde{\chi}^0_{{a}}},0\Big)} \Big({\Gamma^{L*}_{\hat{\tilde{\nu}}^*_{{j}}\tilde{\chi}^0_{{a}}\nu_{{b}}}} {\Gamma^L_{\hat{\tilde{\nu}}^*_{{i}}\tilde{\chi}^0_{{a}}\nu_{{b}}}}   + {\Gamma^{R*}_{\hat{\tilde{\nu}}^*_{{j}}\tilde{\chi}^0_{{a}}\nu_{{b}}}} {\Gamma^R_{\hat{\tilde{\nu}}^*_{{i}}\tilde{\chi}^0_{{a}}\nu_{{b}}}} \Big) \Big]-3 \sum_{a=1}^{6}  \Big[{A_0\Big(m^2_{\tilde{d}_{{a}}}\Big)} {\Gamma_{\hat{\tilde{\nu}}^*_{{i}}\hat{\tilde{\nu}}_{{j}}\tilde{d}^*_{{a}}\tilde{d}_{{a}}}}  \nonumber \\ 
 & \; \;+{A_0\Big(m^2_{\tilde{u}_{{a}}}\Big)} {\Gamma_{\hat{\tilde{\nu}}^*_{{i}}\hat{\tilde{\nu}}_{{j}}\tilde{u}^*_{{a}}\tilde{u}_{{a}}}} \Big] +\sum_{b=1}^{3}{\Gamma^*_{\hat{\tilde{\nu}}^*_{{j}}Z\tilde{\nu}_{{b}}}} {\Gamma_{\hat{\tilde{\nu}}^*_{{i}}Z\tilde{\nu}_{{b}}}} {F_0\Big(m^2_{\tilde{\nu}_{{b}}},m^2_{Z}\Big)}  +\sum_{a,b=1}^{3}{B_0\Big(m^2_{h_{{a}}},m^2_{\tilde{\nu}_{{b}}}\Big)} {\Gamma^*_{\hat{\tilde{\nu}}^*_{{j}}h_{{a}}\tilde{\nu}_{{b}}}} {\Gamma_{\hat{\tilde{\nu}}^*_{{i}}h_{{a}}\tilde{\nu}_{{b}}}}\nonumber \\ 
 & \; \;+\sum_{b=1}^{6}{\Gamma^*_{\hat{\tilde{\nu}}^*_{{j}}W^+\tilde{e}_{{b}}}} {\Gamma_{\hat{\tilde{\nu}}^*_{{i}}W^+\tilde{e}_{{b}}}} {F_0\Big(m^2_{\tilde{e}_{{b}}},m^2_W\Big)} -\frac{1}{2} \sum_{a=1}^{3}{A_0\Big(m^2_{h_{{a}}}\Big)} {\Gamma_{\hat{\tilde{\nu}}^*_{{i}}\hat{\tilde{\nu}}_{{j}}h_{{a}}h_{{a}}}}   
\end{align} 

\subsection{Self-energy of down-squarks}
\label{app:dSquarksSelf}
\begin{align} 
& \Pi_{ij}(p^2) \, = \, +2 {A_0\Big(m^2_{Z}\Big)} {\Gamma_{\check{\tilde{d}}^*_{{i}}\check{\tilde{d}}_{{j}}ZZ}} +4 {A_0\Big(m^2_{W^-}\Big)} {\Gamma_{\check{\tilde{d}}^*_{{i}}\check{\tilde{d}}_{{j}}W^+W^-}} - \sum_{a=1}^{2}{A_0\Big(m^2_{H^-_{{a}}}\Big)} {\Gamma_{\check{\tilde{d}}^*_{{i}}\check{\tilde{d}}_{{j}}H^+_{{a}}H^-_{{a}}}}  \nonumber \\ 
 & \; \;-2 \sum_{a=1}^{2}m_{\tilde{\chi}^-_{{a}}} \sum_{b=1}^{3}{B_0\Big(m^2_{\tilde{\chi}^-_{{a}}},m^2_{u_{{b}}}\Big)} m_{u_{{b}}} \Big({\Gamma^{L*}_{\check{\tilde{d}}^*_{{j}}\tilde{\chi}^-_{{a}}u_{{b}}}} {\Gamma^R_{\check{\tilde{d}}^*_{{i}}\tilde{\chi}^-_{{a}}u_{{b}}}}  + {\Gamma^{R*}_{\check{\tilde{d}}^*_{{j}}\tilde{\chi}^-_{{a}}u_{{b}}}} {\Gamma^L_{\check{\tilde{d}}^*_{{i}}\tilde{\chi}^-_{{a}}u_{{b}}}} \Big)  \nonumber \\ 
 & \; \;+\sum_{a=1}^{2}\sum_{b=1}^{3}{G_0\Big(m^2_{\tilde{\chi}^-_{{a}}},m^2_{u_{{b}}}\Big)} \Big({\Gamma^{L*}_{\check{\tilde{d}}^*_{{j}}\tilde{\chi}^-_{{a}}u_{{b}}}} {\Gamma^L_{\check{\tilde{d}}^*_{{i}}\tilde{\chi}^-_{{a}}u_{{b}}}}  + {\Gamma^{R*}_{\check{\tilde{d}}^*_{{j}}\tilde{\chi}^-_{{a}}u_{{b}}}} {\Gamma^R_{\check{\tilde{d}}^*_{{i}}\tilde{\chi}^-_{{a}}u_{{b}}}} \Big) - \sum_{a=1}^{3}{A_0\Big(m^2_{\tilde{\nu}_{{a}}}\Big)} {\Gamma_{\check{\tilde{d}}^*_{{i}}\check{\tilde{d}}_{{j}}\tilde{\nu}^*_{{a}}\tilde{\nu}_{{a}}}}\nonumber \\ 
 & \; \;+\sum_{a=1}^{2}\sum_{b=1}^{6}{B_0\Big(m^2_{H^-_{{a}}},m^2_{\tilde{u}_{{b}}}\Big)} {\Gamma^*_{\check{\tilde{d}}^*_{{j}}H^-_{{a}}\tilde{u}_{{b}}}} {\Gamma_{\check{\tilde{d}}^*_{{i}}H^-_{{a}}\tilde{u}_{{b}}}} -\frac{1}{2} \sum_{a=1}^{3}{A_0\Big(m^2_{A^0_{{a}}}\Big)} {\Gamma_{\check{\tilde{d}}^*_{{i}}\check{\tilde{d}}_{{j}}A^0_{{a}}A^0_{{a}}}}  \nonumber \\ 
 & \; \;  +\sum_{a=1}^{3}\sum_{b=1}^{6}{B_0\Big(m^2_{A^0_{{a}}},m^2_{\tilde{d}_{{b}}}\Big)} {\Gamma^*_{\check{\tilde{d}}^*_{{j}}A^0_{{a}}\tilde{d}_{{b}}}} {\Gamma_{\check{\tilde{d}}^*_{{i}}A^0_{{a}}\tilde{d}_{{b}}}} +\sum_{a=1}^{3}\sum_{b=1}^{6}{B_0\Big(m^2_{h_{{a}}},m^2_{\tilde{d}_{{b}}}\Big)} {\Gamma^*_{\check{\tilde{d}}^*_{{j}}h_{{a}}\tilde{d}_{{b}}}} {\Gamma_{\check{\tilde{d}}^*_{{i}}h_{{a}}\tilde{d}_{{b}}}} \nonumber \\ 
 & \; \;-\frac{8}{3} m_{\tilde{g}} \sum_{a=1}^{3}{B_0\Big(m^2_{d_{{a}}},m^2_{\tilde{g}}\Big)} m_{d_{{a}}} \Big({\Gamma^{L*}_{\check{\tilde{d}}^*_{{j}}d_{{a}}\tilde{g}_{{1}}}} {\Gamma^R_{\check{\tilde{d}}^*_{{i}}d_{{a}}\tilde{g}_{{1}}}}  + {\Gamma^{R*}_{\check{\tilde{d}}^*_{{j}}d_{{a}}\tilde{g}_{{1}}}} {\Gamma^L_{\check{\tilde{d}}^*_{{i}}d_{{a}}\tilde{g}_{{1}}}} \Big)  -\frac{1}{2} \sum_{a=1}^{3}{A_0\Big(m^2_{h_{{a}}}\Big)} {\Gamma_{\check{\tilde{d}}^*_{{i}}\check{\tilde{d}}_{{j}}h_{{a}}h_{{a}}}}  \nonumber \\ 
 & \; \;+\frac{4}{3} \sum_{a=1}^{3}{G_0\Big(m^2_{d_{{a}}},m^2_{\tilde{g}}\Big)} \Big({\Gamma^{L*}_{\check{\tilde{d}}^*_{{j}}d_{{a}}\tilde{g}_{{1}}}} {\Gamma^L_{\check{\tilde{d}}^*_{{i}}d_{{a}}\tilde{g}_{{1}}}}  + {\Gamma^{R*}_{\check{\tilde{d}}^*_{{j}}d_{{a}}\tilde{g}_{{1}}}} {\Gamma^R_{\check{\tilde{d}}^*_{{i}}d_{{a}}\tilde{g}_{{1}}}} \Big) - \sum_{a=1}^{6}{A_0\Big(m^2_{\tilde{e}_{{a}}}\Big)} {\Gamma_{\check{\tilde{d}}^*_{{i}}\check{\tilde{d}}_{{j}}\tilde{e}^*_{{a}}\tilde{e}_{{a}}}} \nonumber \\ 
 & \; \;-2 \sum_{a=1}^{5}m_{\tilde{\chi}^0_{{a}}} \sum_{b=1}^{3}{B_0\Big(m^2_{\tilde{\chi}^0_{{a}}},m^2_{d_{{b}}}\Big)} m_{d_{{b}}} \Big({\Gamma^{L*}_{\check{\tilde{d}}^*_{{j}}\tilde{\chi}^0_{{a}}d_{{b}}}} {\Gamma^R_{\check{\tilde{d}}^*_{{i}}\tilde{\chi}^0_{{a}}d_{{b}}}}  + {\Gamma^{R*}_{\check{\tilde{d}}^*_{{j}}\tilde{\chi}^0_{{a}}d_{{b}}}} {\Gamma^L_{\check{\tilde{d}}^*_{{i}}\tilde{\chi}^0_{{a}}d_{{b}}}} \Big)  \nonumber \\ 
 & \; \;+\sum_{a=1}^{5}\sum_{b=1}^{3}{G_0\Big(m^2_{\tilde{\chi}^0_{{a}}},m^2_{d_{{b}}}\Big)} \Big({\Gamma^{L*}_{\check{\tilde{d}}^*_{{j}}\tilde{\chi}^0_{{a}}d_{{b}}}} {\Gamma^L_{\check{\tilde{d}}^*_{{i}}\tilde{\chi}^0_{{a}}d_{{b}}}}  + {\Gamma^{R*}_{\check{\tilde{d}}^*_{{j}}\tilde{\chi}^0_{{a}}d_{{b}}}} {\Gamma^R_{\check{\tilde{d}}^*_{{i}}\tilde{\chi}^0_{{a}}d_{{b}}}} \Big)- C \sum_{a=1}^{6}{A_0\Big(m^2_{\tilde{d}_{{a}}}\Big)} {\Gamma_{\check{\tilde{d}}^*_{{i}}\check{\tilde{d}}_{{j}}\tilde{d}^*_{{a}}\tilde{d}_{{a}}}}\nonumber \\ 
 & \; \;+\frac{4}{3} \sum_{b=1}^{6}{\Gamma^*_{\check{\tilde{d}}^*_{{j}}g_{{1}}\tilde{d}_{{b}}}} {\Gamma_{\check{\tilde{d}}^*_{{i}}g_{{1}}\tilde{d}_{{b}}}} {F_0\Big(m^2_{\tilde{d}_{{b}}},0\Big)}  +\sum_{b=1}^{6}{\Gamma^*_{\check{\tilde{d}}^*_{{j}}Z\tilde{d}_{{b}}}} {\Gamma_{\check{\tilde{d}}^*_{{i}}Z\tilde{d}_{{b}}}} {F_0\Big(m^2_{\tilde{d}_{{b}}},m^2_{Z}\Big)}  - C \sum_{a=1}^{6}{A_0\Big(m^2_{\tilde{u}_{{a}}}\Big)} {\Gamma_{\check{\tilde{d}}^*_{{i}}\check{\tilde{d}}_{{j}}\tilde{u}^*_{{a}}\tilde{u}_{{a}}}}\nonumber \\ 
 & \; \;+\sum_{b=1}^{6}{\Gamma^*_{\check{\tilde{d}}^*_{{j}}W^-\tilde{u}_{{b}}}} {\Gamma_{\check{\tilde{d}}^*_{{i}}W^-\tilde{u}_{{b}}}} {F_0\Big(m^2_{\tilde{u}_{{b}}},m^2_{W^-}\Big)}   +\sum_{b=1}^{6}{\Gamma^*_{\check{\tilde{d}}^*_{{j}}\gamma\tilde{d}_{{b}}}} {\Gamma_{\check{\tilde{d}}^*_{{i}}\gamma\tilde{d}_{{b}}}} {F_0\Big(m^2_{\tilde{d}_{{b}}},0\Big)} 
\end{align}

\subsection{Self-energy of up-squarks}
\label{app:uSquarksSelf}
\begin{align} 
& \Pi_{ij}(p^2) \, = \, +2 {A_0\Big(m^2_{Z}\Big)} {\Gamma_{\check{\tilde{u}}^*_{{i}}\check{\tilde{u}}_{{j}}ZZ}} +4 {A_0\Big(m^2_{W^-}\Big)} {\Gamma_{\check{\tilde{u}}^*_{{i}}\check{\tilde{u}}_{{j}}W^+W^-}} - \sum_{a=1}^{2}{A_0\Big(m^2_{H^-_{{a}}}\Big)} {\Gamma_{\check{\tilde{u}}^*_{{i}}\check{\tilde{u}}_{{j}}H^+_{{a}}H^-_{{a}}}}  \nonumber \\ 
 & \; \;-2 \sum_{a=1}^{2}m_{\tilde{\chi}^+_{{a}}} \sum_{b=1}^{3}{B_0\Big(m^2_{\tilde{\chi}^-_{{a}}},m^2_{d_{{b}}}\Big)} m_{d_{{b}}} \Big({\Gamma^{L*}_{\check{\tilde{u}}^*_{{j}}\tilde{\chi}^+_{{a}}d_{{b}}}} {\Gamma^R_{\check{\tilde{u}}^*_{{i}}\tilde{\chi}^+_{{a}}d_{{b}}}}  + {\Gamma^{R*}_{\check{\tilde{u}}^*_{{j}}\tilde{\chi}^+_{{a}}d_{{b}}}} {\Gamma^L_{\check{\tilde{u}}^*_{{i}}\tilde{\chi}^+_{{a}}d_{{b}}}} \Big)  \nonumber \\ 
 & \; \;+\sum_{a=1}^{2}\sum_{b=1}^{3}{G_0\Big(m^2_{\tilde{\chi}^-_{{a}}},m^2_{d_{{b}}}\Big)} \Big({\Gamma^{L*}_{\check{\tilde{u}}^*_{{j}}\tilde{\chi}^+_{{a}}d_{{b}}}} {\Gamma^L_{\check{\tilde{u}}^*_{{i}}\tilde{\chi}^+_{{a}}d_{{b}}}}  + {\Gamma^{R*}_{\check{\tilde{u}}^*_{{j}}\tilde{\chi}^+_{{a}}d_{{b}}}} {\Gamma^R_{\check{\tilde{u}}^*_{{i}}\tilde{\chi}^+_{{a}}d_{{b}}}} \Big)- \sum_{a=1}^{3}{A_0\Big(m^2_{\tilde{\nu}_{{a}}}\Big)} {\Gamma_{\check{\tilde{u}}^*_{{i}}\check{\tilde{u}}_{{j}}\tilde{\nu}^*_{{a}}\tilde{\nu}_{{a}}}}\nonumber \\ 
 & \; \;+\sum_{a=1}^{2}\sum_{b=1}^{6}{B_0\Big(m^2_{H^-_{{a}}},m^2_{\tilde{d}_{{b}}}\Big)} {\Gamma^*_{\check{\tilde{u}}^*_{{j}}H^+_{{a}}\tilde{d}_{{b}}}} {\Gamma_{\check{\tilde{u}}^*_{{i}}H^+_{{a}}\tilde{d}_{{b}}}} -\frac{1}{2} \sum_{a=1}^{3}{A_0\Big(m^2_{A^0_{{a}}}\Big)} {\Gamma_{\check{\tilde{u}}^*_{{i}}\check{\tilde{u}}_{{j}}A^0_{{a}}A^0_{{a}}}} \nonumber \\ 
 & \; \;+\sum_{a=1}^{3}\sum_{b=1}^{6}{B_0\Big(m^2_{A^0_{{a}}},m^2_{\tilde{u}_{{b}}}\Big)} {\Gamma^*_{\check{\tilde{u}}^*_{{j}}A^0_{{a}}\tilde{u}_{{b}}}} {\Gamma_{\check{\tilde{u}}^*_{{i}}A^0_{{a}}\tilde{u}_{{b}}}} +\sum_{a=1}^{3}\sum_{b=1}^{6}{B_0\Big(m^2_{h_{{a}}},m^2_{\tilde{u}_{{b}}}\Big)} {\Gamma^*_{\check{\tilde{u}}^*_{{j}}h_{{a}}\tilde{u}_{{b}}}} {\Gamma_{\check{\tilde{u}}^*_{{i}}h_{{a}}\tilde{u}_{{b}}}} \nonumber \\ 
 & \; \;-\frac{8}{3} m_{\tilde{g}} \sum_{a=1}^{3}{B_0\Big(m^2_{u_{{a}}},m^2_{\tilde{g}}\Big)} m_{u_{{a}}} \Big({\Gamma^{L*}_{\check{\tilde{u}}^*_{{j}}u_{{a}}\tilde{g}_{{1}}}} {\Gamma^R_{\check{\tilde{u}}^*_{{i}}u_{{a}}\tilde{g}_{{1}}}}  + {\Gamma^{R*}_{\check{\tilde{u}}^*_{{j}}u_{{a}}\tilde{g}_{{1}}}} {\Gamma^L_{\check{\tilde{u}}^*_{{i}}u_{{a}}\tilde{g}_{{1}}}} \Big) \nonumber \\ 
 & \; \;+\frac{4}{3} \sum_{a=1}^{3}{G_0\Big(m^2_{u_{{a}}},m^2_{\tilde{g}}\Big)} \Big({\Gamma^{L*}_{\check{\tilde{u}}^*_{{j}}u_{{a}}\tilde{g}_{{1}}}} {\Gamma^L_{\check{\tilde{u}}^*_{{i}}u_{{a}}\tilde{g}_{{1}}}}  + {\Gamma^{R*}_{\check{\tilde{u}}^*_{{j}}u_{{a}}\tilde{g}_{{1}}}} {\Gamma^R_{\check{\tilde{u}}^*_{{i}}u_{{a}}\tilde{g}_{{1}}}} \Big)  -\frac{1}{2} \sum_{a=1}^{3}{A_0\Big(m^2_{h_{{a}}}\Big)} {\Gamma_{\check{\tilde{u}}^*_{{i}}\check{\tilde{u}}_{{j}}h_{{a}}h_{{a}}}}  \nonumber \\ 
 & \; \;-2 \sum_{a=1}^{5}m_{\tilde{\chi}^0_{{a}}} \sum_{b=1}^{3}{B_0\Big(m^2_{\tilde{\chi}^0_{{a}}},m^2_{u_{{b}}}\Big)} m_{u_{{b}}} \Big({\Gamma^{L*}_{\check{\tilde{u}}^*_{{j}}\tilde{\chi}^0_{{a}}u_{{b}}}} {\Gamma^R_{\check{\tilde{u}}^*_{{i}}\tilde{\chi}^0_{{a}}u_{{b}}}}  + {\Gamma^{R*}_{\check{\tilde{u}}^*_{{j}}\tilde{\chi}^0_{{a}}u_{{b}}}} {\Gamma^L_{\check{\tilde{u}}^*_{{i}}\tilde{\chi}^0_{{a}}u_{{b}}}} \Big)- \sum_{a=1}^{6}{A_0\Big(m^2_{\tilde{e}_{{a}}}\Big)} {\Gamma_{\check{\tilde{u}}^*_{{i}}\check{\tilde{u}}_{{j}}\tilde{e}^*_{{a}}\tilde{e}_{{a}}}}  \nonumber \\ 
 & \; \;+\sum_{a=1}^{5}\sum_{b=1}^{3}{G_0\Big(m^2_{\tilde{\chi}^0_{{a}}},m^2_{u_{{b}}}\Big)} \Big({\Gamma^{L*}_{\check{\tilde{u}}^*_{{j}}\tilde{\chi}^0_{{a}}u_{{b}}}} {\Gamma^L_{\check{\tilde{u}}^*_{{i}}\tilde{\chi}^0_{{a}}u_{{b}}}}  + {\Gamma^{R*}_{\check{\tilde{u}}^*_{{j}}\tilde{\chi}^0_{{a}}u_{{b}}}} {\Gamma^R_{\check{\tilde{u}}^*_{{i}}\tilde{\chi}^0_{{a}}u_{{b}}}} \Big)- C \sum_{a=1}^{6}{A_0\Big(m^2_{\tilde{d}_{{a}}}\Big)} {\Gamma_{\check{\tilde{u}}^*_{{i}}\check{\tilde{u}}_{{j}}\tilde{d}^*_{{a}}\tilde{d}_{{a}}}}\nonumber \\ 
 & \; \;+\sum_{b=1}^{6}{\Gamma^*_{\check{\tilde{u}}^*_{{j}}\gamma\tilde{u}_{{b}}}} {\Gamma_{\check{\tilde{u}}^*_{{i}}\gamma\tilde{u}_{{b}}}} {F_0\Big(m^2_{\tilde{u}_{{b}}},0\Big)} +\frac{4}{3} \sum_{b=1}^{6}{\Gamma^*_{\check{\tilde{u}}^*_{{j}}g_{{1}}\tilde{u}_{{b}}}} {\Gamma_{\check{\tilde{u}}^*_{{i}}g_{{1}}\tilde{u}_{{b}}}} {F_0\Big(m^2_{\tilde{u}_{{b}}},0\Big)}  - C \sum_{a=1}^{6}{A_0\Big(m^2_{\tilde{u}_{{a}}}\Big)} {\Gamma_{\check{\tilde{u}}^*_{{i}}\check{\tilde{u}}_{{j}}\tilde{u}^*_{{a}}\tilde{u}_{{a}}}} \nonumber \\ 
 & \; \;+\sum_{b=1}^{6}{\Gamma^*_{\check{\tilde{u}}^*_{{j}}Z\tilde{u}_{{b}}}} {\Gamma_{\check{\tilde{u}}^*_{{i}}Z\tilde{u}_{{b}}}} {F_0\Big(m^2_{\tilde{u}_{{b}}},m^2_{Z}\Big)}  +\sum_{b=1}^{6}{\Gamma^*_{\check{\tilde{u}}^*_{{j}}W^+\tilde{d}_{{b}}}} {\Gamma_{\check{\tilde{u}}^*_{{i}}W^+\tilde{d}_{{b}}}} {F_0\Big(m^2_{\tilde{d}_{{b}}},m^2_{W^-}\Big)} 
\end{align} 

\subsection{Self-energy of up-quarks}
\label{app:SEup}
\begin{align} 
&  \Sigma^S_{ij}(p^2) \, = \, +\sum_{b=1}^{3}\Big[\sum_{a=1}^{2}{B_0\Big(m^2_{d_{{b}}},m^2_{H^-_{{a}}}\Big)} {\Gamma^{L*}_{\hat{\bar{u}}_{{j}}H^+_{{a}}d_{{b}}}} m_{d_{{b}}} {\Gamma^R_{\hat{\bar{u}}_{{i}}H^+_{{a}}d_{{b}}}} -\frac{16}{3} {B_0\Big(m^2_{u_{{b}}},0\Big)} {\Gamma^{R*}_{\hat{\bar{u}}_{{j}}G_{{1}}u_{{b}}}} m_{u_{{b}}} {\Gamma^L_{\hat{\bar{u}}_{{i}}G_{{1}}u_{{b}}}} \Big] \nonumber \\ 
 & \; \;+\sum_{a,b=1}^{3} m_{u_{{b}}}\Big[{B_0\Big(m^2_{u_{{b}}},m^2_{A^0_{{a}}}\Big)} {\Gamma^{L*}_{\hat{\bar{u}}_{{j}}A^0_{{a}}u_{{b}}}} {\Gamma^R_{\hat{\bar{u}}_{{i}}A^0_{{a}}u_{{b}}}}+{B_0\Big(m^2_{u_{{b}}},m^2_{h_{{a}}}\Big)} {\Gamma^{L*}_{\hat{\bar{u}}_{{j}}h_{{a}}u_{{b}}}} {\Gamma^R_{\hat{\bar{u}}_{{i}}h_{{a}}u_{{b}}}}\Big] \nonumber \\ 
 & \; \;-4 \sum_{b=1}^{3}\Big[{B_0\Big(m^2_{u_{{b}}},0\Big)} {\Gamma^{R*}_{\hat{\bar{u}}_{{j}}\gamma u_{{b}}}} m_{u_{{b}}} {\Gamma^L_{\hat{\bar{u}}_{{i}}\gamma u_{{b}}}} +{B_0\Big(m^2_{u_{{b}}},m^2_{Z}\Big)} {\Gamma^{R*}_{\hat{\bar{u}}_{{j}}Zu_{{b}}}} m_{u_{{b}}} {\Gamma^L_{\hat{\bar{u}}_{{i}}Zu_{{b}}}}\nonumber \\ 
 & \; \;+{B_0\Big(m^2_{d_{{b}}},m^2_{W^-}\Big)} {\Gamma^{R*}_{\hat{\bar{u}}_{{j}}W^+d_{{b}}}} m_{d_{{b}}} {\Gamma^L_{\hat{\bar{u}}_{{i}}W^+d_{{b}}}}\Big]+\sum_{a=1}^{6}\sum_{b=1}^{2}{B_0\Big(m^2_{\tilde{\chi}^+_{{b}}},m^2_{\tilde{d}_{{a}}}\Big)} {\Gamma^{L*}_{\hat{\bar{u}}_{{j}}\tilde{d}_{{a}}\tilde{\chi}^-_{{b}}}} m_{\tilde{\chi}^-_{{b}}} {\Gamma^R_{\hat{\bar{u}}_{{i}}\tilde{d}_{{a}}\tilde{\chi}^-_{{b}}}}  \nonumber \\ 
 & \; \; +\sum_{a=1}^{6}\Big[\sum_{b=1}^{5}{B_0\Big(m^2_{\tilde{\chi}^0_{{b}}},m^2_{\tilde{u}_{{a}}}\Big)} {\Gamma^{L*}_{\hat{\bar{u}}_{{j}}\tilde{u}_{{a}}\tilde{\chi}^0_{{b}}}} m_{\tilde{\chi}^0_{{b}}} {\Gamma^R_{\hat{\bar{u}}_{{i}}\tilde{u}_{{a}}\tilde{\chi}^0_{{b}}}} +\frac{4}{3} m_{\tilde{g}} {B_0\Big(m^2_{\tilde{g}},m^2_{\tilde{u}_{{a}}}\Big)} {\Gamma^{L*}_{\hat{\bar{u}}_{{j}}\tilde{u}_{{a}}\tilde{g}_{{1}}}} {\Gamma^R_{\hat{\bar{u}}_{{i}}\tilde{u}_{{a}}\tilde{g}_{{1}}}} \Big]  \\ 
&  \Sigma^R_{ij}(p^2) \, = \, +\sum_{b=1}^{3}\Big[\sum_{a=1}^{2}{B_0\Big(m^2_{d_{{b}}},m^2_{H^-_{{a}}}\Big)} {\Gamma^{L*}_{\hat{\bar{u}}_{{j}}H^+_{{a}}d_{{b}}}} m_{d_{{b}}} {\Gamma^R_{\hat{\bar{u}}_{{i}}H^+_{{a}}d_{{b}}}} -\frac{16}{3} {B_0\Big(m^2_{u_{{b}}},0\Big)} {\Gamma^{R*}_{\hat{\bar{u}}_{{j}}G_{{1}}u_{{b}}}} m_{u_{{b}}} {\Gamma^L_{\hat{\bar{u}}_{{i}}G_{{1}}u_{{b}}}}\Big] \nonumber \\ 
 & \; \;+\sum_{a,b=1}^{3}m_{u_{{b}}}\Big[{B_0\Big(m^2_{u_{{b}}},m^2_{A^0_{{a}}}\Big)} {\Gamma^{L*}_{\hat{\bar{u}}_{{j}}A^0_{{a}}u_{{b}}}}  {\Gamma^R_{\hat{\bar{u}}_{{i}}A^0_{{a}}u_{{b}}}} +{B_0\Big(m^2_{u_{{b}}},m^2_{h_{{a}}}\Big)} {\Gamma^{L*}_{\hat{\bar{u}}_{{j}}h_{{a}}u_{{b}}}}  {\Gamma^R_{\hat{\bar{u}}_{{i}}h_{{a}}u_{{b}}}} \Big] \nonumber \\ 
 & \; \;-4\Big[ \sum_{b=1}^{3}{B_0\Big(m^2_{u_{{b}}},0\Big)} {\Gamma^{R*}_{\hat{\bar{u}}_{{j}}\gamma u_{{b}}}} m_{u_{{b}}} {\Gamma^L_{\hat{\bar{u}}_{{i}}\gamma u_{{b}}}}  +{B_0\Big(m^2_{u_{{b}}},m^2_{Z}\Big)} {\Gamma^{R*}_{\hat{\bar{u}}_{{j}}Zu_{{b}}}} m_{u_{{b}}} {\Gamma^L_{\hat{\bar{u}}_{{i}}Zu_{{b}}}}  \nonumber \\ 
 & \; \;+{B_0\Big(m^2_{d_{{b}}},m^2_{W^-}\Big)} {\Gamma^{R*}_{\hat{\bar{u}}_{{j}}W^+d_{{b}}}} m_{d_{{b}}} {\Gamma^L_{\hat{\bar{u}}_{{i}}W^+d_{{b}}}}\Big]+\sum_{a=1}^{6}\sum_{b=1}^{2}{B_0\Big(m^2_{\tilde{\chi}^+_{{b}}},m^2_{\tilde{d}_{{a}}}\Big)} {\Gamma^{L*}_{\hat{\bar{u}}_{{j}}\tilde{d}_{{a}}\tilde{\chi}^-_{{b}}}} m_{\tilde{\chi}^-_{{b}}} {\Gamma^R_{\hat{\bar{u}}_{{i}}\tilde{d}_{{a}}\tilde{\chi}^-_{{b}}}}    \nonumber \\ 
 & \; \;+\sum_{a=1}^{6}\Big[\sum_{b=1}^{5}{B_0\Big(m^2_{\tilde{\chi}^0_{{b}}},m^2_{\tilde{u}_{{a}}}\Big)} {\Gamma^{L*}_{\hat{\bar{u}}_{{j}}\tilde{u}_{{a}}\tilde{\chi}^0_{{b}}}} m_{\tilde{\chi}^0_{{b}}} {\Gamma^R_{\hat{\bar{u}}_{{i}}\tilde{u}_{{a}}\tilde{\chi}^0_{{b}}}} +\frac{4}{3} m_{\tilde{g}} {B_0\Big(m^2_{\tilde{g}},m^2_{\tilde{u}_{{a}}}\Big)} {\Gamma^{L*}_{\hat{\bar{u}}_{{j}}\tilde{u}_{{a}}\tilde{g}_{{1}}}} {\Gamma^R_{\hat{\bar{u}}_{{i}}\tilde{u}_{{a}}\tilde{g}_{{1}}}} \Big]  \\ 
&  \Sigma^L_{ij}(p^2) \, = \, +\sum_{b=1}^{3}\Big[\sum_{a=1}^{2}{B_0\Big(m^2_{d_{{b}}},m^2_{H^-_{{a}}}\Big)} {\Gamma^{L*}_{\hat{\bar{u}}_{{j}}H^+_{{a}}d_{{b}}}} m_{d_{{b}}} {\Gamma^R_{\hat{\bar{u}}_{{i}}H^+_{{a}}d_{{b}}}}  -\frac{16}{3} {B_0\Big(m^2_{u_{{b}}},0\Big)} {\Gamma^{R*}_{\hat{\bar{u}}_{{j}}G_{{1}}u_{{b}}}} m_{u_{{b}}} {\Gamma^L_{\hat{\bar{u}}_{{i}}G_{{1}}u_{{b}}}} \Big] \nonumber \\ 
 & \; \;+\sum_{a,b=1}^{3} m_{u_{{b}}}\Big[{B_0\Big(m^2_{u_{{b}}},m^2_{A^0_{{a}}}\Big)} {\Gamma^{L*}_{\hat{\bar{u}}_{{j}}A^0_{{a}}u_{{b}}}} {\Gamma^R_{\hat{\bar{u}}_{{i}}A^0_{{a}}u_{{b}}}} +{B_0\Big(m^2_{u_{{b}}},m^2_{h_{{a}}}\Big)} {\Gamma^{L*}_{\hat{\bar{u}}_{{j}}h_{{a}}u_{{b}}}} {\Gamma^R_{\hat{\bar{u}}_{{i}}h_{{a}}u_{{b}}}}\Big] \nonumber \\ 
 & \; \; -4 \sum_{b=1}^{3}\Big[{B_0\Big(m^2_{u_{{b}}},0\Big)} {\Gamma^{R*}_{\hat{\bar{u}}_{{j}}\gamma u_{{b}}}} m_{u_{{b}}} {\Gamma^L_{\hat{\bar{u}}_{{i}}\gamma u_{{b}}}}  +{B_0\Big(m^2_{u_{{b}}},m^2_{Z}\Big)} {\Gamma^{R*}_{\hat{\bar{u}}_{{j}}Zu_{{b}}}} m_{u_{{b}}} {\Gamma^L_{\hat{\bar{u}}_{{i}}Zu_{{b}}}} \nonumber \\ 
 & \; \;+{B_0\Big(m^2_{d_{{b}}},m^2_{W^-}\Big)} {\Gamma^{R*}_{\hat{\bar{u}}_{{j}}W^+d_{{b}}}} m_{d_{{b}}} {\Gamma^L_{\hat{\bar{u}}_{{i}}W^+d_{{b}}}}\Big]+\sum_{a=1}^{6}\sum_{b=1}^{2}{B_0\Big(m^2_{\tilde{\chi}^+_{{b}}},m^2_{\tilde{d}_{{a}}}\Big)} {\Gamma^{L*}_{\hat{\bar{u}}_{{j}}\tilde{d}_{{a}}\tilde{\chi}^-_{{b}}}} m_{\tilde{\chi}^-_{{b}}} {\Gamma^R_{\hat{\bar{u}}_{{i}}\tilde{d}_{{a}}\tilde{\chi}^-_{{b}}}}   \nonumber \\ 
 & \; \;+\sum_{a=1}^{6}\Big[\sum_{b=1}^{5}{B_0\Big(m^2_{\tilde{\chi}^0_{{b}}},m^2_{\tilde{u}_{{a}}}\Big)} {\Gamma^{L*}_{\hat{\bar{u}}_{{j}}\tilde{u}_{{a}}\tilde{\chi}^0_{{b}}}} m_{\tilde{\chi}^0_{{b}}} {\Gamma^R_{\hat{\bar{u}}_{{i}}\tilde{u}_{{a}}\tilde{\chi}^0_{{b}}}}   +\frac{4}{3} m_{\tilde{g}} {B_0\Big(m^2_{\tilde{g}},m^2_{\tilde{u}_{{a}}}\Big)} {\Gamma^{L*}_{\hat{\bar{u}}_{{j}}\tilde{u}_{{a}}\tilde{g}_{{1}}}} {\Gamma^R_{\hat{\bar{u}}_{{i}}\tilde{u}_{{a}}\tilde{g}_{{1}}}} \Big]
\end{align} 

\subsection{Self-energy of down-quarks}
\label{app:SEdown}
\begin{align} 
&  \Sigma^S_{ij}(p^2) \, = \, +\sum_{b=1}^{3}\Big[\sum_{a=1}^{2}{B_0\Big(m^2_{u_{{b}}},m^2_{H^-_{{a}}}\Big)} {\Gamma^{L*}_{\hat{\bar{d}}_{{j}}H^-_{{a}}u_{{b}}}} m_{u_{{b}}} {\Gamma^R_{\hat{\bar{d}}_{{i}}H^-_{{a}}u_{{b}}}}  -\frac{16}{3} {B_0\Big(m^2_{d_{{b}}},0\Big)} {\Gamma^{R*}_{\hat{\bar{d}}_{{j}}G_{{1}}d_{{b}}}} m_{d_{{b}}} {\Gamma^L_{\hat{\bar{d}}_{{i}}G_{{1}}d_{{b}}}}\Big] \nonumber \\ 
 & \; \;+\sum_{a,b=1}^{3}m_{d_{{b}}}\Big[{B_0\Big(m^2_{d_{{b}}},m^2_{A^0_{{a}}}\Big)} {\Gamma^{L*}_{\hat{\bar{d}}_{{j}}A^0_{{a}}d_{{b}}}}  {\Gamma^R_{\hat{\bar{d}}_{{i}}A^0_{{a}}d_{{b}}}} +{B_0\Big(m^2_{d_{{b}}},m^2_{h_{{a}}}\Big)} {\Gamma^{L*}_{\hat{\bar{d}}_{{j}}h_{{a}}d_{{b}}}}  {\Gamma^R_{\hat{\bar{d}}_{{i}}h_{{a}}d_{{b}}}} \Big] \nonumber \\ 
 & \; \; -4 \sum_{b=1}^{3}\Big[{B_0\Big(m^2_{d_{{b}}},0\Big)} {\Gamma^{R*}_{\hat{\bar{d}}_{{j}}\gamma d_{{b}}}} m_{d_{{b}}} {\Gamma^L_{\hat{\bar{d}}_{{i}}\gamma d_{{b}}}} +{B_0\Big(m^2_{u_{{b}}},m^2_{W^-}\Big)} {\Gamma^{R*}_{\hat{\bar{d}}_{{j}}W^-u_{{b}}}} m_{u_{{b}}} {\Gamma^L_{\hat{\bar{d}}_{{i}}W^-u_{{b}}}} \nonumber \\ 
 & \; \;+{B_0\Big(m^2_{d_{{b}}},m^2_{Z}\Big)} {\Gamma^{R*}_{\hat{\bar{d}}_{{j}}Zd_{{b}}}} m_{d_{{b}}} {\Gamma^L_{\hat{\bar{d}}_{{i}}Zd_{{b}}}}\Big]+\sum_{a=1}^{6}\sum_{b=1}^{2}{B_0\Big(m^2_{\tilde{\chi}^+_{{b}}},m^2_{\tilde{u}_{{a}}}\Big)} {\Gamma^{L*}_{\hat{\bar{d}}_{{j}}\tilde{u}_{{a}}\tilde{\chi}^+_{{b}}}} m_{\tilde{\chi}^+_{{b}}} {\Gamma^R_{\hat{\bar{d}}_{{i}}\tilde{u}_{{a}}\tilde{\chi}^+_{{b}}}}   \nonumber \\
& \; \; +\sum_{a=1}^{6}\Big[\sum_{b=1}^{5}{B_0\Big(m^2_{\tilde{\chi}^0_{{b}}},m^2_{\tilde{d}_{{a}}}\Big)} {\Gamma^{L*}_{\hat{\bar{d}}_{{j}}\tilde{d}_{{a}}\tilde{\chi}^0_{{b}}}} m_{\tilde{\chi}^0_{{b}}} {\Gamma^R_{\hat{\bar{d}}_{{i}}\tilde{d}_{{a}}\tilde{\chi}^0_{{b}}}}+\frac{4}{3} m_{\tilde{g}} {B_0\Big(m^2_{\tilde{g}},m^2_{\tilde{d}_{{a}}}\Big)} {\Gamma^{L*}_{\hat{\bar{d}}_{{j}}\tilde{d}_{{a}}\tilde{g}_{{1}}}} {\Gamma^R_{\hat{\bar{d}}_{{i}}\tilde{d}_{{a}}\tilde{g}_{{1}}}}\Big]  \\ 
&  \Sigma^R_{ij}(p^2)\, = \, +\sum_{b=1}^{3}\Big[\sum_{a=1}^{2}{B_0\Big(m^2_{u_{{b}}},m^2_{H^-_{{a}}}\Big)} {\Gamma^{L*}_{\hat{\bar{d}}_{{j}}H^-_{{a}}u_{{b}}}} m_{u_{{b}}} {\Gamma^R_{\hat{\bar{d}}_{{i}}H^-_{{a}}u_{{b}}}} -\frac{16}{3} {B_0\Big(m^2_{d_{{b}}},0\Big)} {\Gamma^{R*}_{\hat{\bar{d}}_{{j}}G_{{1}}d_{{b}}}} m_{d_{{b}}} {\Gamma^L_{\hat{\bar{d}}_{{i}}G_{{1}}d_{{b}}}} \Big] \nonumber \\ 
 & \; \;+\sum_{a,b=1}^{3}m_{d_{{b}}}\Big[{B_0\Big(m^2_{d_{{b}}},m^2_{A^0_{{a}}}\Big)} {\Gamma^{L*}_{\hat{\bar{d}}_{{j}}A^0_{{a}}d_{{b}}}}  {\Gamma^R_{\hat{\bar{d}}_{{i}}A^0_{{a}}d_{{b}}}} +{B_0\Big(m^2_{d_{{b}}},m^2_{h_{{a}}}\Big)} {\Gamma^{L*}_{\hat{\bar{d}}_{{j}}h_{{a}}d_{{b}}}} {\Gamma^R_{\hat{\bar{d}}_{{i}}h_{{a}}d_{{b}}}}\Big] \nonumber \\ 
 & \; \; -4 \sum_{b=1}^{3}\Big[{B_0\Big(m^2_{d_{{b}}},0\Big)} {\Gamma^{R*}_{\hat{\bar{d}}_{{j}}\gamma d_{{b}}}} m_{d_{{b}}} {\Gamma^L_{\hat{\bar{d}}_{{i}}\gamma d_{{b}}}}  +{B_0\Big(m^2_{u_{{b}}},m^2_{W^-}\Big)} {\Gamma^{R*}_{\hat{\bar{d}}_{{j}}W^-u_{{b}}}} m_{u_{{b}}} {\Gamma^L_{\hat{\bar{d}}_{{i}}W^-u_{{b}}}} \nonumber \\ 
 & \; \;+{B_0\Big(m^2_{d_{{b}}},m^2_{Z}\Big)} {\Gamma^{R*}_{\hat{\bar{d}}_{{j}}Zd_{{b}}}} m_{d_{{b}}} {\Gamma^L_{\hat{\bar{d}}_{{i}}Zd_{{b}}}}\Big]  +\sum_{a=1}^{6}\sum_{b=1}^{2}{B_0\Big(m^2_{\tilde{\chi}^+_{{b}}},m^2_{\tilde{u}_{{a}}}\Big)} {\Gamma^{L*}_{\hat{\bar{d}}_{{j}}\tilde{u}_{{a}}\tilde{\chi}^+_{{b}}}} m_{\tilde{\chi}^+_{{b}}} {\Gamma^R_{\hat{\bar{d}}_{{i}}\tilde{u}_{{a}}\tilde{\chi}^+_{{b}}}}  \nonumber \\ 
 & \; \;+\sum_{a=1}^{6}\Big[\sum_{b=1}^{5}{B_0\Big(m^2_{\tilde{\chi}^0_{{b}}},m^2_{\tilde{d}_{{a}}}\Big)} {\Gamma^{L*}_{\hat{\bar{d}}_{{j}}\tilde{d}_{{a}}\tilde{\chi}^0_{{b}}}} m_{\tilde{\chi}^0_{{b}}} {\Gamma^R_{\hat{\bar{d}}_{{i}}\tilde{d}_{{a}}\tilde{\chi}^0_{{b}}}} +\frac{4}{3} m_{\tilde{g}} {B_0\Big(m^2_{\tilde{g}},m^2_{\tilde{d}_{{a}}}\Big)} {\Gamma^{L*}_{\hat{\bar{d}}_{{j}}\tilde{d}_{{a}}\tilde{g}_{{1}}}} {\Gamma^R_{\hat{\bar{d}}_{{i}}\tilde{d}_{{a}}\tilde{g}_{{1}}}} \Big] \\ 
&  \Sigma^L_{ij}(p^2)\, = \,\sum_{b=1}^{3}\Big[\sum_{a=1}^{2}{B_0\Big(m^2_{u_{{b}}},m^2_{H^-_{{a}}}\Big)} {\Gamma^{L*}_{\hat{\bar{d}}_{{j}}H^-_{{a}}u_{{b}}}} m_{u_{{b}}} {\Gamma^R_{\hat{\bar{d}}_{{i}}H^-_{{a}}u_{{b}}}} -\frac{16}{3} {B_0\Big(m^2_{d_{{b}}},0\Big)} {\Gamma^{R*}_{\hat{\bar{d}}_{{j}}G_{{1}}d_{{b}}}} m_{d_{{b}}} {\Gamma^L_{\hat{\bar{d}}_{{i}}G_{{1}}d_{{b}}}}\Big]  \nonumber \\ 
 & \; \;+\sum_{a,b=1}^{3} m_{d_{{b}}}\Big[{B_0\Big(m^2_{d_{{b}}},m^2_{A^0_{{a}}}\Big)} {\Gamma^{L*}_{\hat{\bar{d}}_{{j}}A^0_{{a}}d_{{b}}}} {\Gamma^R_{\hat{\bar{d}}_{{i}}A^0_{{a}}d_{{b}}}}+{B_0\Big(m^2_{d_{{b}}},m^2_{h_{{a}}}\Big)} {\Gamma^{L*}_{\hat{\bar{d}}_{{j}}h_{{a}}d_{{b}}}}{\Gamma^R_{\hat{\bar{d}}_{{i}}h_{{a}}d_{{b}}}}\Big] \nonumber \\ 
 & \; \; -4 \sum_{b=1}^{3}\Big[{B_0\Big(m^2_{d_{{b}}},0\Big)} {\Gamma^{R*}_{\hat{\bar{d}}_{{j}}\gamma d_{{b}}}} m_{d_{{b}}} {\Gamma^L_{\hat{\bar{d}}_{{i}}\gamma d_{{b}}}} +{B_0\Big(m^2_{u_{{b}}},m^2_{W^-}\Big)} {\Gamma^{R*}_{\hat{\bar{d}}_{{j}}W^-u_{{b}}}} m_{u_{{b}}} {\Gamma^L_{\hat{\bar{d}}_{{i}}W^-u_{{b}}}} \nonumber \\ 
 & \; \;  +{B_0\Big(m^2_{d_{{b}}},m^2_{Z}\Big)} {\Gamma^{R*}_{\hat{\bar{d}}_{{j}}Zd_{{b}}}} m_{d_{{b}}} {\Gamma^L_{\hat{\bar{d}}_{{i}}Zd_{{b}}}}\Big] +\sum_{a=1}^{6}\sum_{b=1}^{2}{B_0\Big(m^2_{\tilde{\chi}^+_{{b}}},m^2_{\tilde{u}_{{a}}}\Big)} {\Gamma^{L*}_{\hat{\bar{d}}_{{j}}\tilde{u}_{{a}}\tilde{\chi}^+_{{b}}}} m_{\tilde{\chi}^+_{{b}}} {\Gamma^R_{\hat{\bar{d}}_{{i}}\tilde{u}_{{a}}\tilde{\chi}^+_{{b}}}}   \nonumber \\ 
 & \; \; +\sum_{a=1}^{6}\Big[\sum_{b=1}^{5}{B_0\Big(m^2_{\tilde{\chi}^0_{{b}}},m^2_{\tilde{d}_{{a}}}\Big)} {\Gamma^{L*}_{\hat{\bar{d}}_{{j}}\tilde{d}_{{a}}\tilde{\chi}^0_{{b}}}} m_{\tilde{\chi}^0_{{b}}} {\Gamma^R_{\hat{\bar{d}}_{{i}}\tilde{d}_{{a}}\tilde{\chi}^0_{{b}}}} +\frac{4}{3} m_{\tilde{g}}{B_0\Big(m^2_{\tilde{g}},m^2_{\tilde{d}_{{a}}}\Big)} {\Gamma^{L*}_{\hat{\bar{d}}_{{j}}\tilde{d}_{{a}}\tilde{g}_{{1}}}} {\Gamma^R_{\hat{\bar{d}}_{{i}}\tilde{d}_{{a}}\tilde{g}_{{1}}}} \Big]
\end{align} 

\subsection{Self-energy of leptons}
\label{app:SElep}
\begin{align} 
&  \Sigma^S_{ij}(p^2) \, = \, +\sum_{a=1}^{2}\sum_{b=1}^{3}\Big[{B_0\Big(m^2_{\nu_{{b}}},m^2_{H^-_{{a}}}\Big)} {\Gamma^{L*}_{\hat{\bar{e}}_{{j}}H^-_{{a}}\nu_{{b}}}} m_{\nu_{{b}}} {\Gamma^R_{\hat{\bar{e}}_{{i}}H^-_{{a}}\nu_{{b}}}} +{B_0\Big(m^2_{\tilde{\chi}^+_{{a}}},m^2_{\tilde{\nu}_{{b}}}\Big)} {\Gamma^{L*}_{\hat{\bar{e}}_{{j}}\tilde{\nu}_{{b}}\tilde{\chi}^+_{{a}}}} m_{\tilde{\chi}^+_{{a}}} {\Gamma^R_{\hat{\bar{e}}_{{i}}\tilde{\nu}_{{b}}\tilde{\chi}^+_{{a}}}} \Big] \nonumber \\ 
 & \; \;+\sum_{a,b=1}^{3}m_{e_{{b}}}\Big[{B_0\Big(m^2_{e_{{b}}},m^2_{A^0_{{a}}}\Big)} {\Gamma^{L*}_{\hat{\bar{e}}_{{j}}A^0_{{a}}e_{{b}}}}  {\Gamma^R_{\hat{\bar{e}}_{{i}}A^0_{{a}}e_{{b}}}}+{B_0\Big(m^2_{e_{{b}}},m^2_{h_{{a}}}\Big)} {\Gamma^{L*}_{\hat{\bar{e}}_{{j}}h_{{a}}e_{{b}}}} {\Gamma^R_{\hat{\bar{e}}_{{i}}h_{{a}}e_{{b}}}}\Big]  \nonumber \\ 
 & \; \;-4 \sum_{b=1}^{3}\Big[{B_0\Big(m^2_{e_{{b}}},0\Big)} {\Gamma^{R*}_{\hat{\bar{e}}_{{j}}\gamma e_{{b}}}} m_{e_{{b}}} {\Gamma^L_{\hat{\bar{e}}_{{i}}\gamma e_{{b}}}}  +{B_0\Big(m^2_{\nu_{{b}}},m^2_{W^-}\Big)} {\Gamma^{R*}_{\hat{\bar{e}}_{{j}}W^-\nu_{{b}}}} m_{\nu_{{b}}} {\Gamma^L_{\hat{\bar{e}}_{{i}}W^-\nu_{{b}}}}  \nonumber \\ 
 & \; \;+{B_0\Big(m^2_{e_{{b}}},m^2_{Z}\Big)} {\Gamma^{R*}_{\hat{\bar{e}}_{{j}}Ze_{{b}}}} m_{e_{{b}}} {\Gamma^L_{\hat{\bar{e}}_{{i}}Ze_{{b}}}}\Big]+\sum_{a=1}^{6}\sum_{b=1}^{5}{B_0\Big(m^2_{\tilde{\chi}^0_{{b}}},m^2_{\tilde{e}_{{a}}}\Big)} {\Gamma^{L*}_{\hat{\bar{e}}_{{j}}\tilde{e}_{{a}}\tilde{\chi}^0_{{b}}}} m_{\tilde{\chi}^0_{{b}}} {\Gamma^R_{\hat{\bar{e}}_{{i}}\tilde{e}_{{a}}\tilde{\chi}^0_{{b}}}} \\ 
&  \Sigma^R_{ij}(p^2) \, = \, +\sum_{a=1}^{2}\sum_{b=1}^{3}\Big[{B_0\Big(m^2_{\nu_{{b}}},m^2_{H^-_{{a}}}\Big)} {\Gamma^{L*}_{\hat{\bar{e}}_{{j}}H^-_{{a}}\nu_{{b}}}} m_{\nu_{{b}}} {\Gamma^R_{\hat{\bar{e}}_{{i}}H^-_{{a}}\nu_{{b}}}}+{B_0\Big(m^2_{\tilde{\chi}^+_{{a}}},m^2_{\tilde{\nu}_{{b}}}\Big)} {\Gamma^{L*}_{\hat{\bar{e}}_{{j}}\tilde{\nu}_{{b}}\tilde{\chi}^+_{{a}}}} m_{\tilde{\chi}^+_{{a}}} {\Gamma^R_{\hat{\bar{e}}_{{i}}\tilde{\nu}_{{b}}\tilde{\chi}^+_{{a}}}}\Big]  \nonumber \\ 
 & \; \;+\sum_{a,b=1}^{3}m_{e_{{b}}}\Big[{B_0\Big(m^2_{e_{{b}}},m^2_{A^0_{{a}}}\Big)} {\Gamma^{L*}_{\hat{\bar{e}}_{{j}}A^0_{{a}}e_{{b}}}}  {\Gamma^R_{\hat{\bar{e}}_{{i}}A^0_{{a}}e_{{b}}}} +{B_0\Big(m^2_{e_{{b}}},m^2_{h_{{a}}}\Big)} {\Gamma^{L*}_{\hat{\bar{e}}_{{j}}h_{{a}}e_{{b}}}}  {\Gamma^R_{\hat{\bar{e}}_{{i}}h_{{a}}e_{{b}}}}\Big] \nonumber \\ 
 & \; \;-4 \sum_{b=1}^{3}\Big[{B_0\Big(m^2_{e_{{b}}},0\Big)} {\Gamma^{R*}_{\hat{\bar{e}}_{{j}}\gamma e_{{b}}}} m_{e_{{b}}} {\Gamma^L_{\hat{\bar{e}}_{{i}}\gamma e_{{b}}}}+{B_0\Big(m^2_{\nu_{{b}}},m^2_{W^-}\Big)} {\Gamma^{R*}_{\hat{\bar{e}}_{{j}}W^-\nu_{{b}}}} m_{\nu_{{b}}} {\Gamma^L_{\hat{\bar{e}}_{{i}}W^-\nu_{{b}}}}  \nonumber \\ 
 & \; \;+{B_0\Big(m^2_{e_{{b}}},m^2_{Z}\Big)} {\Gamma^{R*}_{\hat{\bar{e}}_{{j}}Ze_{{b}}}} m_{e_{{b}}} {\Gamma^L_{\hat{\bar{e}}_{{i}}Ze_{{b}}}}\Big]+\sum_{a=1}^{6}\sum_{b=1}^{5}{B_0\Big(m^2_{\tilde{\chi}^0_{{b}}},m^2_{\tilde{e}_{{a}}}\Big)} {\Gamma^{L*}_{\hat{\bar{e}}_{{j}}\tilde{e}_{{a}}\tilde{\chi}^0_{{b}}}} m_{\tilde{\chi}^0_{{b}}} {\Gamma^R_{\hat{\bar{e}}_{{i}}\tilde{e}_{{a}}\tilde{\chi}^0_{{b}}}}   \\ 
&  \Sigma^L_{ij}(p^2) \, = \, +\sum_{a=1}^{2}\sum_{b=1}^{3}\Big[{B_0\Big(m^2_{\nu_{{b}}},m^2_{H^-_{{a}}}\Big)} {\Gamma^{L*}_{\hat{\bar{e}}_{{j}}H^-_{{a}}\nu_{{b}}}} m_{\nu_{{b}}} {\Gamma^R_{\hat{\bar{e}}_{{i}}H^-_{{a}}\nu_{{b}}}} +{B_0\Big(m^2_{\tilde{\chi}^+_{{a}}},m^2_{\tilde{\nu}_{{b}}}\Big)} {\Gamma^{L*}_{\hat{\bar{e}}_{{j}}\tilde{\nu}_{{b}}\tilde{\chi}^+_{{a}}}} m_{\tilde{\chi}^+_{{a}}} {\Gamma^R_{\hat{\bar{e}}_{{i}}\tilde{\nu}_{{b}}\tilde{\chi}^+_{{a}}}}\Big]  \nonumber \\ 
 & \; \;+\sum_{a,b=1}^{3}m_{e_{{b}}}\Big[{B_0\Big(m^2_{e_{{b}}},m^2_{A^0_{{a}}}\Big)} {\Gamma^{L*}_{\hat{\bar{e}}_{{j}}A^0_{{a}}e_{{b}}}}  {\Gamma^R_{\hat{\bar{e}}_{{i}}A^0_{{a}}e_{{b}}}} +{B_0\Big(m^2_{e_{{b}}},m^2_{h_{{a}}}\Big)} {\Gamma^{L*}_{\hat{\bar{e}}_{{j}}h_{{a}}e_{{b}}}}  {\Gamma^R_{\hat{\bar{e}}_{{i}}h_{{a}}e_{{b}}}}\Big] \nonumber \\ 
 & \; \;-4 \sum_{b=1}^{3}\Big[{B_0\Big(m^2_{e_{{b}}},0\Big)} {\Gamma^{R*}_{\hat{\bar{e}}_{{j}}\gamma e_{{b}}}} m_{e_{{b}}} {\Gamma^L_{\hat{\bar{e}}_{{i}}\gamma e_{{b}}}}  +{B_0\Big(m^2_{\nu_{{b}}},m^2_{W^-}\Big)} {\Gamma^{R*}_{\hat{\bar{e}}_{{j}}W^-\nu_{{b}}}} m_{\nu_{{b}}} {\Gamma^L_{\hat{\bar{e}}_{{i}}W^-\nu_{{b}}}}  \nonumber \\ 
 & \; \;+{B_0\Big(m^2_{e_{{b}}},m^2_{Z}\Big)} {\Gamma^{R*}_{\hat{\bar{e}}_{{j}}Ze_{{b}}}} m_{e_{{b}}} {\Gamma^L_{\hat{\bar{e}}_{{i}}Ze_{{b}}}} \Big]   +\sum_{a=1}^{6}\sum_{b=1}^{5}{B_0\Big(m^2_{\tilde{\chi}^0_{{b}}},m^2_{\tilde{e}_{{a}}}\Big)} {\Gamma^{L*}_{\hat{\bar{e}}_{{j}}\tilde{e}_{{a}}\tilde{\chi}^0_{{b}}}} m_{\tilde{\chi}^0_{{b}}} {\Gamma^R_{\hat{\bar{e}}_{{i}}\tilde{e}_{{a}}\tilde{\chi}^0_{{b}}}}  
\end{align} 

\subsection{Self-energy of gluino}
\label{app:SEglu}
\begin{align} 
\Sigma^S(p^2) & \; \;= +\sum_{a=1}^{6}\sum_{b=1}^{3}\Big[{B_0\Big(m^2_{d_{{b}}},m^2_{\tilde{d}_{{a}}}\Big)} {\Gamma^{L*}_{\tilde{g}_{{j}}\tilde{d}^*_{{a}}d_{{b}}}} m_{d_{{b}}} {\Gamma^R_{\tilde{g}_{{i}}\tilde{d}^*_{{a}}d_{{b}}}} +{B_0\Big(m^2_{u_{{b}}},m^2_{\tilde{u}_{{a}}}\Big)} {\Gamma^{L*}_{\tilde{g}_{{j}}\tilde{u}^*_{{a}}u_{{b}}}} m_{u_{{b}}} {\Gamma^R_{\tilde{g}_{{i}}\tilde{u}^*_{{a}}u_{{b}}}}\Big]\nonumber \\ 
 & \; \; -12 {B_0\Big(m^2_{\tilde{g}},0\Big)} {\Gamma^{R*}_{\tilde{g}_{{j}}G_{{1}}\tilde{g}_{{1}}}} m_{\tilde{g}} {\Gamma^L_{\tilde{g}_{{i}}G_{{1}}\tilde{g}_{{1}}}} \\ 
\Sigma^R(p^2) & \; \;= +\sum_{a=1}^{6}\sum_{b=1}^{3}\Big[{B_0\Big(m^2_{d_{{b}}},m^2_{\tilde{d}_{{a}}}\Big)} {\Gamma^{L*}_{\tilde{g}_{{j}}\tilde{d}^*_{{a}},d_{{b}}}} m_{d_{{b}}} {\Gamma^R_{\tilde{g}_{{i}},\tilde{d}^*_{{a}},d_{{b}}}} +{B_0\Big(m^2_{u_{{b}}},m^2_{\tilde{u}_{{a}}}\Big)} {\Gamma^{L*}_{\tilde{g}_{{j}},\tilde{u}^*_{{a}},u_{{b}}}} m_{u_{{b}}} {\Gamma^R_{\tilde{g}_{{i}},\tilde{u}^*_{{a}},u_{{b}}}} \Big]\nonumber \\ 
 & \; \; -12 {B_0\Big(m^2_{\tilde{g}},0\Big)} {\Gamma^{R*}_{\tilde{g}_{{j}},G_{{1}},\tilde{g}_{{1}}}} m_{\tilde{g}} {\Gamma^L_{\tilde{g}_{{i}},G_{{1}},\tilde{g}_{{1}}}} \\ 
\Sigma^L(p^2) & \; \;= +\sum_{a=1}^{6}\sum_{b=1}^{3}\Big[{B_0\Big(m^2_{d_{{b}}},m^2_{\tilde{d}_{{a}}}\Big)} {\Gamma^{L*}_{\tilde{g}_{{j}},\tilde{d}^*_{{a}},d_{{b}}}} m_{d_{{b}}} {\Gamma^R_{\tilde{g}_{{i}},\tilde{d}^*_{{a}},d_{{b}}}} +{B_0\Big(m^2_{u_{{b}}},m^2_{\tilde{u}_{{a}}}\Big)} {\Gamma^{L*}_{\tilde{g}_{{j}},\tilde{u}^*_{{a}},u_{{b}}}} m_{u_{{b}}} {\Gamma^R_{\tilde{g}_{{i}},\tilde{u}^*_{{a}},u_{{b}}}} \Big]\nonumber \\ 
 & \; \; -12 {B_0\Big(m^2_{\tilde{g}},0\Big)} {\Gamma^{R*}_{\tilde{g}_{{j}},G_{{1}},\tilde{g}_{{1}}}} m_{\tilde{g}} {\Gamma^L_{\tilde{g}_{{i}},G_{{1}},\tilde{g}_{{1}}}} 
\end{align}

\newpage

\addcontentsline{toc}{chapter}{References}
\providecommand{\href}[2]{#2}\begingroup\raggedright\endgroup

%
\newpage

\listoftables
\addcontentsline{toc}{chapter}{List of tables}
 
\newpage
 
\listoffigures
\addcontentsline{toc}{chapter}{List of figures}

%

\cleardoublepage
\end{appendix}
\end{document}